\documentclass[review]{elsarticle}

\usepackage[colorlinks,citecolor=blue,linktoc=all,linkcolor=cyan]{hyperref}
\usepackage{graphicx}

\usepackage[T1]{fontenc}
\usepackage{dsfont}               % use mathds instead of mathbb for outline fonts
\usepackage{mathrsfs}             % provides mathscr without overwriting mathcal
\usepackage{slashed}              % For Dirac slash notation.
\usepackage{amsmath}
\usepackage{amssymb}
\usepackage{amsbsy}
\usepackage{amsfonts}
\usepackage{tabularx}
\usepackage{multirow,makecell}
\usepackage{subcaption}
\usepackage{gensymb}
\usepackage{braket}
\usepackage{threeparttable}
\usepackage{array}

\usepackage{xcolor}
\usepackage{academicons}
\definecolor{orcidlogocol}{HTML}{A6CE39}

\newcommand{\orcid}[1]{%
  \href{https://orcid.org/#1}{%
    \includegraphics[height=1.8ex]{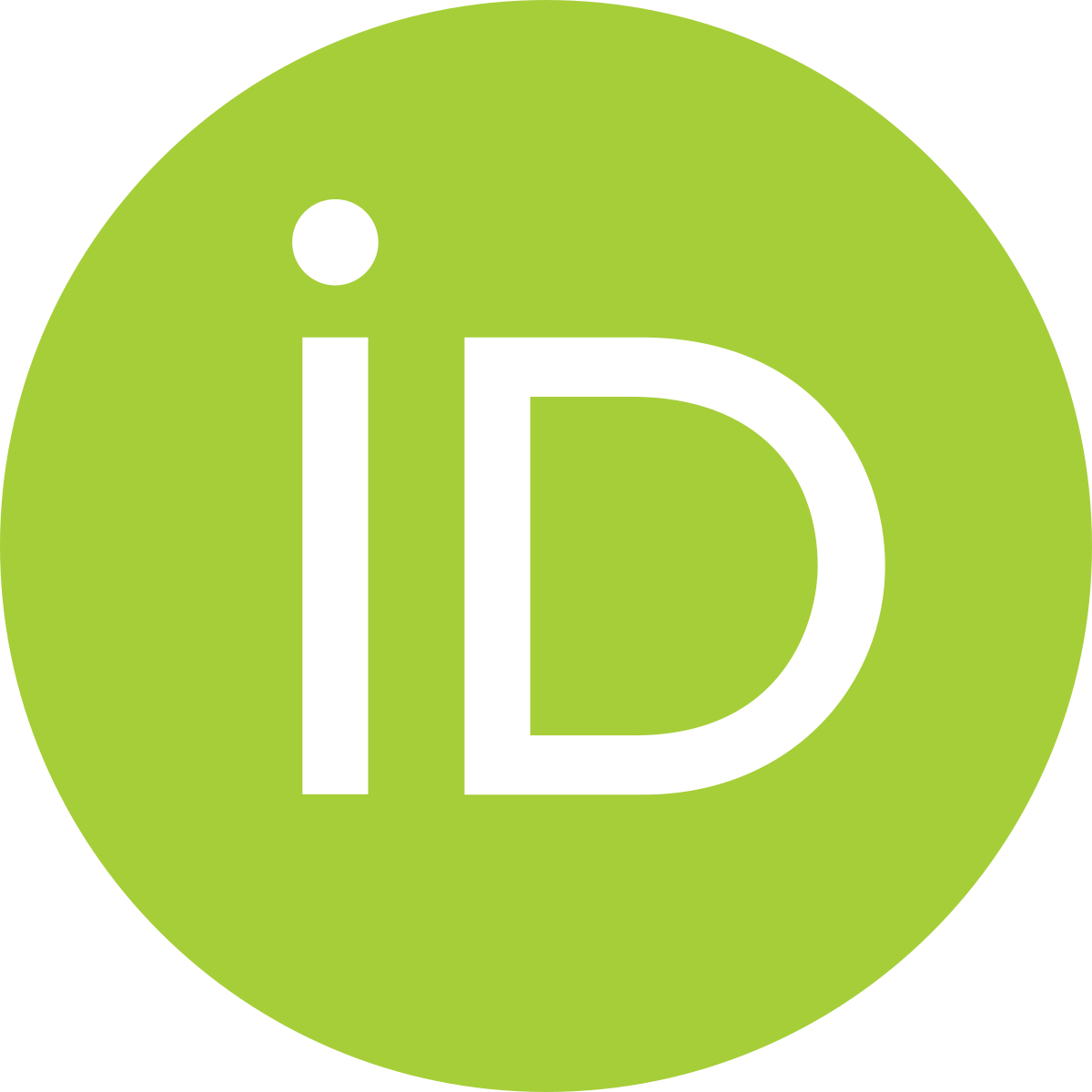}%
  }%
}
\numberwithin{equation}{section}
\numberwithin{table}{section}
\numberwithin{figure}{section}

\journal{Progress in Particle and Nuclear Physics}

\usepackage{titlesec}
\usepackage{sectsty}
\titleformat{\section}{\normalfont\Large\bfseries}{\thesection}{1em}{}
\titleformat{\subsection}{\normalfont\large\bfseries}{\thesubsection}{1em}{}
\titleformat{\subsubsection}{\normalfont\normalsize\bfseries}{\thesubsubsection}{1em}{}
\biboptions{sort&compress}
\begin{document}
	
	\begin{frontmatter}
		\title{Electromagnetic Production of Kaons on the Nucleon}

        %authors, affiliations, corresponding author mention 
        \author[mymainaddress]{Terry Mart\corref{mycorrespondingauthor}\orcid{0000-0003-4628-2245}}
        \cortext[mycorrespondingauthor]{Corresponding author}
        \ead{terry.mart@sci.ui.ac.id}
        \author[mymainaddress]{Jovan Alfian Djaja\orcid{0009-0008-4495-3918}}
        \author[address2]{Daniel S. Carman\orcid{0000-0002-1280-0983}}

        \address[mymainaddress]{Departemen Fisika, FMIPA, Universitas Indonesia, Depok 16424, Indonesia}
        \address[address2]{Thomas Jefferson National Accelerator Facility, Newport News, Virginia 23606, USA}

\begin{abstract} 
Studies of the electromagnetic production of strange quarks began in the 1950s as something of a curiosity that puzzled experimentalists and theorists alike. Eventually, a nascent understanding of these processes began to take shape through the first pioneering experiments dedicated to explore photo- and electroproduction that were carried out in the period from the 1950s to the 1980s. As the datasets increased, concomitant advances in theoretical models were realized. However, these initial studies also made clear that more precise data was essential to continue to move forward. A paradigm shift occurred in the 1990s with the development of second-generation facilities at ELSA, MAMI, SPring-8, and JLab. High-intensity, high duty-factor accelerators, coupled with novel detector systems and advances in computing and readout electronics, brought nuclear physics experiments forward by orders of magnitude in counting statistics compared to the first-generation efforts. This was an utter boon to strangeness physics investigations, and to date, more than 50 dedicated experiments in kaon photo- and electroproduction have been completed at facilities around the world, leading to a host of experimental observables that have enabled significant advances in the exploration of strongly interacting systems that decay via $s\bar{s}$ quark pair creation. These data have proven to be an essential complementary pathway to study the spectrum and structure of the excited states of the nucleon, and the search for missing and exotic baryon configurations. As well, investigations in these channels are requisite for exploring hypernuclear production as a probe of the $YN$ interaction and for studies of the electromagnetic form factors of strange mesons. This review was designed to provide the first-ever in-depth overview of both the experimental and theoretical progress in the field of the electromagnetic production of strangeness. This work looks back over 70 years of past developments, discusses ongoing work and near-term plans, and details future possibilities being considered for third-generation facilities. Extensive lists of the available datasets and theoretical models are provided, together with a comprehensive supporting bibliography of the field. Throughout this work, the primary impacts of these explorations are highlighted, along with connections to a wide range of related phenomenological applications. An important goal of this review is to provide a complete, (reasonably) self-contained guide into this field prepared at a level that is relevant for both new and seasoned scientists, whether experimentalists, phenomenologists, or theorists, to better understand what has been accomplished by so many dedicated folks-each building on what has come before-and to appreciate the exciting future potential for continued studies in this area.\\[1ex]
Preprint number: JLAB-PHY-26-4599
\end{abstract}
		
		
	\end{frontmatter}
	
	\newpage
	
	\thispagestyle{empty}
	\tableofcontents
	
	%to begin the line numbers: 
	%\linenumbers

	%beginning of the core of the manuscript
	\newpage

%=== INTRODUCTION ================================
%  \input{introduction}
\section {Introduction}
\label{sec:intro}

Experiments to measure kaon-hyperon ($KY$) final states through photo- and electroproduction processes on both proton and quasi-free neutron targets have a history spanning back over more than 70 years. What began as an observational puzzle, particles that were plentifully produced yet had anomalously long decay times given their sizable mass, ultimately led to the concept of strangeness and the development of the groundbreaking eight-fold way by Gell-man and Ne'eman~\cite{Gell-Mann:2000mam}. This scheme to organize the proverbial zoo of baryons and mesons, ultimately led to the development of the quark model and set the stage for the development of the theory of strongly interacting particles, Quantum Chromodynamics (QCD), which describes the interactions between colored quarks and gluons~\cite{Gross:1996dla,Gross:2022hyw}. 

Studies of the non-strange final states probed first through $\pi N$ scattering and later by $\gamma N$ photoproduction reactions~\cite{Hey:1982aj,Ireland:2019uwn, Burkert:2025coj} dominated the experimental landscape from the 1950s to 1980s. These programs led to significant advancements in the understanding of baryon resonances and the light baryon spectrum. Over these decades, insights were gleaned that made it possible to begin to understand the relevant low energy degrees of freedom in strongly interacting systems, the structure of the ground state nucleon and its excited resonance states, and hints of possible exotic configurations of both baryons and mesons. These experimental programs went hand-in-hand with phenomenological and theoretical advancements that led to the development of reaction models, both single-channel and coupled-channel approaches, in concert with the development of QCD and predictions from lattice-regularized and QCD-kindred calculations that today are beginning to unify low-energy nuclear with high-energy particle physics, spanning the full regime from perturbative to non-perturbative dynamics and distance scales.

\begin{figure*}[htbp]
\centering
\includegraphics[width=0.7\textwidth]{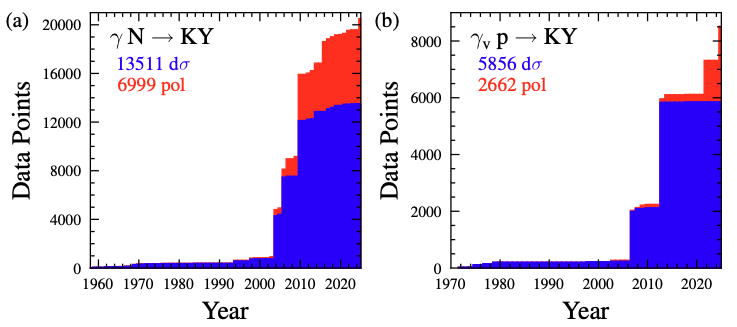}
\caption{(a) Stacked histogram of the full experimental $\gamma N \to KY$ photoproduction database including $\gamma p$ and $\gamma n$ data as a function of year. (b) Stacked histogram of the full experimental $e p \to e'KY$ electroproduction database as function of year. The blue histograms represent the cross section measurements and the red histograms represent the polarization measurements. The total number of data points for each type of measurement is listed on each plot.}
\label{data-points-year} 
\end{figure*}

Due to the much smaller cross sections for $KY$ production compared to $\pi N$ (or $\pi \pi N$) final states, realizing the potential for the relevance and importance of strangeness physics studies necessarily had to wait for the advent of the modern, high-intensity electron machines that came on the scene in the 1990s in the United States (U.S.) and in Europe. Figure~\ref{data-points-year} provides a history of the available data for $KY$ photo- and electroproduction from the first measurements until today. The number of experimental data points remained in the low-statistics, large-bin-size, proof-of-principle realm until about 2005 when there was a sudden opening of the proverbial floodgates with the release of the first results from SAPHIR at Bonn and CLAS at Jefferson Laboratory (JLab). These experimental datasets finally made detailed investigations of $KY$ final states relevant and important complementary channels to their non-strange counterparts in the investigations 
of strongly interacting systems.

%%%%%%%%%%%%%%%%%%%%%%%%%%%%%%%%%%%%%%%%%%%%%%%%%%%%%%%%%%%%%%%%%%%%%%%%%%%%%%%%%%%%%%%%%%%%%%%%%%%%%%%%%%%%%%%%%%%%%%%%%%%%%%%%%%%%%%%%%%%%%%%%%%%%%%%%%%%%%%%%
%%%%%%%%%%%%%%%%%%%%%%%%%%%%%%%%%%%%%%%%%%%%%%%%%%%%%%%%%%%%%%%%%%%%%%%%%%%%%%%%%%%%%%%%%%%%%%%%%%%%%%%%%%%%%%%%%%%%%%%%%%%%%%%%%%%%%%%%%%%%%%%%%%%%%%%%%%%%%%%%

The most recent review on the electromagnetic production of kaons off the nucleon was published over 20 years ago as part of a conference proceedings~\cite{Bennhold:1999nd}. That review was actually prepared before the release of the 2005 experimental wave of data. Over the past three decades, however, the operation of CEBAF at JLab in the U.S., alongside other modern electron accelerators such as ELSA in Bonn, Germany, MAMI in Mainz, Germany, and SPring-8 in Hyogo, Japan, has enabled more than 50 experiments over several generations of data-taking sequences to measure observables in kaon photo- and electroproduction. The study of $KY$ final states is proving more relevant today than ever before. These final states are of considerable theoretical importance, not only as a key input for predicting hypernuclear production via electromagnetic probes, but also for exploring related phenomenological issues such as missing resonances, studies of the structure of nucleon excited states, and the electromagnetic form factors of strange baryons and mesons.

This review was designed to systematically summarize both the experimental and theoretical developments in kaon production induced by electromagnetic probes, i.e., photon and electron beams, made in this field since its commencement. We begin with the most recent review on the elementary process, provided in Ref.~\cite{Bennhold:1999nd}, and build upon it with insights from key literature (including a compilation of nearly 700 citations). This work will offer the most complete and up-to-date review of the subject available. It begins in Section~\ref{theory-formalism} with an introduction of the relevant kinematics, followed by an overview of the scattering amplitudes and observables commonly used to describe the process. Section~\ref{expt-measurements} provides a comprehensive overview of the experimental facilities and datasets in the U.S., Europe, and Asia, spanning over the full range of the history of strangeness photo- and electroproduction. This section includes a detailed overview of the current 12-GeV CEBAF machine and its experimental program and the current plans for a possible future 22-GeV upgrade of this facility. Section~\ref{nstar-studies} details the importance of the $KY$ photo- and electroproduction data for the studies of both the spectrum and structure of excited nucleon states and the importance of these efforts to understand the underpinnings of the strong interaction. A complete layout of the existing theoretical models and their status is provided in Section~\ref{sec:models}. Section~\ref{sec:phenomenlogy} explores a wide range of phenomenological applications, including the search for missing and narrow resonances, the potential existence of pentaquark states, and investigations into hadronic and electromagnetic coupling constants and form factors. Section~\ref{sec:unsettled} provides somewhat of a ``reality check'' discussing various unsettled problems in the field, which serves to provide some possible directions for the future of both the experimental and theoretical efforts in strangeness physics. This includes the role of lattice calculations in the electromagnetic production of kaons. This review concludes in Section~\ref{sec:conclusion} with a brief summation from both the experimental and theoretical viewpoints.

In preparing such a review spanning decades of work, we have labored to provide a balanced discussion of the subject both from the experimental and theoretical points of view. Inevitably, our selections and our focuses reflect our own personal opinions and biases. No doubt, too, there are some omissions that stem from oversights, our own ignorance, and misunderstandings in our archaeology of the available work. Furthermore, to put this current summary review focusing on the electromagnetic production of kaons using beams of photons and electrons as a tool for understanding hadronic structure and strongly interacting systems in its approprite context, it is essential to understand that this is an arena that provides complementary investigations to that provided by experiments employing hadronic probes. Ultimately, the different facilities provide different facets to our understanding of strongly interacting systems from the non-perturbative to the perturbative regimes spanning a broad range of distance scales to explore QCD. To provide a bit of context to complement this review, we provide a {\em very} brief sketch of the major facilities and their primary hadron structure focus along with one or two primary references:

\begin{itemize}
\item J-PARC (Japan) - The Japan Proton Accelerator Research Complex specializes in high-intensity, low-to-medium-energy proton and secondary meson beams (kaons and pions). Its primary contributions to hadron structure center on the role of strangeness and heavy-flavor environments. Focus: hypernuclei, strangeness, exotic bound states \cite{Naruki:2012eka}.
\item RHIC (U.S.) - The Relativistic Heavy Ion Collider operated from 2000 until 2026 when its operations officially ended. The machine provided proton-proton and nucleus-nucleus collisions to create a plasma of quarks and gluons. The focus of its program was to study the internal dynamics of the proton. Focus: Gluon, spin, quark orbital angular motion through transverse momentum dependent distributions (TMDs) \cite{Aschenauer:2015eha}.
\item FNAL (U.S.) - The Fermi National Accelerator Laboratory focuses on high-energy fixed-target experiments using intense muon and proton beams. An important aspect of the program has been the high-precision mapping of collinear parton distributions. Focus: Flavor asymmetry of the sea quarks $\bar{d}/\bar{u}$ \cite{Reimer:2007iy,SeaQuest:2021zxb}.
\item GSI (Germany) - The Helmholtz Center for Heavy Ion Research focuses on the structure of hadrons embedded within high-density, relatively low-temperature nuclear matter using the HADES detector and FAIR (Facility for Antiproton and Ion Research)-- a project now nearing completion. Focus: In-medium properties, charmonium, dense QCD \cite{HADES:2009aat,Gianotti:2010zza}.
\item CERN (Switzerland/France) - The European Organization for Nuclear Research explores hadron structure across the highest energy regimes using the Large Hadron Collider (LHC) and a dedicated fixed-target program (COMPASS/AMBER). Focus: small-$x$ gluons, multi-quark exotics, spatial maps through Generalized Parton Distribution functions (GPDs) \cite{Gao:2017yyd, COMPASS:2007rjf}.
\end{itemize}

The experiments from the hadron facilities, together with the experiments to study photo- and electroproduction processes, allow for the picture of how quarks and gluons build the Universe to be filled out across the entire QCD phase diagram.
%=================================================

%=== THEORETICAL FORMALISM =======================
%  \input{theoretical_formalism}
\section{Theoretical Formalism of the Elementary Operator}
\label{theory-formalism}

\begin{figure}[t]
    \centering
    \includegraphics[width=0.85\columnwidth]{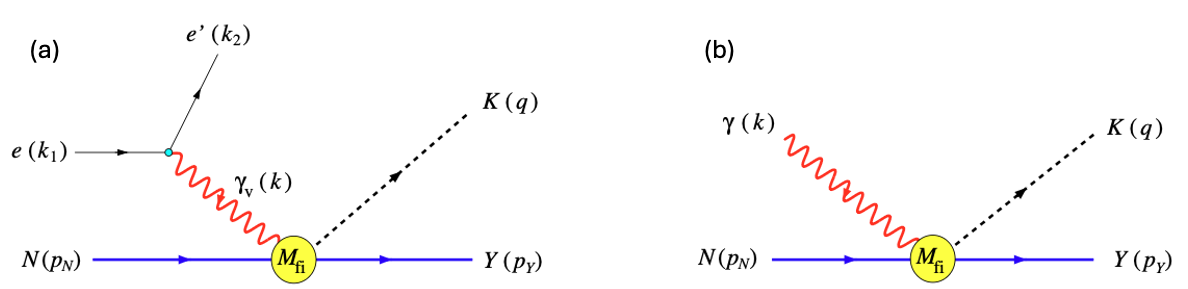} 
    \caption{Kinematic variables for (a) kaon electroproduction on the nucleon and (b) kaon photoproduction on the nucleon. In panel (b), the photon $\gamma$ may be either real or virtual. In the latter case, the process is referred to as virtual photoproduction, which corresponds to the electroproduction process shown in panel (a). }
    \label{electro_photo_kin}
\end{figure}

Since photoproduction is a special case of electroproduction, we will focus on the electroproduction process to describe the general formalism. The reaction under consideration is the inelastic scattering of an electron $e$ and a nucleon $N$ into a final state consisting of a scattered electron $e'$, a kaon $K$, and a hyperon $Y$, with the corresponding four-momenta defined as
\begin{equation}
\label{eq:electro_process}
    e(k_1)+N(p_N) \to e'(k_2)+ K(q) + Y(p_Y) .
\end{equation}
The process given by Eq.~(\ref{eq:electro_process}) is schematically shown in Fig.~\ref{electro_photo_kin}(a) under the so-called one-photon exchange approximation. Given that the electromagnetic coupling constant is given by $\alpha = e^2/4\pi \approx 1/137$, the one-photon exchange approximation is generally valid to within about 1\% accuracy. Note that throughout the discussion, we have the squared mass of the virtual photon $k^2=(k_1-k_2)^2\leq 0$. The photoproduction formalism can then be recovered by taking the virtual photon momentum transfer squared to zero, i.e., $k^2 = 0$. Furthermore,  for convenience, especially when we discuss the electromagnetic form factors,  we also define $Q^2=-k^2\geq 0$.

By considering the conservation of relevant quantum numbers, the process given by Eq.~(\ref{eq:electro_process}) yields two isoscalar channels
\begin{subequations}
\begin{align}
    \label{eq:electro_isoscalar1}
e(k_1)+p(p_p) ~&\to~ e'(k_2)+ K^+(q) + \Lambda(p_\Lambda),\\
\label{eq:electro_isoscalar2}
e(k_1)+n(p_n) ~&\to~ e'(k_2)+ K^0(q) + \Lambda(p_\Lambda),
\end{align}
\end{subequations}
which are analogous to $\eta$ electroproduction, as the $\Lambda$ is an isoscalar baryon. In addition, the process also implies four isovector channels
\begin{subequations}
\begin{align}
\label{eq:electro_isovector1}
e(k_1)+p(p_p) ~&\to~ e'(k_2)+ K^+(q) + \Sigma^0(p_\Sigma),\\
\label{eq:electro_isovector2}
e(k_1)+p(p_p) ~&\to~ e'(k_2)+ K^0(q) + \Sigma^+(p_\Sigma),\\
\label{eq:electro_isovector3}
e(k_1)+n(p_n) ~&\to~ e'(k_2)+ K^+(q) + \Sigma^-(p_\Sigma),\\
\label{eq:electro_isovector4}
e(k_1)+n(p_n) ~&\to~ e'(k_2)+ K^0(q) + \Sigma^0(p_\Sigma),
\end{align}
\end{subequations}
which are comparable to the four isospin channels observed in pion electroproduction, where the $\Sigma$ hyperons form an isospin triplet similar to the pions.

It is also well known that the electron kinematics can be factored out from the expressions for the cross section and polarization observables. Therefore, it is very convenient to work within the framework of virtual photoproduction shown in Fig.~\ref{electro_photo_kin}(b), which is justified by the fact that the reaction amplitude $\mathcal{M}_{\rm fi}$ remains unaffected as long as $k^2\neq 0$. Note that in this framework, Eq.~(\ref{eq:electro_process}) can be written as
\begin{equation}
% Note : eq:electro_process is doubly defined, so this eq is relabeled
\label{eq:virtual_photo_process}
    \gamma_v(k)+N(p_N) \to K(q) + Y(p_Y) ,
\end{equation}
where the virtual photon momentum is defined by $k\equiv k_1-k_2$. 

%%%%%%%FIGURE 6%%%%%
%%%%%%%%%%%%%%%%%%%%

\subsection{Kinematics}
\label{subsec:kinematics}

In the virtual photoproduction framework, Eq.~(\ref{eq:virtual_photo_process}), we consider four four-momenta
\begin{eqnarray}
    \label{eq:momenta_four_particle}
    k = (k_0,\boldsymbol{k}),~~~~
    p_N = (E_N,\boldsymbol{p}_N),~~~~
    q = (E_K,\boldsymbol{q}),~~~~
    p_Y = (E_Y,\boldsymbol{p}_Y).
\end{eqnarray}
Due to momentum conservation
\begin{equation}
    \label{eq:mom_conserv}
    k+p_N=q+p_Y,
\end{equation}
only three of these quantities are independent variables at the interaction vertex. In what follows, we choose $k,~p_N$ and $q$ as the independent momenta. From these we can further define the Mandelstam variables
\begin{equation}
\label{eq:mandelstam}
    s = (k+p_N)^2,~~~~ t = (k-q)^2,~~~~ u = (k-p_Y)^2,
\end{equation}
which satisfy
\begin{equation}
    s+t+u=k^2+m_N^2+m_K^2+m_Y^2.
\end{equation}
Since all three hadrons are on-shell, the electroproduction process involves only three independent scalar variables: the momentum transfer squared $k^2$ and any two of the three Mandelstam variables given in Eq.~(\ref{eq:mandelstam}). We note that the Mandelstam variable $t$ can be used to define the center-of-momentum (c.m.) frame scattering angle as
\begin{equation}
    \cos{\theta}=\frac{\boldsymbol{k}\cdot\boldsymbol{q}}{|\boldsymbol{k}||\boldsymbol{q}|},
\end{equation}
since from Eq.~(\ref{eq:mandelstam}) we have
\begin{equation}
    t= k^2+m_K^2-2k_0E_K+2\boldsymbol{k}\cdot\boldsymbol{q}.
\end{equation}
In the photon-nucleon (or kaon-hyperon) c.m. frame, the four-momenta given in Eq.~(\ref{eq:momenta_four_particle}) take the form
\begin{eqnarray}
    \label{eq:momenta_cm}
    k = (k_0,\boldsymbol{k}),~~~~
    p_N = (E_N,-\boldsymbol{k}),~~~~
    q = (E_K,\boldsymbol{q}),~~~~
    p_Y = (E_Y,-\boldsymbol{q}).
\end{eqnarray}
In this frame, it is customary to define the total c.m. energy as
\begin{equation}
    \label{eq:cm_defined}
    W \equiv k_0+E_N = \sqrt{s}.
\end{equation}
%and the Mandelstam variable $t$ can be used to define the c.m. scattering angle as
%\begin{equation}
%    \cos \theta = \frac{\boldsymbol{k}\cdot\boldsymbol{q}}{|\boldsymbol{k}||\boldsymbol{q}|},
%\end{equation}
%since from Eq.~(\ref{eq:mandelstam}) we have
%\begin{equation}
%    t= k^2+m_K^2-2k_0E_K+2\boldsymbol{k}\cdot\boldsymbol{q}.
%\end{equation}
%The energies of the four particles in Eq.~(\ref{eq:momenta_cm}) can be written as
Equation~(\ref{eq:cm_defined}) allows us to express the energies of the four particles in Eq.~(\ref{eq:momenta_cm}) as
\begin{equation}
    k_0 = \frac{s-m_N^2+k^2}{2W},~~~
    E_N = \frac{s+m_N^2-k^2}{2W},~~~
    E_K = \frac{s-m_Y^2+m_K^2}{2W},~~~
    E_Y = \frac{s+m_Y^2-m_K^2}{2W},
\end{equation}
which is particularly useful for evaluating the amplitudes and observables in kaon electroproduction.

In the laboratory frame, where the nucleon target is at rest, the four-momenta of the particles are defined as
\begin{eqnarray}
    \label{eq:momenta_four_particle_lab}
    \tilde{k} = (\tilde{k}_0,\tilde{\boldsymbol{k}}),~~~~
    \tilde{p}_N = (m_N,0),~~~~
    \tilde{q} = (\tilde{E}_K,\tilde{\boldsymbol{q}}),~~~~
    \tilde{p}_Y = (\tilde{E}_Y,\tilde{\boldsymbol{p}}_Y).
\end{eqnarray}
In this frame, the Mandelstam variable $s$ is given by 
\begin{equation}
    \label{eq:s_in_lab_frame}
    s=k^2+m_N^2+2\tilde{k}_0 m_N.
\end{equation}
From this expression, the relation between the photon energies in the two frames can be written as
\begin{equation}
    \tilde{k}_0=k_0\frac{W}{m_N}-\frac{k^2}{m_N}.
\end{equation}
Note that in the case of photoproduction ($k^2=0$), this simplifies to $\tilde{k}_0=k_0{W}/{m_N}$. At threshold, where both the kaon and hyperon in the final state are at rest, we have $s=(m_K+m_Y)^2$. Substituting this into Eq.~(\ref{eq:s_in_lab_frame}) yields the threshold photon energy for kaon production in the laboratory frame,
\begin{equation}
    \label{eq:threshold_energies_elementary}
    \tilde{k}_0^{\rm thr}=\frac{(m_Y+m_K)^2-m_N^2-k^2}{2m_N}.
\end{equation}
For photoproduction, the threshold energies for the six isospin channels are listed in Table~\ref{tab:kaon_thresholds}.

\begin{table}[h!]
\setlength{\tabcolsep}{6pt} % Default value: 6pt
\renewcommand{\arraystretch}{0.8} % Default value: 1
\centering
\caption{Threshold photon energies ($\tilde{k}_0^{\rm thr}$) and corresponding total c.m.\ energies ($W^{\rm thr}$) for the six isospin channels of kaon photoproduction in the laboratory frame (nucleon at rest).}
\label{tab:kaon_thresholds}
\begin{tabularx}{\textwidth}{XXc}
\hline\hline
Channel & $\tilde{k}_0^{\rm thr}$ [MeV] & $W^{\rm thr}$ [MeV] \\[0.1ex]
\hline
$\gamma p \to K^+ \Lambda$ & ~\,911.1 & 1609.4  \\
$\gamma n \to K^0 \Lambda$ & ~\,915.3 & 1613.3  \\
$\gamma p \to K^+ \Sigma^0$ & 1046.2 & 1686.3 \\
$\gamma p \to K^0 \Sigma^+$ & 1047.4 & 1687.0 \\
$\gamma n \to K^0 \Sigma^0$ & 1050.6 & 1690.3  \\
$\gamma n \to K^+ \Sigma^-$ & 1052.1 & 1691.1 \\
\hline\hline
\end{tabularx}
\end{table}

\subsection{The Production Amplitude}
\label{sec:prodamp}

In general, the Lorentz-invariant amplitude for the kaon electroproduction process is given by \cite{Berends:1967vi}
\begin{equation}
    \label{eq:transition_amplitude}
    {\cal M}_{\rm fi}=\epsilon_{\mu}J^{\mu},
\end{equation}
where $\epsilon_{\mu}$ is the lepton electromagnetic current defined as
\begin{equation}
    \label{eq:lepton_current}
    \epsilon_\mu={\bar u}({\boldsymbol k}_2)\,\gamma_\mu\, u({\boldsymbol k}_1)/k^2,
\end{equation}
and $J^\mu$ is the nucleon electromagnetic current defined as
\begin{equation}
    \label{eq:nucleon_current}
    J^\mu=\langle p_Y, q |\, j^\mu\, | p_N\rangle,
\end{equation}
with $j^\mu$ the nucleon current operator. Note that the lepton current of Eq.~(\ref{eq:lepton_current}) is replaced by the real photon polarization in the case of photoproduction, which makes it possible to discuss kaon photo- and electroproduction simultaneously within a single unified formulation. It is clear that in both cases we have $k\cdot\epsilon=0$, since in photoproduction the real photon polarization vector satisfies the Lorentz gauge condition, whereas in electroproduction the contraction of $k_\mu=(k_1-k_2)_\mu$ with Eq.~(\ref{eq:lepton_current}) vanishes. 

The most general form for the pseudovector nucleon current operator $j^\mu$ can be expanded in terms of eight pseudovector matrices $N_i^\mu$ with amplitudes $B_i(s,t,u,k^2)$ as \cite{Berends:1967vi} 
\begin{equation}
    j^\mu=\sum_{i=1}^{8} B_i(s,t,u,k^2)\, N_i^\mu,
\end{equation}
where
\begin{eqnarray*}
    N_1^\mu &=& i\gamma_5\,\gamma^\mu\,\gamma\cdot k,\\
    N_2^\mu &=& 2i\gamma_5\,P^\mu,\\
    N_3^\mu &=& 2i\gamma_5\,q^\mu,\\
    N_4^\mu &=& 2i\gamma_5\,k^\mu,\\
    N_5^\mu &=& \gamma_5\,\gamma^\mu,\\
    N_6^\mu &=& \gamma_5\,\gamma\cdot k\,P^\mu,\\
    N_7^\mu &=& \gamma_5\,\gamma\cdot k\,q^\mu,\\
    N_8^\mu &=& \gamma_5\,\gamma\cdot k\,k^\mu,
\end{eqnarray*}
with $P=\frac{1}{2}\left(p_N+p_Y\right)$. By using the current conservation relation
\begin{equation}
    k_\mu\,j^\mu=0,
\end{equation}
we can eliminate two of these matrices and construct six new explicitly gauge- and Lorentz-invariant matrices $M_i$ with
\begin{equation}
    \label{eq:Ai_Mi}
    \epsilon_\mu\,j^\mu=\sum_{i=1}^{6}A_i(s,t,u,k^2)\, M_i,
\end{equation}
where $A_i(s,t,u,k^2)$ are the new amplitudes. 
There are a number of variances in writing the matrices $M_i$ in the literature. In the following, we will use that of Refs.~\cite{Deo:1974ik,Mart:1996gx,Mart:2008gq}, that is,
\begin{subequations}
\begin{align}
    M_1 ~&=~ {\textstyle\frac{1}{2}}\gamma_5\left( \slashed{\epsilon}\slashed{k}-\slashed{k}\slashed{\epsilon} \right),\\
    M_2 ~&=~ \gamma_5\bigl[ \left( 2q-k\right)\cdot\epsilon\, P\cdot k- \left( 2q-k\right)\cdot k\, P\cdot\epsilon\bigr],\\
    M_3 ~&=~ \gamma_5\left( q\cdot k\,\slashed{\epsilon}-q\cdot \epsilon\,\slashed{k}\right),\\
    M_4 ~&=~ i\varepsilon_{\mu\nu\rho\sigma}\,\gamma^\mu q^\nu \epsilon^\rho k^\sigma,\\
    M_5 ~&=~ \gamma_5\left( q\cdot \epsilon\, k^2-q\cdot k \, k\cdot\epsilon\right),\\
    M_6 ~&=~ \gamma_5\left( k\cdot \epsilon\,\slashed{k}-k^2 \,\slashed{\epsilon}\right).
\end{align}
\end{subequations}
It is worth mentioning that both $M_5$ and $M_6$ do not exist in the case of photoproduction. For the purpose of multipole decomposition of the amplitudes, as well as calculations of cross section and polarization observables, it is important to express the matrices in terms of the Pauli matrices and spinors in the c.m. system, i.e.,
\begin{eqnarray}
    {\bar u}({\boldsymbol p}_Y)\,\sum_{i=1}^{6}A_i(s,t,u,k^2)\, M_i\, u({\boldsymbol p}_N) &=& \chi^\dagger_{\rm f}~{\cal F}~\chi_{\rm i},
\end{eqnarray}
where the matrix ${\cal F}$ is given by \cite{Dennery:1961zz}
\begin{eqnarray}
\label{eq:F_in_terms_of_Fi}
    {\cal F} &=& i\boldsymbol{\sigma\cdot\epsilon}\, F_1+\boldsymbol{\sigma\cdot{\hat q}}\,\boldsymbol{\sigma\cdot({\hat k}\times\epsilon)}\,F_2+i\boldsymbol{\sigma\cdot{\hat k}}\,\boldsymbol{{\hat q}\cdot\epsilon}\,F_3 + i\boldsymbol{\sigma\cdot{\hat q}}\,\boldsymbol{{\hat q}\cdot\epsilon}\,F_4 \nonumber\\
    &&+~ i\boldsymbol{\sigma\cdot{\hat k}}\,\boldsymbol{{\hat k}\cdot\epsilon}\,F_5 + i\boldsymbol{\sigma\cdot{\hat q}}\,\boldsymbol{{\hat k}\cdot\epsilon}\,F_6 -i\boldsymbol{\sigma\cdot{\hat q}} \,\epsilon_0\,F_7-i\boldsymbol{\sigma\cdot{\hat k}} \,\epsilon_0\,F_8. 
\end{eqnarray}
We note that in expressing the production amplitude in terms of Pauli matrices and spinors, there are different normalizations used in the literature. For instance, Kn\"ochlein et al. \cite{Knochlein:1995qz} used
\begin{eqnarray}
    \label{eq:Ai_to_Fi}
    {\bar u}({\boldsymbol p}_{N'})\,\sum_{i=1}^{6}A_i(s,t,u,k^2)\, M_i\, u({\boldsymbol p}_N) &=& \frac{4\pi W}{m_N} \chi^\dagger_{\rm f}~{\cal F}~\chi_{\rm i},
\end{eqnarray}
for $\eta$ electroproduction on the nucleon, whereas Chew et al. \cite{Chew:1957tf} (later known as CGLN) used the same normalization for meson photoproduction. 

As in the case of the amplitudes $A_i$ given in Eq.~(\ref{eq:Ai_Mi}), we can also eliminate two amplitudes $F_i$ in Eq.~(\ref{eq:F_in_terms_of_Fi}) by invoking current conservation, i.e., ${\cal F}$ vanishes when substituting $\epsilon^\mu\to k^\mu$. There are two options to this end, i.e., eliminating the Pauli amplitudes $F_7$ and $F_8$, so that we obtain \cite{Dennery:1961zz}
\begin{eqnarray}
\label{eq:F_in_terms_of_F1_F6}
    {\cal F} &=& i\boldsymbol{\sigma\cdot a}\, F_1+\boldsymbol{\sigma\cdot{\hat q}}\,\boldsymbol{\sigma\cdot({\hat k}\times a)}\,F_2+i\boldsymbol{\sigma\cdot{\hat k}}\,\boldsymbol{{\hat q}\cdot a}\,F_3 \nonumber\\&& +~ i\boldsymbol{\sigma\cdot{\hat q}}\,\boldsymbol{{\hat q}\cdot a}\,F_4 
    + i\boldsymbol{\sigma\cdot{\hat k}}\,\boldsymbol{{\hat k}\cdot a}\,F_5 + i\boldsymbol{\sigma\cdot{\hat q}}\,\boldsymbol{{\hat k}\cdot a}\,F_6 ,
\end{eqnarray}
with 
\begin{equation}
    a_\mu = \epsilon_\mu - \epsilon_0 k_\mu /k_0,
\end{equation}
or eliminating $F_5$ and $F_6$, which leads to \cite{Berends:1967vi}
\begin{eqnarray}
\label{eq:F_in_terms_of_F7_F8}
    {\cal F} &=& i\boldsymbol{\sigma\cdot b}\, F_1+\boldsymbol{\sigma\cdot{\hat q}}\,\boldsymbol{\sigma\cdot({\hat k}\times b)}\,F_2+i\boldsymbol{\sigma\cdot{\hat k}}\,\boldsymbol{{\hat q}\cdot b}\,F_3 \nonumber\\&& +~ i\boldsymbol{\sigma\cdot{\hat q}}\,\boldsymbol{{\hat q}\cdot b}\,F_4 
    - i\boldsymbol{\sigma\cdot{\hat q}}\,b_0\,F_7 - i\boldsymbol{\sigma\cdot{\hat k}}\,b_0\,F_8 ,
\end{eqnarray}
with 
\begin{equation}
    b_\mu = \epsilon_\mu - \boldsymbol{{\hat k}\cdot\epsilon}\, k_\mu /|\boldsymbol{\hat k}| .
\end{equation}
However, using $F_7$ and $F_8$ as in Eq.~(\ref{eq:F_in_terms_of_F7_F8}) is advantageous, as they allow for a simpler angular decomposition and help avoid the spurious singularities that arise when employing $F_5$ and $F_6$ \cite{Berends:1967vi,Zagury:1966}. We note that Kn\"ochlein et al. \cite{Knochlein:1995qz} employed a similar matrix ${\cal F}$ as in Eq.~(\ref{eq:F_in_terms_of_F7_F8}), but with the definitions 
\begin{equation}
    \boldsymbol{\tilde{\sigma}}\equiv \boldsymbol{\sigma}-(\boldsymbol{\sigma\cdot \hat{k}})\boldsymbol{\hat{k}}, ~~~~ \boldsymbol{\tilde{q}}\equiv \boldsymbol{q}-(\boldsymbol{q\cdot \hat{k}})\boldsymbol{\hat{k}},
\end{equation}
in place of $b_\mu$. The same convention is also used in Ref.~\cite{Drechsel:1992pn}. In the following, we will use Eq.~(\ref{eq:F_in_terms_of_F7_F8}). %but we will replace $F_7$ and $F_8$ with $F_5$ and $F_6$, respectively, for convenience. 

It is also important to introduce the helicity amplitudes $H_i$, through which the differential cross section, e.g., for meson photoproduction, can be expressed in a more efficient form as \cite{Walker:1968xu,Levy:1973aq}
\begin{equation}
    \label{eq:diff_cs_photo_helicity}
    \frac{d\sigma}{d\Omega}=\frac{1}{2}\frac{|\boldsymbol{q}|}{|\boldsymbol{k}|}\sum_{i=1}^{4} |\,H_i\,|^2 .
\end{equation}
%where we have defined the photon equivalent energy $k_{\rm e}=(s-m_N^2)/W$, which is the real photon laboratory energy required to excite a nucleon with total c.m. energy $W$. 
In this notation, the relations between the helicity amplitudes $H_i$ and the Pauli amplitudes of Eqs.~(\ref{eq:F_in_terms_of_F1_F6}) and (\ref{eq:F_in_terms_of_F7_F8}) read \cite{Levy:1973zz}
\begin{subequations}
    \label{eq:helicity_amplitudes}
\begin{align}
    H_1 ~&=~ -\frac{1}{\sqrt{2}}\sin{\theta}\cos{{\textstyle\frac{1}{2}}\theta} \left( F_3+F_4\right),\\
    H_2 ~&=~ -\sqrt{2}\cos{{\textstyle\frac{1}{2}}\theta} \left( F_1-F_2\right) + H_3,\\
    H_3 ~&=~ \frac{1}{\sqrt{2}}\sin{\theta}\sin{{\textstyle\frac{1}{2}}\theta} \left( F_3-F_4\right),\\
    H_4 ~&=~ \sqrt{2}\sin{{\textstyle\frac{1}{2}}\theta} \left( F_1+F_2\right) - H_1,\\
    H_5 ~&=~ -\cos{{\textstyle\frac{1}{2}}\theta} \left( F_5'+F_6'\right),\\
    H_6 ~&=~ -\sin{{\textstyle\frac{1}{2}}\theta} \left( F_5'-F_6'\right),
\end{align}
\end{subequations}
with
\begin{subequations}
\begin{align}
    F_5' ~&=~ F_1+F_3\cos{\theta}+F_5 = \frac{k_0}{|\boldsymbol{k}|} F_8,\\
    F_6' ~&=~ F_4\cos{\theta}+F_6 = \frac{k_0}{|\boldsymbol{k}|} F_7.
\end{align}
\end{subequations}

Finally, we can use Eq.~(\ref{eq:Ai_to_Fi}) to relate the Pauli amplitudes $F_i$ to the amplitudes $A_i$ defined in Eq.~(\ref{eq:Ai_Mi}), yielding
\begin{subequations}
\begin{align}
    F_{1,2} ~&=~ \frac{1}{8\pi W} \left[\left(E_N\pm m_N\right)\left(E_Y\pm m_Y\right)\right]^{1/2} \Bigl[ \pm(W\mp m_N) A_1+ k\cdot q\, (A_3-A_4) \nonumber\\
    ~&~ +(W\mp m_N)(W\mp m_Y) A_4 -k^2 A_6 \Bigr],\\
    F_{3,4} ~&=~ \frac{|\boldsymbol{k}| |\boldsymbol{q}|}{8\pi W} 
    \left(\frac{E_Y\pm m_Y}{E_N\pm m_N}\right)^{1/2} \Bigl[ (s-m_N)^2\, A_2 \mp
    {\textstyle \frac{1}{2}} k^2(A_2-A_5) + (W\pm m_N) (A_3-A_4) \Bigr].
    \\
    F_{5,6}' ~&=~ \frac{k_0}{8\pi W} \left[\left(E_N\pm m_N\right)\left(E_Y\pm m_Y\right)\right]^{1/2} \Bigl[ \pm A_1 + (W\mp m_Y) A_4 - (W\mp m_N) A_6
    \nonumber\\ ~&~ -\frac{1}{E_N\pm m_N}\, \Bigl\{ \pm W(|\boldsymbol{k}|^2-2 \boldsymbol{k\cdot q}) A_2 \pm (k\cdot q\, k_0-E_K k^2) (A_5-{\textstyle \frac{3}{2}} A_2)\nonumber\\
    ~&~ - [E_K(W\pm m_N)-k\cdot q](A_3-A_4) \Bigr\} \Bigr].
\end{align}
\end{subequations}

Besides using helicity amplitudes, one can also write the cross section and polarization observables in terms of transversity amplitudes $b_i$. As in the case of helicity amplitudes, the photoproduction cross section can be simply written as
\begin{equation}
    \frac{d\sigma}{d\Omega}=\frac{1}{2}\frac{|\boldsymbol{q}|}{|\boldsymbol{k}|}\sum_{i=1}^{4} |\,b_i\,|^2,
\end{equation}
where the relations between the transversity and Pauli amplitudes are given by \cite{Barker:1975bp,Adelseck:1990ch,Jacob:1959at,Bussey:1979ju}
\begin{subequations}
    \begin{align}
        b_1 ~&=~ -\frac{i}{\sqrt{2}}\left(F_1e^{i\theta/2}-F_2e^{-i\theta/2}\right),\\
        b_2 ~&=~ +\frac{i}{\sqrt{2}}\left(F_1e^{-i\theta/2}-F_2e^{i\theta/2}\right),\\
        b_3 ~&=~ -b_1-\frac{1}{\sqrt{2}}\sin{\theta}\left(F_3e^{i\theta/2}+F_4e^{-i\theta/2}\right),\\
        b_4 ~&=~ -b_2-\frac{1}{\sqrt{2}}\sin{\theta}\left(F_3e^{-i\theta/2}+F_4e^{i\theta/2}\right).
    \end{align}
\end{subequations}
The relations above do not include the longitudinal terms that arise in the case of a virtual photon and are therefore insufficient to fully define the electroproduction observables. Consequently, two additional longitudinal transversity amplitudes are required, i.e., \cite{Mart:1996gx}
\begin{subequations}
    \begin{align}
        b_5 ~&=~ -\frac{1}{2}\left(F_5e^{i\theta/2}+F_6e^{-i\theta/2}\right),\\
        b_6 ~&=~ -\frac{1}{2}\left(F_5e^{-i\theta/2}+F_6e^{i\theta/2}\right).
    \end{align}
\end{subequations}

It is straightforward to relate the transversity amplitudes to the helicity amplitudes. The relations are given by \cite{Mart:1996gx}
\begin{subequations}
    \begin{align}
        b_1 ~&=~ \frac{1}{2}\left[\left(H_1+H_4\right)+ i\left(H_2-H_3\right) \right],\\
        b_2 ~&=~ \frac{1}{2}\left[\left(H_1+H_4\right)-i\left(H_2-H_3\right) \right],\\
        b_3 ~&=~ \frac{1}{2}\left[\left(H_1-H_4\right)-i\left(H_2+H_3\right) \right],\\
        b_4 ~&=~ \frac{1}{2}\left[\left(H_1-H_4\right)+i\left(H_2+H_3 \right) \right],\\
        b_5 ~&=~ \frac{1}{2}\left(H_5+iH_6\right),\\
        b_6 ~&=~ \frac{1}{2}\left(H_5-iH_6\right).
    \end{align}
\end{subequations}

\subsection{Multipoles Decomposition of the Amplitudes}

In quantum mechanics, it is well established that the transition amplitudes of processes involving angular variables can be expanded in terms of partial waves characterized by the orbital angular momentum $\ell$. This provides the foundation for the multipole decomposition in kaon photo- and electroproduction. Let $\ell$ denote the orbital angular momentum of the final state, i.e., the orbital motion of the kaon relative to the hyperon. The Pauli amplitudes in Eqs.~(\ref{eq:F_in_terms_of_F1_F6}) and (\ref{eq:F_in_terms_of_F7_F8}) can then be expanded into a complete set of states with orbital angular momentum $\ell$. Depending on their transformation properties, these states correspond to electric radiation (transverse vectors with parity $(-1)^\ell$), magnetic radiation (transverse pseudovectors with parity $(-1)^{\ell+1}$), or longitudinal radiation (longitudinal vectors with parity $(-1)^\ell$) \cite{Dennery:1961zz}. Since the hyperon has spin $1/2$, the total angular momentum of the final state is obtained by coupling $\ell$ with the spin of the hyperon, giving $J = \ell \pm 1/2 \equiv \ell\pm$. By conservation of angular momentum, this must equal the total angular momentum of the initial state, $J = L \pm 1/2$, where $L$ denotes the orbital angular momentum of the photon coupled to its intrinsic spin.

Thus, there are six multipole amplitudes in electroproduction that characterize the possible electromagnetic transitions: the electric $E_{\ell\pm}$, magnetic $M_{\ell\pm}$, and scalar $S_{\ell\pm}$ (or, equivalently, longitudinal $L_{\ell\pm}$) multipoles. A brief derivation of these amplitudes can be found, e.g., in Refs.~\cite{Berends:1967vi,Dennery:1961zz}. By comparing the Pauli amplitudes in the angular momentum basis and those in  Eq.~(\ref{eq:F_in_terms_of_F7_F8}), one obtains \cite{Berends:1967vi}
\begin{eqnarray}
    \label{eq:F_in_terms_of_multipoles}
    \left[ \begin{array}{c}
        F_1\\F_2\\F_3\\F_4\\F_5\\F_6 
    \end{array} \right]  &=& \sum_{\ell\geq 0}
    \left[ \begin{array}{cccccc}
        P_{\ell+1}'& P_{\ell-1}' & \ell P_{\ell+1}' & (\ell+1)P_{\ell-1}'&0&0\\ 
        0& 0 & (\ell+1) P_{\ell}' & \ell P_{\ell}'&0&0\\ 
        P_{\ell+1}''& P_{\ell-1}'' & -P_{\ell+1}'' & P_{\ell-1}''&0&0\\ 
        -P_{\ell}''& -P_{\ell}'' & P_{\ell}'' & -P_{\ell}''&0&0\\ 
        0 & 0 & 0 & 0 & -(\ell+1) P_{\ell}' & \ell P_{\ell}'\\
        0 & 0 & 0 & 0 & (\ell+1) P_{\ell+1}' & -\ell P_{\ell-1}' 
    \end{array} \right] ~
    \left[ \begin{array}{c}
        E_{\ell+}\\E_{\ell-}\\M_{\ell+}\\M_{\ell-}\\S_{\ell+}\\S_{\ell-} 
    \end{array} \right] ,
\end{eqnarray}
where $P_\ell\equiv P_\ell(\cos{\theta})=P_\ell(\boldsymbol{\hat{k}\cdot \hat{q}})$ are the Legendre polynomials. Furthermore, the scalar multipole amplitude $S_{\ell\pm}$ is related to the longitudinal amplitude \cite{Dennery:1961zz} through $L_{\ell\pm}=(k_0/|\boldsymbol{k}|) S_{\ell\pm}$. Because Eq. (\ref{eq:F_in_terms_of_multipoles}) involves derivatives of Legendre polynomials, not all values of $\ell$ contribute. The allowed values of $\ell$ and the corresponding types of multipoles are summarized in Table \ref{tab:multipoles}.

\begin{table}[t]
\centering
\caption{Electric, magnetic, and scalar multipoles in kaon electroproduction, their total angular momentum $J$, photon angular momentum $L$, parity, the lowest value of $\ell$, and the lowest contributing  multipoles \cite{Berends:1967vi}.}
\label{tab:multipoles}
\begin{tabularx}{\textwidth}{XXXXrcc}
%\begin{tabular}{lcccccc}
\hline\hline
Multipole &Transition& $J$ & $L$ &Parity& ~~~~$\ell_{\min}$~~~~ & Lowest multipole\\
\hline
$E_{\ell+}$ &Electric& $\ell + \tfrac{1}{2}$ &$\ell+1$&$(-1)^L$&  $0$ &$E_{0+}$\\
$E_{\ell-}$ &Electric& $\ell - \tfrac{1}{2}$ &$\ell-1$&$(-1)^L$&  $2$ &$E_{2-}$\\
$M_{\ell+}$ &Magnetic& $\ell + \tfrac{1}{2}$ &$\ell$     &$-(-1)^L$&  $1$ &$M_{1+}$\\
$M_{\ell-}$ &Magnetic& $\ell - \tfrac{1}{2}$ &$\ell$     &$-(-1)^L$&  $1$ &$M_{1-}$\\
$S_{\ell+}$ &Scalar  & $\ell + \tfrac{1}{2}$ &$\ell+1$&$(-1)^L$&  $0$ &$S_{0+}$\\
$S_{\ell-}$ &Scalar  & $\ell - \tfrac{1}{2}$ &$\ell-1$&$(-1)^L$&  $1$ &$S_{1-}$\\
\hline\hline
%\end{tabular}
\end{tabularx}
\end{table}

Equation~(\ref{eq:F_in_terms_of_multipoles}) can be inverted to extract the multipole amplitudes by exploiting the orthogonality of the Legendre polynomials. This procedure is particularly useful when the amplitudes are derived within a covariant formalism, such as the Feynman diagrammatic approach employed in the effective Lagrangian model Kaon-MAID~\cite{kaonmaid} (see Section~\ref{sec:isobar}). The resulting relations are \cite{Berends:1967vi}
\begin{eqnarray}
    \label{eq:multipoles_in_terms_of_F}
    \left[ \begin{array}{c}
        E_{\ell+}(s)\\E_{\ell-}(s)\\M_{\ell+}(s)\\M_{\ell-}(s)\\S_{\ell+}(s)\\S_{\ell-}(s) 
    \end{array} \right] 
    &=& \int_{-1}^{+1} dx
    \left[ \begin{array}{cccccc}
        \frac{1}{2(\ell+1)}\Bigl\{P_{\ell},& -P_{\ell+1}, & \frac{\ell}{2\ell+1}\bigl(P_{\ell-1}-P_{\ell+1}\bigr), & \frac{\ell+1}{2\ell+3}\bigl(P_{\ell}-P_{\ell+2}\bigr)\Bigr\}&0&0\\ 
        \frac{1}{2\ell}\Bigl\{P_{\ell},& -P_{\ell-1}, & \frac{\ell+1}{2\ell+1}\bigl(P_{\ell+1}-P_{\ell-1}\bigr), & \frac{\ell}{2\ell-1}\bigl(P_{\ell}-P_{\ell-2}\bigr)\Bigr\}&0&0\\ 
        \frac{1}{2(\ell+1)}\Bigl\{P_{\ell},& -P_{\ell+1}, & \frac{1}{2\ell+1}\bigl(P_{\ell+1}-P_{\ell-1}\bigr), & 0\Bigr\}&0&0\\ 
        \frac{1}{2\ell}\Bigl\{-P_{\ell},& P_{\ell-1}, & \frac{1}{2\ell+1}\bigl(P_{\ell-1}-P_{\ell+1}\bigr), & 0\Bigr\}&0&0\\ 
        0 & 0 & 0 & 0 & \frac{1}{2(\ell+1)}\Bigl\{ P_{\ell+1}, & P_{\ell}\Bigr\}\\
        0 & 0 & 0 & 0 & \frac{1}{2\ell}\Bigl\{ P_{\ell-1}, & P_{\ell}\Bigr\}
    \end{array} \right]
    \nonumber\\
    && \times\left[ \begin{array}{c}
        F_1(s,t)\\F_2(s,t)\\F_3(s,t)\\F_4(s,t)\\F_5(s,t)\\F_6(s,t) 
    \end{array} \right],
\end{eqnarray}
where $t$ is a function of $x=\cos{\theta}$.

\subsection{Isospin Amplitudes}
\label{sec:isospin}

As mentioned earlier, the $K\Lambda$ electroproduction channels are isoscalar and thus resemble $\eta$ electroproduction, while the $K\Sigma$ channels are isovector in nature, making them analogous to pion electroproduction. However, the isospin amplitudes in the kaon case are not identical to those in pion electroproduction, since the final baryon is a hyperon, i.e., an isodoublet in the $K\Lambda$ case and an isotriplet in the $K\Sigma$ case, rather than a nucleon. 

To analyze all isospin channels in kaon electroproduction in a unified way, Ref.~\cite{Mart:1996gx} employed SU(3) symmetry to relate the corresponding coupling constants. For the isoscalar $K\Lambda$ channels, the relations are
\begin{eqnarray}
    g_{K^+\Lambda p} &=& g_{K^0\Lambda n},
\end{eqnarray}
which also hold for the excited kaons,
\begin{eqnarray}
    g_{K^{*+}\Lambda p}^{\rm V,T} &=& g_{K^{*0}\Lambda n}^{\rm V,T}.
\end{eqnarray}
For the isovector $K\Sigma$ channels, the coupling constants satisfy
\begin{eqnarray}
    \label{eq:coupl_const_relation_nucl}
    g_{K^+\Sigma^0 p} &=& g_{K^0\Sigma^+ p}/\sqrt{2} ~=~ g_{K^+\Sigma^- n}/\sqrt{2} ~=~ -g_{K^0\Sigma^0 n},
\end{eqnarray}
for isospin-1/2 and
\begin{eqnarray}
    \label{eq:coupl_const_relation_delta}
    g_{K^+\Sigma^0 \Delta^+} &=& -\sqrt{2}\, g_{K^0\Sigma^+ \Delta^+} ~=~ \sqrt{2}\, g_{K^+\Sigma^- \Delta^0} ~=~ g_{K^0\Sigma^0 \Delta^0},
\end{eqnarray}
for isospin-3/2. It is important to note that the isospin-3/2 states ($\Delta$s) contribute only in $K\Sigma$ production.

Nevertheless, an equivalent result can be derived by adopting the same procedure as in pion electroproduction, where the proton and neutron amplitudes with total isospin-1/2 are defined as $A_p^{1/2}$ and $A_n^{1/2}$, respectively \cite{Drechsel:1998hk}. Accordingly, the isospin amplitudes for the isovector $K\Sigma$ electroproduction take the form \cite{Mart:2014eoa}
\begin{subequations}
    \label{eq:sigma_iso_amplitudes}
    \begin{align}
    A(\gamma_v+p\to K^++\Sigma^0)~&=~ A_p^{(1/2)}+{\textstyle\frac{2}{3}} A^{(3/2)},\\
    A(\gamma_v+p\to K^0+\Sigma^+)~&=~ \sqrt{2}\left\{A_p^{(1/2)}-{\textstyle\frac{1}{3}} A^{(3/2)}\right\},\\
    A(\gamma_v+n\to K^++\Sigma^-)~&=~ \sqrt{2}\left\{A_n^{(1/2)}+{\textstyle\frac{1}{3}} A^{(3/2)}\right\},\\
    A(\gamma_v+n\to K^0+\Sigma^0)~&=~ -A_n^{(1/2)}+{\textstyle\frac{2}{3}} A^{(3/2)}. 
    \end{align}
\end{subequations}
By grouping the coefficients of the proton and neutron amplitudes in Eq.~(\ref{eq:sigma_iso_amplitudes}), one obtains Eq.~(\ref{eq:coupl_const_relation_nucl}), whereas collecting the coefficients of the isospin-3/2 amplitudes leads to Eq.~(\ref{eq:coupl_const_relation_delta}). We note that Eq.~(\ref{eq:sigma_iso_amplitudes}) provides a convenient framework for analyzing kaon photo- and electroproduction within the multipole approach \cite{Mart:2014eoa}.

\subsection{Differential Cross Section and Polarization Observables}
\label{dcs-formalism}

\begin{figure}[t]
    \centering
    \includegraphics[width=0.55\columnwidth]{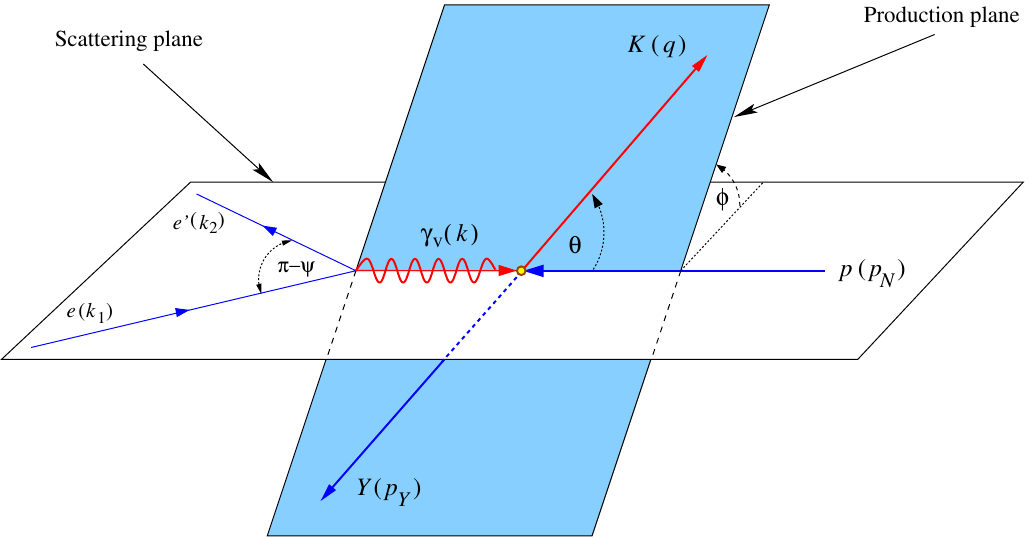} 
    \caption{Kinematics for kaon electroproduction on the nucleon in the kaon-hyperon c.m. frame.}
    \label{fig:electro_kinematics}
    \end{figure}

For an unpolarized experiment, the differential cross section of kaon electroproduction on the nucleon can be written as \cite{Bjorken:1965sts,Halzen:1984mc}
\begin{equation}
    d\sigma = 
    \frac{(2\pi)^4\delta^{(4)}(p_Y+k_2+q-k_1-p_N)}{4\bigl\{ (k_1\cdot p_N)^2-m_em_N \bigr\}^{1/2}}\left| \bar{\cal M}_{\rm fi} \right|^2 \frac{d^3\boldsymbol{p}_Y}{(2\pi)^32E_Y} \frac{d^3\boldsymbol{k}_2}{(2\pi)^32E_2} \frac{d^3\boldsymbol{q}}{(2\pi)^32E_K},
\end{equation}
where $|\bar{\cal M}_{\rm fi}|^2$ denotes the spin-averaged squared transition amplitude ${\cal M}_{\rm fi}$ defined in Eq.~(\ref{eq:transition_amplitude}). Calculation of $|\bar{\cal M}_{\rm fi}|^2$ is straightforward but tedious. This is given, e.g., in Refs.~\cite{Berends:1967vi,Donnachie:1972aa}, where the calculation is carried out in Pauli space. Explicitly, the differential cross section is written as
\begin{equation}
    \frac{d\sigma}{dE_2 d\Omega_e d\Omega_K} = \Gamma_v \frac{d\sigma}{d\Omega_v},
\end{equation}
with the flux of the virtual photon field
\begin{equation}
    \Gamma = \frac{\alpha}{2\pi^2}\frac{E_2}{E_1}\frac{W^2-m_N^2}{2m_N}\frac{1}{Q^2}\frac{1}{1-\epsilon},
\end{equation}
and the degree of transverse polarization of the virtual photon
\begin{equation}
\label{defeps}
    \epsilon = \left( 1+\frac{2|\boldsymbol{q}|^2}{Q^2} \tan^2\frac{\psi}{2} \right)^{-1}.
\end{equation}
The virtual photoproduction differential cross section is given by
\begin{eqnarray}
    \label{eq:virtual_diff_cs}
    \frac{d\sigma_v}{d\Omega_K} &=& \frac{d\sigma_{\rm T}}{d\Omega_K} + 
    \epsilon_{\rm L}\frac{d\sigma_{\rm L}}{d\Omega_K}+
    \epsilon\frac{d\sigma_{\rm TT}}{d\Omega_K}\cos{2\phi}
    +[2\epsilon_{\rm L}(1+\epsilon)]^{1/2}\frac{d\sigma_{\rm TL}}{d\Omega_K}\cos\phi,    
\end{eqnarray}
with $\epsilon_{\rm L}=(Q^2/|\boldsymbol{k}|^2) \epsilon$. The electron and kaon scattering angles $\psi$, $\theta$, and $\phi$ are defined in Fig.~\ref{fig:electro_kinematics}. The subscripts T, L, TT, and TL denote the transverse, longitudinal, transverse-transverse, and transverse-longitudinal interference contributions to the cross section, respectively. These components are often referred to as structure functions or response functions, since they encode the dynamical information of the target's response to the virtual photon. In terms of the helicity amplitudes defined in Eq.~(\ref{eq:helicity_amplitudes}), the individual cross sections on the right hand side of Eq.~(\ref{eq:virtual_diff_cs}) can be written as \cite{Deo:1974ik,Mart:1996gx,Knochlein:1995qz,Levy:1973zz}
\begin{subequations}
    \label{eq:virtual_diff_cs_indiv}
    \begin{align}
        \frac{d\sigma_{\rm T}}{d\Omega_K} ~&=~ \frac{|\boldsymbol{q}|}{2K_\gamma}\left(|H_1|^2+|H_2|^2+|H_3|^2+|H_4|^2\right),    \label{eq:virtual_diff_cs_indiv_a}\\
        \frac{d\sigma_{\rm L}}{d\Omega_K} ~&=~ \frac{|\boldsymbol{q}|}{K_\gamma}\left(|H_5|^2+|H_6|^2\right),\\
        \frac{d\sigma_{\rm TT}}{d\Omega_K} ~&=~ \frac{|\boldsymbol{q}|}{K_\gamma}~{\rm Re}\left(H_2^*H_3-H_1^*H_4\right),\\
        \frac{d\sigma_{\rm TL}}{d\Omega_K} ~&=~ \frac{|\boldsymbol{q}|}{\sqrt{2}K_\gamma}~{\rm Re}\bigl\{\left(H_5^*(H_4-H_1)-H_6^*(H_2+H_3\right)\bigr\},
    \end{align}
\end{subequations}
where we have defined the photon equivalent energy $K_\gamma=(s-m_N^2)/2W$, which in the special case of photoproduction ($k^2=0$) reduces to $K_\gamma=|\boldsymbol{k}|$. In this limit, Eq.~(\ref{eq:virtual_diff_cs_indiv_a}) is identical to Eq.~(\ref{eq:diff_cs_photo_helicity}), thus establishing their consistency for real photons.

%%%%%%%FIGURE 1%%%%%
%%%%%%%%%%%%%%%%%%%%

\begin{figure}[t]
    \centering
    \includegraphics[width=0.3\columnwidth]{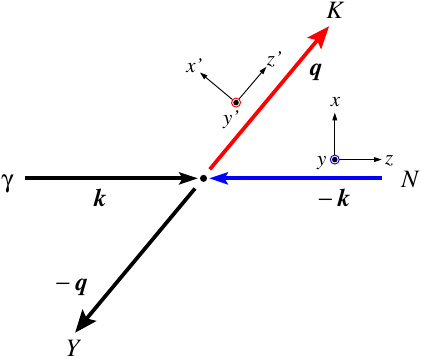} 
    \caption{Target $(x,y,z)$ and recoil $(x',y',z')$ coordinate frames used to define the directions of the target and recoil polarizations \cite{Knochlein:1995qz}. }
    \label{fig:frames_convention}    
    \end{figure}

In experiments with complete polarization, the corresponding cross section is more involved since it includes additional contributions arising from the polarizations of the electron beam, the target nucleon, and the recoiling hyperon. Although slightly different definitions can be found in the literature, in the present discussion, we adopt the convention of Kn\"ochlein et al.~\cite{Knochlein:1995qz}. Within this convention, the target and recoil polarizations are expressed in their respective reference frames, as illustrated in Fig.~\ref{fig:frames_convention}. The target nucleon polarizations are defined in the $(x,y,z)$ frame, where the $z$-axis is directed along the photon momentum vector $\boldsymbol{k}$. For the recoil polarization, the corresponding frame is denoted as $(x',y',z')$, with the $z'$-axis aligned along the kaon momentum vector.

%%%%%%%FIGURE 1%%%%%
%%%%%%%%%%%%%%%%%%%%

Denoting the target and recoil polarizations by $\boldsymbol{P}=(P_x,P_y,P_z)$ and $\boldsymbol{P}'=(P_x',P_y',P_z')$, respectively, the most general differential cross section for kaon electroproduction with polarized beam, target, and recoil, can be written as \cite{Knochlein:1995qz,Drechsel:1992pn}
\begin{eqnarray}
    \label{eq:most_general_cs}
    \frac{d\sigma_v}{d\Omega_K} &=& \frac{|\boldsymbol{q}|}{K_\gamma}P_\mu P_\nu \,\Bigl\{ R_{\rm T}^{\nu\mu}+\epsilon_{\rm L}R_{\rm L}^{\nu\mu} + [2\epsilon_{\rm L}(1+\epsilon)]^{1/2} \bigl(\, {^{\rm c}\! R}_{\rm TL}^{\nu\mu}\cos{\phi_K} + {^{\rm s}\! R}_{\rm TL}^{\nu\mu}\sin{\phi_K} \bigr) \nonumber\\
    && + \epsilon\bigl(\, {^{\rm c}\! R}_{\rm TT}^{\nu\mu}\cos{2\phi_K} + {^{\rm s}\! R}_{\rm TT}^{\nu\mu}\sin{2\phi_K} \bigr) + h[2\epsilon_{\rm L}(1-\epsilon)]^{1/2} \bigl(\, {^{\rm c}\! R}_{\rm TL'}^{\nu\mu}\cos{\phi_K} + {^{\rm s}\! R}_{\rm TL'}^{\nu\mu}\sin{\phi_K} \bigr) \nonumber\\
    && + h(1-\epsilon^2)^{1/2} ~ {^c\! R}_{\rm TT'}^{\nu\mu}\Bigr\},    
\end{eqnarray}
where $R_i^{\nu\mu}$ are the response functions, $h$ is the helicity of the incoming electron, $P_\mu=(1,\boldsymbol{P})$, and $P_\nu=(1,\boldsymbol{P}')$. It is straightforward to prove that Eq.~(\ref{eq:virtual_diff_cs}) can be obtained from the most general differential cross section by setting $\mu=\nu=0$ and the photon helicity $h=0$. The complete set of response functions $R_i^{\nu\mu}$ is listed in Appendix~\ref{app:response_function}. It should be emphasized that not all of the response functions in Eq.~(\ref{eq:most_general_cs}) are independent, and their interrelations are also provided in Appendix~\ref{app:response_function}. Here, we only want to note that certain well-known photoproduction observables can be expressed in terms of the response functions $R_i^{\nu\mu}$ \cite{Knochlein:1995qz}, i.e.,

\begin{center}
\begin{tabular*}{0.75\textwidth}{llll}
    $d\sigma/d\Omega_K=({|\boldsymbol{q}|}/{K_\gamma}) R_{\rm T}^{00}$, ~~~~ & $\Sigma = {^{\rm c}\! R}_{\rm T}^{0y}/R_{\rm T}^{00}$,~~~~ & $T = {R}_{\rm T}^{00}/R_{\rm T}^{00}$,~~~~ &  $P = R_{\rm T}^{y'0}/R_{\rm T}^{00}$, \\
    $E=-{R}_{\rm TT'}^{0z}/R_{\rm T}^{00}$,  & $F = {R}_{\rm TT'}^{0x}/R_{\rm T}^{00}$,~~~~ & $G = -{^{\rm s}\! R}_{\rm TT}^{0z}/R_{\rm T}^{00}$,~~~~ &  $H = {^{\rm s}\! R}_{\rm TT}^{0x}/R_{\rm T}^{00}$,\\
    $O_{x'}={^{\rm s}\! R}_{\rm TT}^{x'0}/R_{\rm T}^{00}$,  & $O_{z'} = {^{\rm s}\! R}_{\rm TT}^{z'0}/R_{\rm T}^{00}$,~~~~ &  $C_{x'}=-{R}_{\rm TT'}^{x'0}/R_{\rm T}^{00}$,  & $C_{z'} = -{R}_{\rm TT'}^{z'0}/R_{\rm T}^{00}$,\\
    $T_{x'}={R}_{\rm T}^{x'x}/R_{\rm T}^{00}$,  & $T_{z'} = {R}_{\rm TT}^{z'x}/R_{\rm T}^{00}$,~~~~ &  $L_{x'}=-{R}_{\rm T}^{x'z}/R_{\rm T}^{00}$,  & $L_{z'} = {R}_{\rm T}^{z'z}/R_{\rm T}^{00}$.
\end{tabular*}
\end{center}

\subsection{Hadronic Form Factor and Problems with Gauge Invariance}

\begin{figure}[t]
    \centering
    \includegraphics[width=0.75\columnwidth]{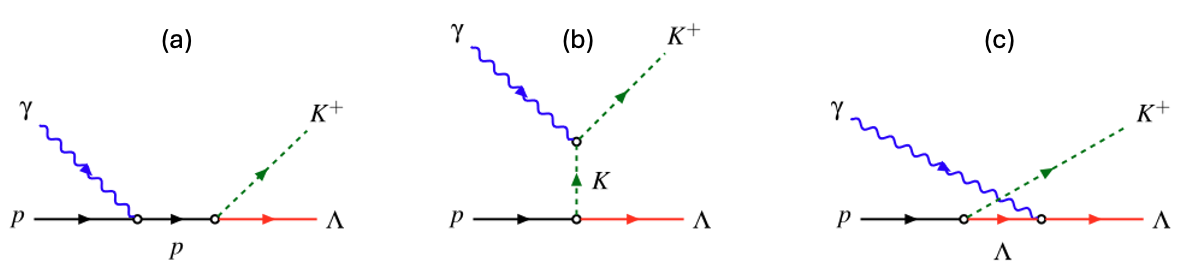} 
    \caption{Feynman diagrams for the Born terms contributing to kaon photoproduction on the nucleon, $\gamma+p \to K^+ + \Lambda$: (a) $s$-channel with an intermediate proton, (b) $t$-channel with an intermediate kaon, and (c) $u$-channel with an intermediate $\Lambda$ hyperon.}
    \label{fig:Born_diagrams} 
    \end{figure}

A quick inspection of the Feynman diagrams for the Born terms in kaon photoproduction, shown in Fig.~\ref{fig:Born_diagrams}, immediately reveals  their similarity to those of pion photoproduction. Furthermore, SU(3) flavor symmetry predicts that the leading coupling constants relevant to kaon photoproduction are of the same order as the $\pi N$ coupling constant. In particular, SU(3) symmetry implies that \cite{Adelseck:1990ch,deSwart:1963pdg}
\begin{equation}
    g_{K\Lambda N} = -\frac{1}{\sqrt{3}}(3-2\alpha)\, g_{\pi NN}, ~~~~~~~~~~~ g_{K\Sigma N} = +(2\alpha-1)\, g_{\pi NN},
\end{equation}
with $\alpha\approx 0.644$ \cite{Donoghue:1982hs}. Accordingly, at low energies, where contributions from baryon resonances can be neglected, the kaon photoproduction cross section is expected to be of the same order of magnitude as that of pion photoproduction. However, experimentally the total cross section for pion photoproduction is found to be nearly three orders of magnitude larger than that for kaon photoproduction. This striking discrepancy underscores the essential role of hadronic form factors, which is required to suppress the scattering amplitude and avoid divergences at higher energies~\cite{Mart:1995wu}. In contrast, such a mechanism is far less significant in the case of pion photoproduction. Nevertheless, introducing hadronic form factors in the Born terms gives rise to another fundamental problem, namely the violation of gauge invariance in the amplitude. 

%%%%%%%FIGURE 1%%%%%
%%%%%%%%%%%%%%%%%%%%

Let us consider again Fig.~\ref{fig:Born_diagrams}. The transition amplitude obtained from the $u$-channel diagram is independently gauge invariant. In contrast, the amplitude obtained from the $s$-channel within the pseudoscalar coupling theory is given by \cite{Mart:1996gx}
\begin{equation}
    \label{eq:s-channel-contrib}
    {\cal M}_p = {\bar u}(p_\Lambda)\, \frac{ieg_{K\Lambda N}\gamma_5}{s-m_p^2}\left[ \frac{\kappa_p}{m_p}(p_p\cdot k\, \slashed{\epsilon}-p_p\cdot\epsilon\,\slashed{k})-(1+\kappa_p)\slashed{\epsilon}\slashed{k}-\frac{4}{t-m_K^2} p_p\cdot\epsilon\, q\cdot k
    \right]\, u(p_p),
\end{equation}
where the last term is explicitly not gauge invariant, since it does not vanish under the substitution $\epsilon \to k$. The amplitude obtained from Fig.~\ref{fig:Born_diagrams}(b) reads
\begin{equation}
    \label{eq:t-channel-contrib}
    {\cal M}_{K^+}={\bar u}(p_\Lambda)\, \frac{ieg_{K\Lambda N}\gamma_5}{t-m_K^2}\left[ \frac{4}{s-m_p^2} q\cdot\epsilon\, p_p\cdot k
    \right] \, u(p_p).
\end{equation}
Obviously, the combined contributions from the $s$- and $t$-channel diagrams yield a gauge-invariant transition amplitude.
However, this result holds only under the assumption of point-like hadrons. In reality, hadrons are extended objects, and the hadronic vertex must therefore be modified by a form factor that accounts for the internal constituent distribution underlying the strong interaction. Naturally, this form factor depends on the momenta of all particles participating in the vertex. Since two of these correspond to on-shell particles, the dependence reduces to a single momentum variable associated with the off-shell intermediate state \cite{Haberzettl:1998aqi, Haberzettl:2001kr, Haberzettl:2006bn, Haberzettl:2021wcz}.

The inclusion of hadronic form factors $F_{\rm h}(p_p^2,q^2,p_\Lambda^2)$ at the hadronic vertices introduces an additional non-gauge-invariant contribution arising from the combination of Eqs.~(\ref{eq:s-channel-contrib}) and (\ref{eq:t-channel-contrib}), i.e.,
\begin{equation}
    \label{eq:excess-non-gi}
    \Delta{\cal M}={\bar u}(p_\Lambda)\, \frac{ieg_{K\Lambda N}\gamma_5}{s-m_p^2}\frac{4}{t-m_K^2}\left[  q\cdot\epsilon\, p_p\cdot k \, F_{\rm h}(m_p^2,t,m_\Lambda^2)-q\cdot k\, p_p\cdot\epsilon \, F_{\rm h}(s,m_K^2,m_\Lambda^2)
    \right] \, u(p_p),
\end{equation}
which evidently does not vanish under the substitution $\epsilon \to k$.

In order to restore gauge invariance, we note that Ohta \cite{Ohta:1989ji} introduced an additional amplitude, derived by applying the minimal substitution prescription. In our convention, the amplitude is given by \cite{Haberzettl:1998aqi}
\begin{eqnarray}
    \label{eq:Ohta_receipe}
    \Delta{\cal M}^{\rm Ohta}&=&{\bar u}(p_\Lambda)\, \frac{ieg_{K\Lambda N}\gamma_5}{s-m_p^2}\frac{4}{t-m_K^2}\Bigl[  q\cdot\epsilon\, p_p\cdot k \, \Bigl\{F_{\rm h}(m_p^2,m_K^2,m_\Lambda^2) - F_{\rm h}(m_p^2,t,m_\Lambda^2) \Bigr\} \nonumber\\ &&
    - q\cdot k\, p_p\cdot\epsilon \, \Bigl\{F_{\rm h}(m_p^2,m_K^2,m_\Lambda^2)-F_{\rm h}(s,m_K^2,m_\Lambda^2) \Bigr\}
    \Bigr] \, u(p_p),
\end{eqnarray}
where the form factor is normalized to unity when all three particles are on-shell, i.e., $F_{\rm h}(m_p^2,m_K^2,m_\Lambda^2)=1$. Incorporating this amplitude into Eq.~(\ref{eq:excess-non-gi}) restores gauge invariance, but simultaneously eliminates the off-shell hadronic form factors in the non-gauge-invariant term. In our convention, Ohta's prescription removes the form factor from the $A_2^{\rm Born}$ amplitude in Eq.~(\ref{eq:Ai_Mi}).

The removal of the form factor is problematic, as it eliminates all degrees of freedom associated with the compositeness of hadrons and contradicts our primary motivation, i.e., to tame the excessive growth of the amplitude at higher energies. Recognizing this limitation, Haberzettl \cite{Haberzettl:1997jg} proposed an alternative approach that preserves gauge invariance while retaining the physical role of hadronic form factors, by making use of the freedom to multiply the $A_2^{\rm Born}$ amplitude with an appropriately chosen form factor $\hat{F}$. Specifically, the form factor is constructed in a more ``democratic'' manner, 
\begin{equation}
    \label{eq:Fhat_haberzettl}
    \hat{F}(s,t,u)=a_1F_1(s)+a_2F_2(t)+a_3F_3(u),
\end{equation} 
with 
\begin{equation}
    F_1(s)=F(s,m_K^2,m_\Lambda^2), ~~~ F_2(t)=F(m_p^2,t,m_\Lambda^2), ~~~ F_3(u)=F(m_p^2,m_K^2,u),
\end{equation} 
and $a_1+a_2+a_3=1$ to ensure the correct limit at $k=0$. At this stage, it is important to note that Ohta's prescription is equivalent to choosing $\hat{F}=1$.

Haberzettl's prescription is clearly more preferred than Ohta's, as it provides a better description of the data by  yielding a lower $\chi^2$ in the fitting process. Qualitatively, the use of Haberzettl prescription tends to enhance the leading coupling constants toward the SU(3) limits while simultaneously reducing the extracted $\chi^2$ \cite{Haberzettl:1998aqi}. Nevertheless, the use of Haberzettl's democratic form factor, given by Eq.~(\ref{eq:Fhat_haberzettl}), has been criticized by Davidson and Workman \cite{Davidson:2001rk}, as it still generates poles and fails to satisfy crossing symmetry in pion photoproduction. These shortcomings can be resolved by adopting a specific, though not unique, form factor,
\begin{equation}\label{eq:Davidson_Workman_HFF}
    \hat{F}(s,u,t)= F_1(s) + F_1(u) + F_3(t) - F_1(s)F_1(u) - F_1(s)F_3(t) - F_1(u)F_3(t) + F_1(s)F_1(u)F_3(t),
\end{equation}
which simultaneously eliminates the spurious poles and restores crossing symmetry. 

\subsection{Self-Analyzing $\Lambda$ Polarization}

\begin{figure}[t]
    \centering
    \includegraphics[width=0.65\columnwidth]{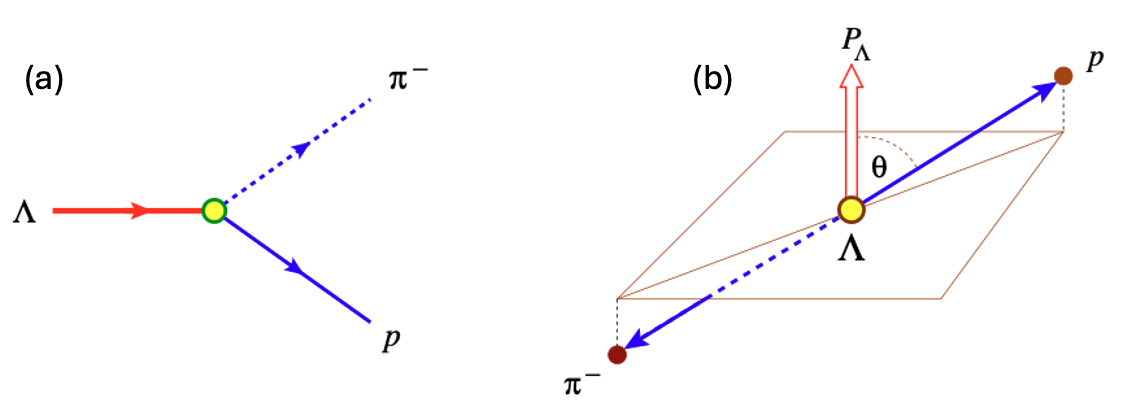} 
    \caption{(a) Feynman diagram for the weak decay of the $\Lambda$ hyperon into a proton and a pion, $\Lambda \to p\pi^-$. (b) Kinematics of the decay in the $\Lambda$ rest frame, where $P_\Lambda$ denotes the $\Lambda$ polarization and $\theta$ is the angle between the polarization direction and the proton momentum. The plane shown is perpendicular to the $\Lambda$ polarization direction.}
    \label{fig:Lambda_decay_p_pi-} 
    \end{figure}

Another unique feature of the electromagnetic production of kaons is that the polarization of the recoiling $\Lambda$ hyperon (or $\Sigma^0$ via its decay into $\Lambda \gamma$ with 100\% branching ratio) can be determined from its self-analyzing weak decay, thereby eliminating the need for an external polarimeter. This characteristic constitutes a significant advantage over pion production. Consider the Feynman diagram of the $\Lambda$ weak decay into a proton and a pion shown in Fig.~\ref{fig:Lambda_decay_p_pi-}(a). The corresponding decay amplitude is given by \cite{Donoghue:2022wrw}
\begin{equation}
\label{eq:Lambda_to_p_pi-}
{\cal M}_{\Lambda\to p+\pi^-} = \bar{u}_p\,(A-B\gamma_5)\,u_\Lambda,
\end{equation}
where $A$ and $B$ denote the parity-conserving and parity-violating amplitudes, respectively.

Since our focus is on the $\Lambda$ polarization, it is convenient to recast the amplitude in terms of Pauli spinors in the $\Lambda$ rest frame, as illustrated in Fig.~\ref{fig:Lambda_decay_p_pi-}(b), i.e.,
\begin{equation}
u_\Lambda = \begin{pmatrix} \chi_\Lambda \\ 0 \end{pmatrix},
\qquad
\bar{u}_p=\Big(\chi_p^\dagger\, ,\, -\chi_p^\dagger \frac{\boldsymbol{\sigma\cdot p}}{E_p+m_p}\Big).
\end{equation}
Substituting these expressions into Eq.~(\ref{eq:Lambda_to_p_pi-}) yields
\begin{equation}
\label{eq:Lambda_to_p_pi-_Pauli}
{\cal M}_{\Lambda\to p+\pi^-} = \chi_p^\dagger \,\big(a_s + a_p,\boldsymbol{\sigma\cdot\hat{n}}\big)\, \chi_\Lambda,
\end{equation}
with
%\begin{equation}
$\boldsymbol{\hat{n}}={\boldsymbol{p}_p}/{|\boldsymbol{p}_p|},~
 a_s=A, ~{\rm and}
~ a_p=B\, {|\boldsymbol{p}_p|}/{(E_p+m_p)}$.
%\end{equation}
By squaring the amplitude in Eq.~(\ref{eq:Lambda_to_p_pi-_Pauli}) and employing the projection identity
\begin{equation*}
\chi^{(s_i)}\chi^{(s_i)\dagger}=\tfrac{1}{2}\big(1+\boldsymbol{\sigma\cdot\hat{s}}_i\big),
\end{equation*}
where $\boldsymbol{\hat{s}}_i$ is a unit vector along the spin-polarization direction $\boldsymbol{s}_i$, one arrives at the expression for the $\Lambda$ decay rate in the form of
\begin{equation}
    \label{eq:dGamma-dOmega-complete}
    \frac{d\Gamma}{d\Omega}\propto 1+\gamma \boldsymbol{\hat{s}}_p\boldsymbol{\cdot\,{\hat{s}}}_\Lambda + (1-\gamma) (\boldsymbol{\hat{s}}_p\boldsymbol{\cdot\,{\hat{n}}})\,(\boldsymbol{\hat{s}}_\Lambda\boldsymbol{\cdot\,{\hat{n}}}) + \alpha(\boldsymbol{\hat{s}}_p\boldsymbol{\cdot\,{\hat{n}}}+\boldsymbol{\hat{s}}_\Lambda\boldsymbol{\cdot\,{\hat{n}}})+\beta(\boldsymbol{\hat{s}}_p\boldsymbol{\times\hat{s}}_\Lambda)\boldsymbol{\cdot\,{\hat{n}}},
\end{equation}
where $\boldsymbol{\hat{s}}_p$ and $\boldsymbol{\hat{s}}_\Lambda$ are the spin polarizations of the proton and $\Lambda$ hyperon, respectively, and 
\begin{eqnarray*}
    \alpha = 2{\rm Re}(a_s^*a_p)/(|a_s^2|+|a_p^2|),~~
    \beta = 2{\rm Im}(a_s^*a_p)/(|a_s^2|+|a_p^2|),~~
    \gamma = (|a_s^2|-|a_p^2|)/(|a_s^2|+|a_p^2|).
\end{eqnarray*}

%%%%%%%FIGURE 1%%%%%
%%%%%%%%%%%%%%%%%%%%

\begin{figure}[t]
    \centering
    \includegraphics[width=0.65\columnwidth]{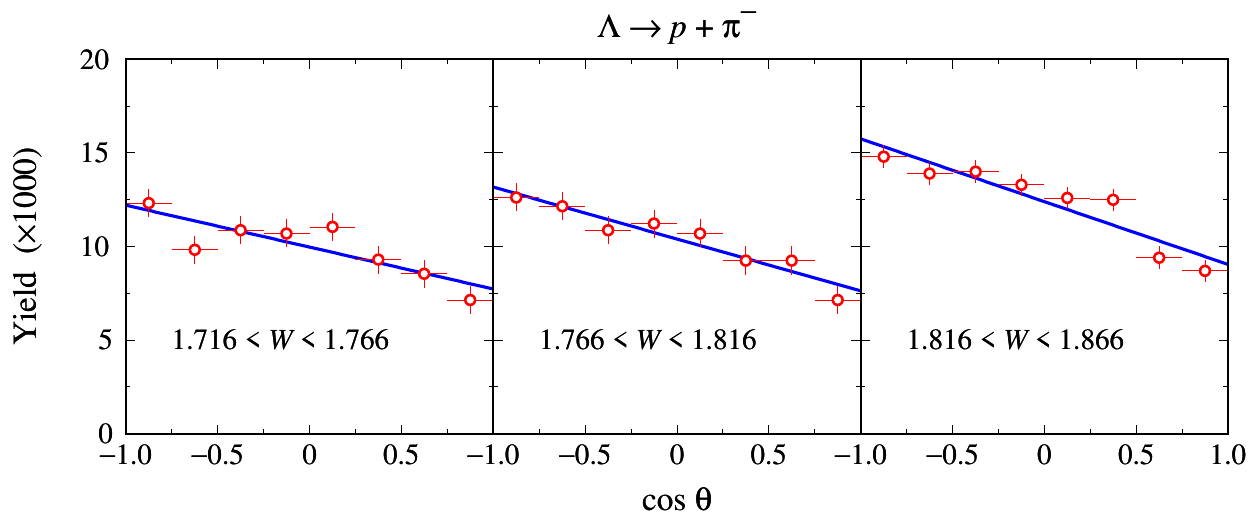} 
    \caption{Angular distributions of the $\Lambda$ decay to a proton and a $\pi^-$ for three different invariant mass bins (GeV) measured by CLAS. Figure adapted from Ref.~\cite{McAleer:2002gaa}.}
    \label{fig:Lambda_polarization} 
    \end{figure}

If the proton polarization is not measured, the average over $\boldsymbol{\hat{s}}_p$ vanishes, and Eq.~(\ref{eq:dGamma-dOmega-complete}) simplifies to
\begin{equation}
    \label{eq:dGamma-dOmega-no-proton}
    \frac{d\Gamma}{d\Omega}\propto 1+ \alpha\, \boldsymbol{\hat{s}}_\Lambda\boldsymbol{\cdot\,{\hat{n}}} = 1 +\alpha\, P_\Lambda\cos{\theta}\,,
\end{equation}
where we have defined $P_\Lambda$ as the $\Lambda$ polarization. Figure~\ref{fig:Lambda_polarization} shows three representative $W$ bins of the angular distributions in the hyperon decay frame from measurements performed at CLAS~\cite{McAleer:2002gaa}. The linear dependence predicted by Eq.~(\ref{eq:dGamma-dOmega-no-proton}) is clearly supported by the experimental results.

%The decay of the $\Lambda$ hyperon into a proton and a pion was studied using the CLAS detector at JLab. Here, $\theta$ is the angle of the proton relative to the $\Lambda$ spin direction. However, since this direction could not be directly determined, the angle was measured relative to the hyperon production plane~\cite{McAleer:2002gaa}. Consequently, Eq.~(\ref{eq:dGamma-dOmega-no-proton}) takes the form 
%\begin{equation}
%    \label{eq:dGamma-dOmega-JLAB}
%    \frac{d\Gamma}{d\Omega}\propto  1 -\alpha\, P_\Lambda\cos{\theta}.
%\end{equation}

%Figure~\ref{fig:Lambda_polarization} compares Eq.~(\ref{eq:dGamma-dOmega-JLAB}) with three representative data bins from measurements performed at JLab by the CLAS Collaboration \cite{McAleer:2002gaa}. The linear dependence predicted by Eq.~(\ref{eq:dGamma-dOmega-JLAB}) is clearly supported by the experimental results. The $\Lambda$ polarization $P_\Lambda$ is extracted from the slopes obtained through linear regression of the data. Using the updated value $\alpha = 0.746$ from the 2025 Particle Data Group (PDG) estimate~\cite{ParticleDataGroup:2024cfk}, the extracted $\Lambda$ polarizations for the three bins are $P_\Lambda = 0.301$, $0.357$, and $0.362$, respectively.

%%%%%%%FIGURE 1%%%%%
%%%%%%%%%%%%%%%%%%%%%=================================================

%=== EXPERIMENTAL MEASUREMENTS ===================
%  \input{experimental_measurements}
\section{Experimental Measurements}
\label{expt-measurements}

Studies of kaon photo- and electroproduction on both proton and quasi-free neutron targets now stretch back over many decades. The first photoproduction datasets were collected starting in the late 1950s and the first electroproduction datasets were collected starting in the early 1970s. With each passing decade the experimental facilities and detector systems evolved, enabling deeper and deeper insights into processes involving the electromagnetic production of strange quarks.

This section provides a brief tour along this historical path to provide background into how the experimental situation advanced and developed. The data collected in the processes $\gamma N \to K Y$ and $\gamma_v N \to K Y$ provide important opportunities for understanding the spectroscopy of excited nucleon and hyperon states, allow access to unravel the structure of these baryons including configurations beyond conventional $qqq$ states, furnish a laboratory to study the $YN$ and $YNN$ interactions through spectroscopy of hypernuclei, and enable studies of the form factor of the associated kaon produced in the reaction. Together these areas of strangeness study represent an important avenue with which to understand strongly interacting systems in a manner complementary to studies involving only light $u$ and $d$ quark baryons and mesons to gain further insight into QCD. Studies of photo- and electroproduction of strangeness are more important today than ever as precision data now become available to provide detailed constraints and checks of theory. Several surveys of the experimental landscape have been published recently that are relevant for further, complementary perspectives~\cite{Ireland:2019uwn,Crede:2024hur,Gross:2022hyw}.

This section is organized as follows: Section~\ref{early-meas} provides an overview of the first introduction of the notion of ``strange'' particles and their initial experimental investigations. The period of early measurements here is defined to span from the late 1950s through the 1980s. Sections~\ref{sec:elsa} -- \ref{sec:jlab6gev} provide an overview of the second generation studies of the photo- and electroproduction measurements of $KY$ final states in facilities around the world. This period extends from the 1990s to the 2010s. The present day measurements from the last decade of the electromagnetic production of strangeness are mainly concentrated at JLab and are detailed in Section~\ref{jlab:12gev}. Finally, Section~\ref{jlab:22gev} looks to the future of a possible further energy-upgraded JLab facility.
    
\subsection{Early Measurements}
\label{early-meas}

The history of strangeness physics effectively began in the 1940s with the experimental observation of the $K^+$ meson in experiments detecting cosmic rays~\cite{Rochester:1947mi}. In the mid-1950s the field advanced still further with initial evidence for detection of the $V_1^0$ ($\Lambda$), $V^+$ ($\Sigma^+$), $V^-$ ($\Sigma^-$), and $V^0$ ($\Sigma^0$) \cite{PhysRev.80.1099,PhysRev.90.167,Armenteros01061952,Anderson1953,PhysRev.98.1407}. The curious experimental observation of particles that were copiously produced in beam-target collisions yet had long decay lengths given their mass had yet to be understood. A significant theoretical advance was put forward in the mid-1950s by Gell-Mann and Pais that introduced the concept of ``strangeness'' \cite{Gell-Mann:1955ipe}. The first tentative theoretical understanding of particles created ``strongly'' that decayed ``weakly'' set the stage for the rapid experimental developments from the late 1950s through the 1980s. This period before the advent of advanced, next-generation accelerator facilities with high-luminosity operations that arrived in the 1990s are set apart here under this section termed ``early measurements''. An informative seminar by Venanzoni provides a much more complete narrative of this fascinating early measurement period in the history of nuclear physics \cite{venanzoni24}. 

In many ways the development of particle accelerators, detector technologies, experimental setups, high-speed electronics, data acquisition systems, and analysis tools were groundbreaking achievements that truly represent tour-de-force efforts of many scientists, engineers, and technicians at facilities around the world. In this review section a full and complete history cannot be attempted. However, a brief survey of the timeline and key results has been provided. Several of the experimental layouts are featured to give an indication of the tools developed by these pioneering initiatives. Finally, several selected measurements from this early period are featured and highlighted. The first painstaking experiments of photo- and electroproduction of strangeness provided datasets amounting to roughly 1000 $K^+\Lambda$ and $K^+\Sigma^0$ events with sizable bin widths in the key kinematic variables. Today's most recent experimental results are based on datasets 1000 times larger than the first efforts. However, only from the humble beginnings of the programs in the 1950s have come the precision experiments possible today and the exciting plans that are being considered for the future.

An attempt to capture the developments of the emergent field of strangeness photo- and electroproduction experiments from their earliest days is provided in Tables~\ref{early-photo} and \ref{early-electro}, respectively. Here, the key early measurements up through the 1980s are provided. Some notable milestones along this path include

\vskip 0.5cm

\noindent
\underline{Photoproduction -- Selected Key Results Timeline}

\begin{itemize}
\item 1958 -- First $K^+\Lambda$ and $K^+\Sigma^0$ differential cross sections at the Cornell Synchrotron
\item 1960 -- First $\Lambda$ recoil polarization measurement at Cornell Synchrotron
\item 1969 -- First studies of $K^*Y$ and $KY^*$ production cross sections at DESY Synchrotron
\item 1978 -- First measurements of polarized target spin asymmetry for $K^+\Lambda$ at Bonn Synchrotron
\end{itemize}

\noindent
\underline{Electroproduction -- Selected Key Results Timeline}

\begin{itemize}
\item 1972 -- First $K^+\Lambda$ and $K^+\Sigma^0$ differential cross sections at Cambridge Electron Accelerator
\item 1974 -- First extraction of interference structure functions $\sigma_{\rm LT}$ and $\sigma_{\rm TT}$ at Cornell Synchrotron
\item 1975 -- First studies of $KY^*$ production cross sections at DESY Synchrotron
\item 1977 -- First studies of $\sigma_\mathrm{L}$ and $\sigma_\mathrm{T}$ for $K^+\Lambda$ and $K^+\Sigma^0$ at Cornell Synchrotron
\end{itemize}

The experimental equipment developed for these investigations included magnetic spectrometers and tracking detectors for momentum analysis of charged tracks, bubble chambers for precise event reconstruction, scintillator hodoscopes for energy loss and timing measurements, Cherenkov counters for charged particle identification, and calorimeters for electron and neutral particle detection. The 1950s also saw the development of the first liquid-hydrogen cryotargets for nuclear and particle physics experiments~\cite{Mulholland:2004pt}.

\begin{figure*}[htbp]
\centering
\includegraphics[width=0.85\textwidth]{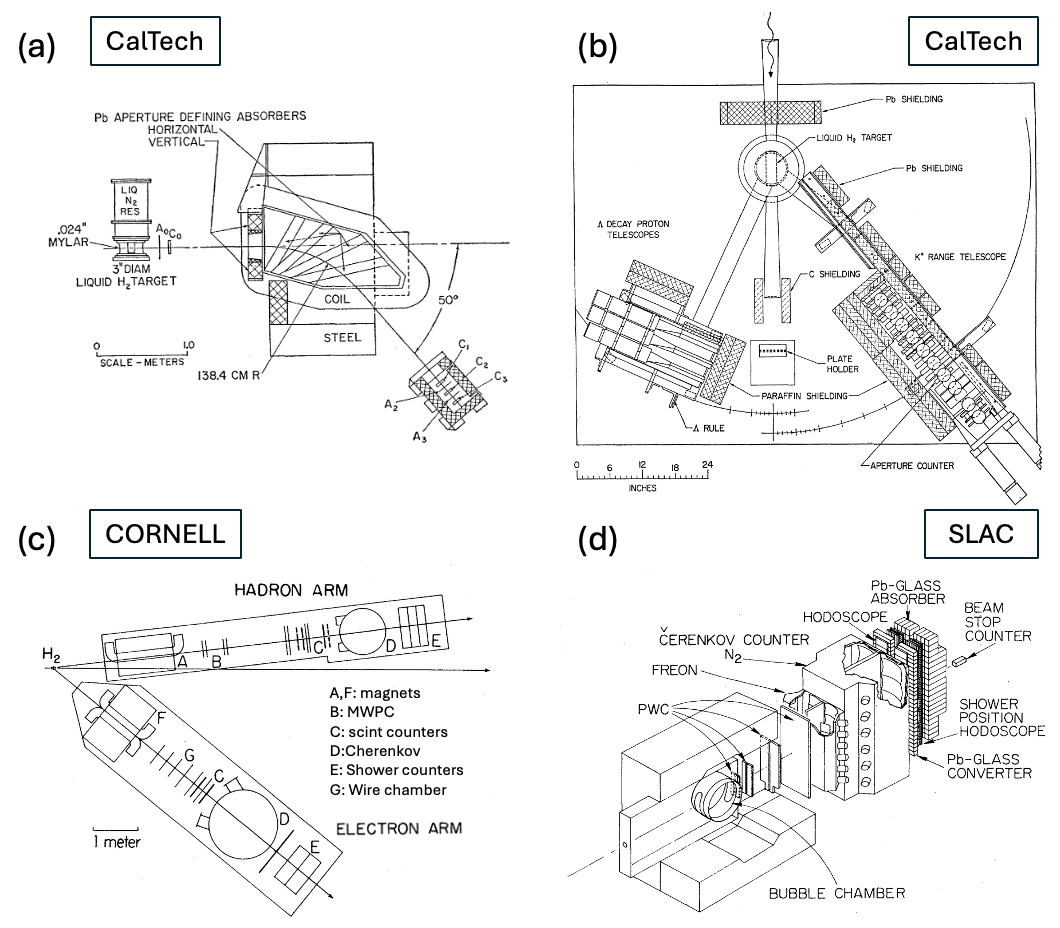}
\caption{Schematics of representative experimental setups for strangeness physics studies in the period from the 1950s to the 1980s. (a) Spectrometer for the study of kaon photoproduction from the CalTech 1.1-GeV electron synchrotron (1958) - Figure from Ref.~\cite{Donoho58}. (b) Two arm telescopes for $K^+Y$ photoproduction studies from the CalTech 1.1-GeV electron synchrotron (1967) - Figure from Ref.~\cite{Groom:1967zz}. (c) Two arm spectrometer system used for $e'K^+$ detection in electroproduction at the Cornell electron synchrotron (1977) - Figure adapted from Ref.~\cite{Bebek:1977bv}. (d) Bubble chamber system at the SLAC electron accelerator hybrid facility (1985) - Figure from Ref.~\cite{Abe:1985nbw}.}
\label{early-setups}
\end{figure*}

To capture a sense of the experimental setups across the decades of this period, Fig.~\ref{early-setups} shows several examples: CalTech electron synchrotron (1950s) \cite{Donoho58}, CalTech electron synchrotron (1960s) \cite{Groom:1967zz}, Cornell electron synchrotron (1970s) \cite{Bebek:1977bv}, and SLAC linear accelerator (1980s) \cite{Abe:1985nbw}. The photon fluxes produced by the bremsstrahlung technique or later by laser backscattering rapidly increased in intensity during this period. However, even at the peak of development, the practical photon beam intensity on target was still at least 3 to 4 orders of magnitude below what is typical in present-day experiments. For the electron beam experiments during this period, the beam-target luminosities were no more than $\sim$$10^{30}$~cm$^{-2}$s$^{-1}$, again many orders of magnitude below what is typical in today's experiments.

\begin{figure*}[htbp]
\centering
\includegraphics[width=0.8\textwidth]{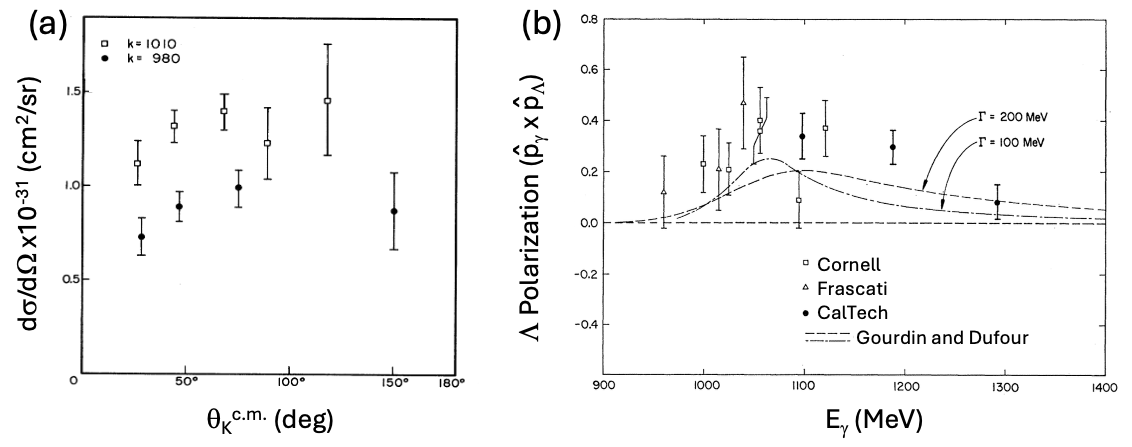}
\caption{Representative key early results from $KY$ photoproduction studies. (a) Differential cross section from Cornell (1958) for $K^+\Lambda$ at $E_\gamma^{\rm lab}$ = 980~MeV and 1010~MeV as a function of $\theta_K^{\mathrm{c.m.}}$. Figure from Ref.~\cite{PhysRevLett.1.109}. (b) $\Lambda$ recoil polarization $P$ for exclusive $K^+\Lambda$ as a function of $E_\gamma^{\rm lab}$ for $\theta_K^{\mathrm{c.m.}} \approx 0^\circ$ from Cornell (1963) (open squares)~\cite{Thom:1963zz}, Frascati (1964) (open triangles)~\cite{Borgia:1964mza}, and CalTech (1967) (solid circles)~\cite{Groom:1967zz}. The curves represent predictions from early development of an isobar model constrained by the available data~\cite{Gourdin1963}. Figure adapted from Ref.~\cite{Groom:1967zz}.}
\label{early-gp-data}
\end{figure*}

\begin{figure*}[htbp]
\centering
\includegraphics[width=0.95\textwidth]{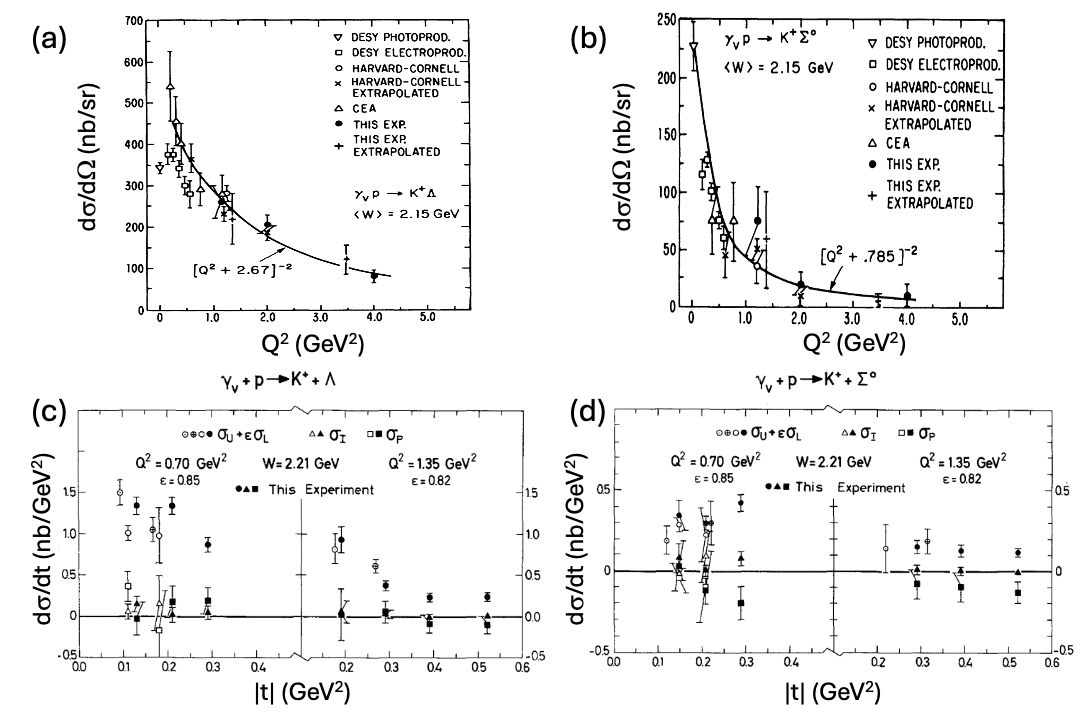}
\caption{Representative key early results from $KY$ electroproduction studies. Differential cross sections for (a) $K^+\Lambda$ and (b) $K^+\Sigma^0$ vs. $Q^2$ for data at an average $W$ of 2.15~GeV selected for $\theta_K^{\mathrm{c.m.}} < 25^\circ$. The solid line is a dipole fit to the available data. $\cos \Phi$ moment analysis of differential cross sections for (c) $K^+\Lambda$ and (d) $K^+\Sigma^0$ to separate $\sigma_U$ ($=\sigma_{\rm T} + \epsilon \sigma_{\rm L}$), $\sigma_{\rm LT}$ ($\sigma_{\rm I}$), and $\sigma_{\rm TT}$ ($\sigma_{\rm P}$) vs. four-momentum transfer squared $\vert t \vert$ ($\vert t \vert = (p_{\gamma_v} - p_{K^+})^2)$. The data from Cornell (1977) are shown for average $Q^2$ values of 0.70~GeV$^2$ and 1.35~GeV$^2$ at an average $W$ of 2.21~GeV. Figures from Ref.~\cite{Bebek:1976qg}.}
\label{early-ep-data}
\end{figure*}

\begin{figure*}[htbp]
\centering
\includegraphics[width=0.45\textwidth]{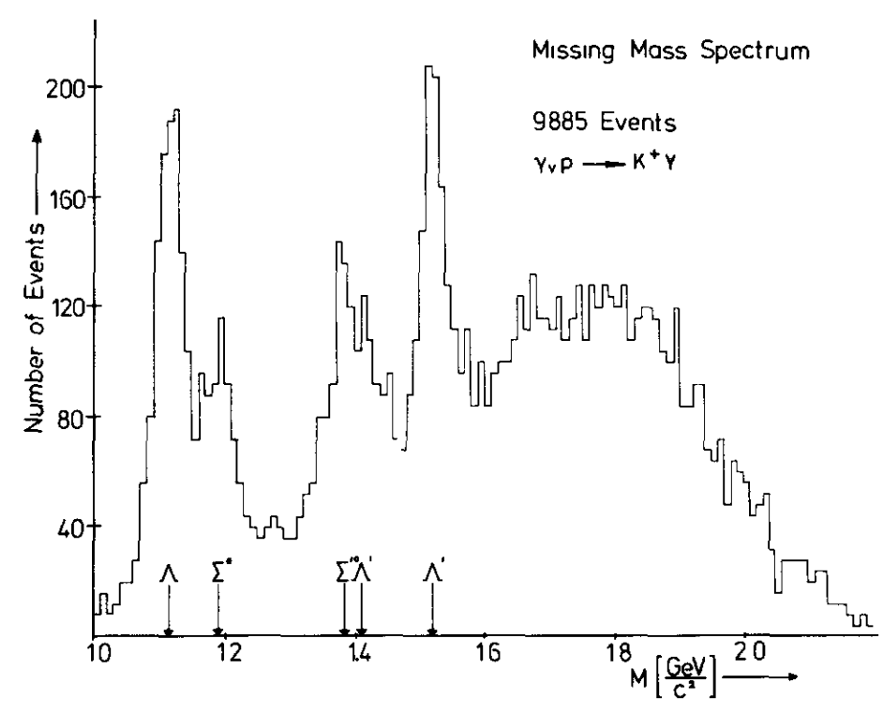}
\caption{Missing mass $MM(e'K^+)$ spectrum from an early electroproduction experiment showing contributions from the ground state $\Lambda$ and $\Sigma^0$, as well as the hyperon excited states $\Sigma(1385)$, $\Lambda(1405)$, and $\Lambda(1520)$. This spectrum from DESY (1975) shows the total dataset for final analysis with 9885 events. Figure from Ref.~\cite{Azemoon:1974dt}. Reprinted with permission from Elsevier.}
\label{early-ep-mm}
\end{figure*}

As the accelerator capabilities, experimental equipment, and analysis techniques advanced over the period from the 1950s to 1980s, so too did the quality and precision of the experimental observables. Several representative results of this evolution are shown in Fig.~\ref{early-gp-data} for photoproduction experiments and in Figs.~\ref{early-ep-data} and \ref{early-ep-mm} for electroproduction. The key aspects of these studies were at first to quantify the production levels of the strange particles and then to understand their production mechanism, both for ground and excited state hyperons. Ultimately, connections were drawn from the structures seen in the photo- and electroproduction energy spectra in the region below invariant mass $W \approx 2.5$~GeV to the nucleon excited states known from production with pion beams.

%%%%%%%%%%%%%%%%%%%%%%%%%%%%%%%%%%%%%%%%%%%%%%%%%%%%%%%%%%%%%%%%%%%%%%%%%%%%%%%%%%%%%%%%%%%%%%%%%%%%%%%%%%%%%%%%%%%%%%%%%%%%%%%%%%%%%%%%%%%%%%%%%%%%%%%%%%%%%%%
\begin{table}[tbh]
\setlength{\tabcolsep}{6pt} % Default value: 6pt
\renewcommand{\arraystretch}{0.8} % Default value: 1
\begin{center}
\caption{Summary of the key strangeness photoproduction measurements from experiments in the early measurement period from the 1950s-1980s. The column labeled $N_{\mathrm{bin}}$ indicates the number of kinematic bins included in the analysis.}
\begin{tabular}{cccccccc} \hline \hline
Observable                  & Final State(s)                        & $W$ (GeV)     & $\cos \theta_K^{\mathrm{c.m.}}$ & $N_{\mathrm{bin}}$              & Facility  & Year & Ref. \\ \hline
\multirow{15}{*}{$d\sigma$} & $K^+\Lambda$, $K^+\Sigma^0$           & $[1.62:1.78]$ & $[-0.9:0.9]$           & 14 $\Lambda$, 1 $\Sigma^0$               & Cornell   & 1958 & \cite{PhysRevLett.1.109} \\
                            & $K^+\Lambda$, $K^+\Sigma^0$           & $[1.62:1.72]$ & $[-0.7:0.9]$           & 31 $\Lambda$, 3 $\Sigma^0$               & Cornell   & 1959 & \cite{McDaniel:1959zz} \\ 
                            & $K^+\Lambda$, $K^+\Sigma^0$           & $[1.62:1.75]$ & $[-0.7:0.9]$           & 25 $\Lambda$, 8 $\Sigma^0$               & Cornell   & 1962 & \cite{Anderson:1962za} \\ 
                            & $K^+\Lambda$                          & $[1.64:1.69]$ & $[-0.35:0.74]$         & 14                                       & CalTech   & 1958 & \cite{Donoho58} \\
                            & $K^+\Lambda$                          & 1.8           & $[0.1:1.0]$            & 12                                       & CalTech   & 1964 & \cite{Peck:1964zz} \\ 
                            & $K^+\Lambda$                          & $[1.72:1.82]$ & 0.0                    & 3                                        & CalTech   & 1967 & \cite{Groom:1967zz} \\ 
                            & $K^+\Lambda$, $K^+\Sigma^0$           & $[1.6:3.2]$   & $[-1.0:1.0]$           & 21 $\Lambda$, 12 $\Sigma^0$              & DESY      & 1967 & \cite{ABBHHM:1967maf} \\
                            & $K^+\Lambda$, $K^+\Sigma^0$, $K\pi Y$ & $[1.62:3.2]$  & $[-1.0:1.0]$           & 20 $\Lambda$, 16 $\Sigma^0$, 4 $\pi Y^*$ & DESY      & 1969 & \cite{ABBHHM:1969pjo} \\
                            & $K^+\Lambda$ $K^+\Sigma^0$            & $[3.2:5.6]$   & $[0.0:1.0]$            & 50 $\Lambda$, 50 $\Sigma^0$              & SLAC      & 1969 & \cite{Boyarski:1969iy} \\ 
                            & $K^+\Lambda$ $K^+\Sigma^0$, $KY^*$    & 4.64          & Forward                & 6 $\Lambda$, 6 $\Sigma^0$, 6 $Y^*$       & SLAC      & 1971 & \cite{Boyarski:1970yc} \\
                            & $K^+\Lambda$, $K^+\Sigma^0$           & 2.9, 3.5      & $[-0.3:0.7]$           & 12 $\Lambda$, 11 $\Sigma^0$              & SLAC      & 1976 & \cite{Anderson:1976ph} \\
                            & $K^0_S X$, $\Lambda X$                & 6.2           & $[-1.0:1.0]$           & 19 $K^0_S$, 16 $\Lambda$                 & SLAC      & 1984 & \cite{SLACHFP:1983yyn} \\
                            & $K^+\Lambda$, $K^+\Sigma^0$           & $[1.8:1.9]$   & $[0.0:1.0]$            & 18 $\Lambda$, 12 $\Sigma^0$              & Bonn      & 1970 & \cite{Bleckmann:1970kb} \\
                            & $K^+\Lambda$, $K^+\Sigma^0$           & $[1.7:2.2]$   & 0.9                    & 10 $\Lambda$, 7 $\Sigma^0$               & Bonn      & 1972 & \cite{Feller:1972ph} \\
                            & $K^+\Lambda^*$                        & $[2.8:4.8]$   & Forward                & 10                                       & Daresbury & 1980 & \cite{Barber:1980zv} \\ \hline
\multirow{6}{*}{$P$}        & \multirow{6}{*}{$K^+\Lambda$}         & 1.66          & 0.85                   & 1                                        & Cornell   & 1960 & \cite{McDaniel:1960} \\
                            &                                       & $[1.66:1.73]$ & 0.0                    & 6                                        & Cornell   & 1963 & \cite{Thom:1963zz} \\ 
                            &                                       & $[1.63:1.69]$ & $[-0.1:0.1]$           & 3                                        & Frascati  & 1964 & \cite{Borgia:1964mza} \\ 
                            &                                       & $[1.63:1.69]$ & 0.0, 0.5               & 4                                        & Frascati  & 1965 & \cite{Grilli:1965jia} \\ 
                            &                                       & $[1.72:1.82]$ & 0.0                    & 3                                        & CalTech   & 1967 & \cite{Groom:1967zz} \\
                            &                                       & $[1.82:2.02]$ & 0.0, 0.77              & 4                                        & Bonn      & 1978 & \cite{Haas:1978qv} \\ \hline 
$T$                         & $K^+\Lambda$                          & $[1.72:1.82]$ & 0.0                    & 3                                        & Bonn      & 1978 & \cite{Althoff:1978qw} \\ \hline\hline 
\end{tabular}
\label{early-photo}
\end{center}
\end{table}
%%%%%%%%%%%%%%%%%%%%%%%%%%%%%%%%%%%%%%%%%%%%%%%%%%%%%%%%%%%%%%%%%%%%%%%%%%%%%%%%%%%%%%%%%%%%%%%%%%%%%%%%%%%%%%%%%%%%%%%%%%%%%%%%%%%%%%%%%%%%%%%%%%%%%%%%%%%%%%%%

%%%%%%%%%%%%%%%%%%%%%%%%%%%%%%%%%%%%%%%%%%%%%%%%%%%%%%%%%%%%%%%%%%%%%%%%%%%%%%%%%%%%%%%%%%%%%%%%%%%%%%%%%%%%%%%%%%%%%%%%%%%%%%%%%%%%%%%%%%%%%%%%%%%%%%%%%%%%%%%
\begin{table}[tbh]
\setlength{\tabcolsep}{6pt} % Default value: 6pt
\renewcommand{\arraystretch}{0.8} % Default value: 1
\begin{center}
\caption{Summary of the key strangeness electroproduction measurements from experiments in the early measurement period from the 1950s-1980s. The column labeled $N_{\mathrm{bin}}$ indicates the number of kinematic bins included in the analysis.}
\begin{tabular}{ccccccccc} \hline\hline
Observables                          & Final States(s)             & $Q^2$ (GeV$^2$) & $W$ (GeV)    & $\cos \theta_K^{\mathrm{c.m.}}$ & $N_{\mathrm{bin}}$                   & Facility  & Year & Ref. \\ \hline
$d\sigma$                            & $K^+\Lambda$, $K^+\Sigma^0$ & $[0.18:1.2]$    & $[1.85:2.6]$ & $[0.88:1.0]$           & 15 $\Lambda$, 15 $\Sigma^0$ & Cambridge & 1972 & \cite{Brown:1972pf} \\ \hline
                                     & $K^+\Lambda$, $K^+\Sigma^0$ & $[0.62:2.0]$    & $[2.2:2.7]$  & $[0.97:1.0]$           & 4  $\Lambda$, 4 $\Sigma^0$  & Cornell   & 1974 & \cite{Bebek:1974bt} \\
$\sigma_{\rm U}$, $\sigma_{\rm LT}$, & $K^+\Lambda$, $K^+\Sigma^0$ & $[1.2:4.0]$     & $[2.15:3.1]$ & $[0.87:1.0]$           & 11 $\Lambda$, 11 $\Sigma^0$ & Cornell   & 1977 & \cite{Bebek:1976qg} \\ 
$\sigma_{\rm TT}$                    & $K^+\Sigma^-$ ($n$ target)  & $[1.2:4.0]$     & $[2.15:3.1]$ & $[0.87:1.0]$           & 6                           & Cornell   & 1977 & \cite{Bebek:1976qg} \\ 
                                     & $K^+\Lambda$, $K^+\Sigma^0$ & $[0.1:0.6]$     & $[1.9:2.8]$  & Forward                & 41 $\Lambda$, 39 $\Sigma^0$ & DESY      & 1975 & \cite{Azemoon:1974dt} \\
                                     & $KY^*$                      & $[0.1:0.6]$     & $[1.9:2.8]$  & Forward                & 31                          & DESY      & 1975 & \cite{Azemoon:1974dt} \\
                                     & $K^+\Lambda$, $K^+\Sigma^0$ & $[0.06:1.35]$   & $[1.9:2.5]$  & Forward                & 27 $\Lambda$, 26 $\Sigma^0$ & DESY      & 1979 & \cite{Brauel:1979zk} \\ \hline
$\sigma_{\rm L}/\sigma_{\rm T}$      & $K^+\Lambda$, $K^+\Sigma^0$ & $[1.2:3.3]$     & $[2.15:3.1]$ & $[0.87:1.0]$           & 6  $\Lambda$, 6 $\Sigma^0$  & Cornell   & 1977 & \cite{Bebek:1977bv} \\ \hline\hline
\end{tabular}
\label{early-electro}
\end{center}
\end{table}
%%%%%%%%%%%%%%%%%%%%%%%%%%%%%%%%%%%%%%%%%%%%%%%%%%%%%%%%%%%%%%%%%%%%%%%%%%%%%%%%%%%%%%%%%%%%%%%%%%%%%%%%%%%%%%%%%%%%%%%%%%%%%%%%%%%%%%%%%%%%%%%%%%%%%%%%%%%%%%%%

%%%%%%%%%%%%%%%%%%%%%%%%%%%%%%%%%%%%%%%%%%%%%%%%%%%%%%%%%%%%%%%%%%%%%%%%%%%%%%%%%%%%%%%%%%%%%%%%%%%%%%%%%%%%%%%%%%%%%%%%%%%%%%%%%%%%%%%%%%%%%%%%%%%%%%%%%%%%%%%%
%%%%%%%%%%%%%%%%%%%%%%%%%%%%%%%%%%%%%%%%%%%%%%%%%%%%%%%%%%%%%%%%%%%%%%%%%%%%%%%%%%%%%%%%%%%%%%%%%%%%%%%%%%%%%%%%%%%%%%%%%%%%%%%%%%%%%%%%%%%%%%%%%%%%%%%%%%%%%%%%

%%%%%%%%%%%%%%%%%%%%%%%%%%%%%%%%%%%%%%%%%%%%%%%%%%%%%%%%%%%%%%%%%%%%%%%%%%%%%%%%%%%%%%%%%%%%%%%%%%%%%%%%%%%%%%%%%%%%%%%%%%%%%%%%%%%%%%%%%%%%%%%%%%%%%%%%%%%%%%%%
%%%%%%%%%%%%%%%%%%%%%%%%%%%%%%%%%%%%%%%%%%%%%%%%%%%%%%%%%%%%%%%%%%%%%%%%%%%%%%%%%%%%%%%%%%%%%%%%%%%%%%%%%%%%%%%%%%%%%%%%%%%%%%%%%%%%%%%%%%%%%%%%%%%%%%%%%%%%%%%%

%%%%%%%%%%%%%%%%%%%%%%%%%%%%%%%%%%%%%%%%%%%%%%%%%%%%%%%%%%%%%%%%%%%%%%%%%%%%%%%%%%%%%%%%%%%%%%%%%%%%%%%%%%%%%%%%%%%%%%%%%%%%%%%%%%%%%%%%%%%%%%%%%%%%%%%%%%%%%%%%
%%%%%%%%%%%%%%%%%%%%%%%%%%%%%%%%%%%%%%%%%%%%%%%%%%%%%%%%%%%%%%%%%%%%%%%%%%%%%%%%%%%%%%%%%%%%%%%%%%%%%%%%%%%%%%%%%%%%%%%%%%%%%%%%%%%%%%%%%%%%%%%%%%%%%%%%%%%%%%%%

%%%%%%%%%%%%%%%%%%%%%%%%%%%%%%%%%%%%%%%%%%%%%%%%%%%%%%%%%%%%%%%%%%%%%%%%%%%%%%%%%%%%%%%%%%%%%%%%%%%%%%%%%%%%%%%%%%%%%%%%%%%%%%%%%%%%%%%%%%%%%%%%%%%%%%%%%%%%%%%%
%%%%%%%%%%%%%%%%%%%%%%%%%%%%%%%%%%%%%%%%%%%%%%%%%%%%%%%%%%%%%%%%%%%%%%%%%%%%%%%%%%%%%%%%%%%%%%%%%%%%%%%%%%%%%%%%%%%%%%%%%%%%%%%%%%%%%%%%%%%%%%%%%%%%%%%%%%%%%%%%

\subsection{Experiments at ELSA}
\label{sec:elsa}

The Electron Stretcher Accelerator (ELSA) facility at the University of Bonn in Germany was constructed in the period from 1982 to 1987. The accelerator relies on a stretcher ring to convert the pulsed output of its synchrotron into a continuous electron beam with a 100\% duty factor. The facility began operations for physics experiments in 1989. ELSA consists of three stages: two injector linear accelerators (LINACs), a 0.5 to 1.6~GeV booster synchrotron, and a 0.5 to 3.5~GeV stretcher ring~\cite{Hillert:2006yb}. Either unpolarized or polarized electron beams can be delivered. Beams of real photons are produced via the bremsstrahlung process. Linearly polarized photon beams are available using coherent bremsstrahlung from a precisely oriented diamond radiator. A tagging facility detects the scattered electrons with a tagging hodoscope to precisely measure their energy and timing to reconstruct the photon energy at the experimental target.

Several detector systems have been incorporated at ELSA for hadron physics experiments, each based on large acceptance detector configurations: SAPHIR (1993--2005), CB-ELSA (1996--2015), and BGOOD (2016--2025). These facilities are further detailed in Sections~\ref{sec:saphir}, \ref{sec:cbelsa}, and \ref{sec:bgood}, respectively. The next generation facility for ELSA is the new INSIGHT experiment now under construction~\cite{Jude:2025bzt}. The current schedule plans for beam commissioning to commence in 2029 and operations to begin in 2030. Further details are included in Section~\ref{sec:insight}.

%%%%%%%%%%%%%%%%%%%%%%%%%%%%%%%%%%%%%%%%%%%%%%%%%%%%%%%%%%%%%%%%%%%%%%%%%%%%%%%%%%%%%%%%%%%%%%%%%%%%%%%%%%%%%%%%%%%%%%%%%%%%%%%%%%%%%%%%%%%%%%%%%%%%%%%%%%%%%%%%
\begin{figure*}[ht]
\centering
\includegraphics[width=0.95\textwidth]{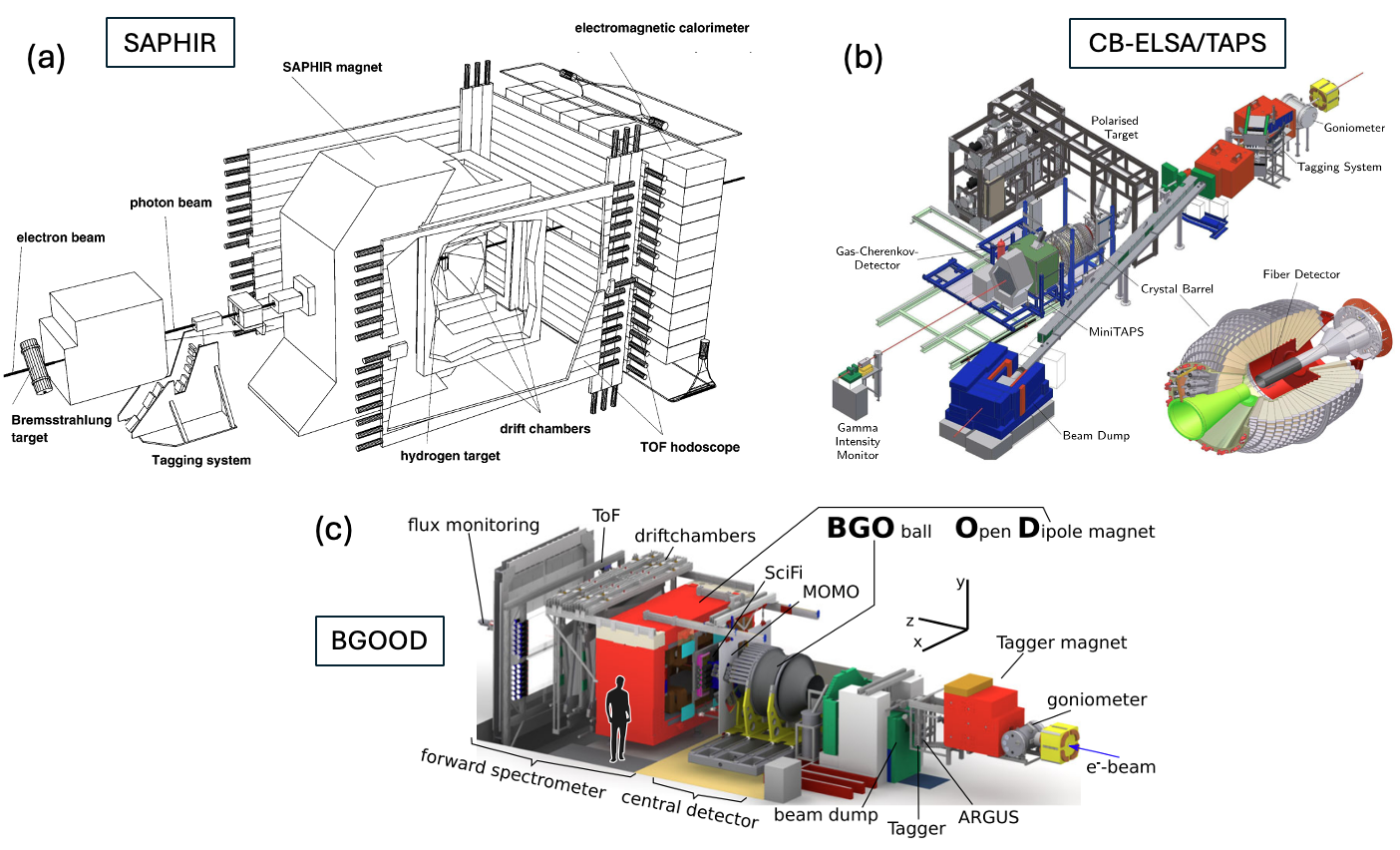}
\caption{Schematic drawings of the three large acceptance hadronic physics installations at ELSA: (a) SAPHIR -- showing the photon tagger, spectrometer dipole, tracking systems, time-of-flight, and forward calorimeter \cite{Glander:2003jw} (length along beamline $\sim$4~m), (b) CB-ELSA/TAPS -- showing the Crystal Barrel calorimeter (CB) and particle identification systems in the central detector region along with the TAPS calorimeter in the forward detector~\cite{TAPS:2022ijf} (length along beamline $\sim$2~m), (c) BGOOD -- showing the BGO ball calorimeter in the central detector region, and the open dipole magnet with charged particle tracking systems and time-of-flight systems in the forward direction~\cite{BGOOD:2019utx} (length along beamline $\sim$15~m). Figures used with kind permission of The European Physical Journal (EPJ).}
%Could replace TAPS:2022ijf with afzal2024
\label{elsa-layout}
\end{figure*}
%%%%%%%%%%%%%%%%%%%%%%%%%%%%%%%%%%%%%%%%%%%%%%%%%%%%%%%%%%%%%%%%%%%%%%%%%%%%%%%%%%%%%%%%%%%%%%%%%%%%%%%%%%%%%%%%%%%%%%%%%%%%%%%%%%%%%%%%%%%%%%%%%%%%%%%%%%%%%%%%
 
\subsubsection{SAPHIR}
\label{sec:saphir}

The SAPHIR (Spectrometer Arrangement for Photon Induced Reactions) detector~\cite{Schwille:1994vg} was a large acceptance spectrometer built for the study of photon-induced reactions on nucleons and light nuclei. It was designed to detect multi-particle final states using photon beams from 0.87 to 2.65~GeV with a tagged photon rate up to $5 \times 10^6$~$\gamma$/s. The spectrometer (shown in Fig.~\ref{elsa-layout}(a)) was based on a 0.8~T dipole magnet. The photon beam was incident upon the target located in the magnet gap through an opening in the dipole yoke. The region around the target was instrumented with a central drift chamber tracker. This tracker was augmented with additional planar drift chambers on the sides of the magnet and in the forward/beam direction. The particle identification system included three planes of scintillator hodoscopes behind the planar chambers and a downstream electromagnetic calorimeter. The detector spanned polar angles from 1$^\circ$ to 155$^\circ$ and the full azimuthal angle range.

The strangeness program at SAPHIR (detailed in Table~\ref{modern-nonjlab-photo}) consisted of measurements of differential cross sections and hyperon recoil polarization measurements for exclusive $K^+\Lambda$ and $K^+\Sigma^0$ photoproduction reactions on an unpolarized proton target~\cite{Glander:2003uf}. These data are included in the comparative plots in Section~\ref{clas-gp-program} with the differential cross section comparisons in Figs.~\ref{clas-kl-dcs} and \ref{clas-ks-dcs} and the recoil hyperon polarizations in Figs.~\ref{clas-kl-p} and \ref{clas-ks-p} for $K^+\Lambda$ and $K^+\Sigma^0$, respectively.

%%%%%%%%%%%%%%%%%%%%%%%%%%%%%%%%%%%%%%%%%%%%%%%%%%%%%%%%%%%%%%%%%%%%%%%%%%%%%%%%%%%%%%%%%%%%%%%%%%%%%%%%%%%%%%%%%%%%%%%%%%%%%%%%%%%%%%%%%%%%%%%%%%%%%%%%%%%%%%%%
\begin{figure*}[htbp]
\centering
\includegraphics[width=0.8\textwidth]{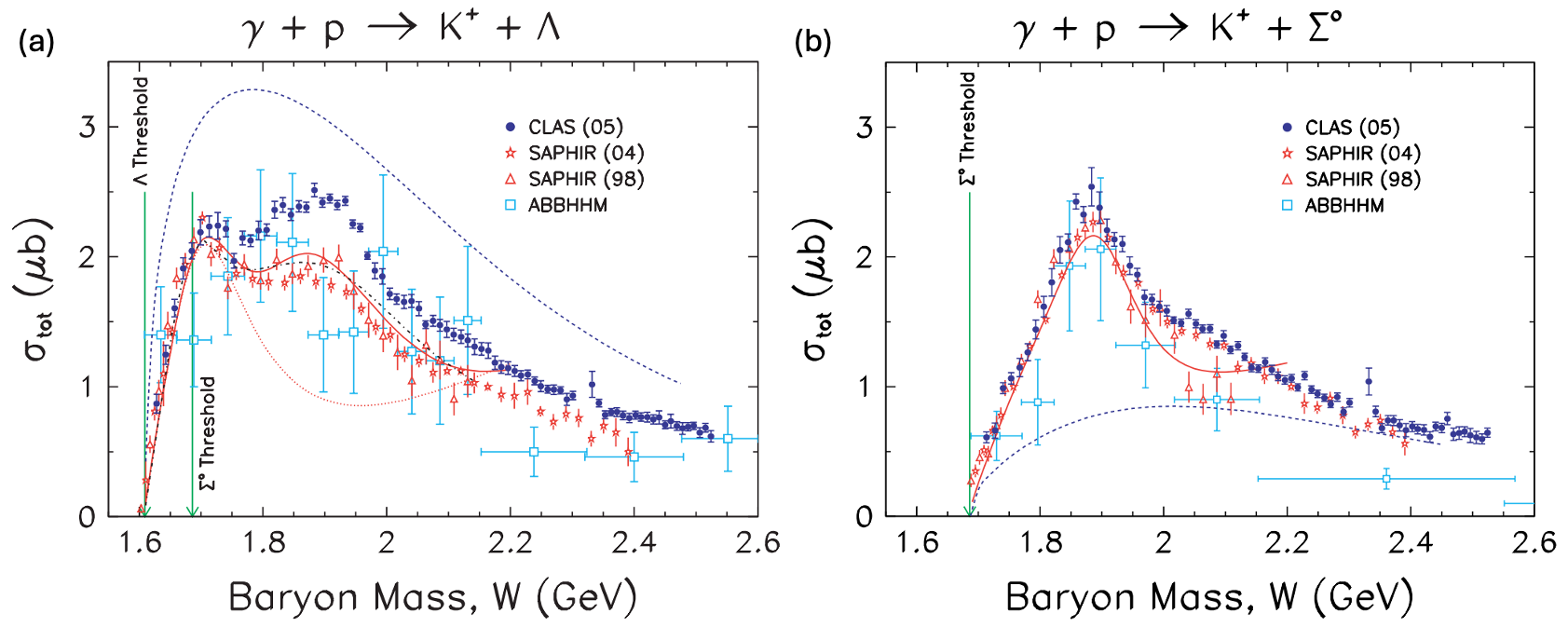}
\caption{Total cross section for (a) $\gamma p \to K^+ \Lambda$ and (b) $\gamma p \to K^+ \Sigma^0$ from SAPHIR (red stars, red triangles)~\cite{Glander:2003jw,SAPHIR:1998fev}, CLAS (blue circles)~\cite{CLAS:2005lui}, and the ABBHHM Collaboration (light blue squares) using data collected at DESY~\cite{ABBHHM:1969pjo}. The curves are from the isobar models Kaon-MAID (solid red, dotted red)~\cite{Mart:1999ed} and Saclay-Lyon (dot-dashed black) \cite{David:1995pi}, as well as from a pure Regge model (dashed blue)~\cite{Guidal:1999qi}. Figures from Ref.~\cite{CLAS:2005lui}.}
\label{ky-sigtot}
\end{figure*}
%%%%%%%%%%%%%%%%%%%%%%%%%%%%%%%%%%%%%%%%%%%%%%%%%%%%%%%%%%%%%%%%%%%%%%%%%%%%%%%%%%%%%%%%%%%%%%%%%%%%%%%%%%%%%%%%%%%%%%%%%%%%%%%%%%%%%%%%%%%%%%%%%%%%%%%%%%%%%%%%

A byproduct of the differential cross section analysis was the determination of the $K^+\Lambda$ and $K^+\Sigma^0$ total cross sections. Comparative plots are shown in Fig.~\ref{ky-sigtot}. The results for $K^+\Lambda$ show a prominent peak centered near 1.9~GeV that is reflective of the expected contribution from $s$-channel nucleon resonances in this mass range. The peak near 1.7~GeV is consistent with contributions from the $N(1710)1/2^+$ and $N(1720)3/2^+$ resonances (notation giving the mass in MeV followed by the spin-parity $J^P$). In the case of $K^+\Sigma^0$, $\sigma_{\rm tot}$ shows a prominent peak centered at $\sim$1.9~GeV, along with a subtle shoulder at about 2.05~GeV. The location of the dominant peak is consistent with the contributions of several well-known $N^*$ and $\Delta^*$ resonances that may contribute to an isospin-3/2 final state. An explanation for the shoulder has not yet been explored within available models. These measurements were important at this time to demonstrate the sensitivity of the $KY$ channels to explorations of $N^*$ and $\Delta^*$ $s$-channel resonance contributions (generically referred to as ``$N^*$'' states). In subsequent years as other datasets became available, discrepancies were revealed among the different datasets outside of the assigned systematic uncertainties that proved problematic for constraining the reaction models that were being developed. Sections~\ref{sec:models}, \ref{sec:phenomenlogy}, and \ref{sec:unsettled} provide further details and discussion on the consistency issues among the $\gamma p$ datasets.

\subsubsection{CB--ELSA}
\label{sec:cbelsa}

The CB-ELSA facility was configured around the large acceptance Crystal Barrel calorimeter based on CsI(Tl) crystals that was originally used at the LEAR (Low Energy Antiproton Ring) experiment at CERN \cite{CrystalBarrel:1992qav}. The measurement program was based on the detection of multi-photon final states~\cite{Afzal:2020geq}. Measurements were carried out at typical photon fluxes of $1 - 2 \times 10^7$~$\gamma$/s on unpolarized liquid-hydrogen targets, as well as both longitudinally and transversely polarized butanol and deuterated-butanol targets with photon beams up to $\sim$2.3~GeV. 

The detector underwent several configuration changes during its lifetime to support its physics program. In its final configuration, known as CB-ELSA/TAPS, the detector was based on three calorimeter systems. The Crystal Barrel calorimeter in the central region, and in the forward region, the MiniTAPS BaF$_2$ crystal calorimeter that was originally part of the TAPS (Two Arm Photon Spectrometer) at MAMI~\cite{Novotny:1991ht} and the Forward Plug CsI(Tl) calorimeter about the beamline. Together, these detectors spanned the full azimuthal range and cover polar angles from 1$^\circ$ to 156$^\circ$. The forward angle crystals ($\lesssim 30^\circ$) were covered by plastic scintillators for charged particle identification capabilities. In addition, a multi-layer scintillation fiber detector was installed around the target to extend the charged particle identification capabilities of the detector~\cite{Hartmann:2016xtu}. Figure~\ref{elsa-layout}(b) shows the layout of the detector systems.

The output of the strangeness physics program is detailed in Table~\ref{modern-nonjlab-photo} and includes measurements of differential cross sections, as well as the linearly polarized beam and recoil hyperon single spin asymmetries $\Sigma$ and $P$ for $K^0_S \Sigma^+$ exclusive photoproduction from a proton target. These data represent some of the first in the $K^0\Sigma^+$ channel, an isospin partner to the $K^+\Sigma^0$ final state, which have proven important to better understand the associated reaction mechanism~\cite{CBELSATAPS:2011gly,CBELSATAPS:2014ibm}. Compared to $K^+\Sigma^0$ photoproduction, the study of the exclusive $K^0\Sigma^+$ channel has many fewer available measurements. From the result summaries in Table~\ref{modern-nonjlab-photo} and Table~\ref{hallb-clas-lista}, the ratio of the available cross section measurements of $K^+\Sigma^0/K^0\Sigma^+$ is $\sim$20 and the ratio of available polarization measurements is $\sim$4. Of course, these ratios are misleading without also appreciating that the measurements of the $K^+\Sigma^0$ channel require detection of only the $K^+$ and possibly the proton from the $\Sigma^0 \to \Lambda \gamma \to p \pi^- \gamma$ decay. For the $K^0\Sigma^+$ studies, the $K^0$ was detected via $K^0_S \to \pi^0\pi^0 \to 4\gamma$ and the $\Sigma^+$ is detected via $\Sigma^+ \to p \pi^0$. Therefore, when accounting for detector acceptance differences for the neutral and charged $\Sigma$ hyperon studies, the statistical quality and bin size choices are markedly superior for the $K^+\Sigma^0$ channel compared to the $K^0\Sigma^+$ channel.

However, $K^0$ photoproduction has some notable advantages in the study of $s$-channel nucleon resonance excitations compared to $K^+$ photoproduction that make these studies relevant for consideration. The main difference in the $K^0$ reaction process is that photons cannot directly couple to the charge of the meson. In this case, the $t$-channel kaon exchange diagram cannot contribute to the production. Since this process is an important part of the $K^+\Sigma^0$ reaction dynamics, the $K^0$ channel can provide a cleaner probe to access the $s$-channel resonance excitations. 

While the nominal $\gamma K^0 K^0$ $t$-channel coupling is absent, in reality, the $t$-channel process does not entirely vanish in $K^0$ photoproduction as photon coupling at the $\gamma K^0 K^{*0}$ vertex can contribute. This makes study of the $K^0\Sigma^+$ channel an important process to access explicit meson-baryon dynamics. If $K^*Y$ dynamics play a significant role in the reaction mechanism, then $K^*$ production may be different above and below the $K^{*0}$ threshold. Hints at such a signature have already been provided in the CB-ELSA/TAPS cross section data \cite{CBELSATAPS:2011gly}. Whether the effects seen in the data are associated with the formation of $K^*Y$ quasi-bound ``molecular''-type states still requires understanding from advanced reaction models.

The CB-ELSA/TAPS facility effectively ended operations by 2015 with analysis of the collected data extending for several years beyond that point. Attention at ELSA for hadron physics then shifted focus to the BGOOD facility detailed in Section~\ref{sec:bgood}.

\subsubsection{BGOOD}
\label{sec:bgood}

The BGOOD experiment was designed to study exclusive photoproduction processes detecting both charged and neutral particles in the final state~\cite{BGOOD:2019utx}. The focus of the program was nominally on $t$-channel processes at low momentum transfer~\cite{Jude:2025bzt}. The facility consisted of two main subsystems. The central detector contained a large acceptance BGO crystal calorimeter spanning polar angles from 10$^\circ$ to 155$^\circ$, which was formerly part of the GRAAL facility (see Section~\ref{sec:graal}). A segmented barrel scintillator and multiple layers of cylindrical wire chambers were positioned within the calorimeter to provide tracking and charged particle identification. In the forward direction, spanning polar angles from 2$^\circ$ to 12$^\circ$, was a large aperture magnetic spectrometer with a 0.5~T open dipole magnet. This system included tracking detectors before and after the magnet, with a scintillating fiber detector before the magnet. The final downstream element was a multi-layer scintillator hodoscope for precise timing information. The photon flux for operations with photon beam energies up to $\sim$3~GeV was up to $1 - 2 \times 10^7$~$\gamma$/s. Figure~\ref{elsa-layout}(c) shows a schematic of the facility.

\begin{figure*}[htbp]
\centering
\includegraphics[width=0.45\textwidth]{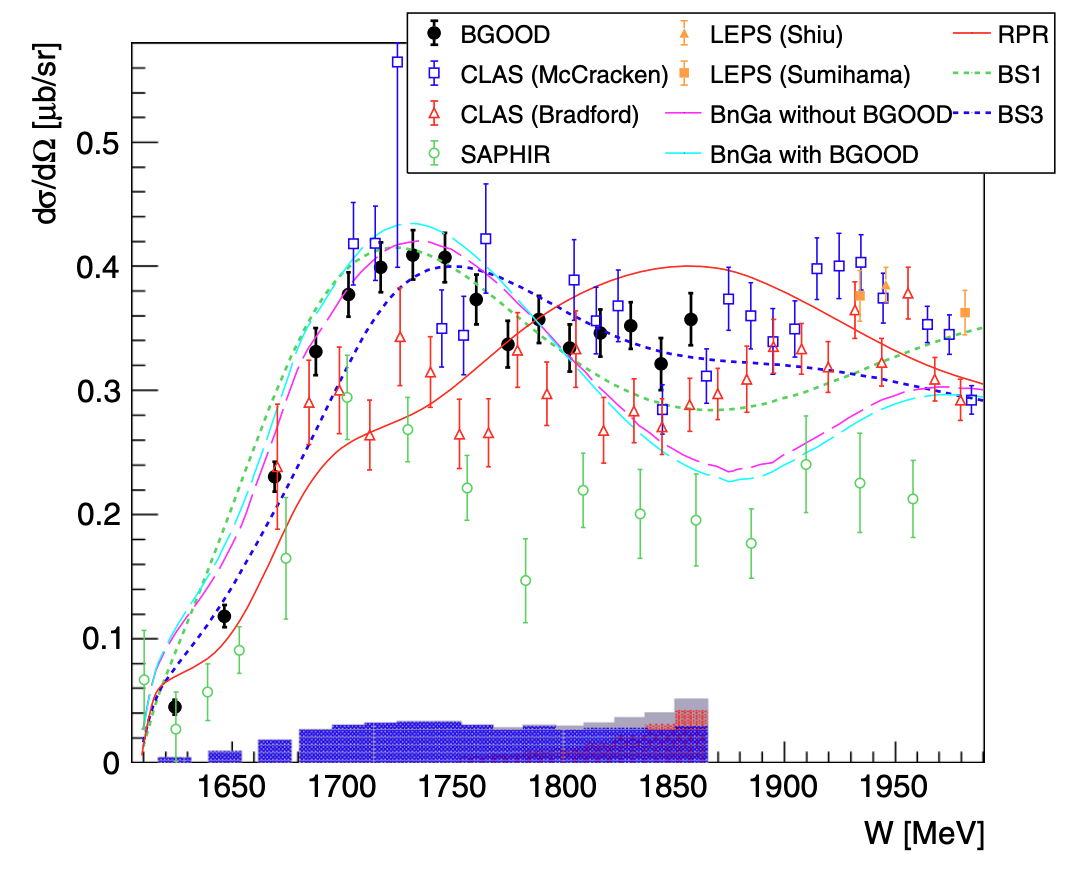}
\caption{Differential cross section of $\gamma p \to K^+ \Lambda$ for $\cos \theta_K^{\mathrm{c.m.}} > 0.90$ from BGOOD (black filled circles). Other data shown in the figure are from CLAS (red open triangles, blue open squares) \cite{CLAS:2005lui,CLAS:2009rdi}, SAPHIR (green open diamonds)~\cite{Glander:2003jw}, and LEPS (orange filled triangle, orange filled squares)~\cite{LEPS:2017pzl,LEPS:2005hji}. The Regge plus resonance model \cite{Bydzovsky:2019hgn} and isobar models BS1 and BS3~\cite{Skoupil:2016ast,Skoupil:2018vdh} of Skoupil and Byd\v{z}ovsk\'{y} are the solid red, dotted green, and dotted blue lines, respectively. The Bonn-Gatchina partial wave analysis~\cite{CBELSATAPS:2019ylw} solutions with and without the inclusion of the BGOOD data are shown by the dashed cyan and magenta lines, respectively. More details on the models are given in Section~\ref{sec:models}. Figure from Ref.~\cite{Alef:2020yul} used with kind permission of The European Physical Journal (EPJ).}
\label{bgood-dcs}
\end{figure*}

An important part of the BGOOD physics program was the study of strangeness photoproduction~\cite{Jude:2019qqd}. Table~\ref{modern-nonjlab-photo} highlights the program output to date, which includes forward-angle measurements up to $W = 1.9$~GeV of differential cross sections for $K^+\Lambda$ and $K^+\Sigma^0$~\cite{Alef:2020yul,Jude:2020byj}. Figure~\ref{bgood-dcs} highlights the kinematic coverage of the BGOOD detector for forward angle studies compared to other facilities. The program has also provided data for differential cross sections of $K^+\Lambda(1405)$ and $K^+\Lambda(1520)$. The studies of the $\Lambda(1405)$ hint at the important role of the $N(2030)5/2^+$ \cite{BGOOD:2021sog}. These data have also allowed for extended kinematic reach to forward angles to study the $\Lambda(1405)$ lineshape in its decay branch to $\Sigma^0\pi^0$. The measurements of the $\Lambda(1520)$ final state~\cite{Rosanowski:2024rww} are very near the reaction threshold and allow for a detailed characterization of the excited state hyperon production mechanism at low-momentum transfer.

Operations at the BGOOD facility effectively ended in 2025. At that point the BGO calorimeter was designated to be removed for reuse at the planned Beam Dump Experiment (BDX) at JLab~\cite{celentano:2022ehx}.

%%%%%%%%%%%%%%%%%%%%%%%%%%%%%%%%%%%%%%%%%%%%%%%%%%%%%%%%%%%%%%%%%%%%%%%%%%%%%%%%%%%%%%%%%%%%%%%%%%%%%%%%%%%%%%%%%%%%%%%%%%%%%%%%%%%%%%%%%%%%%%%%%%%%%%%%%%%%%%%%
%%%%%%%%%%%%%%%%%%%%%%%%%%%%%%%%%%%%%%%%%%%%%%%%%%%%%%%%%%%%%%%%%%%%%%%%%%%%%%%%%%%%%%%%%%%%%%%%%%%%%%%%%%%%%%%%%%%%%%%%%%%%%%%%%%%%%%%%%%%%%%%%%%%%%%%%%%%%%%%%

\subsubsection{INSIGHT}
\label{sec:insight}

\begin{figure*}[htbp]
\centering
\includegraphics[width=0.55\textwidth]{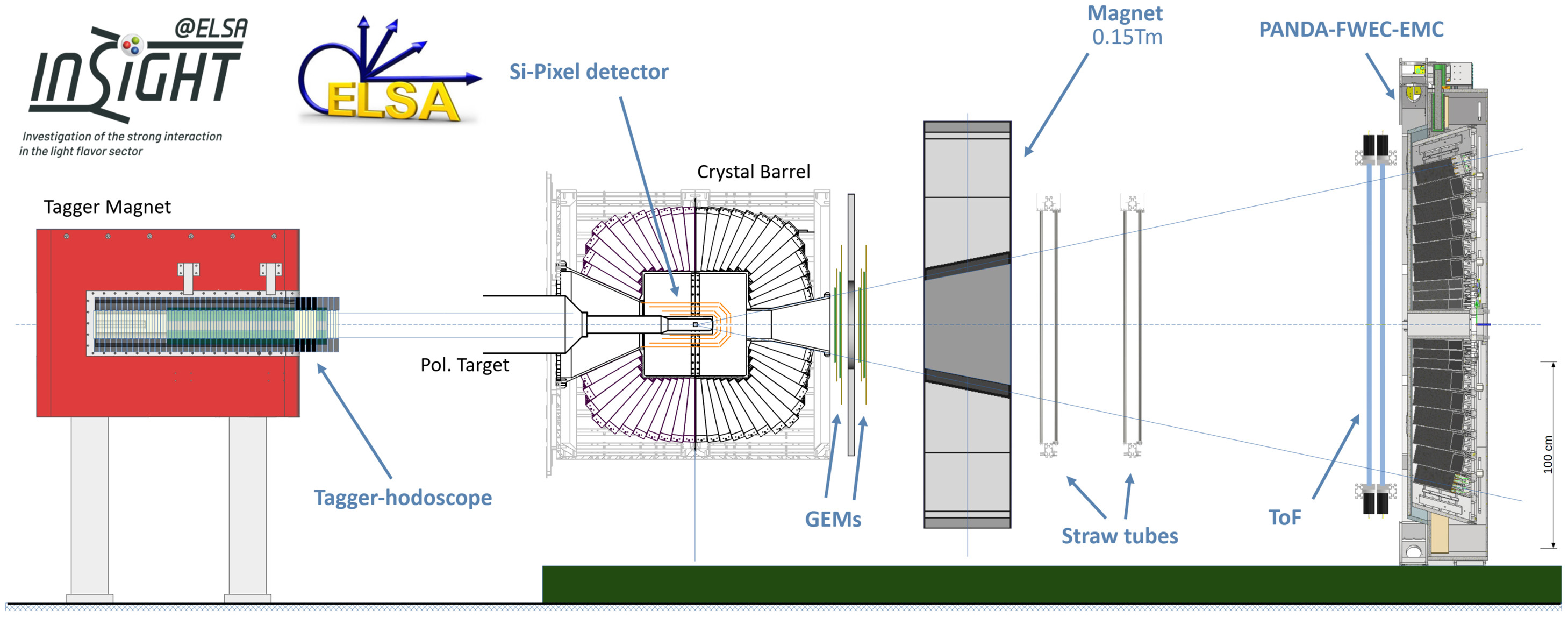}
\caption{Layout of the new INSIGHT facility at ELSA now under construction. Figure from Ref.~\cite{Jude:2025bzt}.}
\label{insight-det}
\end{figure*}

The future of hadron physics at ELSA will continue beyond SAPHIR, CB-ELSA/TAPS, and BGOOD with a major new initiative underway to construct the INSIGHT experiment \cite{Jude:2025bzt}. This new photoproduction facility is based on a detector that includes nearly complete $4\pi$ angular coverage for high resolution photon measurements and detection of charged particles. The facility will operate with both polarized beams and polarized targets. The planned layout will consist of a central detector that will reuse the CB-ELSA/TAPS calorimeter with a new charged particle pixel vertex detector. In the forward region, it will include a magnetic spectrometer with trackers, time-of-flight, and an electromagnetic calorimeter (see Fig.~\ref{insight-det}).

The developing physics program includes experiments on strangeness photoproduction. The initial focus will be on high-precision studies of $K\Lambda^*$ and $K\Sigma^*$. These final states can be used to extend the searches for exotic baryon configurations in the strangeness sector and to continue past and ongoing $s$-channel spectroscopic studies of the $N^*$ and $\Delta^*$ resonances that decay to $KY$~\cite{Jude:2025bzt}. The planned program will complement and extend the established hadron physics program at ELSA, as well as the programs at JLab, SPring-8, and MAMI.

%%%%%%%%%%%%%%%%%%%%%%%%%%%%%%%%%%%%%%%%%%%%%%%%%%%%%%%%%%%%%%%%%%%%%%%%%%%%%%%%%%%%%%%%%%%%%%%%%%%%%%%%%%%%%%%%%%%%%%%%%%%%%%%%%%%%%%%%%%%%%%%%%%%%%%%%%%%%%%%%
%%%%%%%%%%%%%%%%%%%%%%%%%%%%%%%%%%%%%%%%%%%%%%%%%%%%%%%%%%%%%%%%%%%%%%%%%%%%%%%%%%%%%%%%%%%%%%%%%%%%%%%%%%%%%%%%%%%%%%%%%%%%%%%%%%%%%%%%%%%%%%%%%%%%%%%%%%%%%%%%

\subsection{Experiments at ESRF}
\label{sec:graal}

The GRAAL (Gamma Ray Advanced Array for the Laser) facility was located at the European Synchrotron Radiation Facility (ESRF) in Grenoble, France and was in operation from 1996 to 2008~\cite{Bocquet:1997mx}. GRAAL produced circularly and linearly polarized photon beams by Compton backscattering of laser photons from the 6-GeV electrons circulated in the ESRF ring. A photon tagger measured the energy and timing of the scattered electrons. The photons arriving at the target were in the energy range from 0.4 to 1.5~GeV, and the tagged photon flux was $\approx 2 \times 10^6$~$\gamma$/s.

\begin{figure*}[htbp]
\centering
\includegraphics[width=0.55\textwidth]{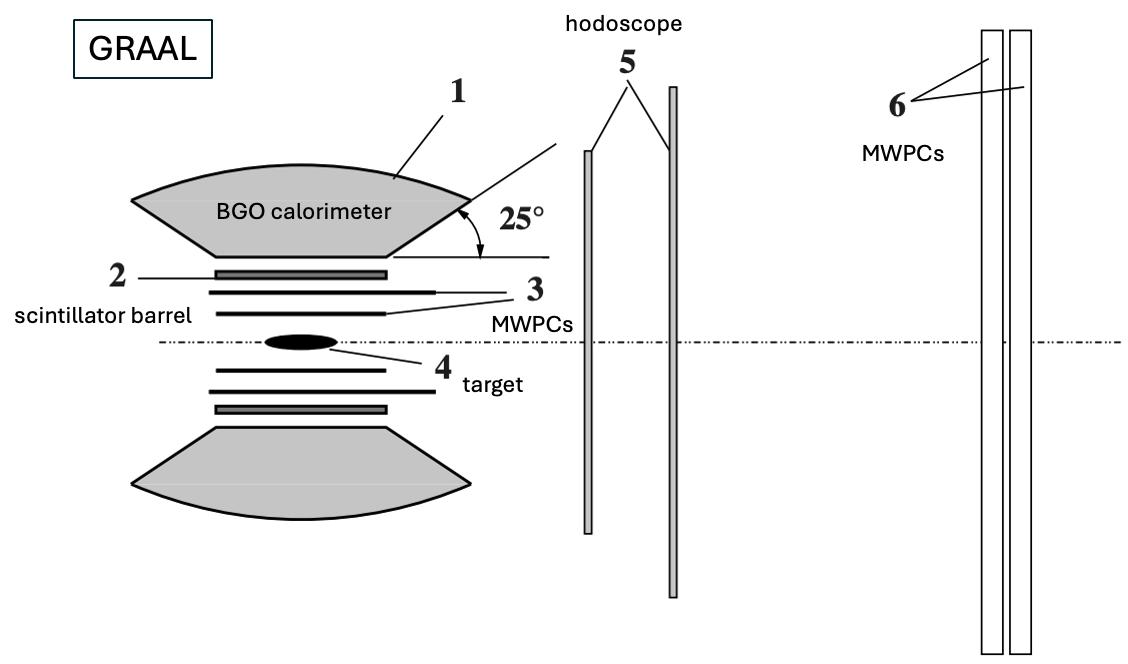}
\caption{Schematic view of the LAGRANGE detector at GRAAL: (1) BGO calorimeter, (2) plastic scintillator barrel, (3) cylindrical MWPCs, (4) target, (5) planar MWPCs, and (6) double plastic scintillator hodoscope. The beam axis is along the horizontal line through the target center with the photon beam incident from the left. Figure adapted from Ref.~\cite{GRAAL:2008jrm} and used with kind permission of The European Physical Journal (EPJ).}
\label{graal-layout}
\end{figure*}

Located 30~m downstream of the ESRF interaction region, the large acceptance LAGRANGE detector (so-named due its use of the Lagrangian particle tracking algorithm~\cite{lagrange:tracking}) was installed (see Fig.~\ref{graal-layout}) \cite{Ghio:1997yw}. This detector system was designed and optimized to reconstruct mesons that decayed to photons. However, it also had efficiency to reconstruct charged hadrons. The central components of the LAGRANGE detector surrounding the target consisted of a BGO crystal calorimeter that spanned polar angles from 25$^\circ$ to 155$^\circ$ (that was later installed at the ELSA BGOOD facility -- see Section~\ref{sec:bgood}), cylindrical multi-wire proportional chambers (MWPCs) for charged particle tracking, and plastic scintillators for energy loss ($dE/dx$) and timing information. In the forward direction, LAGRANGE included multiple MWPC tracking layers, a scintillator hodoscope, and a shower detector that extended the polar angle range down to 1$^\circ$. The detector spanned the full azimuthal angle range.

The measurements in the $KY$ sector included data for the recoil hyperon $P$, target $T$, and beam $\Sigma$ single spin asymmetries for $K^+\Lambda$ and $K^+\Sigma^0$~\cite{Lleres:2007tx,GRAAL:2008jrm} and for the linearly polarized beam-recoil hyperon double spin asymmetries $O_x$ and $O_z$~\cite{GRAAL:2008jrm}. The measurements of the target spin asymmetry $T$ were completed with an unpolarized liquid-hydrogen target but were extracted relying on constraints known as the Fierz identities~\cite{Chiang:1996em,Sandorfi:2010uv}
\begin{equation}
O_x^2 + O_z^2 + C_x^2 + C_z^2 + \Sigma^2 - T^2 + P^2 =1.
\end{equation}

\noindent
Due to the maximum energy of the photon beam, the measurement program was limited to $W_{\rm max}$ = 1.9~GeV. The recoil polarization data from GRAAL are included in the comparative plots of $P_\Lambda$ in Section~\ref{clas-gp-program} (see Figs.~\ref{clas-kl-p} and \ref{clas-ks-p}). Table~\ref{modern-nonjlab-photo} highlights the output of the strangeness physics program and a summary is provided in Ref.~\cite{Bocquet:2001ur}.

%%%%%%%%%%%%%%%%%%%%%%%%%%%%%%%%%%%%%%%%%%%%%%%%%%%%%%%%%%%%%%%%%%%%%%%%%%%%%%%%%%%%%%%%%%%%%%%%%%%%%%%%%%%%%%%%%%%%%%%%%%%%%%%%%%%%%%%%%%%%%%%%%%%%%%%%%%%%%%%%
%%%%%%%%%%%%%%%%%%%%%%%%%%%%%%%%%%%%%%%%%%%%%%%%%%%%%%%%%%%%%%%%%%%%%%%%%%%%%%%%%%%%%%%%%%%%%%%%%%%%%%%%%%%%%%%%%%%%%%%%%%%%%%%%%%%%%%%%%%%%%%%%%%%%%%%%%%%%%%%%
    
\subsection{Experiments at SPring-8}

\subsubsection{LEPS}

The Super Photon Ring-8~GeV (SPring-8) in Hyogo, Japan is an 8~GeV electron synchrotron. One of its beamlines is associated with the LEPS (Laser-Electron-Photon at SPring-8) facility for hadronic physics. The photon beam for LEPS is generated by laser backscattering from the primary electron beam. The generated photon beam for the original LEPS operations had an intensity of $1 - 2 \times 10^6$~$\gamma$/s in the tagged energy range from 1.5 to 2.4~GeV. The first photon beams from this facility were produced in 1999 and data taking for physics began in 2000 \cite{Fujiwara:1998vd}. Photon beams of both circular and linear polarization are available.

\begin{figure*}[htbp]
\centering
\includegraphics[width=0.90\textwidth]{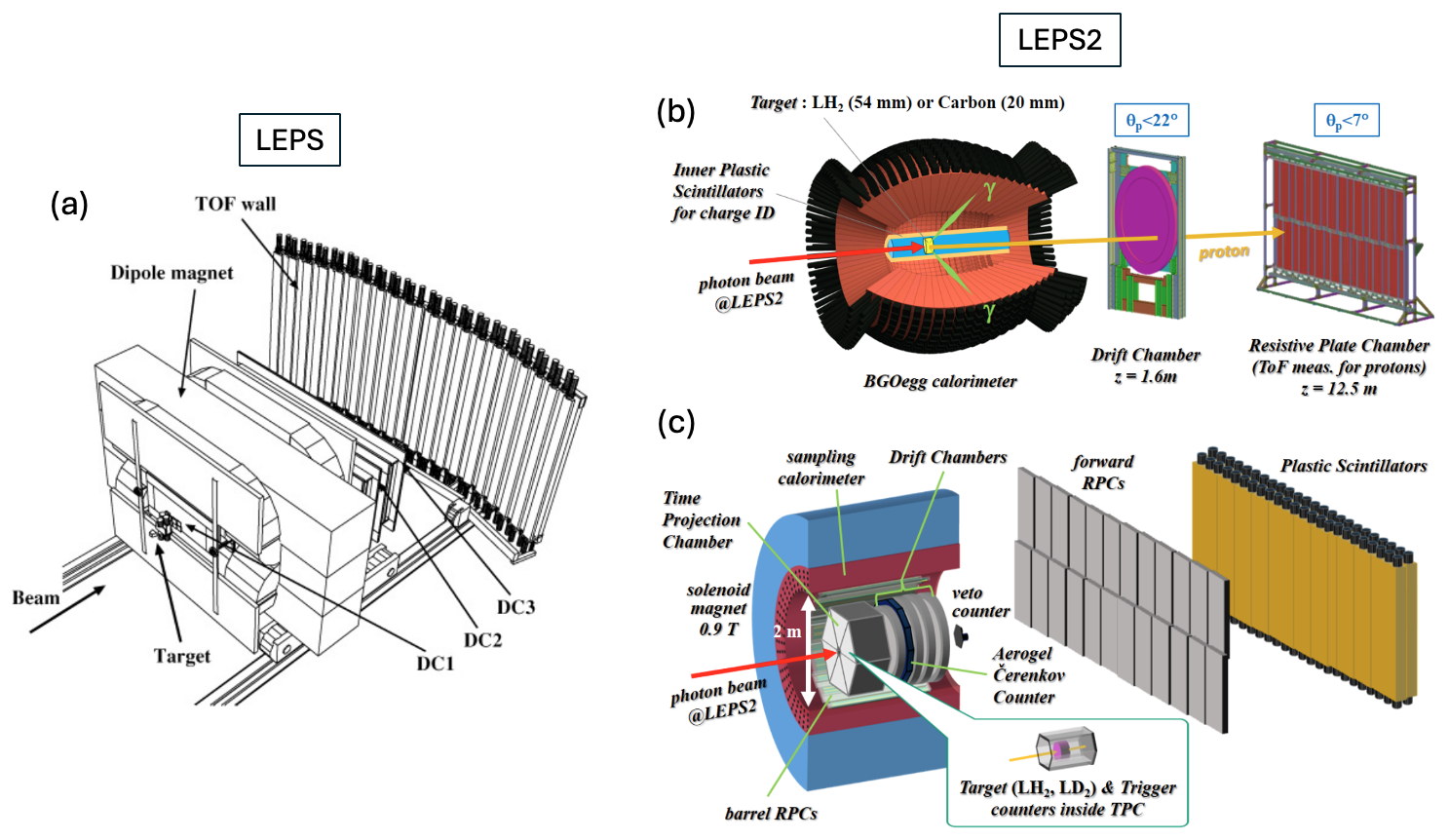}
\caption{Schematic of (a) the LEPS detector at SPring-8 (Figure from Ref.~\cite{LEPS:2005hji}) and the two detector setups for LEPS2 based on (b) the BGOEgg and (c) the LEPS2 solenoid. Figures (b) and (c) from Ref.~\cite{Muramatsu:2024bjh} used with kind permission of The European Physical Journal (EPJ).}
\label{LEPS-layout}
\end{figure*}

The LEPS spectrometer~\cite{Nakano:2000ku} is based on a dipole magnet, a silicon vertex tracker, and multiple layers of drift chambers before and after the magnet for charged particle tracking and momentum analysis. The last element along the beamline is a time-of-flight hodoscope for precise timing information for particle identification. The LEPS spectrometer is designed to span forward angles with an angular coverage of $\pm 23^\circ$ in the horizontal and $\pm 12^\circ$ in the vertical directions. See Fig.~\ref{LEPS-layout}(a) for a schematic of this facility. The measurements in the strangeness physics program are detailed in Table~\ref{modern-nonjlab-photo}.

%%%%%%%%%%%%%%%%%%%%%%%%%%%%%%%%%%%%%%%%%%%%%%%%%%%%%%%%%%%%%%%%%%%%%%%%%%%%%%%%%%%%%%%%%%%%%%%%%%%%%%%%%%%%%%%%%%%%%%%%%%%%%%%%%%%%%%%%%%%%%%%%%%%%%%%%%%%%%%%%
%%%%%%%%%%%%%%%%%%%%%%%%%%%%%%%%%%%%%%%%%%%%%%%%%%%%%%%%%%%%%%%%%%%%%%%%%%%%%%%%%%%%%%%%%%%%%%%%%%%%%%%%%%%%%%%%%%%%%%%%%%%%%%%%%%%%%%%%%%%%%%%%%%%%%%%%%%%%%%%%

The $KY$ measurement program has predominantly focused on measurements of $K^+\Lambda$ and $K^+\Sigma^0$ differential cross sections~\cite{LEPS:2005hji,Kohri:2006yx,PhysRevC.76.042201} and beam single spin asymmetries $\Sigma$ using a linearly polarized beam~\cite{LEPS:2003buk,Kohri:2006yx,LEPS:2005hji,PhysRevC.76.042201,LEPS:2017pzl}. By design of the LEPS spectrometer, the angular coverage spans $\cos \theta_K^{\mathrm{c.m.}} > 0.6$. Figure~\ref{bgood-dcs} and the plots in Section~\ref{clas-gp-program} (see Figs.~\ref{clas-kl-dcs} and \ref{clas-ks-dcs}) show comparative plots of the LEPS differential cross sections to those from other facilities. The program has also provided measurements of differential cross sections and the beam single spin asymmetry $\Sigma$ using an effective neutron target from experiments with an LD$_2$ cryotarget for the $K^+\Sigma^-$ final state~\cite{Kohri:2006yx}. Other investigations include a search for a kaonic nucleus from a $K^-pp$ bound state in $\gamma d$ reactions~\cite{LEPS:2013dqu} and studies of cross sections and spin density matrix elements for the $K^{*0}\Sigma^+$ final state to probe meson exchanges in the $t$-channel \cite{Hwang:2013usa,Hwang:2012zza}. For completeness in this section, it must be mentioned that the LEPS Collaboration has also spent significant effort to search for pentaquark states in $\gamma d$ experiments that recoil against a detected final state $K^+$. This topic is reviewed in Refs.~\cite{Nakano:2010zz,ActaPenta2025} and further information is provided in Section~\ref{sec:penta-narrow} with no further discussion here. 

\subsubsection{LEPS2}

A second hadronic beamline at SPring-8 called the LEPS2 facility was commissioned in 2018 and 2019, with first beam for physics in 2021. LEPS2 builds on the vast experience with the design and operation gained from 20 years of beam operations at LEPS. The LEPS2 facility provides for higher intensity photon beams up to $\approx 10^7$~$\gamma$/s and large acceptance detector systems to expand the LEPS program~\cite{Muramatsu:2024bjh}. The LEPS2 experiment incorporates two separately interchangeable and complementary detector configurations. One based on the BGOegg electromagnetic calorimeter and a second based on a solenoid spectrometer. The ability to switch between the two detector setups allows for a broad and diverse physics program. See Figs.~\ref{LEPS-layout}(b) and (c) for schematics of both LEPS2 detector options.

The first LEPS2 detector system is optimized for the detection of photons and is based on an egg-shaped calorimeter called the BGOegg~\cite{Matsumura:2016roc}. This calorimeter has full coverage of the azimuth and spans polar angles from $24^\circ$ to 144$^\circ$. To enable detection of charged particles in this acceptance regime, a multi-layer cylindrical drift chamber is installed within the calorimeter for charged particle tracking followed by scintillation counters for energy loss and timing information to enable particle identification. The second LEPS2 detector system is based on a solenoid spectrometer that is designed to detect both charged and neutral particles~\cite{Niiyama:2017hjo}. Charged particle tracking is done by a central time projection chamber (TPC) followed by multiple layers of drift chambers in the forward direction. The tracking system resides within the 0.9~T solenoid field. Surrounding the TPC and drift chambers is a barrel resistive plate chamber with good timing resolution for charged particle identification. Outside of this chamber is a barrel electromagnetic calorimeter for detection of neutral particles. This calorimeter spans the full azimuth and the polar angle range from 40$^\circ$ to 110$^\circ$. The particle identification system includes a Cherenkov detector downstream of the TPC within the solenoid. In the forward direction outside of the solenoid is a resistive plate chamber followed by a scintillator hodoscope for neutron detection. One of the flagship programs of strangeness with LEPS2 is the upcoming study of the $\Lambda(1405)$ via the reaction $\gamma p \to K^+ \Lambda(1405)$. This experiment will study the internal structure of this excited hyperon state through extractions of the spin density matrix elements~\cite{Niiyama:2017hjo}.
    
\subsection{Experiments at MAMI}
\label{mami-program}

The Mainz Microtron MAMI located at the University of Mainz in Germany is a continuous wave (100\% duty factor) electron accelerator. This facility was conceived in the mid-1970s and has developed considerably over the decades from MAMI-A -- a 14~MeV machine in 1979, to MAMI-B -- an 850~MeV machine in 1991, to the current facility MAMI-C -- a 1.5~GeV machine. MAMI consists of three stages of racetrack microtrons and one stage formed of a double-sided harmonic microtron~\cite{Kaiser:2008zza}. The current accelerator delivers an electron beam with currents up to 100~$\mu$A. 

\begin{figure*}[htbp]
\centering
\includegraphics[width=0.75\textwidth]{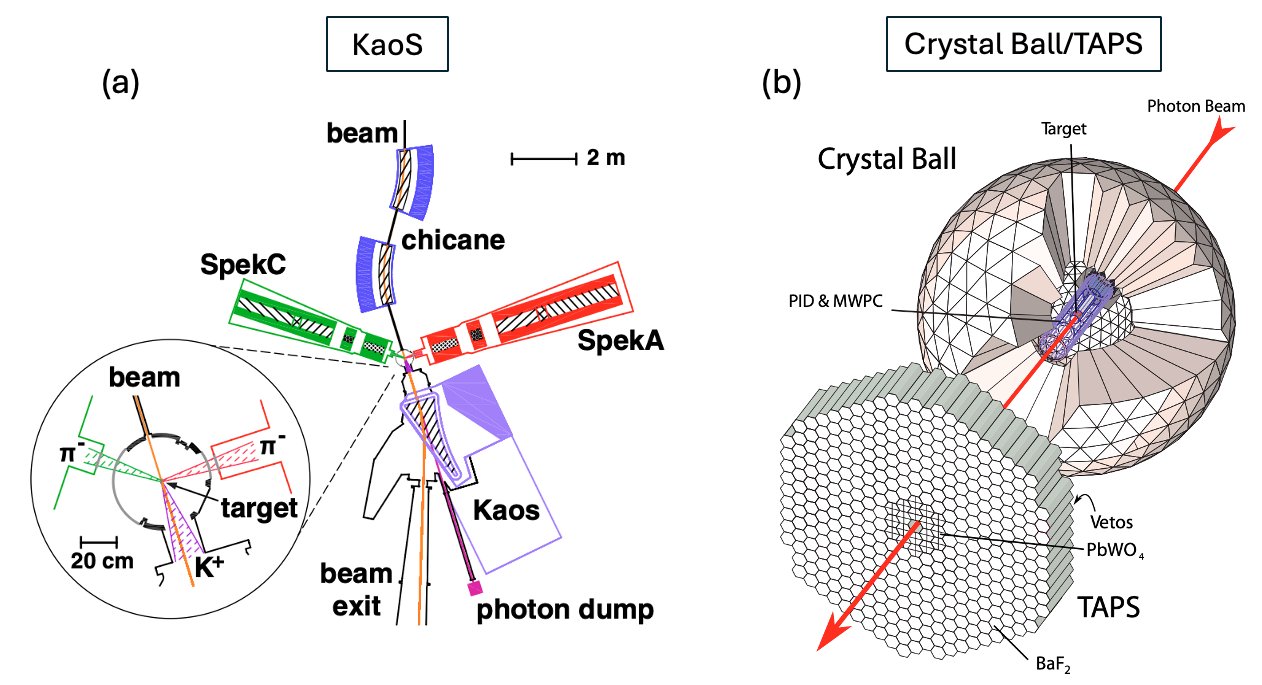}
\caption{Schematic of the experimental layouts at MAMI. (a) Configuration of the electron beam experiments in the A1 area including the two high resolution spectrometers for detection of scattered electrons and the KaoS spectrometer for $K^+$ detection - Figure from Ref.~\cite{A1:2015isi}. This configuration was in place for the $(e,e'K^+)$ program. (b) Configuration of the photon beam experiments in the A2 area showing the Crystal Ball/TAPS detector - Figure from Ref.~\cite{A2:2016sjm}. Note that the original TAPS calorimeter was split into two systems, one detailed in Section~\ref{sec:cbelsa} at CB-ELSA and the other in the MAMI A2 setup.}
\label{mami-layout}
\end{figure*}

There are two experimental areas where studies of the electromagnetic production of kaons are carried out. The A1 beamline is dedicated to electron scattering experiments at beam-target luminosities of up to $\sim$$5 \times 10^{37}$~cm$^{-2}$s$^{-1}$. The experimental hall is currently outfitted with three high resolution spectrometers~\cite{Blomqvist:1998xn} and the short orbit KaoS spectrometer (see Fig.~\ref{mami-layout}(a)) \cite{Esser:2013aya}. KaoS was specifically designed for the detection of charged kaons. A key feature of each of these MAMI A1 spectrometers is their high relative momentum resolution along with their comparatively large angle and momentum acceptances. The single dipole configuration of the KaoS spectrometer allows it to be positioned at very forward angles relative to the beamline, which is an important feature for hypernuclear studies since their production cross sections peak at forward angles. 

The main aspects of the program in strangeness physics are the spectroscopic studies of hypernuclei either tagged with the $(e,e'K^+)$ final state~\cite{Esser:2013aya,Achenbach:2013alv,A1:2015isi} or reconstructed from decay-pion spectroscopy through processes such as $(e,K^+\pi^-)$ where the $K^+$ tags the $\Lambda$ in the production process and the $\pi^-$ comes from the $\Lambda$ decay within the hypernucleus~\cite{A1:2024edr}. Studies of associated production of the exclusive $K^+\Lambda$ and $K^+\Sigma^0$ final states have also been carried out, including several exploratory measurements of differential cross sections and the $\sigma_{\rm LT'}$ interference structure function have been published at an average $Q^2$ of 0.055~GeV$^2$ for $W$ up to 1.725~GeV. These limited measurements are presented in Refs.~\cite{Achenbach:2010gns,A1:2013hzm,A1:2017fvu}. 

The second experimental area at MAMI called A2 is dedicated to experiments with photon beams from the tagged bremsstrahlung technique. The facility can produce beams of circularly and linearly polarized photons at typical photon fluxes of $\sim$$5 \times 10^7$~$\gamma$/s. The scattered electrons are detected in a tagging electrometer. The photons at the target span 5\% to 93\% of the full electron beam energy range. The detector system installed in the A2 beamline is a combination of the Crystal Ball and TAPS detectors~\cite{Neiser:2015rma}. The Crystal Ball detector based on NaI(Tl) crystals was originally built for use at the Stanford accelerator SLAC in the mid-1970s~\cite{Bloom:1983pc}. In its center is a barrel of scintillation counters surrounding the target for $dE/dx$ measurements for charged particle identification. These counters also serve as a charged particle veto for neutral particle reconstruction. The crystals span 94\% of $4 \pi$. The forward angle range up to $\theta \approx 20^\circ$ is covered by the TAPS detector, a BaF$_2$ crystal calorimeter for neutral particle identification. The Crystal Ball/TAPS detector can be used with both unpolarized and polarized targets. See Fig.~\ref{mami-layout}(b) for a schematic of the detector.

%%%%%%%%%%%%%%%%%%%%%%%%%%%%%%%%%%%%%%%%%%%%%%%%%%%%%%%%%%%%%%%%%%%%%%%%%%%%%%%%%%%%%%%%%%%%%%%%%%%%%%%%%%%%%%%%%%%%%%%%%%%%%%%%%%%%%%%%%%%%%%%%%%%%%%%%%%%%%%%%
%%%%%%%%%%%%%%%%%%%%%%%%%%%%%%%%%%%%%%%%%%%%%%%%%%%%%%%%%%%%%%%%%%%%%%%%%%%%%%%%%%%%%%%%%%%%%%%%%%%%%%%%%%%%%%%%%%%%%%%%%%%%%%%%%%%%%%%%%%%%%%%%%%%%%%%%%%%%%%%%

%%%%%%%%%%%%%%%%%%%%%%%%%%%%%%%%%%%%%%%%%%%%%%%%%%%%%%%%%%%%%%%%%%%%%%%%%%%%%%%%%%%%%%%%%%%%%%%%%%%%%%%%%%%%%%%%%%%%%%%%%%%%%%%%%%%%%%%%%%%%%%%%%%%%%%%%%%%%%%%%
\begin{figure*}[htbp]
\centering
\includegraphics[width=0.70\textwidth]{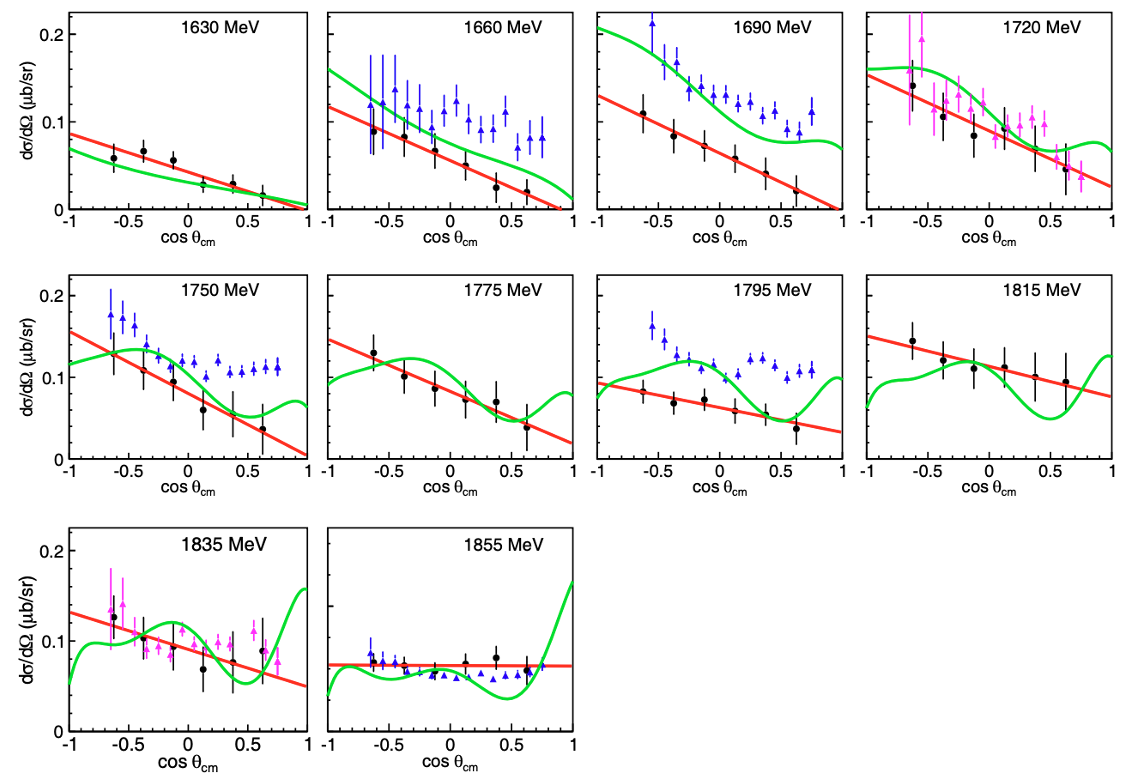}
\caption{Differential cross sections for $\gamma n \to K^0\Lambda$ from MAMI A2 (black circles) plotted as a function of $\cos \theta_K^{\mathrm{c.m.}}$ for bins in $W$. The data are compared to results from CLAS (solid magenta and solid blue triangles)~\cite{CLAS:2017gsu}. The green curves are a prediction from the partial wave analysis of Ref.~\cite{Hunt:2018mrt} and the red line is a linear fit. Figure from Ref.~\cite{A2:2018doh} and used with kind permission of The European Physical Journal (EPJ).}
\label{mami-k0-dcs1}
\end{figure*}
%%%%%%%%%%%%%%%%%%%%%%%%%%%%%%%%%%%%%%%%%%%%%%%%%%%%%%%%%%%%%%%%%%%%%%%%%%%%%%%%%%%%%%%%%%%%%%%%%%%%%%%%%%%%%%%%%%%%%%%%%%%%%%%%%%%%%%%%%%%%%%%%%%%%%%%%%%%%%%%%

\begin{figure*}[htbp]
\centering
\includegraphics[width=0.70\textwidth]{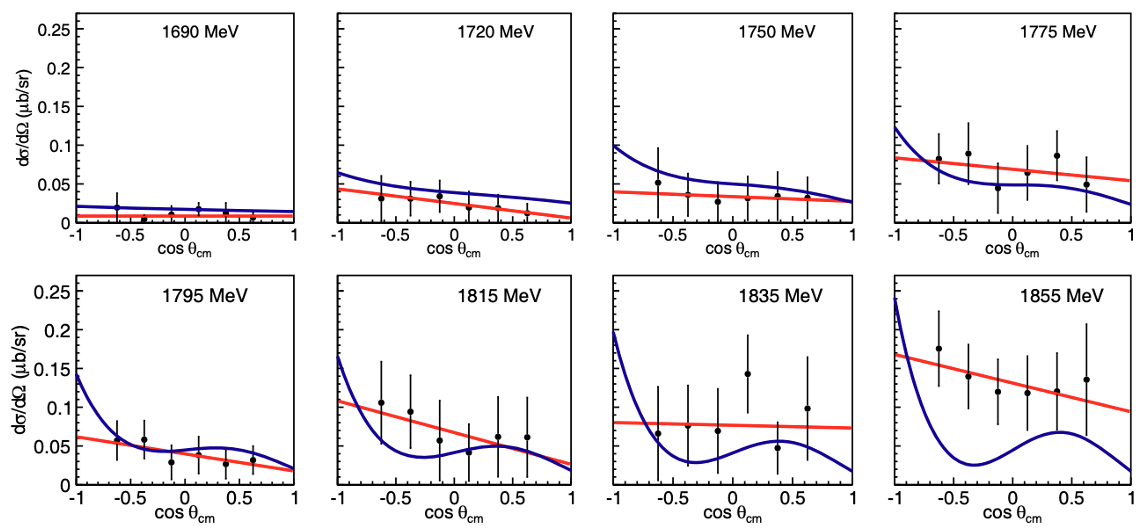}
\caption{Differential cross sections for $\gamma n \to K^0\Sigma^0$ from MAMI A2 plotted as a function of $\cos \theta_K^{\mathrm{c.m.}}$ for bins in $W$~\cite{A2:2018doh}. The blue curve is from the isobar model of Ref.~\cite{Mart:2014eoa} and the red line is a linear fit. Figure from Ref.~\cite{A2:2018doh} and used with kind permission of The European Physical Journal (EPJ).}
\label{mami-k0-dcs2}
\end{figure*}

The photoproduction measurements in the strangeness physics program are detailed in Table~\ref{modern-nonjlab-photo}. They include measurements of $K^+ \Lambda$, $K^+ \Sigma^0$, and $K^0 \Sigma^+$ cross sections from a proton target~\cite{A2:2018doh,A2:2013cqk,CrystalBall:2013iig} and $K^0 \Lambda$ and $K^0 \Sigma^0$ from a neutron target~\cite{A2:2018doh}. Due to the maximum energy of the electron beam, the photoproduction measurements are limited to $W \approx 1.9$~GeV. The available data in these channels amounts to $\sim$$1 \times 10^5$ events. Examples of the data results for the differential cross sections of the neutral particle final state in measurements from an effective neutron (LD$_2$) target in the $\gamma n \to K^0 \Lambda$ and $\gamma n \to K^0 \Sigma^0$ channels are shown in the comparative plots of Figs.~\ref{mami-k0-dcs1} and \ref{mami-k0-dcs2}, respectively. Such measurements represent a significant experimental challenge due to large combinatoric backgrounds, the Fermi motion of the associated quasi-free neutron target, and the non-trivial impact of final state interactions. In these final states the $K^0$ is detected through its $\pi^0\pi^0 \to 4\gamma$ decay mode, the $\Lambda$ through its neutral decay mode $\pi^0 n \to 2\gamma n$, and the $\Sigma^0$ via its $\Lambda \gamma \to n \pi^0 \gamma \to 3\gamma n$ branch. Of the six elementary kaon photoproduction reactions listed in Table~\ref{tab:kaon_thresholds}, this is by far the most challenging to measure experimentally. However, data from each process is ultimately required to fully unravel the $KY$ reaction mechanism and channel couplings, as well as to have maximal sensitivity to the contributing $s$-channel nucleon excited states. The different elementary processes are not related via simple Clebsch-Gordan relations as the electromagnetic interaction does not conserve isospin, so measurements on both proton and neutron targets are necessary. But it must be stressed that an advanced reaction model is essential to fully understand the reaction dynamics of these processes.

%%%%%%%%%%%%%%%%%%%%%%%%%%%%%%%%%%%%%%%%%%%%%%%%%%%%%%%%%%%%%%%%%%%%%%%%%%%%%%%%%%%%%%%%%%%%%%%%%%%%%%%%%%%%%%%%%%%%%%%%%%%%%%%%%%%%%%%%%%%%%%%%%%%%%%%%%%%%%%%%
%%%%%%%%%%%%%%%%%%%%%%%%%%%%%%%%%%%%%%%%%%%%%%%%%%%%%%%%%%%%%%%%%%%%%%%%%%%%%%%%%%%%%%%%%%%%%%%%%%%%%%%%%%%%%%%%%%%%%%%%%%%%%%%%%%%%%%%%%%%%%%%%%%%%%%%%%%%%%%%%

%%%%%%%%%%%%%%%%%%%%%%%%%%%%%%%%%%%%%%%%%%%%%%%%%%%%%%%%%%%%%%%%%%%%%%%%%%%%%%%%%%%%%%%%%%%%%%%%%%%%%%%%%%%%%%%%%%%%%%%%%%%%%%%%%%%%%%%%%%%%%%%%%%%%%%%%%%%%%%%
\begin{table}[htbp]
\setlength{\tabcolsep}{6pt} % Default value: 6pt
\renewcommand{\arraystretch}{0.8} % Default value: 1
\begin{center}
\caption{Summary of the key strangeness photoproduction measurements from experiments since the 1990s at facilities in Europe and Asia. The column labeled $N_\mathrm{bin}$ indicates the number of kinematic bins included in the analysis.}
\begin{tabular}{cccccccc} \hline\hline
Observable                  & Final State(s)                           & $W$ (GeV)     & $\cos \theta_K^{\mathrm{c.m.}}$   & $N_{\mathrm{bin}}$                     & Facility & Year & Ref. \\ \hline
\multirow{25}{*}{$d\sigma$} & $K^+\Lambda$, $K^+\Sigma^0$              & $[1.6:1.9]$   & $[-1.0:1.0]$             & 100 $\Lambda$, 100 $\Sigma^0$ & SAPHIR   & 1994 & \cite{Bockhorst:1994jf} \\
                            & $K^+\Lambda$, $K^+\Sigma^0$              & $[1.6:2.15]$  & $[-1.0:1.0]$             & 90 $\Lambda$, 70 $\Sigma^0$   & SAPHIR   & 1998 & \cite{SAPHIR:1998fev} \\
                            & $K^0\Sigma^+$                            & $[1.72:1.92]$ & $[-1.0:1.0]$             & 17                            & SAPHIR   & 1999 & \cite{SAPHIR:1999wfu} \\
                            & $K^+\Lambda$, $K^+\Sigma^0$              & $[1.6:2.4]$   & $[-1.0:1.0]$             & 720 $\Lambda$, 660 $\Sigma^0$ & SAPHIR   & 2004 & \cite{Glander:2003jw} \\ 
                            & $K^0\Sigma^+$                            & $[1.7:2.4]$   & $[-1.0:1.0]$             & 120                           & SAPHIR   & 2005 & \cite{Lawall:2005np} \\ 
                            & $K^0\Sigma^+$                            & $[1.69:2.26]$ & $[-0.8:0.8]$             & 108                           & CB-ELSA  & 2012 & \cite{CBELSATAPS:2011gly} \\
                            & $K^0\Sigma^+$                            & $[1.73:1.88]$ & $[-1.0:1.0]$             & 42 $\Lambda$, 42 $\Sigma^0$   & MAMI     & 2013 & \cite{A2:2013cqk} \\ 
                            & $K^+\Lambda$, $K^+\Sigma^0$              & $[1.7:1.9]$   & $[-0.7:0.1]$             & 310 $\Lambda$, 280 $\Sigma^0$ & MAMI     & 2014 & \cite{CrystalBall:2013iig} \\ 
                            & $K^0\Sigma^+$                            & $[1.7:1.9]$   & $[-0.6:0.6]$             & 35                            & MAMI     & 2019 & \cite{A2:2018doh} \\ 
                            & $K^0\Lambda$, $K^0\Sigma^0$ ($n$ target) & $[1.7:1.9]$   & $[-0.6:0.6]$             & 60 $\Lambda$, 60 $\Sigma^0$   & MAMI     & 2019 & \cite{A2:2018doh} \\ 
                            & $K^+\Sigma^-$ ($n$ target)               & $[1.92:2.32]$ & $[0.6:1.0]$              & 72                            & LEPS     & 2006 & \cite{Kohri:2006yx} \\ 
                            & $K^+\Lambda$, $K^+\Sigma^0$              & $[1.92:2.32]$ & $[0.6:1.0]$              & 54 $\Lambda$, 54 $\Sigma^0$   & LEPS     & 2006 & \cite{LEPS:2005hji} \\
                            & $K^+\Lambda$                             & $[1.92:2.32]$ & $[-1.0:-0.8]$            & 12                            & LEPS     & 2007 & \cite{PhysRevC.76.042201} \\
                            & $K^+\Lambda(1520)$                       & $[1.92:2.32]$ & $[-1.0:-0.5]$            & 6                             & LEPS     & 2009 & \cite{Muramatsu:2009zp} \\
                            & $K^0\Lambda(1520)$ ($n$ target)          & $[1.92:2.32]$ & $[-1.0:-0.5]$            & 6                             & LEPS     & 2009 & \cite{Muramatsu:2009zp} \\
                            & $K^+\Lambda(1520)$                       & $[1.92:2.32]$ & $[0.6:1.0]$              & 60                            & LEPS     & 2010 & \cite{LEPS:2009isz} \\
                            & $K^{*0}\Sigma^+$                         & $[2.32:2.53]$ & $[0.8:1.0]$              & 12                            & LEPS     & 2013 & \cite{Hwang:2013usa} \\
                            & $K^+\pi^-$                               & $[1.92:2.32]$ & $[0.8:1.0]$              & 45                            & LEPS     & 2014 & \cite{LEPS:2013dqu} \\
                            & $K^+\Lambda$, $K^+\Sigma^0$              & $[1.92:2.55]$ & $[0.6:1.0]$              & 60 $\Lambda$, 60 $\Sigma^0$   & LEPS     & 2018 & \cite{LEPS:2017pzl} \\
                            & $K^+\Lambda$                             & $[1.62:1.86]$ & $[0.9:1.0]$              & 80                            & BGOOD    & 2021 & \cite{Alef:2020yul} \\
                            & $K^+\Sigma^0$                            & $[1.62:1.86]$ & $[0.9:1.0]$              & 22                            & BGOOD    & 2021 & \cite{Jude:2020byj} \\
                            & $K^+ \Lambda(1405)$                      & $[1.95:2.5]$  & $[-0.7:0.7]$             & 42                            & BGOOD    & 2022 & \cite{BGOOD:2021sog} \\
                            & $K^0\Sigma^0$ ($n$ target)               & $[1.75:2.4]$  & $[-0.7:0.5]$             & 40                            & BGOOD    & 2023 & \cite{BGOOD:2021oxp} \\
                            & $K^+ \Lambda(1520)$                      & $[2.0:2.1]$   & $[0.9:1.0]$              & 11                            & BGOOD    & 2025 & \cite{Rosanowski:2024rww} \\
                            & $K^0\Lambda$ ($n$ target)                & $[1.6:1.66]$  & $[0.9:1.0]$              & 22                            & NKS      & 2010 & \cite{NKS:2010dav} \\ \hline
\multirow{8}{*}{$P$}        & $K^+\Lambda$, $K^+\Sigma^0$              & $[1.6:1.9]$   & $[-1.0:1.0]$             & 18 $\Lambda$, 12 $\Sigma^0$   & SAPHIR   & 1994 & \cite{Bockhorst:1994jf} \\
                            & $K^+\Lambda$, $K^+\Sigma^0$              & $[1.6:2.15]$  & $[-1.0:1.0]$             & 12 $\Lambda$, 12 $\Sigma^0$   & SAPHIR   & 1998 & \cite{SAPHIR:1998fev} \\
                            & $K^0\Sigma^+$                            & $[1.72:1.92]$ & $[-1.0:1.0]$             & 4                             & SAPHIR   & 1999 & \cite{SAPHIR:1999wfu} \\ 
                            & $K^+\Lambda$, $K^+\Sigma^0$              & $[1.6:2.4]$   & $[-1.0:1.0]$             & 30 $\Lambda$, 16 $\Sigma^0$   & SAPHIR   & 2004 & \cite{Glander:2003jw} \\ 
                            & $K^0\Sigma^+$                            & $[1.7:2.4]$   & $[-1.0:1.0]$             & 10                            & SAPHIR   & 2005 & \cite{Lawall:2005np} \\ 
                            & $K^0\Sigma^+$                            & $[1.74:2.0]$  & $[-0.8:0.8]$             & 18                            & CB-ELSA  & 2014 & \cite{CBELSATAPS:2014ibm} \\
                            & $K^0\Sigma^+$                            & $[1.73:1.88]$ & $[-1.0:1.0]$             & 42                            & MAMI     & 2013 & \cite{A2:2013cqk} \\  
                            & $K^+\Lambda$, $K^+\Sigma^0$              & $[1.61:1.91]$ & $[-0.75:0.85]$           & 72 $\Lambda$, 8 $\Sigma^0$    & GRAAL    & 2007 & \cite{Lleres:2007tx} \\ \hline
\multirow{9}{*}{$\Sigma$}   & $K^0\Sigma^+$                            & $[1.74:2.0]$  & $[-0.8:0.8]$             & 15                            & CB-ELSA  & 2014 & \cite{CBELSATAPS:2014ibm} \\
                            & $K^+\Lambda$, $K^+\Sigma^0$              & $[1.61:1.91]$ & $[-0.75:0.85]$           & 72 $\Lambda$, 72 $\Sigma^0$   & GRAAL    & 2007 & \cite{Lleres:2007tx} \\
                            & $K^+\Lambda$, $K^+\Sigma^0$              & $[1.92:2.32]$ & $[0.6:1.0]$              & 45 $\Lambda$, 45 $\Sigma^0$   & LEPS     & 2003 & \cite{LEPS:2003buk} \\
                            & $K^+\Sigma^0$                            & $[1.92:2.32]$ & $[0.6:1.0]$              & 72                            & LEPS     & 2006 & \cite{Kohri:2006yx} \\
                            & $K^+\Sigma^-$ ($n$ target)               & $[1.92:2.32]$ & $[0.6:1.0]$              & 72                            & LEPS     & 2006 & \cite{Kohri:2006yx} \\
                            & $K^+\Lambda$, $K^+\Sigma^0$              & $[1.92:2.32]$ & $[0.6:1.0]$              & 30 $\Lambda$, 30 $\Sigma^0$   & LEPS     & 2006 & \cite{LEPS:2005hji} \\
                            & $K^+\Lambda$                             & $[1.92:2.32]$ & $[-1.0:-0.8]$            & 4                             & LEPS     & 2007 & \cite{PhysRevC.76.042201} \\
                            & $K^+\Lambda(1520)$                       & $[1.92:2.32]$ & $[0.6:1.0]$              & 7                             & LEPS     & 2010 & \cite{LEPS:2009isz} \\
                            & $K^+\Lambda$, $K^+\Sigma^0$              & $[1.92:2.55]$ & $[0.6:1.0]$              & 18 $\Lambda$, 18 $\Sigma^0$   & LEPS     & 2018 & \cite{LEPS:2017pzl} \\ \hline
$T$                         & $K^+\Lambda$                             & $[1.61:1.91]$ & $[-0.75:0.85]$           & 66                            & GRAAL    & 2009 & \cite{GRAAL:2008jrm} \\ \hline
$O_x$, $O_z$                & $K^+\Lambda$                             & $[1.61:1.91]$ & $[-0.75:0.85]$           & 66                            & GRAAL    & 2009 & \cite{GRAAL:2008jrm} \\ \hline\hline
\end{tabular}
\label{modern-nonjlab-photo}
\end{center}
\end{table}
%%%%%%%%%%%%%%%%%%%%%%%%%%%%%%%%%%%%%%%%%%%%%%%%%%%%%%%%%%%%%%%%%%%%%%%%%%%%%%%%%%%%%%%%%%%%%%%%%%%%%%%%%%%%%%%%%%%%%%%%%%%%%%%%%%%%%%%%%%%%%%%%%%%%%%%%%%%%%%%%

\subsection{Experiments at JLab -- 6-GeV Era}
\label{sec:jlab6gev}

The electron accelerator at Jefferson Laboratory was constructed in the period from 1987 to 1993. The accelerator, known as CEBAF (Continuous Electron Beam Accelerator Facility), is based on superconducting radiofrequency (RF) technology. The machine parameters allow for up to 200~$\mu$A (with $\approx$150~$\mu$A a more practical limit) of circulated beam with a 100\% duty factor. It is comprised of two anti-parallel LINACs connected through recirculation arcs. The machine began operations for physics in 1995, ultimately delivering highly polarized electrons up to 4~GeV to the three experimental end stations, Halls A, B, and C, with simultaneous beam delivery to all three halls as part of routine operations by 1997. By the year 2000, due to continuous improvements in the associated accelerator technologies, mainly in terms of increased operating gradients in the LINAC cryomodules, the machine was able to provide electron beams up to 6~GeV for physics~\cite{Leemann:2001dg}.

Each of the three experimental endstations included a unique complement of equipment and detector packages to support their scientific programs. The baseline equipment in Halls A and C included dual spectrometer systems of small acceptance for high luminosity operations (see Section~\ref{halla-c-6gev}) and Hall~B housed a large acceptance detector operated at lower luminosities (see Sections~\ref{clas-gp-program} and \ref{clas-ep-program}). The experimental program as part of the 6-GeV era extended until 2012 when it was concluded to begin preparations for the JLab 12-GeV upgrade (see Section~\ref{jlab:12gev}). 

\subsubsection{Halls A and C}
\label{halla-c-6gev}

The baseline equipment in experimental Halls A and C was based on dual-arm spectrometer systems. Hall A included two high resolution spectrometers (HRS)~\cite{Alcorn:2004sb} and Hall~C included a high momentum spectrometer (HMS) and a short orbit spectrometer (SOS)~\cite{Domingo:1992ux}. 

\begin{figure*}[htbp]
\centering
\includegraphics[width=0.9\textwidth]{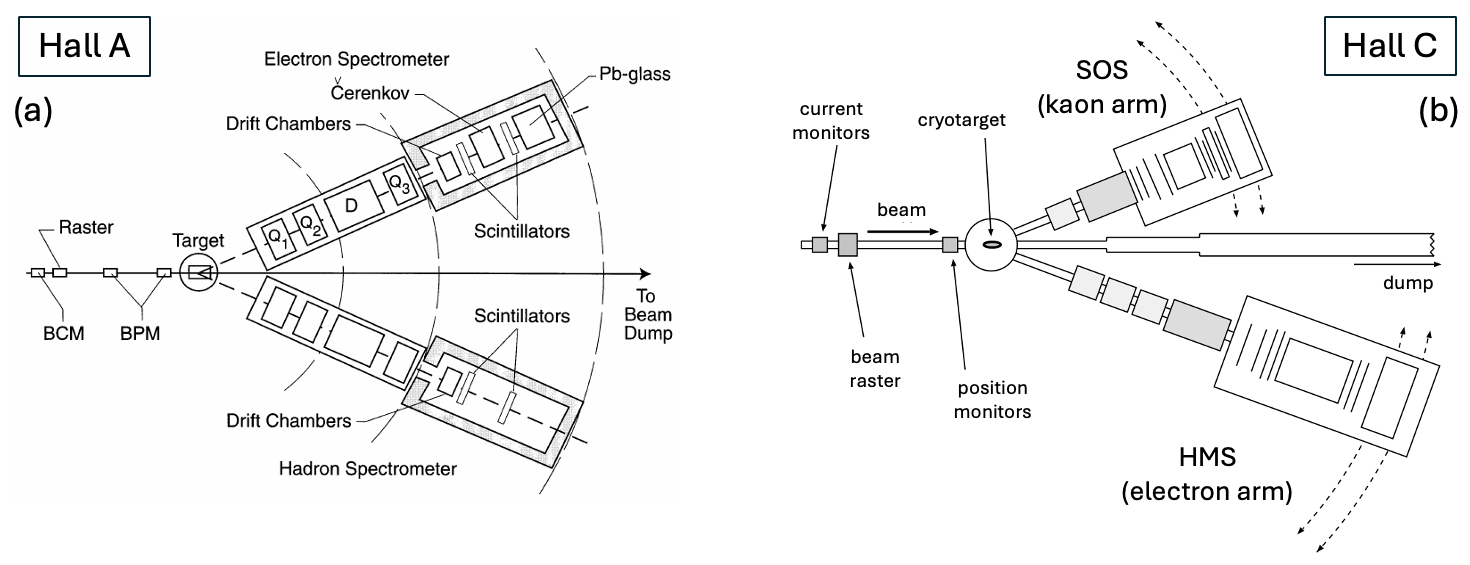}
\caption{(a) Schematic of the Hall~A spectrometer setup showing the two identical HRS high resolution spectrometers. In this figure the spectrometer on beam left is configured as the electron arm and the spectrometer on beam right is configured as the hadron arm. The length of each spectrometer is $\sim$20~m. (b) The Hall~C spectrometer setup showing the Short-Orbit Spectrometer (SOS) on beam left and the High Momentum Spectrometer (HMS) on beam right. The length of the HMS spectrometer is $\sim$18~m. See Ref.~\cite{Domingo:1992ux} for more details.}
\label{halla_c-layout}
\end{figure*}

The Hall A physics program was designed to study electroproduction reactions at luminosities up to $\sim$$10^{39}$~cm$^{-2}$s$^{-1}$ with spectrometers of excellent momentum ($\Delta p/p < 2 \times 10^{-4}$) and angular ($\Delta \theta, \Delta \phi \sim 1-2$~mrad) resolution at a maximum central momentum of 4~GeV. In typical experiments, one spectrometer was configured as the electron arm and the other as the hadron arm as shown in Fig.~\ref{halla_c-layout}(a). Each HRS spectrometer has a design solid angle of 6~msr.

%%%%%%%%%%%%%%%%%%%%%%%%%%%%%%%%%%%%%%%%%%%%%%%%%%%%%%%%%%%%%%%%%%%%%%%%%%%%%%%%%%%%%%%%%%%%%%%%%%%%%%%%%%%%%%%%%%%%%%%%%%%%%%%%%%%%%%%%%%%%%%%%%%%%%%%%%%%%%%%%
%%%%%%%%%%%%%%%%%%%%%%%%%%%%%%%%%%%%%%%%%%%%%%%%%%%%%%%%%%%%%%%%%%%%%%%%%%%%%%%%%%%%%%%%%%%%%%%%%%%%%%%%%%%%%%%%%%%%%%%%%%%%%%%%%%%%%%%%%%%%%%%%%%%%%%%%%%%%%%%%

The Hall C physics program was focused on complementary studies compared to that in Hall~A. The HMS spectrometer was designed to measure electrons up to $\sim$7~GeV with moderate ($\Delta p/p \sim 10^{-3}$) momentum resolution and a solid angle of 8.1~msr. The SOS spectrometer was designed with a short optical path to the detector focal plane to measure particles with a short lifetime. The SOS, which had a solid angle of 9~msr and a momentum resolution $< 8 \times 10^{-4}$, was designed to detect hadrons with momenta below 2~GeV. Figure~\ref{halla_c-layout}(b) shows the Hall~C spectrometer layout. In addition to the baseline equipment, other spectrometer and detector packages were temporarily installed as standalone systems for specific experiments or to be used in conjunction with the HMS or SOS. Like operations in Hall~A, Hall~C was designed to study electroproduction reactions at luminosities up to $\sim$$10^{39}$~cm$^{-2}$s$^{-1}$. 

A summary of the measurements completed in Halls A and C for the $ep\to e'K^+\Lambda$ and $ep\to e'K^+\Sigma^0$ exclusive processes is given in Table~\ref{halla-c-list}. These measurements focused on differential cross sections at multiple beam energies to enable separation of the transverse $\sigma_{\rm T}$ and longitudinal $\sigma_{\rm L}$ structure functions in bins of $Q^2$, $W$, and $\cos \theta_K^{\mathrm{c.m.}}$ via the Rosenbluth separation technique \cite{Rosenbluth:1950yq}. The measurements were carried out only for a few specific kinematic points given the nature of the experiments with small aperture spectrometers. The structure functions were compared to their predictions from single-channel isobar models whose parameters, e.g. coupling constants, form factors, and specific exchanges in the $s$-, $t$-, and $u$-reaction channels, were constrained based on SU(3) predictions and by phenomenological fits to the existing world data available at the time of their publication. Two of the early isobar models that were used for comparison, WJC~\cite{Williams:1992tp} and Saclay-Lyon (SL)~\cite{David:1995pi}, are discussed in Section~\ref{sec:isobar}.

%%%%%%%%%%%%%%%%%%%%%%%%%%%%%%%%%%%%%%%%%%%%%%%%%%%%%%%%%%%%%%%%%%%%%%%%%%%%%%%%%%%%%%%%%%%%%%%%%%%%%%%%%%%%%%%%%%%%%%%%%%%%%%%%%%%%%%%%%%%%%%%%%%%%%%%%%%%%%%%
\begin{table}[tbh]
\setlength{\tabcolsep}{6pt} % Default value: 6pt
\renewcommand{\arraystretch}{0.8} % Default value: 1
\begin{center}
\caption{Summary of $ep\to e'K^+\Lambda$ and $ep\to e'K^+\Sigma^0$ measurements in Halls A and C from the 6-GeV era experiments at JLab. For the entries in the first three rows the cross sections were evolved to $\cos \theta_K^{\mathrm{c.m.}}=1$. For the entry in the last row the structure functions were extracted at kinematics of $-t$=0.4~GeV$^2$ and $x$=0.3. The column labeled $N_{\mathrm{bin}}$ indicates the number of kinematic bins included in the analysis.}
\begin{tabular}{ccccccccc} \hline\hline
Observables                                                            & Final State(s)                               & $Q^2$ (GeV$^2$)        & $W$ (GeV)    & $\cos \theta_K^{\mathrm{c.m.}}$ & $N_{\mathrm{bin}}$                 & Facility & Year  & Ref. \\ \hline
\multirow{3}{*}{$\sigma_{\rm U}$, $\sigma_{\rm L}$, $\sigma_{\rm T}$}  & \multirow{3}{*}{$K^+\Lambda$, $K^+\Sigma^0$} & 0.52, 0.75, 1.00, 2.00 & 1.84         & 1                      & 4 $\Lambda$, 4 $\Sigma^0$ & Hall C   &  1998 & \cite{Niculescu:1998zj} \\
                                                                       &                                              & 0.52, 0.75, 1.00, 2.00 & 1.84         & 1                      & 4 $\Lambda$, 4 $\Sigma^0$ & Hall C   &  2003 & \cite{E93018:2002cpu} \\
                                                                       &                                              & 1.90, 2.35             & $[1.8:2.14]$ & 1                      & 8 $\Lambda$, 8 $\Sigma^0$ & Hall A   &  2010 & \cite{Coman:2009dot} \\ \hline 
$\sigma_{\rm L}$, $\sigma_{\rm T}$                                     & $K^+\Lambda$, $K^+\Sigma^0$                  & 1.00, 1.36, 2.00       & --           & --                     & 3 $\Lambda$, 3 $\Sigma^0$ & Hall C   &  2018 & \cite{Carmignotto:2018uqj} \\ \hline\hline
\end{tabular}
\label{halla-c-list}
\end{center}
\end{table}
%%%%%%%%%%%%%%%%%%%%%%%%%%%%%%%%%%%%%%%%%%%%%%%%%%%%%%%%%%%%%%%%%%%%%%%%%%%%%%%%%%%%%%%%%%%%%%%%%%%%%%%%%%%%%%%%%%%%%%%%%%%%%%%%%%%%%%%%%%%%%%%%%%%%%%%%%%%%%%%%

%%%%%%%%%%%%%%%%%%%%%%%%%%%%%%%%%%%%%%%%%%%%%%%%%%%%%%%%%%%%%%%%%%%%%%%%%%%%%%%%%%%%%%%%%%%%%%%%%%%%%%%%%%%%%%%%%%%%%%%%%%%%%%%%%%%%%%%%%%%%%%%%%%%%%%%%%%%%%%%%
\begin{figure*}[htbp]
\centering
\includegraphics[width=0.35\textwidth]{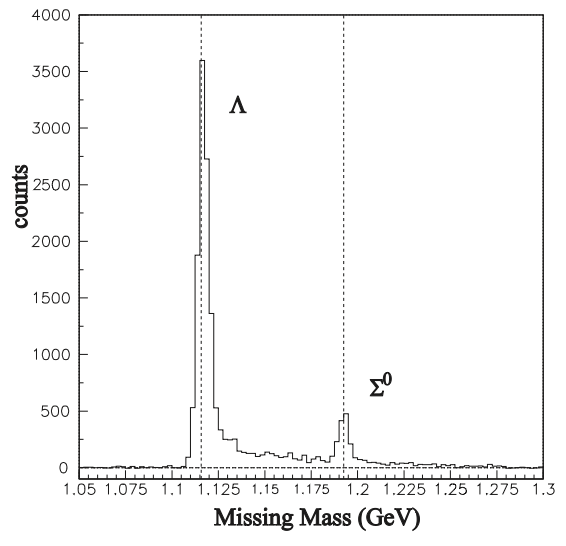} 
\caption{An example $MM(e'K^+)$ spectrum from a Hall~C measurement showing well separated $\Lambda$ and $\Sigma^0$ hyperon peaks. The vertical lines define the hyperon masses. This spectrum has been background subtracted to remove accidental coincidences and cryotarget window contributions. Figure from Ref.~\cite{E93018:2002cpu}.}
\label{hallc-mm}
\end{figure*}
%%%%%%%%%%%%%%%%%%%%%%%%%%%%%%%%%%%%%%%%%%%%%%%%%%%%%%%%%%%%%%%%%%%%%%%%%%%%%%%%%%%%%%%%%%%%%%%%%%%%%%%%%%%%%%%%%%%%%%%%%%%%%%%%%%%%%%%%%%%%%%%%%%%%%%%%%%%%%%%%

\begin{figure*}[htbp]
\centering
\includegraphics[width=0.6\textwidth]{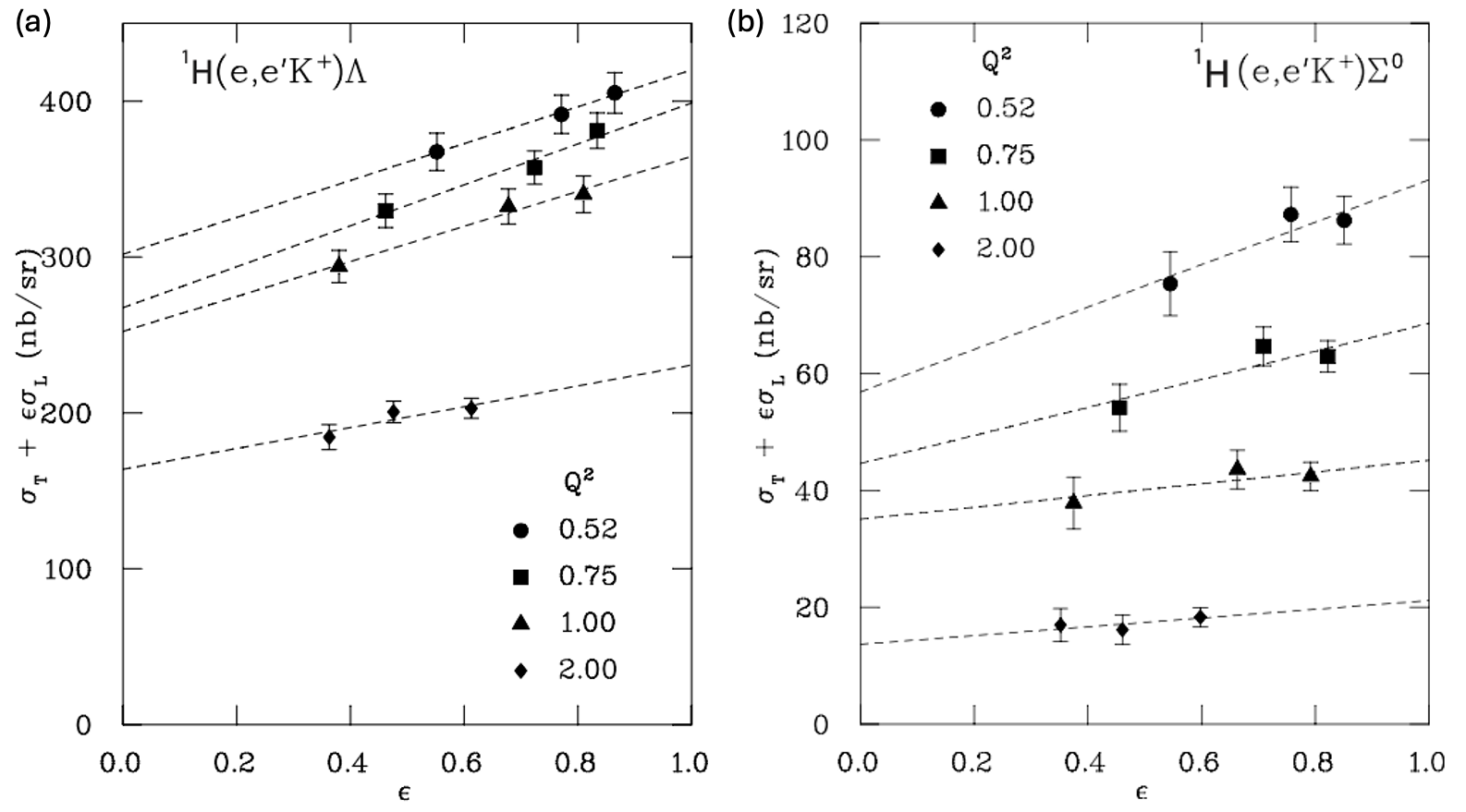}
\caption{Differential cross sections as a function of the transverse virtual photon polarization parameter $\epsilon$ (see Eq.~(\ref{defeps})) for the exclusive electroproduction of (a) $K^+\Lambda$ and  (b) $K^+\Sigma^0$ final states for four different values of $Q^2$ (in GeV$^2$). These data are from a reanalysis of the experimental data used as the basis for Ref.~\cite{Niculescu:1998zj} and correspond to an average $W$ of 1.84 GeV. These Hall~C data enabled separate extraction of $\sigma_{\rm T}$ and $\sigma_{\rm L}$ evolved to $\cos \theta_K^{\mathrm{c.m.}} = 1$ from a linear fit of $\sigma_{\rm U}$ vs. $\epsilon$. Figures from Ref.~\cite{E93018:2002cpu}.}
\label{mohring-data}
\end{figure*}

An example of the quality of the reconstructed hyperon spectrum from the Hall~C measurements is shown in Fig.~\ref{hallc-mm}. Here the $e'K^+$ missing mass distribution is provided, showing well separated $\Lambda$ and $\Sigma^0$ peaks with minimal underlying background. This highlights the spectrometer response function with very good particle identification capabilities (high efficiency and high purity) for the momentum range of the final state particles at beam energies up to 4~GeV. Figure~\ref{mohring-data} highlights the structure functions included in Ref.~\cite{E93018:2002cpu}, demonstrating the quality of the extractions possible in these analyses that combined absolute cross section data from multiple beam energy datasets. These pioneering measurements were from the first year of 4-GeV operations in Hall~C.

%%%%%%%%%%%%%%%%%%%%%%%%%%%%%%%%%%%%%%%%%%%%%%%%%%%%%%%%%%%%%%%%%%%%%%%%%%%%%%%%%%%%%%%%%%%%%%%%%%%%%%%%%%%%%%%%%%%%%%%%%%%%%%%%%%%%%%%%%%%%%%%%%%%%%%%%%%%%%%%%
%%%%%%%%%%%%%%%%%%%%%%%%%%%%%%%%%%%%%%%%%%%%%%%%%%%%%%%%%%%%%%%%%%%%%%%%%%%%%%%%%%%%%%%%%%%%%%%%%%%%%%%%%%%%%%%%%%%%%%%%%%%%%%%%%%%%%%%%%%%%%%%%%%%%%%%%%%%%%%%%

The extraction of $\sigma_{\rm L}$ from the unseparated differential cross sections was also used to attempt an initial determination of the $K^+$ electromagnetic form factor in Ref.~\cite{Carmignotto:2018uqj}. The extraction of $F_K$ was carried out by comparing the measured longitudinal structure function $\sigma_{\rm L}$ to that of a Regge model based on $K$ and $K^*$ exchange with the form factor parameterized by a monopole form, $F_K = (1 + Q^2/\Lambda_K^2)^{-1}$, where the Regge model parameterization of $\sigma_{\rm L}$ is given by~\cite{Vanderhaeghen:1997ts}

\begin{equation}
\label{eq:siglff}
\sigma_{\rm L} \approx \frac{-2 t Q^2}{(t-m_K^2)^2} g_{KYN}^2(t) F_K^2(Q^2,t).
\end{equation}

\begin{figure*}[htbp]
\centering
\includegraphics[width=0.4\textwidth]{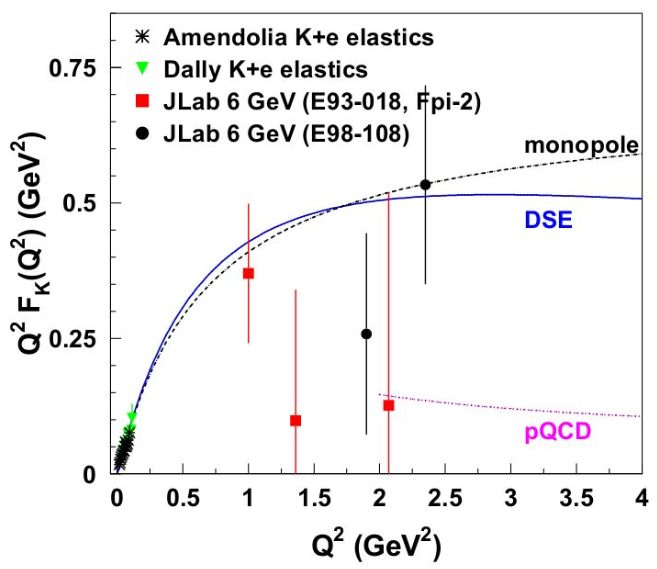}
\caption{Charged kaon form factor vs. $Q^2$ from JLab Hall~C measurements of $\sigma_{\rm L}$ (red squares and black circles). The dashed black curve shows the monopole form factor distribution~\cite{Amendolia:1986ui}; the solid blue line shows the form factor from a Dyson-Schwinger model and the dotted pink curve corresponds to a leading-order pQCD calculation~\cite{Gao:2017mmp}. Figure from Ref.~\cite{Carmignotto:2018uqj}.}
\label{kaonff-hallc}
\end{figure*}

\noindent
Assuming the Regge trajectory cut-off parameters $\Lambda_K = \Lambda_{K^*}$, $F_K$ was determined from a least squares fit of the Regge model prediction to the data with the coupling constant $g_{KYN}$ based on SU(3) constraints. Figure~\ref{kaonff-hallc} shows the extracted $K^+$ form factor vs. $Q^2$ compared to several predictions. These first extractions of $F_K$ from electroproduction data served as a proof of principle for the development of a second-generation $K^+$ form factor measurement in Hall C in the JLab 12-GeV era~\cite{kaonlt} (see Section~\ref{jlab:12gev}).

%%%%%%%%%%%%%%%%%%%%%%%%%%%%%%%%%%%%%%%%%%%%%%%%%%%%%%%%%%%%%%%%%%%%%%%%%%%%%%%%%%%%%%%%%%%%%%%%%%%%%%%%%%%%%%%%%%%%%%%%%%%%%%%%%%%%%%%%%%%%%%%%%%%%%%%%%%%%%%%%
%%%%%%%%%%%%%%%%%%%%%%%%%%%%%%%%%%%%%%%%%%%%%%%%%%%%%%%%%%%%%%%%%%%%%%%%%%%%%%%%%%%%%%%%%%%%%%%%%%%%%%%%%%%%%%%%%%%%%%%%%%%%%%%%%%%%%%%%%%%%%%%%%%%%%%%%%%%%%%%%

\begin{figure*}[htbp]
\centering
\includegraphics[width=0.60\textwidth]{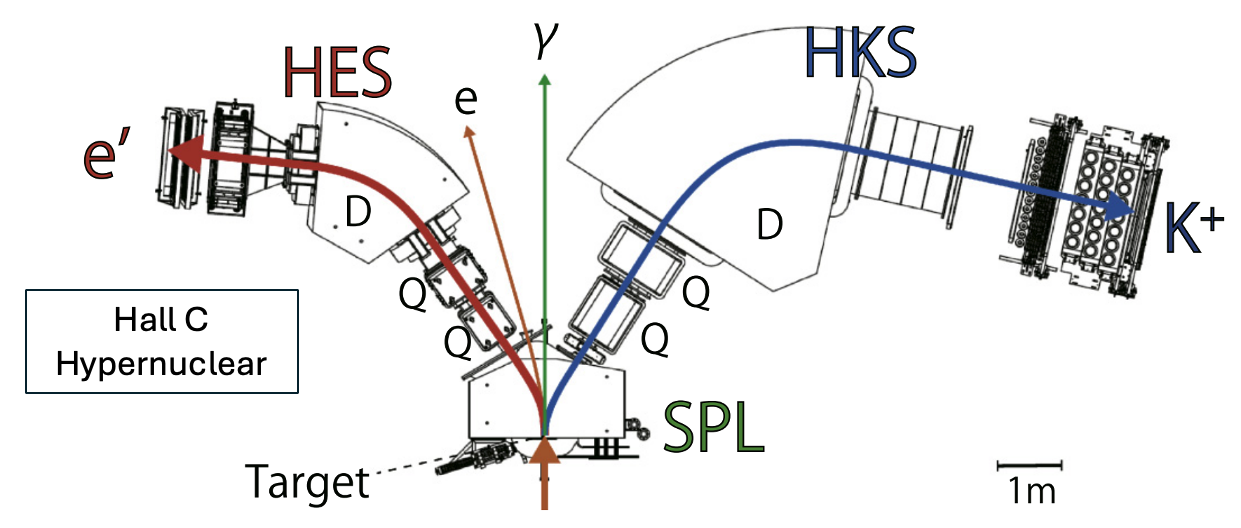}
\caption{Schematic of the JLab hypernuclear spectroscopy configuration in Hall~C. The setup consists of the HES electron arm, the HKS kaon arm, and a dipole (SPL) for charge separation at the very forward angle positioning of the spectrometers. Figure from Ref.~\cite{Gogami:2015tvu}.}
\label{hallc_hyper}
\end{figure*}

Another aspect of the strangeness physics program focuses on hypernuclear studies, which have been carried out in both Halls A and C, tagging production via the $(e,e'K^+)$ missing mass with both spectrometers positioned at very forward angles. The Hall A setup relied on the two HRS spectrometers to measure the scattered electron and kaon. The setup used in Hall C has evolved through multiple generations of spectrometer configurations based on experience gained through each subsequent study. The initial Hall C experiment was configured with an ENGE split-pole electron spectrometer (the HMS spectrometer was not used) and the SOS as a kaon spectrometer \cite{HNSS:2002max,HNSS:2004rep}. The data rate was limited by the accidental backgrounds from bremsstrahlung and the dominant production of M{\o}ller electrons at forward angles. The final optimized configuration employed a new high resolution electron spectrometer (HES) and a new high resolution kaon spectrometer (HKS)~\cite{HKS:2014nmp} (with neither the HMS nor SOS used). Figure~\ref{hallc_hyper} shows the final 6-GeV era hypernuclear configuration installed in Hall~C. The measured binding energy spectra ultimately achieved sub-MeV resolution and data were taken on targets with mass number $A$ = 7 to 52. 

The key motivation of the hypernuclear program at JLab is to understand the behavior of hyperons within (effective) nuclear matter to study the two-body $YN$ interaction, which can be probed through the measurement of the hypernuclear binding energies. These data are also efficacious to gain information about the three-body $YNN$ interaction, which is required to determine the equation of state of dense nuclear matter that governs neutron stars~\cite{Le:2024rkd}. Additional information is included in Section~\ref{sec:hypernuclei}.

%%%%%%%%%%%%%%%%%%%%%%%%%%%%%%%%%%%%%%%%%%%%%%%%%%%%%%%%%%%%%%%%%%%%%%%%%%%%%%%%%%%%%%%%%%%%%%%%%%%%%%%%%%%%%%%%%%%%%%%%%%%%%%%%%%%%%%%%%%%%%%%%%%%%%%%%%%%%%%%%
%%%%%%%%%%%%%%%%%%%%%%%%%%%%%%%%%%%%%%%%%%%%%%%%%%%%%%%%%%%%%%%%%%%%%%%%%%%%%%%%%%%%%%%%%%%%%%%%%%%%%%%%%%%%%%%%%%%%%%%%%%%%%%%%%%%%%%%%%%%%%%%%%%%%%%%%%%%%%%%%

\begin{figure*}[htbp]
\centering
\includegraphics[width=0.9\textwidth]{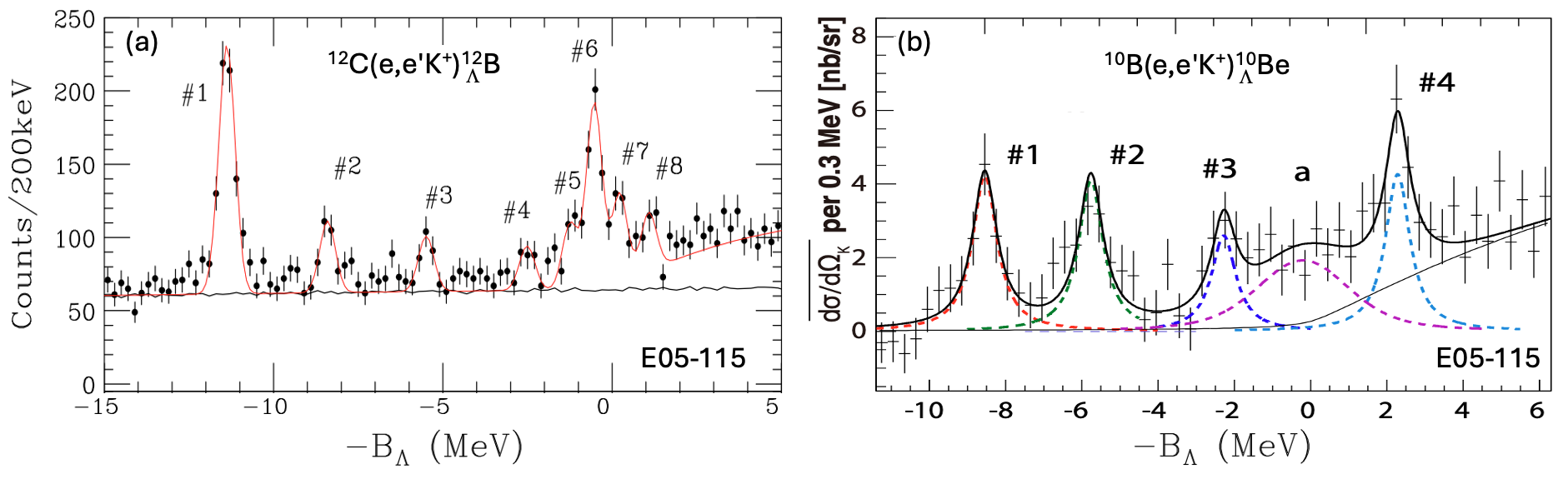}
\caption{Binding energy spectra for (a) $^{12}_{~\Lambda}$B and (b) $^{10}_{~\Lambda}$Be from the final JLab Hall~C 6-GeV era experiments of the hypernuclear program. The fit functions are shown on each plot. Figures from Refs.~\cite{HKS:2014nmp} (a) and \cite{Gogami:2015tvu} (b).}
\label{hyper-plots}
\end{figure*}

Several representative spectra from the JLab hypernuclear spectroscopy program are highlighted in Fig.~\ref{hyper-plots}. Here the $^{12}_{~\Lambda}$B spectrum is shown as it represents the most characteristic $p$-shell hypernucleus and is now commonly used for calibration, along with the $^{10}_{~\Lambda}$Be spectrum. In these plots the $\Lambda$ binding energy $B_\Lambda$ is computed as
\begin{eqnarray}
-B_\Lambda &=& M_H - M_\Lambda - M_{\rm core}, \nonumber \\
M_H &=& \sqrt{(E_e + M_{\rm target} - E_K - E_{e'})^2 - (\vec{p}_e - \vec{p}_K - \vec{p}_{e'})^2},
\end{eqnarray}

\noindent
with $M_{\rm core}$ the core nucleus mass of the hypernucleus.

%%%%%%%%%%%%%%%%%%%%%%%%%%%%%%%%%%%%%%%%%%%%%%%%%%%%%%%%%%%%%%%%%%%%%%%%%%%%%%%%%%%%%%%%%%%%%%%%%%%%%%%%%%%%%%%%%%%%%%%%%%%%%%%%%%%%%%%%%%%%%%%%%%%%%%%%%%%%%%%%
%%%%%%%%%%%%%%%%%%%%%%%%%%%%%%%%%%%%%%%%%%%%%%%%%%%%%%%%%%%%%%%%%%%%%%%%%%%%%%%%%%%%%%%%%%%%%%%%%%%%%%%%%%%%%%%%%%%%%%%%%%%%%%%%%%%%%%%%%%%%%%%%%%%%%%%%%%%%%%%%

\subsubsection{Hall B -- CLAS Photoproduction Program}
\label{clas-gp-program}

\begin{figure*}[htbp]
\centering
\includegraphics[width=0.4\textwidth]{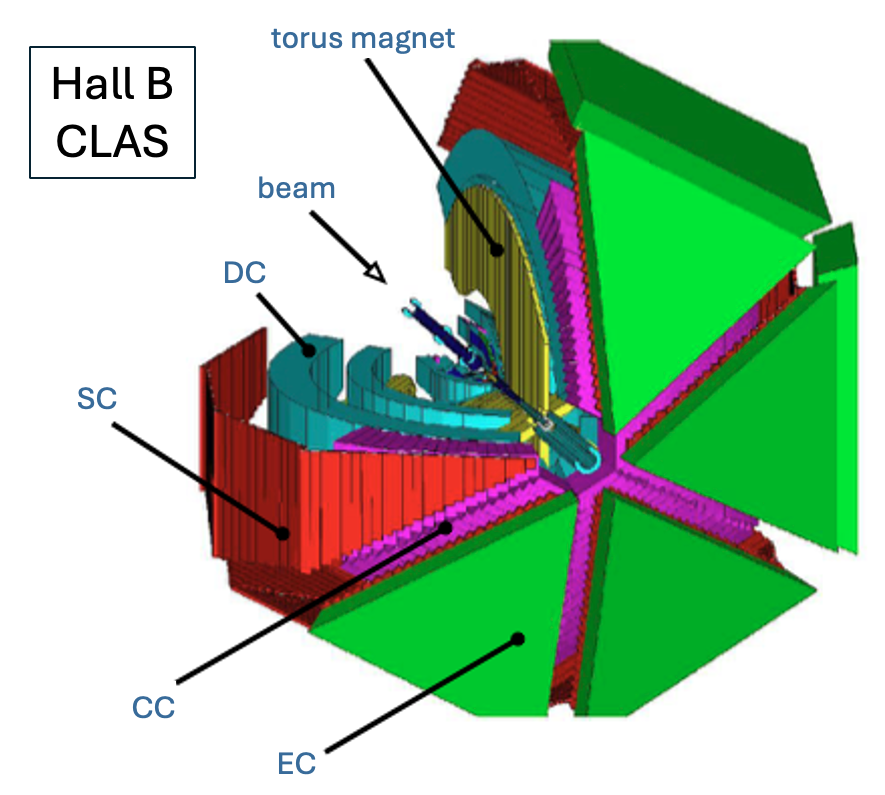}
\caption{Schematic view of the CLAS detector with the different subsystems labeled (DC: drift chamber, SC: scintillation counters, CC: Cherenkov counters, EC: electromagnetic calorimeter). A single sector of the detector has been cut away to enable a view of the inner subsystems and torus magnet. The diameter of the CLAS detector was $\sim$10 m. Figure adapted from Ref.~\cite{CLAS:2009sbn}.}
\label{hallb-equipment}
\end{figure*}

The large acceptance CLAS spectrometer in Hall B~\cite{CLAS:2003umf} was operational in the period from 1997 to 2012. The spectrometer was constructed around six superconducting coils that generated a toroidal magnetic field to momentum analyze charged particles. The torus field had $\int B d\ell = 2.5$~Tm in the forward direction dropping to 0.6~Tm at 90$^\circ$. The detector spanned polar angles from 8$^\circ$ to 142$^\circ$, while covering nearly 80\% of the azimuth. The momentum resolution was $\Delta p/p \sim 0.5\%$ with angular resolution $\Delta \theta, \Delta \phi \sim 2$~mrad. The facility was designed to operate with polarized electron and photon beams with both unpolarized and polarized targets. With electron beam operations the nominal beam-target luminosity was $\sim$$1\times10^{34}$~cm$^{-2}$s$^{-1}$, while with photon beam operations the tagged photon rate was $\sim$$1\times10^7$~$\gamma$/s. For electron beam experiments, a normal-conducting mini-torus was installed outside of the target and before the first layer of tracking detectors to serve as a magnetic shield from the intense M{\o}ller background created from beam interactions with the atomic electrons of the target. For photon beam experiments, the mini-torus was replaced with a segmented scintillation start counter. Figure~\ref{hallb-equipment} shows a model of the CLAS detector.

%%%%%%%%%%%%%%%%%%%%%%%%%%%%%%%%%%%%%%%%%%%%%%%%%%%%%%%%%%%%%%%%%%%%%%%%%%%%%%%%%%%%%%%%%%%%%%%%%%%%%%%%%%%%%%%%%%%%%%%%%%%%%%%%%%%%%%%%%%%%%%%%%%%%%%%%%%%%%%%%
%%%%%%%%%%%%%%%%%%%%%%%%%%%%%%%%%%%%%%%%%%%%%%%%%%%%%%%%%%%%%%%%%%%%%%%%%%%%%%%%%%%%%%%%%%%%%%%%%%%%%%%%%%%%%%%%%%%%%%%%%%%%%%%%%%%%%%%%%%%%%%%%%%%%%%%%%%%%%%%%

The baseline equipment in Hall B for photoproduction experiments included a facility upstream of CLAS that tagged photons with energies between 20\% and 95\% of the incident electron beam energy~\cite{Sober:2000we}. Unpolarized, circularly polarized, and linearly polarized beams were available. The CLAS photoproduction program focusing on strangeness production included measurements at beam energies up to 6~GeV. The different experiments measured both cross sections and polarization observables and are cataloged in Tables~\ref{hallb-clas-lista}--\ref{hallb-clas-listc}.

%%%%%%%%%%%%%%%%%%%%%%%%%%%%%%%%%%%%%%%%%%%%%%%%%%%%%%%%%%%%%%%%%%%%%%%%%%%%%%%%%%%%%%%%%%%%%%%%%%%%%%%%%%%%%%%%%%%%%%%%%%%%%%%%%%%%%%%%%%%%%%%%%%%%%%%%%%%%%%%
\begin{table}[tbh]
\setlength{\tabcolsep}{6pt} % Default value: 6pt
\renewcommand{\arraystretch}{0.8} % Default value: 1
\begin{center}
\caption{Summary of $\gamma p \to K\Lambda$ and $\gamma p \to K\Sigma$ photoproduction measurements in Hall B with CLAS from the 6-GeV era experiments at JLab. The column labeled $N_{\mathrm{bin}}$ indicates the number of kinematic bins included in the analysis.}
\begin{tabular}{cccccccc} \hline\hline
Observable                    & Final State(s)              & $W$ (GeV)     & $\cos \theta_K^{\mathrm{c.m.}}$   & $N_{\mathrm{bin}}$                        & Year & Ref. \\ \hline
\multirow{4}{*}{$d\sigma$}    & $K^+\Lambda$, $K^+\Sigma^0$ & $[1.6:2.3]$   & $[-0.9:0.9]$             & 1140 $\Lambda$, 1020 $\Sigma^0$  & 2004 & \cite{CLAS:2003zrd} \\
                              & $K^+\Lambda$, $K^+\Sigma^0$ & $[1.6:2.5]$   & $[-0.9:0.9]$             & 1494 $\Lambda$, 1386 $\Sigma^0$  & 2006 & \cite{CLAS:2005lui} \\
                              & $K^+\Lambda$                & $[1.6:2.84]$  & $[-0.9:0.9]$             & 2076                             & 2010 & \cite{CLAS:2009rdi} \\
                              & $K^+\Sigma^0$               & $[1.7:2.5]$   & $[-0.9:0.9]$             & 2089                             & 2010 & \cite{CLAS:2010aen} \\ \hline
\multirow{5}{*}{$P$}          & $K^+\Lambda$, $K^+\Sigma^0$ & $[1.6:2.3]$   & $[-0.9:0.9]$             & 220 $\Lambda$, 70 $\Sigma^0$     & 2004 & \cite{CLAS:2003zrd} \\
                              & $K^+\Lambda$                & $[1.6:2.84]$  & $[-0.9:0.9]$             & 1707                             & 2010 & \cite{CLAS:2009rdi} \\
                              & $K^+\Sigma^0$               & $[1.7:2.5]$   & $[-0.9:0.9]$             & 455                              & 2010 & \cite{CLAS:2010aen} \\ 
                              & $K^0_S\Sigma^+$             & $[1.75:2.73]$ & $[-0.9:0.7]$             & 113                              & 2013 & \cite{CLAS:2013owj} \\
                              & $K^0_S \Sigma^+$            & $[1.8:2.1]$   & $[-0.8:0.7]$             & 21                               & 2025 & \cite{CLAS:2024bzi} \\ \hline 
\multirow{2}{*}{$C_x$, $C_z$} & $K^+\Lambda$, $K^+\Sigma^0$ & $[1.6:2.5]$   & $[-0.9:0.9]$             & 196 $\Lambda$, 81 $\Sigma^0$     & 2006 & \cite{CLAS:2006pde} \\
                              & $K^+\Lambda$                & $[1.7:3.3]$   & $[-0.85:0.95]$           & 410                              & 2025 & \cite{CLAS:2025azh} \\ \hline 
\multirow{2}{*}{$O_x$, $O_z$} & $K^+\Lambda$, $K^+\Sigma^0$ & $[1.7:2.2]$   & $[-0.75:0.85]$           & 323 $\Lambda$, 120 $\Sigma^0$    & 2016 & \cite{CLAS:2016wrl} \\ 
                              & $K^0_S \Sigma^+$            & $[1.8:2.1]$   & $[-0.8:0.7]$             & 21                               & 2025 & \cite{CLAS:2024bzi} \\ \hline 
\multirow{2}{*}{$\Sigma$}     & $K^+\Lambda$, $K^+\Sigma^0$ & $[1.7:2.2]$   & $[-0.75:0.85]$           & 323 $\Lambda$, 120 $\Sigma^0$    & 2016 & \cite{CLAS:2016wrl} \\ 
                              & $K^0_S \Sigma^+$            & $[1.8:2.1]$   & $[-0.8:0.7]$             & 21                               & 2025 & \cite{CLAS:2024bzi} \\ \hline 
\multirow{2}{*}{$T$}          & $K^+\Lambda$, $K^+\Sigma^0$ & $[1.7:2.2]$   & $[-0.75:0.85]$           & 323 $\Lambda$, 120 $\Sigma^0$    & 2016 & \cite{CLAS:2016wrl} \\ 
                              & $K^0_S \Sigma^+$            & $[1.8:2.1]$   & $[-0.8:0.7]$             & 21                               & 2025 & \cite{CLAS:2024bzi} \\ \hline\hline 
\end{tabular}
\label{hallb-clas-lista}
\end{center}
\end{table}
%%%%%%%%%%%%%%%%%%%%%%%%%%%%%%%%%%%%%%%%%%%%%%%%%%%%%%%%%%%%%%%%%%%%%%%%%%%%%%%%%%%%%%%%%%%%%%%%%%%%%%%%%%%%%%%%%%%%%%%%%%%%%%%%%%%%%%%%%%%%%%%%%%%%%%%%%%%%%%%%

%%%%%%%%%%%%%%%%%%%%%%%%%%%%%%%%%%%%%%%%%%%%%%%%%%%%%%%%%%%%%%%%%%%%%%%%%%%%%%%%%%%%%%%%%%%%%%%%%%%%%%%%%%%%%%%%%%%%%%%%%%%%%%%%%%%%%%%%%%%%%%%%%%%%%%%%%%%%%%%
\begin{table}[tbh]
\setlength{\tabcolsep}{6pt} % Default value: 6pt
\renewcommand{\arraystretch}{0.8} % Default value: 1
\begin{center}
\caption{Summary of $\gamma n \to K\Lambda$ and $\gamma n \to K\Sigma$ photoproduction measurements in Hall B with CLAS from the 6-GeV era experiments at JLab. The column labeled $N_{\mathrm{bin}}$ indicates the number of kinematic bins included in the analysis.}
\begin{tabular}{cccccccc} \hline\hline
Observable                 & Final State                  & $W$ (GeV)     & $\cos \theta_K^{\mathrm{c.m.}}$ & $N_{\mathrm{bin}}$                 & Year & Ref. \\ \hline
\multirow{2}{*}{$d\sigma$} & $K^+\Sigma^-$                & $[1.67:2.75]$ & $[-0.85:0.85]$         & 396                       & 2010 & \cite{CLAS:2009fmu} \\
                           & $K^0\Lambda$                 & $[1.64:2.34]$ & $[-0.7:0.8]$           & 225                       & 2017 & \cite{CLAS:2017gsu} \\ \hline
$\Sigma$                   & $K^+\Sigma^-$                & $[1.7:2.3]$   & $[-0.8:0.8]$           & 246                       & 2022 & \cite{CLAS:2021hex} \\ \hline
\multirow{2}{*}{$E$}       & $K^0\Lambda$, $K^0 \Sigma^0$ & $[1.7:2.34]$  & $[-0.7:0.7]$           & 6 $\Lambda$, 6 $\Sigma^0$ & 2018 & \cite{CLAS:2018gxz} \\
                           & $K^+\Sigma^-$                & $[1.7:2.3]$   & $[-0.6:0.6]$           & 24                        & 2020 & \cite{CLAS:2020spy} \\ \hline\hline
\end{tabular}
\label{hallb-clas-listb}
\end{center}
\end{table}
%%%%%%%%%%%%%%%%%%%%%%%%%%%%%%%%%%%%%%%%%%%%%%%%%%%%%%%%%%%%%%%%%%%%%%%%%%%%%%%%%%%%%%%%%%%%%%%%%%%%%%%%%%%%%%%%%%%%%%%%%%%%%%%%%%%%%%%%%%%%%%%%%%%%%%%%%%%%%%%%

%%%%%%%%%%%%%%%%%%%%%%%%%%%%%%%%%%%%%%%%%%%%%%%%%%%%%%%%%%%%%%%%%%%%%%%%%%%%%%%%%%%%%%%%%%%%%%%%%%%%%%%%%%%%%%%%%%%%%%%%%%%%%%%%%%%%%%%%%%%%%%%%%%%%%%%%%%%%%%%
\begin{table}[tbh]
\setlength{\tabcolsep}{6pt} % Default value: 6pt
\renewcommand{\arraystretch}{0.8} % Default value: 1
\begin{center}
\caption{Summary of $\gamma p \to K^*Y$ and $\gamma p \to KY^*$ photoproduction measurements in Hall B with CLAS from the 6-GeV era experiments at JLab. The column labeled $N_{\mathrm{bin}}$ indicates the number of kinematic bins included in the analysis.}
\begin{tabular}{cccccccc} \hline\hline
Observable                 & Final State                       & $W$ (GeV)     & $\cos \theta_K^{\mathrm{c.m.}}$   & $N_{\mathrm{bin}}$                   & Year & Ref. \\ \hline
\multirow{7}{*}{$d\sigma$} & $K^{0*}\Sigma^+$                  & $[2.0:2.5]$   & $[-0.75:0.8]$            & 54                          & 2007 & \cite{CLAS:2007kab} \\
                           & $K^{*+}\Lambda$, $K^{*+}\Sigma^0$ & $[2.0:2.9]$   & $[-0.9:0.9]$             & 191 $\Lambda$, 177 $\Sigma$ & 2013 & \cite{CLAS:2013qgi} \\
                           & $K^+\Sigma^0(1385)$               & $[2.0:2.9]$   & $[-0.8:0.8]$             & 144                         & 2013 & \cite{CLAS:2013rxx} \\
                           & $K^+\Lambda(1405)$                & $[2.0:2.9]$   & $[-0.8:0.8]$             & 144                         & 2013 & \cite{CLAS:2013rxx} \\   
                           & $K^+\Lambda(1520)$                & $[2.0:2.9]$   & $[-0.8:0.8]$             & 144                         & 2013 & \cite{CLAS:2013rxx} \\   
                           & $K^{*+}\Lambda$                   & $[2.0:2.8]$   & $[-0.9:0.9]$             & 38                          & 2017 & \cite{CLAS:2017sgi} \\
                           & $K^+\Lambda(1520)$                & $[2.25:3.25]$ & $[-0.8:0.8]$             & 63                          & 2021 & \cite{CLAS:2021osv} \\ \hline
$P$                        & $K^{*+}\Lambda$                   & $[2.0:2.8]$   & $[-0.17:0.84]$           & 21                          & 2017 & \cite{CLAS:2017sgi} \\ \hline\hline
\end{tabular}
\label{hallb-clas-listc}
\end{center}
\end{table}
%%%%%%%%%%%%%%%%%%%%%%%%%%%%%%%%%%%%%%%%%%%%%%%%%%%%%%%%%%%%%%%%%%%%%%%%%%%%%%%%%%%%%%%%%%%%%%%%%%%%%%%%%%%%%%%%%%%%%%%%%%%%%%%%%%%%%%%%%%%%%%%%%%%%%%%%%%%%%%%%

Pseudoscalar meson photoproduction is described by four independent amplitudes. All possible observables are formulated in terms of bilinear combinations of these amplitudes (see Section~\ref{dcs-formalism}). This results in the 16 observable quantities that are shown in Table~\ref{gp-observables}. These observables include the cross section ($\sigma$), three asymmetries that enter into the cross section to leading order scaled by a single polarization observable of either beam, target, or recoil hyperon ($\Sigma$, $T$, $P$), and three sets of four asymmetries that enter into the cross section to leading order involving observables of beam-target polarization ($E$, $G$, $F$, $H$), beam-recoil hyperon polarization ($C_{x'}$, $C_{z'}$, $O_{x'}$, $O_{z'}$), or target-recoil hyperon polarization ($L_{x'}$, $L_{z'}$, $T_{x'}$, $T_{z'}$). The $(x,y,z)$ and $(x',y',z')$ coordinate systems are defined in Fig.~\ref{fig:frames_convention}. Note that with simple rotations, the double-polarization observables can be presented in either the primed or unprimed coordinate systems.

The ultimate experimental outcome for a given reaction channel would be a detailed, finely binned mapping of the 16 pseudoscalar meson photoproduction observables vs. $W$ and $\cos \theta_K^{\mathrm{c.m.}}$. This would enable (in principle) an unambiguous determination of the reaction amplitudes without the need to rely on theoretical models. With the $K^+ \Lambda$ and $K^+ \Sigma^0$ channels from measurements with a proton target, this is becoming closer to reality. There is also hope as part of the existing programs within Hall~B that extensive measurements from an effective neutron target (e.g. $^2$H or $^3$He) will enable the isospin dependence of the $KY$ channels to be studied in detail. Of course, this is a longer-term prospect given the still limited availability of data from a neutron target, the complexities of the data analysis, and the need to properly account for target nuclei Fermi motion and final state interactions. Further discussion on the notion of ``complete'' experiments of this sort is provided in Ref.~\cite{Sandorfi:2010uv}.

%%%%%%%%%%%%%%%%%%%%%%%%%%%%%%%%%%%%%%%%%%%%%%%%%%%%%%%%%%%%%%%%%%%%%%%%%%%%%%%%%%%%%%%%%%%%%%%%%%%%%%%%%%%%%%%%%%%%%%%%%%%%%%%%%%%%%%%%%%%%%%%%%%%%%%%%%%%%%%%
\begin{table}[tbh]
\begin{center}
\caption{Listing of the 16 observables defined for pseudoscalar meson photoproduction. The target polarization components are defined in the $(x,y,z)$ system and the recoil polarization components are nominally defined in the $(x',y',z')$ system. See Section~\ref{dcs-formalism} and Appendix~\ref{app:response_function} for details.}
\begin{tabular}{c|c|ccc|ccc|cccc} \hline \hline
Beam                  &          & \multicolumn{3}{c|}{Target} & \multicolumn{3}{c|}{Recoil}  & \multicolumn{4}{c}{Target + Recoil} \\ \hline
                      & -        & -   & -   & -              & $x'$     & $y'$ & $z'$      & $x'$     & $x'$     & $z'$     & $z'$ \\
                      & -        & $x$ & $y$ & $z$            & -        & -    & -         & $x$      & $z$      & $x$      & $z$ \\ \hline
unpolarized           & $\sigma$ &     & $T$ &                &          & $P$  &           & $T_{x'}$ & $L_{x'}$ & $T_{z'}$ & $L_{z'}$ \\
circularly polarized  &          & $F$ &     & $E$            & $C_{x'}$ &      & $C_{z'}$  &          &          &          & \\
linear polarized      & $\Sigma$ & $H$ &     & $G$            & $O_{x'}$ &      & $O_{z'}$  &          &          &          & \\ \hline \hline
\end{tabular}
\label{gp-observables}
\end{center}
\end{table}
%%%%%%%%%%%%%%%%%%%%%%%%%%%%%%%%%%%%%%%%%%%%%%%%%%%%%%%%%%%%%%%%%%%%%%%%%%%%%%%%%%%%%%%%%%%%%%%%%%%%%%%%%%%%%%%%%%%%%%%%%%%%%%%%%%%%%%%%%%%%%%%%%%%%%%%%%%%%%%%%

\begin{figure}[htbp]
\centering
\includegraphics[width=0.6\columnwidth]{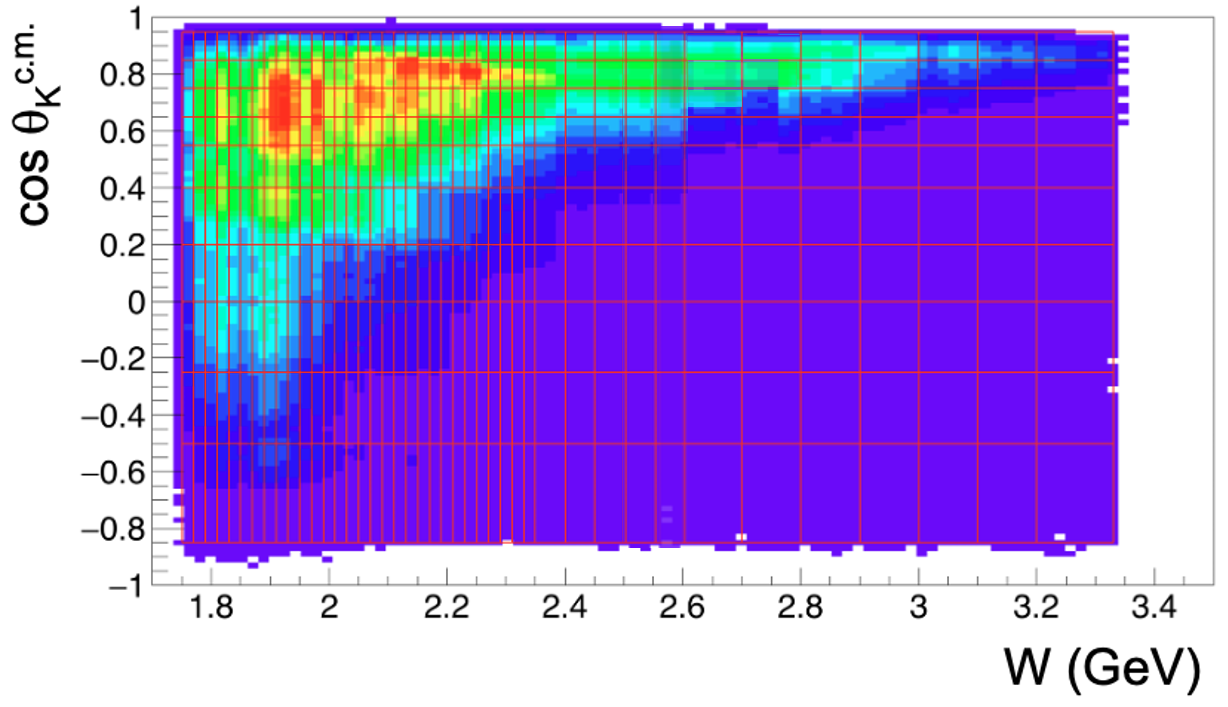}
\caption{Experimental yield distribution in terms of $\cos\theta_K^{\mathrm{c.m.}}$ vs.~$W$ for $K^+\Lambda$ exclusive photoproduction data from the final CLAS photoproduction run using a liquid-hydrogen target. The binning limits of that analysis are overlaid. Figure from Ref.~\cite{CLAS:2025azh}.}
\label{g12:bin}
\end{figure}

The 6-GeV era CLAS $KY$ photoproduction program is extensive, providing measurements of $K^+\Lambda$ and $K^+\Sigma^0$ off the proton over a broad range of $W$ up to 3~GeV and spanning essentially the full range of $\cos \theta_K^{\mathrm{c.m.}}$ (see Fig.~\ref{g12:bin}). These observables fully dominate the available world data and have proven valuable in constraining the reaction mechanism for strangeness production and accounting for and revealing the resonant and non-resonant contributions. 

%%%%%%%%%%%%%%%%%%%%%%%%%%%%%%%%%%%%%%%%%%%%%%%%%%%%%%%%%%%%%%%%%%%%%%%%%%%%%%%%%%%%%%%%%%%%%%%%%%%%%%%%%%%%%%%%%%%%%%%%%%%%%%%%%%%%%%%%%%%%%%%%%%%%%%%%%%%%%%%%
%%%%%%%%%%%%%%%%%%%%%%%%%%%%%%%%%%%%%%%%%%%%%%%%%%%%%%%%%%%%%%%%%%%%%%%%%%%%%%%%%%%%%%%%%%%%%%%%%%%%%%%%%%%%%%%%%%%%%%%%%%%%%%%%%%%%%%%%%%%%%%%%%%%%%%%%%%%%%%%%

\begin{figure}[htbp]
\centering
\includegraphics[width=0.8\columnwidth]{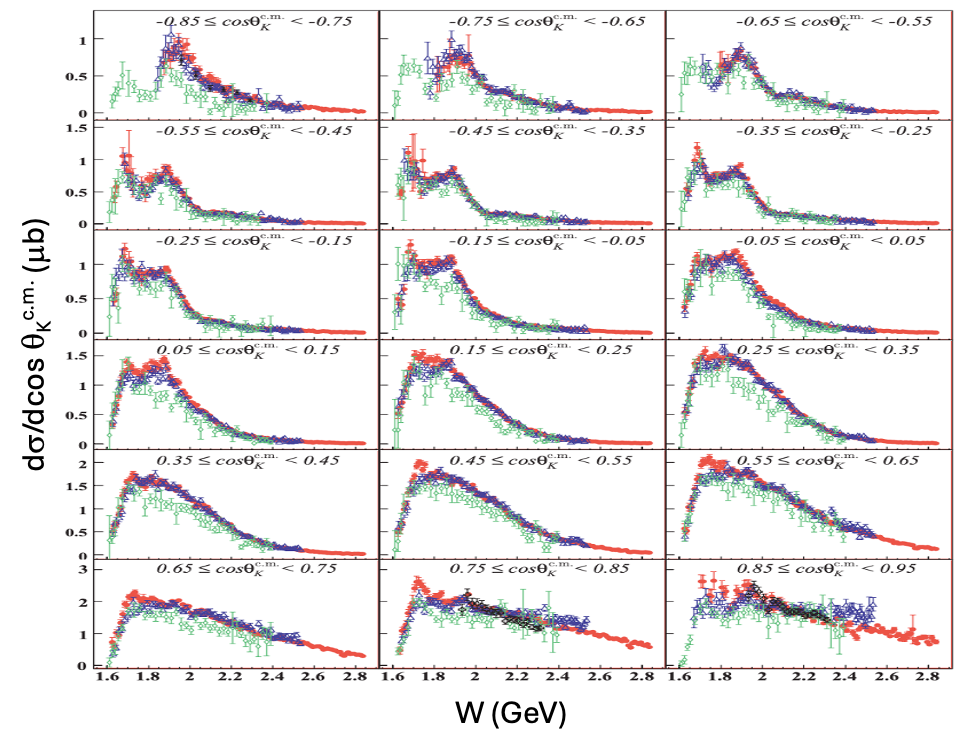}
\caption{Differential cross sections for $K^+\Lambda$ photoproduction vs. $W$ binned in $\cos \theta_K^{\mathrm{c.m.}}$. The plot includes data from CLAS~\cite{CLAS:2005lui,CLAS:2009rdi} (open blue triangles, red circles), SAPHIR~\cite{Glander:2003jw} (open green diamonds), and LEPS~\cite{Sumihama:2005mt,PhysRevC.76.042201} (open black circles). Figure adapted from Ref.~\cite{CLAS:2009rdi}.}
\label{clas-kl-dcs}
\end{figure}

\begin{figure}[htbp]
\centering
\includegraphics[width=0.8\columnwidth]{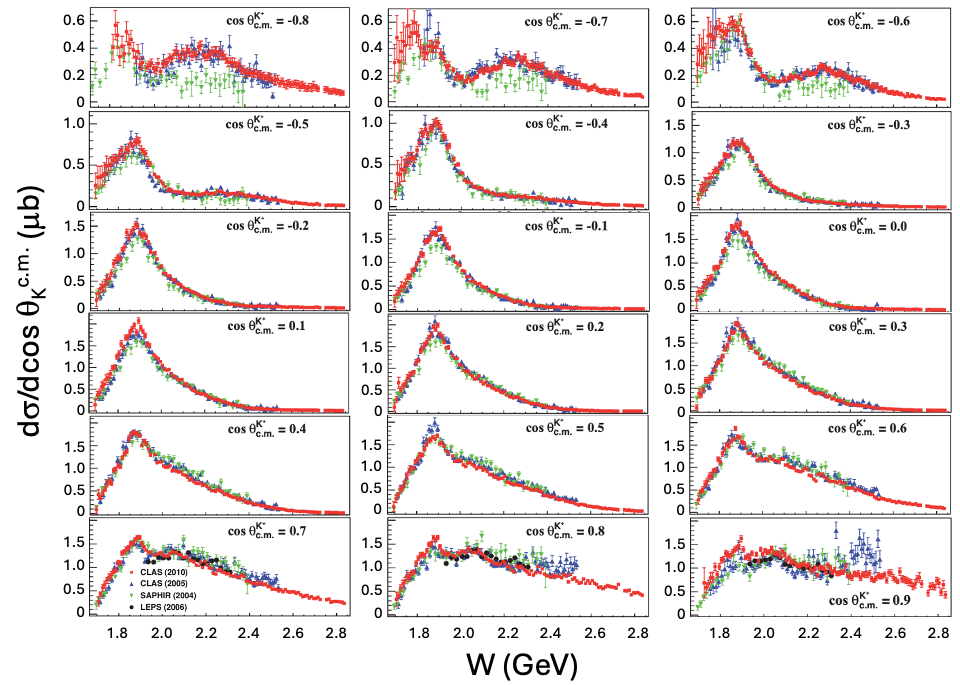}
\caption{Differential cross sections for $K^+\Sigma^0$ photoproduction vs. $W$ binned in $\cos \theta_K^{\mathrm{c.m.}}$. The plot includes data from CLAS~\cite{CLAS:2005lui,CLAS:2010aen} (blue up-triangles, red squares), SAPHIR~\cite{Glander:2003jw} (green down-triangles), and LEPS~\cite{Kohri:2006yx} (black circles). Figure adapted from Ref.~\cite{CLAS:2010aen}.}
\label{clas-ks-dcs}
\end{figure}

Given the significant number of CLAS publications of different observables in the $KY$ sector, only a few key results are highlighted here that have proven essential for the development of reaction models. Figures~\ref{clas-kl-dcs} and \ref{clas-ks-dcs} show comparative plots of the available CLAS $K^+\Lambda$ and $K^+\Sigma^0$ photoproduction differential cross sections from Refs.~\cite{CLAS:2009rdi} and \cite{CLAS:2010aen}, respectively, that were finely binned in $W$ ($\Delta W = 10$~MeV) and $\cos \theta_K^{\mathrm{c.m.}}$ ($\Delta \cos \theta_K^{\mathrm{c.m.}} = 0.1$). These data span the full nucleon resonance region where new $N^*$ states could be discovered. Note, in general, when referring to excited nucleon states, $N^*$ can represent either isospin-1/2 or isospin-3/2 states.

%%%%%%%%%%%%%%%%%%%%%%%%%%%%%%%%%%%%%%%%%%%%%%%%%%%%%%%%%%%%%%%%%%%%%%%%%%%%%%%%%%%%%%%%%%%%%%%%%%%%%%%%%%%%%%%%%%%%%%%%%%%%%%%%%%%%%%%%%%%%%%%%%%%%%%%%%%%%%%%%
%%%%%%%%%%%%%%%%%%%%%%%%%%%%%%%%%%%%%%%%%%%%%%%%%%%%%%%%%%%%%%%%%%%%%%%%%%%%%%%%%%%%%%%%%%%%%%%%%%%%%%%%%%%%%%%%%%%%%%%%%%%%%%%%%%%%%%%%%%%%%%%%%%%%%%%%%%%%%%%%

%%%%%%%%%%%%%%%%%%%%%%%%%%%%%%%%%%%%%%%%%%%%%%%%%%%%%%%%%%%%%%%%%%%%%%%%%%%%%%%%%%%%%%%%%%%%%%%%%%%%%%%%%%%%%%%%%%%%%%%%%%%%%%%%%%%%%%%%%%%%%%%%%%%%%%%%%%%%%%%%
%%%%%%%%%%%%%%%%%%%%%%%%%%%%%%%%%%%%%%%%%%%%%%%%%%%%%%%%%%%%%%%%%%%%%%%%%%%%%%%%%%%%%%%%%%%%%%%%%%%%%%%%%%%%%%%%%%%%%%%%%%%%%%%%%%%%%%%%%%%%%%%%%%%%%%%%%%%%%%%%

The $K^+\Lambda$ cross section data show clear evidence of resonance-like contributions at $W \approx 1.7$~GeV and 1.9~GeV in the backward angle range revealing further details than shown in the total cross sections of Fig.~\ref{ky-sigtot}. At forward angles the cross section increases rapidly with diminishing resonance signatures due to the dominance of $t$-channel $K$ and $K^*$ exchanges. The $K^+\Sigma^0$ cross section data also show strong evidence of resonance contributions at $W \approx 1.9$~GeV and $W \approx 2.25$~GeV. As $s$-channel contributions are much more important in this channel compared to $K^+\Lambda$, the strong forward peaking of the cross sections seen in the $K^+ \Lambda$ channel is not present. With the broad angular coverage of the data spanning the full range of $\cos \theta_K^{\mathrm{c.m.}}$, these data, in conjunction with a reaction model that describes the data, allow for separation of the contributing $s$-, $t$-, and $u$-channel processes necessary to understand the associated $KY$ reaction mechanism.

\begin{figure}[htbp]
\centering
\includegraphics[width=0.75\columnwidth]{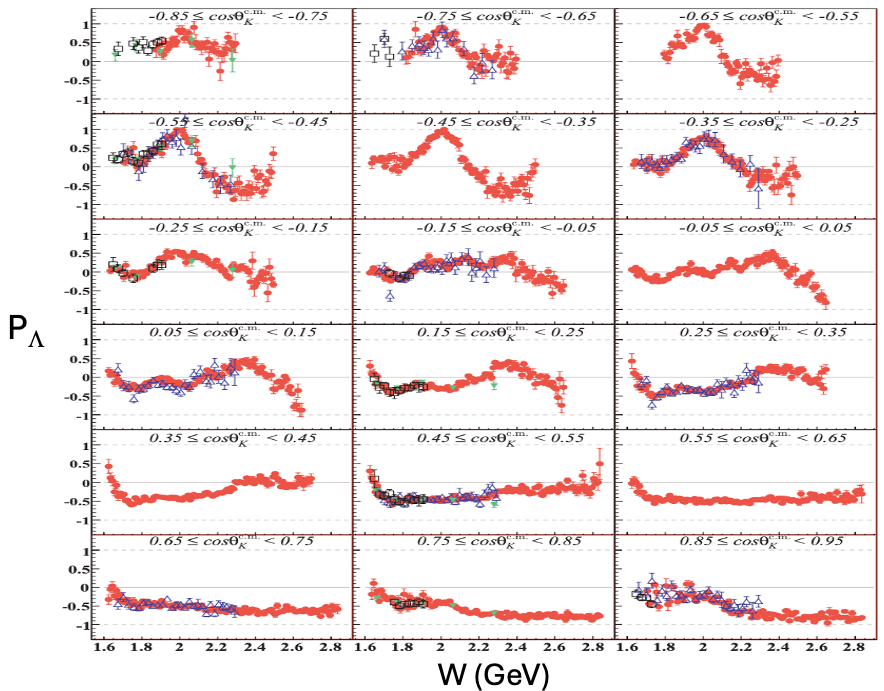}
\caption{Recoil polarization $P$ vs. $W$ binned in $\cos \theta_K^{\mathrm{c.m.}}$ for $K^+\Lambda$ photoproduction from a proton from CLAS~\cite{CLAS:2003zrd,CLAS:2009rdi} (blue triangles, red circles), SAPHIR~\cite{Glander:2003jw} (green triangles), and GRAAL~\cite{Lleres:2007tx} (black squares). Figure adapted from Ref.~\cite{CLAS:2009rdi}.}
\label{clas-kl-p}
\end{figure}

\begin{figure}[htbp]
\centering
\includegraphics[width=0.75\columnwidth]{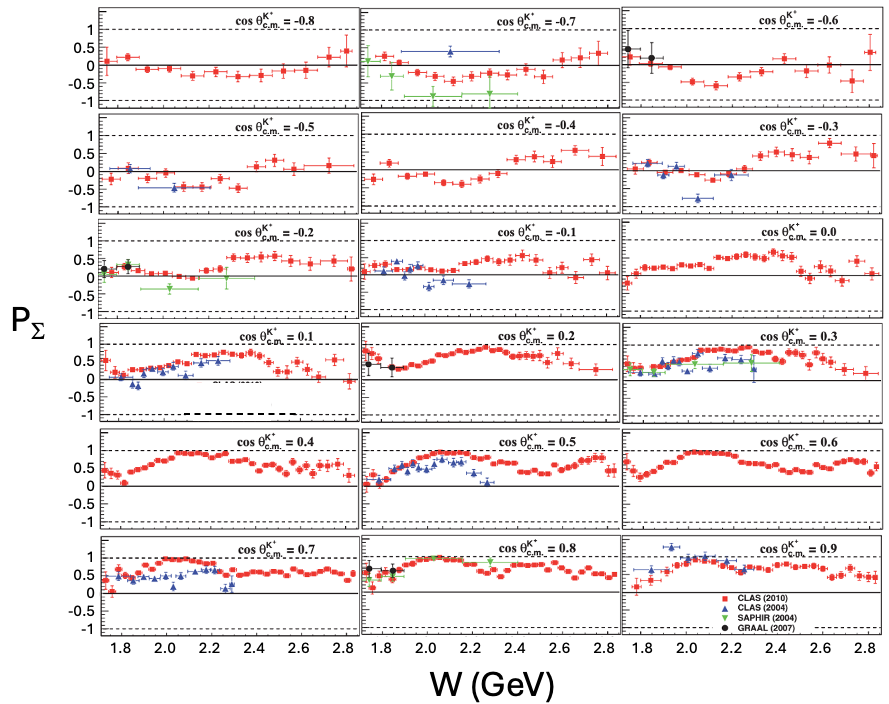}
\caption{Recoil polarization $P$ vs. $W$ binned in $\cos \theta_K^{\mathrm{c.m.}}$ for $K^+\Sigma^0$ photoproduction from a proton from CLAS~\cite{CLAS:2003zrd,CLAS:2010aen} (blue up-triangles, red circles), SAPHIR~\cite{Glander:2003jw} (green down-triangles), and GRAAL~\cite{Lleres:2007tx} (black circles). Figure adapted from Ref.~\cite{CLAS:2010aen}.}
\label{clas-ks-p}
\end{figure}

\begin{figure*}[htbp]
\centering
\includegraphics[width=0.65\textwidth]{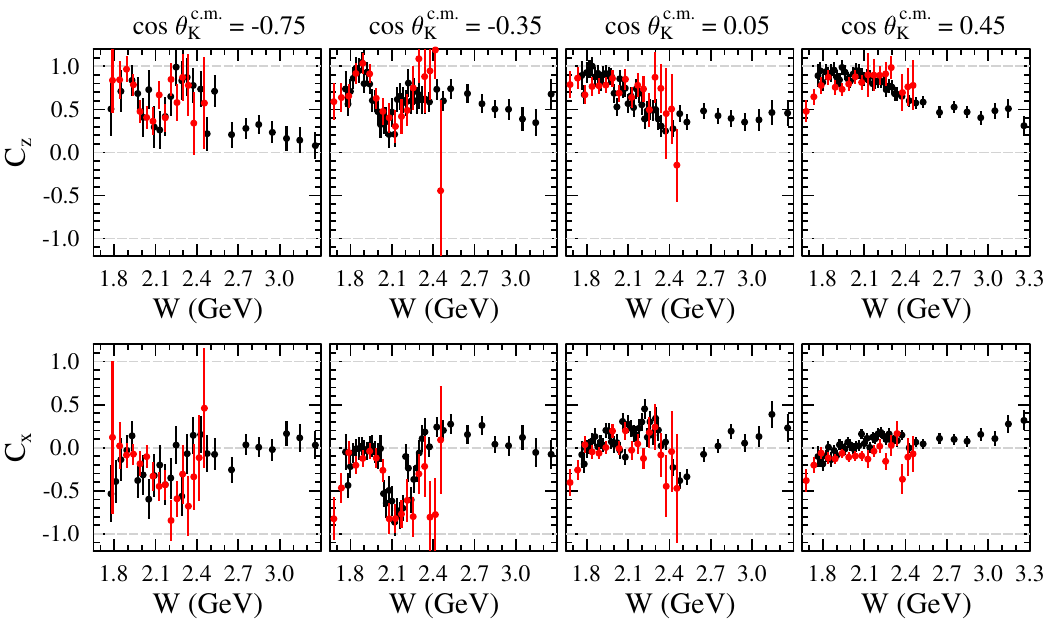}
\caption{$C_z$ (top) and $C_x$ (bottom) beam-recoil hyperon transferred polarization for exclusive $K^+\Lambda$ production extracted from the first generation CLAS dataset (red)~\cite{CLAS:2006pde} compared to second-generation CLAS results (black)~\cite{CLAS:2025azh} as a function of $W$ for representative $\cos \theta_K^{\mathrm{c.m.}}$ bins as shown. Figure from Ref.~\cite{CLAS:2025azh}.}
\label{clas-kl-cxcz}
\end{figure*}

Figures~\ref{clas-kl-p} and \ref{clas-ks-p} show comparative plots of the available CLAS recoil hyperon polarization $P$ (also called the induced hyperon polarization) for $K^+ \Lambda$ from Refs.~\cite{CLAS:2003zrd,CLAS:2009rdi} and $K^+\Sigma^0$ from Refs.~\cite{CLAS:2003zrd,CLAS:2010aen}. Figure~\ref{clas-kl-cxcz} shows a sample of the available beam-recoil hyperon transferred polarization observables $C_x$ and $C_z$ for $K^+\Lambda$ from a second-generation CLAS experiment~\cite{CLAS:2025azh} compared to the first-generation CLAS experiment~\cite{CLAS:2006pde} that highlights the evolution of the program. These observables require measurements with a circularly polarized photon beam. It is important to note that many resonances in the mass range $W > 1.7$~GeV have total decay widths exceeding 250-300~MeV. As a result, they may not manifest as distinct peaks in the $W$-dependence of the different experimental observables. However, their contributions can still be identified in analyses that combine the $KY$ differential cross sections and polarization observables. In fact, only by accounting for these data will reaction models be able to provide detailed information on the reaction mechanism in the $s$-, $t$-, and $u$-channels and separate the resonant and non-resonant contributions to the observables in the $W$ range above 1.6~GeV (i.e. above the $K\Lambda$ production threshold).

%%%%%%%%%%%%%%%%%%%%%%%%%%%%%%%%%%%%%%%%%%%%%%%%%%%%%%%%%%%%%%%%%%%%%%%%%%%%%%%%%%%%%%%%%%%%%%%%%%%%%%%%%%%%%%%%%%%%%%%%%%%%%%%%%%%%%%%%%%%%%%%%%%%%%%%%%%%%%%%%
%%%%%%%%%%%%%%%%%%%%%%%%%%%%%%%%%%%%%%%%%%%%%%%%%%%%%%%%%%%%%%%%%%%%%%%%%%%%%%%%%%%%%%%%%%%%%%%%%%%%%%%%%%%%%%%%%%%%%%%%%%%%%%%%%%%%%%%%%%%%%%%%%%%%%%%%%%%%%%%%

%%%%%%%%%%%%%%%%%%%%%%%%%%%%%%%%%%%%%%%%%%%%%%%%%%%%%%%%%%%%%%%%%%%%%%%%%%%%%%%%%%%%%%%%%%%%%%%%%%%%%%%%%%%%%%%%%%%%%%%%%%%%%%%%%%%%%%%%%%%%%%%%%%%%%%%%%%%%%%%%
%%%%%%%%%%%%%%%%%%%%%%%%%%%%%%%%%%%%%%%%%%%%%%%%%%%%%%%%%%%%%%%%%%%%%%%%%%%%%%%%%%%%%%%%%%%%%%%%%%%%%%%%%%%%%%%%%%%%%%%%%%%%%%%%%%%%%%%%%%%%%%%%%%%%%%%%%%%%%%%%

%%%%%%%%%%%%%%%%%%%%%%%%%%%%%%%%%%%%%%%%%%%%%%%%%%%%%%%%%%%%%%%%%%%%%%%%%%%%%%%%%%%%%%%%%%%%%%%%%%%%%%%%%%%%%%%%%%%%%%%%%%%%%%%%%%%%%%%%%%%%%%%%%%%%%%%%%%%%%%%%
%%%%%%%%%%%%%%%%%%%%%%%%%%%%%%%%%%%%%%%%%%%%%%%%%%%%%%%%%%%%%%%%%%%%%%%%%%%%%%%%%%%%%%%%%%%%%%%%%%%%%%%%%%%%%%%%%%%%%%%%%%%%%%%%%%%%%%%%%%%%%%%%%%%%%%%%%%%%%%%%

\begin{figure*}[htbp]
\centering
\includegraphics[width=1.0\textwidth]{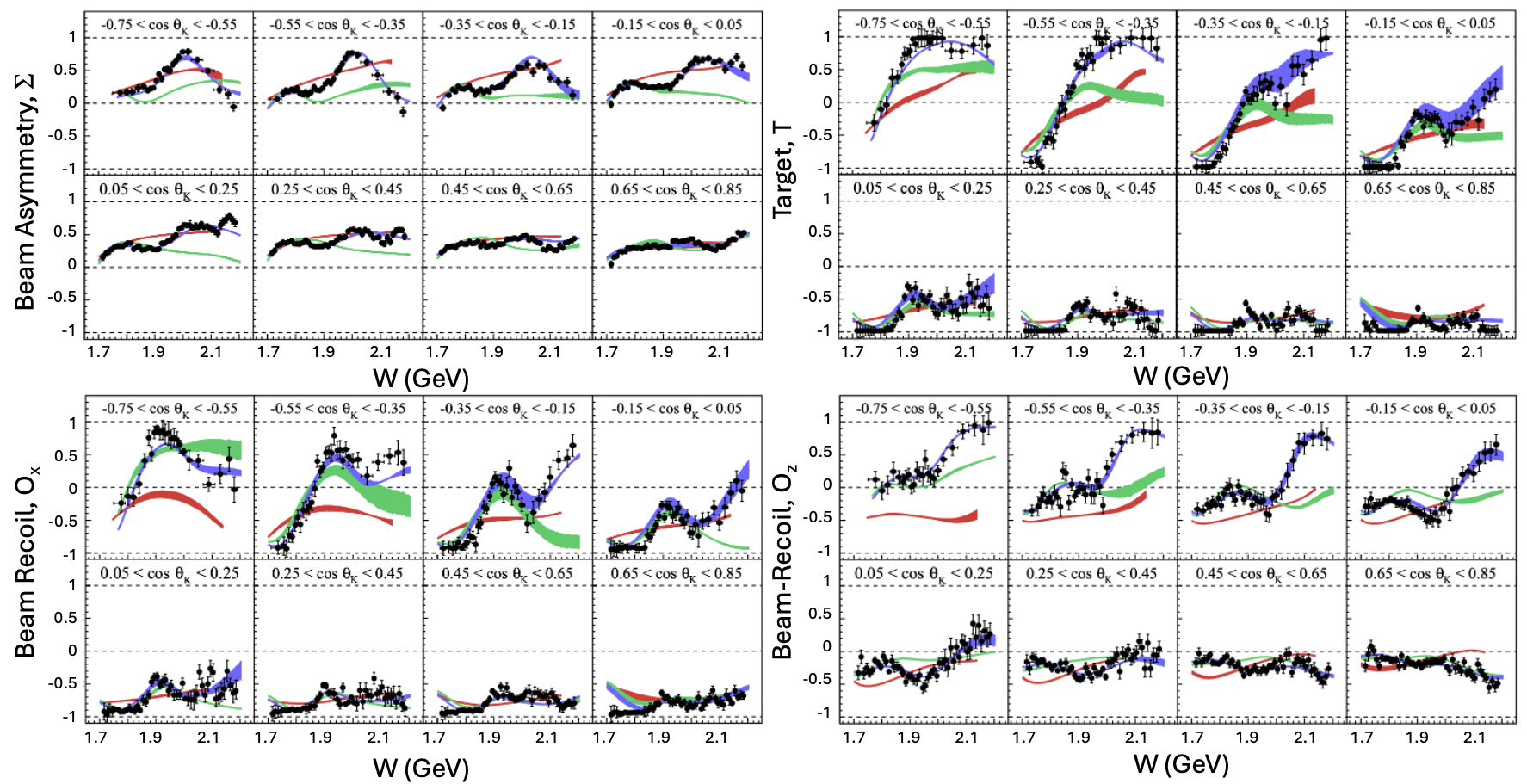}
\caption{The energy dependence of the beam asymmetry $\Sigma$, the target asymmetry $T$, and the beam-recoil hyperon transferred polarizations $O_x$ and $O_z$ as a function of $\cos \theta_K^{\mathrm{c.m.}}$ for CLAS $K^+\Lambda$ photoproduction from a proton target \cite{CLAS:2016wrl}. Red curves: ANL-Osaka coupled-channel model~\cite{Kamano:2013iva}; green curves: predictions from the 2014 solution of the Bonn-Gatchina partial wave analysis~\cite{CBELSATAPS:2014wvh}; blue curves: Bonn-Gatchina calculations after a refit including these data, which include additional $N^*$ states~\cite{Sarantsev:2016fac}. Figures adapted from Ref.~\cite{CLAS:2016wrl}.}
\label{clas-kl-g8}
\end{figure*}

\begin{figure*}[htbp]
\centering
\includegraphics[width=1.0\textwidth]{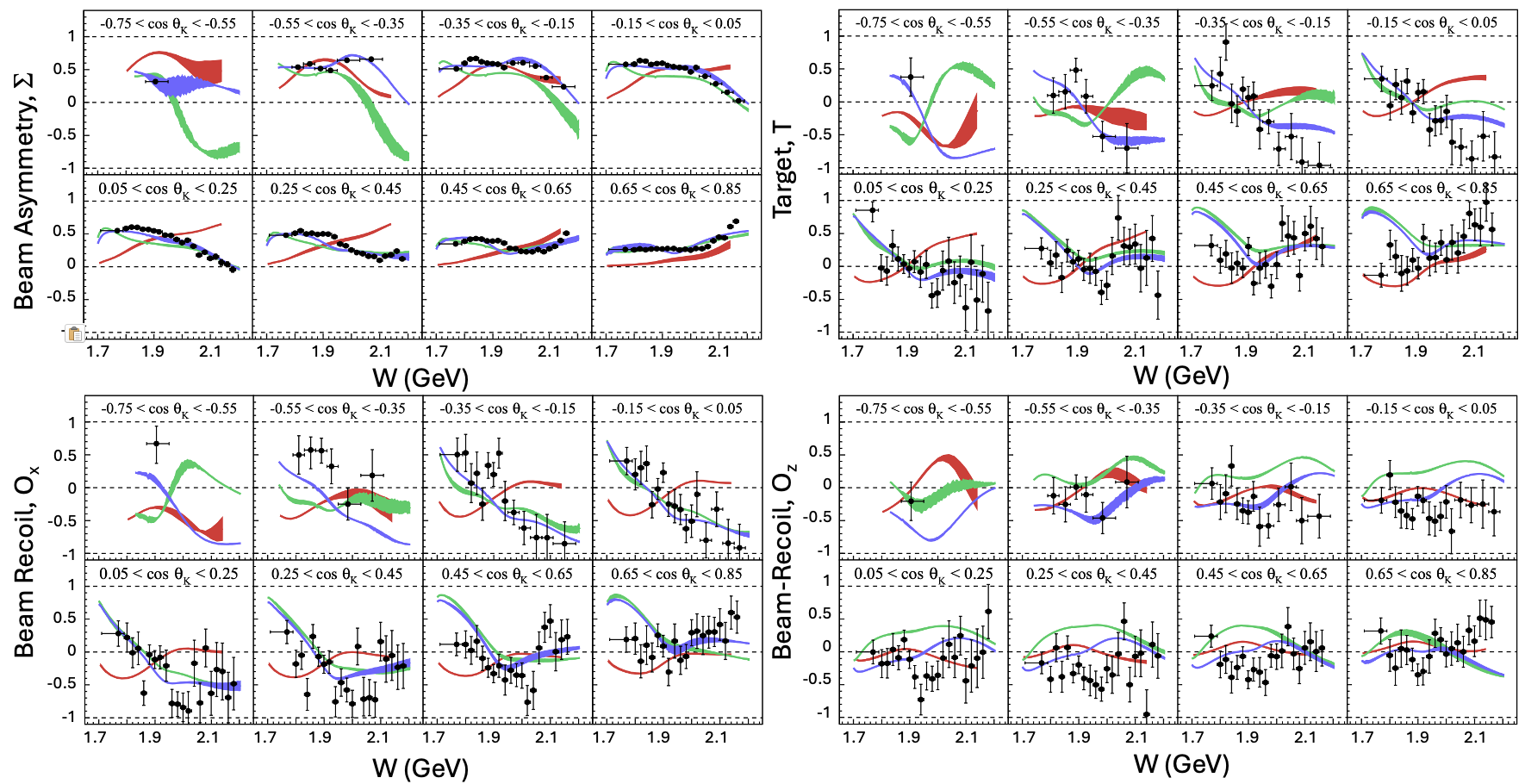}
\caption{The energy dependence of the beam asymmetry $\Sigma$, the target asymmetry $T$, and the beam-recoil hyperon transferred polarizations $O_x$ and $O_z$ as a function of $\cos \theta_K^{\mathrm{c.m.}}$ for CLAS $K^+\Sigma^0$ photoproduction from a proton target~\cite{CLAS:2016wrl}. The different curves are detailed in the caption of Fig.~\ref{clas-kl-g8}. Figures adapted from Ref.~\cite{CLAS:2016wrl}.}
\label{clas-ks-g8}
\end{figure*}

The available hyperon polarization observables from CLAS are complemented through measurements with a linearly polarized photon beam. Figures~\ref{clas-kl-g8} and \ref{clas-ks-g8} show the available data for $K^+\Lambda$ and $K^+\Sigma^0$ for the beam and target single-spin asymmetries $\Sigma$ and $T$ and the beam-recoil hyperon double-spin asymmetries $O_x$ and $O_z$ as a function of $W$ for bins in $\cos \theta_K^{\mathrm{c.m.}}$ \cite{CLAS:2016wrl}. Note that these data were taken with a unpolarized liquid-hydrogen target. The single spin asymmetry $T$ was extracted as discussed in Section~\ref{sec:graal}. Data in terms of the polarization observables ($\Sigma$, $P$, $T$, $O_x$, $O_z$) have also become available for the reaction $\gamma p \to K^0 \Sigma^+$ in the nucleon resonance region~\cite{CLAS:2024bzi}. Comparison with advanced coupled-channels models demonstrate that even though the measurements are provided for only a few kinematic points, they still carry sufficient information to be sensitive to the contributions of $s$-channel baryon resonances \cite{Ronchen:2022hqk,CLAS:2024bzi}. Since the energy dependence of a partial wave amplitude for any particular channel is influenced by other channels due to unitarity constraints, the coupled-channel approach is important to properly describe the energy dependence of the different production amplitudes. In recent years, such coupled-channel approaches that account for the $KY$ photoproduction processes have been developed by groups at Bonn-Gatchina~\cite{Anisovich:2011fc}, JLab~\cite{Julia-Diaz:2007mae}, ANL-Osaka \cite{Kamano:2013iva}, J{\"u}lich-Bonn \cite{Ronchen:2014cna,Ronchen:2018ury}, and elsewhere.

%%%%%%%%%%%%%%%%%%%%%%%%%%%%%%%%%%%%%%%%%%%%%%%%%%%%%%%%%%%%%%%%%%%%%%%%%%%%%%%%%%%%%%%%%%%%%%%%%%%%%%%%%%%%%%%%%%%%%%%%%%%%%%%%%%%%%%%%%%%%%%%%%%%%%%%%%%%%%%%%
%%%%%%%%%%%%%%%%%%%%%%%%%%%%%%%%%%%%%%%%%%%%%%%%%%%%%%%%%%%%%%%%%%%%%%%%%%%%%%%%%%%%%%%%%%%%%%%%%%%%%%%%%%%%%%%%%%%%%%%%%%%%%%%%%%%%%%%%%%%%%%%%%%%%%%%%%%%%%%%%

%%%%%%%%%%%%%%%%%%%%%%%%%%%%%%%%%%%%%%%%%%%%%%%%%%%%%%%%%%%%%%%%%%%%%%%%%%%%%%%%%%%%%%%%%%%%%%%%%%%%%%%%%%%%%%%%%%%%%%%%%%%%%%%%%%%%%%%%%%%%%%%%%%%%%%%%%%%%%%%%
%%%%%%%%%%%%%%%%%%%%%%%%%%%%%%%%%%%%%%%%%%%%%%%%%%%%%%%%%%%%%%%%%%%%%%%%%%%%%%%%%%%%%%%%%%%%%%%%%%%%%%%%%%%%%%%%%%%%%%%%%%%%%%%%%%%%%%%%%%%%%%%%%%%%%%%%%%%%%%%%

The final category of photoproduction measurements off the proton from CLAS for ground state hyperon production is from the $K^{*+}\Lambda$ and $K^{*+}\Sigma^0$ channels involving production of vector mesons in the final state. These reactions were analyzed detecting the three-pion decay mode $K^{*+} \to K^0_S \pi^+ \to \pi^+\pi^+\pi^-$, and reconstructing the associated hyperon by missing mass. One of the motivations for the study of $K^*$ photoproduction is to investigate the role of the $K^{*+}(800)$ (also called the $\kappa$) through $t$-channel exchange~\cite{CLAS:2013qgi}. This meson is expected to reside within the $J^{PC} = 0^{++}$ scalar nonet that includes the $f_0(500)$. The $\kappa^+$ can enter into $t$-channel exchange for $K^{*+}Y$ production but not for $K^+Y$ because the photon cannot couple to the $\kappa^+ - K^+$ vertex due to conservation of $G$-parity~\cite{Tang:2012cta}.

The $K^{*+} \Sigma^0$ channel also has promise in the search for $s$-channel excited nucleon states, including those with isospin-3/2. In this channel diffractive processes do not contribute~\cite{Burkert:2019kxy}, which may allow for access to contributing $N^*$ and $\Delta^*$ $s$-channel resonances in a manner in which their production is enhanced relative to the non-resonant backgrounds. However, due to the increased number of amplitudes that contribute to this process compared to $KY$, it is more complex to understand theoretically.

A number of measurements of cross sections and polarization observables from CLAS analysis are also available off effective neutron targets for the $K^0 \Lambda$, $K^0 \Sigma^0$, and $K^+\Sigma^-$ exclusive final states (see Table~\ref{hallb-clas-listb}). These measurements of isospin-partner channels will be important to isolate the contributions of specific baryon resonances by separating processes with different isospin properties. However, as seen in comparing Tables~\ref{hallb-clas-lista} and \ref{hallb-clas-listb}, the number of $\gamma n \to KY$ data points is only $\sim$7\% that for $\gamma p \to KY$ in terms of both cross sections and polarization observables, so photoproduction studies from a neutron target should still be considered lacking. However, as noted in Section~\ref{mami-program} and earlier in this section, understanding the $\gamma n$ process and the contributing resonant and non-resonant contributions must necessarily involve an accurate reaction model and proper treatment of target Fermi motion and final state interactions.

\begin{figure*}[htbp]
\centering
\includegraphics[width=0.5\textwidth,height=0.35\textheight]{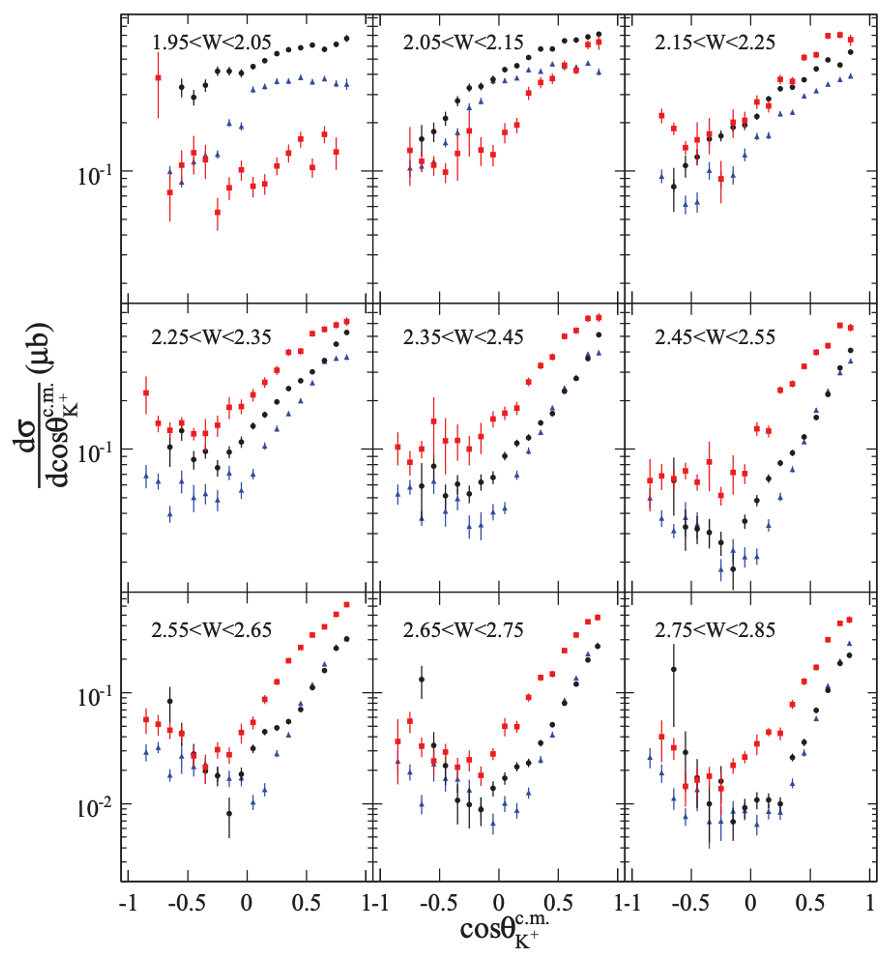}
\caption{Comparison of differential cross sections from CLAS data for the $K^+\Sigma^0(1385)$ (black circles), $K^+\Lambda(1405)$ (blue triangles), and $K^+\Lambda(1520)$ (red squares) as a function of $\cos \theta_K^{\mathrm{c.m.}}$ for bins in $W$. Figure from Ref.~\cite{CLAS:2013rxx}.}
\label{clas-kystar-dcs}
\end{figure*}

Measurements of the CLAS photoproduction data have provided cross sections for the excited hyperon states $\Sigma^0(1385)$, $\Lambda(1405)$, and $\Lambda(1520)$~\cite{CLAS:2013rxx,CLAS:2021osv}. Figure~\ref{clas-kystar-dcs} shows a direct comparison of data for these different hyperon states as a function of $\cos \theta_K^{\mathrm{c.m.}}$ for bins in $W$ from 1.95~GeV to 2.85~GeV. Qualitatively, the cross sections show the strong forward peaking that is indicative of the importance of diffractive production mechanisms in the $t$-channel. The $\Sigma^0(1385)$ was measured through its dominant $\Lambda \pi^0$ decay mode, while the $\Lambda(1405)$ and $\Lambda(1520)$ were measured through each of their $\Sigma \pi$ decay modes. Comparisons of the data for $\Sigma^0(1385)$ and $\Lambda(1520)$ photoproduction to available models show reasonable agreement and provide for an opportunity to study the role of $t$-channel $K^*$ exchange and various $s$-channel $N^*$ contributions in the production. However, studies of the $\Lambda(1405)$ show that detailed understanding of its production mechanism is still lacking~\cite{CLAS:2013rxx}.  

%%%%%%%%%%%%%%%%%%%%%%%%%%%%%%%%%%%%%%%%%%%%%%%%%%%%%%%%%%%%%%%%%%%%%%%%%%%%%%%%%%%%%%%%%%%%%%%%%%%%%%%%%%%%%%%%%%%%%%%%%%%%%%%%%%%%%%%%%%%%%%%%%%%%%%%%%%%%%%%%
%%%%%%%%%%%%%%%%%%%%%%%%%%%%%%%%%%%%%%%%%%%%%%%%%%%%%%%%%%%%%%%%%%%%%%%%%%%%%%%%%%%%%%%%%%%%%%%%%%%%%%%%%%%%%%%%%%%%%%%%%%%%%%%%%%%%%%%%%%%%%%%%%%%%%%%%%%%%%%%%

The CLAS $\gamma p$ studies of the $\Lambda(1405)$ have provided a detailed examination of the $\Sigma^+ \pi^-$, $\Sigma^0 \pi^0$, and $\Sigma^-\pi^+$ lineshapes~\cite{CLAS:2013rjt}. The differential cross section in each decay mode can be written in terms of the contributions from the different complex $I=0$ and $I=1$ isospin amplitudes $T^{(I)}$ as \cite{Nacher:1998mi}

\begin{eqnarray}
\label{dsig-l1405}
\frac{d\sigma(\pi^+ \Sigma^-)}{dM_I} &\propto& \frac{1}{3} \vert T^{(0)} \vert^2 + \frac{1}{2} \vert T^{(1)} \vert^2 + \frac{2}{\sqrt{6}} {\rm Re}(T^{(0)}T^{(1)*}) \\ \nonumber
\frac{d\sigma(\pi^- \Sigma^+)}{dM_I} &\propto& \frac{1}{3} \vert T^{(0)} \vert^2 + \frac{1}{2} \vert T^{(1)} \vert^2 - \frac{2}{\sqrt{6}} {\rm Re}(T^{(0)}T^{(1)*}) \\ \nonumber
\frac{d\sigma(\pi^0 \Sigma^0)}{dM_I} &\propto& \frac{1}{3} \vert T^{(0)} \vert^2.
\end{eqnarray}

\noindent
Using a model of the mass distribution for the $\Sigma \pi$ lineshape, it was shown that the isospin-0 amplitude is consistent with the contribution of two poles modeled as Breit-Wigner resonances with large channel coupling to $N\bar{K}$. In addition, an isospin-1 component of the reaction mechanism has been shown to account for the variation among the different $\Sigma \pi$ channels~\cite{Schumacher:2013vma}. These data also allowed for a determination of the spin and party of the $\Lambda(1405)$ in its $\Sigma^+\pi^-$ decay branch~\cite{CLAS:2014tbc}, showing that the decays are $s$-wave, with the $\Sigma^+$ polarized such that the $\Lambda(1405)$ can be assigned $J^P = 1/2^-$ as expected by most theories. 

For completeness in this brief survey of the CLAS $KY^*$ photoproduction data, several technically challenging but precise measurements of radiative decays of the $\Sigma^0(1385)$, $\Sigma^+(1385)$, and $\Lambda(1520)$ have been completed~\cite{CLAS:2005bgo,CLAS:2011vzg,CLAS:2011iuw}. These measurements provide a unique avenue with which to probe the wavefunctions of these excited state hyperons. The strange quark baryons have an additional degree of freedom that aids in the study of multiplet mixing and non-three quark contributions.

The CLAS photoproduction program via the $\gamma p \to K^+Y$ exclusive channel also included a series of analyses in the $S = -2$ sector that have spanned several generations of experimental datasets, each building on the previous experience. The motivation for the measurements was to expand our knowledge about the production of doubly strange hyperons. Even today the Particle Data Group (PDG) listings for cascade hyperons show only 3 states with known spin-parity $J^P$: $\Xi(1321)1/2^+$, $\Xi(1530)3/2^+$, and $\Xi(1820)3/2^-$~\cite{ParticleDataGroup:2024cfk}. An additional eight more candidate states have been reported with masses up to 2.5~GeV, but knowledge is still quite limited. SU(3) flavor symmetry, however, implies the existence of a $\Xi^*$ state for every $N^*$ state and for every $\Delta^*$ state. At present the PDG includes 27 $N^*$ states with masses up to 2.7~GeV and 21 $\Delta^*$ states with masses up to 2.95~GeV \cite{ParticleDataGroup:2024cfk}. The discrepancy between experiment and theoretical expectations remains an open question~\cite{CLAS:2004gjf}.

Studies of $\Xi$ production with photon beams were initially carried out using inclusive scattering processes, $\gamma p \to \Xi^- X$, with a search for the $\Xi$ through its decays products, $\Xi \to \pi^- \Lambda \to \pi^- \pi^- p$ at CERN~\cite{Aston:1981wxh} and SLAC~\cite{Abe:1985nbw}. These first generation searches provided initial cross sections based on data samples with less than 100 events with sizable statistical and systematic uncertainties due to the significant backgrounds beneath the signal peaks in the mass spectra.

\begin{figure*}[htbp]
\centering
\includegraphics[width=1.0\textwidth]{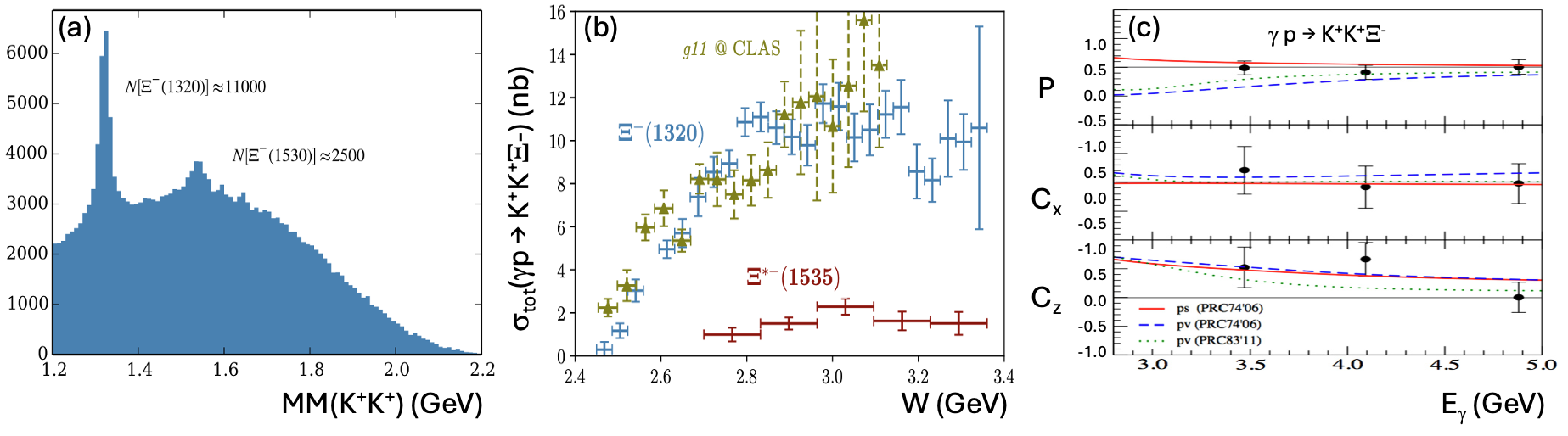}
\caption{Key results from the CLAS $\Xi$ photoproduction analyses. (a) Distribution of the $MM(K^+K^+)$ events showing the exclusive reaction yields for the ground state $\Xi^-(1321)$ and first excited state $\Xi^-(1530)$. (b) Total cross section for the ground $\Xi$ and first excited state as a function of $W$. (c) First measurement of the $\Xi^-(1321)$ recoil and beam-recoil hyperon transferred polarization components as a function of photon energy. The curves are from the phenomenological meson-exchange model including either pseudoscalar or pseudovector exchanges \cite{Man:2011np}. Figures from Ref.~\cite{CLAS:2018kvn} (a,b) and adapted from Ref.~\cite{CLAS:2018xbd} (c).}
\label{clas-cascade-plots}
\end{figure*}

The CLAS photoproduction studies focused instead on the exclusive topologies $\gamma p \to K^+ K^+ X$ and $\gamma p \to K^+ K^+ p X$. The first experiments in this program provided measurements of cross sections for the $\Xi^-(1321)$ ground state based on a sample of $\sim$500 events and a sample of $\sim$150 events for the $\Xi^-(1530)$ first excited state~\cite{CLAS:2004gjf}. The final measurements in the program provided cross sections and the first extraction of the $\Xi$ polarization observables $P$, $C_x$, and $C_z$ based on a data sample of $\sim$5000 events~\cite{Guo:2007dw}. The $S = -2$ photoproduction studies of $\Xi$ hyperons from CLAS are summarized in Table~\ref{clas-cascade}. Figure~\ref{clas-cascade-plots} shows several key results from these measurements.

%%%%%%%%%%%%%%%%%%%%%%%%%%%%%%%%%%%%%%%%%%%%%%%%%%%%%%%%%%%%%%%%%%%%%%%%%%%%%%%%%%%%%%%%%%%%%%%%%%%%%%%%%%%%%%%%%%%%%%%%%%%%%%%%%%%%%%%%%%%%%%%%%%%%%%%%%%%%%%%
\begin{table}[tbh]
\setlength{\tabcolsep}{6pt} % Default value: 6pt
\renewcommand{\arraystretch}{0.8} % Default value: 1
\begin{center}
\caption{Summary of $\gamma p \to K^+K^+\Xi^{(*)}$ photoproduction measurements in Hall B with CLAS from the 6-GeV era experiments at JLab. The column labeled $N_{\mathrm{bin}}$ indicates the number of kinematic bins included in the analysis.}
\begin{tabular}{cccccccc} \hline \hline
Observable                          & Final State         & $W$ (GeV)                    & $\cos \theta_\Xi^{\rm c.m.}$      & $N_{\mathrm{bin}}$          & Year                  & Ref. \\ \hline
\multirow{2}{*}{Mass Spectrum}      & $K^+K^+\Xi^-(1321)$ & \multirow{2}{*}{$[2.6:3.3]$} & \multirow{2}{*}{$[-1.0:1.0]$} & \multirow{2}{*}{-} & \multirow{2}{*}{2005} & \multirow{2}{*}{\cite{CLAS:2004gjf}} \\
                                    & $K^+K^+\Xi^-(1530)$ &                              &                               &                    &                       &  \\ \hline
\multirow{2}{*}{$d\sigma$}          & $K^+K^+\Xi^-(1321)$ & $[2.5:2.9]$                  & \multirow{2}{*}{$[-1.0:1.0]$} & 88                 & \multirow{2}{*}{2007} & \multirow{2}{*}{\cite{Guo:2007dw}} \\
                                    & $K^+K^+\Xi^-(1530)$ & $[2.7:3.1]$                  &                               & 8                  &                       & \\ \hline
$P$, $C_x$, $C_z$                   & $K^+K^+\Xi^-(1321)$ & $[2.5:3.3]$                  & $[-1.0:1.0]$                  & 3                  & 2018                  & \cite{CLAS:2018xbd} \\ \hline
\multirow{2}{*}{$\sigma_{\rm tot}$} & $K^+K^+\Xi^-(1321)$ & \multirow{2}{*}{$[2.5:3.3]$} & \multirow{2}{*}{$[-1.0:1.0]$} & 23 $\Xi^-(1321)$   & \multirow{2}{*}{2018} & \multirow{2}{*}{\cite{CLAS:2018kvn}} \\
                                    & $K^+K^+\Xi^-(1530)$ &                              &                               & 5 $\Xi^-(1530)$    &                       &  \\ \hline\hline
\end{tabular}
\label{clas-cascade}
\end{center}
\end{table}
%%%%%%%%%%%%%%%%%%%%%%%%%%%%%%%%%%%%%%%%%%%%%%%%%%%%%%%%%%%%%%%%%%%%%%%%%%%%%%%%%%%%%%%%%%%%%%%%%%%%%%%%%%%%%%%%%%%%%%%%%%%%%%%%%%%%%%%%%%%%%%%%%%%%%%%%%%%%%%%%

%%%%%%%%%%%%%%%%%%%%%%%%%%%%%%%%%%%%%%%%%%%%%%%%%%%%%%%%%%%%%%%%%%%%%%%%%%%%%%%%%%%%%%%%%%%%%%%%%%%%%%%%%%%%%%%%%%%%%%%%%%%%%%%%%%%%%%%%%%%%%%%%%%%%%%%%%%%%%%%%
%%%%%%%%%%%%%%%%%%%%%%%%%%%%%%%%%%%%%%%%%%%%%%%%%%%%%%%%%%%%%%%%%%%%%%%%%%%%%%%%%%%%%%%%%%%%%%%%%%%%%%%%%%%%%%%%%%%%%%%%%%%%%%%%%%%%%%%%%%%%%%%%%%%%%%%%%%%%%%%%

For completeness in this section, it must be mentioned that the CLAS Collaboration published several papers in the search for pentaquark states in $\gamma d$ experiments that recoil against a detected final state $K^+$. This topic is discussed in Refs.~\cite{CLAS:2003yuj,CLAS:2003wfm,CLAS:2005koo,CLAS:2006anj,CLAS:2006rru,ActaPenta2025} and further information is provided in Section~\ref{sec:penta-narrow} with no further discussion here. 

\subsubsection{Hall B -- CLAS Electroproduction Program}
\label{clas-ep-program}

\begin{figure*}[htbp]
\centering
\includegraphics[width=0.75\textwidth]{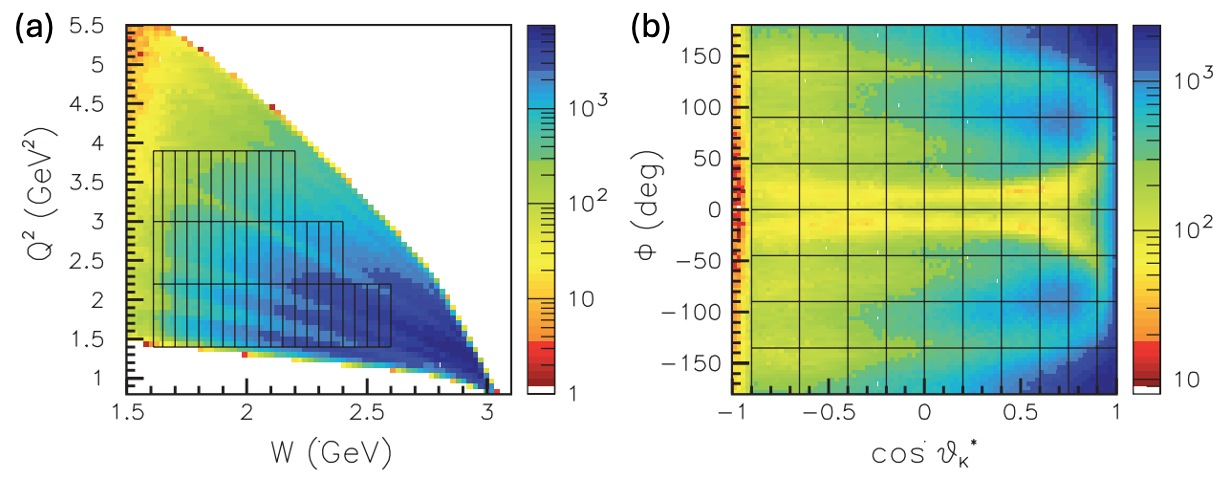}
\caption{Kinematic coverage of a CLAS 6-GeV $KY$ electroproduction dataset in terms of (a) $Q^2$ (GeV$^2$) vs. $W$ (GeV) and (b) $\cos \theta_K^{\mathrm{c.m.}}$ vs. $\Phi$. The plots are overlaid with the binning choices from the analysis. Figure from Ref.~\cite{Carman:2012qj}.}
\label{clas-electro-kin}
\end{figure*}

\begin{figure}[htbp]
\centering
\includegraphics[width=0.4\textwidth]{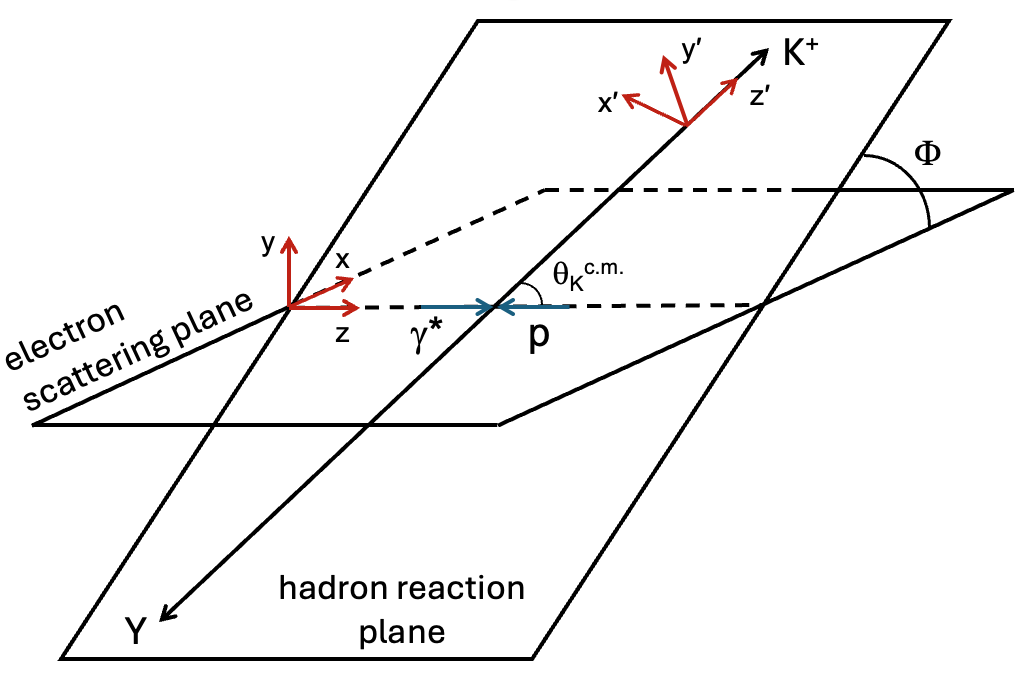}
\caption{Kinematics for $K^+Y$ electroproduction defining the c.m. angles and coordinate systems used to express the formalism and to present the hyperon polarization components. Here the primed coordinate system is connected to the $K^+$ with the $z'$ axis along its direction and the $y'$ axis normal to the hadronic reaction plane. The unprimed coordinate system is connected with the virtual photon with the $z$ along along its direction and the $y$ axis normal to the electron scattering plane. Figure from Ref.~\cite{CLAS:2025vkn}.}
\label{coor4}
\end{figure}

The CLAS program of the JLab 6-GeV era has provided the most extensive datasets presently available for $KY$ electroproduction across the nucleon resonance region. These measurements have included the separated structure functions $\sigma_{\rm T}$, $\sigma_{\rm L}$, $\sigma_{\rm U} = \sigma_{\rm T} + \epsilon \sigma_{\rm L}$, $\sigma_{\rm LT}$, $\sigma_{\rm TT}$, and $\sigma_{\rm LT'}$ for $K^+\Lambda$ and $K^+\Sigma^0$ \cite{Raue:2004us,CLAS:2006ogr,CLAS:2008agj,Carman:2012qj}, recoil hyperon polarization for $K^+\Lambda$~\cite{CLAS:2014udv}, and beam-recoil hyperon transferred polarization for $K^+\Lambda$ and $K^+\Sigma^0$ \cite{CLAS:2002zlc,CLAS:2009sbn}. These observables span $Q^2$ from 0.5 to 4.5~GeV$^2$, $W$ from 1.6 to 3.0~GeV, and the full center-of-mass angular range of the $K^+$. Tables~\ref{hallb-clas-list1} and \ref{hallb-clas-list2} provide an overview of the electroproduction measurements completed with CLAS for exclusive $K^+\Lambda$ and $K^+\Sigma^0$ production from a proton target. Figure~\ref{clas-electro-kin} shows the broad kinematic coverage of the CLAS detector in terms of the relevant kinematic variables $Q^2$, $W$, $\cos \theta_K^{\mathrm{c.m.}}$, and $\Phi$, where $\Phi$ is angle between the electron scattering plane and the hadronic production plane shown in Fig.~\ref{coor4} (this angle is introduced as $\phi$ in Section~\ref{dcs-formalism}, Eq.~(\ref{eq:virtual_diff_cs})). Overviews of the program are included in Refs.~\cite{Carman:2005ms,Carman:2008hb,Carman:2012zz,Carman:2018fsn,Carman:2019lkk}.

%%%%%%%%%%%%%%%%%%%%%%%%%%%%%%%%%%%%%%%%%%%%%%%%%%%%%%%%%%%%%%%%%%%%%%%%%%%%%%%%%%%%%%%%%%%%%%%%%%%%%%%%%%%%%%%%%%%%%%%%%%%%%%%%%%%%%%%%%%%%%%%%%%%%%%%%%%%%%%%%
%%%%%%%%%%%%%%%%%%%%%%%%%%%%%%%%%%%%%%%%%%%%%%%%%%%%%%%%%%%%%%%%%%%%%%%%%%%%%%%%%%%%%%%%%%%%%%%%%%%%%%%%%%%%%%%%%%%%%%%%%%%%%%%%%%%%%%%%%%%%%%%%%%%%%%%%%%%%%%%%

%%%%%%%%%%%%%%%%%%%%%%%%%%%%%%%%%%%%%%%%%%%%%%%%%%%%%%%%%%%%%%%%%%%%%%%%%%%%%%%%%%%%%%%%%%%%%%%%%%%%%%%%%%%%%%%%%%%%%%%%%%%%%%%%%%%%%%%%%%%%%%%%%%%%%%%%%%%%%%%%
%%%%%%%%%%%%%%%%%%%%%%%%%%%%%%%%%%%%%%%%%%%%%%%%%%%%%%%%%%%%%%%%%%%%%%%%%%%%%%%%%%%%%%%%%%%%%%%%%%%%%%%%%%%%%%%%%%%%%%%%%%%%%%%%%%%%%%%%%%%%%%%%%%%%%%%%%%%%%%%%

%%%%%%%%%%%%%%%%%%%%%%%%%%%%%%%%%%%%%%%%%%%%%%%%%%%%%%%%%%%%%%%%%%%%%%%%%%%%%%%%%%%%%%%%%%%%%%%%%%%%%%%%%%%%%%%%%%%%%%%%%%%%%%%%%%%%%%%%%%%%%%%%%%%%%%%%%%%%%%%
\begin{table}[tbh]
\setlength{\tabcolsep}{6pt} % Default value: 6pt
\renewcommand{\arraystretch}{0.8} % Default value: 1
\begin{center}
\caption{Summary of $ep\to e'K^+\Lambda$ electroproduction measurements in Hall B with CLAS from the 6-GeV era experiments at JLab. The column labeled $N_{\mathrm{bin}}$ indicates the number of kinematic bins included in the analysis.}
\begin{tabular}{ccccccc} \hline\hline
Observables                                               & $Q^2$ (GeV$^2$) & $W$ (GeV)      & $\cos \theta_K^{\mathrm{c.m.}}$ & $N_{\rm bin}$ & Year & Ref. \\ \hline
$\sigma_{\rm U}$, $\sigma_{\rm LT}$, $\sigma_{\rm TT}$    & $[0.65:2.55]$   & $[1.65:2.35]$  & $[-0.80:1.0]$          & 318       & 2007 & \cite{CLAS:2006ogr} \\ \hline
$\sigma_{\rm L}$, $\sigma_{\rm T}$                        & 1.0             & $[1.65:1.95]$  & $[-0.80:1.0]$          & 24        & 2007 & \cite{CLAS:2006ogr} \\ \hline
$\sigma_{\rm LT'}$                                        & 0.65, 1.00      & $[1.60:2.10]$  & $[-0.80:1.0]$          & 96        & 2008 & \cite{CLAS:2008agj} \\ \hline
$\sigma_{\rm U}$, $\sigma_{\rm LT}$, $\sigma_{\rm TT}$, $\sigma_{\rm LT'}$  & $[1.80:3.45]$   & $[1.63:2.575]$ & $[-0.9:1.0]$           & 480       & 2013 & \cite{Carman:2012qj} \\ \hline
\multirow{2}{*}{$\sigma_{\rm L}/\sigma_{\rm T}$}          & $[0.3:1.5]$     & $[1.6:2.15]$   & \multirow{2}{*}{1.0}   & 3         & 2005 & \cite{Raue:2004us} \\
                                                          & 1.5, 2.5        & $[1.7:2.2]$    &                        & 6         & 2009 & \cite{CLAS:2009sbn} \\ \hline
${\cal P}^0_{y'}$                                         & 1.9             & $[1.6:2.7]$    & $[-1.0:1.0]$           & 224       & 2014 & \cite{CLAS:2014udv} \\ \hline 
${\cal P}'_x$, ${\cal P}'_z$,                             & $[0.3:1.5]$     & $[1.6:1.25]$   & $[-1.0:1.0]$           & 38        & 2003 & \cite{CLAS:2002zlc} \\ 
${\cal P}'_{x'}$, ${\cal P}'_{z'}$                        & 2.5             & $[1.75:2.31]$  & $[-0.8:0.9]$           & 76        & 2009 & \cite{CLAS:2009sbn} \\ \hline\hline 
\end{tabular}
\label{hallb-clas-list1}
\end{center}
\end{table}
%%%%%%%%%%%%%%%%%%%%%%%%%%%%%%%%%%%%%%%%%%%%%%%%%%%%%%%%%%%%%%%%%%%%%%%%%%%%%%%%%%%%%%%%%%%%%%%%%%%%%%%%%%%%%%%%%%%%%%%%%%%%%%%%%%%%%%%%%%%%%%%%%%%%%%%%%%%%%%%%

%%%%%%%%%%%%%%%%%%%%%%%%%%%%%%%%%%%%%%%%%%%%%%%%%%%%%%%%%%%%%%%%%%%%%%%%%%%%%%%%%%%%%%%%%%%%%%%%%%%%%%%%%%%%%%%%%%%%%%%%%%%%%%%%%%%%%%%%%%%%%%%%%%%%%%%%%%%%%%%
\begin{table}[tbh]
\setlength{\tabcolsep}{6pt} % Default value: 6pt
\renewcommand{\arraystretch}{0.8} % Default value: 1
\begin{center}
\caption{Summary of $ep\to e'K^+\Sigma^0$ electroproduction measurements in Hall B with CLAS from the 6-GeV era experiments at JLab. The column labeled $N_{\rm bin}$ indicates the number of kinematic bins included in the analysis for each channel.}
\begin{tabular}{ccccccc} \hline\hline
Observables                                                                 & $Q^2$ (GeV$^2$) & $W$ (GeV)       & $\cos \theta_K^{\mathrm{c.m.}}$ & $N_{\rm bin}$ & Year & Ref. \\ \hline
$\sigma_{\rm U}$, $\sigma_{\rm LT}$, $\sigma_{\rm TT}$                      & $[0.65:2.55]$   & $[1.725:2.35]$  & $[-0.80:1.0]$          & 240       & 2007 & \cite{CLAS:2006ogr} \\ \hline
$\sigma_{\rm L}$, $\sigma_{\rm T}$                                          & 1.0             & $[1.75:1.95]$   & $[-0.80:1.0]$          & 18        & 2007 & \cite{CLAS:2006ogr} \\ \hline
$\sigma_{\rm U}$, $\sigma_{\rm LT}$, $\sigma_{\rm TT}$, $\sigma_{\rm LT'}$  & $[1.80:3.45]$   & $[1.695:2.575]$ & $[-0.9:1.0]$           & 450       & 2013 & \cite{Carman:2012qj} \\ \hline
${\cal P}'_x$, ${\cal P}'_z$, ${\cal P}'_{x'}$, ${\cal P}'_{z'}$            & 2.5             & 2.1             & $[-0.4:0.9]$           & 15        & 2009 & \cite{CLAS:2009sbn} \\ \hline\hline 
\end{tabular}
\label{hallb-clas-list2}
\end{center}
\end{table}
%%%%%%%%%%%%%%%%%%%%%%%%%%%%%%%%%%%%%%%%%%%%%%%%%%%%%%%%%%%%%%%%%%%%%%%%%%%%%%%%%%%%%%%%%%%%%%%%%%%%%%%%%%%%%%%%%%%%%%%%%%%%%%%%%%%%%%%%%%%%%%%%%%%%%%%%%%%%%%%%

As discussed in Section~\ref{clas-gp-program}, the full differential cross section for pseudoscalar meson photoproduction with purely transverse real photons is described by four independent, complex amplitudes that give rise to a total of 16 experimental observables. However, as detailed in Section~\ref{dcs-formalism}, the situation is more complex in the electroproduction case due to the fact that the virtual photon probe has both transverse and longitudinal polarization components. Appendix~\ref{app:response_function} shows the full set of 36 response functions, $R_{ij}^{\alpha \beta}(Q^2,W,\cos \theta_K^{\mathrm{c.m.}})$ (or $R_{ij}^{\alpha \beta}(Q^2,W,t)$), necessary for a complete description of the process. This increased number of contributing response functions, for the case of the unpolarized cross section, single polarization observables (for beam, target, or hyperon), or double polarization observables with two reaction products in the initial and final state polarized, makes ``complete'' electroproduction experiments unfeasible at the current time. However, the CLAS $KY$ electroproduction program has made a start in providing data that give access to a ``reasonable'' subset of the contributing response functions over a broad range of $Q^2$, $W$, and $\cos \theta_K^{\mathrm{c.m.}}$. 

The measured observables from the CLAS $K^+Y$ electroproduction program provide access to the following response functions

\begin{itemize}
\item Unpolarized: $\sigma_{\rm U}$ : $R_{\rm T}^{00}$, $R_{\rm L}^{00}$ ~$\vert$~ $\sigma_{\rm L}$ : $R_{\rm L}^{00}$ ~$\vert$~ $\sigma_{\rm T}$ : $R_{\rm T}^{00}$ ~$\vert$~ $\sigma_{\rm LT}$ : $R_{\rm LT}^{00}$ ~$\vert$~ $\sigma_{\rm TT}$ : $R_{\rm TT}^{00}$
\item Beam single polarization : $\sigma_{\rm LT'}$ : $R_{\rm LT'}^{00}$
\item Hyperon single polarization : ${\cal P}^0_y$ : $R_{\rm LT}^{x'0}$, $R_{\rm LT}^{y'0}$, $R_{\rm LT}^{z'0}$ ~$\vert$~ ${\cal P}^0_{y'}$ : $R_{\rm T}^{y'0}$, $R_{\rm L}^{y'0}$
\item Beam-recoil hyperon double polarization: ${\cal P}'_x$ : $R_{\rm LT'}^{x'0}$, $R_{\rm LT'}^{y'0}$, $R_{\rm LT'}^{z'0}$ ~$\vert$~ ${\cal P}'_z$ : $R_{\rm TT'}^{x'0}$, $R_{\rm TT'}^{z'0}$ ~$\vert$~ 
${\cal P}'_{x'}$ : $R_{\rm TT'}^{x'0}$ ~$\vert$~ ${\cal P}'_{z'}$ : $R_{\rm TT'}^{z'0}$.
\end{itemize}
\noindent
Note that the hyperon recoil polarization and the beam-recoil hyperon transferred polarization components from CLAS data~\cite{CLAS:2002zlc,CLAS:2009sbn,CLAS:2014udv} are provided integrating over all angles $\Phi$ to accommodate finite bin sizes and to improve statistical precision. These integrated polarization components are designed with the ${\cal P}$ notation. Note also that these hyperon polarization components are presented in both the primed and unprimed coordinate systems. The components in these two different coordinate systems are sensitive to different combinations of the contributing response functions as given above. See Ref.~\cite{CLAS:2009sbn} for details on the complete formalism.

It is necessary to make clear when comparing the experimental observables to theoretical models that the so-called ``pre-factor'' terms containing the virtual photon polarization parameter $\epsilon$ multiplying the structure functions in the electroproduction cross section of Section~\ref{dcs-formalism} (see Eq.~(\ref{eq:most_general_cs})) are not unique in the literature. Some authors use a pre-factor for the $\sigma_{\rm L}$ ($\sigma_{\rm LT}$) term of $\epsilon_{\rm L}$ ($\sqrt{2\epsilon_{\rm L}(1 + \epsilon)}$), where $\epsilon_{\rm L} = \epsilon Q^2/E_{\gamma_v}^2$ parameterizes the longitudinal polarization of the virtual photon. In other work, including that from the CLAS $KY$ measurements, the use of the non-Lorentz invariant (or frame-dependent) term $\epsilon_L$ is strictly avoided and replaced with $\epsilon$ and, furthermore, factors of $\sqrt{2}$ are sometimes absorbed into the interference response functions in the observable extraction compared with the pre-factors defined in Ref.~\cite{Knochlein:1995qz} and discussed in Section~\ref{dcs-formalism}. Finally, in some work a $\sin \theta_K^{\mathrm{c.m.}}$ ($\sin^2 \theta_K^{\mathrm{c.m.}}$) term is taken out of the definition of $\sigma_{\rm LT}$ ($\sigma_{\rm TT}$). ({\it Caveat emptor!})

%%%%%%%%%%%%%%%%%%%%%%%%%%%%%%%%%%%%%%%%%%%%%%%%%%%%%%%%%%%%%%%%%%%%%%%%%%%%%%%%%%%%%%%%%%%%%%%%%%%%%%%%%%%%%%%%%%%%%%%%%%%%%%%%%%%%%%%%%%%%%%%%%%%%%%%%%%%%%%%%
\begin{figure*}[htbp]
\centering
\includegraphics[width=0.9\textwidth]{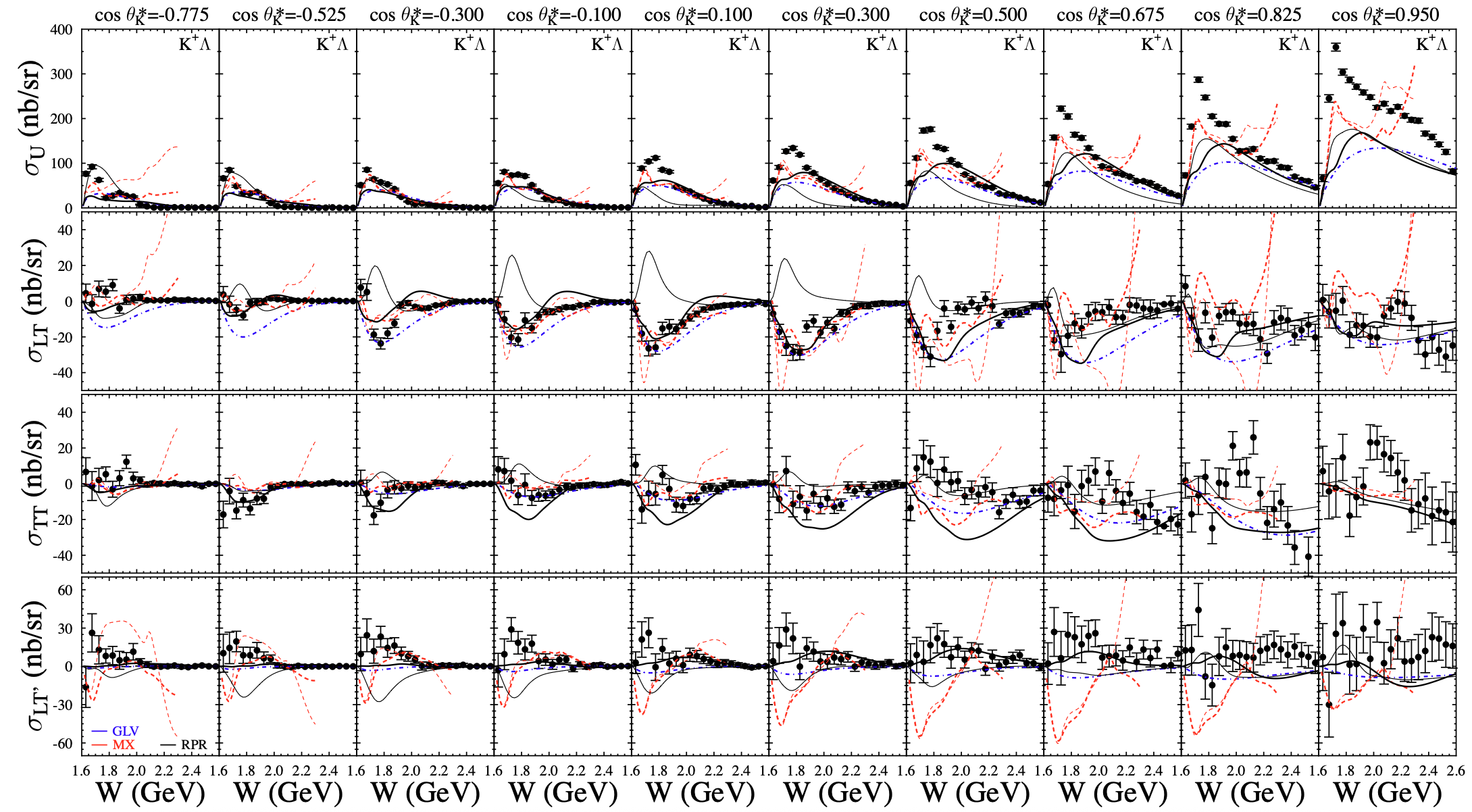}
\caption{Structure functions $\sigma_{\rm U} = \sigma_{\rm T} + \epsilon \sigma_{\rm L}$, $\sigma_{\rm LT}$, $\sigma_{\rm TT}$, and $\sigma_{\rm LT'}$ (nb/sr) from CLAS data for $K^+\Lambda$ production vs. $W$ (GeV) for $E_{\rm beam}=5.5$~GeV for $Q^2=1.80$~GeV$^2$ and $\cos \theta_K^{\mathrm{c.m.}}$ values as shown. The error bars represent the statistical uncertainties only. The red curves are from the hadrodynamic $KY$ model of Maxwell~\cite{Maxwell:2012zz}, the blue curves are from the hybrid RPR-2011 $KY$ model from Ghent~\cite{DeCruz:2012bv}, and the black curves are from the GLV Regge model~\cite{Guidal:1999qi}. Figure from Ref.~\cite{Carman:2012qj}.}
\label{lam_q1_w} 
\end{figure*}
%%%%%%%%%%%%%%%%%%%%%%%%%%%%%%%%%%%%%%%%%%%%%%%%%%%%%%%%%%%%%%%%%%%%%%%%%%%%%%%%%%%%%%%%%%%%%%%%%%%%%%%%%%%%%%%%%%%%%%%%%%%%%%%%%%%%%%%%%%%%%%%%%%%%%%%%%%%%%%%%

%%%%%%%%%%%%%%%%%%%%%%%%%%%%%%%%%%%%%%%%%%%%%%%%%%%%%%%%%%%%%%%%%%%%%%%%%%%%%%%%%%%%%%%%%%%%%%%%%%%%%%%%%%%%%%%%%%%%%%%%%%%%%%%%%%%%%%%%%%%%%%%%%%%%%%%%%%%%%%%%
\begin{figure*}[htbp]
\centering
\includegraphics[width=0.9\textwidth]{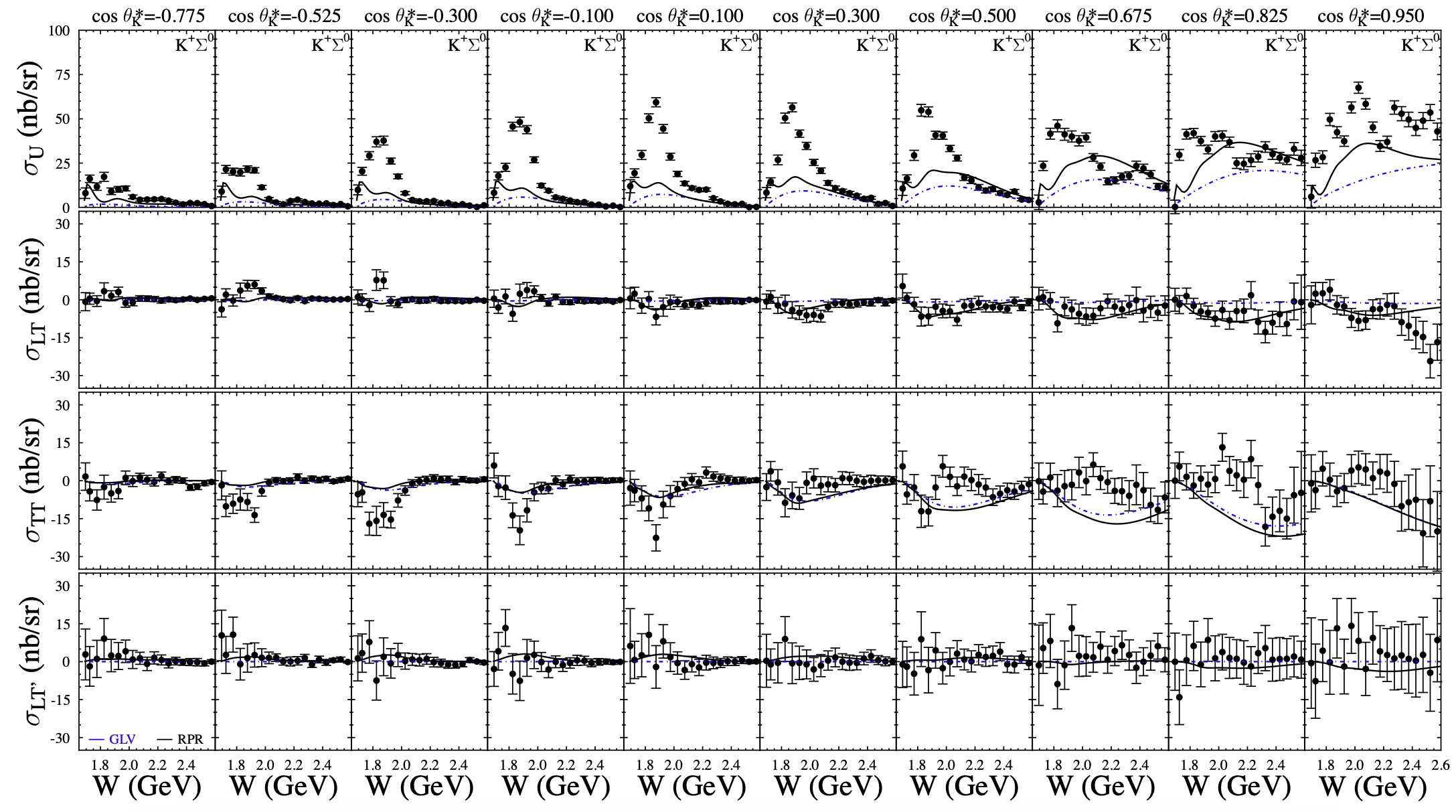}
\caption{Structure functions $\sigma_{\rm U} = \sigma_{\rm T} + \epsilon \sigma_{\rm L}$, $\sigma_{\rm LT}$, $\sigma_{\rm TT}$, and $\sigma_{\rm LT'}$ (nb/sr) from CLAS data for $K^+\Sigma^0$ production vs. $W$ (GeV) for $E_{\rm beam}=5.5$~GeV for $Q^2=1.80$~GeV$^2$ and $\cos \theta_K^{\mathrm{c.m.}}$ values as shown. The error bars represent the statistical uncertainties only. The blue curves are from the hybrid RPR-2007 $KY$ model from Ghent \cite{Corthals:2007kc} and the black curves are from the GLV Regge model~\cite{Guidal:1999qi}. Figure from Ref.~\cite{Carman:2012qj}.}
\label{sig_q1_w} 
\end{figure*}
%%%%%%%%%%%%%%%%%%%%%%%%%%%%%%%%%%%%%%%%%%%%%%%%%%%%%%%%%%%%%%%%%%%%%%%%%%%%%%%%%%%%%%%%%%%%%%%%%%%%%%%%%%%%%%%%%%%%%%%%%%%%%%%%%%%%%%%%%%%%%%%%%%%%%%%%%%%%%%%%

Figures~\ref{lam_q1_w} and \ref{sig_q1_w} show one representative $Q^2$ bin at 1.80~GeV$^2$ from the available CLAS data (other bins are centered at 2.60~GeV$^2$ and 3.45~GeV$^2$) for the $K^+\Lambda$ and $K^+\Sigma^0$ structure functions $\sigma_{\rm U}$, $\sigma_{\rm LT}$, $\sigma_{\rm TT}$, and $\sigma_{\rm LT'}$~\cite{Carman:2012qj}, illustrating its broad kinematic coverage in $W$ and $\cos \theta_K^{\mathrm{c.m.}}$, as well as its statistical precision. The helicity-dependent polarized structure function $\sigma_{\rm LT'}$ is intrinsically different from the structure functions $\sigma_{\rm U}$, $\sigma_{\rm TT}$, $\sigma_{\rm LT}$ of the unpolarized cross section. This term is generated by the imaginary part of terms involving the interference between longitudinal and transverse components of the hadronic and leptonic currents, in contrast to $\sigma_{\rm LT}$, which is generated by the real part of the same interference. $\sigma_{\rm LT'}$ is non-vanishing only if the hadronic tensor is antisymmetric, which occurs in the presence of rescattering effects, interference contributions between multiple resonances, interferences between resonant and non-resonant processes, or even between non-resonant processes alone~\cite{Boffi:1993gs}. In fact, $\sigma_{\rm LT'}$ could be non-zero even when $\sigma_{\rm LT}$ is zero. When the reaction proceeds through a channel in which a single amplitude dominates, the longitudinal-transverse response will be real and $\sigma_{\rm LT'}$ will vanish. Extractions of both $\sigma_{\rm LT}$ and $\sigma_{\rm LT'}$ are required to unravel the full longitudinal-transverse response of the $K^+Y$ electroproduction reactions. It should be made clear in terms of notation of the response functions provided in Table~\ref{tab:response_functions} of Appendix~\ref{dcs-formalism} that $\sigma_{\rm LT}$ ($\sigma_{\rm LT'}$) can equivalently be written as $\sigma_{\rm TL}$ ($\sigma_{\rm TL'}$).

%%%%%%%%%%%%%%%%%%%%%%%%%%%%%%%%%%%%%%%%%%%%%%%%%%%%%%%%%%%%%%%%%%%%%%%%%%%%%%%%%%%%%%%%%%%%%%%%%%%%%%%%%%%%%%%%%%%%%%%%%%%%%%%%%%%%%%%%%%%%%%%%%%%%%%%%%%%%%%%%
\begin{figure*}[htbp]
\centering
\includegraphics[width=0.65\textwidth]{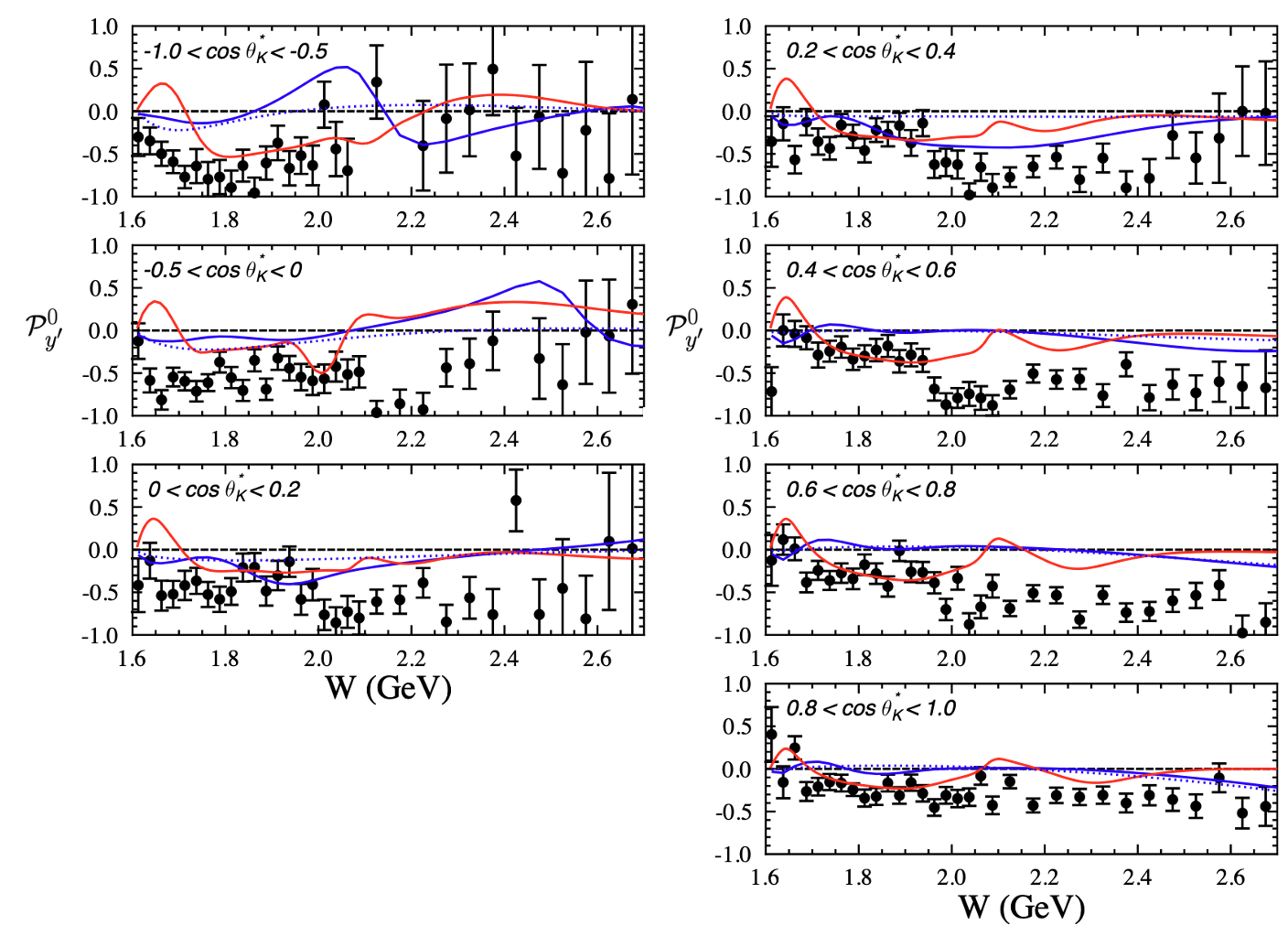}
\caption{Recoil hyperon polarization ${\cal P}^0_{y'}$ in the reaction $ep \to e'K^+\Lambda$ from CLAS data vs. $W$ at an average $Q^2=1.9$~GeV$^2$ for $E_{\rm beam}=5.5$~GeV and $\cos \theta_K^{\mathrm{c.m.}}$ ranges as shown. The error bars represent the statistical uncertainties only. The red curves are from the isobar model of Maxwell~\cite{Maxwell:2012zz} and the blue curves are from the hybrid RPR-2011 $KY$ model from Ghent~\cite{DeCruz:2012bv}. Figure adapted from Ref.~\cite{CLAS:2014udv}.}
\label{ipol-kl} 
\end{figure*}
%%%%%%%%%%%%%%%%%%%%%%%%%%%%%%%%%%%%%%%%%%%%%%%%%%%%%%%%%%%%%%%%%%%%%%%%%%%%%%%%%%%%%%%%%%%%%%%%%%%%%%%%%%%%%%%%%%%%%%%%%%%%%%%%%%%%%%%%%%%%%%%%%%%%%%%%%%%%%%%%

\begin{figure*}[htbp]
\centering
\includegraphics[width=0.5\textwidth]{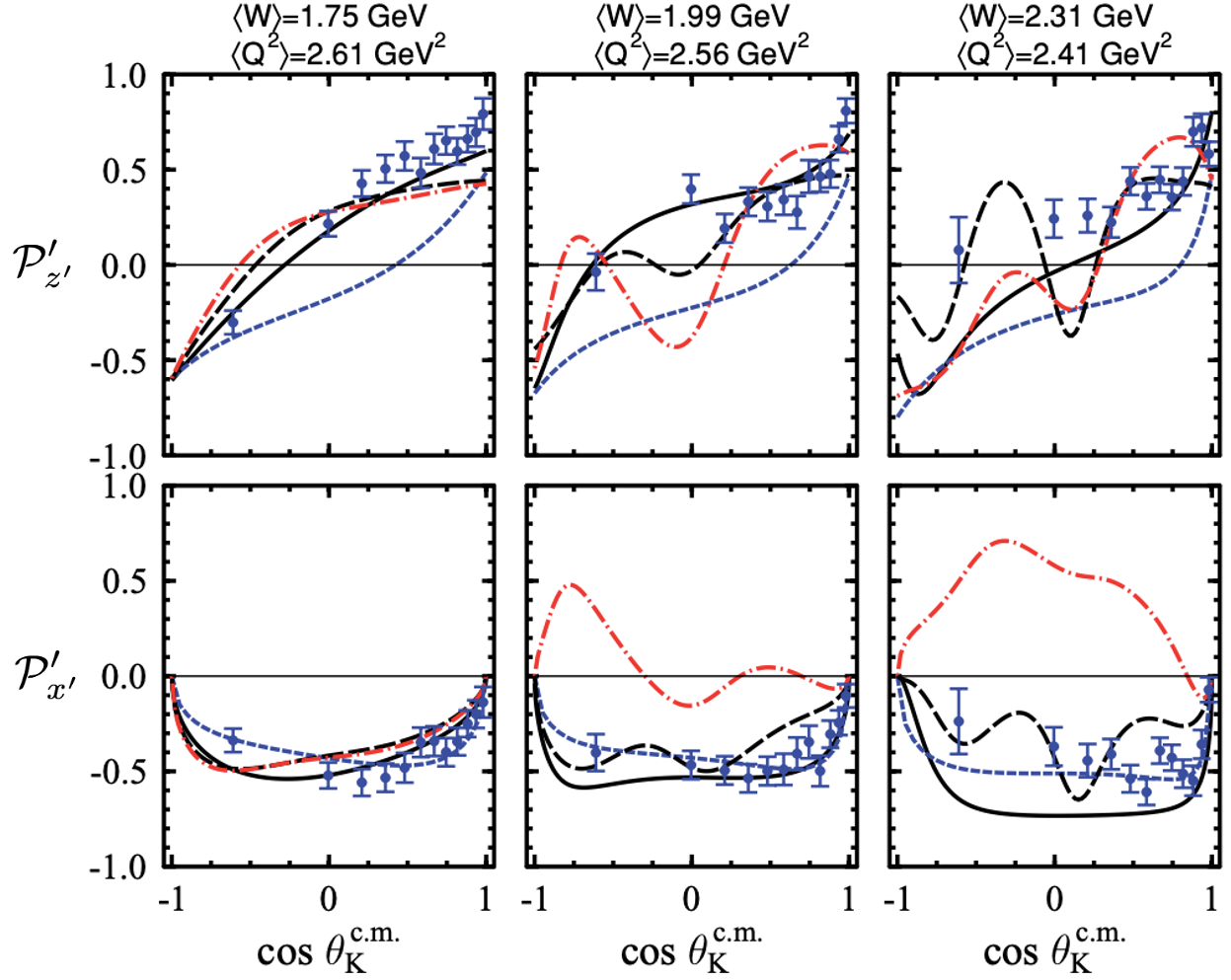}
\caption{Beam-recoil hyperon transferred polarization components ${\cal P}'$ with respect to the $(x',z')$ axes in the reaction $ep \to e'K^+\Lambda$ from CLAS data vs. $\cos \theta_K^{\mathrm{c.m.}}$ for three bin-averaged $W/Q^2$ values as indicated for a beam energy of 5.754~GeV. The curves are calculations from the MB isobar model~\cite{Mart:2006dk} (solid, black), the GLV Regge model~\cite{Guidal:1999qi} (short dash, blue), and the RPR model variant including an $N(1900)1/2^+$ state (dot-dash, red) and an $N(1900)3/2^-$ state (long dash, black)~\cite{Corthals:2007kc}. Figure from Ref.~\cite{CLAS:2009sbn}.}
\label{tpol-kl} 
\end{figure*}

Figure~\ref{ipol-kl} shows the $W$ dependence of the recoil $\Lambda$ polarization normal to the hadronic reaction plane ${\cal P}^0_{y'}$ at an average $Q^2$ of 1.9~GeV$^2$ for different angular ranges for $\cos \theta_K^{\mathrm{c.m.}}$ \cite{CLAS:2014udv}. Finally, Fig.~\ref{tpol-kl} shows the beam-recoil transferred $\Lambda$ polarization components ${\cal P}'$ within the hadronic reaction plane at an average $Q^2$ of 2.5~GeV$^2$ for bins in $W$ from 1.75~GeV to 2.31~GeV. The recoil and beam-recoil hyperon polarization components are mainly dependent on the ${\rm LT'}$ and ${\rm TT'}$ interference structure functions and are sensitive to the resonant amplitudes and the resonant/non-resonant interference amplitudes to the reaction mechanism. Extending this statement a bit further, polarization observables, in general, allow access to interference terms between the production amplitudes, while the unpolarized cross sections are only sensitive to the squared absolute values of the amplitudes. Section~\ref{nstar-spectrum} details the importance of the hyperon polarization observables to the spectroscopy of $s$-channel nucleon resonances that couple to the $KY$ final states.

These polarization data also provide insight into the nature of quark-pair production. It has long been suggested that the appropriate degrees of freedom to describe the phenomenology of hadronic decays are constituent quarks bound through a gluonic flux tube~\cite{Isgur:1984bm}. The non-perturbative nature of this flux tube gives rise to the well-known linear potential of heavy-quark confinement \cite{Bali:1994de}. Other properties of the flux tube can be probed through processes involving $q\bar{q}$ quark pair production, since this is expected to produce the color field neutralization that breaks the flux tube. Since the 1970s it has been suggested that a quark pair with vacuum quantum numbers is responsible for breaking the color flux tube, e.g. the $^3P_0$ model detailed in Ref.~\cite{LeYaouanc:1972vsx}. The most sensitive experiments to test this model have measured ratios in certain meson decays of strong amplitudes differing in their orbital angular momenta \cite{Geiger:1994kr}. Since the $^3P_0$ operator has $S = 1$ and $L = 1$, it implies a different amplitude ratio than, e.g., a $^3S_1$ operator with $S = 1$ and $L = 0$, corresponding to one gluon exchange. The results of Ref.~\cite{CLAS:2002zlc} are of interest as it seems that the spin properties of the quark-pair creation operator might also be responsible for the trends seen in the $\Lambda$ polarization, which indicate that the relevant operator dominating this process produces the $s\bar{s}$ pair with spins anti-aligned~\cite{cern_2003,cern_2007}. If this is a true reflection of the dynamics, it brings into question the universal applicability of the $^3P_0$ model, which has important implications since many, if not most, calculations of hadronic spectroscopy use the $^3P_0$ operator to calculate the transition to the final-state particles~\cite{Barnes:2002pw}.

%%%%%%%%%%%%%%%%%%%%%%%%%%%%%%%%%%%%%%%%%%%%%%%%%%%%%%%%%%%%%%%%%%%%%%%%%%%%%%%%%%%%%%%%%%%%%%%%%%%%%%%%%%%%%%%%%%%%%%%%%%%%%%%%%%%%%%%%%%%%%%%%%%%%%%%%%%%%%%%%
%%%%%%%%%%%%%%%%%%%%%%%%%%%%%%%%%%%%%%%%%%%%%%%%%%%%%%%%%%%%%%%%%%%%%%%%%%%%%%%%%%%%%%%%%%%%%%%%%%%%%%%%%%%%%%%%%%%%%%%%%%%%%%%%%%%%%%%%%%%%%%%%%%%%%%%%%%%%%%%%

There has been considerable effort during the past three decades to develop models for $KY$ photo- and electroproduction (see Section~\ref{sec:models}). However, model fits to the cross section data are generally obtained at the expense of many free parameters, which makes it difficult to provide precise constraints given the datasets used so far in the model fits. Moreover, cross section data alone are not sufficient to fully understand the reaction mechanism, as they represent only a portion of the full amplitude response (see the response functions listed in Table~\ref{tab:response_functions} of Appendix~\ref{dcs-formalism}). In this regard, measurements of spin observables are essential for continued improvement and refinement in the reaction models used to describe the $KY$ process. Fits to only limited subsets of the available data lead to ambiguities and model dependence in interpreting the results. Only by fitting all available cross section and polarization observables can the reaction models be developed to provide for improved constraints on their parameters, increasing their discriminatory power and allowing for a quantitative measure of whether or not new resonance states are required to explain the data. One central issue involves discriminating $s$-channel resonance states from the non-resonant background, from effects caused by final-state interactions or channel couplings, or from incomplete treatment of Fermi motion of the target nucleon for the $\gamma_v n$ observables.

Figures~\ref{lam_q1_w} to \ref{tpol-kl} include predictions from the single channel reaction models available at the time of publication. The models include the MX isobar model from Maxwell~\cite{Maxwell:2012zz}, the hybrid Regge-plus-Resonance RPR model from Ghent~\cite{Corthals:2007kc,DeCruz:2012bv}, the MB isobar model from Mart-Bennhold~\cite{Mart:2006dk}, and the GLV Regge model~\cite{Guidal:1999qi} (see Section~\ref{sec:models} for details). The available single-channel isobar models that have been developed to date have been constrained by fits to only a limited portion of the available $KY$ photoproduction data. However, since the time the models were developed, they have not been refit to the full set of available photoproduction cross section and polarization observables. The models at least (more or less) faithfully describe the specific observables in the kinematic ranges for which they have been constrained by the data in their fitting database. Yet, it is abundantly clear that none of the available isobar models are able to reproduce the kinematic dependence seen in the $KY$ electroproduction cross sections and polarization observables with their current parameters. Given that they were based on fits to the $KY$ photoproduction data, it should not be expected that they will be able to explain or predict the electroproduction data. Therefore, it must be stated that their current level of discrepancy compared to the data is not (yet) an indictment of the models. Only by including the electroproduction data in their databases and simultaneously fitting both the photo- and electroproduction observables will it be possible to shed light on their ability to first describe the electroproduction data and second to gain insights into the presence of additional mechanisms that are relevant for $Q^2 >0$. These mechanisms may gradually emerge with increasing $Q^2$ or be related to the contribution from the amplitudes for longitudinally polarized photons that are absent in photoproduction. Further tests turning individual $N^*$ states on and off within the single-channel isobar models could also provide for insight into how individual states affect the observables as a function of $Q^2$, $W$, and $\cos \theta_K^{\mathrm{c.m.}}$. Section~\ref{sec:models} contains an essential overview of the data included in the fits of the available models.

Ultimately, however, it will be necessary to move beyond the single-channel models to include the full dynamics from coupled-channel approaches that make possible a combined global analysis of all available data on exclusive meson photo-, electro-, and hadroproduction. Such efforts are advancing by groups at ANL-Osaka~\cite{Kamano:2013iva}, Bonn-Gatchina~\cite{Sarantsev:2016fac}, J{\"u}lich-Bonn \cite{Ronchen:2014cna,Ronchen:2018ury}, J{\"u}lich-Bonn-Washington~\cite{Mai:2023cbp}, and JLab \cite{Julia-Diaz:2007mae}. It is also important to note that the $\pi N$ and $\pi\pi N$ electroproduction channels represent the two dominant exclusive channels in the resonance region. Knowledge of the electroproduction mechanisms for these channels is directly relevant for $N^*$ studies in channels with smaller cross sections such as $K^+\Lambda$ and $K^+\Sigma^0$ production, as they can be significantly affected in leading order by coupled-channel effects produced by their hadronic interactions in the pion channels.

The final area of study from the CLAS strangeness program in the electroproduction sector of the JLab 6-GeV era includes measurements of the excited hyperon states $\Lambda(1405)$ and $\Lambda(1520)$. These measurements are detailed in Table~\ref{hallb-clas-list3}. The CLAS electroproduction measurement of the $\Lambda(1405)$ focused on the study of the $Q^2$ evolution of its lineshape in the decay branch $\Lambda(1405) \to \Sigma^+ \pi^- \to p \pi^0 \pi^-$~\cite{CLAS:2013zie}. This work served as a complementary analysis to the companion photoproduction studies from CLAS mentioned in Section~\ref{clas-gp-program}. The mass distribution again revealed evidence that the $\Lambda(1405)$ cannot be described by a simple Breit-Wigner lineshape. Instead, it is better described by a two-pole structure in agreement with chiral unitary models \cite{Hyodo:2011ur}. Furthermore, the relative strength of the two contributions to the $\Lambda(1405)$ seems to evolve with $Q^2$. The $\Lambda(1520) \to K^- p$ decay was studied through the kinematic dependence of its differential cross sections vs. $Q^2$, $W$, $\theta_{K^+}^{c.m.}$, and $\Phi$~\cite{Clas:2001hvx}. This analysis probed the spin-parity $J^P = 3/2^-$ $\Lambda(1520)$ production mechanism to show that it is dominated by longitudinal $t$-channel exchange contributions as $Q^2$ increases away from the photon point. This analysis also measured the $\Lambda(1520)$ daughter decay angular distributions in the hyperon decay frame to reveal a dominance of the contributions of the $m_z = \pm 3/2$ spin projections relative to $m_z = \pm 1/2$ in its decay. 

%%%%%%%%%%%%%%%%%%%%%%%%%%%%%%%%%%%%%%%%%%%%%%%%%%%%%%%%%%%%%%%%%%%%%%%%%%%%%%%%%%%%%%%%%%%%%%%%%%%%%%%%%%%%%%%%%%%%%%%%%%%%%%%%%%%%%%%%%%%%%%%%%%%%%%%%%%%%%%%
\begin{table}[tbh]
\setlength{\tabcolsep}{6pt} % Default value: 6pt
\renewcommand{\arraystretch}{0.8} % Default value: 1
\begin{center}
\caption{Summary of $ep\to e'K^+\Lambda^*$ electroproduction measurements measurements in Hall B with CLAS from the 6-GeV era experiments at JLab. The column labeled $N_{\rm bin}$ indicates the number of kinematic bins included in the analysis.}
\begin{tabular}{cccccccc} \hline\hline
Observable & Final State        & $Q^2$ (GeV$^2$) & $W$ (GeV)     & $\cos \theta_K^{\mathrm{c.m.}}$ & $N_{\rm bin}$ & Year & Ref. \\ \hline
Lineshape  & $K^+\Lambda(1405)$ & $[1.0:3.0]$     & $[1.5:3.5]$   & $[-1.0:1.0]$             & -       & 2013 & \cite{CLAS:2013zie} \\ \hline 
$d\sigma$  & $K^+\Lambda(1520)$ & $[0.9:2.4]$     & $[1.95:2.65]$ & $[-1.0:1.0]$             & 96      & 2001 & \cite{Clas:2001hvx} \\ \hline\hline
\end{tabular}
\label{hallb-clas-list3}
\end{center}
\end{table}
%%%%%%%%%%%%%%%%%%%%%%%%%%%%%%%%%%%%%%%%%%%%%%%%%%%%%%%%%%%%%%%%%%%%%%%%%%%%%%%%%%%%%%%%%%%%%%%%%%%%%%%%%%%%%%%%%%%%%%%%%%%%%%%%%%%%%%%%%%%%%%%%%%%%%%%%%%%%%%%%

\subsection{Experiments at JLab -- 12-GeV Era}
\label{jlab:12gev}

After supporting a broad program for fixed-target experiments with beams of electrons and photons at energies up to 6~GeV in the period from 1997-2012, Jefferson Lab transitioned to a new phase of operations. As part of the 12-GeV upgrade project, the CEBAF accelerator energy was doubled due to advances made possible in superconducting RF technology that enabled significantly increased accelerating gradients~\cite{Adderley:2024czm}. This project included significant equipment upgrades in the three existing experimental endstations, Halls A, B, and C, and the construction of a new experimental Hall D devoted to photoproduction studies.

The 12-GeV research program was developed around five primary subject areas:

\begin{itemize}
\item Investigations of baryon and meson spectroscopy and structure including the search for non-$qqq$ baryon and non-$q\bar{q}$ meson hybrid configurations;
\item Precision tests of the Standard Model and fundamental symmetries using parity violating electron scattering;
\item Mapping out the spin and flavor dependence of valence proton distributions;
\item Unraveling the three-dimensional structure of the nucleon via femtographic measurements of the Generalized Parton Distributions;
\item Exploration of how valence quark structure is modified in a dense nuclear medium.
\end{itemize}

The 12-GeV experimental program at JLab began in 2016/2017 with beam operations in all four experimental halls. See Refs.~\cite{Burkert:2018nvj,Arrington:2021alx} for an overview of the full scientific program. In the following subsections a summary of the ongoing program in strangeness physics is provided.

\subsubsection{Hall C Hypernuclear Program}

The hypernuclear program that commenced during the 6-GeV era at JLab (see Section~\ref{halla-c-6gev}) will continue through a fourth generation of experiments that are scheduled to begin in the next several years \cite{Achenbach:2024lpw}. Due to the long-term staging of new experiments in Hall~A, the entirety of the upcoming hypernuclear running has been designed to be accommodated in Hall~C. Studies of hypernuclei tagged through the $(e,e'K^+)$ reaction with the forward-angle of detection of the electron in the HES spectrometer and the kaon in the HKS spectrometer (see Fig.~\ref{hallc_hyper}) are planned at beam energies up to 4~GeV. The detector packages in the spectrometers, the beamline, and the septum magnets to separate the forward-scattered electrons and kaons have been upgraded from their third-generation counterparts used during the JLab 6-GeV era experiments to enable higher-luminosity operations, improved resolution in the measured binding energies, and reduced experimental backgrounds.

The approved program consists of several complementary experiments. The first experiments are focused on what is termed the ``hyperon puzzle'' \cite{Bombaci:2016xzl} that has arisen due to the observation of neutron stars larger than two solar masses, which has challenged present understanding of the interaction among baryons in such conditions. Within the interior of neutron stars, nuclear matter is highly asymmetric, with nucleons embedded in an environment at significantly increased densities compared to standard nuclei. In such systems, the formation of hyperons is energetically favored, since their presence reduces the Fermi energy of the nucleon sea. However, in present models, this energy reduction leads to neutron stars smaller than 1.5 solar masses, leading to a tension with astronomical observations. It is expected that improved understanding of the $\Lambda N$ and $\Lambda NN$ interactions in a variety of light to heavy hypernuclei can help to address this puzzle. The planned experiments will provide a spectroscopic study of the binding energy of the levels in the hypernuclei $^{40}_{~\Lambda}$K, $^{48}_{~\Lambda}$K, and $^{208}_{~\Lambda}$Pb to enable reliable predictions on the structure and dynamics of neutron stars~\cite{nakamura-hyper,garibaldi-hyper}. The program will also include an experiment to study light hypernuclei through decay pion spectroscopy (including a third spectrometer for the detection of the $\pi^-$) based on the approach pioneered at MAMI~\cite{A1:2024edr} (see Section~\ref{mami-program})~\cite{Achenbach:2024lpw,JLabhypernuclear:2025vfp}.

A second hypernuclear experiment focuses on investigation of charge symmetry breaking in the $\Lambda N$ interaction. Evidence of charge symmetry breaking was found in measurements of the binding-energy difference for the $s$-shell isodoublet hypernuclear systems, $^4_\Lambda$He and $^4_\Lambda$H~\cite{A1:2015isi,A1:2016nfu,J-PARCE13:2015uwb}. The experimental goal is to advance and extend such investigations with the measurement of binding energies in the $p$-shell hypernuclei $^6_\Lambda$He, $^9_\Lambda$Li, and $^{11}_{~\Lambda}$Be~\cite{gogami-hyper}. The final experiment in the program is designed to use a $\Lambda$ hyperon to probe the collective deformation in $^{26}$Mg through the study of the reaction $^{27}$Al$(e,e'K^+)$$^{27}_\Lambda$Mg~\cite{nakamura-hyper2}.

\subsubsection{Hall C Kaon Form Factor Studies}
\label{hallc12-program}

Investigations of the exclusive $ep\to e'K^+\Lambda$ and $ep\to e'K^+\Sigma^0$ reactions in the 12-GeV era in Hall~C are presently underway based on data taken in 2018/2019. These investigations serve as an extension of the measurements of the separated $\sigma_{\rm T}$ and $\sigma_{\rm L}$ structure functions carried out as the first experiment in Hall~C in 1998 (see Section~\ref{halla-c-6gev}), and are designed to measure forward $K^+$ electroproduction, detecting the kaon in the Super High Momentum Spectrometer (built to upgrade the Hall~C experimental equipment as part of the 12-GeV upgrade~\cite{Ali:2025dan}) in coincidence with the scattered electron in the HMS (see Fig.~\ref{halla_c-layout}). Differential cross sections will ultimately be provided at multiple beam energies and, hence, different values of the transverse virtual photon polarization $\epsilon$. In overlapping bins of ($Q^2$, $W$, $\cos \theta_K^{\mathrm{c.m.}}$), the separated transverse and longitudinal structure functions will be extracted using the Rosenbluth approach. Measurements in non-parallel kinematics, i.e. $\vert \cos \theta_K^{\mathrm{c.m.}} \vert \ne 1$, will allow for simultaneous extraction of the interference terms $\sigma_{\rm TT}$ and $\sigma_{\rm LT}$, and measurements of the $-t$ dependence of the $K^+$ cross section will be carried out~\cite{kaonlt}.

There are two main physics goals from these studies. The first is an investigation of scaling and factorization to higher $Q^2$ than the available Hall~C data. The separated $\sigma_{\rm T}$ and $\sigma_{\rm L}$ structure functions of the exclusive $K^+\Lambda$ and $K^+\Sigma^0$ channels allow investigations of the transition from hadronic to partonic degrees of freedom as a function of increasing $Q^2$ or decreasing distance scale. Recent analysis of the related $ep\to e'\pi^+p$ channel from JLab suggests that the power law behavior expected from the hard scattering mechanism is consistent with the evolution of $\sigma_{\rm L}$ as a function of $Q^2$~\cite{Yao:2024drm}. The $Q^2$-dependence of the pion electromagnetic form factor is also consistent with the $Q^2$ scaling expectation in the range of $Q^2 > 1$~GeV$^2$, although the observed magnitude of the charged pion form factor $F_\pi$ is larger than the hard QCD prediction. It is possible that QCD factorization is not applicable in this regime or that understanding of the soft contributions from the wavefunction in meson production is incomplete. A direct comparison of the scaling properties of the $K^+Y$ cross sections as a function of increasing $Q^2$ using precision data from this experiment could provide insights into the onset of factorization in the transition from the hadronic to the partonic regimes, and provide a study of scaling in a strange system for the first time.

This measurement will provide high quality $\sigma_{\rm T}$ and $\sigma_{\rm L}$ structure functions in kinematics above the nucleon resonance region for the first time, which is essential for a better understanding of the $K^+$ reaction mechanism. If these studies indicate that the $K^+$ pole term dominates $\sigma_{\rm L}$ at low $-t$, they could be used to extract the $K^+$ form factor using the Chew-Low approach in which $F_K$ is determined by extrapolating the measured $t$-dependence of the longitudinal cross section $\sigma_{\rm L}$ to the kaon pole relying on the dependence shown in Eq.~(\ref{eq:siglff})~\cite{Chew:1956zz,Ghahramany:2004sc}. A theoretical discussion of the kaon form factor is provided in Section~\ref{subsec:Kaon_form_factor}.

\subsubsection{Hall D GlueX Photoproduction}

The gluonic excitations (GlueX) experiment in Hall~D at JLab was designed to study the light quark meson spectrum through photoproduction processes. The initial motivation for the GlueX program was the search for exotic hybrid mesons with glue as an active degree of freedom~\cite{Dzierba:2003fe}. Since these $q\bar{q}g$ states are expected to decay into charged and neutral final state particles, the GlueX spectrometer was designed to be able to detect and reconstruct both types of reaction products, spanning nearly the full azimuthal angle range for polar angles from $2^\circ - 150^\circ$ surrounding the beam-target interaction point. The practical momentum resolution of GlueX is in the range from 2\%--7\% with angular resolution $\Delta \theta, \Delta \phi \sim 2-5$~mrad, where the resolution is dependent on particle momentum and polar angle. However, as the event reconstruction for exclusive final states relies on constrained kinematic fitting, the effective momentum and angular resolution are significantly better than these values.

The 12-GeV electron beam from the CEBAF accelerator is used in Hall~D to create a high-intensity, tagged, linearly polarized photon beam by coherent bremsstrahlung off a precisely aligned diamond radiator. The polarization approaches 40\% in the region of the coherent peak, from 8.2 to 8.8~GeV. The scattered electrons are directed into the Hall~D photon tagger, which allows a measurement of the energy of the produced photons to 0.1\% precision within the region of the coherent peak. The photon beam passes through a collimator in order to suppress the incoherent contribution, a triplet polarimeter, and a pair spectrometer, which provide a continuous, non-invasive measurement of the photon beam polarization and the relative flux, respectively, before reaching the liquid-hydrogen target positioned within the 2~T solenoid of the GlueX spectrometer~\cite{GlueX:2020idb}. The GlueX solenoid was originally built in the 1970s as part of the LASS (Large Aperture Superconducting Solenoid) installation at SLAC~\cite{Aston:1987uc}. It was then refurbished as part of the MEGA (Muon-to-Electron Gamma) experiment at Los Alamos (LAMPF) in the 1980s/1990s~\cite{Barakat:1994bu} before being made available to JLab for Hall~D. 

\begin{figure}[ht]
\centering
\includegraphics[width=0.6\columnwidth]{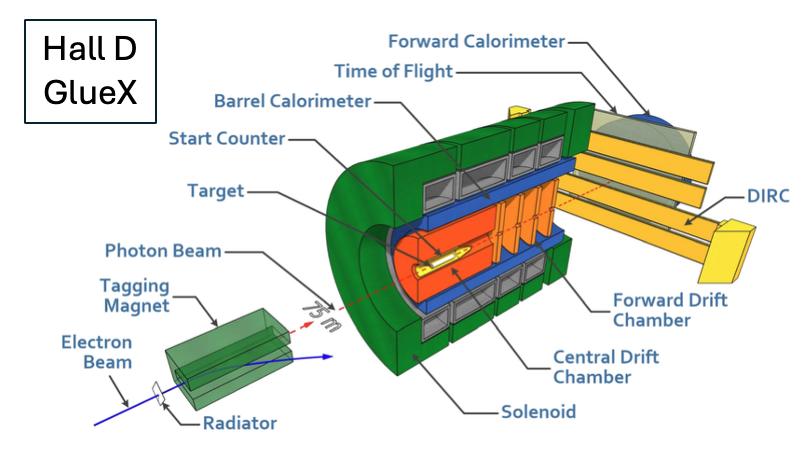} 
\caption{Model representation of the GlueX tagger hall and spectrometer in Hall D. The detector subsystems are labeled and outlined in the text. Note that the above model is not drawn to scale. The length of the solenoid magnet is 4.8~m. Figure adapted from Ref.~\cite{GlueX:2021myx}.}
\label{gluex-model}
\end{figure}

The GlueX target cell (nominal length of 30~cm) is surrounded by a scintillator start counter, a straw-tube central drift chamber, and a lead $+$ scintillating-fiber barrel calorimeter. The forward detector systems include multiple layers of drift chambers inside the solenoid, and downstream of the solenoid, a scintillator hodoscope, a Cherenkov detector for $\pi/K$ separation, and a forward calorimeter. The drift chambers provide measurements of momentum and $dE/dX$ energy loss for charged particle identification, while the calorimeters provide energy and position measurements for reconstruction of both charged and neutral particles. Time-of-flight measurements for particle identification are provided by the start counter, the calorimeters, and the time-of-flight wall. Figure~\ref{gluex-model} shows a schematic model of the Hall~D photon tagger and the GlueX spectrometer.

%%%%%%%%%%%%%%%%%%%%%%%%%%%%%%%%%%%%%%%%%%%%%%%%%%%%%%%%%%%%%%%%%%%%%%%%%%%%%%%%%%%%%%%%%%%%%%%%%%%%%%%%%%%%%%%%%%%%%%%%%%%%%%%%%%%%%%%%%%%%%%%%%%%%%%%%%%%%%%%%%%%%%%%%%
%%%%%%%%%%%%%%%%%%%%%%%%%%%%%%%%%%%%%%%%%%%%%%%%%%%%%%%%%%%%%%%%%%%%%%%%%%%%%%%%%%%%%%%%%%%%%%%%%%%%%%%%%%%%%%%%%%%%%%%%%%%%%%%%%%%%%%%%%%%%%%%%%%%%%%%%%%%%%%%%%%%%%%%%

The initial phase of GlueX data taking (called GlueX-I) was completed in 2018. This period was dedicated to detector commissioning and initial measurement cross-checks with operations at beam photon rates up to $\sim$$2 \times 10^7$~$\gamma$/s in the coherent peak. The second phase of the GlueX experiment (called GlueX-II) that took place from 2020 to 2023 operated at $\sim$$5 \times 10^7$~$\gamma$/s in the coherent peak and was the first beam run to include the DIRC-type Cherenkov detector downstream of the solenoid for enhanced $\pi/K$ separation~\cite{Ali:2022amh}. The third phase of the GlueX experiment (called GlueX-III) includes an upgraded forward calorimeter system to provide improved pixel granularity for photon detection for $\pi^0$ and $\eta$ meson decays~\cite{Somov:2025uiw}. Data taking will occur at a photon flux of $1 \times 10^8$~$\gamma$/s through 2026. Beyond this time frame a new $K_L$ facility will be setup in Hall~D~\cite{Dobbs:2022agy}. This experiment will create an intense secondary $K_L$ beam that will be directed onto the target cell within the GlueX spectrometer.

The strangeness physics program at GlueX is focused on three primary research areas that extend measurements previously studied with the CLAS 6-GeV era photoproduction program (see Section~\ref{clas-gp-program}). These include studies of ground state $\Lambda$ and $\Sigma$ hyperon production, studies of $\Lambda$ excited states with a focus on the $\Lambda(1405)$ and $\Lambda(1520)$, and $S = -2$ spectroscopy of $\Xi$ hyperons. An overview of this program is provided in Ref.~\cite{Pauli:2022ehd} and discussed briefly below. A summary of the published results is included in Table~\ref{halld-gluex-list}.

%%%%%%%%%%%%%%%%%%%%%%%%%%%%%%%%%%%%%%%%%%%%%%%%%%%%%%%%%%%%%%%%%%%%%%%%%%%%%%%%%%%%%%%%%%%%%%%%%%%%%%%%%%%%%%%%%%%%%%%%%%%%%%%%%%%%%%%%%%%%%%%%%%%%%%%%%%%%%%%
\begin{table}[tbh]
\setlength{\tabcolsep}{6pt} % Default value: 6pt
\renewcommand{\arraystretch}{0.8} % Default value: 1
\begin{center}
\caption{Summary of strangeness photoproduction measurements in Hall D with GlueX from the 12-GeV era experiments at JLab. The column labeled $N_{\rm bin}$ indicates the number of kinematic bins included in the analysis.}
\begin{tabular}{cccccccc} \hline\hline
Observable & Final State        & $-t$ (GeV$^2$) & $E_\gamma^{\rm lab}$ (GeV) & $N_{\rm bin}$ & Year & Ref. \\ \hline
$\Sigma$   & $K^+\Sigma^0$      & $[0.1:1.4]$    & $[4.0:4.2]$                & 4             & 2020 & \cite{GlueX:2020qat} \\ \hline 
SDMEs      & $K^+\Lambda(1520)$ & $[0.1:2.0]$    & $[4.0:4.2]$                & 8             & 2022 & \cite{GlueX:2021pcl} \\ \hline\hline
\end{tabular}
\label{halld-gluex-list}
\end{center}
\end{table}
%%%%%%%%%%%%%%%%%%%%%%%%%%%%%%%%%%%%%%%%%%%%%%%%%%%%%%%%%%%%%%%%%%%%%%%%%%%%%%%%%%%%%%%%%%%%%%%%%%%%%%%%%%%%%%%%%%%%%%%%%%%%%%%%%%%%%%%%%%%%%%%%%%%%%%%%%%%%%%%%

\begin{figure}[ht]
\centering
\includegraphics[width=0.4\columnwidth]{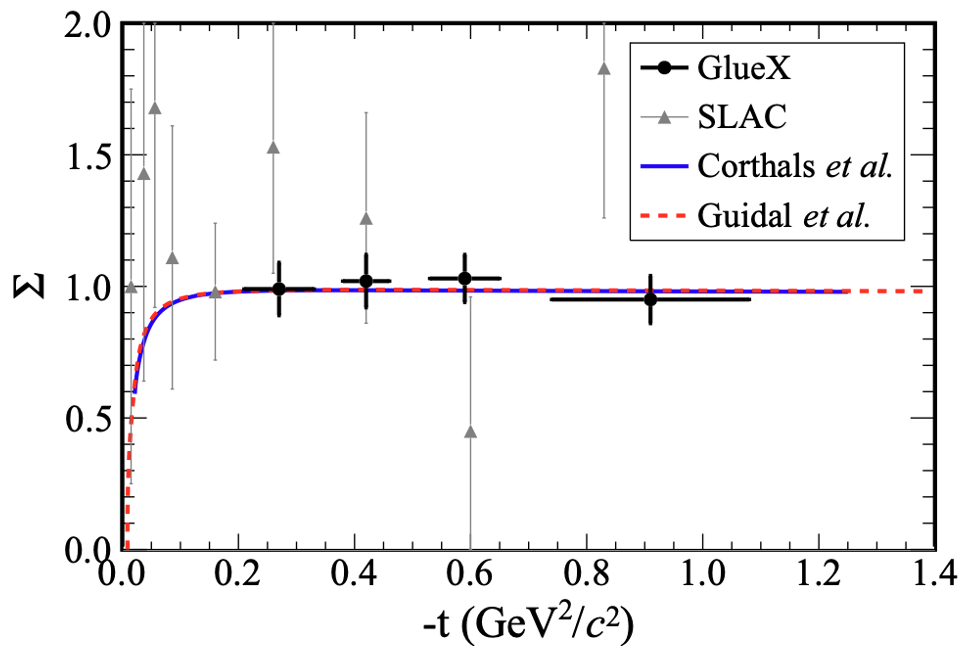} 
\caption{The linearly polarized beam spin asymmetry $\Sigma$ for exclusive $\gamma p \to K^+\Sigma^0$ photoproduction from GlueX as a function of four-momentum transfer squared $-t$ (solid circles) compared to previous SLAC results $E_\gamma^{\rm lab} = 16$~GeV (triangles)~\cite{Quinn:1979zp}. The curves represent predictions from the hybrid Regge-plus-Resonance model (solid blue)~\cite{Corthals:2005ce} and a pure Regge model (dashed red) at $E_\gamma^{\rm lab} = 8.5$~GeV~\cite{Guidal:1999qi}. Figure from Ref.~\cite{GlueX:2020qat}.}
\label{gluex-sigma-bsa}
\end{figure}

The linearly polarized beam spin asymmetry $\Sigma$ for ground state $\Sigma^0$ production in the exclusive process $\gamma p \to K^+ \Sigma^0$ was measured detecting the $K^+$ and the $\Sigma^0$ through its decay branch $\Sigma^0 \to \Lambda \gamma \to p \pi^- \gamma$. The beam asymmetry is formed from the cross sections with the photon beam polarized perpendicular and parallel to the reaction plane and was determined as a function of the four-momentum transfer squared $-t$ in the range 0.1 -- 1.4~GeV$^2$ for the tagged photon energy range from 8.2 -- 8.8~GeV. The measured beam asymmetry is close to unity over this range up to $\sim$1~GeV$^2$ (see Fig.~\ref{gluex-sigma-bsa})~\cite{GlueX:2020qat}. This result agrees with models that describe the reaction via the natural-parity exchange of the $K^*(892)$ Regge trajectory. 

%%%%%%%%%%%%%%%%%%%%%%%%%%%%%%%%%%%%%%%%%%%%%%%%%%%%%%%%%%%%%%%%%%%%%%%%%%%%%%%%%%%%%%%%%%%%%%%%%%%%%%%%%%%%%%%%%%%%%%%%%%%%%%%%%%%%%%%%%%%%%%%%%%%%%%%%%%%%%%%%%%%%%%%%%
%%%%%%%%%%%%%%%%%%%%%%%%%%%%%%%%%%%%%%%%%%%%%%%%%%%%%%%%%%%%%%%%%%%%%%%%%%%%%%%%%%%%%%%%%%%%%%%%%%%%%%%%%%%%%%%%%%%%%%%%%%%%%%%%%%%%%%%%%%%%%%%%%%%%%%%%%%%%%%%%%%%%%%%%%

Due to a unique low-energy configuration of the accelerator planned at JLab for 2026 configured for only 700~MeV/pass, the Hall~D maximum energy will be limited to 4~GeV. An opportunistic experiment will take data on a liquid-deuterium target using an elliptically polarized photon beam, i.e. with both linearly and circularly polarized components, to measure the $\gamma d \to K^0_S \Lambda$ exclusive channel~\cite{dalton:2025prop}. The $K^0_S$ will be detected from its $\pi^+\pi^-$ decay branch and the $\Lambda$ from its $p\pi^-$ decay branch. The experiment will extract differential cross sections, as well as a slate of polarization observables, $P$, $T$, $C_{x'}$, $C_{z'}$, $O_{x'}$, and $O_{z'}$, over the full range of $\cos \theta_{K^0_S}^{\rm c.m.}$. The primary focus of the study is to further explore evidence for a narrow $N(1685)1/2^+$ state that has been hinted at through multiple previous experiments \cite{GRAAL:2006gzl,CBELSA:2008epm,A2:2013tbo,A2:2016bij} (see Section~\ref{sec:penta-narrow} for further discussion). However, the experiment will span a broad range of $W$ up to $\approx$ 3~GeV with high statistics and narrow binning in energy and angles, so it will ultimately bring a wealth of data for this important channel. It should also be mentioned that this same experiment will collect a comparable dataset on a liquid-hydrogen target and will complete an extraction of the weak decay asymmetry parameter $\alpha$ with an expected absolute uncertainty of $\approx$ 0.021 (or $< 3\%)$ (see Refs.~\cite{BESIII:2018cnd,Ireland:2019uja} for details on recent extractions).

The GlueX studies of the $\Lambda(1405)$ and $\Lambda(1520)$ are focused on measurements of the $t$-dependence of the differential cross sections~\cite{Wickramaarachchi:2022mhi} in the region of the coherent photon energy peak. The $\Lambda(1405)$ is detected through its decay branch $\Lambda(1405) \to \Sigma^0 \pi^0$ with $\Sigma^0 \to \Lambda \gamma \to p \pi^- \gamma$ and $\pi^0 \to \gamma \gamma$. The $\Sigma^0 \pi^0$ decay branch of the $\Lambda(1405)$ is preferred for study of this state since it has isospin $I=0$ and is therefore free of contamination from the isospin-1 $\Sigma(1385)$ (see Eq.~(\ref{dsig-l1405})), which significantly overlaps the $\Lambda(1405)$ in the $K^+$ missing mass distribution. This analysis, which is still preliminary and ongoing, also allows for studies of the $t$-dependence of the $\Lambda(1405)$ lineshape to extend the CLAS results of Refs.~\cite{CLAS:2013rjt,Schumacher:2013vma}. These studies aim to better understand the nature of the $\Lambda(1405)$ as a simple quark model resonance or whether it is better classified as an $N\bar{K}$ bound state or $\Sigma \pi$ continuum resonance.

\begin{figure}[ht]
\centering
\includegraphics[width=0.55\columnwidth]{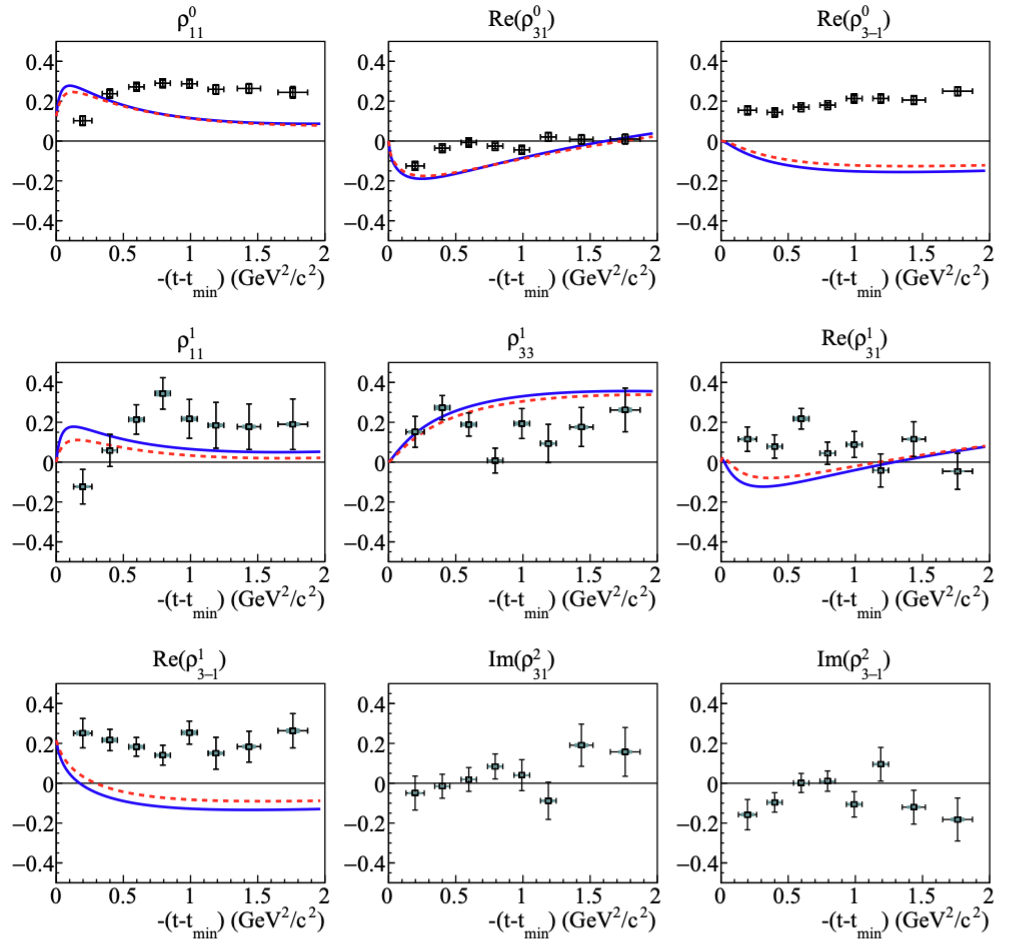} 
\caption{Spin density matrix elements $\rho$ ($\rho^0$ unpolarized, $\rho^{1,2}$ polarized) as a function of four-momentum transfer squared $t$ measured at GlueX for exclusive photoproduction of $\Lambda(1520)$ \cite{GlueX:2021pcl} compared to predictions from a Regge approach \cite{Yu:2017kng} constrained by data from CLAS~\cite{CLAS:2013rxx} and LEPS~\cite{Muramatsu:2009zp,LEPS:2009isz} (blue solid) and from LAMP2 (the Large Aperture Magnetic Spectrometer at Daresbury~\cite{Barber:1978hq})~\cite{Barber:1980zv} and SLAC~\cite{Boyarski:1970yc} (red dashed). The SDMEs are denoted as $\rho^\alpha_{2 \lambda_{\Lambda}, 2 \lambda_{\Lambda'}}$, where $\lambda_\Lambda$, $\lambda_{\Lambda'}$ refer to the spin projections of the $\Lambda(1520)$ on the photon direction in the c.m. frame and $\alpha$ refers to the photon polarization ($\alpha$=0 unpolarized beam, $\alpha=1,2$ two directions of linear polarization). Figure from Ref.~\cite{GlueX:2021pcl}.}
\label{gluex_l1520_sdme}
\end{figure}

The $\Lambda(1520)$ is reconstructed through its charged $K^-p$ branch. Measurements for this state include extraction of precision cross sections and determination of the hyperon spin density matrix elements (SDMEs). The spin density matrix $\rho$ quantifies the spin polarization of the $\Lambda(1520)$ and parametrizes the angular distribution of its daughter particles in its decay frame to enable insights into the production mechanism. Assuming $t$-channel exchange as the dominant production mode, the naturality $\eta = P(-1)^J$ of the exchanged particle with spin-parity $J^P$ can be studied. Utilizing the linear polarization of the GlueX photon beam, the unpolarized and polarized SDMEs were extracted as a function of $t$ in the range of $-(t-t_{\rm min})$ from 0.1 to 2.0~GeV$^2$. The analysis showed that $\Lambda(1520)$ production in the region of the coherent peak is dominated  by natural parity exchanges, i.e. by vector or tensor mesons~\cite{GlueX:2021pcl}. Figure~\ref{gluex_l1520_sdme} shows the six polarized and three unpolarized SDMEs extracted from the GlueX analysis (of the full $4 \times 4$ spin density matrix for a spin-3/2 particle) compared to predictions from a Regge model constrained by existing data. The extraction of the cross sections is in progress with preliminary results included in Ref.~\cite{Pauli:2022ehd}.

%%%%%%%%%%%%%%%%%%%%%%%%%%%%%%%%%%%%%%%%%%%%%%%%%%%%%%%%%%%%%%%%%%%%%%%%%%%%%%%%%%%%%%%%%%%%%%%%%%%%%%%%%%%%%%%%%%%%%%%%%%%%%%%%%%%%%%%%%%%%%%%%%%%%%%%%%%%%%%%%%%%%%%%%%
%%%%%%%%%%%%%%%%%%%%%%%%%%%%%%%%%%%%%%%%%%%%%%%%%%%%%%%%%%%%%%%%%%%%%%%%%%%%%%%%%%%%%%%%%%%%%%%%%%%%%%%%%%%%%%%%%%%%%%%%%%%%%%%%%%%%%%%%%%%%%%%%%%%%%%%%%%%%%%%%%%%%%%%%%

The final GlueX measurements in the strangeness sector focus on $S = -2$ spectroscopy to search for low-lying $\Xi$ excited states. As detailed in Section~\ref{clas-ep-program}, theoretical expectations call for a much larger number of $\Xi^*$ states than have been seen to date experimentally. The production mechanism for $\Xi$ baryons should be different compared to their $S = -1$ $\Lambda$ and $\Sigma$ hyperon counterparts. Instead of being formed in direct $t$-channel exchange processes like the $\Lambda$ and $\Sigma^0$, which can only generate one unit of strangeness, $\Xi$ production must proceed instead through the creation of an excited state $S = -1$ hyperon, which then decays into $K \Xi$ to add another unit of strangeness. Therefore, reactions are expected to be observed through $\gamma p \to K^+ (\Lambda^*/\Sigma^*) \to K^+K^{+/0}\Xi^{-/0}$. The initial phase of data studies with GlueX in this regard have focused on ``bump'' hunts in the reconstructed mass spectra for specific low-lying states. The ultimate plan is to provide production cross sections, angular distributions, and cross section upper limits.

\begin{figure}[ht]
\centering
\includegraphics[width=0.55\columnwidth]{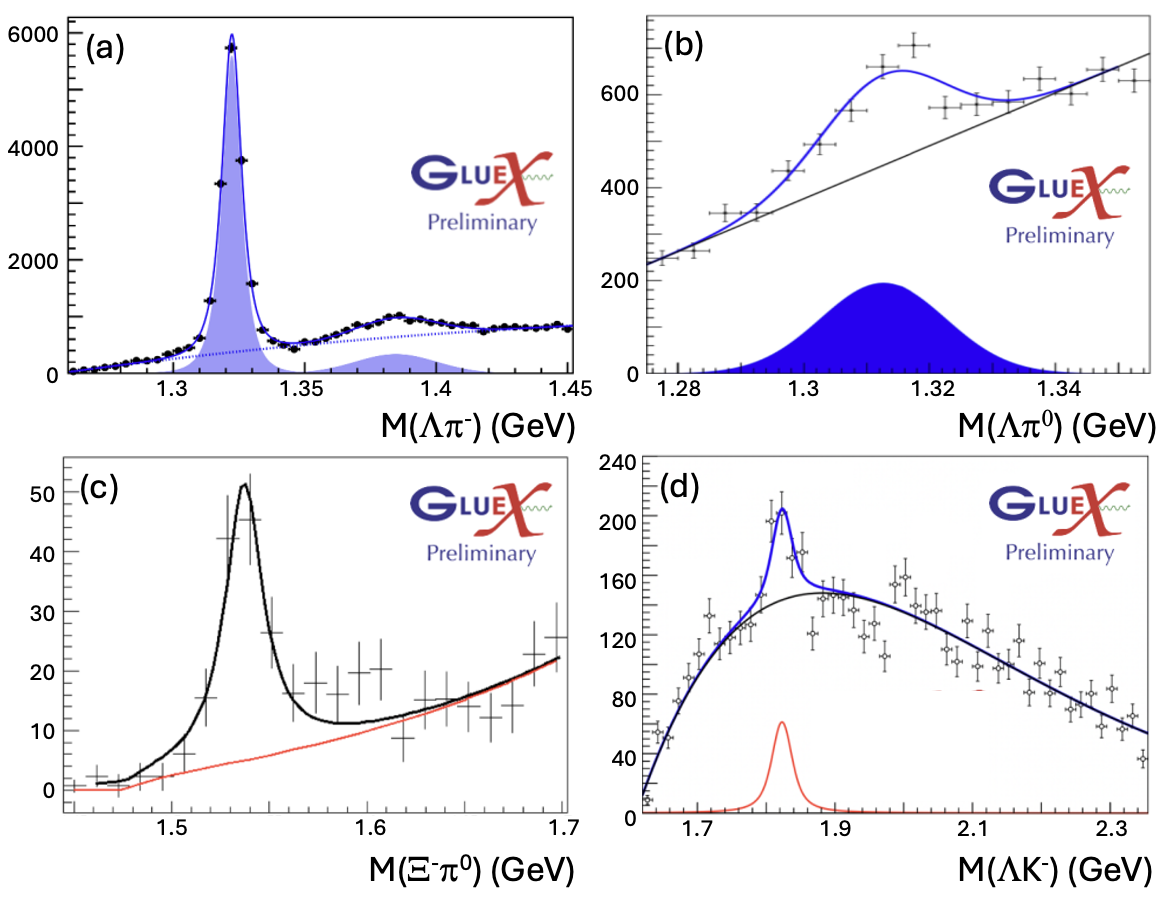} 
\caption{Spectroscopy studies at GlueX for various $\Xi$ states: (a) $\Xi^-(1321)$, (b) $\Xi^0(1315)$, (c) $\Xi^{*-}(1530)$, (d) $\Xi^{*-}(1820)$. See text for details on the different final state topologies selected in each of these analyses. The curves on the plots show the signal and background fit functions. Figures adapted from Ref.~\cite{Pauli:2022ehd} and used with kind permission of The European Physical Journal (EPJ).}
\label{gluex_cascades}
\end{figure}

The ground state cascade doublet states decay with a branching fraction of $\sim$100\% to $\Lambda \pi$. Therefore, when searching for the charged ground state cascade, the reaction of interest is $\gamma p \to K^+K^+\Xi^- \to K^+K^+\Lambda \pi^-$. GlueX studies of the $\Lambda \pi^-$ mass spectrum for this final state show a clean, narrow peak at the mass of the $\Xi^-(1321)$. When searching for the neutral ground state cascade $\Xi^0(1315)$, the reaction 
\begin{equation}
    \gamma p \to K^+K^0_S\Xi^0 \to K^+{\pi^+\pi^-}\pi^0\Lambda \to K^+{\pi^+\pi^-}\gamma\gamma \pi^-p,
\end{equation}
was studied. The $\Xi^0(1315)$ is seen for this channel in the reconstructed $\Lambda \pi^0$ mass spectrum. The excited $\Xi^{*-}(1530)$ state has a known decay to $\pi^0 \Xi^-$ and was searched for through the reaction 
\begin{equation}
    \gamma p \to K^+K^+\Sigma^{*-} \to K^+K^+\pi^0\Xi^- \to K^+K^+\gamma\gamma \pi^-\Lambda \to K^+K^+\gamma\gamma\pi^-\pi^-p.
\end{equation}
The highest mass $\Xi^*$ state seen in GlueX so far is the $\Xi^{*-}(1820)$, which was identified in the reaction
\begin{equation}
   \gamma p \to K^+K^+\Xi^{*-} \to K^+K^+K^-\Lambda \to K^+K^+K^-\pi^-p, 
\end{equation}
where a peak is seen in the $K^-\Lambda$ mass spectrum. Figure~\ref{gluex_cascades} shows a collection of selected GlueX mass plots for selected $S = -2$ spectra provided in Ref.~\cite{Pauli:2022ehd}. These painstaking analyses of high multiplicity, mixed charge, low cross section processes are quite an impressive demonstration of the capabilities of the GlueX detector, collected datasets, and analysis algorithms.

%%%%%%%%%%%%%%%%%%%%%%%%%%%%%%%%%%%%%%%%%%%%%%%%%%%%%%%%%%%%%%%%%%%%%%%%%%%%%%%%%%%%%%%%%%%%%%%%%%%%%%%%%%%%%%%%%%%%%%%%%%%%%%%%%%%%%%%%%%%%%%%%%%%%%%%%%%%%%%%%%%%%%%%%%
%%%%%%%%%%%%%%%%%%%%%%%%%%%%%%%%%%%%%%%%%%%%%%%%%%%%%%%%%%%%%%%%%%%%%%%%%%%%%%%%%%%%%%%%%%%%%%%%%%%%%%%%%%%%%%%%%%%%%%%%%%%%%%%%%%%%%%%%%%%%%%%%%%%%%%%%%%%%%%%%%%%%%%%%%

\subsubsection{Hall B CLAS12 Electroproduction}
\label{clas12-program}

The CLAS spectrometer was decommissioned in 2012 and the large acceptance CLAS12 spectrometer was installed in its place in Hall~B~\cite{Burkert:2020akg}. This new spectrometer was optimized for electron beam operations up to 11~GeV, the maximum beam energy deliverable to Hall~B as part of the JLab 12-GeV upgrade project. The physics program with CLAS12 focuses on a diverse array of topics, including femtographic imaging of quark distributions in the nucleon, investigations of the spectrum and structure of excited baryons, and studies of nucleon correlations within nuclei~\cite{Burkert:2018nvj}. The approved experiments include studies with longitudinally polarized electron beams on both unpolarized and (longitudinally and transversely) polarized hydrogen and deuterium targets, as well as nuclear targets, at beam-target luminosities of $\sim 1 \times 10^{35}$~cm$^{-2}$s$^{-1}$. The CLAS12 spectrometer provides for a broad kinematic coverage in invariant mass $W$ up to 4~GeV, four-momentum transfer squared $Q^2$ from 0.05 to 12~GeV$^2$, and nearly complete angular coverage of the final state reaction phase space through studies of inclusive, semi-inclusive, and exclusive reaction processes.

%%%%%%%%%%%%%%%%%%%%%%%%%%%%%%%%%%%%%%%%%%%%%%%%%%%%%%%%%%%%%%%%%%%%%%%%%%%%%%%%%%%%%%%%%%%%%%%%%%%%%%%%%%%%%%%%%%%%%%%%%%%%%%%%%%%%%%%%%%%%%%%%%%%%%%%%%%%%%%%%%%%%%%%%%
\begin{figure}[ht]
\centering
\includegraphics[width=0.6\columnwidth]{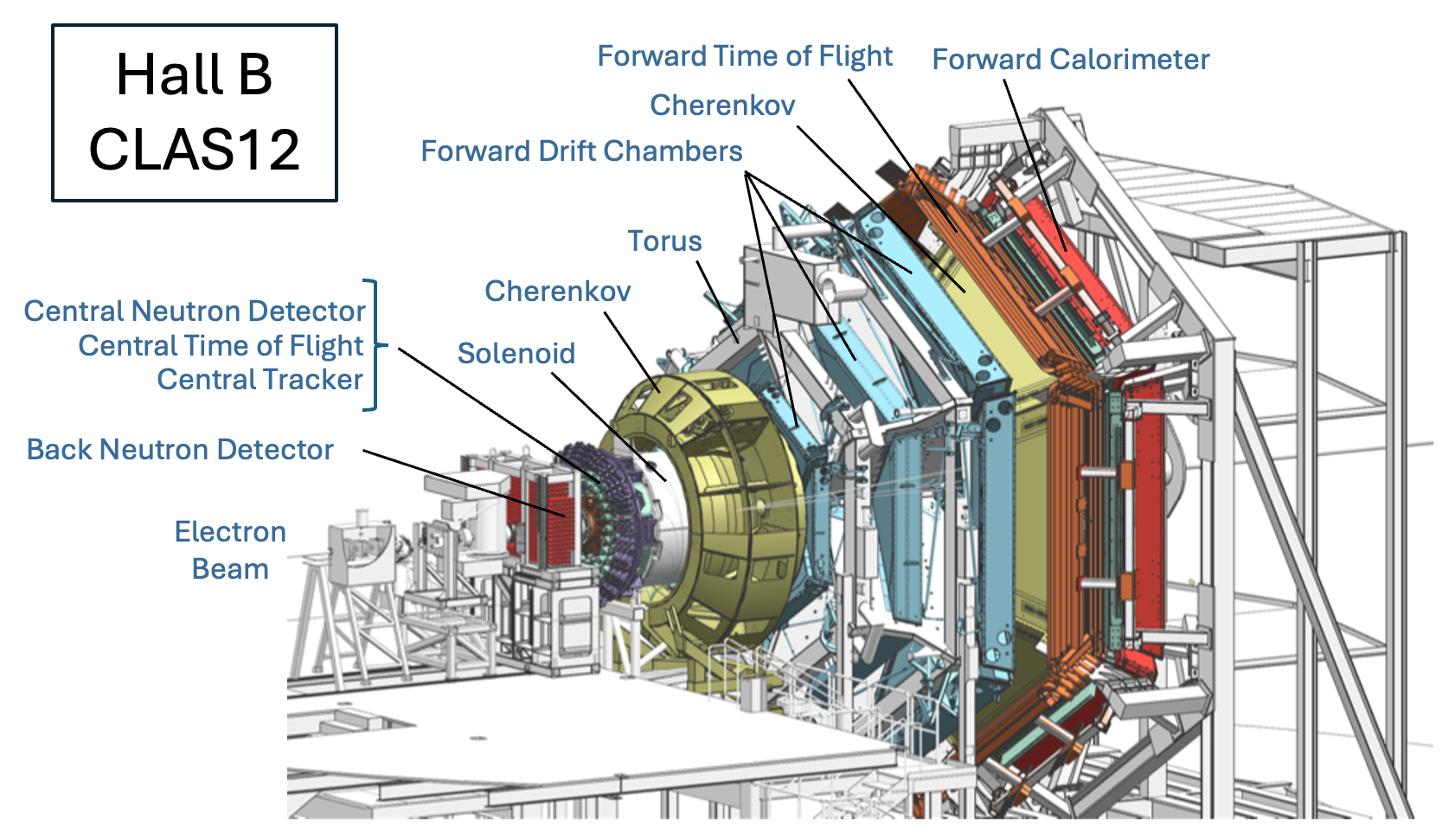} 
\caption{Model representation of the CLAS12 spectrometer in Hall~B. The detector subsystems are labeled and outlined in the text. The overall detector length along the beamline is $\sim$20~m. Figure adapted from Ref.~\cite{CLAS:2025vkn}.}
\label{clas12-model}
\end{figure}
%%%%%%%%%%%%%%%%%%%%%%%%%%%%%%%%%%%%%%%%%%%%%%%%%%%%%%%%%%%%%%%%%%%%%%%%%%%%%%%%%%%%%%%%%%%%%%%%%%%%%%%%%%%%%%%%%%%%%%%%%%%%%%%%%%%%%%%%%%%%%%%%%%%%%%%%%%%%%%%%%%%%%%%%

\begin{figure}[ht]
\centering
\includegraphics[width=0.4\columnwidth]{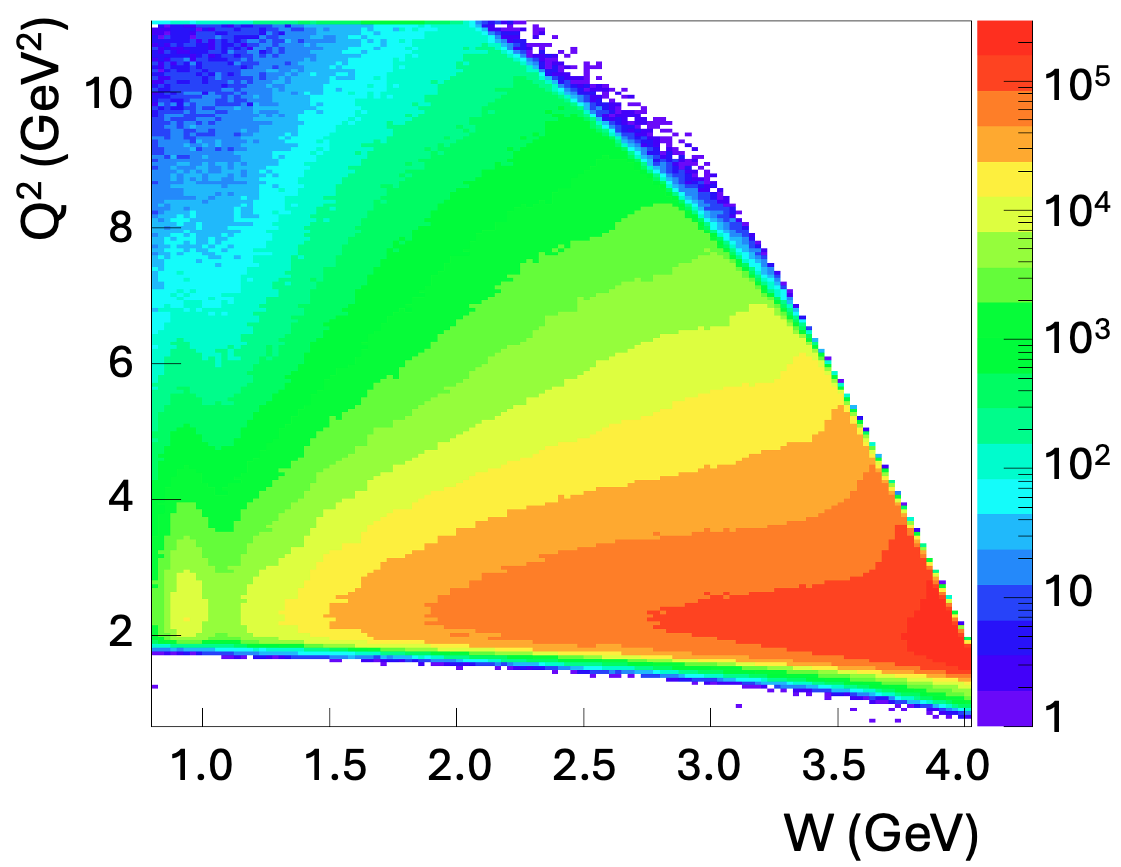} 
\caption{Kinematic coverage of CLAS12 from data with an 11~GeV electron on a liquid-hydrogen target in terms of $Q^2$ vs. $W$ from a 2018 data run with the torus polarity set to bend negatively charged particles toward the beamline. This plot was based on reconstructed electrons detected in the forward electromagnetic calorimeter. Figure from Ref.~\cite{CLAS:2025zup}.}
\label{clas12-kincov}
\end{figure}

CLAS12 is based on two superconducting magnets, a 5~T solenoid in the central region around the target and a toroid at forward angles. The CLAS12 torus magnet, like the original CLAS torus magnet, has a six-fold symmetry that divides the forward azimuthal acceptance in the polar angle range from 5$^\circ$ to 35$^\circ$ into six 60$^\circ$-wide sectors. The torus produces a field primarily in the azimuthal direction. Its strength in terms of $\int B d\ell$ varies from 2.78~Tm at 5$^\circ$ to 0.54~Tm at 40$^\circ$. A set of three multi-layer drift chambers in each sector (before the field, within the field, and after the field) and a forward vertex tracker are used for charged particle tracking to measure momenta and to define the event vertex. Downstream of the torus each sector is instrumented with a Cherenkov counter for hadron identification, scintillation counters for charged particle timing, and an electromagnetic calorimeter for electron and neutral particle identification. Just upstream of the first set of drift chambers is a large-volume, light-weight, high-threshold gas Cherenkov counter for electron/pion discrimination and a forward tagging system to detect electrons and photons at polar angles below $5^\circ$. The solenoid, which spans the angular range from 35$^\circ$ to 125$^\circ$, serves to focus the low-energy M{\o}ller background down the beampipe and to provide a magnetic field for charged particle momentum analysis in the central region. The detectors mounted within the solenoid include a thicker scintillation counter for neutron identification, a barrel of thinner scintillation counters for charged particle timing measurements, and a set of silicon strip/micromegas tracking detectors around the target. Upstream of the solenoid is a back angle scintillation counter wall for neutron detection at angles above 155$^\circ$. Figure~\ref{clas12-model} shows a model representation of CLAS12 to highlight its overall layout and scale. CLAS12 was installed and instrumented in Hall~B in the period from 2012 to 2017, and began operations for physics in 2018. In the forward region it provides $\Delta p/p \sim 0.5\%$ and $\Delta \theta, \Delta \phi \sim 1$~mrad. In the central region it provides $\Delta p/p \sim 5\%$ and $\Delta \theta, \Delta \phi \sim 1$~mrad. Figure~\ref{clas12-kincov} shows the broad $Q^2$ vs. $W$ coverage of this detector system from an experiment with incident 11~GeV electrons on a liquid-hydrogen target. This data was collected for a polarity of the CLAS12 torus set to bend negatively charged particles in toward the electron beamline. Data collected reversing the magnet polarity allows coverage to lower $Q^2$. 

%%%%%%%%%%%%%%%%%%%%%%%%%%%%%%%%%%%%%%%%%%%%%%%%%%%%%%%%%%%%%%%%%%%%%%%%%%%%%%%%%%%%%%%%%%%%%%%%%%%%%%%%%%%%%%%%%%%%%%%%%%%%%%%%%%%%%%%%%%%%%%%%%%%%%%%%%%%%%%%%%%%%%%%%%
%%%%%%%%%%%%%%%%%%%%%%%%%%%%%%%%%%%%%%%%%%%%%%%%%%%%%%%%%%%%%%%%%%%%%%%%%%%%%%%%%%%%%%%%%%%%%%%%%%%%%%%%%%%%%%%%%%%%%%%%%%%%%%%%%%%%%%%%%%%%%%%%%%%%%%%%%%%%%%%%%%%%%%%%

The strangeness physics program in Hall B with CLAS12 is focused on an extension of the $N^*$ program started with the CLAS 6-GeV era program (see Section~\ref{clas-ep-program}). It is designed to provide measurements on exclusive electroproduction in the $KY$, $K^*Y$, and $KY^*$ channels in the $Q^2$ range beyond the CLAS 6-GeV limit of 5~GeV$^2$ up to 10-12~GeV$^2$ that are possible with the 11-GeV electron beam in Hall~B. The approved experiments employ beam energies of 6.6 GeV, 8.8 GeV, and 11 GeV. The measurements include precision extractions of differential cross sections, separated structure functions $\sigma_{\rm U}$, $\sigma_{\rm LT}$, $\sigma_{\rm TT}$, and $\sigma_{\rm LT'}$, and beam and beam-recoil hyperon polarization measurements on an unpolarized liquid-hydrogen target. The kinematic coverage of CLAS12 in these experiments overlaps that of the existing CLAS results, however, given that the nominal beam-target luminosity of CLAS12 is an order of magnitude larger than that for the CLAS program, much improved statistical precision is possible with finer bin sizes in the relevant kinematic variables $Q^2$, $W$, $\cos \theta_K^{\mathrm{c.m.}}$, and $\Phi$ in the overlap region. As well, measurements at different beam energies and, hence, different values of the virtual photon polarization parameter $\epsilon$, will enable Rosenbluth separations of $\sigma_{\rm U}$ into $\sigma_{\rm T}$ and $\sigma_{\rm L}$ from overlapping bins of $Q^2$, $W$, and $\cos \theta_K^{\mathrm{c.m.}}$ to complement those planned in Hall~C (see Section~\ref{hallc12-program}). 

The study of the spectrum and structure of excited nucleon states (known as the $N^*$ program) represents one of the foundations for the measurement program in Hall~B with the CLAS spectrometer. To date, measurements with CLAS in the 6-GeV era of experiments have provided a significant amount of precision data (cross sections and polarization observables) for a number of different exclusive final states for $Q^2$ from 0 to 4.5~GeV$^2$, $W$ up to 2.5~GeV, and the full center-of-mass angular range of the final state decay products~\cite{Burkert:2025coj,Proceedings:2020fyd} (see Section~\ref{clas-ep-program}). The continuing $N^*$ program with CLAS12 will extend the studies from CLAS up to $Q^2$ of 12~GeV$^2$, the highest photon virtualities ever probed in exclusive reactions in the resonance region. A central focus in this program is the study of the $Q^2$ evolution of the electroexcitation amplitudes (or $\gamma_vpN^*$ electrocouplings) for the $s$-channel resonances that decay through the different exclusive reaction channels~\cite{Mokeev:2022xfo} (see Section~\ref{nstar-structure}). The momentum dependence of the underlying degrees of freedom shapes the structure of the $N^*$ states and is reflected in the $Q^2$ evolution of these electrocouplings. These quantities, directly related to the helicity amplitudes or to the transition form factors for the process $\gamma_v p \to N^*,\Delta^*$ represent the only source of information on many facets of the non-perturbative strong interaction in the generation of excited nucleon states of different quantum numbers and their emergence from QCD.

From the $\pi N$, $\eta N$, and $\pi \pi N$ data, the electrocouplings of most $N^*$ states up to $\sim$1.8~GeV have been extracted for the first time from data collected with CLAS~\cite{Carman:2023zke}. A powerful cross-check of these quantities and their assigned systematic uncertainties comes from comparisons of these extractions from independent analyses of different exclusive final states. With the development and refinement of reaction models that accurately describe the extensive CLAS $K^+\Lambda$ and $K^+\Sigma^0$ electroproduction data over its full available kinematic phase space, the CLAS and CLAS12 data from the strangeness channels is expected to provide an important complement to study the electrocouplings determined for higher-lying $N^*$ resonances with masses above 1.6~GeV and will serve as an important cross-check of the corresponding analyses from the dominant $\pi \pi N$ channel. These studies, in concert with theoretical developments, will allow for insight into the strong interaction dynamics of dressed quarks and their confinement in baryons of different quantum numbers over a broad $Q^2$ range. The data will allow for further insights into the Standard Model relating to the nature of hadron mass and quark-gluon confinement~\cite{Achenbach:2025kfx}.

\begin{figure}[ht]
\centering
\includegraphics[width=0.7\columnwidth]{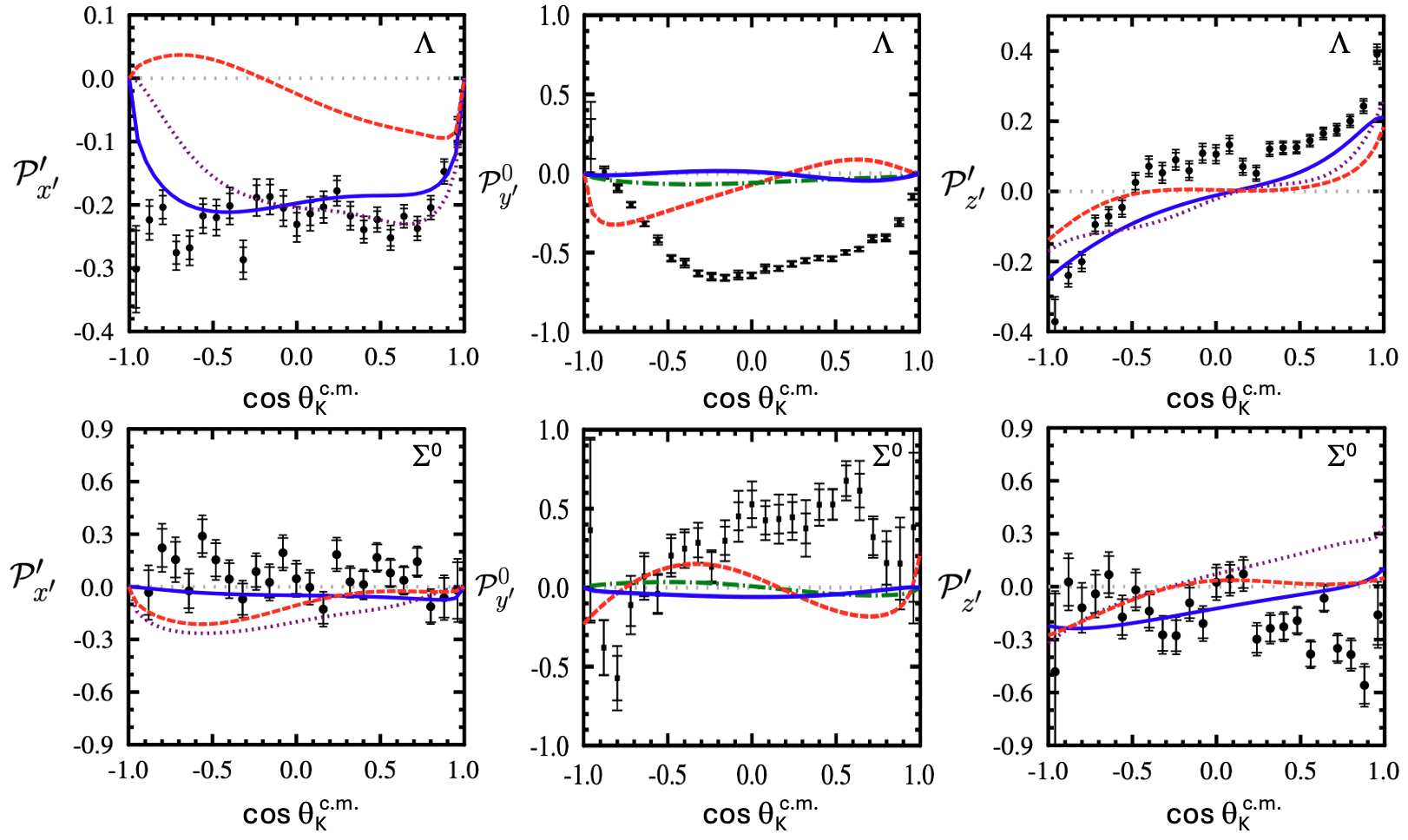} 
\caption{Representative $K^+Y$ polarization results from CLAS12 at 6.535~GeV for the recoil polarization ${\cal P}^0$ (middle column)~\cite{CLAS:2025vkn} and for the beam-recoil transferred polarization ${\cal P}'$ (outer columns)~\cite{CLAS:2022yzd} with respect to the $(x',y',z')$ coordinate system defined in Fig.~\ref{coor4}. The top row is for the $\Lambda$ and the bottom row is for the $\Sigma^0$. The data shown here are averaged over the $Q^2$ and $W$ ranges shown in Table~\ref{hallb-clas12}. The curves are calculations from the isobar models of Saclay-Lyon (SL) \cite{David:1995pi,Mizutani:1997sd} (green dot-dash), Kaon-MAID (KM) \cite{Mart:1999ed,Mart:2000jv} (solid blue), and the Czech group (BS) \cite{Skoupil:2018vdh,Petrellis:2024ybj} (red dashed), and from the Ghent Regge-plus-Resonance (RPR) model \cite{Corthals:2007kc} (dotted purple). Figure adapted from Refs.~\cite{CLAS:2022yzd,CLAS:2025vkn}.}
\label{clas12-ky-pol}
\end{figure}

To date only $\sim$50\% of the approved $N^*$ experimental program with CLAS12 has been completed in terms of data collection. Additional beam time will be scheduled in the years ahead, however, the analysis of the data already collected is in progress. Two analyses from a first short beam run at 6.5 and 7.5~GeV in 2018 related to the recoil and beam-recoil transferred polarizations for $\Lambda$ and $\Sigma^0$ hyperons from exclusive $KY$ electroproduction have been completed~\cite{CLAS:2022yzd,CLAS:2025vkn}, greatly expanding the available data. These measurements are detailed in Table~\ref{hallb-clas12} and representative plots are shown in Fig.~\ref{clas12-ky-pol}. This limited duration run with CLAS12 already accounts for five times the size of any of the individual $KY$ datasets acquired using CLAS (in terms of collected charge). However, it represents only $< 10\%$ of the full approved beam time. The ultimate goal of the CLAS12 $KY$ electroproduction program is to supply a dataset up to $Q^2 \approx 2 - 3$~GeV$^2$ with statistics and binning comparable to the available CLAS photoproduction datasets detailed in Section~\ref{clas-gp-program}, while extending the $Q^2$ coverage up to 10-12 GeV$^2$ for the purpose of nucleon resonance spectrum and structure studies as detailed in Section~\ref{nstar-structure}. Table~\ref{clas12-data} provides an overview of the datasets collected with CLAS12 so far related to the approved strangeness physics program. Analyses of cross sections and multi-dimensional polarization extractions for $K^+\Lambda$ and $K^+\Sigma^0$ from the remaining datasets are in progress.

%%%%%%%%%%%%%%%%%%%%%%%%%%%%%%%%%%%%%%%%%%%%%%%%%%%%%%%%%%%%%%%%%%%%%%%%%%%%%%%%%%%%%%%%%%%%%%%%%%%%%%%%%%%%%%%%%%%%%%%%%%%%%%%%%%%%%%%%%%%%%%%%%%%%%%%%%%%%%%%
\begin{table}[tbh]
\begin{center}
\caption{Summary of $ep\to e'K^+\Lambda$ and $ep\to e'K^+\Sigma^0$ electroproduction measurements in Hall B with CLAS12 from the 12-GeV era experiments at JLab. The column labeled $N_{\rm bin}$ indicates the number of kinematic bins included in the analysis.}
\begin{tabular}{ccccccc} \hline\hline
Observables                         & $Q^2$ (GeV$^2$)              & $W$ (GeV)                    & $\cos \theta_K^{\mathrm{c.m.}}$        & $N_{\rm bin}$                                      & Year                  & Ref. \\ \hline
${\cal P}^0_{y}$, ${\cal P}^0_{y'}$ & $[0.3:4.5]$                  & $[1.6:2.4]$                  & $[-1.0:1.0]$                  & 312 $\Lambda$, 312 $\Sigma^0$                  & 2025                  & \cite{CLAS:2025vkn} \\ \hline 
${\cal P}'_x$, ${\cal P}'_z$,       & \multirow{2}{*}{$[0.3:4.5]$} & \multirow{2}{*}{$[1.6:2.4]$} & \multirow{2}{*}{$[-1.0:1.0]$} & \multirow{2}{*}{312 $\Lambda$, 312 $\Sigma^0$} & \multirow{2}{*}{2022} & \multirow{2}{*}{\cite{CLAS:2022yzd}} \\ 
${\cal P}'_{x'}$, ${\cal P}'_{z'}$  &                              &                              &                               &                                                &                       &  \\ \hline\hline 
\end{tabular}
\label{hallb-clas12}
\end{center}
\end{table}
%%%%%%%%%%%%%%%%%%%%%%%%%%%%%%%%%%%%%%%%%%%%%%%%%%%%%%%%%%%%%%%%%%%%%%%%%%%%%%%%%%%%%%%%%%%%%%%%%%%%%%%%%%%%%%%%%%%%%%%%%%%%%%%%%%%%%%%%%%%%%%%%%%%%%%%%%%%%%%%%

%%%%%%%%%%%%%%%%%%%%%%%%%%%%%%%%%%%%%%%%%%%%%%%%%%%%%%%%%%%%%%%%%%%%%%%%%%%%%%%%%%%%%%%%%%%%%%%%%%%%%%%%%%%%%%%%%%%%%%%%%%%%%%%%%%%%%%%%%%%%%%%%%%%%%%%%%%%%%%%%%%%%%%%%%
%%%%%%%%%%%%%%%%%%%%%%%%%%%%%%%%%%%%%%%%%%%%%%%%%%%%%%%%%%%%%%%%%%%%%%%%%%%%%%%%%%%%%%%%%%%%%%%%%%%%%%%%%%%%%%%%%%%%%%%%%%%%%%%%%%%%%%%%%%%%%%%%%%%%%%%%%%%%%%%%%%%%%%%%

%%%%%%%%%%%%%%%%%%%%%%%%%%%%%%%%%%%%%%%%%%%%%%%%%%%%%%%%%%%%%%%%%%%%%%%%%%%%%%%%%%%%%%%%%%%%%%%%%%%%%%%%%%%%%%%%%%%%%%%%%%%%%%%%%%%%%%%%%%%%%%%%%%%%%%%%%%%%%%%
\begin{table}[htb]
\setlength{\tabcolsep}{6pt} % Default value: 6pt
\renewcommand{\arraystretch}{0.8} % Default value: 1
\begin{center}
\caption{Summary of the CLAS12 datasets on a liquid-hydrogen target taken to date related to the approved strangeness physics program. These data amount to roughly 50\% of the approved experimental beam time.}
\begin{tabular}{ccc} \hline\hline
Dataset           & Beam Energy (GeV) & Collected Charge (mC) \\ \hline
Feb. - May 2018   & 10.6              & 126 \\ \hline
Sep. - Nov. 2018  & 10.6              &  99 \\ \hline
Dec. 2018         & 6.5 / 7.5         &  18 / 15 \\ \hline
Mar. - Apr. 2019  & 10.2              &  58 \\ \hline
Jan. - Mar. 2024  & 6.4 / 8.5         &  91 / 82 \\ \hline\hline
\end{tabular}
\label{clas12-data}
\end{center}
\end{table}
%%%%%%%%%%%%%%%%%%%%%%%%%%%%%%%%%%%%%%%%%%%%%%%%%%%%%%%%%%%%%%%%%%%%%%%%%%%%%%%%%%%%%%%%%%%%%%%%%%%%%%%%%%%%%%%%%%%%%%%%%%%%%%%%%%%%%%%%%%%%%%%%%%%%%%%%%%%%%%%%

Work is presently ongoing to upgrade the CLAS12 detector to enable roughly two times higher luminosity operations using new micro-pattern gas detectors as tracking layers upstream of the first layer of drift chambers in the forward direction~\cite{Gnanvo:2024jag}. The hardware improvements are proceeding in parallel with new tracking algorithmic improvements implementing artificial intelligence/machine learning methods for tracking finding and denoising \cite{Chabanat:2005zz} and hybrid algorithms that extend the Kalman filter to improve track determination~\cite{Fleischmann:2006uq,Gavalian:2024knw}. These algorithms have been implemented for CLAS12 forward track reconstruction and lead to a significant reduction in charged track inefficiency with increasing beam-target luminosity, as well as significant improvements in momentum resolution by removing noise hits that pollute the track fitting. Development of kindred algorithms for the central track reconstruction is in progress. In addition, due to upgrades of the drift chamber high voltage system, operations with increased electric fields have allowed for significant improvements in the track resolution that will benefit future datasets. The improved momentum resolution of charged tracks for CLAS12 is important to enable cleaner separation of the $\Lambda$ and $\Sigma^0$ hyperons in the $MM(e'K^+)$ mass distributions for the higher energy datasets.

The final aspect of the CLAS12 strangeness physics program is focused on studies of $\Xi$ and $\Omega$ baryons. A recent review provides a much more complete worldwide history--past, present, and future--of studies of $S = -2$ and $S = -3$ hyperons~\cite{Crede:2024hur}. The CLAS12 spectrometer includes a forward tagging system about the electron beamline spanning polar angles from $\theta = 2^\circ \to 5^\circ$ that allows for quasi-real photoproduction studies at very low $Q^2$ ($10^{-2} \to 10^{-1}$~GeV$^2$)~\cite{Acker:2020brf} where production rates are maximal. The goal of the measurements is to study the production mechanisms of these baryons produced in exclusive reactions, extending the results provided from the 6-GeV era CLAS data (see Section~\ref{clas-gp-program}). The approved experimental program is expected to yield a total data sample of several million $\Xi$ ($S = -2$) and $\sim$4000 reconstructed $\Omega^-$ ($S = -3$) baryons, based on the predicted cross sections and simulations of the experimental conditions and detector apparatus~\cite{Afanasev:2012fh}. The new data will provide for a significant increase in the available statistics for photoproduction of these hyperons.

The quasi-real photoproduction $\Xi$ data sample will be used to search for evidence of higher-lying excited $\Xi$ states that could allow an opportunity to measure their quantum numbers, as well as the mass splittings of ground and excited state $\Xi$ doublets. The new data will also allow access to the recoil and beam-recoil polarization transfer observables of the ground state $\Xi^-$ in the reaction $\gamma p \to K^+K^+ \Xi^-$ to extend the initial measurements made during the CLAS 6-GeV era~\cite{CLAS:2018xbd}. Analysis of the collected data is currently in progress~\cite{Khanal:2022ynr,carvajal2024}. Using the same dataset, a search is underway for the photoproduction of the $\Omega^-$ baryon, utilizing the reaction $\gamma p \to K^+K^+K^0 \Omega^-$ to enable a search for $\Omega^-$ excited states and provide insight into the associated production mechanism. 

\subsection{Future Possibilities -- JLab at 22-GeV}
\label{jlab:22gev}

The CEBAF accelerator at JLab currently delivers the world's highest intensity and highest precision multi-GeV electron beams. The 12-GeV era (see Section~\ref{jlab:12gev}) is now well underway, with many important experimental results already published, and an exciting experimental program planned over the next decade~\cite{Arrington:2021alx}. However, the CEBAF community is already looking toward its future and the science that could be obtained through a future upgrade to a 22~GeV electron machine~\cite{Benesch:2023fug}. The potential to upgrade CEBAF to higher energies opens a rich and unique experimental nuclear physics program that has been detailed in a JLab 22-GeV White Paper~\cite{Accardi:2023chb}
and a subsequent update~\cite{Accardi:2026slw}. The future realization of such a facility will depend on support within the community, available funding, and resolving various technical issues in its design.

A 22 GeV CEBAF machine would provide a high-precision, high-luminosity measurement program to elucidate the properties of QCD in the valence quark regime (struck quark momentum fraction $x \ge 0.1$). In fact, such a machine would operate with several orders of magnitude higher luminosity than what is planned at the Electron-Ion Collider (EIC) at Brookhaven National Laboratory~\cite{AbdulKhalek:2021gbh}. CEBAF's current and envisioned capabilities enable exciting scientific opportunities  that complement the EIC operational reach, thus giving scientists the full suite of tools necessary to comprehensively understand how QCD builds hadronic matter over the full range of partonic distance scales. In the following, elements of a developing strangeness physics program with a 22-GeV JLab facility across the different experimental halls are highlighted.

\subsubsection{Kaon Form Factor Studies in Hall C}
\label{subsec:Studies-in-Hall-C}
The measurement of the charged kaon electromagnetic form factor has evolved through several generations of experiments in Hall~C, first from the 6-GeV era (see Section~\ref{halla-c-6gev}), and into the current 12-GeV era (see Section~\ref{jlab:12gev}). Considerations of the possibility to extend the program to higher beam energies are ongoing. The goal is to measure the $K^+$ form factor $F_K$ to increasingly higher $Q^2$ to allow for improved separation between the soft and hard contributions in order to provide for cleaner access to information governing hadron mass generation. The measurement would be part of a broader program to also measure $F_\pi$, the charged pion form factor. The $\pi^+$ form factor is the best hope of experimentally observing the QCD transition from soft to hard physics (i.e. bridging the regime from non-perturbative to perturbative dynamics) as it should occur at lower $Q^2$ than for the proton (as the virtual photon momentum sharing in the pion is only between two quarks instead of three). Study of the $K^+$ form factor is complementary to the pion as it allows for a probe of meson structure when an $s$ quark is substituted for a $d$ quark. 

These experiments to measure $e'\pi^+X$ and $e'K^+X$ final states rely on forward angle detection of the scattered electron and electroproduced meson. At small four-momentum transfer squared ($-t$), the meson pole term is expected to dominate the longitudinal cross section $\sigma_{\rm L}$, which is directly proportional to the charged meson form factor $F_{\pi,K}(Q^2,t)$. There are drawbacks to this approach to accessing the form factors that include the difficulty of isolating $\sigma_{\rm L}$ and the assumption that the meson pole term dominates $\sigma_{\rm L}$ that requires theory input.

\begin{figure*}[htbp]
\centering
\includegraphics[width=0.45\textwidth]{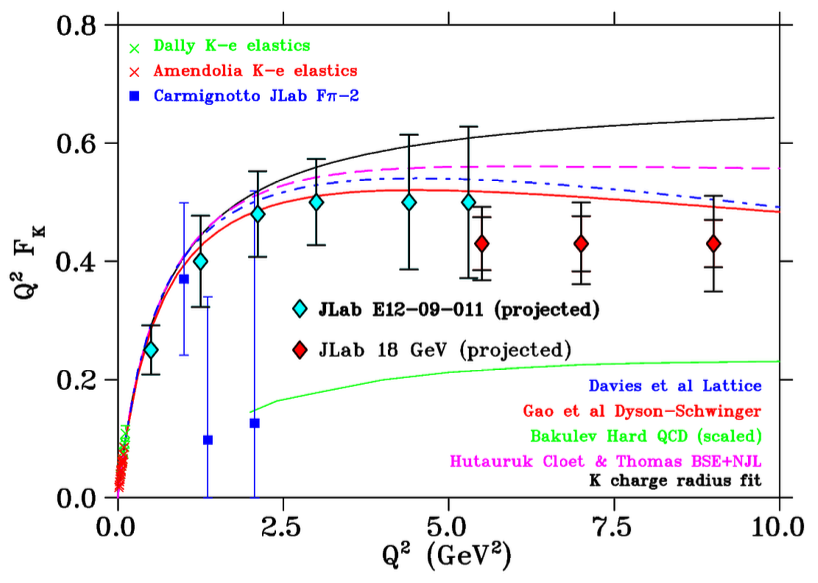}
\caption{Projections of the kinematic reach and uncertainties of the charged kaon form factor vs. $Q^2$ from the ongoing JLab Hall~C 12-GeV era measurement and expectations of what could be possible at an energy-upgraded JLab at 22 GeV. Previous measurements are included from Refs.~\cite{Dally:1980dj,Amendolia:1986ui,Carmignotto:2018uqj} and calculations from different approaches: lattice QCD~\cite{Davies:2018zav}, Dyson-Schwinger~\cite{Gao:2017mmp}, hard QCD~\cite{Bakulev:2009ib}, Bethe-Salpeter~\cite{Hutauruk:2016sug}, and extractions from the kaon charge radius~\cite{Krutov:2016luz}. Figure from Ref.~\cite{ghuber2024}.}
\label{kaonff-proj12} 
\end{figure*}

The experiment is being planned for staging in Hall C and is being considered in two phases. Phase 1 assumes the current HMS for the electron arm and the SHMS for the meson arm (see Section~\ref{hallc12-program}) but is limited to an 18~GeV electron beam (and $Q^2 < 12$~GeV$^2$) due to the momentum limitations of the spectrometers. Phase 2 assumes an upgrade of the HMS to detect higher momentum mesons with the scattered electron then detected in the SHMS to allow for an electron beam energy of 22~GeV (and $Q^2$ up to 15~GeV$^2$). The program relies on measurement of absolute cross sections at different beam energies to perform Rosenbluth $\sigma_{\rm T}$ and $\sigma_{\rm L}$ separations. The key to reduce the experimental uncertainties is to take data at a given $(Q^2,W,t)$ over the broadest possible beam energy range to allow for the largest possible spread in the virtual photon polarization parameter $\epsilon$. Planned measurements of $F_K$ and $F_\pi$ at the EIC~\cite{Aguilar:2019teb} have different systematics as they cannot separate $\sigma_{\rm T}$ and $\sigma_{\rm L}$ as all high energy measurements essentially take place in kinematics with $\epsilon=1$. Figure~\ref{kaonff-proj12} shows a summary plot of the existing measurements from older data deriving $F_K$ from $K-e$ elastic scattering, from JLab 6-GeV era experiments, from expectations of the $F_K$ measurement in the 12-GeV era experiment, and a projection of would could be possible at a 22-GeV energy-upgraded JLab~\cite{ghuber2024}. Also shown are predictions from various theoretical approaches.

%%%%%%%%%%%%%%%%%%%%%%%%%%%%%%%%%%%%%%%%%%%%%%%%%%%%%%%%%%%%%%%%%%%%%%%%%%%%%%%%%%%%%%%%%%%%%%%%%%%%%%%%%%%%%%%%%%%%%%%%%%%%%%%%%%%%%%%%%%%%%%%%%%%%%%%%%%%%%%%
%%%%%%%%%%%%%%%%%%%%%%%%%%%%%%%%%%%%%%%%%%%%%%%%%%%%%%%%%%%%%%%%%%%%%%%%%%%%%%%%%%%%%%%%%%%%%%%%%%%%%%%%%%%%%%%%%%%%%%%%%%%%%%%%%%%%%%%%%%%%%%%%%%%%%%%%%%%%%%%

\subsubsection{Structure of Excited Nucleons and Emergence of Hadron Mass in Hall B}
\label{sec:clas22}

The Standard Model of Particle Physics has one well-known mass-generating mechanism for hadrons, namely, the Higgs boson \cite{Higgs:2014aqa}. However, only 9~MeV of the 940~MeV mass of a nucleon is directly generated by Higgs boson couplings into QCD. The true mass-generating mechanism of visible matter, often referred to as emergent hadron mass (EHM) \cite{Roberts:2020udq}, is responsible for 94\% of $m_N$, with the remaining 5\% generated by constructive interference between EHM and the Higgs boson. This makes studies of the structure of the ground and excited states of the nucleon probed in experiments with electron beams a promising avenue to gain insight into the strong interaction dynamics that underlie the emergence of the dominant part of the visible mass in the Universe~\cite{Carman:2023zke}.

Studies of $N^*$ structure from the data on exclusive meson electroproduction in terms of their electroexcitation amplitudes over a broad range of $Q^2$ are critical in order to explore the evolution of the strong interaction in the transition from the strongly coupled to the perturbative QCD regimes~\cite{Achenbach:2025kfx}. These amplitudes provide the essential experimental input for the development of the QCD-connected theoretical approaches necessary for understanding the structure of the ground and excited nucleon states vs. partonic distance scale.

\begin{figure*}[htbp]
\centering
\includegraphics[width=0.7\textwidth]{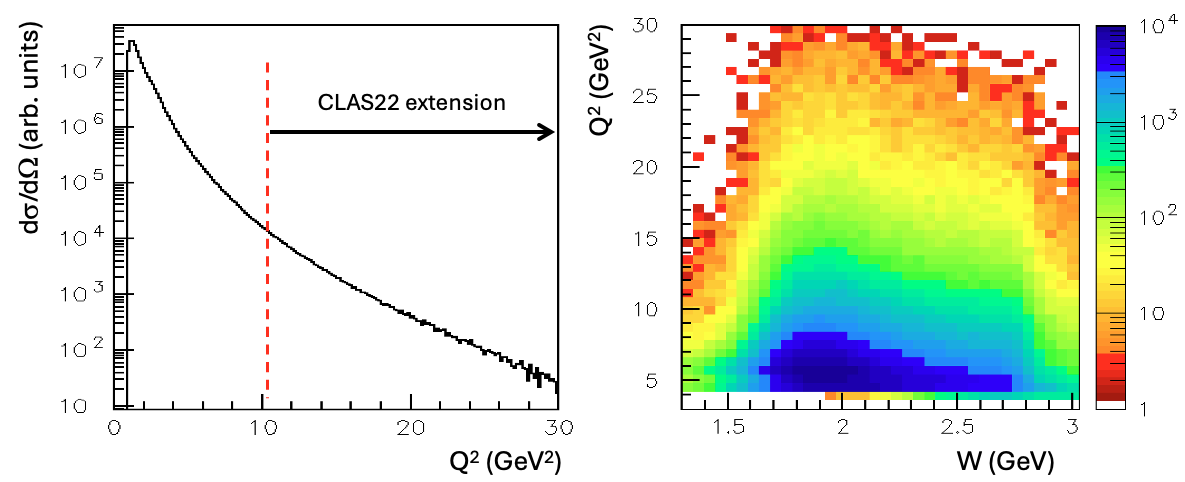}
\caption{(a) Evolution of the $K^+Y$ cross section vs. $Q^2$ (assuming a dipole form factor) to illustrate the kinematic range opened up by the possible energy upgrade of JLab to 22~GeV. This figure shows the rapid fall-off of the cross section at increasing $Q^2$ that must be compensated for with data-taking at higher beam-target luminosities and longer running times. (b) Preliminary simulation results for exclusive $KY$ electroproduction for the $Q^2$ vs. $W$ coverage of a version of CLAS22 at 22~GeV beam energy.}
\label{clas22-proj} 
\end{figure*}

Opportunities are being explored to extend the studies of the $\gamma_vpN^*$ electrocouplings from exclusive meson electroproduction processes initiated with the CLAS detector in Hall~B at beam energies up to 6~GeV and continued with the CLAS12 detector at beam energies up to 11~GeV, to a proposed ``CLAS22'' configuration at beam energies up to 22~GeV. Simulations of the $\pi N$, $\pi^+\pi^-p$, $K\Lambda$, and $K\Sigma$ exclusive channels with a 22~GeV electron beam show that the electrocouplings could be accessed up to $Q^2 \approx 20-30$~GeV$^2$ utilizing the large acceptance spectrometer at luminosities ${\cal L} \gtrsim 2 \times 10^{35}$~cm$^{-2}$s$^{-1}$ with a 1-2 year duration beam run. Figure~\ref{clas22-proj}(b) shows a simulation result of the expected kinematic coverage for $KY$ exclusive production at CLAS22 in terms of $Q^2$ vs. $W$, giving an indication of the kinematic domain that could be accessed in such a measurement program. The push to higher $Q^2$ must face the reality of the rapid drop-off in the cross section as illustrated in Fig.~\ref{clas22-proj}(a) (based on a standard dipole form factor evolution) and set practical limits for the beam-target luminosity and the experiment duration for data collection.

A comparison of the parameters for the available and anticipated facilities for studies of hadron structure with electromagnetic probes in this regime demonstrates that after the CEBAF energy increase to 22~GeV, CLAS22 would be the only facility capable of measuring $\gamma_v p N^*$ electrocouplings for $Q^2 \gtrsim 10$~GeV$^2$. The expected maximum luminosities for the EIC and EIcC Electron--Ion Collider facilities in the U.S. and China are more than an order of magnitude below the levels required for such extractions and do not effectively span the $x \to 1$ regime that covers the nucleon resonance excitation region~\cite{AbdulKhalek:2021gbh}. Such experiments at the highest photon virtualities $Q^2$ ever achieved (10-30~GeV$^2$) in the studies of exclusive meson electroproduction in the nucleon resonance region will allow for the realization of the goal to improve our understanding of the fundamental underpinnings of the mechanism for EHM in these strongly interacting baryon states over the full range of parton momenta where hadron mass is generated~\cite{Achenbach:2025kfx}. The proposed 22-GeV experimental program, along with the associated experiments in JLab Halls~A/C and the planned studies at AMBER@CERN~\cite{Quintans:2022utc}, EIC~\cite{AbdulKhalek:2021gbh}, and EicC~\cite{Chen:2020ijn} focused on the structure of $\pi$ and $K$ mesons \cite{Roberts:2021nhw}, are of particular importance in order to understand the dynamics of the processes that generate the dominant portion of visible hadron mass in the Universe.

%%%%%%%%%%%%%%%%%%%%%%%%%%%%%%%%%%%%%%%%%%%%%%%%%%%%%%%%%%%%%%%%%%%%%%%%%%%%%%%%%%%%%%%%%%%%%%%%%%%%%%%%%%%%%%%%%%%%%%%%%%%%%%%%%%%%%%%%%%%%%%%%%%%%%%%%%%%%%%%
%%%%%%%%%%%%%%%%%%%%%%%%%%%%%%%%%%%%%%%%%%%%%%%%%%%%%%%%%%%%%%%%%%%%%%%%%%%%%%%%%%%%%%%%%%%%%%%%%%%%%%%%%%%%%%%%%%%%%%%%%%%%%%%%%%%%%%%%%%%%%%%%%%%%%%%%%%%%%%%%=================================================

%=== NUCLEON EXCITED STATES ===================
%  \input{nucleon-excited-states}
\section{Nucleon Resonance Spectrum and Structure Studies}
\label{nstar-studies}

\subsection{Spectroscopy Studies}
\label{nstar-spectrum}

The most comprehensive predictions of the nucleon excitation spectrum have come from various implementations of the constituent quark model incorporating broken SU(6) symmetry~\cite{Burkert:2004sk}. Additional dynamical contributions from gluonic excitations in the wavefunction may also play a central role \cite{Dudek:2012ag} and resonances may be dynamically generated through meson-baryon interactions~\cite{Oset:2004dd}. Quark model calculations of the nucleon spectrum have predicted many more states than have been seen experimentally~\cite{Capstick:1998uh,Loring:2001kx}. This long-standing discrepancy has been termed the ``missing'' resonance problem~\cite{Koniuk:1979vy}, and the existence of these predicted states is directly connected with the underlying degrees of freedom of the nucleon that govern hadronic production at moderate energies~\cite{Isgur:2000ad}. Additional information is included in Section~\ref{sec:missing}.

Spectroscopic studies of the nucleon resonance excitation spectrum are essential to gain a deeper understanding into strong interaction dynamics and QCD. This relevance comes not from a mere ``stamp collecting'' point of view. Rather, baryons are the most fundamental three-body systems in nature. If we do not understand how QCD generates these bound states of three dressed quarks, then our understanding of nature is incomplete. Remarkable progress has been achieved in recent decades through the studies of the structure of the ground and excited nucleon states based on the experimental results detailed in Section~\ref{expt-measurements}. These experiments have provided for significant new opportunities for QCD-based hadron structure theory to explore many facets of the strong interaction in the regime of large QCD running coupling, i.e., $\alpha_s/\pi \gtrsim 0.2$--referred to as the strong QCD regime, by providing measurements of the $\gamma_vpN^*$ electrocouplings for numerous $N^*$ states, with different quantum numbers and structural features. However, the ``missing'' resonance problem is still very much an issue that must be addressed by pushing our understanding of higher-lying states beyond $\approx$2~GeV, a boundary that effectively represents a realm of {\it terra incognita} where we still have barely made inroads. From another viewpoint, being mindful of earlier history, understanding the ground state of the hydrogen atom required understanding its full excitation spectrum. Only through this path was it possible to evolve from the Bohr model of the atom to the complete field theory of Quantum Electrodynamics (QED). Likewise, understanding the ground state of the proton requires understanding its full excitation energy spectrum, the so-called $N^*$ states. It is this crucial knowledge that will enable evolution beyond the constituent quark model to a complete understanding of the corresponding field theory of QCD. The study of $N^*$ states is critical to charting the true character of the wavefunction of baryons. Many models have the same ground state, but only the one that captures the intricacies of the full excitation spectrum truly reflects nature.

Another appreciation for effectively cataloging the nucleon resonance excitation spectrum has arisen in recent years connected with events crucial in their significance in the phase transition in the early universe from the quark-gluon plasma of non-interacting colored quarks and gluons to the formation of protons and neutrons. During this transition, a number of dramatic events occurred - chiral symmetry was broken, quarks acquired a dynamically generated mass, baryon resonances occurred abundantly, and colored quarks and gluons became confined. This phase evolution is apparently governed by the formation of excited hadrons~\cite{Chatterjee:2017yhp} and in this process strong QCD was born as the process describing the interaction of colored quarks and gluons. The intimate relationship of the baryon resonance excitation spectrum and the evolution of the early universe makes the experimental search for the ``missing'' resonances imminently compelling~\cite{Burkert:2019kxy}.

It should be expected that QCD could provide a reliable prediction of the nucleon excitation spectrum. However, due to the non-perturbative nature of QCD at these energies, this expectation has not been fully realized. There has been progress in lattice QCD calculations for predictions of the $N^*$ spectrum with dynamical quarks, although with unphysically large pion masses \cite{Bulava:2009jb,Dudek:2012ag}. Calculations in this approach with improved actions, larger volumes, and smaller quark masses continue to progress~\cite{Khan:2020ahz}.

Most of our present knowledge of baryon resonances comes from reactions involving pions in the initial and/or final states. One explanation for the excess of predicted states relative to experimental observation could be that pionic coupling to the intermediate $N^*$ or $\Delta^*$ states is weak. This suggests a search for these states in the strangeness channels with electromagnetic probes. In addition to different coupling constants (i.e., $g_{KNY}$ vs. $g_{\pi NN}$), the study of the exclusive $K^+\Lambda$ and $K^+\Sigma^0$ final states has other advantages in the search for missing quark model resonances. The $KY$ final states, due to the creation of an $s\bar{s}$ quark pair in the intermediate state, are naturally sensitive to coupling to higher-lying $s$-channel resonance states at $W > 1.6$~GeV, the region where our knowledge of the $N^*$ spectrum is the most limited. In addition, baryon resonances have large widths (100-400~MeV) and are often overlapping. Studies of different final state channels can provide for important cross checks to understand the contributing amplitudes in kinematic domains with different ratios of the resonant to non-resonant contributions. Although the two ground-state hyperons have the same valence quark structure ($uds$), they differ in isospin, such that intermediate $N^*$ resonances can decay strongly to $K^+\Lambda$ final states, while both $N^*$ and $\Delta^*$ decays can couple to $K^+\Sigma^0$ final states, so a separate focus on these two channels provides an isospin filter in the analysis (as mentioned in Section~\ref{sec:isospin}).

The majority of the advancements in understanding the excited nucleon spectrum have been provided by analyses of the $\pi N$, $\pi \eta$, and $\pi^+ \pi^- p$ channels~\cite{Carman:2018fsn}. However, with the publication of the high statistics $K^+ \Lambda$ and $K^+ \Sigma^0$ photoproduction data, the potential and importance of the hyperon channels has been realized. The $K^+\Lambda$ and $K^+\Sigma^0$ cross sections and polarization observables have had a significant impact on the discovery of several new $N^*$ states based on partial wave analysis fits~\cite{Anisovich:2007bq,CLAS:2017sgi,Anisovich:2017bsk}. They have also provided new evidence for several candidates states that had been claimed but lacked confirmation~\cite{Burkert:2019kxy}. Table~\ref{pdg-nstar-comp} provides a comparison from the PDG listings \cite{ParticleDataGroup:2024cfk} of the evolution of our understanding of the $N^*$ spectrum over the past 15 years based on significant input from the $KY$ photoproduction data. Returning to the earlier discussion about the role of $N^*$ states and their importance in understanding the phase transition evolution of the early universe, Ref.~\cite{Burkert:2019kxy} notes that including the additional states in Table 4.1 in available evolution models has a strong impact in improving agreement with lattice QCD calculations of so-called ``hot'' QCD.

%%%%%%%%%%%%%%%%%%%%%%%%%%%%%%%%%%%%%%%%%%%%%%%%%%%%%%%%%%%%%%%%%%%%%%%%%%%%%%%%%%%%%%%%%%%%%%%%%%%%%%%%%%%%%%%%%%%%%%%%%%%%%%%%%%%%%%%%%%%%%%%%%%%%%%%%%%%%%%%
\begin{table}[htpb]
\setlength{\tabcolsep}{6pt} % Default value: 6pt
\renewcommand{\arraystretch}{0.8} % Default value: 1
\begin{center}
\caption{PDG listings for a dozen $N^*$ states (masses given in MeV) whose status has been upgraded since the 2010 tables due to inclusion of the recent $KY$ photoproduction data. The * symbols are the measure from the PDG for the existence of a given state (**** $\equiv$ existence to certain to * $\equiv$ evidence of existence is poor). The table also provides the * rating for the decays of the $N^*$ resonances to different final states \cite{ParticleDataGroup:2024cfk}.}
\begin{tabular} {ccccccc} \hline\hline
State               & PDG  & PDG  & \multirow{2}{*}{$\pi N$} & \multirow{2}{*}{$K \Lambda$} & \multirow{2}{*}{$K \Sigma$} & \multirow{2}{*}{$\gamma N$} \\
$N$(mass)$J^P$      & 2010 & 2024 &                          &                              &                             &                             \\ \hline
$N(1710)1/2^+$      & ***  & **** & **** & **  & *   & **** \\ %\hline
$N(1875)3/2^-$      &      & ***  & **   & *   & *   & **   \\ %\hline
$N(1880)1/2^+$      &      & ***  & *    & **  & **  & **   \\ %\hline
$N(1895)1/2^-$      &      & **** & *    & **  & **  & **** \\ %\hline
$N(1900)3/2^+$      & **   & **** & **   & **  & **  & **** \\ %\hline
$N(2000)5/2^+$      & *    & **   & *    &     &     & **   \\ %\hline
$N(2060)5/2^-$      &      & ***  & **   & *   & *   & ***  \\ %\hline
$N(2100)1/2^+$      & *    & ***  & ***  & *   &     & **   \\ %\hline
$N(2120)3/2^-$      &      & ***  & **   & **  & *   & ***  \\ %\hline
$\Delta(1600)3/2^+$ & ***  & **** & ***  &     &     & **** \\ %\hline
$\Delta(1900)1/2^+$ & **   & ***  & ***  &     & **  & ***  \\ %\hline
$\Delta(2200)7/2^-$ & *    & ***  & **   &     & **  & ***  \\ \hline \hline
\end{tabular}                     
\label{pdg-nstar-comp}
\end{center}                      
\end{table}
%%%%%%%%%%%%%%%%%%%%%%%%%%%%%%%%%%%%%%%%%%%%%%%%%%%%%%%%%%%%%%%%%%%%%%%%%%%%%%%%%%%%%%%%%%%%%%%%%%%%%%%%%%%%%%%%%%%%%%%%%%%%%%%%%%%%%%%%%%%%%%%%%%%%%%%%%%%%%%%

It is also important to appreciate that significant amounts of data in terms of cross sections and polarization observables for electroproduction of $K^+ \Lambda$ and $K^+ \Sigma^0$ spanning the full nucleon resonance region and beyond ($W$ up to 2.5~GeV) for $Q^2$ from 0.3 to 5.4~GeV$^2$ are now available (see Section~\ref{clas-ep-program}). These $KY$ data are also valuable as input to spectroscopic studies as they can be used to confirm the signals of new baryon states observed in photoproduction in a complementary fashion. Within each bin of $Q^2$, the contributing states must have the same mass and decay widths. In such studies, the electroproduction data can be used to verify the findings for the states shown in Table~\ref{pdg-nstar-comp} and can also allow searches for higher-lying states that are not revealed in photoproduction processes. In this regard, lattice QCD calculations predict the existence of baryon states with glue as an active structural component~\cite{Dudek:2012ag}. Measurements of the extracted electrocouplings of such candidate states as a function of $Q^2$ is one approach to confirm the existence of such hybrid states as their amplitudes are expected to have a significantly different evolution with distance scale compared to three quark baryons with the same spin and parity~\cite{Li:1991yba,clas12-hybrid}.

\subsection{Structure Studies}
\label{nstar-structure}

Investigations of the spectrum and structure of excited nucleon states have played a crucial role in the development of our understanding of the strong interaction within the light quark sector. As a result of extensive experimental, phenomenological, and theoretical effort over the past 40 years, it is now recognized that the structure of nucleon excited states is much more complex than what can be described in terms of simple constituent quark models. At the typical energy and distance scales relevant for the $N^*$ states, the quark-gluon coupling is large. It is therefore the case that theoretical calculations are confronted with the fact that quark-gluon confinement, hadron mass generation, and the dynamics that give rise to the $N^*$ spectrum, cannot be understood within the framework of perturbative QCD. In order to understand QCD in this domain, studies of $N^*$ structure can provide important insights. Such studies, in fact, represent a necessary step toward understanding how QCD in the regime of large quark-gluon coupling generates mass and how systems of confined quarks and gluons, i.e. mesons and baryons, are formed. These questions remain among the most challenging open problems within the Standard Model of fundamental particles and interactions~\cite{DOE2023LRP}.

\begin{figure*}[htbp]
\centering
\includegraphics[width=0.7\textwidth]{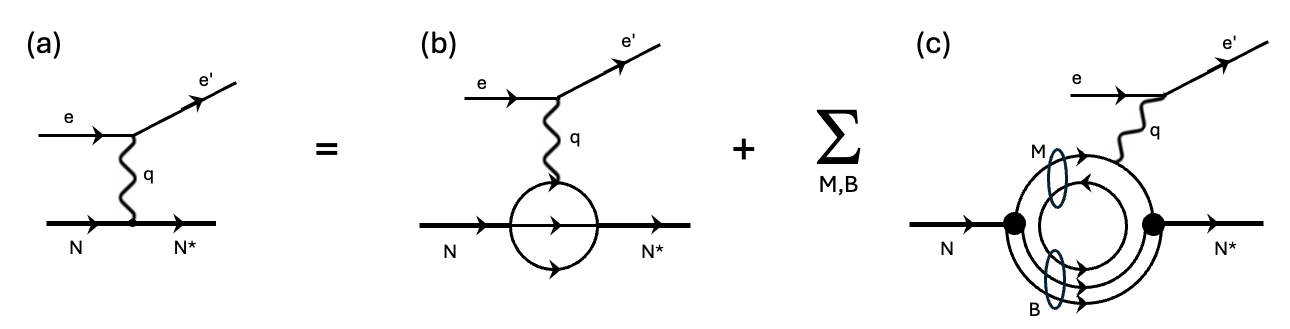}
\caption{Schematic representation of the $\gamma_v N \to N^*$ electroproduction process. (a) The fully dressed $\gamma_vNN^*$ electrocoupling that determines the $N^*$ contribution to the resonant part of the meson electroproduction amplitude. (b) The contribution of the three-quark core. (c) The contribution from the meson-baryon cloud, where the sum is over all intermediate meson and baryon states. Figure motivation from Ref.~\cite{Mokeev:2013kka}.}
\label{electrocoupling} 
\end{figure*}

Studies of low-lying nucleon excited states using electromagnetic probes at $Q^2 < 5$~GeV$^2$ have revealed that the structure of $N^*$ states is a complex interplay between the internal core of three dressed quarks and an external meson-baryon (MB) cloud~\cite{Carman:2023zke} (see Fig.~\ref{electrocoupling}). Excited nucleon states of different quantum numbers have significantly different relative contributions from these two components, demonstrating different manifestations of the non-perturbative strong interaction in their formation. The relative contribution of the quark core increases with $Q^2$ in a gradual transition to a dominance of quark degrees of freedom for $Q^2 > 5$~GeV$^2$. This kinematic regime still remains almost unexplored in exclusive reactions. Studies of the evolution of $N^*$ structure over a broad range of $Q^2$ offer access to the strong interaction between dressed quarks in the non-perturbative regime that is responsible for $N^*$ formation. Such studies are part of the existing CLAS12 program (see Section~\ref{clas12-program}) and the proposed CLAS22 program (see Section~\ref{sec:clas22}).

%%%%%%%%%%%%%%%%%%%%%%%%%%%%%%%%%%%%%%%%%%%%%%%%%%%%%%%%%%%%%%%%%%%%%%%%%%%%%%%%%%%%%%%%%%%%%%%%%%%%%%%%%%%%%%%%%%%%%%%%%%%%%%%%%%%%%%%%%%%%%%%%%%%%%%%%%%%%%%%
%%%%%%%%%%%%%%%%%%%%%%%%%%%%%%%%%%%%%%%%%%%%%%%%%%%%%%%%%%%%%%%%%%%%%%%%%%%%%%%%%%%%%%%%%%%%%%%%%%%%%%%%%%%%%%%%%%%%%%%%%%%%%%%%%%%%%%%%%%%%%%%%%%%%%%%%%%%%%%%

Electroproduction reactions of the form $\gamma_v N \to N^* \to M + B$ provide a tool to probe the inner structure of the contributing $N^*$ resonances through the extraction of the amplitudes that describe the transition between the virtual photon-nucleon initial state and the intermediate excited $N^*$ state. These $\gamma_vNN^*$ electrocouplings are directly related to the structure of their associated $N^*$ states and can be represented by the helicity amplitudes $A_{1/2}(Q^2)$, $A_{3/2}(Q^2)$, and $S_{1/2}(Q^2)$. $A_{1/2}$ and $A_{3/2}$ describe the $N^*$ resonance electroexcitation for transversely polarized virtual photons and a nucleon with helicity parallel or anti-parallel to the photon, while $S_{1/2}$ describes the $N^*$ resonance electroexcitation by longitudinally polarized virtual photons of zero helicity~\cite{Aznauryan:2011qj}. The $Q^2$ evolution of these electrocouplings provides fundamental information on the relevant degrees of freedom that describe the structure of the nucleon as a function of distance scale. These fundamental quantities are now subject to computations starting from theoretical approaches based on the QCD Lagrangian~\cite{Mokeev:2022xfo}.

Reliable information on $KY$ hadronic decays from $N^*$s is not yet available due to the lack of a reaction model that accurately reproduces the available data over its full kinematic range of $Q^2$, $W$, and $\cos \theta_K^{\rm c.m.}$. However, after such a model is developed, the $N^*$ electroexcitation amplitudes for states that couple to $KY$ can be extracted from fits to the extensive existing CLAS $KY$ photo- and electroproduction data over the range $Q^2 \lesssim 5$~GeV$^2$, which should be carried out independently in different bins of $Q^2$ with the same $KY$ hadronic decays, extending the available information on these $N^*$ states. The development of reaction models for the extraction of the $\gamma_vpN^*$ electrocouplings from the $KY$ photo- and electroproduction channels is urgently needed. The development of such models as an extraction tool for the electrocouplings from the available CLAS data at $Q^2 < 5$~GeV$^2$ then sets the stage for extension of the models to $Q^2 > 5$~GeV$^2$ from the forthcoming CLAS12 data.

Electroexcitation amplitudes for most $N^*$ states below 1.8~GeV have been extracted for the first time from analysis of CLAS $ep$ data in the exclusive $\pi^+ n$ and $\pi^0 p$ channels for $Q^2$ up to 5~GeV$^2$, in $\eta p$ for $Q^2$ up to 4~GeV$^2$, and for $\pi^+ \pi^- p$ for $Q^2$ up to 5~GeV$^2$~\cite{Mokeev:2023zhq}. The resonance parameter extractions from the experimental observables represent a complex exercise and involve a level of model and fit uncertainty. Therefore it is highly desirable that the resonance electrocouplings are determined from at least two different final states in order to understand the model-dependent systematic uncertainties.

With a goal to have an independent determination of the electrocouplings for each $N^*$ state from multiple exclusive reaction channels, a natural avenue to investigate for the higher-lying $N^*$ states is the strangeness channels $K^+\Lambda$ and $K^+\Sigma^0$. In fact, data from the $KY$ channels are critical to provide an independent extraction of the electroexcitation amplitudes for the higher-lying $N^*$ states and represent a central part of the $N^*$ program with CLAS (see Section~\ref{clas-ep-program}) and underway with CLAS12 (see Section~\ref{clas12-program}). For most $N^*$ states with coupling to $KY$, the $\pi N$ coupling is larger than the $KY$ coupling. In other words, $g_{\pi NN} > g_{KYN}$. However, there are states for which the $KY$ coupling is predicted to be of the same magnitude (or even larger) than the $\pi N$ coupling~\cite{Capstick:1998uh} (see Section~\ref{sec:missing}). In addition, in cases where the $\pi N$ coupling is weak, it may be that distinguishing $N^*$ states from $\pi N$ elastic scattering experimental data could be quite difficult. Therefore, the study of $KY$ final states should be viewed as complementary to the study of $\pi N$ final states. Also due to the fact that $m_K + m_Y > m_{\pi} + m_N$, the $KY$ decays kinematically favor a two-body decay mode for resonances with masses near 2~GeV. This amounts to a significant experimental advantage for the study of $N^* \to KY$ decays as two-body decay modes are easier to interpret than extracting $N^*$ spectrum and structure information from the reconstruction of a series of sequential non-strange decays. In comparison to the $\pi N$, $\pi\pi N$, and other prominent non-strange channels, the backgrounds for the $KY$ channels are markedly different as are the resonant to non-resonant amplitude contributions. In such a case, the systematics of the extracted electroexcitation amplitudes would be wholly different. Consistency of the extracted amplitudes from the independent analysis of the non-strange and strange final states would then serve to give confidence in the results by quantifying the systematic uncertainties related to the model dependence of the extracted electrocouplings. It is important to note that the statistical quality of the $KY$ data from the existing CLAS measurements is comparable to that of the CLAS $\pi\pi N$ data from which the electrocouplings of many $N^*$ states have already successfully been extracted~\cite{Mokeev:2022xfo}.%=================================================

%=== EXISTING MODELS =============================
%  \input{existing_models}
\section {Existing Models}
\label{sec:models}

The theoretical and phenomenological description of kaon photo- and electroproduction on the nucleon has been under development since the 1950s. To the best of our knowledge, the pioneering works in describing kaon photoproduction on the nucleon were the works of Kawaguchi and Moravcsik \cite{Kawaguchi57} and of Fujii and Marshak \cite{Fujii:1957}, both published in 1957, which were based on perturbation theory. Other notable approaches include dispersion theory \cite{Nelipa:1963ynj,Amati1957,Fayyazuddin64,Pickering:1973yxk} and Partial Conservation of Axial-Vector Current (PCAC) \cite{Basu:1968dfk}. Over the years, a number of other frameworks have been proposed to model the electromagnetic production of kaons on the nucleon, ranging from phenomenological isobar models to more sophisticated coupled-channel analyses. In the following sections, we will discuss these models according to their type to emphasize the conceptual differences between frameworks. Within each model type, we will also review the work of various research groups to further highlight the distinctive features and developments of their models. We note that several other models are not included in the following discussion, either because they are overly simplified, lack sufficient available documentation, or can be confronted with only a very limited set of experimental data~\cite{Ahlig:2000qu,Alkofer:2001qj,Henley:2010yf,Bhat04,Kupsch:1966fbv,Moorhouse66,Mart:2000jv}.

%=============================================
%=============================================
%=============================================

\subsection{Quark Models}

\begin{figure}[t]
\centerline{
\includegraphics[width=0.65\textwidth]{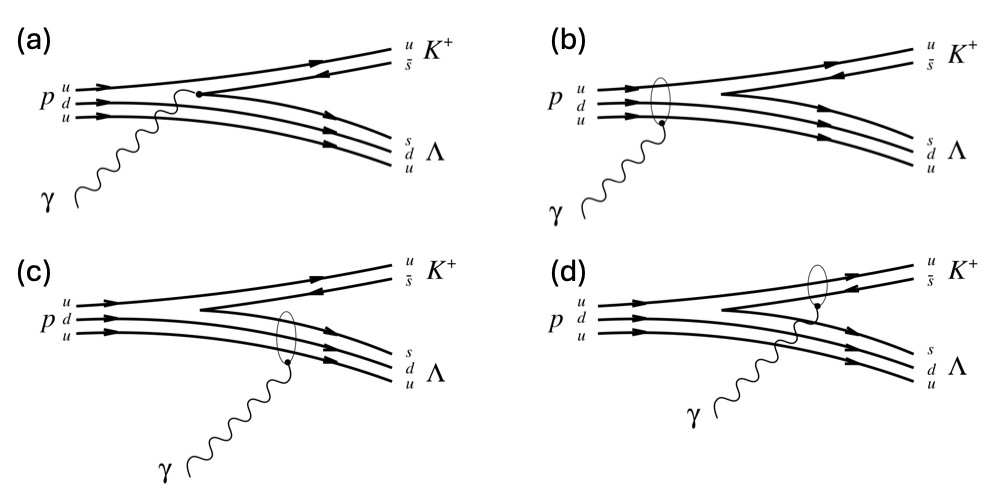}}
\caption{Quark-flow diagrams for $K^+$ photoproduction in the semi-relativistic quark model (SRQM). Panel (a) shows the direct photon-induced creation of an $s{\bar s}$ pair, representing the seagull term, while panels (b), (c), and (d) illustrate indirect production mechanisms corresponding to the $s$-, $u$-, and $t$-channel processes, respectively. Figures adapted from Ref.~\cite{Mart:1996gx}.}
\label{fig:quark_model_diagram} 
\end{figure}

In the mid-1990s, several efforts were made to describe kaon photoproduction within the framework of quark models. Among these was the semi-relativistic quark model (SRQM) developed by Kumar and Onley~\cite{Kumar:1994,Onley:1994}. This model was based on the Quark Pair Creation (QPC) model formulated by Le Yaouanc et al.~\cite{LeYaouanc:1972vsx,LeYaouanc:1973ldf}. The QPC model posits that a quark-antiquark pair is produced during a strong decay process without disturbing the constituent quarks of the initial hadron, analogous to Dirac's concept of electron-positron pair creation from the negative energy sea. Although the QPC model permitted only indirect quark-pair production, it could not generate an $s\bar{s}$ pair directly from the photon, as required by gauge invariance. To resolve this shortcoming, the SRQM constructed its Hamiltonian to allow direct creation of $s\bar{s}$ pairs. In this framework, kaon photoproduction, $\gamma p \to K^+\Lambda$, can be represented by the quark-flow diagrams in Fig.~\ref{fig:quark_model_diagram}. In Fig.~\ref{fig:quark_model_diagram}(a), the photon produces the $s\bar{s}$ pair directly, while the remaining quarks act as spectators and subsequently recombine with the newly created quarks to form the $K^+$ and $\Lambda$. Translating this process into a Feynman diagram yields the well-known seagull term. The photon may also couple to any of the quarks inside the proton, as illustrated in Fig.~\ref{fig:quark_model_diagram}(b). In this case, the photon excites the proton to an intermediate state, which then decays after the $s\bar{s}$ pair is produced. This corresponds to the $s$-channel contribution in the isobar model. Finally, the quark-flow diagrams shown in Figs.~\ref{fig:quark_model_diagram}(c),(d) represent the $u$- and $t$-channel processes, respectively. 

%%%%%%%%%%%%%%%%%%%%%%%%%%%%%%%%%%%%%%%%%%%%%%%%%%%%%%%%%%%%%%%%%%%%%%%%%%%%%%%%%%%%%%%%%%%%%%%%%%%%%%%%%%%%%%%%%%%%%%%%%%%%%%%%%%%%%%%%%%%%%%%%%%%%%%%%%%%%%%%
%%%%%%%%%%%%%%%%%%%%%%%%%%%%%%%%%%%%%%%%%%%%%%%%%%%%%%%%%%%%%%%%%%%%%%%%%%%%%%%%%%%%%%%%%%%%%%%%%%%%%%%%%%%%%%%%%%%%%%%%%%%%%%%%%%%%%%%%%%%%%%%%%%%%%%%%%%%%%%%

The hadron wavefunctions were constructed as products of their flavor, color, spin, and spatial components, with the latter described by harmonic-oscillator wavefunctions. The $u$ and $d$ quark masses were taken to be 330 MeV, while the $s$ quark was assigned a mass of 500 MeV. In the case of $K^+\Lambda$ photoproduction, the model reproduced the experimental data with moderate accuracy. The predictions exhibited strong sensitivity to the assumed proton and kaon radii. Moreover, the model favored mixing between the $N(1535)1/2^-$ and $N(1650)1/2^-$, as well as between the $N(1520)3/2^-$ and $N(1700)3/2^-$. However, for $K^+\Sigma^0$ photoproduction, the SRQM showed a noticeable shortcoming, it failed to reproduce the observed energy dependence of the differential cross sections.

Another quark-based approach to the $\gamma p \to K^+\Lambda$ reaction is the color-dielectric model (CDM)~\cite{Lu:1995bk}, which applies the earlier chiral color-dielectric model (CCDM) developed by the same authors~\cite{Lu:1994qi} to photoproduction processes. In this framework, the kaon is treated as an elementary Goldstone boson that preserves chiral symmetry, whereas baryons are described as non-elementary composites of three relativistic quarks confined by the scalar expectation value of the glueball field.

At the quark level, the CCDM with pseudovector coupling is governed by the Lagrangian
\begin{eqnarray}
    {\cal L} =  {\bar q} \left[ i\gamma^\mu\partial_\mu-\frac{m_q}{\chi}+\frac{i}{2f} \gamma_5\gamma^\mu \lambda_a\cdot \partial_\mu\phi_a \right] q + \frac{\sigma_v^2}{2} \left(\partial_\mu\chi\right)^2-U(\chi) + \frac{1}{2} \left( \partial_\mu \phi_a \right)^2 - \frac{1}{2} m_\phi^2 \phi_a^2 \,,
\end{eqnarray}
with $q$, $\phi_a$, and $\chi$ corresponding to the quark, pseudoscalar meson, and scalar glueball fields, respectively. The matrices $\lambda_a$ are the SU(3) Gell-Mann generators, and $f$ denotes the mean weak decay constant of the pseudoscalar octet. A key component of the model is the dielectric self-interaction field $U(\chi)$, which effectively incorporates the mean gluon field and ensures that quarks are confined through a self-consistent, dynamical mechanism.

\begin{figure}[htb]
\centerline{
\includegraphics[height=0.4\textwidth]{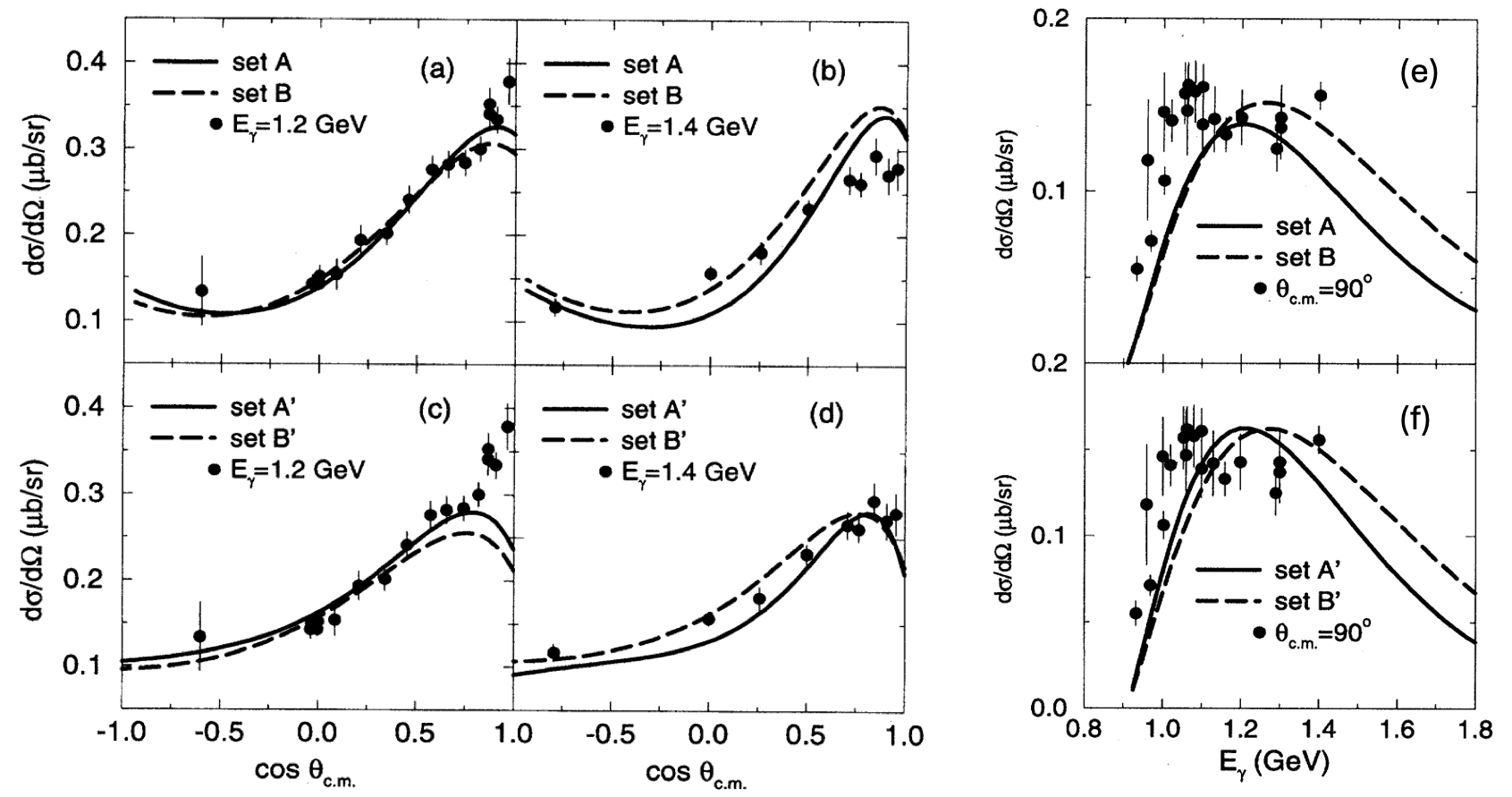}}
\caption{The CDM predictions for the $\gamma p \to K^+\Lambda$ differential cross sections compared with experimental data plotted as a function of $\cos \theta_K^{\mathrm{c.m.}}$ (a)-(d) and the photon laboratory energy (e)-(f). The differences between the solid and dashed curves illustrate the sensitivity of the model to a 10\% variation in the kaon decay constant $f_K$. Figures from Ref.~\cite{Lu:1995bk}.}
\label{fig:Landau_CDM_model} 
\end{figure}

To apply the Lagrangian to photoproduction, the photon-quark interaction is incorporated through the minimal substitution $\partial_\mu \to \partial_\mu - i e Q A_\mu$, where $A_\mu$ is the photon field, $e$ the proton charge, and $Q$ the quark or meson charge operator. This procedure leads to different form factors at each interaction Lagrangian. These vertex-dependent form factors account for the composite nature of the baryons while simultaneously providing effective couplings that ensure the convergence of the amplitude at higher energies. 
Note that only Born terms were included in this model. Several parameters enter the numerical calculations, including those associated with the self-interaction of the scalar field, the quark masses, the glueball mass, and the pion weak decay constant. Comparisons with very limited experimental data indicate good agreement for photon laboratory energies $E_\gamma^{\mathrm{lab}}$ from threshold up to 1.4~GeV (see Fig.~\ref{fig:Landau_CDM_model}), while the description of the recoil polarization $P_\Lambda$ at $E_\gamma^{\mathrm{lab}} = 1.1$~GeV is reasonably fair. The observables were also found to be quite sensitive to variations in the kaon decay constant $f_K$, as illustrated in Fig.~\ref{fig:Landau_CDM_model}.

%%%%%%%%%%%%%%%%%%%%%%%%%%%%%%%%%%%%%%%%%%%%%%%%%%%%%%%%%%%%%%%%%%%%%%%%%%%%%%%%%%%%%%%%%%%%%%%%%%%%%%%%%%%%%%%%%%%%%%%%%%%%%%%%%%%%%%%%%%%%%%%%%%%%%%%%%%%%%%%
%%%%%%%%%%%%%%%%%%%%%%%%%%%%%%%%%%%%%%%%%%%%%%%%%%%%%%%%%%%%%%%%%%%%%%%%%%%%%%%%%%%%%%%%%%%%%%%%%%%%%%%%%%%%%%%%%%%%%%%%%%%%%%%%%%%%%%%%%%%%%%%%%%%%%%%%%%%%%%%

The chiral quark model ($\chi$QM) of Li~\cite{Li:1995si}, developed as an extension of his earlier study of threshold pion photoproduction~\cite{Li:1994cy}, provides better predictive power than the CDM, particularly in applications to kaon photoproduction. In this framework, the kaon is treated as a Goldstone boson, while baryons are described as simple SU(6) three-quark bound states represented by harmonic-oscillator wavefunctions. This differs from the CDM, in which baryons are modeled as three-quark systems embedded in a scalar color-dielectric field that simulates gluonic confinement. In both approaches the kaon couples directly to the constituent quarks, but an important distinction is that the $\chi$QM enforces chiral symmetry at low energies, requiring the quark-meson interaction to take the pseudovector form. The low-energy QCD Lagrangian employed in this model can be written as~\cite{Manohar:1983md}
\begin{equation}
    {\cal L}={\bar\psi} \Bigl[ \gamma_\mu \bigl( i\partial^\mu+V^\mu+A^\mu\bigr) - m  \Bigr] \psi + \cdots ,
\end{equation}
where
\begin{equation}
    V_\mu = \frac{1}{2} \left( \xi^\dagger \partial_\mu \xi + \xi \partial_\mu \xi^\dagger \right), ~~~~~~~
    A_\mu = \frac{i}{2} \left( \xi^\dagger \partial_\mu \xi - \xi \partial_\mu \xi^\dagger \right), ~~~~~~~
    \xi = e^{i\pi/f} ,
\end{equation}
with $\psi$ denoting the quark field, $f$ the meson decay constant, and $\pi$ the Goldstone boson field. As in the CDM, the quark-photon-kaon vertex arises from gauging the axial-vector field $A^\mu$ in the Lagrangian, generating a seagull term relevant for charged-kaon photoproduction. Its non-relativistic reduction reads
\begin{equation}
    H_{K,e}^{\rm NR} = i\sum_j \frac{e}{f_K}\, a_j^\dagger(s)\, a_j(u)\, {\boldsymbol{\sigma}_j\,\boldsymbol{\cdot\epsilon}} \,,
\end{equation}
with $\boldsymbol{\sigma}$ the nucleon spin operator, $\boldsymbol{\epsilon}$ the photon polarization vector, and $a_j^\dagger(s)$ and $a_j(u)$ are the strange- and up-quark creation and annihilation operators, respectively. The remaining interaction terms, including the standard pseudovector and electromagnetic couplings, arise from expanding the nonlinear field $\xi$ in powers of the Goldstone boson field $\pi$.

Form factors emerge naturally in both models. However, in the $\chi$QM they originate from overlaps of the harmonic-oscillator wavefunctions. The resulting Gaussian form provides strong suppression of high-momentum components, yielding improved agreement with experimental data at higher energies. After reducing the transition operators to their non-relativistic forms and evaluating their matrix elements between the baryon wavefunctions, the amplitudes can be expressed in the standard CGLN basis (see Section~\ref{sec:prodamp}), from which the observables are calculated. The model was applied to calculate the $\gamma p\to K^+\Lambda,~K^+\Sigma^0,~K^0\Sigma^+$ cross sections, as well as the recoil polarization in $\gamma p\to K^+\Lambda$. 

\begin{figure}[htb]
\centering
\includegraphics[width=0.65\textwidth]{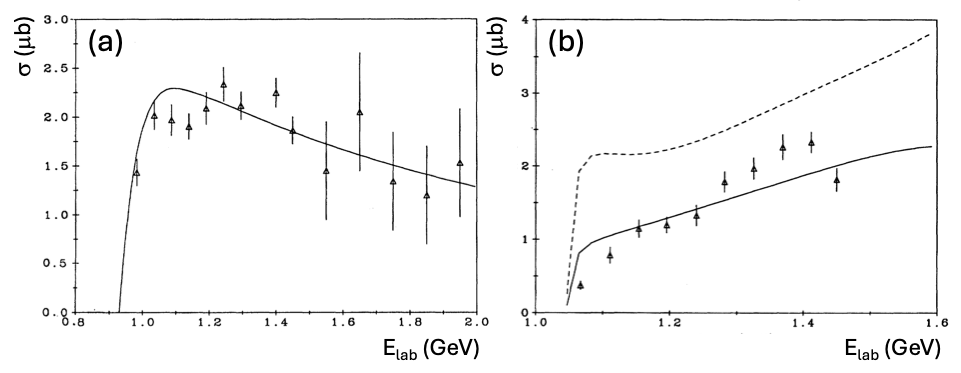}
\caption{Total cross sections vs. $E_\gamma^{\rm lab}$ predicted by the chiral quark model compared to the experimental data for the (a) $\gamma p\to K^+\Lambda$ and (b) $\gamma p\to K^+\Sigma^0$ channels. Figures from Ref.~\cite{Li:1995si}.}
\label{fig:Li_quark_model} 
\end{figure}

Only two parameters were fitted to reproduce the limited experimental data from Refs.~\cite{Adelseck:1990ch,Bockhorst:1994jf}, i.e., the kaon-quark coupling $\alpha_K$ and the $K^*$ coupling constant $g_{K^*}$. All other inputs, including the constituent-quark masses and the harmonic-oscillator parameter, were taken from standard values commonly used in the literature. Good overall agreement between the model predictions and the differential cross section data near 1.2 and 1.4 GeV, together with the total cross section data below 2 GeV, was obtained, as illustrated in Fig.~\ref{fig:Li_quark_model}. For both $K^+\Lambda$ and $K^+\Sigma^0$ photoproduction, the $\chi$QM provides a satisfactory description of the data. In particular, the pronounced forward peaking observed in the $K^+\Lambda$ channel is well reproduced, which is attributed to the dominance of the seagull term at low energies. For the $K^+\Sigma^0$ reaction, sizable contributions are found from the isospin-3/2 resonances $\Delta(1910)1/2^+$, $\Delta(1920)3/2^+$, $\Delta(1950)5/2^+$, and $\Delta(1905)7/2^+$. Moreover, the recoil polarization in the $K^+\Sigma^0$ channel exhibits an opposite sign compared to that of $K^+\Lambda$, in agreement with the predictions of the SRQM.

%%%%%%%%%%%%%%%%%%%%%%%%%%%%%%%%%%%%%%%%%%%%%%%%%%%%%%%%%%%%%%%%%%%%%%%%%%%%%%%%%%%%%%%%%%%%%%%%%%%%%%%%%%%%%%%%%%%%%%%%%%%%%%%%%%%%%%%%%%%%%%%%%%%%%%%%%%%%%%%
%%%%%%%%%%%%%%%%%%%%%%%%%%%%%%%%%%%%%%%%%%%%%%%%%%%%%%%%%%%%%%%%%%%%%%%%%%%%%%%%%%%%%%%%%%%%%%%%%%%%%%%%%%%%%%%%%%%%%%%%%%%%%%%%%%%%%%%%%%%%%%%%%%%%%%%%%%%%%%%

The $\chi$QM was subsequently extended to study $K\Sigma$ photoproduction in all four isospin channels~\cite{Li:1996kv}, yielding better overall agreement with the available data than traditional isobar models. It should be emphasized, however, that the experimental database at that time was still relatively limited. The model was later further developed into a unified framework for pseudoscalar-meson photoproduction, encompassing the $\pi$, $\eta$, and $K$ channels on the nucleon~\cite{Li:1997gd}. A similar technique was later employed to predict the $K^*$ photoproduction cross sections~\cite{Zhao:2001jw}. The quark model is particularly advantageous in this context because the $K^*\Sigma N^*$ couplings are poorly known. Interestingly, the predicted cross section for the $K^{*+}\Sigma^0$ channel is found to be much larger than that for $K^{*0}\Sigma^+$. This difference arises from the strong suppression of the $t$-channel kaon-exchange contribution in $K^{*0}$ photoproduction, whereas it remains comparatively small in the $K^{*+}$ case. 

Within a coupled-channel framework the chiral quark model was also employed to investigate the $\eta n$, $\eta p$, $K^+\Lambda$, and $K^0\Sigma^+$ photoproduction channels simultaneously~\cite{Golli:2016dlj}. In this approach, the Cloudy Bag Model (CBM)~\cite{Theberge:1980ye,Thomas:1981vc} provides the quark wavefunctions, meson-quark interaction vertices, and electromagnetic currents, while the coupled-channel dynamics are implemented through a $K$-matrix formalism in which the resonance pole positions and configuration-mixing angles are constrained solely by the $\pi N$ and $\eta N$ channels. With these inputs fixed, the model predicted the $\pi N\to K\Lambda$, $\pi N\to K\Sigma$, and $\gamma N\to K\Lambda$ processes. Neutral kaon photoproduction was the focus, since the charged kaon channels are dominated by large Born background contributions that mask the resonance signal. The resulting amplitudes reproduce the main qualitative features of $\gamma p\to K^+\Lambda$ and $\gamma p\to K^0\Sigma^+$ photoproduction data.

Another quark model worth mentioning here is the simple spectator quark model (SSQM)~\cite{Keiner:1995rm}. The SSQM is termed “simple” because it includes only the direct creation of the $s\bar{s}$ pair, analogous to the seagull term shown in Fig.~\ref{fig:quark_model_diagram}(a). In this framework, the baryon wavefunctions are constructed as integrals over quark states weighted by a Gaussian profile, while the meson wavefunction is obtained from an integral over quark-antiquark states multiplied by a Gaussian factor. The overall performance of the model, when compared with experimental data, is modest. In certain cases it underestimates the measured differential cross section by roughly 60\%. This shortcoming arises from the highly restricted structure of the model, which treats the quarks inside the nucleon purely as spectators. As a consequence, the model even predicts a vanishing recoil polarization because the amplitude remains real. Introducing resonance contributions is therefore necessary to obtain a non-zero polarization.

%=============================================
%=============================================
%=============================================

\subsection{Chiral Perturbation Theory Models}

Chiral Perturbation Theory (ChPT) is the low-energy effective field theory of QCD, constructed to describe the dynamics of the pseudo-Goldstone bosons ($\pi$, $\eta$, and $K$) that emerge from spontaneous chiral symmetry breaking \cite{Scherer:2012xha}. At energies well below typical hadron masses, quarks and gluons are confined and cannot be treated as explicit perturbative degrees of freedom. Instead, the relevant dynamical variables are mesons, whose interactions reflect both spontaneous and explicit chiral-symmetry breaking in QCD. In this regime, observables are expanded systematically in powers of external momenta and light-quark masses, with meson masses serving as natural expansion parameters through the Gell-Mann-Oakes-Renner relation. When baryons are included, an additional small parameter $p/M$ arises, where $p$ denotes a typical meson momentum and $M$ the baryon mass, enabling a consistent expansion of baryon-meson amplitudes (often formulated in heavy-baryon ChPT). Within this framework, ChPT provides a systematic, model-independent method to compute low-energy hadronic processes order by order, with controlled theoretical uncertainties. Examples include $\pi N$ scattering, nucleon form factors, and kaon photoproduction \cite{Scherer:2012xha,Gasser:1987rb,Scherer:2002tk,Bernard:1995dp,Mai:2009ce}. 

It is obvious that in the case of kaon photoproduction, particular caution is required because the reaction threshold already lies at relatively high energies compared with the typical domain of ChPT validity. At such energies, the relatively large strange quark mass, and therefore the comparatively large kaon mass, significantly slows the convergence of the SU(3) chiral expansion. In addition, the proximity of several baryon resonances in the kaon-hyperon channel further distorts the low-energy expansion, limiting the predictive power of ChPT in this region, as will be illustrated in the following examples.

\begin{figure}[htb]
\centering
\includegraphics[width=0.9\textwidth]{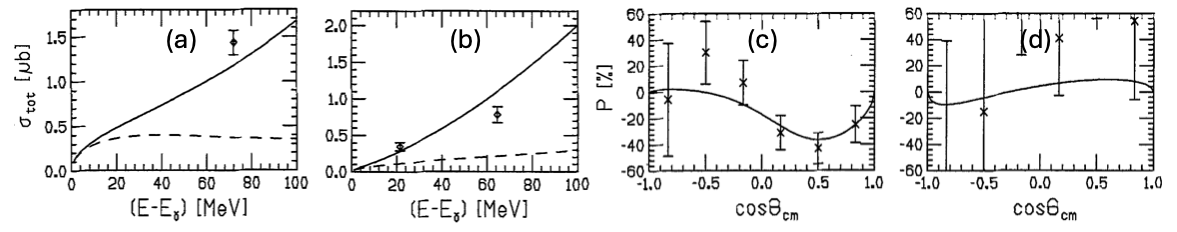}
\caption{Comparison of heavy-baryon ChPT predictions with experimental photoproduction data for the total cross sections of (a) $K^+\Lambda$ and (b) $K^+\Sigma^0$, and the recoil polarizations of (c) $K^+\Lambda$ and (d) $K^+\Sigma^0$. Figures from Ref.~\cite{Steininger:1996xw}. Reprinted with permission from Elsevier.}
\label{fig:steininger} 
\end{figure}

The first application of three-flavor heavy-baryon ChPT to kaon photoproduction on the proton near threshold was carried out by Steininger and Mei{\ss}ner~\cite{Steininger:1996xw}. In their study, the effective meson-baryon Lagrangian was constructed up to third order in the chiral expansion,
\begin{equation}
    {\cal L}_{\rm eff} = {\cal L}_{\rm M} + {\cal L}_{\rm MB} = {\cal L}_{\rm M}^{(2)} + {\cal L}_{\rm MB}^{(1)} + {\cal L}_{\rm MB}^{(2)} + {\cal L}_{\rm MB}^{(3)} + \cdots ,
\end{equation}
where the subscripts M and MB denote the pure-meson and meson-baryon sectors, and the superscripts indicate the chiral expansion order. This expansion incorporates tree-level (Born) terms, loop contributions, and counter terms. At third order, a total of 15 structures contribute to the $S$- and $P$-wave amplitudes. The relevant low-energy constants were fixed partly from known baryon properties and partly estimated via resonance saturation. Very close to threshold, the resulting predictions for the total cross sections and recoil polarizations in both the $K^+\Lambda$ and $K^+\Sigma^0$ channels show reasonable agreement with the available data, as illustrated in Fig.~\ref{fig:steininger}.

%%%%%%%%%%%%%%%%%%%%%%%%%%%%%%%%%%%%%%%%%%%%%%%%%%%%%%%%%%%%%%%%%%%%%%%%%%%%%%%%%%%%%%%%%%%%%%%%%%%%%%%%%%%%%%%%%%%%%%%%%%%%%%%%%%%%%%%%%%%%%%%%%%%%%%%%%%%%%%%
%%%%%%%%%%%%%%%%%%%%%%%%%%%%%%%%%%%%%%%%%%%%%%%%%%%%%%%%%%%%%%%%%%%%%%%%%%%%%%%%%%%%%%%%%%%%%%%%%%%%%%%%%%%%%%%%%%%%%%%%%%%%%%%%%%%%%%%%%%%%%%%%%%%%%%%%%%%%%%%

With the primary motivation to construct a minimal chiral effective approach to meson photoproduction that is exactly unitary and gauge invariant, Borasoy et al.~\cite{Borasoy:2007ku} developed a chiral unitary model based on the Bethe-Salpeter equation using the full off-shell Weinberg-Tomozawa kernel. Gauge invariance was ensured by coupling the photon to all possible hadronic lines. Although the formalism was designed to accommodate electroproduction, the authors performed a $\chi^2$ fit to threshold photon- and pion-induced kaon-production data, i.e., $\gamma p \to K^+\Lambda$, $K^+\Sigma^0$, $K^0\Sigma^+$ and $\pi^- p \to K^0\Lambda$, $K^0\Sigma^0$, to constrain the parameters. For the total cross sections, the model yields good agreement with the experimental data from threshold up to $E_\gamma^{\rm lab} = 1.2~\text{GeV}$ ($W \approx 1.8~\text{GeV}$). In contrast, while the model reproduces the differential cross sections for the $K^+\Lambda$ channel reasonably well, it fails to describe the angular distributions in the $K^+\Sigma^0$ and $K^0\Sigma^+$ channels. This approach has been further developed to next-to-leading order within the SU(3) framework~\cite{Mai:2013cka}, and subsequently applied to the photoproduction of the $\eta$~\cite{Ruic:2011wf} and $\pi$~\cite{Mai:2012wy} off the proton. An extension to the $K^+\pi\Sigma$ photoproduction process, aimed at investigating the effects of meson-baryon rescattering in the final state, has also been carried out~\cite{Bruns:2022sio}.

%\begin{table}[hbt!]
 %   \centering
  %  \caption{Summary of Known Chiral Perturbation Theory Models and the Experimental Data used in the Fitting Process.}
  %  \begin{tabularx}{\textwidth}{XXXc}
  %  \hline \hline
  %  Model     &   Dataset   &    Observable(s)   &  Ref. \\
  %  \hline 
    %Steininger \& Mei{\ss}ner (1997)     & DESY 1969      &    $d\sigma/d\Omega$    &  \cite{ABBHHM:1969pjo} \\
    %&   SAPHIR 1994   &   $d\sigma/d\Omega$, $\sigma$, $P$   &  \cite{Bockhorst:1994jf} \\
%    Borasoy et al. (2007) &   SAPHIR 2003   &   $d\sigma/d\Omega$, $P$    &  \cite{Glander:2003jw} \\
%    &   SAPHIR 2005   &    $d\sigma/d\Omega$, $P$    &  \cite{Lawall:2005np} \\
%    &   LEPS 2003    &   $\Sigma$    &  \cite{LEPS:2003buk} \\
%    &   CLAS 2004   &   $d\sigma/d\Omega$, $P$    &  \cite{CLAS:2003zrd} \\
%    &   CLAS 2006    &   $d\sigma/d\Omega$    &  \cite{CLAS:2005lui} \\
%    \hline \hline
%    \end{tabularx}
%    \label{tab:ChPT_Summary}
%\end{table}

%============================================
%============================================
%============================================

\subsection{Isobar Models}
\label{Subsec:isobar_models}

The central idea behind isobar models is that, in the medium-energy regime, specifically near threshold and within the resonance region corresponding to photon laboratory energies of $E_\gamma^{\mathrm{lab}}=0.91$ - $2.5~\mathrm{GeV}$, the relevant degrees of freedom are hadrons and their excited states \cite{Janssen02,Skoupil10}. Since the properties of these hadronic degrees of freedom are not fully determined from first-principle field theories, they are usually treated as effective degrees of freedom \cite{Corthals07}. Within this framework, interactions are described using effective Lagrangians, and the reaction amplitudes are computed from tree-level Feynman diagrams \cite{Janssen02,Skoupil10,Corthals07} (see Fig.~\ref{fig:s,t,u channels}). However, even when restricted to tree-level processes, the number of diagrams required to adequately describe the reaction mechanism can already be quite large \cite{Skoupil10}.

%%%%%%%%%%%%%%%%%%%%%%%%%%%%%%%%%%%%%%%%%%%%%%%%%%%%%%%%%%%%%%%%%%%%%%%%%%%%%%%%%%%%%%%%%%%%%%%%%%%%%%%%%%%%%%%%%%%%%%%%%%%%%%%%%%%%%%%%%%%%%%%%%%%%%%%%%%%%%%%
\begin{figure}[hbt!]
\centering
\includegraphics[width=0.8\textwidth]{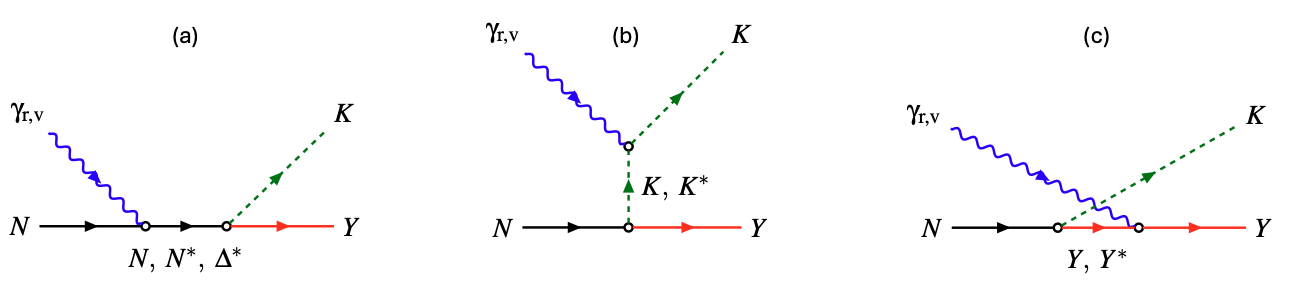}
\caption{The three possible Feynman diagrams for the electromagnetic production of kaons on the nucleon~\mbox{$\gamma_{r,v}+N\to K+Y$}. The subscripts $r$ and $v$ denote real and virtual photons, corresponding to photo- and electroproduction, respectively. The intermediates states in these diagrams are (a) the $s$-channel nucleon, nucleon resonances, and $\Delta$ resonances; (b) the $t$-channel kaon and kaon resonances; and (c) the $u$-channel hyperon and hyperon resonances. We note that $\Delta$ resonances are only present in $K\Sigma$ channel due to isospin conservation.}
\label{fig:s,t,u channels}
\end{figure}
%%%%%%%%%%%%%%%%%%%%%%%%%%%%%%%%%%%%%%%%%%%%%%%%%%%%%%%%%%%%%%%%%%%%%%%%%%%%%%%%%%%%%%%%%%%%%%%%%%%%%%%%%%%%%%%%%%%%%%%%%%%%%%%%%%%%%%%%%%%%%%%%%%%%%%%%%%%%%%%

The simplicity of the tree-level approximation comes at the expense of neglecting final-state interaction effects arising from the rescattering of the final-state particles \cite{Janssen02}. A consistent treatment would require the inclusion of other production channels (e.g., $\pi N$ and $\eta N$), leading to the so-called coupled-channels framework \cite{Skoupil:2018vdh}. While this approach restores unitarity, it also faces challenges, such as the lack of data for some reactions (e.g., $K^+\Lambda \to K^+\Lambda$) and the growing number of parameters that must be introduced \cite{Skoupil:2018vdh,Skoupil12}. Finally, the high production threshold of kaon photoproduction (e.g., $1609~\mathrm{MeV}$ for $\gamma  p \to K^+  \Lambda$ in the c.m. frame) implies that many nucleon and hyperon resonances already contribute to the process near threshold. This results in a highly complex interplay of resonance contributions and an extensive amount of resonance combinations that must be investigated \cite{Mart:1996gx,Skoupil12,Skoupil:2016ast}. We have also summarized the models discussed in this subsection, alongside the datasets used in their fitting process, 
in Tables~\ref{tab:isobar_summary} and \ref{tab:isobar_summary2}. We acknowledge that there are other isobar models by various authors that are not discussed in this paper. Regardless, these works \cite{Deans:1972hh,Deans:1972osv,Egorov:2020ghz,Egorov:2021scy,Kuo63,Kuo:1963zc,Ozaki:2007ka,Ozaki:2008zz,Ozaki09} still provide valuable insights into our understanding of kaon photo- and electroproduction on the nucleon.
    
\subsubsection{Partial Wave Analyses}

The use of isobar models to describe kaon photo- and electroproduction on the nucleon was pioneered by Thom in 1966 \cite{Thom:1966rm}. In his work, Thom derived the background amplitudes from tree-level Feynman diagrams of the standard Born terms and $t$-channel $K^*(892)$ exchange. However, the resonant states were represented by Breit-Wigner electric and magnetic multipoles $E_{\ell\pm}$ and $M_{\ell\pm}$. The electric and magnetic multipoles are parameterized by
\begin{equation}\label{eq:electric multipole}
    E_{\ell\pm}=[qkj_\gamma(j\gamma+1)]^{-1/2}\frac{M_R(\Gamma_E\Gamma_\pm)^{1/2}}{M_R^2-W^2-i\Gamma M_R},
\end{equation}
for $j_\gamma=\ell\pm 1$, and
\begin{equation}\label{eq:magnetic multipole}
    M_{\ell\pm}=[qkj_\gamma(j\gamma+1)]^{-1/2}\frac{M_R(\Gamma_M\Gamma_\pm)^{1/2}}{M_R^2-W^2-i\Gamma M_R},
\end{equation}
for $j_\gamma=\ell$. In Eqs.~(\ref{eq:electric multipole}) and (\ref{eq:magnetic multipole}), $q$ and $k$ are the momentum of the kaon and photon, respectively, in the c.m. frame; $M_R$ and $\Gamma$ are the resonance mass and width; $W$, $\Gamma_{E,M}$, and $\Gamma_\pm$ are the total c.m. energy, the partial decay width of the resonance into the $\gamma N$ final state, and the partial decay width of the resonance into the $K\Lambda$ final state, respectively \cite{Thom:1966rm}. 

The electric and magnetic multipoles $E_{\ell\pm}$ and $M_{\ell\pm}$ can then be expressed in the so-called helicity amplitude $A_{\ell\pm}$, which is given by the Breit-Wigner form \cite{Mart:2006dk}
\begin{equation}\label{eq:Breit_Wigner_Multipole}
	A^R_{\ell\pm}(W)=\bar{A}^R_{\ell\pm}c_{KY}\frac{f_{\gamma R}(W)\Gamma_\mathrm{tot}(W)M_Rf_{KR}(W)}{M_R^2-W^2-iM_R\Gamma_\mathrm{tot}(W)}e^{i\phi},
\end{equation}
where $\bar{A}^R_{\ell\pm}$ is the electric or magnetic photon coupling, $\phi$ is the phase angle, $M_{R}$ is the physical mass of the resonance, $\Gamma_\mathrm{tot}$ is the total width of the resonance, and $f_{\gamma R}$ denotes the $\gamma NR$ vertex. The Breit-Wigner factor $f_{KR}$ is given by
\begin{equation}
	f_{KR}=\left[\frac{1}{(2j+1)\pi}\frac{k_W}{|\boldsymbol{q}|}\frac{m_N}{W}\frac{\Gamma_{KY}}{\Gamma_\mathrm{tot}}\right]^{1/2},~~~k_W=\frac{W^2-m_N^2}{2W},
\end{equation}
where $m_N$ is the mass of the nucleon. The vertex factor $f_{\gamma R}$ is parameterized as
\begin{equation}
	f_{\gamma R}(W)=\left(\frac{k_W}{k_R}\right)^{2\ell'+1}\left(\frac{X^2+k_R^2}{X^2+k_W^2}\right)^{\ell'},
\end{equation} 
where $k_R$ is $k_W$ calculated at $W=M_R$ and $X=500~\mathrm{MeV}$ is the damping factor. We note that the isospin factor $c_{KY}$ depends on the reaction channel. For the $K\Lambda$ channel, which has isospin-1/2,  $c_{K\Lambda}=-1$. However, in the $K\Sigma$ channel, as both nucleon and $\Delta$ resonances can contribute, the case for isospin-1/2 and 3/2 must be considered. Thus, the isospin factor for the $K\Sigma$ channel is given by \cite{Mart:2014eoa,Mart:2019fau}
\begin{equation}
	c_{K\Sigma}=\begin{cases}
		-1/\sqrt{3}~,&~~I=1/2,\\[0.25cm]
		\sqrt{3/2}~,&~~I=3/2.
	\end{cases}
\end{equation}
The total width $\Gamma_\mathrm{tot}$ in Eq.~(\ref{eq:Breit_Wigner_Multipole}) can be expressed as
\begin{equation}
	\Gamma_\mathrm{tot}=\Gamma_{KY}+\Gamma_\mathrm{in},
\end{equation}
where $\Gamma_{KY}$ is the energy-dependent partial-width and $\Gamma_\mathrm{in}$ is the ``inelastic'' width. The widths are given by
\begin{equation}
	\Gamma_{KY}=\beta_K\Gamma_R\left(\frac{|\boldsymbol{q}|}{q_R}\right)^{2\ell+1}\left(\frac{X^2+q_R^2}{X^2+|\boldsymbol{q}|^2}\right)^\ell\frac{W_R}{W},
\end{equation}
\begin{equation}
	\Gamma_\mathrm{in}=(1-\beta_K)\Gamma_R\left(\frac{q_\pi}{q_0}\right)^{2\ell+4}\left(\frac{X^2+q_0^2}{X^2+q_\pi^2}\right)^{\ell+2},
\end{equation}
where $\beta_K$ is the single-kaon branching ratio, $\Gamma_R$ and $q_R$ are the total width and the kaon momentum at $W=M_R$, and $q_\pi$ is the momentum of $\pi$ in the decay process $R\to \pi + N$ in the c.m. frame, assuming dominance of the pion decay channel. Lastly, $q_0$ is $q_\pi$ evaluated at $W=M_R$ \cite{Mart:2006dk}.

The advantage of partial wave analyses is that the electric and magnetic photon couplings $\bar{A}^R_{\ell\pm}$ can be related to the helicity photon couplings $A_{1/2}$ and $A_{3/2}$ \cite{Mart:2017msm}, which are listed in the Review of Particle Physics \cite{ParticleDataGroup:2024cfk}. The electric and magnetic couplings in terms of the helicity photon couplings for $j=\ell+1/2$ are given as \cite{Mart:2006dk}
\begin{equation}
	\bar{E}_{\ell+}=\frac{1}{\ell+1}\left(-A^{\ell+}_{1/2}+\sqrt{\frac{\ell}{\ell+2}}A^{\ell+}_{3/2}\right),~~	\bar{M}_{\ell+}=-\frac{1}{\ell+1}\left(-A^{\ell+}_{1/2}+\sqrt{\frac{\ell+2}{\ell}}A^{\ell+}_{3/2}\right),
\end{equation}
and for $j=\ell-1/2$, 
\begin{equation}
	\bar{E}_{(l+1)-}=-\frac{1}{\ell+1}\left[A^{(l+1)-}_{1/2}+\sqrt{\frac{\ell+2}{\ell}}A^{(l+1)-}_{3/2}\right],~~	\bar{M}_{(l+1)-}=\frac{1}{\ell+1}\left[A^{(l+1)-}_{1/2}-\sqrt{\frac{\ell}{\ell+2}}A^{(l+1)-}_{3/2}\right].
\end{equation}
These electric and magnetic multipoles can then be used to calculate the CGLN amplitudes, as shown in Eq.~(\ref{eq:F_in_terms_of_multipoles}).

As mentioned before, Thom's work pioneered the use of isobar models in describing kaon photo- and electroproduction on the nucleon. In his work, Thom considered many configurations of possible resonant states and their corresponding multipoles (see Table~I of Ref.~\cite{Thom:1966rm}). The model of Thom was fitted to 54 data points, consisting of 46 differential cross section data points and 8 hyperon polarization data points. Unlike most fitting procedures, Thom made his fits based on minimizing the quantity
\begin{equation}
    C^2=\chi_\sigma^2+2\chi^2_P,
\end{equation}
where
\begin{equation}\label{eq:chi2_Thom}
    \chi_\sigma^2=\sum_{i=1}^{46}\left[\frac{\sigma_i(\mathrm{exp})-\sigma_i(\mathrm{cal})}{\Delta\sigma_i(\mathrm{exp})}\right]^2~~~\mathrm{and}~~~\chi_P^2=\sum_{i=1}^{8}\left[\frac{P_i\sigma_i(\mathrm{exp})-P_i\sigma_i(\mathrm{cal})}{\Delta P_i\sigma_i(\mathrm{exp})}\right]^2,
\end{equation}
with $\sigma_i(\mathrm{exp})$ and $\Delta\sigma_i(\mathrm{exp})$ corresponding to the experimental value of the differential cross sections and their corresponding uncertainties, respectively, while $\sigma_i(\mathrm{cal})$ refers to the value of the differential cross section predicted by the model. Similarly for $P_i\sigma_i(\mathrm{exp})$, $\Delta P_i\sigma_i(\mathrm{exp})$, and $P_i\sigma_i(\mathrm{cal})$, where $P_i\sigma_i$ denotes the product of the differential cross section and polarization. The factor of 2 for $\chi_P^2$ was arbitrarily chosen to ensure a good fit to the polarization data. For a detailed explanation on the fitting procedure, see Section~III.B of Ref.~\cite{Thom:1966rm}.

\begin{figure}[hbt!]
\centering
\includegraphics[width=0.7\textwidth]{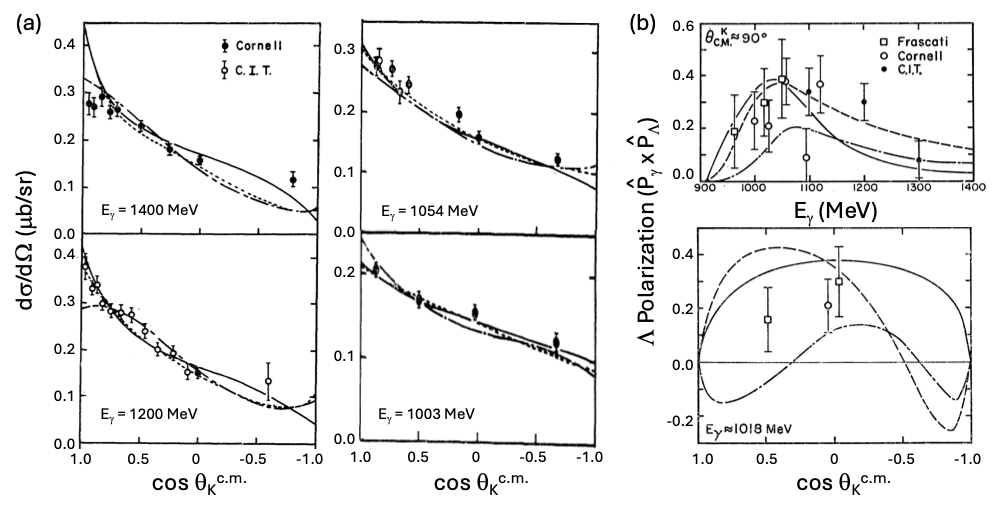}
\caption{(a) Comparison of the angular dependence of the differential cross section data with the fits of Thom. The resonant states in each fit are $P_{1/2}$ (solid curve), $D_{3/2}$ (dashed curve), and $F_{5/2}$ (dash-dotted curve). These fits were referred to as solutions 3, 14a, and 19a, respectively, in Table~I of Ref.~\cite{Thom:1966rm}. Experimental data from Refs.~\cite{Anderson:1962za,Peck:1964zz}. (b) Same as the left panels, but for hyperon polarization. Experimental data from Refs.~\cite{Thom:1963zz,Borgia:1964mza,Grilli:1965jia,Groom:1967zz}. Figures from Ref.~\cite{Thom:1966rm}.}
\label{fig:Thom_Results}
\end{figure}

Figure~\ref{fig:Thom_Results} depicts three of the best fits by Thom--which were referred to as solutions 3, 14a, and 19a in his work--for the differential cross section and hyperon polarization. As seen in Fig.~\ref{fig:Thom_Results}, the three fits reproduced the differential cross section data fairly well. We note that Thom found that even Born amplitudes alone already produced relatively good fits to the data (see Fig.~2 of Ref.~\cite{Thom:1966rm}), indicating that the fits were rather insensitive to the choice of resonant amplitudes.

%%%%%%%%%%%%%%%%%%%%%%%%%%%%%%%%%%%%%%%%%%%%%%%%%%%%%%%%%%%%%%%%%%%%%%%%%%%%%%%%%%%%%%%%%%%%%%%%%%%%%%%%%%%%%%%%%%%%%%%%%%%%%%%%%%%%%%%%%%%%%%%%%%%%%%%%%%%%%%%
%%%%%%%%%%%%%%%%%%%%%%%%%%%%%%%%%%%%%%%%%%%%%%%%%%%%%%%%%%%%%%%%%%%%%%%%%%%%%%%%%%%%%%%%%%%%%%%%%%%%%%%%%%%%%%%%%%%%%%%%%%%%%%%%%%%%%%%%%%%%%%%%%%%%%%%%%%%%%%%

\begin{figure}[hbt!]
    \centering
    \includegraphics[scale=0.6]{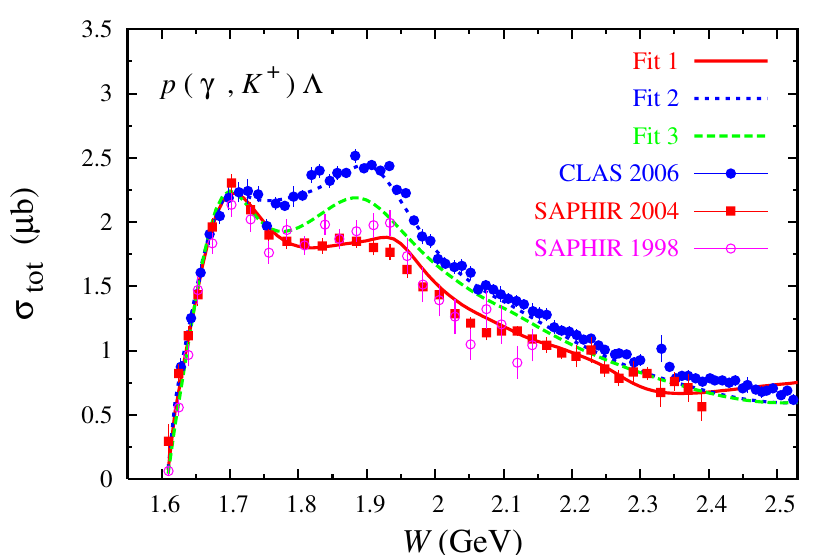}
    \caption{Comparison between the fits and total cross sections of the $\gamma p \to K^+ \Lambda$ data. We note that the data points listed in the figure were not used in the fitting process. Figure from Ref.~\cite{Mart:2006dk}.}
    \label{fig:Mart06_CS}
\end{figure}

Another partial wave analysis model describing the $\gamma p \to K^+\Lambda$ reaction was developed decades after the work of Thom by Mart and Sulaksono \cite{Mart:2006dk}. In contrast to Thom's work, Mart and Sulaksono included multiple nucleon resonances with spins up to 9/2 in their model (see Table~I of Ref.~\cite{Mart:2006dk}). They developed a total of three fits, each fitted to a different dataset (see Table~III in Ref.~\cite{Mart:2006dk}). All fits used dipole hadronic form factors and implemented Davidson and Workman's prescription given in Eq.~(\ref{eq:Davidson_Workman_HFF}) to preserve gauge invariance and satisfy crossing symmetry. A comparison of the three fits with the experimental total cross section data is shown in Fig.~\ref{fig:Mart06_CS}. As seen in the figure, Fit 1 (solid curve) and Fit 2 (dotted curve) reproduce the total cross section data well, indicating that the differential cross section data used in these fits are consistent with the corresponding extracted total cross sections. In contrast, Fit 3--which was obtained by fitting the model simultaneously to both the CLAS and SAPHIR data--is not consistent with either dataset. This issue was also addressed in an earlier study by Byd{\v{z}}ovsk{\'y} and Mart \cite{Bydzovsky:2006wy}. We also note that an updated version of Mart and Sulaksono's model was published around a decade later in a subsequent work by Mart and Sakinah \cite{Mart:2017mwj}. The main difference in this newer model is the significantly larger dataset used in the fitting process, approximately four times larger than the dataset used in the original work by Mart and Sulaksono~\cite{Mart:2006dk}. 

In general, the multipole couplings $\bar{A}_{\ell\pm}^R$ depend on the $Q^2$. Indeed, the couplings can be considered constants for a fixed $Q^2$ analysis, e.g. photoproduction reactions where $Q^2=0$. This dependence on $Q^2$ is captured in the electromagnetic form factor $F(Q^2)$. The multipole couplings as a function of $Q^2$ are given by
\begin{equation}
	\bar{A}^R_{\ell\pm}(Q^2)=\bar{A}^R_{\ell\pm}(0)F(Q^2),
\end{equation}
where $F(Q^2)$ is the electromagnetic form factor. A well-known model that considered the couplings as a function of $Q^2$ is the MAID2007 model for pion electroproduction \cite{TIATOR_2006,Tiator:2009mt}. In their work, Tiator et al. used the electromagnetic form factor
\begin{equation}
    F(Q^2)=(1+a_1Q^2+a_2Q^4+a_3Q^8)\exp(-\gamma Q^2),
\end{equation}
where $a_1$, $a_2$, $a_3$, and $\gamma$ are parameters. A similar prescription was adopted by Respati and Mart in their work, where they used the electromagnetic form factor \cite{Respati:2020fpk,Respati20}
\begin{equation}
\label{eq:FF_multipole_Respati}
    F(Q^2)=(1+a_1Q^2)^n\exp(-\gamma Q^2),~~n=1,2,3.
\end{equation}
Respati and Mart found that utilizing the electromagnetic form factor given in Eq.~(\ref{eq:FF_multipole_Respati}) in a partial wave analysis reproduces the CLAS recoil polarization \cite{Respati:2020fpk} and differential cross section data \cite{Respati20} reasonably well for $n=3$ and $n=1$, respectively.

\subsubsection{Covariant Isobar Models}
\label{sec:isobar}

Another way to model the resonant states in kaon photo- and electroproduction on the nucleon, other than treating them as multipoles, is by treating them as effective fields with their own physical properties \cite{Janssen02}. Interactions between the hadronic fields are then represented by effective Lagrangians \cite{Cauteren:2005iya}. This approach has one clear advantage over partial wave analyses in that it preserves Lorentz invariance \cite{Cauteren:2005iya}. This framework is widely known as a covariant isobar model. Because such models involve many features, especially the choice of resonant states, there have been many distinct isobar models from various groups \cite{Okuyama:2023lxq}.

One of the first covariant isobar models is the work of Renard and Renard in the early 1970s \cite{Renard:1971us}, in which they modeled the photoproduction reactions $\gamma  p \to K^+  \Lambda$ and $\gamma p \to K^+ \Sigma^0$. In addition to the Born terms, the model utilized eight isospin-1/2 resonances with spins up to 5/2 for the $s$-channel exchanges: $N(1525)1/2^-$, $N(1715)1/2^-$, $N(1405)1/2^+$, $N(1785)1/2^+$, $N(1855)3/2^+$, $N(1515)3/2^-$, $N(2030)3/2^-$, and $N(1690)5/2^+$. For $K^+\Sigma^0$ photoproduction, the model also took account of several isospin-3/2 resonances in addition to the isospin-1/2 resonances. The isospin-3/2 resonances used in this model were $\Delta(1630)1/2^-$, $\Delta(1905)1/2^+$, $\Delta(1236)3/2^+$, $\Delta(1690)3/2^+$, $\Delta(1670)3/2^-$, and $\Delta(1880)5/2^+$. The resonance widths were taken to be energy-dependent to partially account for unitarity. Although the model also included the $\Lambda(1405)1/2^-$, $\Sigma(1385)3/2^+$, and $\Lambda(1520)3/2^-$ states for the $u$-channel contributions, it did not consider the exchange of $K(494)$ and $K^*(892)$ in the $t$-channel. The reason is that in their work, Renard and Renard argued that the simultaneous inclusion of $s$-channel and $t$-channel exchanges does not satisfy duality constraints. Thus, the $t$-channel exchange in this model is simulated by a gauge-invariant constraint. 

\begin{figure}[hbt!]
    \centering
%    \includegraphics[scale=0.8]{Renard_DCS_KL_30.pdf}
%    \hspace{1.5cm}
%    \includegraphics{Renard_DCS_KS_95.pdf}
    \includegraphics[width=0.6\textwidth]{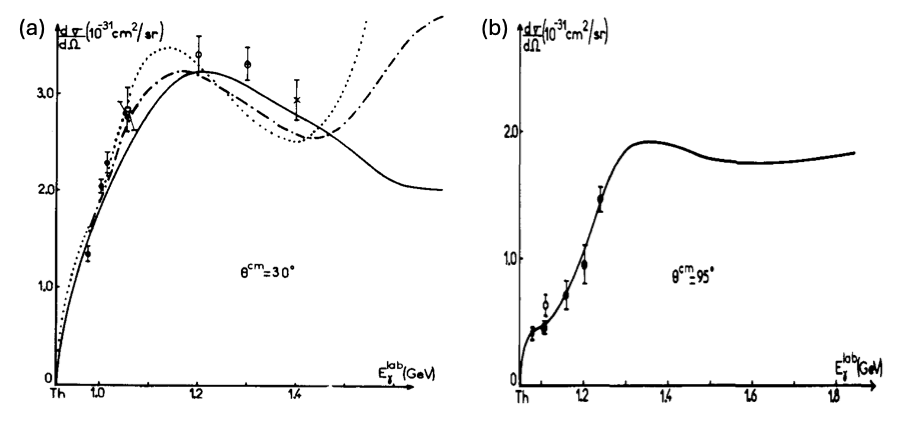}
    \caption{(a) Results of the Renard-Renard model for the differential cross section of the $\gamma p \to K^+ \Lambda$ reaction as a function of beam energy for kaon angle $\theta_K^{\mathrm{c.m.}}=30\degree$. Values of the $g_{KYN}$ coupling constants for each curve are $g_{K\Lambda N}=-8.5$ and $g_{K\Sigma N}=1.3$ (solid curve); $g_{K\Lambda N}=-10.0$ and $g_{K\Sigma N}=1.95$ (dashed curve); $g_{K\Lambda N}=-12.7$ and $g_{K\Sigma N}=1.95$ (dash-dotted curve); $g_{K\Lambda N}=14.0$ and $g_{K\Sigma N}=1.95$ (dotted curve). (b) Same as the left panel, but for the $\gamma p \to K^+ \Sigma^0$ reaction and kaon angle $\theta_K^{\mathrm{c.m.}}=95\degree$. The values of the $g_{KYN}$ coupling constants used in the model are $g_{K\Lambda N}=-12.7$ and $g_{K\Sigma N}=1.95$. Experimental data from Ref.~\cite{Decamp:1970zz}. Figures from Ref.~\cite{Renard:1971us}. Reprinted with permission from Elsevier.}
    \label{fig:Renard_DCS_KL_30}
\end{figure}

For the fitting process, Renard and Renard used the initial values of $-14$ and $1.95$ for the $g_{K\Lambda N}$ and $g_{K\Sigma N}$ coupling constants, respectively. These values were taken from a previous analysis by Kim, which utilized $KN$ forward-dispersion relations to calculate the coupling constants \cite{Kim:1967zzc}. It was found that a good fit could be made when the $g_{KYN}$ coupling constants satisfy the interval $-12.7\lesssim g_{K\Lambda N}\lesssim -8.5$ and $1.3\lesssim g_{K\Sigma N}\lesssim 2.0$. We also note that they did not use hadronic form factors in their model to reduce the background contributions to the cross section. They thus relied solely on the interplay of $s$- and $u$-channel resonances to reduce the contributions of the background terms and form the structures seen in the experimental data. Results of the Renard-Renard model for $K^+\Lambda$ and $K^+\Sigma^0$ photoproduction are shown in Fig.~\ref{fig:Renard_DCS_KL_30}.

%%%%%%%%%%%%%%%%%%%%%%%%%%%%%%%%%%%%%%%%%%%%%%%%%%%%%%%%%%%%%%%%%%%%%%%%%%%%%%%%%%%%%%%%%%%%%%%%%%%%%%%%%%%%%%%%%%%%%%%%%%%%%%%%%%%%%%%%%%%%%%%%%%%%%%%%%%%%%%%
%%%%%%%%%%%%%%%%%%%%%%%%%%%%%%%%%%%%%%%%%%%%%%%%%%%%%%%%%%%%%%%%%%%%%%%%%%%%%%%%%%%%%%%%%%%%%%%%%%%%%%%%%%%%%%%%%%%%%%%%%%%%%%%%%%%%%%%%%%%%%%%%%%%%%%%%%%%%%%%

In the early 1990s, another covariant isobar model to simultaneously describe both the $\gamma p \to K^+\Lambda$ and $K^-p\to \gamma \Lambda$ reactions was developed by Williams, Ji, and Cotanch \cite{Williams:1990hh}. The model is relatively simple as they only included three nucleon and hyperon resonances in their model on top of the Born and extended Born terms. All nucleon and hyperon resonances considered in their work are exclusively spin-1/2 states, namely $N(1440)1/2^+$, $N(1650)1/2^-$, $N(1710)1/2^+$, $\Lambda(1405)1/2^-$, $\Lambda(1670)1/2^-$, and $\Lambda(1800)1/2^-$. 

\begin{figure}[hbt!]
\centering
\includegraphics[width=0.7\textwidth]{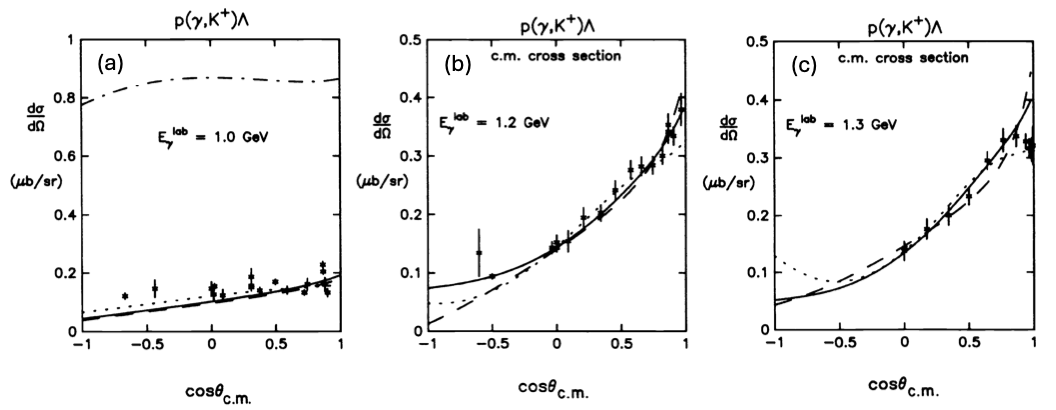}
\caption{Predictions of the WJC model for the angular dependence of the differential cross section of the $\gamma p \to K^+\Lambda$ reaction for several values of beam energy. The C1, C2, and PR4b parameter sets are represented by the solid, dashed, and dotted curves, respectively. Figures from Ref.~\cite{Williams:1990hh}.}
    \label{fig:WJC90_DCS}
\end{figure}

For the sake of comparison, they also investigated four other models from Refs.~\cite{Thom:1966rm}, \cite{Adelseck:1985scp}, \cite{Adelseck:1988yv}, \cite{Bennhold:1989bw}, and \cite{Workman:1987ka}. They constructed a total of five models, named M1 through M5. Some of the models were evaluated with two sets of parameters sourced from different works. A summary of the models and the sources of the parameter sets is given in Ref.~\cite{Williams:1990hh}. We note that model M5 is the original work of Williams, Ji, and Cotanch, and thus the parameter set is derived from the fitting process. Results of the PR4b, C1, and C2 parameterizations are depicted in Fig.~\ref{fig:WJC90_DCS}. As a comparison, the M4 model with the PR5 parameter set is also plotted (dash-dotted curve). A notable feature of the PR5 parameter set is that the prediction far exceeded the data. This is mainly attributed to the fact that the PR5 parameter set was obtained from a radiative capture analysis, whereas the other sets were obtained from fitting the models to the photoproduction data.

%%%%%%%%%%%%%%%%%%%%%%%%%%%%%%%%%%%%%%%%%%%%%%%%%%%%%%%%%%%%%%%%%%%%%%%%%%%%%%%%%%%%%%%%%%%%%%%%%%%%%%%%%%%%%%%%%%%%%%%%%%%%%%%%%%%%%%%%%%%%%%%%%%%%%%%%%%%%%%%
%%%%%%%%%%%%%%%%%%%%%%%%%%%%%%%%%%%%%%%%%%%%%%%%%%%%%%%%%%%%%%%%%%%%%%%%%%%%%%%%%%%%%%%%%%%%%%%%%%%%%%%%%%%%%%%%%%%%%%%%%%%%%%%%%%%%%%%%%%%%%%%%%%%%%%%%%%%%%%%

\begin{figure}[hbt!]
\centering
    \includegraphics[width=0.7\textwidth]{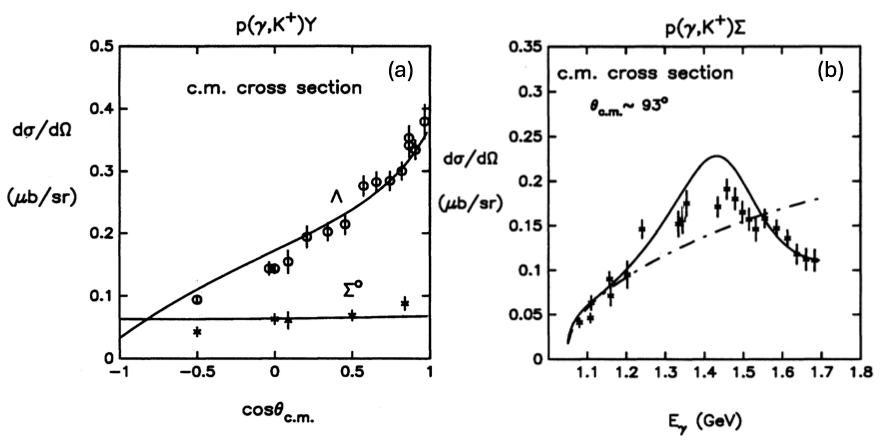}
    \caption{(a) Results of the WJC91 model for the angular distribution of the differential cross section of $K^+\Lambda$ photoproduction at photon energy $E_\gamma^{\mathrm{lab}}=1.2~\mathrm{GeV}$ and $K^+\Sigma^0$ photoproduction at photon energy $E_\gamma^{\mathrm{lab}}=1.1~\mathrm{GeV}$. (b) Results of the WJC91 model for the energy dependence of the differential cross section of $K^+\Sigma^0$ photoproduction at kaon angle $\theta_K^{\mathrm{c.m.}}=93\degree$. The solid and dash-dotted lines represent the full model and the model excluding $\Delta$ resonances, respectively. Figures from Ref.~\cite{Williams:1991tw}.}
    \label{fig:WJC91}
\end{figure}

Soon afterward, Williams, Ji, and Cotanch extended their model (hereinafter referred to as the WJC91 model) in a subsequent work \cite{Williams:1991tw} to describe both kaon photoproduction $\gamma p \to K^+Y$ and kaon radiative capture $K^-p\to \gamma Y$ for $Y=\Lambda,~\Sigma^0,~\Lambda(1405)$ simultaneously. In addition to enforcing crossing relations on the model, they also further constrained the model by imposing a duality argument, similar to the work of Renard and Renard, which also imposed duality constraints \cite{Renard:1971us}. Hence, in this new model, they did not include the exchange of kaon resonances in the $t$-channel. The resonant states in this model were kept to a minimum. In the $s$-channel, only the $N(1650)1/2^-$ and $N(1710)1/2^+$ states were included. For $K^+\Sigma^0$ photoproduction, the $\Delta(1620)1/2^-$ and $\Delta(1910)1/2^+$ were also included. As for hyperon resonances, only the $\Lambda(1405)1/2^-$ was included as a result of that particular resonance dominating the radiative capture width. Results of the WJC91 model are shown in Fig.~\ref{fig:WJC91}. We note that in Fig.~\ref{fig:WJC91}(b), the inclusion of the $\Delta(1620)1/2^-$ and $\Delta(1910)1/2^+$ significantly improved the fit (solid curve). The inclusion of these resonances also formed a structure around $E_\gamma^{\mathrm{lab}}=1.4~\mathrm{GeV}$ or $W\approx 1.9~\mathrm{GeV}$, signifying an important resonance contribution near that energy.

%%%%%%%%%%%%%%%%%%%%%%%%%%%%%%%%%%%%%%%%%%%%%%%%%%%%%%%%%%%%%%%%%%%%%%%%%%%%%%%%%%%%%%%%%%%%%%%%%%%%%%%%%%%%%%%%%%%%%%%%%%%%%%%%%%%%%%%%%%%%%%%%%%%%%%%%%%%%%%%
%%%%%%%%%%%%%%%%%%%%%%%%%%%%%%%%%%%%%%%%%%%%%%%%%%%%%%%%%%%%%%%%%%%%%%%%%%%%%%%%%%%%%%%%%%%%%%%%%%%%%%%%%%%%%%%%%%%%%%%%%%%%%%%%%%%%%%%%%%%%%%%%%%%%%%%%%%%%%%%

Williams, Ji, and Cotanch extended their model (referred to as the WJC92 model in the following) once more not long after to also describe the kaon electroproduction reaction $ep\to e'K^+Y$ for $Y=\Lambda,~\Sigma^0,~\Lambda(1405)$ \cite{Williams:1992tp}. They kept the nucleon and hyperon resonances in the WJC92 model identical to those included in the WJC91 model. However, an additional $\Delta(1900)1/2^-$ resonance was also included in addition to the $\Delta(1620)1/2^-$ and $\Delta(1910)1/2^+$ included in the WJC91 model. Moreover, they also opted to include the $t$-channel kaon resonances meant to emulate the omitted high-spin baryon resonances in the WJC91 model to produce an overall improved behavior for a wider energy range. For the electromagnetic form factors of the particles, they used the Extended Vector Meson Dominance (EVMD) model proposed by Gari and Kr\"{u}mpelmann \cite{Gari:1984ia,Gari:1986rj,Gari:1992tq}. They reasoned that this particular choice of electromagnetic form factor was made because the formulas derived from EVMD explicitly satisfy crossing symmetry, making them valid for space-like and time-like photon momentum transfers \cite{Williams:1992tp,Cotanch:1993ry}. They found that the couplings of the nucleon and hyperon resonances were relatively stable with respect to the inclusion of kaon resonance exchange. For an explicit comparison of the extracted coupling constants, see Table II of Ref.~\cite{Williams:1992tp}.

Another notable covariant isobar model, coined the Saclay-Lyon (SL) model, was developed by David et al. in the late 1990s \cite{David:1995pi}. This model was developed to describe all kaon photo- and electroproduction reaction involving a proton target. Like the Williams-Ji-Cotanch model, David et al. also took the crossing symmetry between kaon photoproduction and kaon radiative capture into account. For the branching ratio of kaon radiative capture, David et al. used the value obtained by Whitehouse et al. \cite{Whitehouse:1989yi}, based on a measurement performed at Brookhaven National Laboratory (BNL).

The model developed by David et al. was an extension of previous work by Adelseck and Saghai \cite{Adelseck:1990ch}, who analyzed the $K^+\Lambda$ photoproduction process from threshold to 1.4 GeV. The main improvements introduced, in addition to accounting for the radiative capture reactions, were the inclusion of spin-3/2 and 5/2 nucleon resonances and the extension of the energy range from $E_\gamma^{\mathrm{lab}}\leq 1.5~\mathrm{GeV}$ to $E_\gamma^{\mathrm{lab}}\leq 2.1~\mathrm{GeV}$. In total, David et al. utilized five nucleon resonances and four hyperon resonances for the $K\Lambda$ channel, and an additional five $\Delta$ resonances for the $K\Sigma$ channels. See Tables~I, VIII, and IX of Ref.~\cite{David:1995pi} for the complete list of the resonances included in the models.

\begin{figure}[hbt!]
    \centering
    \includegraphics[width=0.6\textwidth,height=0.5\textwidth]{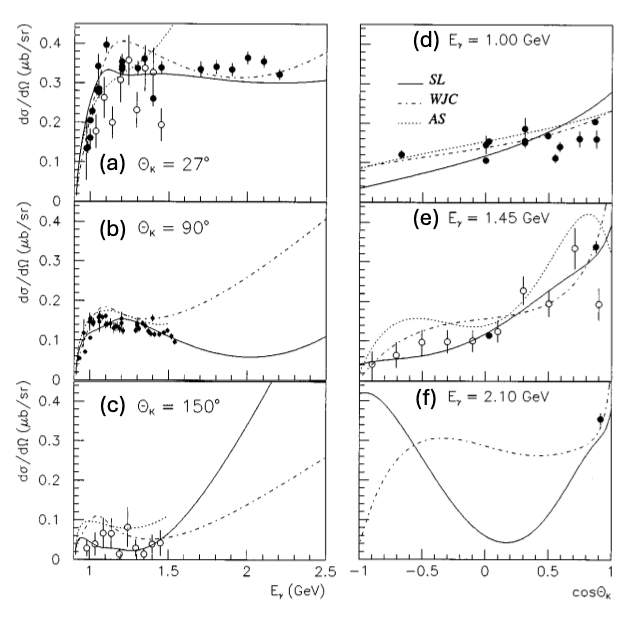}
    \caption{Prediction of the SL model for the differential cross section of the $\gamma p \to K^+\Lambda$ reaction as a function of photon energy $E_\gamma^{\rm lab}$ and kaon angle $\theta_K^{\mathrm{c.m.}}$. Panels (a)-(c) show the differential cross section at $\theta_K^{\mathrm{c.m.}}=27\degree$, $90\degree$, and $150\degree$, respectively, while panels (d)-(f) show the angular dependence of the differential cross section at $E_\gamma^{\rm lab}=1.0~\mathrm{GeV}$, $1.45~\mathrm{GeV}$, and $2.1~\mathrm{GeV}$. The solid, dash-dotted, and dotted curves represent the SL, WJC \cite{Williams:1992tp}, and AS \cite{Adelseck:1990ch,Adelseck:1992ua} models, respectively. Note that the upper limit of the AS model was $E_\gamma^{\rm lab} = 1.5~\mathrm{GeV}$ and thus, the curves are plotted only up to that energy. The experimental data are from Ref.~\cite{Bockhorst:1994jf} (empty circles) and Refs.~\cite{Donoho58,McDaniel:1959zz,Brody:1960zz,Anderson:1962za,Thom:1963zz,Peck:1964zz,Thom:1966rm,Groom:1967zz,Bleckmann:1970kb,Goeing:1971nb,Fujii:1970gn,Feller:1972ph,ABBHHM:1969pjo} (solid circles). Figures from Ref.~\cite{David:1995pi}.}
    \label{fig:David_KL CS}
\end{figure}

As noted above, the SL model was developed to describe all the kaon photo- and electroproduction reactions involving a proton target. For the photoproduction case, this model was used to describe the $\gamma p \to K^+\Lambda$, $\gamma p \to K^+\Sigma^0$, and $\gamma p \to K^0\Sigma^+$ reactions. As a comparison, David et al. compared the results of the SL model with other existing models at the time such as Adelseck-Saghai (AS) \cite{Adelseck:1990ch}, Williams-Ji-Cotanch (WJC) \cite{Williams:1992tp}, and Mart-Bennhold-Hyde (MBH) \cite{Mart:1995wu}. The SL model result for the differential cross section of the $\gamma p \to K^+\Lambda$ reaction is shown in Fig.~\ref{fig:David_KL CS}, where the SL model is seen to have the best agreement with the data compared to the other models.

%%%%%%%%%%%%%%%%%%%%%%%%%%%%%%%%%%%%%%%%%%%%%%%%%%%%%%%%%%%%%%%%%%%%%%%%%%%%%%%%%%%%%%%%%%%%%%%%%%%%%%%%%%%%%%%%%%%%%%%%%%%%%%%%%%%%%%%%%%%%%%%%%%%%%%%%%%%%%%%
%%%%%%%%%%%%%%%%%%%%%%%%%%%%%%%%%%%%%%%%%%%%%%%%%%%%%%%%%%%%%%%%%%%%%%%%%%%%%%%%%%%%%%%%%%%%%%%%%%%%%%%%%%%%%%%%%%%%%%%%%%%%%%%%%%%%%%%%%%%%%%%%%%%%%%%%%%%%%%%

Results of the SL model for the single polarization observables of the $\gamma p \to K^+\Lambda$ reaction can be seen in Fig.~6 of Ref.~\cite{David:1995pi}. Overall, the SL model reproduced the experimental data of the single polarization observables fairly well. Although there were no experimental data for the linearly polarized beam asymmetry $\Sigma$ at that time, we can still infer from Fig.~6(c) of Ref.~\cite{David:1995pi} that the linearly polarized beam asymmetry $\Sigma$ shows high sensitivity to the various reaction mechanisms in the models.

For the electroproduction reactions, David et al. mostly relied on various Vector Meson Dominance (VMD) models for the electromagnetic form factors of the particles. For the nucleon and hyperon resonances, they used the EVMD models of Gari and Kr\"{u}mpelmann that were published in 1985 \cite{Gari:1984ia,Gari:1986rj} and 1992 \cite{Gari:1992tq}. As for the kaon and kaon resonances, David et al. considered several proposed VMD models. For the $K^+$, they adopted the VMD model proposed by Williams, Ji, and Cotanch, as well as the Relativistic Constituent Quark Model (RCQM) introduced by Cardarelli et al~\cite{Cardarelli:1994ix}. For the kaon resonances $K_1$ and $K^*$, they considered the simplified VMD model of Adelseck and Wright \cite{Adelseck:1988yv} and the EVMD model of Williams, Ji, and Cotanch \cite{Williams:1992tp}.

\begin{figure}
    \centering
    \includegraphics[width=0.7\textwidth]{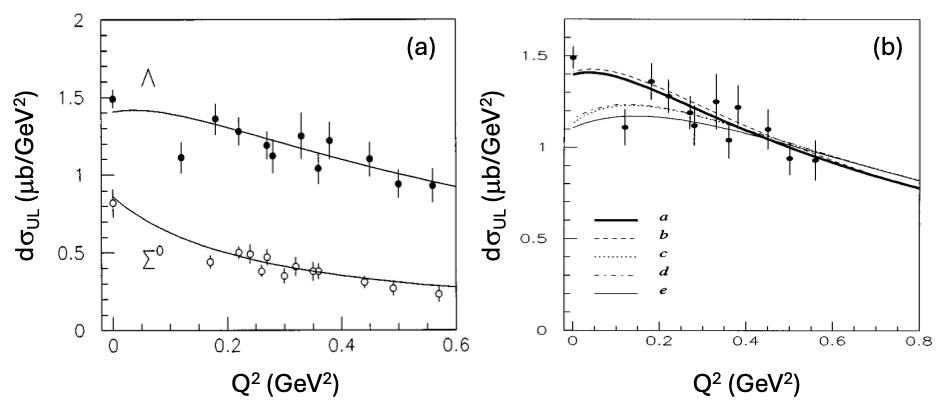}
    \caption{(a) Prediction of the SL model for the structure function $d\sigma_{\rm UL}=d\sigma_{\rm U}+\epsilon_{\rm L}d\sigma_{\rm L}$ (i.e., $\sigma_{\rm T} + \epsilon_{\rm L} \sigma_{\rm L}$ in the notation of Section~\ref{theory-formalism}) for the $ep\to e'K^+\Lambda$ and $ep\to e'K^+\Sigma^0$ reactions as a function of $Q^2$ at $\epsilon=0.72$, $s=5.02~\mathrm{GeV}^2$, and $t=-0.15~\mathrm{GeV}^2$. (b) Same as the left panel, but only for the $ep\to e'K^+\Lambda$ reaction. The five curves correspond to the SL model with different prescriptions for the electromagnetic form factors. The combinations of the electromagnetic form factors are as follows: (a) GK92-WAA, (b) GK92-CAA, (c) GK92-WAW, (d) GK92-WWW, and (e) GK85-WWW. Experimental data from Refs.~\cite{Azemoon:1974dt,Brown:1972pf,Bebek:1976qg,Brauel:1979zk,Bebek:1977bv}. Figures from Ref.~\cite{David:1995pi}.}
    \label{fig:David_Electro}
\end{figure}

Figure~\ref{fig:David_Electro} shows predictions of the SL model for the structure function $d\sigma_{\rm UL}=d\sigma_{\rm U}+\epsilon_{\rm L} d\sigma_{\rm L}$ of the $ep\to e'K^+\Lambda$ and $ep\to e'K^+\Sigma^0$ reactions (see Eq.~(2.24) of Ref.~\cite{David:1995pi} for the definition of the observable). As seen in the figure, the SL model reproduced the experimental data fairly well for both reactions. 

David et al. also investigated the effect of various electromagnetic form factor prescriptions. They considered only the $ep\to e'K^+\Lambda$ reaction for this investigation as it has a simpler reaction mechanism compared to the $K\Sigma$ channel. In total, they constructed five different combinations of electromagnetic form factors, labeled $U-VXY$ in their study, where $U$, $V$, $X$, and $Y$ refers to the choice of electromagnetic form factors of the baryons, kaons, $K^*$, and $K_1$, respectively. The five combinations are GK92-WAA, GK92-CAA, GK92-WAW, GK92-WWW, and GK85-WWW, where W, A, and C denote the parameterizations of Williams \cite{Williams:1992tp}, Adelseck and Saghai \cite{Adelseck:1988yv}, and Cardarelli et al. \cite{Cardarelli:1995hn}, respectively. The results of applying these various combinations are presented in Fig.~\ref{fig:David_Electro}(b), which shows that the structure function $d\sigma_{\rm UL}$ is not significantly sensitive to the various electromagnetic form factors combinations. 

%%%%%%%%%%%%%%%%%%%%%%%%%%%%%%%%%%%%%%%%%%%%%%%%%%%%%%%%%%%%%%%%%%%%%%%%%%%%%%%%%%%%%%%%%%%%%%%%%%%%%%%%%%%%%%%%%%%%%%%%%%%%%%%%%%%%%%%%%%%%%%%%%%%%%%%%%%%%%%%
%%%%%%%%%%%%%%%%%%%%%%%%%%%%%%%%%%%%%%%%%%%%%%%%%%%%%%%%%%%%%%%%%%%%%%%%%%%%%%%%%%%%%%%%%%%%%%%%%%%%%%%%%%%%%%%%%%%%%%%%%%%%%%%%%%%%%%%%%%%%%%%%%%%%%%%%%%%%%%%

Shortly after the publication of the SL model, Mizutani et al. reported an improvement to the model~\cite{Mizutani:1997sd} by taking into account the so-called off-shell effects, which are associated with fermions of spins $\geq 3/2$. It is important to note that Mizutani et al. limited their work to the $K\Lambda$ channels while David et al. considered both the $K\Lambda$ and $K\Sigma$ channels. Mizutani et al. followed the procedure of Benmerrouche et al. in treating a spin-3/2 baryon resonance, which was originally intended for pion and eta photoproduction \cite{Benmerrouche:1989uc}. The main idea came from the different prescription of the effective Lagrangian of spin-3/2 resonances in both the $s$- and $u$-channels. See Eqs.~(2.1)-(2.3) of Ref.~\cite{Mizutani:1997sd} for the effective Lagrangians. These Lagrangians contain extra parameters, often referred to as off-shell parameters (OSP). These OSP were treated as free parameters in the fitting process. For a more comprehensive discussion on off-shell parameters, we refer the reader to Ref.~\cite{Benmerrouche:1989uc}.

A total of four models were developed by Mizutani et al. as a comparison to the SL model developed by David et al. \cite{David:1995pi}. These models were labeled Model A, B, C, and D, respectively. Model A was obtained by removing two nucleon resonances from the SL model, namely the $N(1440)1/2^+$ and $N(1675)5/2^-$, in search of a simpler model. The removal of the $N(1440)1/2^+$ was based on the work of David et al. \cite{David:1995pi}, which found that its coupling constant was effectively zero (see Table~VIII of Ref.~\cite{David:1995pi} and Table~I of Ref.~\cite{Mizutani:1997sd}). This conclusion was supported by the study of Saghai and Tabakin \cite{Saghai:1996hn}, which argued that the dataset available at that time did not require contributions from $P_{11}$ resonances. As for the $N(1675)5/2^-$ state, David et al. found that its contribution to the dynamics of both the $K\Lambda$ and $K\Sigma$ channels is not pivotal (see Table~XI of Ref.~\cite{David:1995pi}). We note that off-shell effects were not taken into account in Model A.

\begin{figure}[hbt!]
    \centering
%    \includegraphics{Mizutani_KL CS.pdf}
%    \hspace{0.5cm}
%    \includegraphics[scale=0.9]{Mizutani_KL Electro.pdf}
    \includegraphics[width=0.8\textwidth]{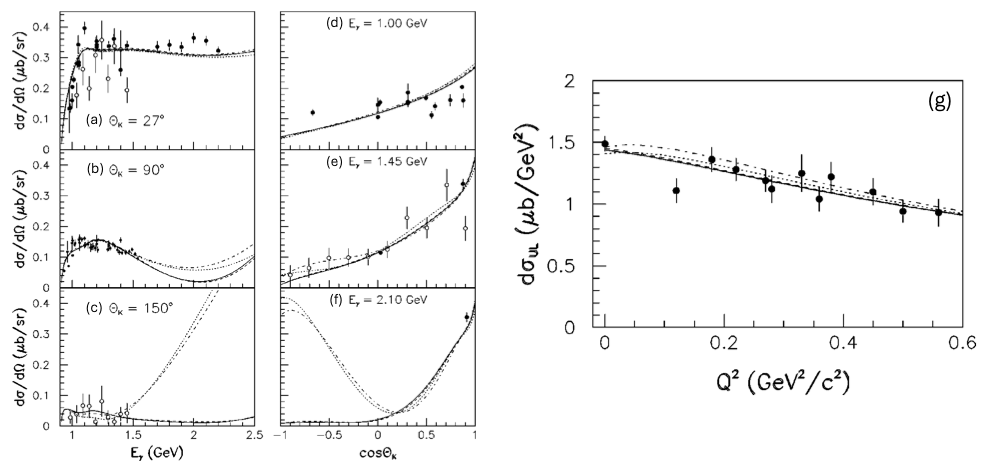}
    \caption{Comparison between the differential cross section data for the $\gamma p \to K^+\Lambda$ reaction and the fits by Mizutani et al. (a)-(f). Dotted, dash-dotted, solid, and dashed curves depict the SL model \cite{David:1995pi}, model A, model B, and model C, respectively. Experimental data are from Ref.~\cite{Bockhorst:1994jf} (empty circles) and Refs.~\cite{Donoho58,Groom:1967zz,McDaniel:1959zz,Anderson:1962za,ABBHHM:1969pjo,Thom:1963zz,Thom:1966rm,Peck:1964zz,Brody:1960zz,Bleckmann:1970kb,Goeing:1971nb,Fujii:1970gn,Feller:1972ph} (solid circles). (g) The structure function $d\sigma_{\rm UL}\equiv d\sigma_{\rm U}+\epsilon_{\rm L} d\sigma_{\rm L}$ data for the $ep\to e'K^+\Lambda$ reaction as a function of $Q^2$ at $\epsilon=0.72$, $s=5.02~\mathrm{GeV}^2$, and $t=-0.15~\mathrm{GeV}^2$. Curves as in the left figure. Experimental data from Refs.~\cite{Azemoon:1974dt,Brown:1972pf,Bebek:1976qg,Brauel:1979zk,Bebek:1977bv}. Figures from Ref.~\cite{Mizutani:1997sd}.}
    \label{fig:Mizutani98_CS}
\end{figure}

Models B and C were obtained by introducing off-shell effects into Model A. Specifically, Model B was obtained by introducing off-shell treatments to the $N(1720)3/2^+$ as the only spin-3/2 nucleon resonance included, while Model~C also took into account the spin-3/2 $\Lambda(1890)3/2^+$ resonance with its off-shell effects. Mizutani et al. found that Model~C had a higher value of $\chi^2/N$ compared to Model B. This was an indication that the addition of the $\Lambda(1890)3/2^+$ was not relevant to the underlying dynamics. See Fig.~\ref{fig:Mizutani98_CS} for comparisons to the data.

Model D was derived from Model B. The only difference in Model D is that the two off-shell parameters $X$ and $Z$ were set as fixed values of $X=-0.5$ and $Z=0$. This was done in an effort to eliminate the undesirable growth of the invariant amplitude as a function of the Mandelstam variable $s$ when $X\neq -0.5$ and $Z\neq 0$. For explicit expressions of the invariant amplitudes in the work of Mizutani et al., see Appendix~B of Ref.~\cite{Mizutani:1997sd}.

Figure~\ref{fig:Mizutani98_CS}(c) shows the distinction between the models that do not incorporate off-shell parameters (SL model and model A) and those that used them (models B and C). Here it is seen that the SL model and model A significantly overpredict the experimental data compared to models B and C. This behavior is also evident in Fig.~\ref{fig:Mizutani98_CS}(f), which shows that the off-shell treatments in models B and C give significantly lower predictions for backward angles.

%%%%%%%%%%%%%%%%%%%%%%%%%%%%%%%%%%%%%%%%%%%%%%%%%%%%%%%%%%%%%%%%%%%%%%%%%%%%%%%%%%%%%%%%%%%%%%%%%%%%%%%%%%%%%%%%%%%%%%%%%%%%%%%%%%%%%%%%%%%%%%%%%%%%%%%%%%%%%%%
%%%%%%%%%%%%%%%%%%%%%%%%%%%%%%%%%%%%%%%%%%%%%%%%%%%%%%%%%%%%%%%%%%%%%%%%%%%%%%%%%%%%%%%%%%%%%%%%%%%%%%%%%%%%%%%%%%%%%%%%%%%%%%%%%%%%%%%%%%%%%%%%%%%%%%%%%%%%%%%

Around the same time that Mizutani et al. developed their models, Mart and Bennhold investigated a resonance structure around $W=1.9~\mathrm{GeV}$, revealed by the new SAPHIR data \cite{SAPHIR:1998fev} in the $\gamma p \to K^+\Lambda$ reaction using a covariant isobar model \cite{Mart:1999ed}. They claimed the structure could be reproduced by including a $D_{13}$ $s$-channel resonance at 1895~MeV. However, there were no three- or four-star isospin-1/2 resonances with mass around 1900 MeV recorded at that time in the Review of Particle Physics. Thus, Mart and Bennhold utilized the constituent quark model of Capstick \cite{Capstick:1992uc} that predicted a number of states around 1900~MeV. One of the predicted states is the $N(1960)3/2^-$, which has a large decay width to the $K\Lambda$ channel and also significant photocouplings \cite{Capstick:1998uh}.

\begin{figure}[hbt!]
	\centering
	\includegraphics[scale=0.4]{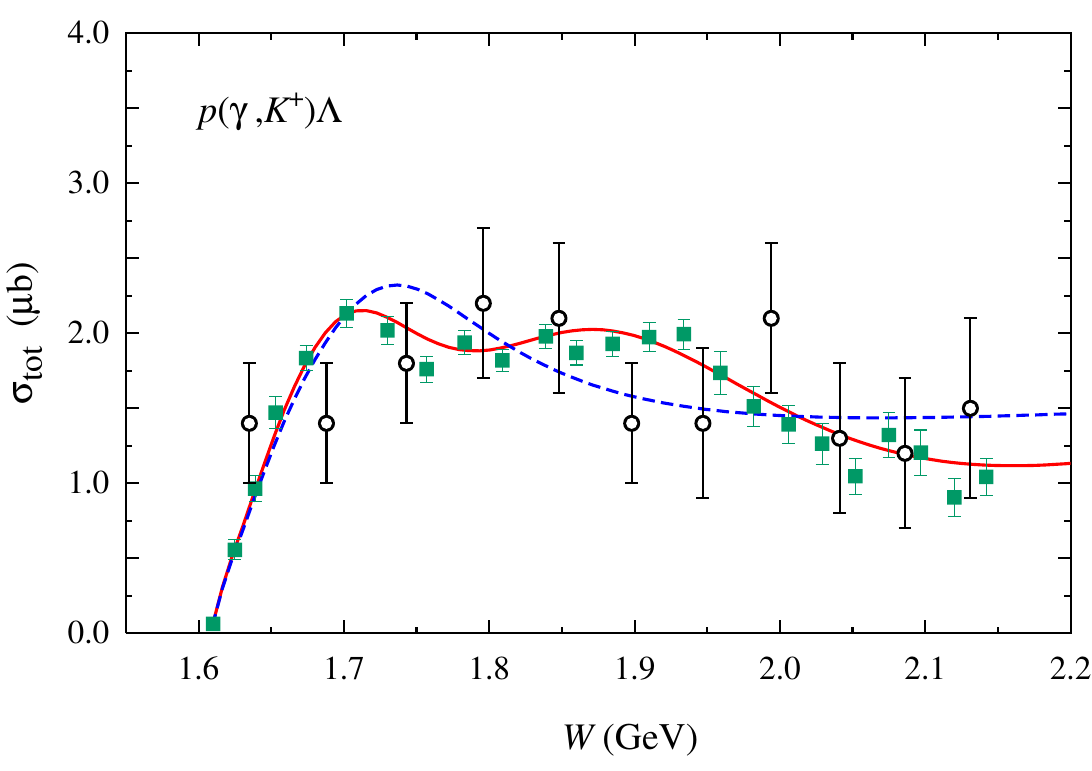}
	\caption{Total cross section data for $K^+\Lambda$ photoproduction on the proton as a function of the c.m. energy $W$. The solid green squares are the SAPHIR data \cite{SAPHIR:1998fev}, while the open circles show older Aachen-Berlin-Bonn-Hamburg-Heidelberg-M\"{u}nchen Collaboration data \cite{ABBHHM:1969pjo}. The dashed blue line displays the model without the $N(1960)3/2^-$. The inclusion of the $N(1960)3/2^-$ leads to the model represented by the solid red line. It is evident that a significant improvement has been achieved after the inclusion of the $N(1960)3/2^-$. Figure from Ref.~\cite{Mart:1999ed}, with color added for clarity.}
    \label{fig:mart99_SAPHIR} 
\end{figure}

The model developed by Mart and Bennhold, which was later named the Kaon-MAID (KM) model, is relatively simple. In addition to the usual Born terms and the $K^*(892)$ and $K_1(1270)$ resonances in the $t$-channel, the model includes only three nucleon resonances, namely the $N(1650)1/2^-$, $N(1710)1/2^+$, and $N(1720)3/2^+$ \cite{Mart:1999ed}. This particular set of resonances was motivated by a previous coupled-channel analysis by Feuster and Mosel \cite{Feuster:1998cj}. Another notable feature of the model is that the resonance widths were taken to be energy-dependent to account approximately for unitarity constraints~\cite{Mart:1999ed}. A further point of interest is that the KM model was among the first covariant isobar models to utilize hadronic form factors to suppress the contribution of the Born terms. Mart and Bennhold followed Haberzettl's gauge prescription, as given in Eq.~(\ref{eq:Fhat_haberzettl}), to include the hadronic form factors in a gauge-invariant fashion. Figure~\ref{fig:mart99_SAPHIR} compares the prediction of the model before (dashed curve) and after (solid curve) the inclusion of the $N(1960)3/2^-$ state. We also note that the KM model is available online via Ref.~\cite{kaonmaid}.

%%%%%%%%%%%%%%%%%%%%%%%%%%%%%%%%%%%%%%%%%%%%%%%%%%%%%%%%%%%%%%%%%%%%%%%%%%%%%%%%%%%%%%%%%%%%%%%%%%%%%%%%%%%%%%%%%%%%%%%%%%%%%%%%%%%%%%%%%%%%%%%%%%%%%%%%%%%%%%%
%%%%%%%%%%%%%%%%%%%%%%%%%%%%%%%%%%%%%%%%%%%%%%%%%%%%%%%%%%%%%%%%%%%%%%%%%%%%%%%%%%%%%%%%%%%%%%%%%%%%%%%%%%%%%%%%%%%%%%%%%%%%%%%%%%%%%%%%%%%%%%%%%%%%%%%%%%%%%%%

Briefly after the publication of the KM model, an isobaric model was also developed for $K\Sigma$ photoproduction \cite{Mart:2000jv,Lee:1999kd}. This model shared the same background, $t$-channel kaon resonances, and nucleon resonances as the model for $K^+\Lambda$ photoproduction with the exception of the $N(1960)3/2^-$ state \cite{Mart:2000jv}. However, since the conservation of isospin allows $\Delta$ resonances to contribute to $K\Sigma$ photoproduction, the model included two spin-1/2 $\Delta$ resonances, namely the $\Delta(1900)1/2^-$ and $\Delta(1910)1/2^+$ \cite{Mart:2000jv,Lee:1999kd}. An overview of the model predictions for the total cross sections of all six kaon photoproduction isospin channels as a function of the total c.m. energy $W$ is shown in Fig.~\ref{fig:GDH-Sumowidagdo}.

%\begin{figure}[hbt!]
%	\centering
%    \includegraphics[scale=0.6]{CS_All Channel.eps}
%    \vspace{-6mm}
%    \caption{Total cross sections for all six isospin channels of kaon photoproduction on the nucleon. Solid curves represent the Set II of Ref.~\cite{Lee:1999kd} while the dashed curves show the older Set I model. The solid squares and open circles are the SAPHIR data \cite{SAPHIR:1998fev} and old ABBHHM data \cite{ABBHHM:1969pjo}, respectively. For the $K^0\Sigma^+$ photoproduction, the solid circles show the data taken from Ref.~\cite{Bennhold98}. Figure from Ref.~\cite{Lee:1999kd}.}
%    \label{fig:KMAID_allCS} 
%\end{figure}

Shortly after the work of Mart and Bennhold \cite{Mart:1999ed}, Janssen et al. from Ghent University proposed an alternative method to explain the second peak around $W=1.9~\mathrm{GeV}$ in the $\gamma p \to K^+\Lambda$ reaction total cross section data, that is, through the inclusion of hyperon resonances \cite{Janssen:2001pe}. To this end, they developed six covariant isobar models that differ in both their resonance content and the imposed lower limit of the cutoff mass for the hadronic form factor. They assumed the commonly used dipole expression for the hadronic form factor, which was implemented using Haberzettl's procedure to preserve gauge invariance. Motivated by the coupled-channel analysis of Feuster and Mosel \cite{Feuster:1998cj}, Janssen et al. constructed an initial set of resonances consisting of the $N(1650)1/2^-$, $N(1710)1/2^+$, and $N(1720)3/2^+$, as well as the kaon resonances $K^*(892)$ and $K_1(1270)$. This set, referred to as the ``basic set'' in their work, acted as the core foundation for the six models. 

%%%%%%%%%%%%%%%%%%%%%%%%%%%%%%%%%%%%%%%%%%%%%%%%%%%%%%%%%%%%%%%%%%%%%%%%%%%%%%%%%%%%%%%%%%%%%%%%%%%%%%%%%%%%%%%%%%%%%%%%%%%%%%%%%%%%%%%%%%%%%%%%%%%%%%%%%%%%%%%
%\begin{table}[hbt!]
%    \centering
%    \caption{Models studied in the work of Janssen et al. \cite{Janssen:2001pe}. We note that the values listed in the cutoff mass column are the imposed lower bounds, while the cutoff masses themselves were allowed to vary freely under this constraint. Table adapted from Ref.~\cite{Janssen:2001pe}.}
%    \begin{tabular}{l @{\hskip 1 cm} c @{\hskip 0.75cm} c}
%    \hline
%    \hline
%       Resonances in the model  &  $\Lambda$ (GeV)  & $\chi^2/N$\\
%    \hline
%        Basic set  & $\geq1.6$ & $10.32$\\
%        Basic set  & $\geq0.4$  & $4.36$\\
%        Basic set + $N(1895)1/2^-$  & $\geq1.6$  & $7.38$\\
%        Basic set + $N(1895)1/2^-$  & $\geq 0.4$  & $2.64$\\
%        Basic set + $\Lambda(1800)1/2^-$, $\Lambda(1810)1/2^+$  & $\geq 1.6$ & $3.43$\\
%        Basic set + $N(1895)1/2^-$, $\Lambda(1800)1/2^-$, $\Lambda(1810)1/2^+$  & $\geq 1.6$ & $2.65$\\
%    \hline
%    \hline
%    \end{tabular}
%    \label{tab:Janssen01_models}
%\end{table}
%%%%%%%%%%%%%%%%%%%%%%%%%%%%%%%%%%%%%%%%%%%%%%%%%%%%%%%%%%%%%%%%%%%%%%%%%%%%%%%%%%%%%%%%%%%%%%%%%%%%%%%%%%%%%%%%%%%%%%%%%%%%%%%%%%%%%%%%%%%%%%%%%%%%%%%%%%%%%%%

The content of the models, along with their corresponding $\chi^2/N$ values, is presented in Table~1 of Ref.~\cite{Janssen:2001pe}. There is it shown that a better fit was achieved with a soft form factor ($\Lambda \sim 0.42~\mathrm{GeV}$) compared to a hard form factor ($\Lambda \geq 1.6~\mathrm{GeV}$) for both the basic set and the extended set, composed of the basic set and the $N(1895)$ state. Obviously, this indicated that the cutoff mass of the hadronic form factor had a significant impact on the model predictions. %We also observe in Fig.~\ref{fig:Janssen_HFF} that a soft cutoff mass value tends to suppress the contribution of the Born terms to the cross section greatly. 
However, while soft cutoff masses seem to improve the model predictions in this case, Janssen et al. argued that introducing a soft value for the cutoff mass of the hadronic form factor was a rather unnatural procedure. 

To resolve this, they proposed a different way to appropriately suppress the contribution of the Born terms, that is, by including hyperon resonances in the model. As demonstrated in Fig.~6 of Ref.~\cite{Janssen:2001pe}, the inclusion of hyperon resonances in the model leads to a proper suppression of the Born terms to the cross section. We also observe that even when all terms but the hyperon resonances were taken into account (dash-dotted curve in Fig.~6 of Ref.~\cite{Janssen:2001pe}), the model still overpredicted the data by an order of magnitude. Thus, they asserted that it is highly likely that hyperon resonances have a significant role in the dynamics of the $\gamma p \to K^+\Lambda$ reaction. 

In a subsequent study, Janssen et al. performed a detailed comparison of the role of hyperon resonances in reducing the strength of the Born terms in the $\gamma  p \to K^+  \Lambda$ reaction with two other plausible approaches, namely, introducing small cutoff masses for the hadronic form factors, and disregarding the constraints of broken SU(3) symmetry on the values of the $g_{KYN}$ coupling constants~\cite{Janssen:2001wk}. Though all three procedures lead to similarly good predictions with $\chi^2/N \sim 2.9$, it was found that the three different procedures of modeling the background contribution influence the extracted information about the resonances quite significantly (see Figs.~2 and 3 of Ref.~\cite{Janssen:2001wk}). Additionally, it was further observed that the gauge recipe used to introduce the hadronic form factors also affects the extracted coupling constants substantially (see Fig.~5 of Ref.~\cite{Janssen:2001wk}). Interestingly, a further study by Janssen, Ryckebusch, and Van Cauteren found that the longitudinal and transverse response functions of $K^+\Lambda$ electroproduction seem to favor the inclusion of hyperon resonances as the method of reducing the strength of the Born terms compared to using soft cutoff masses and relaxing the SU(3) constraints on the $g_{KYN}$ coupling constants \cite{Janssen:2003kk}. 

%\begin{figure}[hbt!]
%    \centering
%    \includegraphics[scale=0.6]{ffdynam.eps}
%    \caption{Behavior of hadronic form factors $F_x(\Lambda)$ for three kaon angles. The cutoff masses are $\Lambda=0.8~\mathrm{GeV}$ and $\Lambda=1.8~\mathrm{GeV}$ for the left and right panels, respectively. Solid, dashed, and dotted curves represent $F_s(\Lambda)$, $F_t(\Lambda)$, and $F_u(\Lambda)$, respectively. Figure from Ref.~\cite{Janssen:2001pe}.}
%    \label{fig:Janssen_HFF}
%\end{figure}

%\begin{figure}[hbt!]
%    \centering
%    \includegraphics[scale=0.4]{totcs.comp.eps}
%    \vspace{-3mm}
%    \caption{Results from Janssen et al. for the total cross section of the $\gamma  p \to K^+  \Lambda$ reaction as a function of the photon energy $\omega_\mathrm{lab}$. Notation of the curves is given in the figure. We note that the curves were evaluated with a cutoff mass of $\Lambda\approx 1.78~\mathrm{GeV}$. Figure from Ref.~\cite{Janssen:2001pe}.}
%    \label{fig:Janssen01_cs}
%\end{figure}

Another notable covariant isobar model that describes the $\gamma p \to K^+\Lambda$ reaction is the work of Han et al. in the early 2000s \cite{Han:1999ck}. This model was developed for photon energies up to $E^{\mathrm{lab}}_\gamma=2.0~\mathrm{GeV}$. The model shared the same dataset with a previous work by Adelseck and Saghai \cite{Adelseck:1990ch} where the same reaction was studied over the energy range up to $E^{\mathrm{lab}}_\gamma=1.4~\mathrm{GeV}$. We note that this dataset also included the branching ratio for the kaon radiative capture reaction. Hence, the model of Han et al. was also constrained by crossing symmetry. 

In total, Han et al. incorporated nine nucleon resonances and eleven hyperon resonances with spins up to 5/2 in their model in addition to the usual Born and extended Born terms (see Table~1 of Ref.~\cite{Han:1999ck} for the complete list of resonances). However, after a rigorous fitting process, many of these resonances were excluded, primarily because their contributions had a negligible effect on the fit $\chi^2$, leaving only seven nucleon resonances and four hyperon resonances included in the analysis. We note that while Han et al. let the coupling constants of the resonances vary in the fitting process, the $g_{K\Lambda N}$ and $g_{K\Sigma N}$ coupling constants were fixed to the SU(3)-predicted values of $g_{K\Lambda N}/\sqrt{4\pi}=-3.74$ and $g_{K\Sigma N}/\sqrt{4\pi}=1.09$ because they found that both varying and fixing their values within the allowed range of broken SU(3) symmetry lead to practically the same value of $\chi^2$.

\begin{figure}[hbt!]
    \centering
    \includegraphics[scale=0.65]{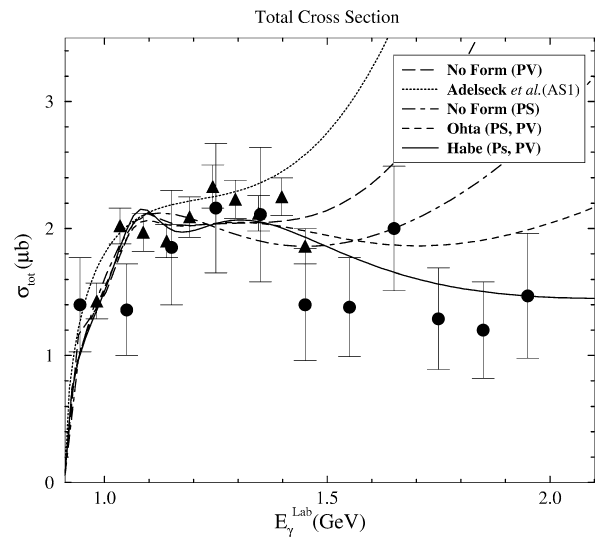}
    \caption{Total cross section of the $\gamma  p \to K^+\Lambda$ reaction for photon energies up to $E_\gamma^{\mathrm{lab}}=2.1$ GeV. Notation of the curves is given in the figure, with the AS1 model from the work of Adelseck and Saghai \cite{Adelseck:1990ch}. We note that the Ohta(PS) and Ohta(PV) models, as well as the Habe(PS) and Habe(PV) models, were each presented as a single model—Ohta(PS, PV) and Habe(PS, PV), respectively—which, in fact, correspond to the Ohta(PV) and Habe(PV) models, since the two coupling schemes show no significant differences. Figure from Ref.~\cite{Han:1999ck}. Reprinted with permission from Elsevier.}
    \label{fig:Han_totcs}
\end{figure}

A key feature of the work by Han et al. was the in-depth comparison of the combination of three aspects: the pseudoscalar (PS) and pseudovector (PV) coupling theories, the gauge restoration procedure, and the use of hadronic form factors. Specifically, they compared the use of Haberzettl's \cite{Haberzettl:1997jg} and Ohta's \cite{Ohta:1989ji} prescriptions on the restoration of gauge invariance. In total, they constructed six models: Habe(PV), Habe(Ps), Ohta(PV), Ohta(PS), Noform(PV), and Noform(PS). The Noform(PV) and Noform(PS) do not use hadronic form factors. As in previous studies, Han et al. also found that the introduction of hadronic form factors improves the quality of the resulting models. This fact is reflected in Fig.~\ref{fig:Han_totcs}, which shows that the Noform(PV) and Noform(PS) models start to diverge and overpredict the data around $E_\gamma^{\mathrm{lab}}=1.5-1.7~\mathrm{GeV}$, whereas only the Habe(PV) model reproduces the experimental data reasonably well up to $E_\gamma^{\mathrm{lab}}=2.1~\mathrm{GeV}$. Interestingly, they reported that the Noform(PS) model produced better results compared to the models utilizing Haberzettl's and Ohta's prescriptions, based on the partial $\chi^2$ value for the target-polarization asymmetry observable. For the full results of all six models, see Tables~6, 7, 8, and 9 in Ref.~\cite{Han:1999ck}. We also note that the results presented in Ref.~\cite{Han:1999ck} show that, when hadronic form factors are included, the pseudoscalar and pseudovector coupling theories yield similar results.

%%%%%%%%%%%%%%%%%%%%%%%%%%%%%%%%%%%%%%%%%%%%%%%%%%%%%%%%%%%%%%%%%%%%%%%%%%%%%%%%%%%%%%%%%%%%%%%%%%%%%%%%%%%%%%%%%%%%%%%%%%%%%%%%%%%%%%%%%%%%%%%%%%%%%%%%%%%%%%%
%%%%%%%%%%%%%%%%%%%%%%%%%%%%%%%%%%%%%%%%%%%%%%%%%%%%%%%%%%%%%%%%%%%%%%%%%%%%%%%%%%%%%%%%%%%%%%%%%%%%%%%%%%%%%%%%%%%%%%%%%%%%%%%%%%%%%%%%%%%%%%%%%%%%%%%%%%%%%%%

From the late 2000s until the mid-2010s, a set of covariant isobar models to describe kaon photo- and electroproduction reactions was intensively developed by Maxwell. In his work, he used an effective Lagrangian model similar to those developed in Refs.~\cite{Adelseck:1985scp,Adelseck:1988yv}. One of Maxwell's first works was an isobar model that describes the $\gamma p \to K^+\Lambda$ reaction over the energy range $E_\gamma^{\rm lab} = 1.0-1.8~\mathrm{GeV}$ \cite{Maxwell:2004ga}. It is important to note that, unlike most covariant isobar models, which utilize well-known amplitudes such as the helicity amplitudes to calculate the differential cross section, Maxwell calculated the differential cross section with the formula
\begin{equation}
    \frac{d\sigma}{d\Omega}=\frac{1}{(2\pi)^2}\frac{m_pm_\Lambda |\boldsymbol{p}_F|}{4E_\gamma s}\frac{1}{4}\sum_\mathrm{spins}\left|\braket{F|\hat{T}|I}\right|^2,
\end{equation}
where $|\boldsymbol{p}_F|$ is the magnitude of the outgoing 3-momentum~\cite{Maxwell:2004ga} and
\begin{equation}\label{eq:Maxwell_Decomposition}
    \bar{u}(\boldsymbol{p}_\Lambda)\hat{T}u(\boldsymbol{p}_p)=\bar{u}(\boldsymbol{p}_\Lambda)\left[\hat{A}+\hat{B}\gamma_5+\hat{C}\gamma^0+\hat{D}\gamma^0\gamma_5\right]u(\boldsymbol{p}_p),
\end{equation}
with $\hat{A}$, $\hat{B}$, $\hat{C}$, and $\hat{D}$ denoting operators whose structure depends on the spin and parity of the individual contributions under consideration. Their explicit expressions are given in the Appendix of Ref.~\cite{Maxwell:2004fs}. In this model, seven nucleon resonances and eleven hyperon resonances with spins up to 3/2 were incorporated (see Table~I of Ref.~\cite{Maxwell:2004ga}). Maxwell developed six fits labeled A to F, where fits A to D utilized the Rarita-Schwinger propagator for spin-3/2 resonances and full energy-dependent width prescription (see Ref.~\cite{Maxwell:2004fs} for the detailed explanation of the width prescription). Meanwhile, fit E was obtained through the simplified width prescription, where on-shell widths in each channel are also used for off-shell widths, while fit F was achieved with the spin-3/2 propagator used by Adelseck, Bennhold, and Wright \cite{Adelseck:1985scp}. We also note that in fits D and F, only spin-1/2 hyperon resonances were included, while in the other fits, both spin-1/2 and 3/2 hyperon resonances were incorporated. For the result of the fits, see Table~IV of Ref.~\cite{Maxwell:2004ga}. 

It is also perhaps worth noting that this model was not developed with the intention to explain the experimental data, but to study model dependence in kaon photoproduction instead. Thus, instead of fitting to the experimental data, Maxwell performed the fitting process by using the SAID cross sections for the $\gamma p \to K^+\Lambda$ reaction as a function of the c.m. scattering angle~\cite{Hyslop:1992}.

Despite resulting in different parameter sets, the six fits to the SAID cross section are relatively similar in quality. However, fits D and F are slightly worse than the rest of the fits due to the absence of spin-3/2 resonances in the $u$-channel (see Figs.~2 and 3 in Ref.~\cite{Maxwell:2004ga}). Although the results for the differential cross section are comparable in quality, the fits produced significantly different predictions for the polarization observables (see Figs.~4, 5, and 6 in Ref.~\cite{Maxwell:2004ga}). This implies that fitting to differential cross section data alone does not constrain the model effectively. In addition, it was also found that the polarization observables are quite sensitive to the prescription for resonance widths, while the form of spin-3/2 propagators has little effect on these observables.

Shortly after the publication of Ref.~\cite{Maxwell:2004ga}, Maxwell extended the photoproduction model to also describe the $ep\to e'K^+\Lambda$ reaction \cite{Maxwell:2007zza}. Maxwell used a slight modification of the EVMD model developed by Gari and Kr\"{u}mpelmann \cite{Gari:1984ia,Gari:1986rj,Gari:1992tq} for the nucleon electromagnetic form factors, whereas the constituent quark model by Cardarelli et al. was used for the kaon electromagnetic form factors (see Ref.~\cite{Maxwell:2007zza} for the detailed discussion). It is important to note that this model was not fitted to any electroproduction data, and, thus the fits did not reproduce the electroproduction data well (see Figs.~7-14 in Ref.~\cite{Maxwell:2007zza}). The electroproduction model was further extended in subsequent studies \cite{Maxwell:2012zz,Maxwell:2012tn,Maxwell:2014txa}.

The photoproduction model in Ref.~\cite{Maxwell:2004ga} was later extended by de la Puente, Maxwell, and Raue in another work \cite{delaPuente:2008bw} by including resonances up to spin-5/2 (see Tables~I and III in Ref.~\cite{delaPuente:2008bw}). However, it was found that the inclusion of some hyperon resonances yielded unphysically large couplings with correspondingly large uncertainties. Thus, the resulting model contains only a few select hyperon resonances (see Table~IV in Ref.~\cite{delaPuente:2008bw}). Overall, this model reproduced the differential cross section and polarization observables data relatively well (see Figs.~2-10 in Ref.~\cite{delaPuente:2008bw}). We note that Maxwell developed a similar model for the $\gamma p \to K^+\Sigma^0$ reaction several years later \cite{Maxwell:2015psa}, which was extended to also describe the $ep\to e'K^+\Sigma^0$ reaction in a subsequent study \cite{Maxwell:2016hdx}.

In the mid-2010s, Byd\v{z}ovsk\'{y} and Skoupil developed two isobar models \cite{Skoupil:2016ast} to describe the new experimental $\gamma p \to K^+\Lambda$ data in the energy range from threshold to $W=2.4~\mathrm{GeV}$ obtained by the CLAS, LEPS, and GRAAL Collaborations \cite{Glander:2003jw,CLAS:2005lui,CLAS:2009rdi,LEPS:2005hji,CLAS:2003zrd}. The two models, labeled BS1 and BS2, do not differ much in terms of their resonance content \cite{Skoupil:2016ast}. The BS1 and BS2 models contain 16 and 15 resonances, respectively, including the kaon resonances $K^*$ and $K_1$ (see Table~II in Ref.~\cite{Skoupil:2016ast} for the detailed list of each model's resonance content and the values of coupling constants used). Although both models contain similar nucleon resonances, they contain significantly different hyperon resonances. The BS1 model contains three spin-3/2 resonances and only one spin-1/2 resonance, whereas the BS2 model contains only one spin-3/2 resonance and three spin-1/2 resonances. 

Another major difference between the two models is the use of different hadronic form factors. Hadronic form factors that are typically used are either a dipole form,
\begin{equation}\label{eq:Dipole_HFF}
    F_{\rm d}(x,m,\Lambda)=\frac{\Lambda^4}{\Lambda^4+(x-m^2)^2},
\end{equation}
or a Gaussian form,
\begin{equation}\label{eq:Gauss_HFF}
    F_{\rm G}(x,m,\Lambda)=\exp\left[-\frac{(x-m^2)^2}{\Lambda^4}\right],
\end{equation}
where $x\equiv \{s,t,u\}$, $m$, and $\Lambda$ denote the Mandelstam variables, mass of the resonance, and cutoff mass of the form factor, respectively. However, a comprehensive study by Vrancx et al. has shown that the use of dipole and Gaussian models for hadronic form factors resulted in a cutoff-dependent resonance structure \cite{Vrancx:2011qv}. To remedy this, Vrancx et al. suggested a spin-dependent hadronic form factor
\begin{equation}
    F_{\rm mdG}(x,m,\Lambda,\Gamma,J)=\left[\frac{m^2\tilde{\Gamma}}{(x-m^2)^2+m^2\tilde{\Gamma}^2}\right]^{J-1/2}\exp\left[-\frac{(x-m^2)^2}{\Lambda^4}\right],
\end{equation}
which was called the multi-dipole-Gauss form factor. A modified decay width $\tilde{\Gamma}$, defined as
\begin{equation}
    \tilde{\Gamma}=\frac{\Gamma}{\sqrt{2^{1/(2J)}-1}},
\end{equation}
was used instead of the natural decay width $\Gamma$ to retain the interpretation of the resonance decay width being the full width at half maximum (FWHM) of the resonance peak \cite{Vrancx:2011qv,Skoupil:2016ast} (see Appendix~C of Ref.~\cite{Vrancx:2011qv} for the explicit derivation of $\tilde{\Gamma}$). Indeed, both Byd\v{z}ovsk\'{y} and Skoupil \cite{Skoupil:2016ast}, as well as Vrancx et al. \cite{Vrancx:2011qv}, had demonstrated that the multi-dipole-Gauss hadronic form factor was able to produce resonance peaks that are virtually independent of the cutoff value (see Fig.~2 of Ref.~\cite{Skoupil:2016ast} and Fig.~6 of Ref.~\cite{Vrancx:2011qv}). An additional spin-dependent hadronic form factor, named the multi-dipole form factor, was also introduced by Byd\v{z}ovsk\'{y} and Skoupil. The multi-dipole form factor is given by
\begin{equation}
    F_{\rm md}(x,m,\Lambda,J)=F_{\rm d}^{J+1/2}(x,m,\Lambda)=\left[\frac{\Lambda^4}{\Lambda^4+(x-m^2)^2}\right]^{J+1/2}.
\end{equation}
In their work, Byd\v{z}ovsk\'{y} and Skoupil used the multi-dipole form factor for the BS1 model and the dipole form factor for the BS2 model. As the multi-dipole model provides greater suppression, it is natural that the fit for the BS1 model leads to a higher cutoff parameter for the resonances  (see Table~II in Ref.~\cite{Skoupil:2016ast}).

\begin{figure}[hbt!]
    \centering
    \includegraphics[scale=0.1]{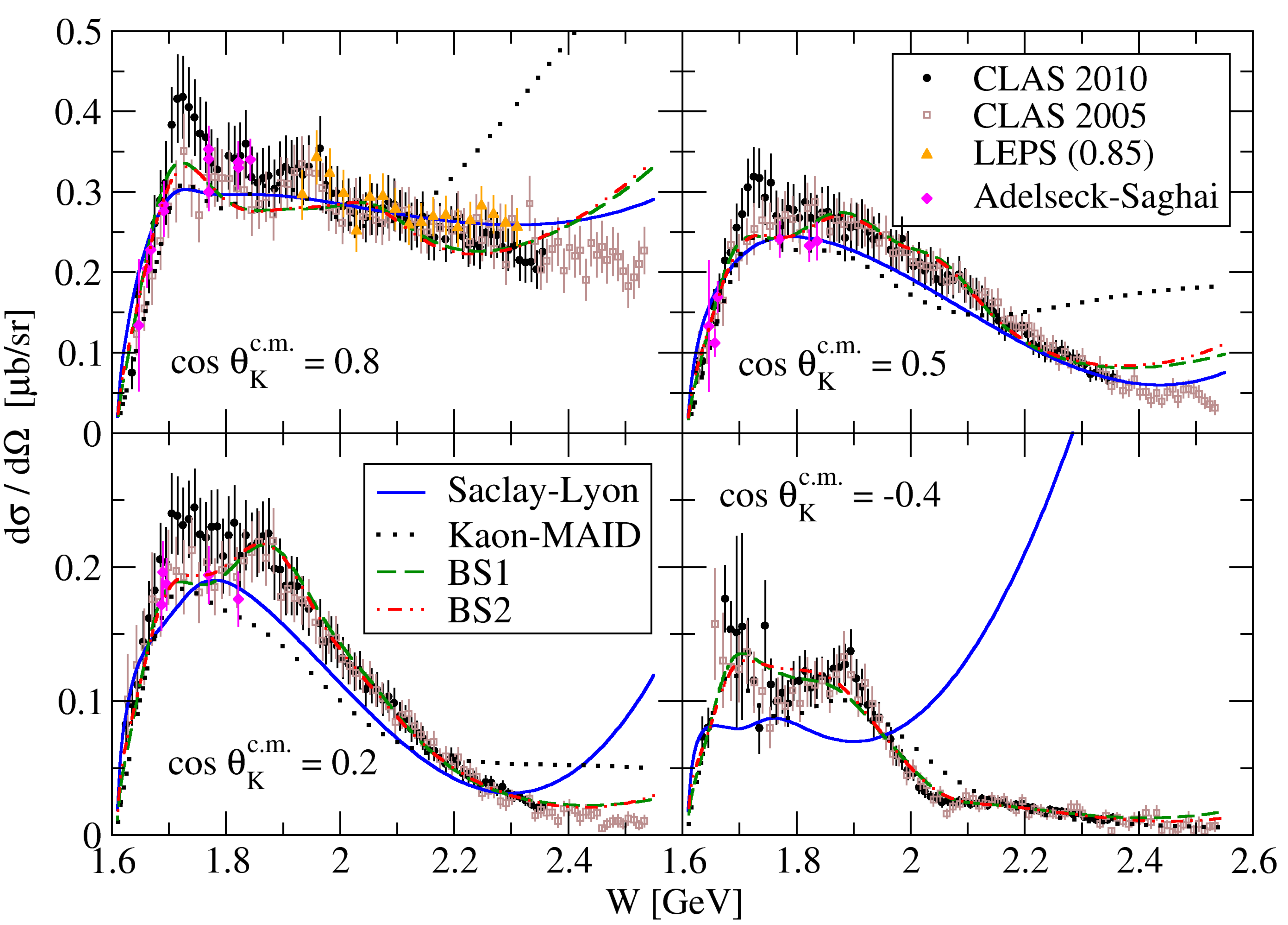}
    \caption{Energy dependence of the differential cross section of the $\gamma p\to K^+\Lambda$ reaction. The experimental data are from CLAS 2005 \cite{CLAS:2005lui,CLAS:2003zrd}, CLAS 2010 \cite{CLAS:2009rdi}, LEPS 2005 (only for $\cos \theta^\mathrm{c.m.}_K=0.85$) \cite{LEPS:2005hji}, and the work of Adelseck and Saghai \cite{Adelseck:1990ch}. It is important to note that Byd\v{z}ovsk\'{y} and Skoupil had modified the mass and width of some resonances in the fitting process, as stated in their work \cite{Skoupil:2016ast}. Figure from Ref.~\cite{Skoupil:2016ast}.}
    \label{fig:skoupil8}
\end{figure}

Results of the BS1 and BS2 models for the differential cross section of the $\gamma p \to K^+\Lambda$ reaction are given in Fig.~\ref{fig:skoupil8}. Both the BS1 and BS2 models are able to reproduce the experimental data relatively well up to $W\approx 2.4~\mathrm{GeV}$, beyond which they start to diverge and overpredict the data. Nevertheless, the BS1 and BS2 models provide a better description of the experimental data compared to the SL model from threshold to the resonance region. 

In addition to the differential cross section, the BS1 and BS2 models are also able to capture the trend of the polarization observables. As shown in Fig.~7 and Fig.~9 of Ref.~\cite{Skoupil:2016ast}, these two models are able to reproduce the overall trend of the hyperon polarization data from CLAS and the beam asymmetry data from LEPS fairly well, while the SL model and the Kaon-MAID model fail to do so. 

%%%%%%%%%%%%%%%%%%%%%%%%%%%%%%%%%%%%%%%%%%%%%%%%%%%%%%%%%%%%%%%%%%%%%%%%%%%%%%%%%%%%%%%%%%%%%%%%%%%%%%%%%%%%%%%%%%%%%%%%%%%%%%%%%%%%%%%%%%%%%%%%%%%%%%%%%%%%%%%
%%%%%%%%%%%%%%%%%%%%%%%%%%%%%%%%%%%%%%%%%%%%%%%%%%%%%%%%%%%%%%%%%%%%%%%%%%%%%%%%%%%%%%%%%%%%%%%%%%%%%%%%%%%%%%%%%%%%%%%%%%%%%%%%%%%%%%%%%%%%%%%%%%%%%%%%%%%%%%%

Shortly after their BS1 and BS2 models, Byd\v{z}ovsk\'{y} and Skoupil extended the models in another work \cite{Skoupil:2018vdh} to account for $K^+\Lambda$ electroproduction. The resulting model was named BS3. One notable feature of this model is the introduction of an energy-dependent decay width (see Eq.~(3) in Ref.~\cite{Skoupil:2018vdh}) for the resonances as an effort to partially restore unitarity. Due to the energy-dependent decay width, the BS3 model predicted a different resonance structure at forward and backward kaon scattering angles. 

The BS3 model shares the majority of its nucleon resonance content with the BS1 and BS2 models, having only extra $N(1710)1/2^+$ and $N(2120)3/2^-$ states compared to the BS1 model. It is also important to note that each nucleon resonance gained an extra parameter, due to the inclusion of longitudinal couplings to describe the electroproduction case. In addition, Byd\v{z}ovsk\'{y} and Skoupil also used Lomon's VMD-inspired parameterization, which was labeled the GKex(02S) model \cite{Lomon:2001,Lomon:2002}, for the electromagnetic form factors of the baryons and their resonances. For the kaon and its resonances, the usual monopole form factor was used.

The BS3 model was fitted to 3554 data points in total, of which 3383 are from photoproduction. The model was first fitted to only the photoproduction data to fix the transverse coupling constants. After a satisfactory result was achieved, the values of the transverse coupling constants were fixed, and the fitting process was redone for both photoproduction and electroproduction data to determine the optimal value for the longitudinal coupling constants of the nucleon resonances. It was also found that the photoproduction data alone were sufficient to constrain the transverse couplings, as these couplings barely changed when the electroproduction data were included.

\begin{figure}[hbt!]
    \centering
    \includegraphics[width=0.8\textwidth]{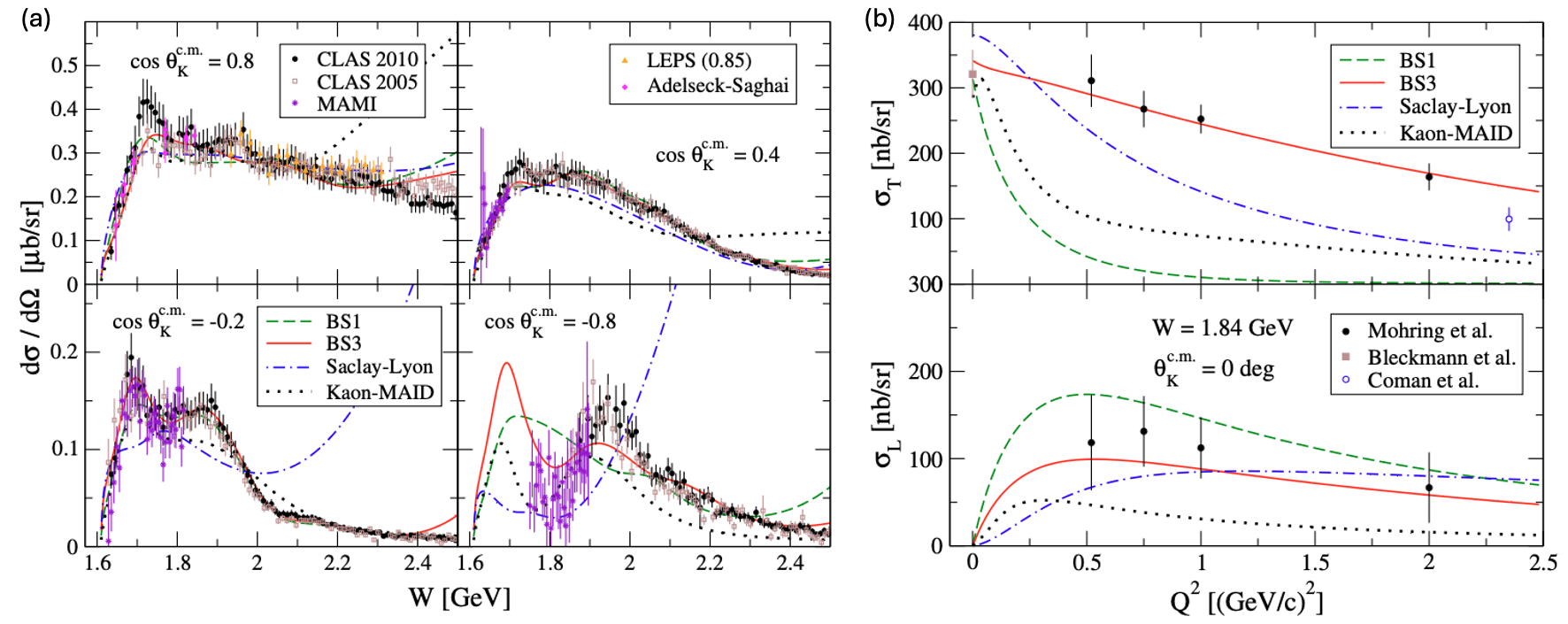}
    \caption{(a) Comparison between the BS3 model and the differential cross section data for $K^+\Lambda$ photoproduction as a function of the total c.m. energy for four values of the kaon angle. The experimental data from CLAS 2005 \cite{CLAS:2006pde}, CLAS 2010 \cite{CLAS:2009rdi}, MAMI 2014 \cite{CrystalBall:2013iig}, and LEPS 2006 \cite{LEPS:2005hji}, and from the work of Adelseck and Saghai \cite{Adelseck:1990ch}. Notation of the curves as in the figure. (b) Same as the left figure, but for the transverse ($\sigma_{\rm T}$) and longitudinal ($\sigma_{\rm L}$) structure functions for $K^+\Lambda$ electroproduction as a function of $Q^2$. Experimental data from JLab \cite{E93018:2002cpu,Coman:2009dot} and Bleckmann et al. \cite{Bleckmann:1970kb}. Figures from Ref.~\cite{Skoupil:2018vdh}.}
    \label{fig:BS3_result}
\end{figure}

Some results of the BS3 model are shown in Fig.~\ref{fig:BS3_result}. In Fig.~\ref{fig:BS3_result}(a), the BS3 model predicted two peaks around $W\approx 1.75~\mathrm{GeV}$ and $W\approx 1.90~\mathrm{GeV}$ for all kaon angles. These structures were mainly attributed to the contributions of the $N(1710)1/2^+$, $N(1900)3/2^+$, and $N(1875)3/2^-$ resonances. We also note that the $N(1900)3/2^+$ and $N(1875)3/2^-$ resonances are of particular importance in the BS3 model, as shown in Table~III and Fig.~1 of Ref.~\cite{Skoupil:2018vdh}. It thus can be inferred that these two resonances make a major contribution to shaping the second peak around $W=1.90~\mathrm{GeV}$. The BS3 model also reproduced the electroproduction data relatively well, as seen in Fig.~\ref{fig:BS3_result}(b). An interesting feature is that the BS1, SL, and Kaon-MAID models all fall too sharply with respect to $Q^2$ to adequately describe the transverse structure function $\sigma_{\rm T}$. In contrast, the BS3 model reproduced the data very well. This difference is mainly attributed to the inclusion of longitudinal couplings in the BS3 model, whereas the other models were developed solely to describe photoproduction and therefore do not include longitudinal couplings. However, the BS1 model is able to describe the longitudinal structure function $\sigma_{\rm L}$ fairly well, indicating that a good description of $\sigma_{\rm L}$ can be attained even without longitudinal couplings in certain kinematical regions.

Inspired by the studies of Landay et al. regarding pion photoproduction and the spectrum of baryon resonances \cite{Landay:2016cjw,Landay:2018wgf}, Byd\v{z}ovsk\'{y} et al. recently developed 
an isobar model for $K^+\Sigma^-$ photoproduction with a novel fitting procedure based on the Least Absolute Shrinkage and Selection Operator (LASSO) method \cite{Bydzovsky:2021rog}. The LASSO method is utilized to prevent overfitting by introducing a penalty term $P(\lambda)$, where $\lambda$ is a regularization parameter. Instead of the usual $\chi^2$, the quantity of interest is $\chi^2_T$, which is defined as
\begin{equation}
    \chi^2_T\equiv \chi^2+P(\lambda).
\end{equation}
For the detailed fitting procedure using the LASSO method, see Ref.~\cite{Bydzovsky:2021rog}. The resulting fit, named fit L, was obtained through the combination of the plain \textsc{MINUIT} procedure and the LASSO method. Despite having larger $\chi^2$ compared to the fit obtained through only the plain \textsc{MINUIT} procedure, fit L proved to give satisfactory agreement with the experimental data even with less parameters. Shortly after, Petrellis and Skoupil also utilized the LASSO method in their work on $K^+\Lambda$ \cite{Petrellis:2022eqw} and $K^+\Sigma^0$ photoproduction \cite{Petrellis:2024ybj}.

Another noteworthy covariant isobar model is the recent work of Fatima, Dar, Athar, and Singh from the Aligarh Muslim University (hereinafter called the AMU model) \cite{Fatima:2020tyh}. Unlike most covariant isobar models, the non-resonant terms of the AMU model were obtained using the non-linear sigma model. The AMU group utilized the non-linear sigma model to derive the meson-baryon interaction Lagrangians that were subsequently used to derive the hadronic currents, which in turn were utilized to calculate the differential cross section. For the explicit formalism used by the AMU group, we refer readers to Section~2.2 of Ref.~\cite{Fatima:2020tyh}.

Another interesting feature of the AMU model is its simplicity. Contrary to most covariant isobar models that incorporated numerous resonances, the AMU group only included 6 nucleon resonances and 2 hyperon resonances with spins up to 3/2 in their model in addition to the $K^*$ and $K_1$ resonances (see Tables~1 and 3 in Ref.~\cite{Fatima:2020tyh} for the complete list of the resonances). We note that the spin-3/2 nucleon resonances in the AMU model were included using consistent propagators and effective Lagrangians, following the prescription of Ref.~\cite{Mizutani:1997sd}.

The AMU model results for the total cross section of the $\gamma p \to K^+\Lambda$ reaction are shown in Fig.~5 of Ref.~\cite{Fatima:2020tyh}. Two versions of the model are shown, one with fixed widths and the other with energy-dependent widths. The two results are consistent with each other up to around $W=1.9~\mathrm{GeV}$. Around $W=1.8~\mathrm{GeV}$, the two results predict a significant dip structure, which is consistent with the SAPHIR 1998 and SAPHIR 2004 data. However, for $W>1.9~\mathrm{GeV}$, the result with energy-dependent widths strongly favors the CLAS 2006 data instead. Overall, both results show fairly good agreement with the experimental data.

%%%%%%%%%%%%%%%%%%%%%%%%%%%%%%%%%%%%%%%%%%%%%%%%%%%%%%%%%%%%%%%%%%%%%%%%%%%%%%%%%%%%%%%%%%%%%%%%%%%%%%%%%%%%%%%%%%%%%%%%%%%%%%%%%%%%%%%%%%%%%%%%%%%%%%%%%%%%%%%
%\begin{figure}[hbt!]
%\centering
%%\includegraphics[scale=0.75]{Fatima_CS_Comparison.pdf}
%\includegraphics[scale=0.4]{Fatima_CS_Comparison.png}
%\caption{Total cross section of the reaction $\gamma p\to K^+\Lambda$ as a function of total c.m. energy $W$. The solid black and dashed-dot blue curves represent the AMU model for energy-dependent widths and constant widths, respectively. The experimental data are from CLAS 2006 (solid circles) \cite{CLAS:2005lui}, SAPHIR 2004 (solid diamonds) \cite{Glander:2003jw}, and SAPHIR 1998 (solid triangles) \cite{SAPHIR:1998fev}. Figure adapted from Ref.~\cite{Fatima:2020tyh}.}
%\label{fig:fatima_cs_comparison}
%\end{figure}
%%%%%%%%%%%%%%%%%%%%%%%%%%%%%%%%%%%%%%%%%%%%%%%%%%%%%%%%%%%%%%%%%%%%%%%%%%%%%%%%%%%%%%%%%%%%%%%%%%%%%%%%%%%%%%%%%%%%%%%%%%%%%%%%%%%%%%%%%%%%%%%%%%%%%%%%%%%%%%%

Before concluding the discussion of isobar models, we note that, for more than two decades, significant efforts have been devoted to achieving a more reliable and accurate Kaon-MAID description. Progress has necessarily been gradual, driven by the steadily increasing volume and precision of experimental data, which in turn demand increasingly sophisticated theoretical treatments. Moreover, given the relatively high threshold energies for kaon photo- and electroproduction and the availability of more precise data in the near-threshold region, it has been essential to first establish a reliable description of kaon photoproduction in this energy domain~\cite{Mart:2010ch,Mart:2011ez,Mart:2014eoa}. For $K^+\Lambda$ photoproduction, experimental data up to 50~MeV above threshold can be well reproduced by including only one nucleon resonance and one hyperon resonance, i.e., the $N(1650)1/2^-$ and $\Lambda(1800)1/2^-$ states. The inclusion of the latter is essential for improving the agreement between the model calculations and the experimental data \cite{Mart:2013hia}. In contrast, for $K\Sigma$ photoproduction within the same energy range, four baryon resonances are required to achieve a satisfactory fit of $\chi^2/N = 1.07$, i.e., the $N(1700)3/2^-$, $N(1710)1/2^+$, $N(1720)3/2^+$, and $\Delta(1700)3/2^-$. In this channel, the $\Lambda(1800)1/2^-$ plays only a minor role, reducing $\chi^2/N$ to 1.05 when included.

For extensions from threshold up to about 3~GeV, two main approaches have been pursued, namely the use of partial-wave amplitudes or the retention of fully covariant amplitudes for the resonance contributions. In both approaches, the non-resonant background terms are described using covariant Feynman diagrammatic techniques. Within the partial-wave framework, the first comprehensive analysis of $K^+\Lambda$ photoproduction was reported in Ref.~\cite{Mart:2006dk}. By including 15 nucleon resonances, approximately 2400 experimental data points, comprising differential cross sections and single-polarization observables over the range from threshold up to $W \simeq 2.5$~GeV, were successfully fitted, yielding $\chi^2/N_{\rm dof}$ values between 0.98 and 1.31, depending on the dataset considered. This analysis was updated more than a decade later in Ref.~\cite{Mart:2017mwj}, where the inclusion of 22 nucleon resonances allowed nearly 9000 data points, consisting of differential cross sections as well as single- and double-polarization observables, to be well reproduced. The analysis indicates that the $N(1650)1/2^-$, $N(1720)3/2^+$, and $N(1900)3/2^+$ provide the dominant contributions to the process. This observation is in qualitative agreement with the previous analysis~\cite{Mart:2006dk}, which, however, also highlighted the relevance of several other resonant states. The extension of this framework to the $K\Sigma$ channels was subsequently carried out by incorporating 19 nucleon resonances and 13 $\Delta$ resonances, covering spins from $1/2$ up to $13/2$~\cite{Mart:2019fau}. The results indicate that, while the $N(1895)1/2^-$ and $\Delta(1900)1/2^-$ provide the dominant contributions in this channel, the $N(1720)3/2^+$ and $N(1900)3/2^+$ remain relevant for $K\Sigma$ photoproduction.

In the case of the fully covariant approach, progress has been considerably slower than for the partial-wave formulation. This is not unexpected, as the covariant treatment is substantially more involved, being affected by off-shell ambiguities and by the propagation of unphysical lower-spin components~\cite{Benmerrouche:1989uc,Vrancx:2011qv} (see Sec.~\ref{sec:unsettled} for further discussion). Consequently, a careful and consistent treatment of both propagators and interaction vertices is required in order to achieve a reliable formulation of this approach. Such efforts were initiated through an analysis of $K^+\Lambda$ photoproduction that included all nucleon resonances listed by the PDG with spins up to $5/2$~\cite{Mart:2015jof}. Despite this nominal spin limitation, a total of 17 nucleon resonances were employed, since states with masses below the reaction threshold were also taken into account. By fitting approximately 7000 experimental data points, it was concluded that a gauge-invariant formulation of spin-$3/2$ and spin-$5/2$ interactions provides the best overall agreement with the data. The model was subsequently extended to include spin-$7/2$ and spin-$9/2$ nucleon resonances~\cite{Clymton:2017nvp}, and was later further generalized by incorporating nucleon resonances with spins up to $15/2$~\cite{Luthfiyah:2020ogb,Luthfiyah:2021yqe}. In the latter work, in addition to the Breit-Wigner parameters, the pole positions of the included resonances were also determined. In contrast to the extracted Breit-Wigner parameters, which depend strongly on the background contributions of the model, the pole positions are largely model independent and can therefore be meaningfully compared with those obtained in other approaches. For $K\Sigma$ photoproduction, the extracted pole positions of the nucleon and $\Delta$ resonances have been reported in Ref.~\cite{Clymton:2021wof}.

Extension to electroproduction is still in progress. In total, nearly 2000 data points, including the newly averaged beam-recoil transferred polarization observables $\mathcal{P}'_{x}$ and $\mathcal{P}'_{x'}$ from CLAS12~\cite{CLAS:2022yzd}, have been fitted using different choices of electromagnetic form factors. Preliminary results have been reported in a conference proceeding~\cite{Djaja:2025hzz}.

\begin{table}[hbt!]
\setlength{\tabcolsep}{6pt} % Default value: 6pt
\renewcommand{\arraystretch}{0.8} % Default value: 1
    \centering
    \caption{Summary of known isobar models and the experimental data used in the fitting process.}
    \begin{threeparttable}
    \begin{tabularx}{\textwidth}{XX wc{3cm} @{\hspace{1cm}} Xc}
    \hline \hline
    Model & Dataset & Final State(s) & Observable(s) & Ref.
    \\ \hline
    %======================
    Thom (1966)   &   Cornell 1962   & $K^+\Lambda$ & $d\sigma/d\Omega$   &   \cite{Anderson:1962za}\\
    %======================
    &   Cornell 1963   & $K^+\Lambda$ &  $P$   &   \cite{Thom:1963zz}\\
    %======================
    &   CalTech 1964   &   $K^+\Lambda$  & $d\sigma/d\Omega$   &   \cite{Peck:1964zz}\\
    %======================
    &   Frascati 1964   & $K^+\Lambda$ &  $P$   &   \cite{Borgia:1964mza}\\
    %======================
    &   Frascati 1965   & $K^+\Lambda$  &  $P$   &   \cite{Grilli:1965jia}\\
    %======================
    &   CalTech 1967   & $K^+\Lambda$ &  $P$   &   \cite{Groom:1967zz}\\ \hline
    %======================
    Renard-Renard (1971)   &   CalTech 1967  & $K^+\Lambda$ &   $P$   &   \cite{Groom:1967zz}\\
    %======================
       &   Old Data 1970\tnote{a}   & $\dagger$ & $\dagger$    &   $\dagger$
    \\ \hline
    %======================
    WJC (1992)   &   SLAC 1971   & $K^+\Lambda$, $K^+\Sigma^0$  & $d\sigma/d\Omega$   &  \cite{Boyarski:1970yc} \\
    %======================
    &   Bonn 1972   & $K^+\Lambda$, $K^+\Sigma^0$ & $d\sigma/d\Omega$   & \cite{Feller:1972ph}   \\
    %======================
    &   DESY 1975   & $e'K^+\Lambda$, $e'K^+\Sigma^0$  & $\sigma_\mathrm{U}$, $\sigma_\mathrm{LT}$, $\sigma_\mathrm{TT}$   &  \cite{Azemoon:1974dt} \\
    %======================
    &   SLAC 1976   & $K^+\Lambda$, $K^+\Sigma^0$  & $d\sigma/d\Omega$   &  \cite{Anderson:1976ph} \\
    %======================
    &   BNL 1989   & $\gamma \Lambda$, $\gamma \Sigma^0$ &  $R_{\gamma \Lambda}$, $R_{\gamma \Sigma^0}$\tnote{b}   &  \cite{Whitehouse:1989yi} \\ \hline
    %======================
    SL (1996)  & DESY 1969   & $K^0\Sigma^+$ & $d\sigma/d\Omega$   &   \cite{ABBHHM:1969pjo}\\
    %======================
    & Old $\gamma N$ Data  & $K^+\Lambda$ & $P$ & \cite{Borgia:1964mza,Grilli:1965jia,Groom:1967zz,Haas:1978qv} \\
    %======================
    & Old $\gamma_v N$ Data  & $e'K^+\Lambda$, $e'K^+\Sigma^0$  & $d\sigma/d\Omega$, $\sigma_\mathrm{U}$, $\sigma_\mathrm{TT}$, $\sigma_\mathrm{LT}$  & \cite{Brown:1972pf,Azemoon:1974dt,Bebek:1976qg,Bebek:1977bv,Brauel:1979zk} \\
    %======================
    &   BNL 1989   & $\gamma \Lambda$, $\gamma \Sigma^0$ & $R_{\gamma \Lambda}$, $R_{\gamma \Sigma^0}$\tnote{b}   &  \cite{Whitehouse:1989yi} \\
    %======================
    & Adelseck-Saghai 1990\tnote{c}  & $K^+\Lambda$, $K^+\Sigma^0$  & $d\sigma/d\Omega$  &   \cite{Adelseck:1990ch}\\
    %======================
    &   SAPHIR 1994   & $K^+\Lambda$, $K^+\Sigma^0$  & $d\sigma/d\Omega$, $\sigma$, $P$   &   \cite{Bockhorst:1994jf}\\
    %======================
    \hline
    %======================
    Kaon-MAID (1999)   &   DESY 1969   & $K^+\Lambda$ & $d\sigma/d\Omega$   &   \cite{ABBHHM:1969pjo}\\
    %======================
     &   SAPHIR 1998   & $K^+\Lambda$ &  $d\sigma/d\Omega$   &   \cite{SAPHIR:1998fev} \\ \hline
    %======================
    Ghent-A,B,C (2001)   &   SAPHIR 1998   &  $K^+\Lambda$ &  $d\sigma/d\Omega$   &   \cite{SAPHIR:1998fev} \\ \hline
    %======================
    HCKC (2001)   &  Adelseck-Saghai 1990\tnote{c}  &  $K^+\Lambda$ & $d\sigma/d\Omega$  &   \cite{Adelseck:1990ch}\\
    %======================
    &   SAPHIR 1994   & $K^+\Lambda$  & $d\sigma/d\Omega$, $\sigma$, $P$   & \cite{Bockhorst:1994jf}
    %======================
    %x   &   x   &  x   &   x   &   x   &   x   &   x   &   x\\
    \\ \hline \hline
    \end{tabularx}
        \begin{tablenotes}
        \footnotesize
            \item[a] We do not have access to some of the experimental references in Ref.~\cite{Renard:1971us}. See Refs.~[4] and [5] in Ref.~\cite{Renard:1971us}. 
            \item[b] Branching ratio in kaon radiative capture reaction, defined as $R_{\gamma Y}\equiv \Gamma(K^-p\to \gamma Y)/\Gamma(K^-p\to \mathrm{all})$ \cite{David:1995pi}.
            \item[c] The work of Adelseck and Saghai \cite{Adelseck:1990ch}. The experimental data listed in this work are not original data, but rather a compilation of older data. See Tables~IX and X in Ref.~\cite{Adelseck:1990ch}.
        \end{tablenotes}
    \end{threeparttable}
    \label{tab:isobar_summary}
\end{table}

\begin{table}[hbt!]
\setlength{\tabcolsep}{6pt} % Default value: 6pt
\renewcommand{\arraystretch}{0.8} % Default value: 1
    \centering
    \caption{Summary of known isobar models and the experimental data used in the fitting process (cont.).}
    \begin{threeparttable}
    \begin{tabularx}{\textwidth}{XX wc{3cm} @{\hspace{1cm}} Xc}
    \hline \hline
    Model & Dataset & Final State(s) & Observable(s) & Ref.
    \\ \hline
    BS1 \& BS2 (2016)   &   Adelseck-Saghai 1990\tnote{a}   &  $K^+\Lambda$ & $d\sigma/d\Omega$  &   \cite{Adelseck:1990ch}\\
    %======================
    &   CLAS 2004   &  $K^+\Lambda$  & $d\sigma/d\Omega$   &   \cite{CLAS:2003zrd}   \\
    %======================
    &   LEPS 2006   & $K^+\Lambda$ & $d\sigma/d\Omega$, $\Sigma$   &   \cite{LEPS:2005hji}   \\
    %======================
    &   CLAS 2006   &  $K^+\Lambda$  & $d\sigma/d\Omega$   &   \cite{CLAS:2005lui}\\
    %======================
    &   CLAS 2010   &  $K^+\Lambda$  & $d\sigma/d\Omega$, $P$   &   \cite{CLAS:2009rdi}\\
    %======================
    \hline
    %======================
    Maxwell (2014) & SAPHIR 1998  & $K^+\Lambda$, $K^+\Sigma^0$  & $d\sigma/d\Omega$  &  \cite{SAPHIR:1998fev} \\
    & SAPHIR 1999  & $K^0\Sigma^+$  & $d\sigma/d\Omega$, $P$  & \cite{SAPHIR:1999wfu}  \\
    & LEPS 2003  & $K^+\Lambda$, $K^+\Sigma^0$  & $\Sigma$  &  \cite{LEPS:2003buk} \\
    & CLAS 2003  & $e'K^+\Lambda$  & $\mathcal{P}'_x$, $\mathcal{P}'_{x'}$, $\mathcal{P}'_z$, $\mathcal{P}'_{z'}$  & \cite{CLAS:2002zlc}  \\
    & CLAS 2004  & $K^+\Lambda$, $K^+\Sigma^0$  & $d\sigma/d\Omega$, $P$  &  \cite{CLAS:2003zrd} \\
    & SAPHIR 2004  & $K^+\Lambda$, $K^+\Sigma^0$  & $d\sigma/d\Omega$, $P$  & \cite{Glander:2003jw}  \\
    & LEPS 2006  & $K^+\Lambda$, $K^+\Sigma^0$  & $\Sigma$  & \cite{LEPS:2005hji}  \\
    & CLAS 2006  &  $K^+\Lambda$, $K^+\Sigma^0$ &  $d\sigma/d\Omega$  & \cite{CLAS:2005lui}  \\
    & CLAS 2007  & $K^+\Lambda$, $K^+\Sigma^0$ & $C_x$, $C_z$  & \cite{CLAS:2006pde}  \\
    & CLAS 2007 & $e'K^+\Lambda$, $e'K^+\Sigma^0$ & $\sigma_\mathrm{U}$, $\sigma_\mathrm{TT}$, $\sigma_\mathrm{LT}$ & \cite{CLAS:2006ogr} \\
    & GRAAL 2007  & $K^+\Lambda$, $K^+\Sigma^0$  & $P$, $\Sigma$  & \cite{Lleres:2007tx,GRAAL:2007tnj}  \\
    & CLAS 2008  & $e'K^+\Lambda$  & $\sigma_\mathrm{LT'}$  &  \cite{CLAS:2008agj} \\
    & GRAAL 2009  & $K^+\Lambda$  & $O_x$, $O_z$, $T$\tnote{b}  & \cite{GRAAL:2008jrm}  \\
    & CLAS 2009  & $e'K^+\Lambda$, $e'K^+\Sigma^0$  & $\mathcal{P}'_{x}$, $\mathcal{P}'_{x'}$, $\mathcal{P}'_{z}$,, $\mathcal{P}'_{z'}$  & \cite{CLAS:2009sbn}  \\
    & CLAS 2014  & $e'K^+\Lambda$  & $\mathcal{P}^0_y$  & \cite{CLAS:2014udv}  \\
    %======================
    \hline
    %======================
    BS3 (2018)   &   Bonn 1970   & $K^+\Lambda$  & $d\sigma/d\Omega$   &   \cite{Bleckmann:1970kb}\\
    %======================
    &   Cambridge 1972   & $e'K^+\Lambda$  & $\sigma_\mathrm{U}$   &   \cite{Brown:1972pf}\\
    %======================
    &   Cornell 1974   & $e'K^+\Lambda$  & $\sigma_\mathrm{U}$   &   \cite{Bebek:1974bt}\\
    %======================
    &   DESY 1975   & $e'K^+\Lambda$ & $\sigma_\mathrm{U}$   &   \cite{Azemoon:1974dt}\\
    %======================
    &   Cornell 1977  & $e'K^+\Lambda$ & $\sigma_\mathrm{U}$   &   \cite{Bebek:1976qg}\\
    %======================
    &   Cornell 1977   & $e'K^+\Lambda$  & $\sigma_{\mathrm{U}}$   &   \cite{Bebek:1977bv}\\
    %======================
    &   Adelseck-Saghai 1990\tnote{a}   & $K^+\Lambda$  & $d\sigma/d\Omega$  &   \cite{Adelseck:1990ch}\\
    %======================
    &   Hall C 2003   & $e'K^+\Lambda$ & $\sigma_\mathrm{U}$, $\sigma_\mathrm{L}$, $\sigma_\mathrm{T}$   &   \cite{E93018:2002cpu}\\
    %======================
    &   LEPS 2006   & $K^+\Lambda$  & $\Sigma$   &   \cite{LEPS:2005hji}\\
    %======================
    &   CLAS 2006   &  $K^+\Lambda$ & $d\sigma/d\Omega$   &   \cite{CLAS:2005lui}\\
    %======================
    &   CLAS 2009   & $e'K^+\Lambda$  & $\sigma_\mathrm{L}$, $\sigma_\mathrm{T}$   &   \cite{CLAS:2009sbn}\\
    %======================
    &   CLAS 2010   & $K^+\Lambda$  & $d\sigma/d\Omega$, $P$   &   \cite{CLAS:2009rdi}\\
    %======================
    &   Hall A 2010   & $e'K^+\Lambda$  & $\sigma_\mathrm{L}$, $\sigma_\mathrm{T}$   &   \cite{Coman:2009dot}\\
    %======================
    &   MAMI 2012   &  $e'K^+\Lambda$ & $d\sigma/d\Omega$   &   \cite{A1:2011eeq} \\
    %======================
    &   MAMI 2017   & $e'K^+\Lambda$  & $\sigma_\mathrm{LT'}$   &  \cite{A1:2017fvu} \\ \hline
    %======================
    AMU (2020)   &   SAPHIR 1998   & $K^+\Lambda$  & $d\sigma/d\Omega$   &   \cite{SAPHIR:1998fev}\\
    %======================
    &   SAPHIR 2004   & $K^+\Lambda$ & $d\sigma/d\Omega$   &   \cite{Glander:2003jw}\\
    %======================
    &   CLAS 2006   & $K^+\Lambda$  & $d\sigma/d\Omega$   &   \cite{CLAS:2005lui}\\
    %======================
    &   CLAS 2010   & $K^+\Lambda$  & $d\sigma/d\Omega$   &   \cite{CLAS:2009rdi}
    %======================
    %x   &   x   &  x   &   x   &   x   &   x   &   x   &   x\\
    \\ \hline \hline
    \end{tabularx}
        \begin{tablenotes}
        \footnotesize
            \item[a] The work of Adelseck and Saghai \cite{Adelseck:1990ch}. The experimental data listed in this work are not original data, but rather a compilation of older data. See Tables~IX and X in Ref.~\cite{Adelseck:1990ch}.
            \item[b] The target asymmetry $T$ was not directly measured, but indirectly extracted from the data \cite{GRAAL:2008jrm}.
        \end{tablenotes}
    \end{threeparttable}
    \label{tab:isobar_summary2}
\end{table}

%=============================================

\subsection{Coupled-Channel Analyses}

Both scattering and decay processes are not independent. They share common dynamics because quantum mechanics allows any excited hadronic system to evolve through all intermediate states permitted by conservation laws. In photo- and electroproduction of kaons this is particularly important, since the electromagnetic probe can excite several hadronic configurations that subsequently re-scatter into the observed final state. Thus, a reaction such as $\gamma p \to K^+\Lambda$, can receive significant contributions from intermediate channels like $\pi N$, $\eta N$, or $\pi\pi N$, all of which carry the same conserved quantum numbers. Because these intermediate states are dynamically connected, a coupled-channels treatment becomes necessary. The observed amplitude in any given reaction contains not only direct production but also multi-step transitions among all accessible hadronic channels. These rescattering effects generate correlations across states with the same baryon number, strangeness, isospin, spin, and parity, meaning that resonance structures or background behaviors in one channel can be strongly altered by processes occurring in others, even those that cannot be directly observed in experiments. Note that neglecting this coupled-channels dynamics inevitably violates unitarity, analyticity, and multi-channel rescattering, leading to amplitudes that misrepresent the reaction mechanism and distort the extracted resonance parameters. Modern approaches, whether based on the $K$-matrix approximation or on the fully dynamical coupled-channels (DCC) scheme derived from the Lippmann-Schwinger equation, address these issues by treating all relevant hadronic and electromagnetic channels simultaneously, thereby ensuring a consistent and physically reliable description of the data.

The availability of high-quality data from various experimental facilities has significantly improved our ability to study the reaction mechanism of kaon photo- and electroproduction on the nucleon. Although direct-channel mechanisms provide valuable insights into the underlying baryon resonances, previous studies have shown that higher-order processes, such as the multi-step reaction $\gamma p \to \pi N \to K^+\Lambda$, can contribute up to 20\% of the total cross section \cite{Chiang:2001pw,Chiang:2004ye}. This highlights the importance of including coupled-channel effects in theoretical models. Accounting for these effects helps to constrain model parameters more reliably and reduces model dependence in the analysis \cite{Usov:2005wy, Bruns:2010sv,Mai:2012dt,Mai:2014xna,Mai:2020ltx}.

\begin{figure}[h]
\centering
\includegraphics[width=0.95\columnwidth]{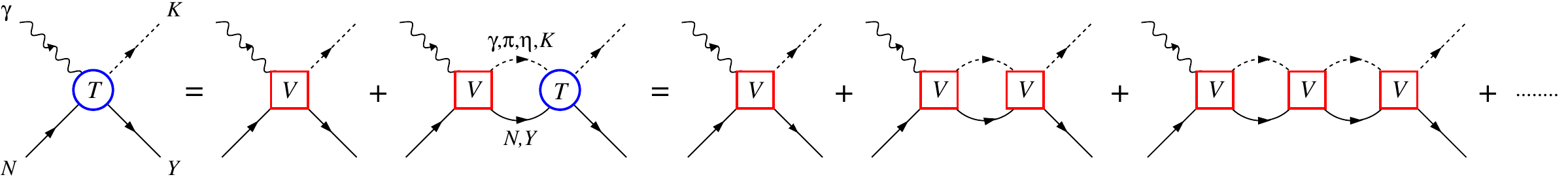}
\caption{Contributions of the rescattering terms in electromagnetic production of kaons, illustrating the Lippmann-Schwinger equation given in Eq.~(\ref{eq:LS_equation}).}
\label{fig:LS_eq_illustrated}
\end{figure}

In multi-channel scattering the central quantity in the scattering formalism is the $S$-matrix, written as
\begin{equation}
    S = 1 + 2iT,
\end{equation}
where the transition matrix $T$ contains all reaction dynamics among the channels $i,j = 1,2,\dots,N$. Because any channel permitted by conservation laws can contribute through intermediate propagation, the multi-channel amplitude satisfies the Lippmann-Schwinger equation
\begin{equation}
    \label{eq:LS_eq_details}
    T_{ij}(E) = V_{ij}(E) + \sum_{k} V_{ik}(E)\, G_k(E)\, T_{kj}(E),
\end{equation}
where $V_{ij}(E)$ denotes the interaction kernel (or driving term) describing the direct transition $i \to j$, and $G_k(E)$ is the propagator (Green's function) for channel $k$, containing the intermediate-state phase space and analytic structure. In compact matrix notation this equation can be written as
\begin{equation}
T = V + VGT = V + VGV + VGVGV + \cdots = (1 - VG)^{-1}V.
\label{eq:LS_equation}
\end{equation}
This constitutes the full DCC formulation, expressed through the Lippmann-Schwinger equation, where the propagator $G$ contains both real and imaginary parts, ensuring the correct analytic structure and dynamical dressing of resonances. In the case of  electromagnetic production of kaons, Eq.~(\ref{eq:LS_equation}) is illustrated in Fig.~\ref{fig:LS_eq_illustrated}.

%%%%%%%%%%%%%%%%%%%%%%%%%%%%%%%%%%%%%%%%%%%%%%%%%%%%%%%%%%%%%%%%%%%%%%%%%%%%%%%%%%%%%%%%%%%%%%%%%%%%%%%%%%%%%%%%%%%%%%%%%%%%%%%%%%%%%%%%%%%%%%%%%%%%%%%%%%%%%%%
%%%%%%%%%%%%%%%%%%%%%%%%%%%%%%%%%%%%%%%%%%%%%%%%%%%%%%%%%%%%%%%%%%%%%%%%%%%%%%%%%%%%%%%%%%%%%%%%%%%%%%%%%%%%%%%%%%%%%%%%%%%%%%%%%%%%%%%%%%%%%%%%%%%%%%%%%%%%%%%

However, the use of DCC can be computationally demanding because it requires evaluating full off-shell propagation and dispersive integrals across all intermediate channels.
To simplify the calculation, one often adopts the $K$-matrix approximation, which follows directly from the unitarity condition. Starting from the unitarity of the $S$-matrix, $S^\dagger S = 1$, and substituting $S = 1 + 2iT$, one obtains the multi-channel unitarity condition
\begin{equation}
\operatorname{Im}\, T^{-1} = -\,1 ,
\end{equation}
with $1$ the identity matrix. By separating its real and imaginary parts, the inverse $T$-matrix can be written as
\begin{equation}
T^{-1} = \operatorname{Re}T^{-1} - i\,1 \;\equiv\; K^{-1} - i\,1 ,
\end{equation}
where the real, Hermitian matrix $K$ is defined by $K^{-1} = \operatorname{Re}T^{-1}$. Inverting the expression gives the standard relation between the $T$ and $K$-matrices,
\begin{equation}
T = (K^{-1} - i\,1)^{-1} = (1 - iK)^{-1}K.
\label{eq:K-matrix-def}
\end{equation}
Thus, the DCC scheme retains the full off-shell and dispersive dynamics generated by $VG$, whereas the $K$-matrix approximation enforces unitarity algebraically by imposing $\operatorname{Im} T^{-1} = -1$ and approximating the real part of $T^{-1}$ with the energy-dependent kernel $K$. Nevertheless, once the $K$-matrix is specified from modeling the physical process, the resulting $S$-matrix is guaranteed to be unitary, as shown above~\cite{Glockle:1983}. 

For illustration, consider imposing the unitarity constraint on single-channel fits within the $K$-matrix formalism. A commonly used construction for the multi-channel $K$ matrix reads~\cite{Workman:2005eu}
\begin{equation}
\label{eq:K-matrix-example}
K =
\begin{pmatrix}
K_{\gamma\gamma} & K_{\gamma\pi} & K_{\gamma\Delta} \\
K_{\gamma\pi}    & K_{\pi\pi}    & K_{\pi\Delta}    \\
K_{\gamma\Delta} & K_{\pi\Delta} & K_{\Delta\Delta}
\end{pmatrix},
\end{equation}
where, for example, $K_{\pi\Delta} \equiv K(\pi N \to \pi\Delta)$.  
Substituting Eq.~(\ref{eq:K-matrix-example}) into the unitarized definition of the $T$ matrix, Eq.~(\ref{eq:K-matrix-def}), yields relations between the $T$- and $K$-matrix elements.  
For the $\gamma N \to \pi N$ channel one obtains~\cite{Workman:2005eu}
\begin{equation}
    T_{\gamma\pi}
    =
    \bigl(1 + i\,T_{\pi\pi}\bigr)
    \left(
      K_{\gamma\pi}
      - \frac{K_{\gamma\Delta}\,K_{\pi\pi}}{K_{\pi\Delta}}
    \right)
    + \frac{K_{\gamma\Delta}}{K_{\pi\Delta}}\, T_{\pi\pi},
\end{equation}
which demonstrates explicitly that, even within this approximation, the single-channel pion photoproduction amplitude is constrained by other photoproduction and meson-baryon scattering processes.

Although the basic scattering equation given in Eq.~(\ref{eq:LS_eq_details}) is employed in all coupled-channels models, its detailed implementation differs from one model to another. In the following, we briefly highlight these differences and examine the performance of the models in describing electromagnetic kaon production data. 

\subsubsection{Dynamical Coupled-Channels Models}

In the literature, two dynamical coupled-channels (DCC) models explicitly include electromagnetic kaon production, namely the Argonne-National-Laboratory-Osaka model (ANL-Osaka, formerly EBAC or JLab)~\cite{Julia-Diaz:2007mae,Kamano:2013iva,Chiang:2001pw,Chiang:2004ye,Matsuyama:2006rp,Julia-Diaz:2006rvy,Julia-Diaz:2007qtz,Kamano:2008gr,Suzuki:2010yn,Kamano:2014zba,Kamano:2015hxa,Kamano:2016bgm} and the J\"ulich-Bonn-Washington (JBW) model~\cite{Mai:2023cbp,Mai:2021vsw,Mai:2021aui,Wang:2024byt}, the latter being an extension of the earlier J\"uBo framework~\cite{Ronchen:2022hqk,Ronchen:2014cna,Ronchen:2012eg}. For a recent summary of developments in DCC modeling, readers may consult Ref.~\cite{Doring:2025sgb}.

The ANL-Osaka DCC model is rooted in the seminal work of Sato and Lee in the late 1990s, who developed a fully dynamical and unitary framework for meson-baryon reactions based on effective Lagrangians and time-ordered perturbation theory~\cite{Sato:1996gk,Sato:2000jf}. Their approach provided a consistent description of pion-induced and pion photoproduction reactions, such as $\pi N \to \pi N$ and $\gamma N \to \pi N$, by explicitly incorporating meson-baryon rescattering effects and dressing the bare baryon states. A key conceptual advance of the Sato-Lee model is the dynamical generation of resonance widths while exactly preserving multi-channel unitarity at the amplitude level. Thus, for example, the pion photoproduction amplitude is written as \cite{Sato:2000jf}
\begin{equation}
    \label{eq:Sato_Lee_amplitude}
    T_{\gamma\pi}(E)=t_{\gamma\pi}(E)+\frac{{\bar \Gamma}_{\Delta\to\pi N}{\bar \Gamma}_{\gamma N\to\Delta}}{E-m_\Delta-\Sigma_\Delta(E)},
\end{equation}
with $\Sigma_\Delta(E)=\Gamma_{\pi N\to\Delta}\, G_{\pi N}(E)\,{\bar\Gamma}_{\Delta\to\pi N}$ is the $\Delta$ self-energy and the non-resonant amplitude $t_{\gamma\pi}$ is calculated from the non-resonant $\gamma N\to \pi N$ transition potential $v_{\gamma\pi}$ and the $\pi N$ free propagator $G_{\pi N}$, 
\begin{equation}
    t_{\gamma\pi}(E)=v_{\gamma\pi}+t_{\pi N}(E)\, G_{\pi N}(E)\, v_{\gamma\pi}.
\end{equation}
The dressed vertices ${\bar \Gamma}_{\Delta\to\pi N}$ and ${\bar \Gamma}_{\gamma N\to\Delta}$ given in Eq.~(\ref{eq:Sato_Lee_amplitude}) are calculated from 
\begin{equation}
    {\bar \Gamma}_{\Delta\to\pi N} = \bigl[1+t_{\pi N}(E) G_{\pi N}(E)\bigr] \Gamma_{\Delta\to\pi N} ~~~~{\rm and}~~~~
    {\bar \Gamma}_{\gamma N\to\Delta} = \Gamma_{\gamma N\to\Delta} +{\bar \Gamma}_{\pi N\to\Delta}\,G_{\pi N}(E)\,v_{\gamma\pi} .
\end{equation}
This formulation explicitly incorporates meson-baryon dressing effects and ensures unitarity through the consistent treatment of background and resonance contributions. As such, it laid the theoretical foundation for subsequent dynamical coupled-channels analyses of nucleon excitations.

Building on this framework, the multi-channel DCC program known as the Excited Baryon Analysis Center (EBAC) at JLab was established. Matsuyama, Sato, and Lee extended the model to include multiple meson-baryon channels, such as $\pi N$, $\eta N$, and $\pi\pi N$, thereby enabling a unified treatment of hadronic scattering and electromagnetic reactions in the nucleon-resonance region~\cite{Matsuyama:2006rp}. Subsequently, Juli\'{a}-D\'{i}az and collaborators carried out extensive studies within this framework, refining the resonance parameterization and demonstrating that coupled-channel effects play a crucial role in the reliable extraction of resonance properties from pion- and photon-induced reactions~\cite{Julia-Diaz:2006ios,Julia-Diaz:2007qtz,Julia-Diaz:2007mae}.

\begin{figure}[h]
\centering
\includegraphics[width=0.95\columnwidth]{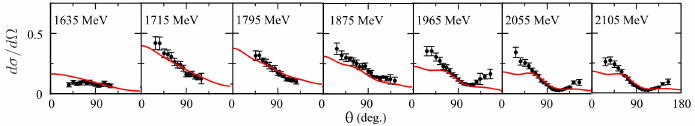}
\caption{Samples of differential cross sections for the $\gamma p \to K^+\Lambda$ reaction calculated within the ANL-Osaka DCC model at different total c.m. energies, as indicated in each panel. The differential cross sections are given in units of $\mu\mathrm{b}/\mathrm{sr}$. Figure adapted from Ref.~\cite{Kamano:2013iva}.}
\label{fig:ANL_cs_KL}
\end{figure}

\begin{figure}[h]
\centering
\includegraphics[width=0.85\columnwidth]{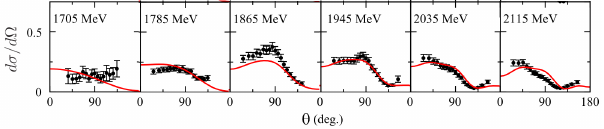}
\caption{As in Fig.~\ref{fig:ANL_cs_KL}, but for the  $\gamma p \to K^+\Sigma^0$ channel. Figure adapted from Ref.~\cite{Kamano:2013iva}.}
\label{fig:ANL_cs_KS}
\end{figure}

The ANL-Osaka DCC model was further improved by Kamano, Nakamura, Lee, and Sato through a combined analysis of the available data for the reactions $\pi N,\gamma N \to \pi N, \eta N, K\Lambda,$ and $K\Sigma$. In particular, very extensive data on $K\Lambda$ and $K\Sigma$ photoproduction, covering c.m. energies up to $W \simeq 2.1$--$2.3~\mathrm{GeV}$, were included for the first time~\cite{Kamano:2008gr,Kamano:2013iva,Kamano:2016bgm}. Within this DCC framework, electromagnetic kaon production has emerged as a particularly stringent constraint on the model. Notably, the model encounters difficulties in accurately reproducing the differential cross sections in the forward and backward angular regions for the $\gamma p \to K^+\Lambda$ reaction~\cite{Kamano:2013iva,Kamano:2015hxa}. These discrepancies become even more pronounced for the $\gamma p \to K^+\Sigma^0$ and $\gamma p \to K^0\Sigma^+$ channels. Figures~\ref{fig:ANL_cs_KL} and~\ref{fig:ANL_cs_KS} compare the experimental differential cross section data for the $K^+\Lambda$ and $K^+\Sigma^0$ channels, respectively, with the corresponding predictions of the ANL-Osaka model, which are shown only at selected values of the c.m. energy $W$.

%%%%%%%%%%%%%%%%%%%%%%%%%%%%%%%%%%%%%%%%%%%%%%%%%%%%%%%%%%%%%%%%%%%%%%%%%%%%%%%%%%%%%%%%%%%%%%%%%%%%%%%%%%%%%%%%%%%%%%%%%%%%%%%%%%%%%%%%%%%%%%%%%%%%%%%%%%%%%%%
%%%%%%%%%%%%%%%%%%%%%%%%%%%%%%%%%%%%%%%%%%%%%%%%%%%%%%%%%%%%%%%%%%%%%%%%%%%%%%%%%%%%%%%%%%%%%%%%%%%%%%%%%%%%%%%%%%%%%%%%%%%%%%%%%%%%%%%%%%%%%%%%%%%%%%%%%%%%%%%

%%%%%%%%%%%%%%%%%%%%%%%%%%%%%%%%%%%%%%%%%%%%%%%%%%%%%%%%%%%%%%%%%%%%%%%%%%%%%%%%%%%%%%%%%%%%%%%%%%%%%%%%%%%%%%%%%%%%%%%%%%%%%%%%%%%%%%%%%%%%%%%%%%%%%%%%%%%%%%%
%%%%%%%%%%%%%%%%%%%%%%%%%%%%%%%%%%%%%%%%%%%%%%%%%%%%%%%%%%%%%%%%%%%%%%%%%%%%%%%%%%%%%%%%%%%%%%%%%%%%%%%%%%%%%%%%%%%%%%%%%%%%%%%%%%%%%%%%%%%%%%%%%%%%%%%%%%%%%%%

The development of the J\"uBo model~\cite{Ronchen:2012eg} builds upon a series of earlier efforts aimed at analyzing multi-channel reaction data within a fully dynamical framework~\cite{Doring:2009bi,Doring:2009yv,Huang:2011as}. In this model, for a given partial wave, the scattering amplitude reads
\begin{eqnarray}
    \label{eq:scat_ampl_JuBo}
    T_{\mu\nu}(p'',p',z) = V_{\mu\nu}(p'',p',z) + \sum_{\kappa}\int_0^\infty dp\, p^2\, \frac{V_{\mu\kappa}(p'',p,z)\, T_{\kappa\nu}(p,p',z)}{z-E_a(p)-E_b(p)+i\epsilon},
\end{eqnarray}
where $p''$ and $p'$ denote the momenta of the outgoing and incoming particles, respectively, $z=\sqrt{s}$ is the total c.m. energy, and $\mu$, $\nu$, and $\kappa$ label the reaction channels. The interaction kernel $V$ consists of non-pole and pole contributions. The non-pole part $V^{\rm NP}$ arises from the sum of the $u$- and $t$-channel exchange diagrams, while the pole part $V^{\rm P}$ originates from $s$-channel resonance exchanges. Together, they define the full interaction kernel $V_{\mu\nu}$ appearing in Eq.~(\ref{eq:scat_ampl_JuBo}), with 
\begin{equation}
    V_{\mu\nu} = V_{\mu\nu}^{\rm NP} + V_{\mu\nu}^{\rm P} \equiv V_{\mu\nu}^{\rm NP} + \sum_{i=0}^n\frac{\gamma_{\mu;i}^{a}\gamma_{\nu;i}^{c}}{z-m_i^b},
\end{equation}
with $\gamma^c (\gamma^a)$ denotes the creation (annihilation) vertex. This separation of the interaction into pole and non-pole terms is advantageous, as it simplifies the numerical implementation and facilitates the extraction of resonance properties within the coupled-channels framework.

The non-pole part of the full scattering $T$-matrix is obtained by unitarizing the non-pole interaction kernel. For the sake of brevity, the momentum integration is omitted, and the resulting expression can be written as
\begin{equation}
    T_{\mu\nu}^{\rm NP} = V_{\mu\nu}^{\rm NP} + \sum_{\kappa} V_{\mu\kappa}^{\rm NP}\, G_\kappa\, T_{\kappa\nu}^{\rm NP},
\end{equation}
where $G_\kappa$ denotes the propagator of the intermediate two- or three-body states in channel $\kappa$. The pole part of the $T$-matrix can then be constructed from the non-pole amplitude $T^{\rm NP}$. To this end, dressed resonance creation and annihilation vertices, as well as the corresponding self-energy, are introduced as
\begin{equation}
    \Gamma^c_{\mu;i} = \gamma^c_{\mu;i} + \sum_\nu \gamma^c_{\nu;i}\, G_\nu\, T_{\nu\mu}^{\rm NP}, ~~~~~
    \Gamma^a_{\mu;i} = \gamma^a_{\mu;i} + \sum_\nu T_{\mu\nu}^{\rm NP}\, G_\nu\,  \gamma^a_{\nu;i}, ~~~~~
    \Sigma_{ij} = \sum_\mu \gamma^c_{\mu;i}\, G_\mu\, \Gamma^a_{j;\mu}.
\end{equation}
In the two-resonance case, the pole part reads~\cite{Doring:2009uc} 
\begin{equation}
    T_{\mu\nu}^{\rm P} = \Gamma_\mu^a\, D^{-1}\, \Gamma_\nu^c, ~~~{\rm with} ~~~ \Gamma_\mu^a = \left( \Gamma_{\mu;1}^a , \Gamma_{\mu;2}^a  \right) , ~~~ \Gamma_\mu^c = \left( \begin{array}{c}
    \Gamma_{\mu;1}^c \\ \Gamma_{\mu;2}^c     \end{array}  \right) , ~~~
    D = \left( \begin{array}{cc}
    z-m_1^b-\Sigma_{11} & -\Sigma_{12} \\ -\Sigma_{21}  & z-m_2^b-\Sigma_{22}   \end{array}  \right).
\end{equation}
The corresponding one-resonance expression follows straightforwardly as a special case. The full scattering amplitude is finally obtained as the sum of pole and non-pole contributions, $T_{\mu\nu}=T_{\mu\nu}^{\rm P}+T_{\mu\nu}^{\rm NP}$. Within this framework, the model provides a unified description of the $\pi N\to \pi N, \eta N, K\Lambda, K\Sigma$ reactions. For this purpose, results from the SAID partial-wave analysis~\cite{Arndt:2006bf} were used as empirical input.

\begin{figure}[htbp]
\centering
\includegraphics[width=1.0\textwidth]{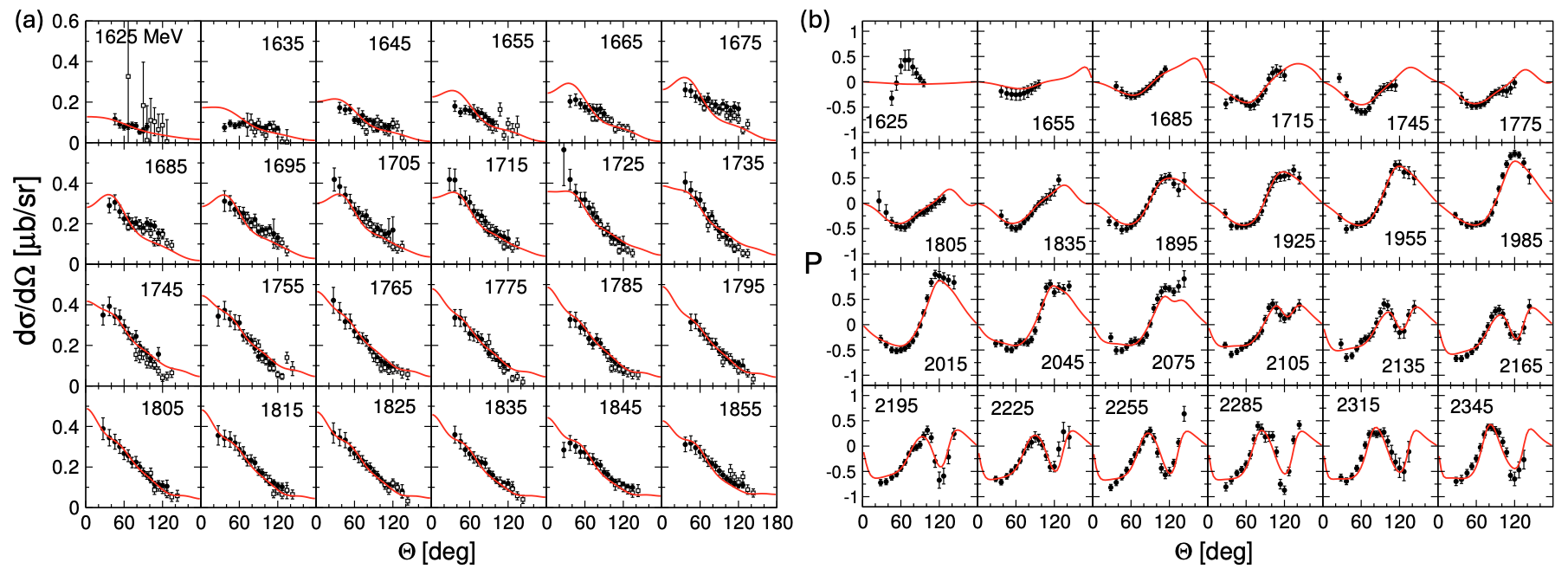}
\caption{(a) Selected differential cross sections for $\gamma p \to K^+\Lambda$ calculated within the J\"uBo DCC model compared with experimental data. (b) Same as the left panel, but for the recoil polarization. Figures from Ref.~\cite{Ronchen:2018ury}. Figures used with kind permission of The European Physical Journal (EPJ).}
\label{fig:Jubo-klam-CS-P}
\end{figure}

The extension to pion photoproduction on the proton was implemented at a later stage~\cite{Ronchen:2014cna}. In analogy to Eq.~(\ref{eq:scat_ampl_JuBo}), the photoproduction multipole amplitude is defined as
\begin{eqnarray}
    \label{eq:multipole_scat_ampl_JuBo}
    M_{\mu\gamma}(q,E) = V_{\mu\gamma}(q,E) + \sum_{\kappa}\int_0^\infty dp\, p^2\, T_{\mu\kappa}(q,p,E)\, G_\kappa (p,E)\, V_{\kappa\gamma}(p,E) \, ,
\end{eqnarray}
with the corresponding photoproduction kernel
\begin{equation}
    \label{eq:kernel_mult_scat_ampl_JuBo}
    V_{\mu\gamma}(p,E) = \alpha_{\mu\gamma}^{\rm NP}(p,E) + \sum_{i}\frac{\gamma_{\mu;i}^{a}(p)\gamma_{\gamma;i}^{c}(E)}{E-m_i^b}\, ,
\end{equation}
where $\alpha_{\mu\gamma}^{\rm NP}(p,E)$ denotes the non-pole contribution, representing the photon couplings to the $t$- and $u$-channel exchange mechanisms, as well as to contact interaction terms. Within this framework, kaon photoproduction was incorporated by including the $\gamma p \to K^+\Lambda$ data. In the energy range from threshold up to $W = 2345~\mathrm{MeV}$, nearly 6000 data points covering differential cross sections as well as single- and double-polarization observables were simultaneously analyzed together with the pion and eta channels within the DCC model~\cite{Ronchen:2018ury}. The quality of the resulting fit is illustrated by representative samples shown in Fig.~\ref{fig:Jubo-klam-CS-P}. The inclusion of the $K^+\Lambda$ channel leaves the pole positions of most established three- and four-star resonances largely unchanged compared to the previous J\"uBo analysis. Notable exceptions are the $N(1710)1/2^+$ and $N(1720)3/2^+$, whose pole positions move closer to the PDG values. Moreover, the analysis indicates that a satisfactory description of the $K^+\Lambda$ data requires the inclusion of the $N(1900)3/2^+$, a feature that is commonly observed in kaon photoproduction studies. 

%%%%%%%%%%%%%%%%%%%%%%%%%%%%%%%%%%%%%%%%%%%%%%%%%%%%%%%%%%%%%%%%%%%%%%%%%%%%%%%%%%%%%%%%%%%%%%%%%%%%%%%%%%%%%%%%%%%%%%%%%%%%%%%%%%%%%%%%%%%%%%%%%%%%%%%%%%%%%%%
%%%%%%%%%%%%%%%%%%%%%%%%%%%%%%%%%%%%%%%%%%%%%%%%%%%%%%%%%%%%%%%%%%%%%%%%%%%%%%%%%%%%%%%%%%%%%%%%%%%%%%%%%%%%%%%%%%%%%%%%%%%%%%%%%%%%%%%%%%%%%%%%%%%%%%%%%%%%%%%

The $K\Sigma$ channels were incorporated at a later stage~\cite{Ronchen:2022hqk}. It was found that the $K^+\Sigma^0$ channel is predominantly governed by $I=3/2$ partial waves, with the notable exception of the $P_{13}$ wave, in which the $N(1900)3/2^+$ provides a qualitative explanation for the cusp-like structure observed in recent BGOOD data~\cite{Jude:2020byj}. It is also noteworthy that the model was employed to clarify the controversy triggered by a recent BESIII measurement~\cite{BESIII:2018cnd}, which reported a value of the decay parameter $\alpha_{-}$ for the parity-violating weak decay $\Lambda\to p\pi^-$ as $0.750 \pm 0.009 \pm 0.004$, significantly larger than the then-established PDG value of $0.642\pm 0.013$. By refitting the J\"uBo model, the best fit was obtained for $\alpha_- = 0.721 \pm 0.006~\text{(statistical)} \pm 0.005~\text{(systematic)}$~\cite{Ireland:2019uja}, clearly favoring the BESIII result.

The extension of the J\"uBo model to the JBW framework is characterized by the inclusion of pion~\cite{Mai:2021vsw} and eta~\cite{Mai:2021aui} electroproduction data in the fitting database. Kaon electroproduction data were subsequently incorporated as well~\cite{Mai:2023cbp}, adding nearly 2800 data points, bringing the total number of fitted data points to more than 110,000. It is important to note that the extension of the model to finite $Q^2$ requires a generalization of the photoproduction kernel in Eq.~(\ref{eq:kernel_mult_scat_ampl_JuBo}) to
\begin{equation}
    \label{eq:kernel_electro_JuBo}
    V_{\mu\gamma}(p,W,Q^2) = \alpha_{\mu\gamma}^{\rm NP}(p,W,Q^2) + \sum_{i}\frac{\gamma_{\mu;i}^{a}(p)\gamma_{\gamma;i}^{c}(W,Q^2)}{W-m_i^b}\, .
\end{equation}
The $Q^2$-dependence of the non-pole photon coupling $\alpha_{\mu\gamma}^{\rm NP}(p,W,Q^2)$ and the electromagnetic creation vertex $\gamma_{\gamma;i}^{c}(W,Q^2)$ is introduced through a phenomenological ansatz,
\begin{equation}
    \alpha_{\mu\gamma}^{\rm NP}(p,W,Q^2) = {\tilde F}_\mu (Q^2)\, \alpha_{\mu\gamma}^{\rm NP}(p,W)  ~~~~~\text{and} ~~~~~
    \gamma_{\gamma;i}^{c}(W,Q^2) = {\tilde F}_i (Q^2) \,\gamma_{\gamma;i}^{c}(W) .
\end{equation}
Consequently, the entire $Q^2$ dependence is absorbed into the channel-dependent form factors ${\tilde F}_\mu (Q^2)$ and the resonance-dependent form factors ${\tilde F}_i (Q^2)$, which are parameterized as
\begin{equation}
    {\tilde F}_\mu (Q^2) = {\tilde F}_D (Q^2)\, e^{\beta_\mu^0 Q^2/m^2}\, P^N(Q^2/m^2,\vec{\beta}_\mu) ~~~~~\text{and} ~~~~~
    {\tilde F}_i (Q^2) = {\tilde F}_D (Q^2)\, e^{\delta_i^0 Q^2/m^2}\, P^N(Q^2/m^2,\vec{\delta}_i) ,
\end{equation}
respectively, where $P^N$ denotes a general polynomial and $\beta_\mu^0$ and $\delta_i^0$ are fit parameters.

\begin{figure}[htbp]
\centering
\includegraphics[width=1.0\textwidth]{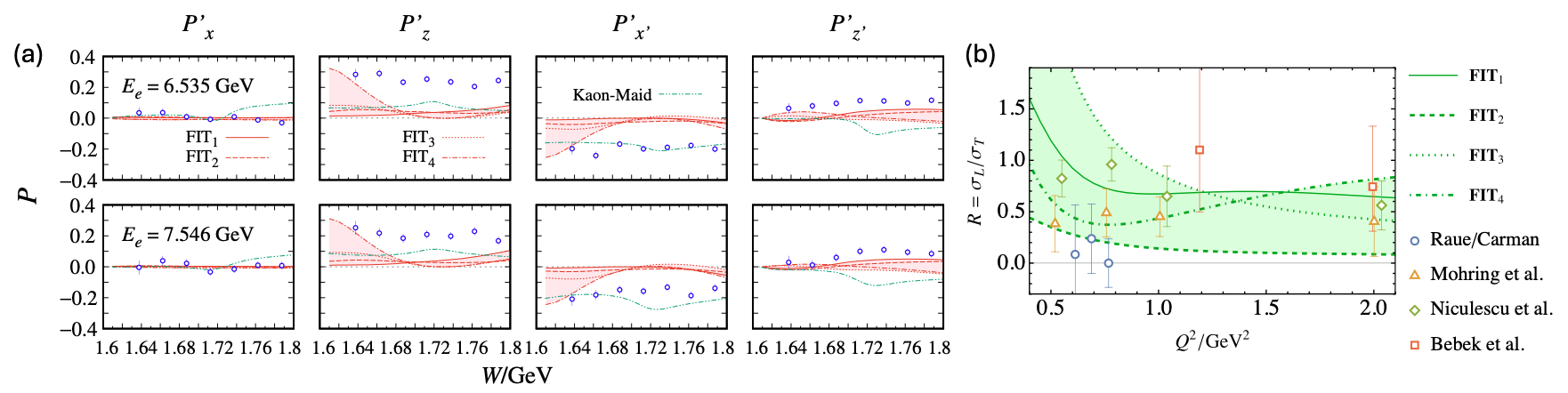}
\caption{(a) Comparison of the transferred $\Lambda$ polarization components $\mathcal{P}'_x$, $\mathcal{P}'_z$, $\mathcal{P}'_{x'}$, and $\mathcal{P}'_{z'}$ with predictions from the JBW DCC model~\cite{Mai:2023cbp} and Kaon-MAID~\cite{kaonmaid}. The shaded bands connect the different solutions denoted by $\mathrm{FIT}_{1,\ldots,4}$ to guide the eye. Predictions from Kaon-Maid are displayed as green dash-dot-dotted lines. The experimental data are from Ref.~\cite{CLAS:2022yzd}. (b) Ratio of the longitudinal to transverse $K^+\Lambda$ cross sections as a function of $Q^2$. Predictions from the JBW DCC model are shown by dark green lines at $\theta = 0^\circ$ and $W = 1.84~\mathrm{GeV}$, with the shaded region connecting different solutions to guide the eye. Figures from Ref.~\cite{Mai:2023cbp}. Reprinted with kind permission of The European Physical Journal (EPJ).}
\label{fig:JBW_2023_electro}
\end{figure}

The $K^+\Lambda$ electroproduction data can be described very well within the JBW framework and were even found to be not sufficiently restrictive, as indicated by $\chi^2$ per datum values ranging between 0.32 and 0.70 for the four solutions presented in Ref.~\cite{Mai:2023cbp}. It should be noted that, in the JBW model, the kinematic coverage is limited to $W < 1.8$~GeV and $Q^2 < 8~\mathrm{GeV}^2$. More recently, new CLAS12 data on the transferred $\Lambda$ polarization components $\mathcal{P}'_x$, $\mathcal{P}'_z$, $\mathcal{P}'_{x'}$, and $\mathcal{P}'_{z'}$ have become available~\cite{CLAS:2022yzd}. Since these observables are provided as acceptance-weighted, bin-averaged quantities extracted in finite kinematic bins of $Q^2$, $W$, and $\cos\theta_K^{\rm c.m.}$ (see Section~\ref{clas12-program}), a two-dimensional integration is required for each data point, which significantly complicates their direct inclusion in the fitting procedure. Figure~\ref{fig:JBW_2023_electro}(a) compares these data with predictions from the JBW and Kaon-MAID models. While both models reproduce several qualitative features of the measurements, noticeable deviations remain in certain kinematic regions. It is also important to note that a recent effort to partially fit these new CLAS12 data within an isobar-model framework has been reported in Ref.~\cite{Djaja:2025hzz}.

On the other hand, the ratio of longitudinal to transverse cross sections in meson electroproduction provides a stringent test of meson pole dominance and meson form-factor extractions~\cite{Vanderhaeghen:1997ts,Horn:2007ug}, the onset of QCD factorization and handbag dynamics~\cite{Collins:1996fb,Guidal:2004nd}, the relative importance of chiral-even versus chiral-odd generalized parton distributions~\cite{Goloskokov:2011rd}, and, within a coupled-channels framework, the helicity structure of nucleon resonance electroexcitation and the interplay between resonant and non-resonant mechanisms~\cite{Drechsel:2007if,Kamano:2013iva,Tiator:2011pw}. The predicted ratios for the $K^+\Lambda$ electroproduction, compared with the available experimental data, are shown in the Fig.~\ref{fig:JBW_2023_electro}(b). While the JBW predictions are consistent with the data within the quoted uncertainties, the relatively large experimental error bars currently limit the discriminating power of this observable.

%%%%%%%%%%%%%%%%%%%%%%%%%%%%%%%%%%%%%%%%%%%%%%%%%%%%%%%%%%%%%%%%%%%%%%%%%%%%%%%%%%%%%%%%%%%%%%%%%%%%%%%%%%%%%%%%%%%%%%%%%%%%%%%%%%%%%%%%%%%%%%%%%%%%%%%%%%%%%%%
%%%%%%%%%%%%%%%%%%%%%%%%%%%%%%%%%%%%%%%%%%%%%%%%%%%%%%%%%%%%%%%%%%%%%%%%%%%%%%%%%%%%%%%%%%%%%%%%%%%%%%%%%%%%%%%%%%%%%%%%%%%%%%%%%%%%%%%%%%%%%%%%%%%%%%%%%%%%%%%

\begin{figure}[htbp]
\centering
\includegraphics[width=0.7\textwidth]{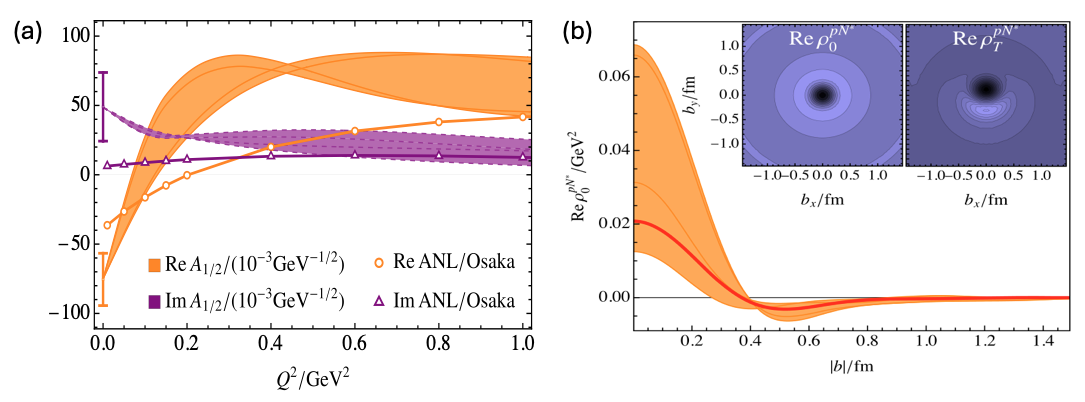}
\caption{(a) Real and imaginary parts of the $p \to N(1440)1/2^+$ transition form factors at low $Q^2$, obtained from the JBW DCC model~\cite{Wang:2024byt} and the ANL-Osaka model~\cite{Kamano:2018sfb}. The two error bars at $Q^2 = 0$ indicate the uncertainties of the photoproduction pole solution. (b) Transverse charge density for the $p \to N(1440)1/2^+$ transition as a function of the transverse distance $b$ in the $xy$ plane. Figures from Ref.~\cite{Wang:2024byt}.}
\label{fig:Wang-2024-TFF}
\end{figure}

Finally, the JBW DCC model was used to extract twelve $N^*$ and $\Delta^*$ transition form factors (TFFs) between ground- and excited-state baryons at the resonance poles~\cite{Wang:2024byt}. This extraction is unique in that, for the first time, it was carried out using more than $10^5$ experimental data points covering most available meson-nucleon scattering and meson electromagnetic production processes. For the $\Delta(1232)3/2^+$ and $N(1440)1/2^+$ resonances, the results are in good agreement with previous studies, whereas for the other excited states this work constitutes the first such effort. For the $N(1440)$ Roper resonance, the extracted TFFs are shown in Fig.~\ref{fig:Wang-2024-TFF}. The JBW results are consistent with previous findings, namely the presence of a zero crossing in the real part of the transition form factor. In the JBW analysis, this zero occurs at a smaller value than in the ANL-Osaka case, from which it is concluded that the $N(1440)$ core can be interpreted as a radial excitation of the nucleon, despite the overall complexity of its structure. The helicity amplitudes determined in the JBW study allow for the determination of the transverse transition charge density $\rho$, shown in Fig.~\ref{fig:Wang-2024-TFF}(b), from which a positively charged central region surrounded by a weakly negatively charged distribution is observed.

%%%%%%%%%%%%%%%%%%%%%%%%%%%%%%%%%%%%%%%%%%%%%%%%%%%%%%%%%%%%%%%%%%%%%%%%%%%%%%%%%%%%%%%%%%%%%%%%%%%%%%%%%%%%%%%%%%%%%%%%%%%%%%%%%%%%%%%%%%%%%%%%%%%%%%%%%%%%%%%
%%%%%%%%%%%%%%%%%%%%%%%%%%%%%%%%%%%%%%%%%%%%%%%%%%%%%%%%%%%%%%%%%%%%%%%%%%%%%%%%%%%%%%%%%%%%%%%%%%%%%%%%%%%%%%%%%%%%%%%%%%%%%%%%%%%%%%%%%%%%%%%%%%%%%%%%%%%%%%%

\subsubsection{$K$-Matrix Models}

In contrast to the DCC models, a larger number of coupled-channel analyses rely on the $K$-matrix formalism. These include the Kent State University (KSU) model~\cite{Hunt:2018mrt,Manley:1992yb,Fernandez-Ramirez:2015tfa,Shrestha:2012ep,Hunt:2018wqz}, the George Washington University Institute for Nuclear Studies (GWU-INS or SAID) model~\cite{Arndt:1994br,Arndt:1995bj,Arndt:1995ak,Arndt:2007qn,Arndt:2006bf,Workman:2011vb,Briscoe:2023gmb}, the MAID model~\cite{Knochlein:1995qz,Mart:1999ed,Kamalov:1999hs,Fix:2005if,Drechsel:2007if,Kamalov:2000en,Chiang:2001as,Hilt:2013fda}, the Giessen model~\cite{Feuster:1998cj,Feuster:1997pq,Penner:2002ma,Shklyar:2004ba,Shklyar:2014kra}, and the Bonn-Gatchina model~\cite{Anisovich:2011fc,Anisovich:2004zz,Sarantsev:2005tg,Anisovich:2006bc}. It is important to note that in both the GWU-INS and MAID models the electromagnetic production of kaons is handled as an independent reaction channel. Consequently, the resulting framework corresponds naturally to the Kaon-Maid isobar model, which has already been discussed in Section~\ref{Subsec:isobar_models}. A related effort was undertaken in Ref.~\cite{Svarc:2021gcs}, where the $K^+\Lambda$ photoproduction data were analyzed using the single-channel, single-energy amplitude (partial-wave) analysis method, previously developed and validated for $\eta$ photoproduction~\cite{Svarc:2020cic}. By examining the experimental data in the energy range $1625<W<2296$ MeV, this study confirmed the sizes and shapes of the multipoles obtained by the Bonn-Gatchina group~\cite{Anisovich:2017bsk,Anisovich:2017ygb}.

The SAID (Scattering Analysis Interactive Dial-in) program, maintained by the GWU-INS, evolved from a long sequence of partial wave analyses (PWA) performed by Arndt and collaborators beginning in the 1980s. The earliest foundations were laid in the development of global comprehensive analyses of $\pi N$ elastic scattering data based on the $K$-matrix parametrization and energy-dependent fits to the world database~\cite{Arndt:1985vj}. These analyses established the methodological framework, unitarized parametrizations and global optimization, which later became the backbone of the SAID project. The latest update of this model was reported in Ref.~\cite{Briscoe:2023gmb}, where it was extended to provide a unified description of single-pion photoproduction together with pion and eta hadroproduction within the Chew-Mandelstam $K$-matrix approach. SAID analyses are widely regarded as an empirical benchmark for further phenomenological and dynamical analyses of meson-baryon interactions and electromagnetic production, as they provide stable, data-driven amplitudes constrained by unitarity and analyticity in well-measured channels. Consequently, many subsequent studies use SAID solutions as reference inputs or consistency checks, while extending beyond them through additional channels, dynamical mechanisms, or alternative unitarization schemes.

The early MAID model, commonly referred to as MAID2000, was developed by Drechsel, Hanstein, Kamalov, and Tiator as a unitary isobar model for pion photo- and electroproduction on the nucleon~\cite{Drechsel:1998hk}. Its original motivation was to provide a unitary, flexible, and computationally efficient framework capable of describing electromagnetic pion production data over a wide energy range, while allowing for extraction of nucleon resonance properties through a $K$-matrix unitarization. The model includes non-resonant Born terms, vector-meson exchange contributions, and explicit nucleon-resonance excitations, i.e., the $\Delta(1232)3/2^+$, $N(1440)1/2^+$, $N(1520)3/2^-$, $N(1535)1/2^-$, $N(1680)5/2^+$, and $\Delta(1700)3/2^-$ states. %Unitarity is imposed to ensure the correct phase behavior of the pion photoproduction multipoles, consistent with elastic $\pi N$ scattering. 
In the background terms, special attention is given to the $E_{0+}$ multipole, which exhibits a relatively large imaginary part even at low pion energies. The origin of this behavior is well known and arises from $\pi N$ rescattering effects. To account for this mechanism, the Fermi-Watson theorem is implemented through a $K$-matrix unitarization of the background contribution, leading to
\begin{equation}
    E_{0+}^{(I)} = E_{0+}^{(I)} ({\rm Born}+\omega,\rho)(1+it^I_{\pi N}),
\end{equation}
where $t^I_{\pi N}=[\eta_I\, \exp(i\delta^I_{\pi N})-1]/2i$ denotes the elastic $\pi N$ scattering amplitude in the isospin channel $I$, expressed in terms of the corresponding phase shift $\delta^I_{\pi N}$ and inelasticity parameter $\eta_I$. The resonance contributions to the multipoles are parameterized by assuming a Breit-Wigner energy dependence,
\begin{equation}
    A_{\ell\pm}={\bar A}_{\ell\pm}f_{\gamma N}(W)\frac{\Gamma_{\rm tot}W_R e^{i\phi}}{W_R^2-W^2-iW_R\Gamma_{\rm tot}}f_{\pi N}(W)C_{\pi N},
\end{equation}
where $f_{\gamma N}(W)$ and $f_{\pi N}(W)$ are the Breit-Wigner factors describing the $\gamma NN^*$ and $\pi NN^*$ vertices, respectively. The unitary phase $\phi$ plays a crucial role in this parameterization: it is chosen such that the phase of the total multipole amplitude, including both background and resonance contributions, reproduces the corresponding $\pi N$ scattering phase $\delta_{\pi N}$, in accordance with the Fermi-Watson theorem. On the other hand, the extension to finite $Q^2$ is achieved by introducing appropriate electromagnetic form factors in the helicity amplitudes. With these ingredients, MAID2000 provides a good overall description of pion photo- and electroproduction observables, including differential cross sections and polarization asymmetries.

The MAID2003 and MAID2007 versions represented successive developments driven by the rapidly increasing precision and volume of photo- and electroproduction data from MAMI, ELSA, and JLab. In MAID2003, the resonance parameters and background amplitudes were refitted over an extended energy range, with particular emphasis on improving the description of electroproduction observables~\cite{Kamalov:2000en}. MAID2007 further advanced the framework by incorporating additional resonances, refining the unitarization procedure to more consistently satisfy Watson's theorem below the two-pion threshold, and performing a comprehensive global fit that included a wide range of new polarization data~\cite{Drechsel:2007if}. As a result, MAID2007 became one of the standard references for unitary isobar analyses of pion photo- and electroproduction.

The KSU model was originally applied only to the $\pi N$ sector~\cite{Manley:1992yb,Fernandez-Ramirez:2015tfa,Shrestha:2012ep}. Within this framework, the unitary and symmetric scattering matrix ${\bf S}$ is constructed as
\begin{eqnarray}
    {\bf S}={\bf B}^{\rm T}\,{\bf R\, B}\,,
\end{eqnarray}
where ${\bf R}$ denotes the resonant contribution to the scattering amplitude and ${\bf B}$ represents the product of unitary and symmetric background matrices ${\bf B}={\bf B}_1{\bf B}_2\cdots{\bf B}_n$. The resonant matrix ${\bf R}$ is defined by
\begin{equation}
    {\bf R} = {\bf I}+2i{\bf T_R} = {\bf I}+2i{\bf K}({\bf I}-i{\bf K})^{-1} = ({\bf I}+i{\bf K})({\bf I}-i{\bf K})^{-1},
\end{equation}
where ${\bf T_R}$ is the resonant transition matrix and ${\bf K}$ is the real and symmetric $K$-matrix. 
In recent years the KSU model has been extended to include the $K^+\Lambda$ channel, using partial wave analysis and experimental data from threshold up to $W = 2200$ MeV~\cite{Hunt:2018mrt,Hunt:2018wqz}. The photoproduction datasets incorporated in this analysis consist of differential cross sections, single-polarization observables ($\Sigma$, $T$, and $P$), and double-polarization observables ($G$, $H$, $F$, $E$, $O$, $C$, $T$, and $L$). The results indicated the presence of the $N(1880)1/2^+$, $N(2120)3/2^-$, and $N(2080)5/2^-$ resonances, along with a possible $J^P=7/2^+$ resonance near 2300 MeV. The latest update of the KSU model was reported in Ref.~\cite{Hunt:2018wqz}.

The foundations of the Giessen model were established in the late 1990s through the work of Feuster and Mosel~\cite{Feuster:1997pq,Feuster:1998cj}, who formulated a multi-channel $K$-matrix framework based on effective hadronic Lagrangians constrained by SU(3) symmetry. Their approach simultaneously considered the $\pi N$, $2\pi N$, $\eta N$, and $K\Lambda/K\Sigma$ channels. The hadronic sector of the model was established in the initial work~\cite{Feuster:1997pq}, while the electromagnetic interaction with $W$ from threshold up to 2 GeV was incorporated in a subsequent study~\cite{Feuster:1998cj}. Since the contribution from Compton scattering can be neglected and, moreover, only two of the four independent Compton amplitudes can be extracted, photon-rescattering effects were omitted in the extraction of photoproduction amplitudes from the data. Schematically, the photoproduction amplitude for a scalar meson $\varphi$ can be written as
\begin{equation}
    T^{I_\gamma}_{\varphi\gamma} = V^{I_\gamma}_{\varphi\gamma} + i\sum_{\varphi'} T^{I_\varphi}_{\varphi\varphi'}V^{I_\gamma}_{\varphi'\gamma} \, , ~~~~{\rm with}~~~~ I_\gamma = 0,\tfrac{1}{2},\tfrac{3}{2} ~~~~ {\rm and} ~~~~ I_\varphi = \tfrac{1}{2},\tfrac{3}{2} ,
\end{equation}
where $I_\gamma$ and $I_\varphi$ denote the isospin of the photon-nucleon and meson-nucleon systems, respectively. For Compton scattering, the corresponding amplitude is given by
\begin{equation}
    T^{p,n}_{\gamma\gamma} = V^{p,n}_{\gamma\gamma} + i\sum_{c} T_{\gamma c}V_{c\gamma} \,  ,  ~~~~{\rm with,~ e.g.,}~~ c=\pi^0 p, \pi^+ n, .~.~. ~~ {\rm for}~ \gamma p\to \gamma p \, .
\end{equation}

The next major step in the development of the Giessen model was the extension of the framework to a comprehensive treatment of vector-meson production, in particular the $\rho N$, $\omega N$, and $\phi N$ channels~\cite{Penner:2002ma,Penner:2002md}. In these works, the emphasis was placed on performing a global coupled-channel analysis of hadronic reactions ($\pi N \to \pi N,\eta N,\pi\pi N$) together with photon-induced processes ($\gamma N \to \rho N,\omega N,\phi N$). The primary goal was to refine the nucleon-resonance spectrum and to constrain non-resonant background mechanisms over a broad kinematic range relevant to the resonance region. Although kaon photoproduction was not incorporated yet, the model's dynamical foundation is strengthened by including higher-mass vector-meson final states.

\begin{figure}[htb]
\centering
\includegraphics[width=0.65\columnwidth]{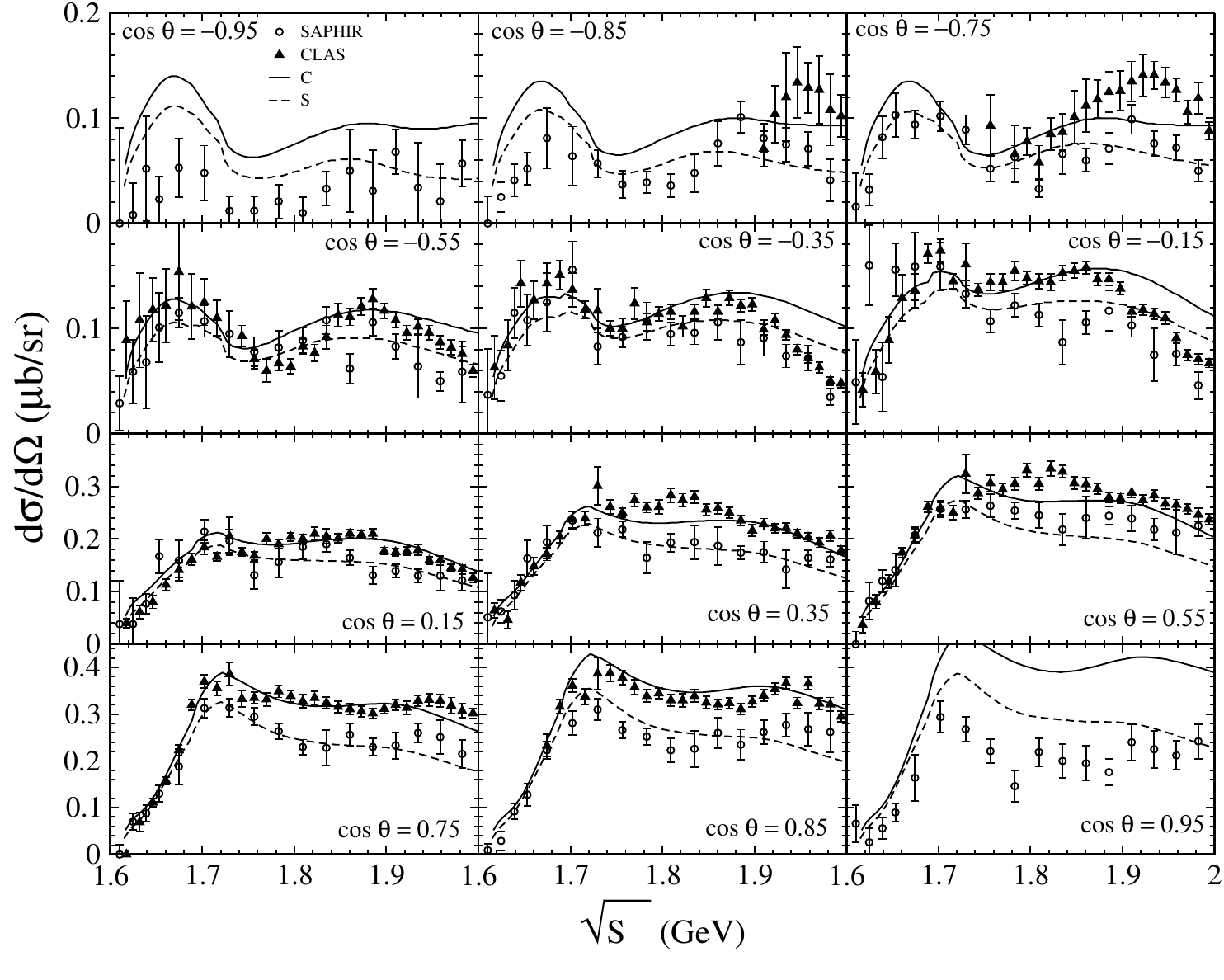}
\caption{Comparison of the $\gamma p \to K^+\Lambda$ differential cross sections obtained from the Giessen model with experimental data from SAPHIR~\cite{Glander:2003jw} and CLAS~\cite{CLAS:2003zrd}. The solid and dashed curves correspond to fits to the CLAS (C parameter set) and SAPHIR (S parameter set) data, respectively. Figure from Ref.~\cite{Shklyar:2005xg}.}
\label{fig:Giessen_Shklyar}
\end{figure}

In Ref.~\cite{Shklyar:2004ba}, the Giessen group carried out a full coupled-channel analysis of $\omega$ photoproduction, demonstrating that multi-channel dynamics, particularly intermediate $\pi N$, $2\pi N$, $\eta N$, and $\rho N$ states, are essential for reproducing the observed structures in the data. The follow-up study~\cite{Shklyar:2005xg} extended the model to investigate kaon production and, in particular, to examine the proposed $N(1895)3/2^-$ “missing” resonance highlighted in Ref.~\cite{Mart:1999ed}. This was achieved through a combined analysis of the $(\pi,\gamma)N\to K\Lambda$ channels from threshold up to $W \approx 2~\text{GeV}$, providing a dynamically consistent description of strangeness photoproduction within the Giessen scheme. The analysis identified the $N(1650)1/2^-$, $N(1720)3/2^+$, and $N(1900)3/2^+$ as providing the dominant contributions, while no evidence was found for the “missing” $N(1895)3/2^-$. Instead, the second peak in the cross section arises from the coherent superposition of resonance and background contributions. Furthermore, the issue of data consistency between the CLAS and SAPHIR photoproduction cross sections was also examined, yielding two distinct solutions associated with the respective datasets as shown in Fig.~\ref{fig:Giessen_Shklyar}. Nevertheless, the conclusion regarding the three dominant resonances is insensitive to these discrepancies.

%%%%%%%%%%%%%%%%%%%%%%%%%%%%%%%%%%%%%%%%%%%%%%%%%%%%%%%%%%%%%%%%%%%%%%%%%%%%%%%%%%%%%%%%%%%%%%%%%%%%%%%%%%%%%%%%%%%%%%%%%%%%%%%%%%%%%%%%%%%%%%%%%%%%%%%%%%%%%%%
%%%%%%%%%%%%%%%%%%%%%%%%%%%%%%%%%%%%%%%%%%%%%%%%%%%%%%%%%%%%%%%%%%%%%%%%%%%%%%%%%%%%%%%%%%%%%%%%%%%%%%%%%%%%%%%%%%%%%%%%%%%%%%%%%%%%%%%%%%%%%%%%%%%%%%%%%%%%%%%

\begin{figure}[htb]
\centering
\includegraphics[width=\columnwidth]{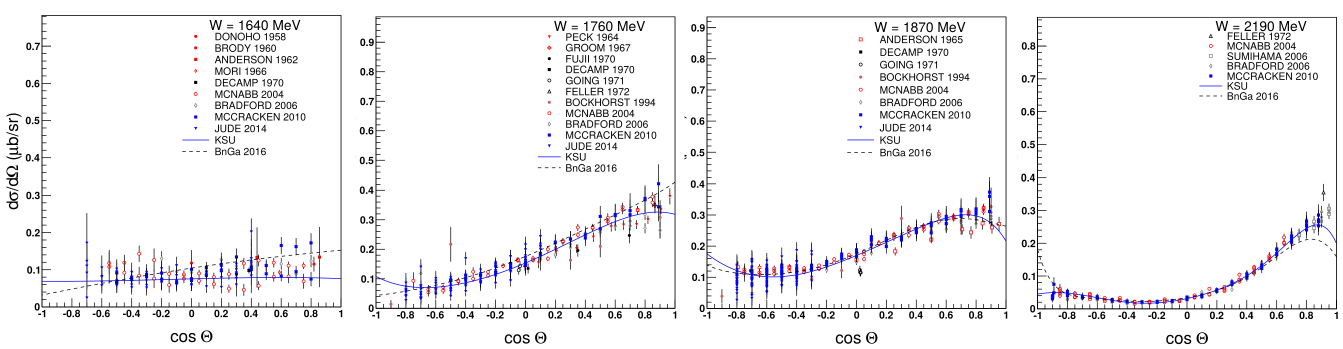}
\caption{Sample of the $\gamma p \to K^+\Lambda$ differential cross sections vs. $\cos \theta_K^{c.m.}$ for different $W$ bins (as labeled) predicted by the KSU (solid lines)~\cite{Hunt:2018mrt} and BnGa 2016 (dashed lines) \cite{Anisovich:2016vzt} models. Figures adapted from Ref.~\cite{Hunt:2018mrt}.}
\label{fig:KSU_BnGA_difCS}
\end{figure}

The Bonn-Gatchina (BnGa) partial wave analysis originated in the early 2000s with studies of pion-induced and pion-photoproduction reactions. It was formulated using a relativistic multi-channel parameterization of the scattering amplitudes based on a $K$-matrix-type formalism~\cite{Anisovich:2004zz,Sarantsev:2005tg,Anisovich:2006bc}. Subsequently, the kaon photoproduction channels, $\gamma p \to K^+\Lambda$ and $\gamma p \to K^+\Sigma^0$, were incorporated into the global database, with analyses extending up to $W \approx 2.3~\mathrm{GeV}$ \cite{Anisovich:2010an,Anisovich:2012ct,Anisovich:2014yza}. With the availability of increasingly precise data from ELSA, CLAS, and SAPHIR, the BnGa model was further developed to include a wide range of single- and double-polarization observables, covering the full angular distributions of the measured reactions~\cite{Nikonov:2007br,Anisovich:2011fc}. A small sample comparing the differential cross sections calculated using the BnGa and KSU models is shown in Fig.~\ref{fig:KSU_BnGA_difCS}. Despite visible differences between the two calculations, both models reproduce the experimental data well within the quoted error bars.

Together with the KSU, GWU-INS, MAID, and Giessen models, the BnGa approach has made a significant contribution to our understanding of nucleon resonances that are strongly coupled to the $K\Lambda$ and $K\Sigma$ channels. In particular, through the use of the kaon photoproduction channel $\gamma p \to K^+\Lambda$, the BnGa model~\cite{Anisovich:2017ygb} has firmly established the $N(1895)1/2^-$ and $N(1900)3/2^+$, and has placed strong constraints on the properties of the $N(1875)3/2^-$ and $N(2060)5/2^-$. These results indicate that these resonances play an important role in this reaction channel. It is noteworthy that the $N(1900)3/2^+$ had already been identified as a contributor to the second peak of the $K^+\Lambda$ cross section in earlier studies~\cite{Nikonov:2007br,Mart:2012fa}.

%%%%%%%%%%%%%%%%%%%%%%%%%%%%%%%%%%%%%%%%%%%%%%%%%%%%%%%%%%%%%%%%%%%%%%%%%%%%%%%%%%%%%%%%%%%%%%%%%%%%%%%%%%%%%%%%%%%%%%%%%%%%%%%%%%%%%%%%%%%%%%%%%%%%%%%%%%%%%%%
%%%%%%%%%%%%%%%%%%%%%%%%%%%%%%%%%%%%%%%%%%%%%%%%%%%%%%%%%%%%%%%%%%%%%%%%%%%%%%%%%%%%%%%%%%%%%%%%%%%%%%%%%%%%%%%%%%%%%%%%%%%%%%%%%%%%%%%%%%%%%%%%%%%%%%%%%%%%%%%

%=============================================
%=============================================
%=============================================

\subsection{Regge and Hybrid Models}

One major problem of isobars model is that they only work reliably within a limited energy domain \cite{Corthals07}. At higher energies, the role of individual resonances diminishes, and the reaction becomes dominated by kaon resonance exchanges in the $t$-channel \cite{Okuyama:2023lxq}. Since this $t$-channel background is built from many resonances, no clear structures appear in that regime. Incorporating all these contributions explicitly would require introducing a copious number of free parameters, making the model unnecessarily complicated. Furthermore, isobar models also violate the Froissart bound, which limits the total scattering cross section to rise no faster than $\left(\log s\right)^2$ \cite{Corthals07,Froissart:1961}. On the other hand, the background terms in isobar models grow with a positive power of $s$, outpacing the logarithmic bound. For a more in-depth discussion of the Froissart bound, we refer readers to Ref.~\cite{Froissart:1961}.

A way to resolve the shortcomings of isobar models was introduced in 1959 by Regge. His idea was to extend the concept of partial wave amplitudes by treating the angular momentum as a complex variable \cite{Regge:1959mz}. In this framework, the poles of the amplitude correspond to resonant states, which naturally arrange themselves into families known as Regge trajectories. The states along a given trajectory share the same internal quantum numbers but differ in spin, and thus can be seen as related excitations of the same underlying structure. In other words, a Regge trajectory represents a whole class of particles with common internal quantum numbers but distinct masses and angular momenta \cite{Corthals07}. Unlike the isobar model, where interactions are described through the exchange of individual hadrons, Regge theory replaces these by the exchange of entire trajectories. For an in-depth treatment of Regge theory and its applications, see Refs.~\cite{Collins:1971ff,Irving:1977ea}. 

Interest in applying Regge theory to kaon photo- and electroproduction has grown steadily since the late 1990s, following reports that Regge-based approaches reproduce high-energy data with great accuracy \cite{Janssen02}. A well-known example is the series of studies by Guidal, Laget, and Vanderhaeghen (GLV) \cite{Guidal:2003qs,Vanderhaeghen:1997ts}. In addition, several other Regge-inspired frameworks, such as Reggeized unitary isobar models \cite{Aznauryan:2002gd,Aznauryan:2003zg} and the quark-gluon string model \cite{Grishina:2005cy}, have also been reported. The models discussed in this subsection have been summarized in Table~\ref{tab:regge_summary}.

\subsubsection{Regge Models}

\begin{figure}[hbt!]
\centering
\includegraphics[scale=0.3]{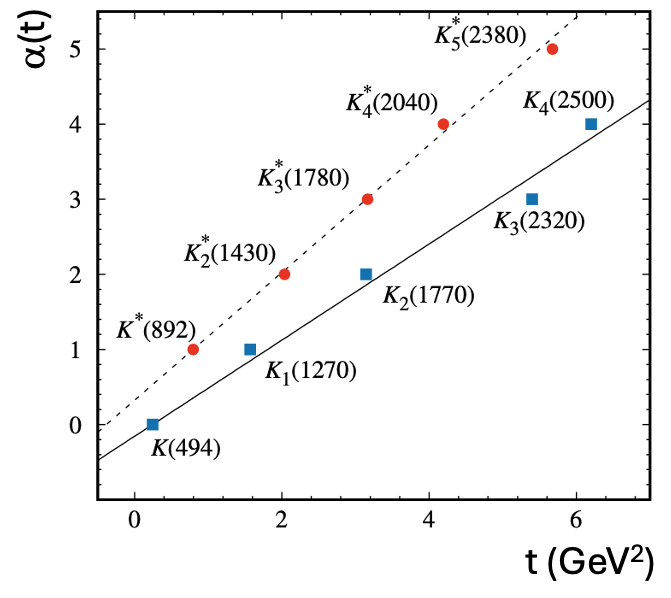}
\caption{The Chew-Frautschi plot for the $K(494)$ and $K^*(892)$ Regge trajectories. The mass of each particle is taken from Ref.~\cite{ParticleDataGroup:2024cfk}. The trajectories satisfy $\alpha_K(t)=0.64(t-m_K^2)$ and $\alpha_{K^*}(t)=1+0.85(t-m_{K^*}^2)$.}
\label{fig:Chew Frautschi Plot}
\end{figure}

Regge theory revolves around the idea that the spin $\alpha$ of a particle is related to its mass squared through Regge trajectories $\alpha(t)$. Concretely, when the spins of a family of resonant states are plotted against their squared masses on what is known as a Chew-Frautschi plot, the points follow an approximately linear relationship given by \cite{Janssen02,Vancraeyveld:2011mta}
\begin{equation}
	\alpha_X(t)=\alpha_{X,0}+\alpha'(t-m_X^2),
\end{equation}
where $\alpha_{X,0}$ and $m_X$ denote the spin and mass of the lightest particle on the trajectory, respectively. This lightest state is referred to as the first materialization of the trajectory \cite{Corthals07,Skoupil11}. The slope parameter $\alpha'$ was found to have an approximate value of $\alpha' \simeq 0.8~\mathrm{GeV}^{-2}$ for almost all known meson and baryon trajectories \cite{Janssen02}. An example of a Chew-Frautschi plot for the $K(494)$ and $K^*(892)$ Regge trajectories is shown in Fig.~\ref{fig:Chew Frautschi Plot}. The Regge trajectories of $K(494)$ and $K^*(892)$ are parameterized as $\alpha_K(t)=0.64(t-m_K^2)$ and $\alpha_{K^*}(t)=1+0.85(t-m_{K^*}^2)$.

%%%%%%%%%%%%%%%%%%%%%%%%%%%%%%%%%%%%%%%%%%%%%%%%%%%%%%%%%%%%%%%%%%%%%%%%%%%%%%%%%%%%%%%%%%%%%%%%%%%%%%%%%%%%%%%%%%%%%%%%%%%%%%%%%%%%%%%%%%%%%%%%%%%%%%%%%%%%%%%
%%%%%%%%%%%%%%%%%%%%%%%%%%%%%%%%%%%%%%%%%%%%%%%%%%%%%%%%%%%%%%%%%%%%%%%%%%%%%%%%%%%%%%%%%%%%%%%%%%%%%%%%%%%%%%%%%%%%%%%%%%%%%%%%%%%%%%%%%%%%%%%%%%%%%%%%%%%%%%%

Regge models for kaon photo- and electroproduction on the nucleon have been developed since the late 1960s. Some of the pioneering works can be found in Refs.~\cite{Ader:1967tqj,Ball:1968zza,Shih:1969wc,Cooper:1969ma}. However, one of the most notable foundational works was the work of Levy, Majerotto, and Read \cite{Levy:1973aq,Levy:1973zz}. They proposed that for sufficiently high energies, the kaon photo- and electroproduction reactions at forward angles are dominated by the $t$-channel $K$ and $K^*$ Regge exchanges (equivalently, $u$-channel exchanges for backward angles). The Regge exchanges were introduced by first considering the reactions through the Feynman diagrammatic technique, and then generalizing it to the exchange of a Regge trajectory by replacing the pole-like Feynman propagator $(t-m^2)^{-1}$ with the so-called Regge propagator, while keeping the vertex structure of the Feynman diagrams associated with the first materialization of the trajectory \cite{Levy:1973aq,Guidal:1997hy}.

The Regge amplitude is given by (see Refs.~\cite{Janssen02,Donnachie:2002en} for the detailed derivation)
\begin{equation}\label{eq:regge_amplitude}
    \mathcal{M}_\mathrm{Regge}^{\zeta=\pm}(s,t)=\pi\alpha'\left(\frac{s}{s_0}\right)^{\alpha(t)}\frac{\beta(t)}{\sin \pi\alpha(t)}\frac{1+\zeta e^{-i\pi \alpha(t)}}{2}\frac{1}{\Gamma(\alpha(t)+1)},
\end{equation}
where $s_0\equiv 1~\mathrm{GeV}^2$ is a scale factor and $\beta(t)$ is a residue function. To determine the Regge propagator,  we first define
\begin{equation}\label{eq:regge_amplitude_propagator}
    \mathcal{M}_\mathrm{Regge}(s,t)=\beta(t)\times\mathcal{P}_\mathrm{Regge}(s,t).
\end{equation}
By comparing Eqs.~(\ref{eq:regge_amplitude}) and (\ref{eq:regge_amplitude_propagator}) we obtain
\begin{equation}\label{eq:regge_propagator}
    \mathcal{P}_\mathrm{Regge}^{\zeta=\pm}(s,t)=\left(\frac{s}{s_0}\right)^{\alpha(t)}\frac{\pi\alpha'}{\sin \pi\alpha(t)}\frac{1+\zeta e^{-i\pi \alpha(t)}}{2}\frac{1}{\Gamma(\alpha(t)+1)}.
\end{equation}
It is important to note that in obtaining Eq.~(\ref{eq:regge_amplitude}), the two signatures $\zeta=+$ and $\zeta=-$ have to be distinguished to satisfy the convergence criteria \cite{Corthals:2005ce}. However, the positive- and negative-signature parts of a trajectory would often coincide. Furthermore, if their residues $\beta(t)$ are identical or at least differ only in their signs, both trajectories are called strongly degenerate and thus, their contributions to the amplitude differ only by a phase \cite{Corthals:2005ce}. Hence, in determining the total amplitude, the Regge propagator in Eq.~(\ref{eq:regge_propagator}) can either be added or subtracted. It can be seen that the propagator can contain either a constant phase or a rotating phase:
\begin{equation}
    \frac{1+e^{-i\pi\alpha(t)}}{2}\pm \frac{1-e^{-i\pi\alpha(t)}}{2}=\begin{cases}
        1~~~~&\mathrm{(constant~phase)}\\
        e^{-i\pi\alpha(t)}~~~~&\mathrm{(rotating~phase)}
    \end{cases}.
\end{equation}
Hence, the Regge propagator for a degenerate trajectory is given by \cite{Corthals:2005ce}
\begin{equation}\label{eq:regge_propagator_phase}
    \mathcal{P}_\mathrm{Regge}(s,t)=\left(\frac{s}{s_0}\right)^{\alpha(t)}\frac{\pi\alpha'}{\sin \pi\alpha(t)}\frac{1}{\Gamma(\alpha(t)+1)}\begin{Bmatrix}
        1\\
        e^{-i\pi\alpha(t)}
    \end{Bmatrix}.
\end{equation}
Depending on the process of interest, one can determine whether the trajectories should be considered degenerate or not. If there are no structures present in the differential cross section data, then the involved trajectories are assumed to be degenerate \cite{Corthals:2005ce}. In contrast to degenerate trajectories, non-degenerate trajectories produce dips in the differential cross section, due to the so-called wrong-signature zeroes \cite{Corthals:2005ce,Collins:1971ff}. As there are no obvious structures present in the $\gamma  p \to K^+  \Lambda$ reaction for $E_\gamma^\mathrm{lab}\gtrsim 4~\mathrm{GeV}$, the $K$ and $K^*$ trajectories are thus assumed to be degenerate.

We note that Eq.~(\ref{eq:regge_propagator_phase}) is only valid for a scalar first materialization. If the first materialization has a non-zero spin $\alpha_0$, Eq.~(\ref{eq:regge_propagator_phase}) is modified with the substitution
\begin{equation}
    \alpha(t)~\to~\alpha(t)-\alpha_0
\end{equation}
in the exponent of $s/s_0$ and the argument of the gamma function \cite{Corthals:2006nz}. Thus, the general form of the Regge propagator reads
\begin{equation}\label{eq:regge_propagator_general}
    \mathcal{P}_\mathrm{Regge}(s,t)=\left(\frac{s}{s_0}\right)^{\alpha(t)-\alpha_0}\frac{\pi\alpha'}{\sin \pi\alpha(t)}\frac{1}{\Gamma(\alpha(t)-\alpha_0+1)}\begin{Bmatrix}
        1\\
        e^{-i\pi\alpha(t)}
    \end{Bmatrix}.
\end{equation}
This substitution ensures the propagator has poles at the physical materialization of the trajectory \cite{Corthals:2006nz}. It is important to note that while there are no strict restrictions in the choice of phase for the Regge propagator, the rotating phase is often preferred because the constant phase produces no imaginary part in the amplitude, and thus, yields zero recoil and target polarization asymmetries \cite{Corthals:2005ce,Janssen02}.

Despite originating in the late 1960s, Regge models for kaon photo- and electroproduction only started gaining traction in the late 1990s through the GLV developments \cite{Guidal:1997hy,Vanderhaeghen:1997ts,Guidal:1999qi,Guidal:2003qs}. While their method of Reggeizing the $t$-channel exchanges is very similar to that of Levy, Majerotto, and Read \cite{Levy:1973aq,Levy:1973zz}, they addressed the problem of gauge invariance, which was not intensively discussed by Levy, Majerotto, and Read. Guidal, Laget, and Vanderhaeghen proposed that the $t$-channel diagrams must be complemented with the $s$-channel nucleon diagram for the total amplitude to be gauge-invariant due to the $K$ exchange breaking gauge invariance \cite{Guidal:1997hy}. The total gauge-invariant amplitude is thus given by \cite{Guidal:1997hy,Corthals07,Corthals:2005ce,Janssen02}
\begin{equation}\label{eq:regge_total_amplitude}
    \mathcal{M}_\mathrm{fi}=\mathcal{M}^K_\mathrm{Regge}+\mathcal{M}^{K^*}_\mathrm{Regge}+\mathcal{M}^p_\mathrm{Feyn}\times\mathcal{P}^K_\mathrm{Regge}\times (t-m_K^2).
\end{equation}
However, a study by Haberzettl, Wang, and He \cite{Haberzettl:2015exa} later criticized Eq.~(\ref{eq:regge_total_amplitude}) for having no physical foundation. Nevertheless, the GLV recipe proved to be fairly successful in describing the experimental data \cite{Guidal:1997hy,Vanderhaeghen:1997ts,Chiang:2002vq,Corthals:2005ce,Corthals:2006nz}. 

\begin{figure}[hbt!]
    \centering
%    \includegraphics[scale=0.65]{GLV97_KL.pdf}
%    \hspace{1cm}
%    \includegraphics[scale=0.65]{GLV97_KS.pdf}
    \includegraphics[width=0.65\textwidth]{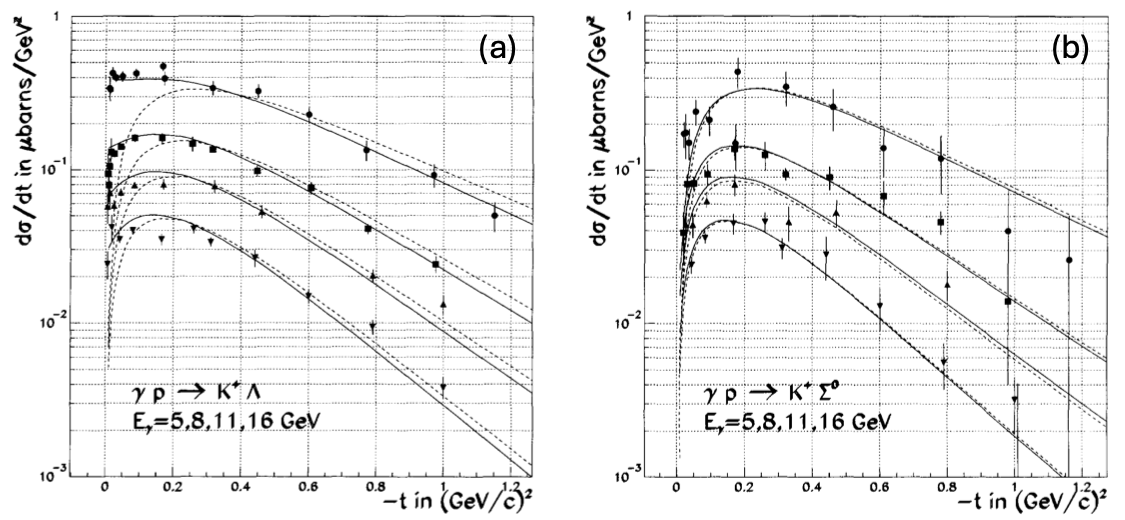}
   \caption{(a) Differential cross section $d\sigma/dt$ for the $\gamma p \to K^+\Lambda$ reaction for four photon energies as a function of the Mandelstam variable $t$. The solid curves show the full Reggeized $K$ and $K^*$ exchanges while the dashed curves show only the contribution from $K^*$ exchanges. Experimental data from Ref.~\cite{Boyarski:1969iy}. (b) Same, but for the $\gamma p \to K^+\Sigma^0$ reaction. Figures from Ref.~\cite{Guidal:1997hy}. Reprinted with permission from Elsevier.}
    \label{fig:GLV97_DCS}
\end{figure}

Results of the GLV model are shown in Fig.~\ref{fig:GLV97_DCS}. The model is constructed from the $K(494)$ and $K^*(892)$ trajectories, which are the dominant trajectories in kaon photoproduction. One notable feature of the results is the presence of a plateau in the differential cross section of the $\gamma p \to K^+\Lambda$ at very forward angles ($|t|\to 0$). This plateau is attributed to the introduction of the $s$-channel electric nucleon term for gauge invariance. Meanwhile, the plateau is not present in the $\gamma p \to K^+\Sigma^0$ reaction due to the small value of the $g_{K\Sigma N}$ coupling constant (see Ref.~\cite{Guidal:1997hy} for a more detailed discussion). Not long after, this model was extended to electroproduction in a series of follow-up works \cite{Vanderhaeghen:1997ts, Guidal:1999qi,Guidal:2003qs}. 

Even though most Regge models include only the $K(494)$ and $K^*(892)$ trajectories, a study by Yu, Choi, and Kim in 2011 \cite{Yu:2011fv} proposed the inclusion of the tensor meson $K^*_2(1430)$ exchange. The inclusion of tensor meson exchange was found to provide a better description of the kaon photoproduction process. 

%%%%%%%%%%%%%%%%%%%%%%%%%%%%%%%%%%%%%%%%%%%%%%%%%%%%%%%%%%%%%%%%%%%%%%%%%%%%%%%%%%%%%%%%%%%%%%%%%%%%%%%%%%%%%%%%%%%%%%%%%%%%%%%%%%%%%%%%%%%%%%%%%%%%%%%%%%%%%%%
%%%%%%%%%%%%%%%%%%%%%%%%%%%%%%%%%%%%%%%%%%%%%%%%%%%%%%%%%%%%%%%%%%%%%%%%%%%%%%%%%%%%%%%%%%%%%%%%%%%%%%%%%%%%%%%%%%%%%%%%%%%%%%%%%%%%%%%%%%%%%%%%%%%%%%%%%%%%%%%

\subsubsection{Regge-Plus-Resonance Models}

\begin{figure}[hbt!]
    \centering
    \includegraphics[scale=0.85]{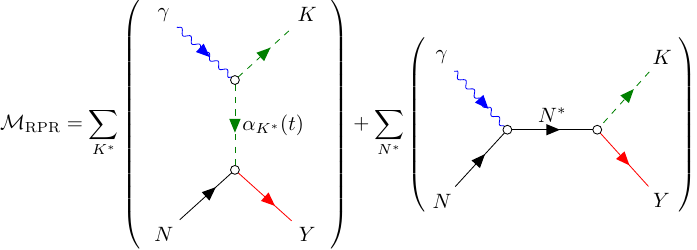}
    \caption{Illustration of the Regge-plus-Resonance approach. The Feynman diagrams in this figure were generated using the {\tt{tikz-feynman}} package \cite{Ellis_2017}. Figure motivation from Ref.~\cite{Corthals:2006nz}.}
    \label{fig:RPP_Amplitude_Illustration}
\end{figure}

As discussed in the previous subsection, Regge theory is able to describe the high energy behavior of kaon photoproduction very well. However, it is well-known that for low energies, the cross section of kaon photoproduction exhibits intricate structures, such as peaks and dips at certain energies. Hence, Reggeized $t$-channel exchanges alone are obviously not sufficient to describe the reaction dynamics in the resonance region \cite{Corthals:2005ce}. To remedy this, a few select dominant resonances are chosen to reproduce the resonance region. These resonances are then complemented with the Reggeized background (see Fig.~\ref{fig:RPP_Amplitude_Illustration}). This approach is often referred to as the Regge-plus-Resonance (RPR) model. 

%%%%%%%%%%%%%%%%%%%%%%%%%%%%%%%%%%%%%%%%%%%%%%%%%%%%%%%%%%%%%%%%%%%%%%%%%%%%%%%%%%%%%%%%%%%%%%%%%%%%%%%%%%%%%%%%%%%%%%%%%%%%%%%%%%%%%%%%%%%%%%%%%%%%%%%%%%%%%%%
%%%%%%%%%%%%%%%%%%%%%%%%%%%%%%%%%%%%%%%%%%%%%%%%%%%%%%%%%%%%%%%%%%%%%%%%%%%%%%%%%%%%%%%%%%%%%%%%%%%%%%%%%%%%%%%%%%%%%%%%%%%%%%%%%%%%%%%%%%%%%%%%%%%%%%%%%%%%%%%

To the best of our knowledge, the most notable RPR models for kaon photo- and electroproduction up to date are the works of the Ghent University group \cite{Corthals:2005ce,Corthals:2006nz,Corthals:2007kc,Vancraeyveld:2009qt,DeCruz:2012bv}. The first models, named RPR-2, RPR-3, and RPR-4, were developed in the mid-2000s by Corthals, Ryckebusch, and Van Cauteren \cite{Corthals:2005ce}. The background was constructed using the same $t$-channel Reggeization method as for the GLV work \cite{Guidal:1997hy}, while the resonances were introduced through the usual $s$-channel Feynman diagrams. Obviously, the contribution of the resonances have to vanish for high energies. Thus, the introduction of hadronic form factors is necessary. To this end, Corthals, Ryckebusch, and Van Cauteren used the Gaussian form factor as in Eq.~(\ref{eq:Gauss_HFF}) because it falls off faster than the usual dipole form. 

Corthals, Ryckebusch, and Van Cauteren considered three possible combinations for the phases of the $K$ and $K^*$ Regge propagators: rotating phases for both $K$ and $K^*$; a constant $K$ phase with a rotating $K^*$ phase; and a rotating $K$ phase with a constant $K^*$ phase. Constant phases for both propagators are not considered as they leave no imaginary parts in the Regge amplitude, which results in zero recoil polarization. However, Corthals, Ryckebusch, and Van Cauteren also found that the choice of constant phase for the $K$ propagator and rotating phase for the $K^*$ propagator resulted in a high value of $\chi^2$, and thus, is deemed unsatisfactory. See Table~I of Ref.~\cite{Corthals:2005ce} for all of the background models, along with the extracted coupling constants.

After investigating all combinations of the background models with various nucleon resonances, Corthals, Ryckebusch, and Van Cauteren found three potential RPR models, labeled RPR-2, RPR-3, and RPR-4, referring to background models 2, 3, and 4 in Table~I of Ref.~\cite{Corthals:2005ce}. The nucleon resonances of interest are the $N(1650)1/2^-$, $N(1710)1/2^+$, $N(1720)3/2^+$, $N(1900)3/2^+$, and $N(1900)1/2^+$. Both the RPR-2 and RPR-3 models incorporated all resonances, differing only in the cutoff for the hadronic form factor. Interestingly, the RPR-4 model, while not incorporating the $N(1900)3/2^+$ and $N(1900)1/2^+$ states, has a similar $\chi^2$ value as for the RPR-2 and RPR-3 models, which contain more parameters from the extra resonances. Results of the RPR-2, RPR-3, and RPR-4 models are presented in Figs.~7, 8, and 9 in Ref.~\cite{Corthals:2005ce}. Overall, all models described the experimental data well. These models were later extended in subsequent studies to describe $K^+\Sigma^0$ and $K^0\Sigma^+$ photoproductions \cite{Corthals:2006nz}, $K^+\Lambda$ and $K^+\Sigma^0$ electroproduction \cite{Corthals:2007kc}, and $K^+\Sigma^-$ photoproduction \cite{Vancraeyveld:2009qt}.

Two other notable RPR models for $K^+\Lambda$ photoproduction was developed by Byd\v{z}ovsk\'{y} and Skoupil in 2013 \cite{Bydzovsky:2012qg}, coined RPR-1 and RPR-2. The nucleon resonances incorporated in these models were inspired by the resonances included in the Ghent group's RPR-2011B model \cite{deCruz:2011}, namely the $N(1650)1/2^-$, $N(1710)1/2^+$, $N(1720)3/2^+$, $N(1900)3/2^+$, $N(1900)1/2^+$, and $N(1895)3/2^-$. The two models differ in the data to which they were fitted. RPR-1 was fitted to experimental data covering the full angular range, whereas RPR-2 was fitted only to forward-angle data. The magnitudes and signs of the coupling constants of the $K$ and $K^*$ trajectories in both models also differ, leading to relatively significant differences in their predictions of the cross sections (see Fig.~4 in Ref.~\cite{Bydzovsky:2012qg}).

\begin{figure}[hbt!]
    \centering
    \includegraphics[width=0.95\textwidth]{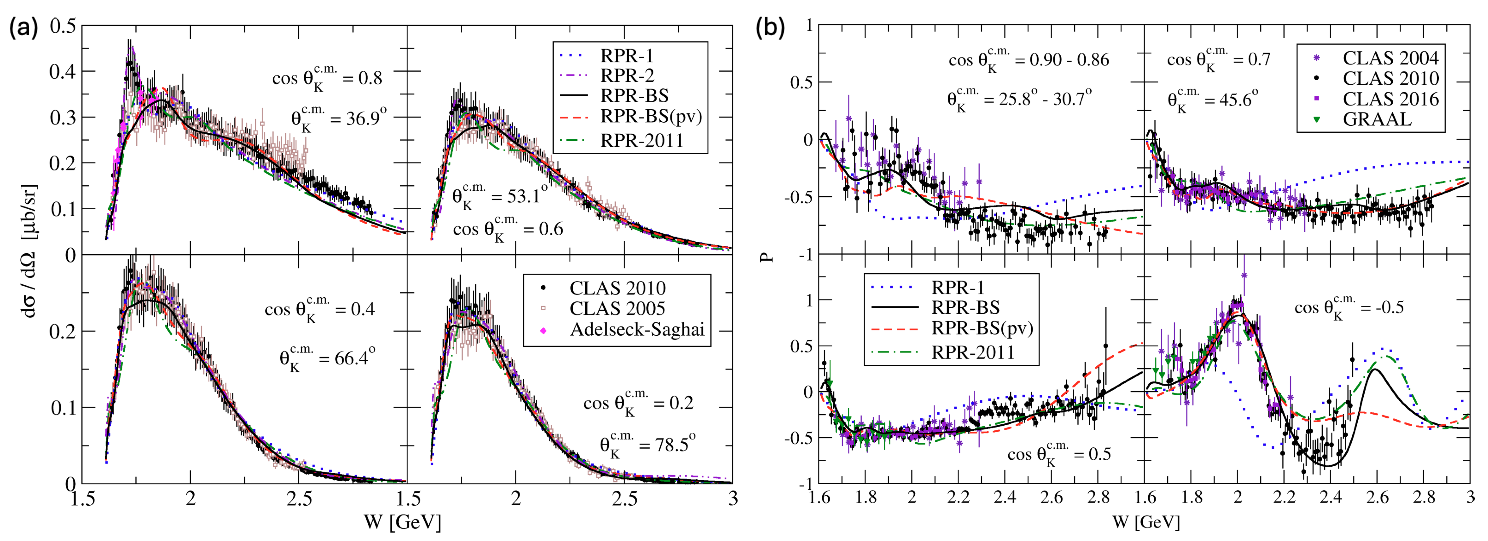}
    \caption{(a) Results of the RPR-BS and RPR-BS(pv) models for the differential cross section of the $\gamma  p \to K^+\Lambda$ reaction for four kaon angles as a function of the total c.m. energy. Experimental data are from the CLAS 2005 \cite{CLAS:2005lui}, CLAS 2010 \cite{CLAS:2009rdi}, LEPS 2006 \cite{LEPS:2005hji}, and LEPS 2018 \cite{LEPS:2017pzl}. (b) Same as the left figure, but for hyperon polarization $P$. Experimental data are from the CLAS \cite{CLAS:2009rdi,CLAS:2016wrl,CLAS:2003zrd} and GRAAL \cite{Lleres:2007tx}. Figures from Ref.~\cite{Bydzovsky:2019hgn}.}
    \label{fig:BS_RPR}
\end{figure}

Byd\v{z}ovsk\'{y} and Skoupil later extended their RPR models to also include spin-5/2 nucleon resonances (see Table~I in Ref.~\cite{Bydzovsky:2019hgn}). The main novelty of this work lies in the use of the gauge restoration recipe of Haberzettl, Wang, and He \cite{Haberzettl:2015exa}, which was based on the generalized Ward-Takahashi identities. Additionally, Byd\v{z}ovsk\'{y} and Skoupil also investigated the effects of pseudoscalar and pseudovector couplings, resulting in two models, named RPR-BS and RPR-BS(pv). Both models describe the differential cross section data well, as presented in Fig.~\ref{fig:BS_RPR}(a). On the other hand, the RPR-BS(pv) model is not able to describe the hyperon polarization data satisfactorily at $\cos \theta_K^\mathrm{c.m.}=-0.5$, as seen in Fig.~\ref{fig:BS_RPR}(b). This is mostly attributed to the $N(2570)5/2^-$ state, which was present in the RPR-BS model, but not in the RPR-BS(pv) model.

%%%%%%%%%%%%%%%%%%%%%%%%%%%%%%%%%%%%%%%%%%%%%%%%%%%%%%%%%%%%%%%%%%%%%%%%%%%%%%%%%%%%%%%%%%%%%%%%%%%%%%%%%%%%%%%%%%%%%%%%%%%%%%%%%%%%%%%%%%%%%%%%%%%%%%%%%%%%%%%
%%%%%%%%%%%%%%%%%%%%%%%%%%%%%%%%%%%%%%%%%%%%%%%%%%%%%%%%%%%%%%%%%%%%%%%%%%%%%%%%%%%%%%%%%%%%%%%%%%%%%%%%%%%%%%%%%%%%%%%%%%%%%%%%%%%%%%%%%%%%%%%%%%%%%%%%%%%%%%%

%%%%%%%%%%%%%%%%%%%%%%%%%%%%%%%%%%%%%%%%%%%%%%%%%%%%%%%%%%%%%%%%%%%%%%%%%%%%%%%%%%%%%%%%%%%%%%%%%%%%%%%%%%%%%%%%%%%%%%%%%%%%%%%%%%%%%%%%%%%%%%%%%%%%%%%%%%%%%%%
\begin{table}[hbt!]
\setlength{\tabcolsep}{6pt} % Default value: 6pt
\renewcommand{\arraystretch}{0.8} % Default value: 1
    \centering
    \caption{Summary of known Regge and Regge-plus-Resonance models and the experimental data used in the fitting process.}
    \label{tab:regge_summary}
    \begin{threeparttable}
    \begin{tabularx}{\textwidth}{XX wc{3cm} @{\hspace{1cm}} Xc}
    \hline \hline
    Model & Dataset & Final State(s) & Observable(s) & Ref. \\ \hline
    %======================
    GLV (1997) & SLAC 1971 & $K^+\Lambda$, $K^+\Sigma^0$  & $d\sigma/d\Omega$ & \cite{Boyarski:1970yc}\\ 
    %======================
    & DESY 1972 & $K^+\Lambda$ & $P$ & \cite{Vogel:1972xh} \\
    %======================
    & SLAC 1979 & $K^+\Lambda$, $K^+\Sigma^0$ & $d\sigma/d\Omega$, $\Sigma$ & \cite{Quinn:1979zp}\\
    %======================
    \hline
    %======================
    RPR-2007 (2007) & SLAC 1971 & $K^+\Lambda$, $K^+\Sigma^0$ & $d\sigma/d\Omega$ & \cite{Boyarski:1970yc}\\ 
    %======================
    & DESY 1972 & $K^+\Lambda$ & $P$ & \cite{Vogel:1972xh}\\
    %======================
    & SLAC 1979 & $K^+\Lambda$, $K^+\Sigma^0$ & $d\sigma/d\Omega$, $\Sigma$ & \cite{Quinn:1979zp}\\
    %======================
    & SAPHIR 2004 & $K^+\Lambda$, $K^+\Sigma^0$ & $d\sigma/d\Omega$, $P$ & \cite{Glander:2003jw} \\
    %======================
    & Hall C 2003 & $e'K^+\Lambda$, $e'K^+\Sigma^0$ & $\sigma_\mathrm{U}$, $\sigma_\mathrm{L}$, $\sigma_\mathrm{T}$ & \cite{E93018:2002cpu} \\
    %======================
    & CLAS 2003 & $e'K^+\Lambda$ & $\mathcal{P}'_x$, $\mathcal{P}'_z$, $\mathcal{P}'_{x'}$, $\mathcal{P}'_{z'}$ & \cite{CLAS:2002zlc} \\
    %======================
    & CLAS 2004 & $K^+\Lambda$, $K^+\Sigma^0$ & $d\sigma/d\Omega$, $P$ & \cite{CLAS:2003zrd} \\
    %======================
    %& SAPHIR 2005 & & $d\sigma/d\Omega$, $P$ & \cite{Lawall:2005np} \\
    %======================
    & GRAAL 2007 & $K^+\Lambda$, $K^+\Sigma^0$ & $P$, $\Sigma$ & \cite{Lleres:2007tx} \\
    %======================
    & CLAS 2007 & $e'K^+\Lambda$, $e'K^+\Sigma^0$ & $\sigma_\mathrm{U}$, $\sigma_\mathrm{TT}$, $\sigma_\mathrm{LT}$ & \cite{CLAS:2006ogr} \\ 
    %======================
    \hline 
    RPR-2011 (2011) & LEPS 2003 & $K^+\Lambda$, $K^+\Sigma^0$ & $\Sigma$ & \cite{LEPS:2003buk} \\
    %======================
    & CLAS 2004 & $K^+\Lambda$, $K^+\Sigma^0$ & $P$ & \cite{CLAS:2003zrd} \\
    %======================
    & CLAS 2006 & $K^+\Lambda$, $K^+\Sigma^0$ & $d\sigma/d\Omega$ & \cite{CLAS:2005lui} \\
    %======================
    & LEPS 2006 & $K^+\Lambda$, $K^+\Sigma^0$ & $\Sigma$ & \cite{LEPS:2005hji} \\
    %======================
    & LEPS 2007 & $K^+\Lambda$ & $d\sigma/d\Omega$, $\Sigma$ & \cite{PhysRevC.76.042201} \\
    %======================
    & CLAS 2007 & $K^+\Lambda$, $K^+\Sigma^0$ & $C_x$, $C_z$ & \cite{CLAS:2006pde} \\
    %======================
    & GRAAL 2007 & $K^+\Lambda$, $K^+\Sigma^0$ & $P$, $\Sigma$ & \cite{Lleres:2007tx} \\
    %======================
    & GRAAL 2009  & $K^+\Lambda$ & $O_x$, $O_z$, $T$\tnote{a} & \cite{GRAAL:2008jrm} \\
    %======================
    & CLAS 2010 & $K^+\Lambda$ & $d\sigma/d\Omega$, $P$ & \cite{CLAS:2009rdi} \\
    %======================
    \hline
    RPR-BS (2019) & Adelseck-Saghai 1990\tnote{b} & $K^+\Lambda$, $K^+\Sigma^0$ & $d\sigma/d\Omega$ & \cite{Adelseck:1990ch}\\
    %======================
    & LEPS 2006 & $K^+\Lambda$, $K^+\Sigma^0$ & $d\sigma/d\Omega$ & \cite{LEPS:2005hji} \\
    %======================
    & CLAS 2006 & $K^+\Lambda$ & $d\sigma/d\Omega$ & \cite{CLAS:2005lui} \\
    %======================
    & CLAS 2010 & $K^+\Lambda$ & $d\sigma/d\Omega$, $P$ & \cite{CLAS:2009rdi} \\
    %======================
    & CLAS 2016 & $K^+\Lambda$ & $P$ & \cite{CLAS:2016wrl} \\
    %======================
    & LEPS 2018 & $K^+\Lambda$ & $d\sigma/d\Omega$ & \cite{LEPS:2017pzl}\\
    %======================
    \hline \hline
    \end{tabularx}
        \begin{tablenotes}
            \item[a] \footnotesize The target asymmetry $T$ was not directly measured, but indirectly extracted from the data \cite{GRAAL:2008jrm}. 
            \item[b] \footnotesize The work of Adelseck and Saghai \cite{Adelseck:1990ch}. The experimental data listed in this work are not original data, but rather a compilation of older data. See Tables~IX and X in Ref.~\cite{Adelseck:1990ch}.
        \end{tablenotes}
    \end{threeparttable}
\end{table}
%%%%%%%%%%%%%%%%%%%%%%%%%%%%%%%%%%%%%%%%%%%%%%%%%%%%%%%%%%%%%%%%%%%%%%%%%%%%%%%%%%%%%%%%%%%%%%%%%%%%%%%%%%%%%%%%%%%%%%%%%%%%%%%%%%%%%%%%%%%%%%%%%%%%%%%%%%%%%%%
%=================================================

%=== PHENOMENOLOGICAL APPLICATIONS ===============
%  \input{phenomenological_applications}
\section {Phenomenological Applications}
\label{sec:phenomenlogy}

Numerous phenomenological applications of kaon-hyperon production have been discussed in the last three decades. This section will review these applications, which are less well known compared to the intensive discussions of the model development given in Section~\ref{sec:models}.

\subsection{Elementary Operator for Use in Nuclear Physics: Production of Hypernuclei}
\label{sec:hypernuclei}

Photoproduction of hypernuclear states, in which the nucleus contains one or more hyperons in addition to protons and neutrons, commonly denoted in spectroscopic notation as $(\gamma,K^+)$, plays an important role in hypernuclear physics. Because of the mass difference between the hyperon and the nucleon, this reaction involves a large momentum transfer and consequently a large recoil momentum. As a result, the associated production cross section is much smaller than those for hadron-induced processes such as $(K^-,\pi^-)$ and $(\pi^+,K^+)$. Furthermore, the $(K^-,\pi^-)$ reaction predominantly excites natural-parity states with low angular momentum, such as $0^+$ and $1^-$, while the $(\pi^+,K^+)$ reaction also favors natural-parity states but with higher angular momentum, e.g., $1^-$ and $2^+$. In contrast, the electromagnetic process $(\gamma,K^+)$ preferentially excites unnatural-parity states with higher angular momentum, such as $2^-$ and $3^+$. Thus, $(\gamma,K^+)$ provides complementary spectroscopic information, and a complete description of hypernuclear structure requires the combined use of all three reactions \cite{Bennhold:1999nd,Bando:1990yi}.

\begin{figure}[hbt!]%[t]
\centering
\includegraphics[width=0.65\columnwidth]{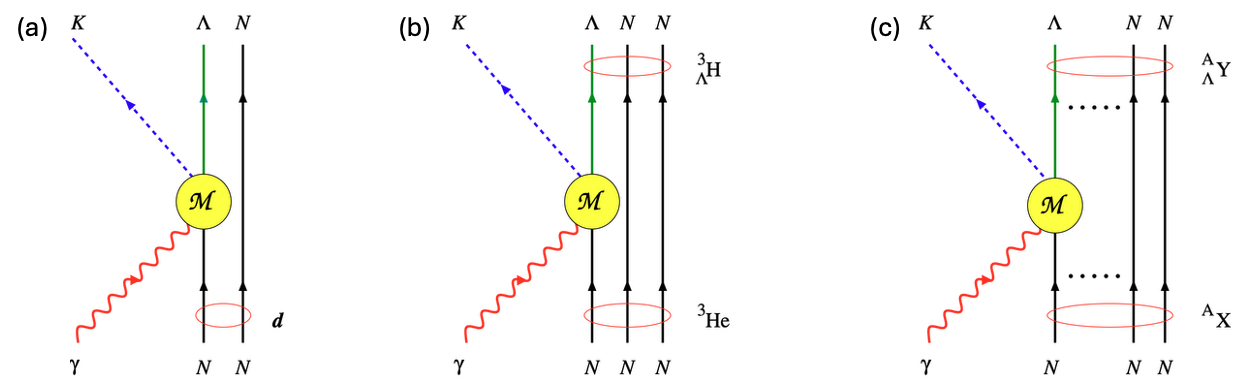} 
\caption{Kaon photoproduction on nuclei within the impulse approximation: (a) kaon photoproduction on the deuteron, $\gamma d\to K^+\Lambda N$; (b) kaon photoproduction on $^3$He with a hypertriton in the final state; and (c) kaon photoproduction on a heavier nucleus with a hypernucleus in the final state. In all panels, the elementary operator ${\mathcal{M}}$ is shown as a yellow circle.}
\label{fig:elementary_operator} 
\end{figure}

To first order, hypernucleus photoproduction can be described within the impulse approximation, as illustrated in Fig.~\ref{fig:elementary_operator}. In this framework, the incoming photon interacts with a single nucleon inside the nucleus, producing a kaon that immediately exits the nucleus together with a hyperon, which may remain free or become bound depending on the interaction potential. Beyond this approximation, corrections may be included, such as initial- and final-state interactions or two-step processes in which the photon-nucleon interaction does not directly produce the kaon-hyperon pair.

%%%%%%%FIGURE 1%%%%%
%%%%%%%%%%%%%%%%%%%%

The earliest elementary operator for this purpose was constructed by Thom~\cite{Thom:1966rm}, where the Born terms were obtained using the Feynman diagrammatic technique, while the resonance contributions were derived from partial-wave amplitudes. Since in nuclear calculations the struck nucleon is not at rest due to Fermi motion, a frame-independent operator is required, which is difficult to achieve with partial-wave amplitudes. Technically, this involves integrating the nucleon momentum from zero to infinity in order to obtain the average contribution, which in turn necessitates an analytical continuation.

A more consistent operator, employing Feynman diagrams for both the Born and resonance terms, was developed by Blomqvist and Laget~\cite{Blomqvist:1977rv}. Although formulated for pion photoproduction on nuclei, the non-relativistic reduction adopted in this work later inspired many studies. Nearly a decade afterward, Adelseck et al.~\cite{Adelseck:1985scp} constructed an elementary operator for kaon photoproduction $\gamma p\to K^+\Lambda$ using the Feynman diagrammatic approach for both the Born and resonance terms. Incorporating seven nucleon and five hyperon resonances, the operator was fit to experimental data on cross sections and polarization observables. Unlike previous works, the reduction of this operator to the non-relativistic limit was carried out in a systematic manner, i.e. \cite{Adelseck:1985scp},
\begin{eqnarray}
\label{eq:F_adelseck_operator}
    \lefteqn{{\bar u}({\boldsymbol p}_\Lambda)\,\sum_{i=1}^{4}A_i(s,t,u)\, M_i\, u({\boldsymbol p}_p) =  \left(\frac{E_p+m_p}{2m_p}\right)^{1/2} \left(\frac{E_\Lambda+m_\Lambda}{2m_\Lambda}\right)^{1/2}}
    \nonumber\\  && \times \,\chi_\Lambda^\dagger\Bigl(F_1\,\boldsymbol{\sigma\cdot\,\epsilon} +F_2\,\boldsymbol{\sigma\cdot{k}}\,\boldsymbol{p}_p\boldsymbol{\cdot\,\epsilon}+F_3\,\boldsymbol{\sigma\cdot{k}}\,\boldsymbol{p}_\Lambda\boldsymbol{\cdot\,\epsilon} + F_4\,\boldsymbol{\sigma\cdot p}_p\,\boldsymbol{p}_p\boldsymbol{\cdot\,\epsilon} + F_5\,\boldsymbol{\sigma\cdot p}_p\,\boldsymbol{p}_\Lambda\boldsymbol{\cdot\,\epsilon}
    \nonumber\\
    &&+~ F_6\,\boldsymbol{\sigma\cdot p}_\Lambda\,\boldsymbol{p}_p\boldsymbol{\cdot\,\epsilon} + F_7\,\boldsymbol{\sigma\cdot p}_\Lambda\,\boldsymbol{p}_\Lambda\boldsymbol{\cdot\,\epsilon} + F_8\, \boldsymbol{\sigma\cdot \epsilon}\,\boldsymbol{\sigma\cdot k}\,\boldsymbol{\sigma\cdot p}_p + F_9\, \boldsymbol{\sigma\cdot p}_\Lambda\,\boldsymbol{\sigma\cdot k}\,\boldsymbol{\sigma\cdot\,\epsilon} \nonumber\\
    &&+~ F_{10}\, \boldsymbol{\sigma\cdot p}_\Lambda\,\boldsymbol{\sigma\cdot\,\epsilon}\,\boldsymbol{\sigma\cdot p}_p + F_{11}\, \boldsymbol{\sigma\cdot p}_\Lambda\,\boldsymbol{\sigma\cdot k}\,\boldsymbol{\sigma\cdot p}_p
    \,\boldsymbol{p}_p\boldsymbol{\cdot\,\epsilon} + F_{12}\, \boldsymbol{\sigma\cdot p}_\Lambda\,\boldsymbol{\sigma\cdot k}\,\boldsymbol{\sigma\cdot p}_p
    \,\boldsymbol{p}_\Lambda\boldsymbol{\cdot\,\epsilon} \Bigr) \,\chi_p. 
\end{eqnarray}
The amplitudes $F_1, F_2,$ and $F_3$ provide the dominant contributions, whereas $F_4,\ldots,F_{12}$ are suppressed because their denominators contain additional kinematical factors, i.e., $E_p+m_p$ for $F_4,F_5,$ and $F_8$, $E_\Lambda+m_\Lambda$ for $F_6,F_7,$ and $F_9$, and the product of both for $F_{10},F_{11},$ and $F_{12}$. In the non-relativistic limit, $F_{10},F_{11},$ and $F_{12}$ can therefore be safely neglected. One might also be tempted to omit the smaller amplitudes $F_4,\ldots,F_9$, since in pion photoproduction near threshold the so-called Kroll-Ruderman term, corresponding to $F_1\,\boldsymbol{\sigma\cdot\epsilon}$ in Eq.~(\ref{eq:F_adelseck_operator}), already provides an adequate description of the data. The situation is different in kaon photoproduction, since its threshold is much higher than that of the pion. Consequently, the reaction involves larger momentum transfers, stronger recoil effects, and more significant relativistic corrections. As a result, the smaller amplitudes $F_4,\ldots,F_9$ cannot be neglected, and a reliable approximation requires keeping all amplitudes up to $F_9$ \cite{Adelseck:1985scp}. 

The non-relativistic photoproduction operator in Eq.~(\ref{eq:F_adelseck_operator}) can be extended to the general case, i.e., electroproduction, by including the longitudinal contributions in the matrices $M_i$ ($i=1,\ldots,6$). This extension yields eight additional amplitudes beyond those in Eq.~(\ref{eq:F_adelseck_operator}). The complete operator is then given by \cite{Mart:2008gq}
\begin{eqnarray}
\lefteqn{ \overline{u}({\boldsymbol p}_{Y})\,\sum_{i=1}^{6}\, 
A_{i}M_{i}\, u({\boldsymbol p}_{N})
 ~=~ \left(\frac{E_{N} + m_{N}}{2m_{N}} \right)^{\frac{1}{2}}
        \left(\frac{E_{Y} + m_{Y}}{2m_{Y}} 
        \right)^{\frac{1}{2}}}\nonumber\\ && \times\, \chi_{Y}^{\dagger}
        \, \Bigl(\, {\cal{F}}_{1}
\,{\boldsymbol{\sigma} \,\cdot\,} {\boldsymbol{\epsilon}}
+ {\cal{F}}_{2}\,  {\boldsymbol{\sigma}}{\boldsymbol \,\cdot\,} {\boldsymbol k}\, \epsilon_{0}
+ {\cal{F}}_{3}\, {\boldsymbol{\sigma}}\boldsymbol{\, \cdot \,}{\boldsymbol k}\, 
{\boldsymbol k}\boldsymbol{\, \cdot \,} {\boldsymbol{\epsilon}} + {\cal{F}}_{4}\, 
{\boldsymbol{\sigma}}\boldsymbol{\, \cdot \,}{\boldsymbol k}\, {\boldsymbol p}_{N}\boldsymbol{\, \cdot \,}
{\boldsymbol{\epsilon}}  + {\cal{F}}_{5}\, {\boldsymbol{\sigma}\,\cdot\,} {\boldsymbol k}\, {\boldsymbol p}_{Y}\boldsymbol{\, \cdot \,}{\boldsymbol{\epsilon}}
+ {\cal{F}}_{6}\, {\boldsymbol{\sigma}\,\cdot\,} {\boldsymbol p}_{N}\,\epsilon_{0} \nonumber\\ 
&& 
+\, {\cal{F}}_{7}\, {\boldsymbol{\sigma}} 
\cdot {\boldsymbol p}_{N}\, {\boldsymbol k}\boldsymbol{\, \cdot \,}{\boldsymbol{\epsilon}}
+ {\cal{F}}_{8}\, {\boldsymbol{\sigma}}\boldsymbol\, \cdot \,{\boldsymbol p}_{N}\, {\boldsymbol p}_{N} \,{\boldsymbol \cdot}\, 
{\boldsymbol {\epsilon}} + {\cal{F}}_{9}\, 
{\boldsymbol{\sigma}}\boldsymbol{\, \cdot \,}{\boldsymbol p}_{N}\, 
{\boldsymbol p}_{Y}\boldsymbol{\, \cdot \,} {\boldsymbol{\epsilon}}+ {\cal{F}}_{10}\, {\boldsymbol{\sigma}}\boldsymbol{\, \cdot \,}{\boldsymbol p}_{Y}\, 
\epsilon_{0} 
+ {\cal{F}}_{11}\, {\boldsymbol{\sigma}}\boldsymbol{\, \cdot \,}{\boldsymbol p}_{Y}\, {\boldsymbol k}\boldsymbol{\, \cdot \,}
{\boldsymbol{\epsilon}} 
\nonumber\\ 
&& 
+\, {\cal{F}}_{12}\, {\boldsymbol{\sigma}}
\boldsymbol{\, \cdot \,}{\boldsymbol p}_{Y}\, {\boldsymbol p}_{N}\boldsymbol{\, \cdot \,}{\boldsymbol{\epsilon}} + 
{\cal{F}}_{13}\, {\boldsymbol{\sigma}}\boldsymbol{\, \cdot \,}{\boldsymbol p}_{Y}\, 
{\boldsymbol p}_{Y}\boldsymbol{\, \cdot \,} {\boldsymbol{\epsilon}}
 + {\cal{F}}_{14}\, {\boldsymbol{\sigma}}\boldsymbol{\, \cdot \,}
{\boldsymbol{\epsilon}}\, {\boldsymbol{\sigma}}
\boldsymbol{\, \cdot \,}{\boldsymbol k}\, {\boldsymbol{\sigma}}\boldsymbol{\, \cdot \,}{\boldsymbol p}_{N}+ {\cal{F}}_{15}\, {\boldsymbol{\sigma}}\boldsymbol{\, \cdot \,}{\boldsymbol p}_{Y}\, 
{\boldsymbol{\sigma}}\boldsymbol{\, \cdot \,}{\boldsymbol{\epsilon}}\, {\boldsymbol{\sigma}}\boldsymbol{\, \cdot \,}{\boldsymbol k}
\nonumber\\ 
&& 
 + {\cal{F}}_{16}\, {\boldsymbol{\sigma}}\boldsymbol{\, \cdot \,}{\boldsymbol p}_{Y}\, 
{\boldsymbol{\sigma}}\boldsymbol{\, \cdot \,}{\boldsymbol{\epsilon}}\, {\boldsymbol{\sigma}}\boldsymbol{\, \cdot \,}{\boldsymbol p}_{N} 
+ {\cal{F}}_{17}\, {\boldsymbol{\sigma}}\boldsymbol{\, \cdot \,}{\boldsymbol p}_{Y}\, 
 {\boldsymbol{\sigma}}\boldsymbol{\, \cdot \,}
{\boldsymbol k}\, {\boldsymbol{\sigma}}\boldsymbol{\, \cdot \,}{\boldsymbol p}_{N}\, \epsilon_{0} 
+ {\cal{F}}_{18}\, {\boldsymbol{\sigma}}\boldsymbol{\, \cdot \,}{\boldsymbol p}_{Y}\, 
{\boldsymbol{\sigma}}\boldsymbol{\, \cdot \,}{\boldsymbol k}\, {\boldsymbol{\sigma}\,\cdot\, } 
{\boldsymbol p}_{N}\, {\boldsymbol k}\boldsymbol{\, \cdot \,}{\boldsymbol{\epsilon}}
\nonumber\\ 
 & &  + {\cal{F}}_{19}\, {\boldsymbol{\sigma}}\boldsymbol{\, \cdot \,}{\boldsymbol p}_{Y}\, 
{\boldsymbol{\sigma}}\boldsymbol{\, \cdot \,}{\boldsymbol k}\, {\boldsymbol{\sigma}\,
\cdot\,} {\boldsymbol p}_{N}\, {\boldsymbol p}_{N}\boldsymbol{\, \cdot \,} {\boldsymbol{\epsilon}}
 +\, {\cal{F}}_{20}\, {\boldsymbol{\sigma}}\boldsymbol{\, \cdot \,}{\boldsymbol p}_{Y}\, 
{\boldsymbol{\sigma}}\boldsymbol{\, \cdot \,}{\boldsymbol k}\, {\boldsymbol{\sigma}} 
\cdot {\boldsymbol p}_{N}\, {\boldsymbol p}_{Y}\boldsymbol{\, \cdot \,} {\boldsymbol{\epsilon}}
\, \Bigr) \, \chi_{N},
\label{eq:gen_nonrel_op}
\end{eqnarray}
where we have also generalized the notation to include all six isospin channels in kaon electroproduction. Note that the amplitudes $F_i$ can be related to the amplitudes $A_i$ in Eq.~(\ref{eq:Ai_Mi}) by using the lengthy equations given in Appendix \ref{app:non-rel-amplitudes} and the conventions for all momenta and energies are defined in Section~\ref{subsec:kinematics}. 

The operator in Eq.~(\ref{eq:gen_nonrel_op}) can be also written in terms of the spin-non-flip and spin-flip amplitudes, $L$ and $\boldsymbol{K}$, respectively, 
\begin{equation}\label{eq:L_and_K}
    \overline{u}({\boldsymbol p}_{Y})\,\sum_{i=1}^{6}\, A_{i}M_{i}\, u({\boldsymbol p}_{N}) \equiv  \chi_{Y}^{\dagger}~ t_{\gamma K}~\chi_{N} = \chi_{Y}^{\dagger}~ i\bigl(L+i\boldsymbol{\sigma\cdot K}\bigr)~\chi_{N} . 
\end{equation}
This representation is especially useful, since it separates the spin-independent ($L$) and spin-dependent ($\boldsymbol{K}$) parts of the operator, and has therefore been widely adopted in nuclear calculations. 

It should be noted that both the $L$ and $\boldsymbol{K}$ amplitudes depend on the photon polarization vector $\boldsymbol{\epsilon}$. This dependence can make nuclear physics calculations more involved, because the polarization is usually defined in the nuclear frame of reference. To simplify the treatment and make the operator more suitable for frame-independent calculations, the amplitude in Eq.~(\ref{eq:L_and_K}) can be rewritten as \cite{Miyagawa:2006kj}
\begin{eqnarray} 
\label{eq:t_gamma_K} 
t_{\gamma K} &=& i\,(1,i\sigma_x,i\sigma_y,i\sigma_z) \,
\left( \begin{array}{ccc}
{\cal L}_x & {\cal L}_y & {\cal L}_z \\ {\cal K}_{xx} & {\cal K}_{xy} & {\cal K}_{xz} \\ 
{\cal K}_{yx} & {\cal K}_{yy} & {\cal K}_{yz} \\ 
{\cal K}_{zx} & {\cal K}_{zy} & {\cal K}_{zz} \end{array}\right) ~
\left( \begin{array}{c}
\epsilon_x \\ \epsilon_y \\ \epsilon_z  \end{array}\right).
\end{eqnarray}
Equation (\ref{eq:t_gamma_K}) demonstrates that the non-relativistic operator is expressed as a $4\times 3$ matrix, fully independent of the reference frame in which the spin operator $\boldsymbol{\sigma}$ and the photon polarization vector $\boldsymbol{\epsilon}$ are defined. The components of the matrix in Eq.~(\ref{eq:t_gamma_K}) are given by \cite{Miyagawa:2006kj}
\begin{eqnarray} 
\boldsymbol{\cal L} &=& N \Bigl[{\cal F}_{14}\,\boldsymbol{k} \times \boldsymbol{p}_N +
{\cal F}_{15}\,\boldsymbol{k} \times \boldsymbol{p}_Y +{\cal F}_{16}\,\boldsymbol{p}_N
 \times \boldsymbol{p}_Y + \boldsymbol{p}_Y\boldsymbol{\,\cdot\, k} \times \boldsymbol{p}_N 
\bigl({\cal F}_{18}\, \boldsymbol{k} +
{\cal F}_{19}\, \boldsymbol{p}_N + {\cal F}_{20}\, \boldsymbol{p}_Y\bigr)\Bigr]
\end{eqnarray}
and
\begin{eqnarray} 
{\cal K}_{ij} &=& \delta_{ij}\,A + p_{\gamma ,i}\,{B}_j + 
p_{N,i}\,{C}_j + p_{Y,i}\,{D}_j, ~~~~ i,j=x,y,z,
\end{eqnarray}
with 
\begin{eqnarray} 
A &=& -N \left[{\cal F}_1+{\cal F}_{14}\,\boldsymbol{p}_N\boldsymbol{\,\cdot\, k} -
{\cal F}_{15}\,\boldsymbol{p}_Y\boldsymbol{\,\cdot\,k}-{\cal F}_{16}\,\boldsymbol{p}_N
\boldsymbol{\,\cdot\,p}_Y\right],\\
\boldsymbol{B} &=& -N \left[({\cal F}_4-{\cal F}_{14}-{\cal F}_{19}\,
\boldsymbol{p}_N\boldsymbol{\,\cdot\, p}_Y)\,\boldsymbol{p}_N 
+ ({\cal F}_{5}+{\cal F}_{15}-{\cal F}_{20}\,\boldsymbol{p}_N\boldsymbol{\,\cdot\, p}_Y)\,
\boldsymbol{p}_Y 
\right],\\
\boldsymbol{C} &=& -N \left[({\cal F}_8+{\cal F}_{19}\,\boldsymbol{p}_Y\boldsymbol{\,\cdot\, p}
_\gamma)\,\boldsymbol{p}_N 
+ ({\cal F}_{9}+{\cal F}_{16}+{\cal F}_{20}\,\boldsymbol{p}_Y
\boldsymbol{\,\cdot\, k})\,\boldsymbol{p}_Y \right],
\\
\boldsymbol{D} &=& -N \left[({\cal F}_{12}+{\cal F}_{16}+{\cal F}_{19}\,\boldsymbol{p}_N
\boldsymbol{\,\cdot\, k})\,\boldsymbol{p}_N 
+ ({\cal F}_{13}+{\cal F}_{20}\,\boldsymbol{p}_N\boldsymbol{\,\cdot\,k})\,\boldsymbol{p}_Y
 \right],
\end{eqnarray}
and
\begin{equation} 
N= \left(\frac{E_N+m_N}{2m_N}\right)^{\frac{1}{2}}
\left(\frac{E_Y+m_Y}{2m_Y}\right)^{\frac{1}{2}}
\sqrt{\frac{m_Y}{E_Y}}
\sqrt{\frac{m_N}{E_N}}.
\end{equation}

Using this elementary operator, kaon photoproduction on the deuteron has been investigated in both the quasi-free scattering (QFS) region, where one nucleon acts as a spectator with negligible momentum, and in the kinematical region where the $YN$ final-state interaction (FSI) is significant. Therefore, the study of kaon production on the deuteron has a two-fold purpose. In the QFS region the reaction provides access to the elementary operator, particularly important in the case of kaon production on the neutron. Since no free neutron target is available, the deuteron is used, with kinematics chosen such that the proton remains a spectator to minimize nuclear distortion. In the region where both nucleons participate, the reaction offers a means to test the available $YN$ potential models. 

\begin{figure}[t]
\centering
\includegraphics[width=0.55\columnwidth]{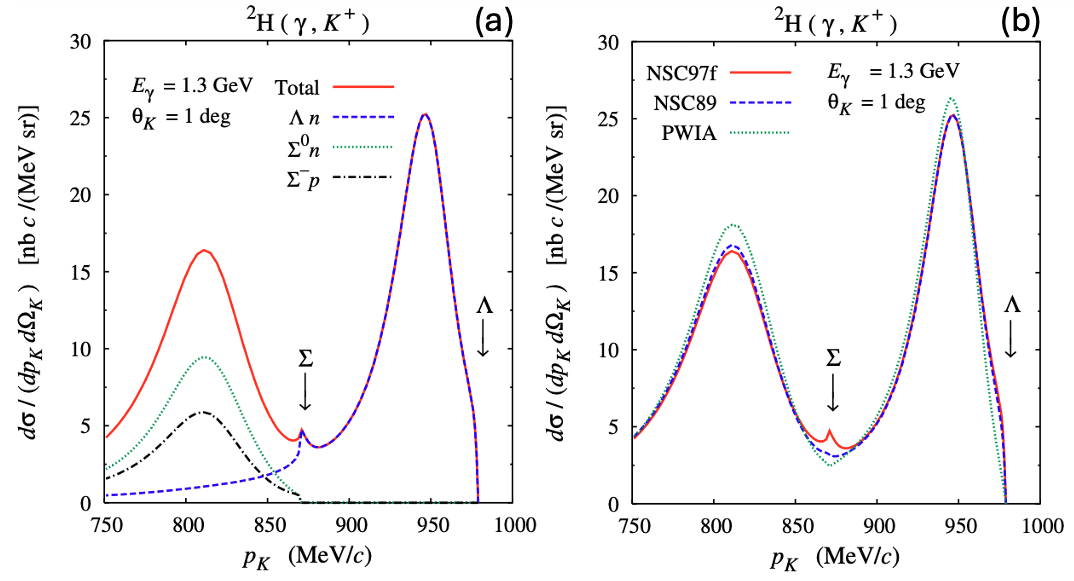} 
\caption{Inclusive cross section for kaon photoproduction on the deuteron as a function of kaon laboratory momentum at $\theta=1^\circ$ and $E_\gamma^{\rm lab} = 1300$ MeV: (a) contributions from the $\Lambda n$, $\Sigma^0 n$, and $\Sigma^- p$ channels of the elementary process calculated with the NSC97f $YN$ potential; (b) effects of the final-state interaction for different $YN$ interactions (NSC97f, NSC89, and partial-wave impulse approximation). Figures adapted from Ref.~\cite{Miyagawa:2006kj}, with color added for clarity.}
\label{fig:inclusive_deut_cs} 
\end{figure}

Figure~\ref{fig:inclusive_deut_cs} shows the inclusive cross sections for $\gamma d\to K^+YN$, illustrating the two-fold purpose discussed above~\cite{Miyagawa:2006kj}. In Fig.~\ref{fig:inclusive_deut_cs}(a) two pronounced peaks appear at $p_K=810$ and $950$ MeV, which correspond to the QFS regions. The first peak originates mainly from the $\Lambda n$ and $\Sigma^0 n$ channels, indicating that the elementary operator for these processes can be extracted from the nuclear cross section. Nevertheless, some FSI effects are visible, especially at $p_K=810$ MeV, as shown in Fig.~\ref{fig:inclusive_deut_cs}(b), where the impulse approximation is compared with results obtained using the NSC89~\cite{Maessen:1989sx} and NSC97f~\cite{Rijken:1998yy} $YN$ potentials. It is important to note that NSC97f represents an improvement over NSC89.

%%%%%%%FIGURE 1%%%%%
%%%%%%%%%%%%%%%%%%%%

Beyond the two peak regions, Fig.~\ref{fig:inclusive_deut_cs}(b) also reveals FSI effects near the $\Lambda N$ and $\Sigma N$ thresholds. In particular, around the $\Sigma N$ threshold the effect is significantly enhanced, where the NSC97f potential exhibits a prominent cusp-like structure, while NSC89 shows only minor deviations from the impulse approximation. Therefore, experimental data on the inclusive cross section are essential for testing the validity of the potential models. In the exclusive cross section, the differences among these three interaction assumptions become large for $\theta \geq 20^\circ$ \cite{Miyagawa:2006kj}. In summary, the elementary operator discussed here has proven very useful for nuclear calculations of kaon photoproduction on the deuteron, enabling simultaneous investigation of both the operator itself and the $YN$ interaction. 

Previous calculations employing a similar operator used simple $YN$ potentials to examine their effects on the inclusive and exclusive cross sections \cite{Adelseck:1989zt,Li:1991qa,Li:1992tza}. These studies showed that the FSI have a substantial impact on the cross section only near production thresholds. They also demonstrated that corrections to the $s$-state wavefunction are insufficient to account for the FSI, in contrast to the conclusions of much earlier works \cite{Renard:1967frj,Renard:1967ruy}. 

Using the same technique, a more elaborate calculation incorporating meson rescattering contributions to kaon photoproduction on the deuteron, $\gamma d\to K^+ \Lambda n$, was later carried out \cite{Maxwell:2004fs}. It was shown that rescattering effects can be significant across a wide kinematical range, except in kinematical regions where the spectator neutron has a small momentum. This study was subsequently extended to investigate the sensitivity of the observables to the choice of elementary operator, final-state interactions, and the deuteron wavefunction \cite{Maxwell:2004ga}. The analysis showed that deuteron cross sections and polarization observables are largely insensitive to these model ingredients, including the choice of deuteron wavefunction, whereas the corresponding proton observables are much more sensitive.

Still within the QFS kinematical region, a general formalism for kaon photoproduction on heavier nuclei, A$(\gamma,KY)$B, has been developed within the distorted-wave impulse approximation (DWIA) framework \cite{Lee:1999kd}. For certain kinematical conditions, the calculated observables exhibit sensitivity to the hyperon-nucleus interaction, making them well suited for testing available hyperon-nucleus optical potentials. On the other hand, some polarization observables are found to be relatively insensitive to distortion effects and, therefore, provide an excellent tool for studying medium modifications of the elementary amplitude.

Another application of the elementary operator is in the photoproduction~\cite{Mart:1996ay,Mart:1997dc} and electroproduction~\cite{Mart:2008gq} of the hypertriton, as illustrated in Fig.~\ref{fig:elementary_operator}(b). Unlike production on the deuteron, the final state in this case is the bound hypertriton. Therefore, the calculation can be carried out entirely with the few-body wavefunctions of $^3$He and the hypertriton. It should be noted that experimental data are scarce, with only three points available for hypertriton electroproduction \cite{JeffersonLabE91-016:2004qei}, measured more than two decades ago, while no experimental measurement has yet been performed for photoproduction, even though the latter is both theoretically and experimentally simpler.

The wavefunctions of both $^3$He and hypertriton can be written as
\begin{eqnarray}
\label{eq:wfhe3}
\Psi(\boldsymbol{p},\boldsymbol{q}) & = & \sum_{\alpha=(LSJljT)} 
\phi_{\alpha}(p,q)~ \left|\,\{(LS)J,(l {\textstyle \frac{1}{2}})
j\} {\textstyle \frac{1}{2}} M_{\rm i}\ \right\rangle ~ \left|\,
(T{\textstyle \frac{1}{2}}) {\textstyle \frac{1}{2}} M_{t}\right\rangle
\nonumber\\
&=&\sum_{\alpha =(LSJljT)}
\sum_{\left. \begin{array}{c} \\[-6ex]
m_{L}m_{S}m_{l}\\[-2ex] m_{s}m_{J}m_{j} \end{array} \right.}\phi_{\alpha}
(p,q)~ (Lm_{L}Sm_{S}|J m_{J})~ 
(l m_{l}{\textstyle \frac{1}{2}} m_{s}|j m_{j}) \nonumber\\
&& \hspace{2.5cm} \times (J m_{J} j m_{j}
|{\textstyle \frac{1}{2}} M_{\rm i}) ~ 
Y^{L}_{m_{L}} (\boldsymbol{\hat{p}})~Y^{l}_{m_{l}}(\boldsymbol{\hat{q}})
~\chi^{S}_{m_{S}}~\chi_{m_{s}}^{\frac{1}{2}}~ \left|\,
(T{\textstyle \frac{1}{2}}){\textstyle \frac{1}{2}} M_{t}\right\rangle,
\end{eqnarray}
where the notation of Ref.~\cite{deShalit:1963} for the Clebsch-Gordan coefficients is adopted. The three-body Jacobi momenta are defined as $\boldsymbol{p}=\frac{1}{2}(\boldsymbol{k}_2-\boldsymbol{k}_3)$ and $\boldsymbol{q}=\boldsymbol{k}_1$, where $\boldsymbol{k}_2$ and $\boldsymbol{k}_3$ denote the momenta of the spectator nucleons with spin and orbital angular momenta $S$ and $L$, respectively, while $\boldsymbol{k}_1$ is the momentum of the active nucleon interacting with the photon, carrying spin $1/2$ and orbital angular momentum $l$. The number of included partial waves is controlled by $\alpha$. In total, the $^3$He wavefunction contains 34 partial waves \cite{Stoks:1994wp}, while the hypertriton has 16 \cite{Miyagawa:1993rd}. A more detailed discussion can be found in Ref.~\cite{Mart:2008gq}.

The elementary operator of Eq.~(\ref{eq:t_gamma_K}) can be slightly modified to $t_{\gamma K}=\epsilon_\mu\, J^\mu$, with
\begin{equation}
    J^\mu = j^\mu_0 + \sigma_x\, j^\mu_x + \sigma_y\, j^\mu_y + \sigma_z\, j^\mu_z = \sum_{n=0,1}\;\sum_{m_n=-n}^{+n}\, (-1)^{m_n}\,\sigma^{(n)}_{-m_n}\, [\,j^{\mu}\,]^{(n)}_{m_n},
\end{equation}
where the elementary operator $[j^\mu]^{(n)}_{m_n}$ is completely frame independent, since it does not depend on the reference frame in which $\epsilon^\mu$ and $\sigma^{(n)}$ are defined. The transition matrix can then be expressed as
\begin{eqnarray}
  \langle\, {\rm f} \left| \, J^\mu \, \right| {\rm i} \,\rangle &=& 
  \sqrt{6}\, \sum_{\alpha , \alpha '}~
  \sum_{{\sf m},{\sf m}'}~  \sum_{n,m_{n}}
        \left(Lm_{L}Sm_{S}|Jm_{J}\right) \, 
        \left(Lm_{L}Sm_{S}|J'm_{J'}\right) \, 
        \left(lm_{l}{\textstyle \frac{1}{2}}m_{s}|jm_{j}\right)
   \nonumber\\
&& \times\,\left(l'm_{l'}{\textstyle \frac{1}{2}}m_{s'}|j'm_{j'}\right) \, 
        \left(Jm_{J}jm_{j}|{\textstyle \frac{1}{2}}M_{\rm i}\right) \, 
        \left(J'm_{J'}j'm_{j'}|{\textstyle \frac{1}{2}}M_{\rm f}\right) \,
        \left({\textstyle \frac{1}{2}}-\! m_{s'}
              {\textstyle \frac{1}{2}}m_{s} | n m_{n} \right)
        \nonumber\\
&& \times\, (-1)^{n-\frac{1}{2}-m_{s'}}
            \delta_{LL'}\,\delta_{m_{L}m_{L'}}\,
            \delta_{SS'}\,\delta_{m_{S}m_{S'}}\,\delta_{T0}
   \nonumber\\
&& \times\,\int p^{2}dp~d^{3}\boldsymbol{q}\;\phi_{\alpha '}(p,q')\;
   \phi_{\alpha}(p,q)\; Y^{l'}_{m_{l'}}(\boldsymbol{\hat q}')
   \, Y^{l}_{m_{l}}(\boldsymbol{\hat q})
   \; \left[\,j^{\mu}\,\right]^{(n)}_{m_n},
\label{eq:trans_matrix_hyp}
\end{eqnarray}
from which the cross section follows \cite{Mart:2008gq}.

%%%%%%%FIGURE 1%%%%%
    \begin{figure}[htb]
    \centering
    \includegraphics[width=0.6\columnwidth]{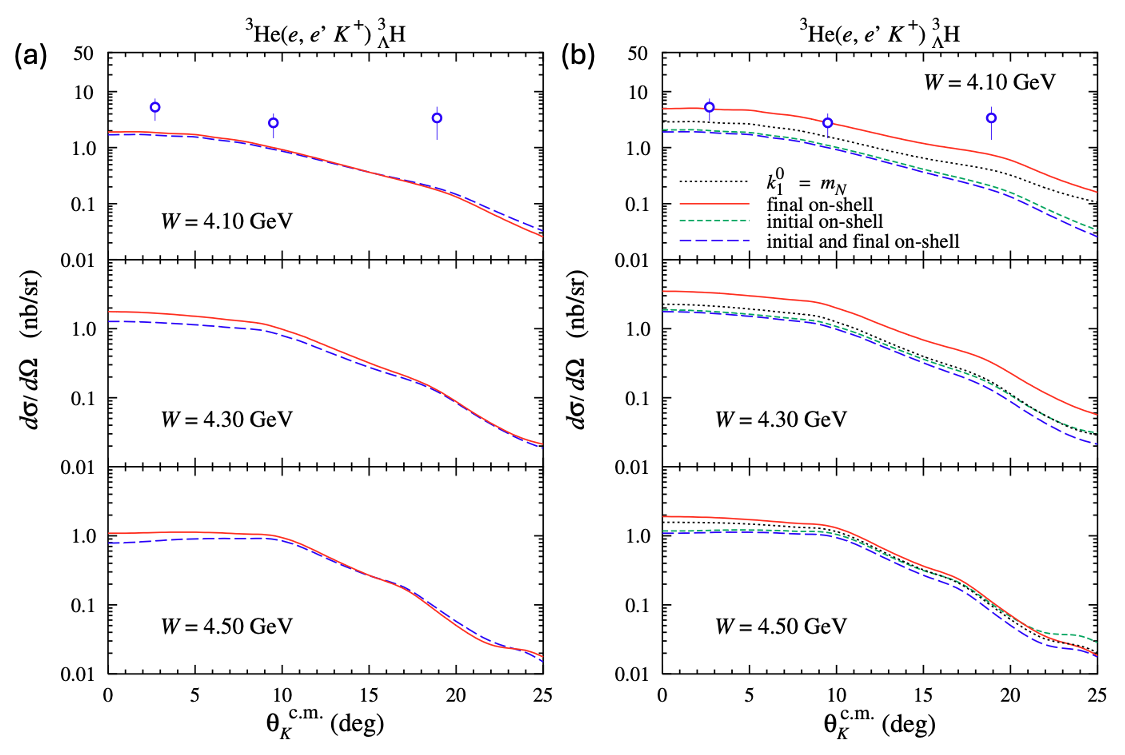} 
    \caption{Comparison of model calculations with experimental data \cite{JeffersonLabE91-016:2004qei} for the differential cross section of hypertriton electroproduction as a function of the kaon scattering angle. (a) Effect of the $s$-wave approximation: the dashed curves are obtained using the $s$-wave contribution only, while the solid curves include all partial waves. (b) Effect of different off-shell assumptions for the active nucleon in the initial nucleus and the hyperon in the final nucleus. Figures from Ref.~\cite{Mart:2008gq}.}
    \label{fig:hyp_el_dCS} 
    \end{figure}
%%%%%%%%%%%%%%%%%%%%

It is well known that for $^3$He most of the nuclear contribution arises from the two $s$-wave components, whereas the hypertriton has only a single $s$-wave component due to its isospin being zero. As a result, the cross section for hypertriton production is much smaller than that for pion production \cite{Tiator:1980jw}. A comparison between the calculated cross sections and experimental data is shown in Fig.~\ref{fig:hyp_el_dCS}. Figure~\ref{fig:hyp_el_dCS}(a) demonstrates that the contribution from higher partial waves is relatively small. Only at larger total c.m. energies $W$ and near forward angles, where the momentum transfer is sizable, do these effects become significant. This behavior reflects the limited number of partial waves and the single $s$-wave component in the hypertriton. 

Figure~\ref{fig:hyp_el_dCS}(b) illustrates the effect of different off-shell assumptions for the active nucleon in the initial and final state nuclei. Remarkably, assuming that the nucleon in the hypertriton is on-shell yields good agreement with the experimental data. Although this behavior differs from the phenomenon observed in pion photoproduction on $^3$He, it can be understood as a consequence of the weak binding of the $\Lambda$ to the deuteron in the $^3_\Lambda$H. Accordingly, bringing the hyperon mass closer to its on-shell value makes the model more realistic \cite{Mart:2008gq}. 

Considerable effort has also been devoted to studying the electromagnetic production of heavier hypernuclei by employing different nuclear-structure approaches while still using similar elementary operators. For example, the production of $^{12}_{~\Lambda}\mathrm{B}$ through the photoproduction process $^{12}\mathrm{C}(\gamma,K^+)^{12}_{~\Lambda}\mathrm{B}$ has been investigated with various approaches \cite{Bennhold:1989bw,Sotona:1994st,Motoba:1994sr,vanderVentel:2011nk}. More recently, calculations based on the equation-of-motion phonon method (EMPM) have been performed for both light and medium-mass hypernuclei such as $^{12}_{~\Lambda}\mathrm{B}$ and $^{16}_{~\Lambda}\mathrm{N}$, the heavier systems $^{28}_{~\Lambda}\mathrm{Al}$, $^{40}_{~\Lambda}\mathrm{K}$, and $^{48}_{~\Lambda}\mathrm{K}$, as well as the very heavy nucleus $^{208}_{~~\Lambda}\mathrm{Tl}$ \cite{Bydzovsky:2023osl,Bydzovsky:2025nnt}.

Finally, it is important to note that a new elementary operator for kaon photoproduction in all six isospin channels has recently become available \cite{Mart:2025ufa}. This state-of-the-art operator incorporates 26 nucleon resonances in the $K\Lambda$ channels and an additional 17 $\Delta$ resonances in the $K\Sigma$ channels. It was fitted to nearly 17,000 experimental data points, including differential cross sections, as well as single- and double-polarization observables from the latest measurements. The operator has also been applied to calculations of hypertriton photoproduction \cite{Mart:2025ufa} and kaon photoproduction on the deuteron in quasi-free kinematics \cite{Mutoharoh:2026rxn}, where it revealed a pronounced forward-peaking behavior in the differential cross section that was absent in previous calculations \cite{Mart:2008gq}. 

\subsection{Missing Nucleon Resonances}
\label{sec:missing}

In this section, it is not our intention to provide a comprehensive discussion of the progress in missing $s$-channel resonance searches, as this is a broad topic with numerous findings that cannot be covered here in detail (for more recent and comprehensive reviews, see, e.g., Refs.~\cite{Burkert:2025coj,Doring:2025sgb,Thiel:2022xtb,Crede:2013kia}). Our purpose is merely to highlight the small part of this effort that is directly relevant to kaon photoproduction, one of the important processes in which the missing resonances are more likely to be observed. Additional information is included in Section~\ref{nstar-spectrum}.

The term ``missing resonances'' refers to resonant states that are predicted by quark and other theoretical models but have not yet been confirmed experimentally and therefore do not appear in the PDG listings. The reason was straightforward, as the PDG listings had primarily relied on $\pi N$ scattering and pion-induced electromagnetic production, for which extensive experimental data and theoretical analyses were available. However, the situation began to change in the late 1990s with the operation of modern detectors and continuous-wave electron accelerators at facilities such as JLab, MAMI, and SPring-8 (see Section~\ref{expt-measurements} for details). These advances enabled precise measurements of reactions involving strange particles, such as kaons and hyperons, which in turn attracted greater theoretical attention. Since then, the investigation of missing resonances has intensified, leading to the discovery and confirmation of numerous nucleon resonances that had long been predicted but were absent from earlier compilations. 

On the theoretical side, Isgur and Karl developed a relativized quark model for baryons in the late 1970s, predicting a rich spectrum of $N^*$ and $\Delta^*$ states. Many of these states were not observed in $\pi N$ scattering data, revealing a significant discrepancy between theory and experiment~\cite{Isgur:1978xj}. Building on this framework, Koniuk and Isgur incorporated chromodynamic effects and calculated the decay properties of these states, emphasizing that several of the predicted resonances might be ``missing'' due to their weak couplings to the pion channels traditionally used in experiments~\cite{Koniuk:1979vy}. It was in this context that the term ``missing resonance'' was first introduced. Subsequent refinements by Capstick, Roberts, Lee, and \v{S}varc employed relativized quark models in conjunction with partial-wave analyses that included not only the $\pi N$ but also the $\eta N$ channel. Their investigation demonstrated that one of the missing $P_{11}$ states could be revealed through the $\eta N$ data~\cite{Capstick:1999dg}. These theoretical advances established a solid foundation for the concept of missing resonances and motivated searches in alternative reaction channels.

\begin{figure}[htb]
\centering  
\includegraphics[width=1.0\textwidth]{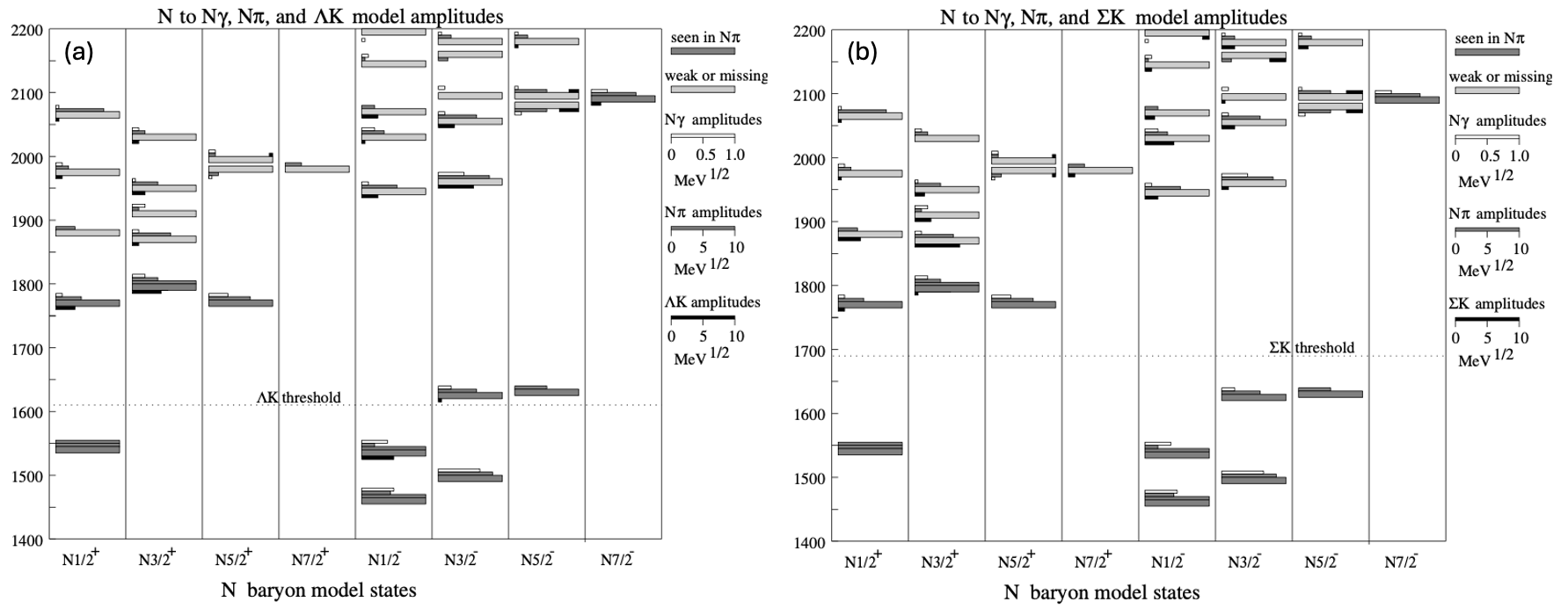}    
\caption{(a) Predicted masses and decay amplitudes to the $\gamma N$, $\pi N$, and $K \Lambda$ channels for nucleon resonances up to 2200 MeV, shown for different spin and parity states along the horizontal axis. The thick bars represent the predicted resonance masses, while the thin bars denote the strengths of the corresponding decay amplitudes. Resonances with large $\gamma N$, $\pi N$, and $K \Lambda$ amplitudes make significant contributions to the $\gamma N\,(\pi N)\to K\Lambda$ process. (b) Same as in the left panel, but for the predicted $\gamma N$, $\pi N$, and $K \Sigma$ amplitudes. Figures from Ref.~\cite{Capstick:1998uh}.}
\label{fig:capstick_1998}
\end{figure}

Capstick and Roberts have also shown that a number of resonances predicted by the relativized quark model couple only weakly to the $\pi N$ channel, but exhibit sizable decay amplitudes into strange final states~\cite{Capstick:1998uh}. In particular, their results for the partial decay amplitudes to $K\Lambda$ and $K\Sigma$, depicted in Fig.~\ref{fig:capstick_1998}, indicate that several higher-mass $N^*$ states possess strong couplings to these channels, in some cases comparable to or larger than their non-strange decay modes. Since the $K\Lambda$ final state also acts as an isospin filter, restricting contributions to $N^*$ resonances only, it provides an especially clean environment for studying states that remain unobserved in $\pi N$ scattering. These findings emphasize that many of the missing resonances are more likely to be revealed in kaon photoproduction and electroproduction, thereby motivating the focus on strangeness channels in modern experimental programs.

Based on the results of the quark model shown in Fig.~\ref{fig:capstick_1998}, Mart and Bennhold in 1999 compared the fits to the older data with those obtained using the published SAPHIR data \cite{Mart:1999ed}. As shown in Fig.~\ref{fig:mart99_SAPHIR}, the older data exhibit no clear structure up to $W = 2.2$ GeV, mainly due to their large uncertainties. Fitting the isobar model to these data yields a single broad peak in the total cross section, represented by the dashed curve. In contrast, the SAPHIR data \cite{SAPHIR:1998fev} revealed a distinct structure near $W \approx 2$ GeV that is absent in the earlier measurements. To reproduce this feature, Mart and Bennhold tested four missing resonances predicted by the quark model of Capstick and Roberts, whose properties are summarized in Table~\ref{tab:fit_to_mis_res}. For comparison, the fitted resonance parameters obtained from the old (pre-SAPHIR) and the 1998 SAPHIR data are also listed in the same table. It is obvious that the best agreement is achieved when the $N(1960)3/2^-$ state from the quark model is included. The corresponding total cross section calculated with this state is shown by the solid red line in Fig.~\ref{fig:mart99_SAPHIR}, demonstrating excellent agreement with the experimental data.

\begin{table}[htbp]
\setlength{\tabcolsep}{6pt} % Default value: 6pt
\renewcommand{\arraystretch}{0.8} % Default value: 1
    \centering
    \caption{Properties of the missing resonances obtained from fits to the old (pre-SAPHIR) and the 1998 SAPHIR data (Fit), compared with those predicted by the quark model (QM). The QM photocouplings are taken from Ref.~\cite{Capstick:1992uc}, while the $K\Lambda$ decay widths are from Ref.~\cite{Capstick:1998uh}. Table adapted from Ref.~\cite{Mart:1999ed}.}
    \label{tab:fit_to_mis_res}
    %\begin{tabular}{lcccc}
    \begin{tabularx}{\textwidth}{XXXXr} \hline \hline
        Missing Resonance & Model & $m_{N^*}$ (MeV)& $\Gamma_{N^*}$ (MeV) & $\sqrt{\Gamma_{N^*N\gamma}\Gamma_{N^*K\Lambda}}/\Gamma_{N^*}~ (10^{-3})$ \\
        \hline
        $N(1945)1/2^-$ & Fit & $1847$ & $258$ & $-10.370\pm 0.875$ \\
               & QM  & $1945$ & $595$ & $0.298\pm 0.349$ \\ [0.5ex]
        $N(1975)1/2^+$ & Fit & $1935$ & $131$ & $9.623\pm 0.789$ \\
               & QM  & $1975$ & $ 45$ & $1.960\pm 0.535$ \\ [0.5ex]
        $\boldsymbol{N(1960)3/2^-}$ & {\bf Fit} & ${\bf 1895}$ & ${\bf 372}$ & ${\bf 2.292^{+0.722}_{-0.204}}$ \\
        & {\bf QM}  & ${\bf 1960}$ & ${\bf 535}$ &${\bf -2.722\pm 0.729}$ \\ [0.5ex]
        $N(1950)3/2^+$ & Fit & $1853$ & $189$ & $1.097^{+0.011}_{-0.010}$ \\
        & QM  & $1950$ & $140$ & $-0.334\pm 0.070$\\ [0.5ex]
    \hline \hline
    \end{tabularx}
\end{table}

\begin{figure}[htb]
    \centering
    \includegraphics[width=0.5\columnwidth]{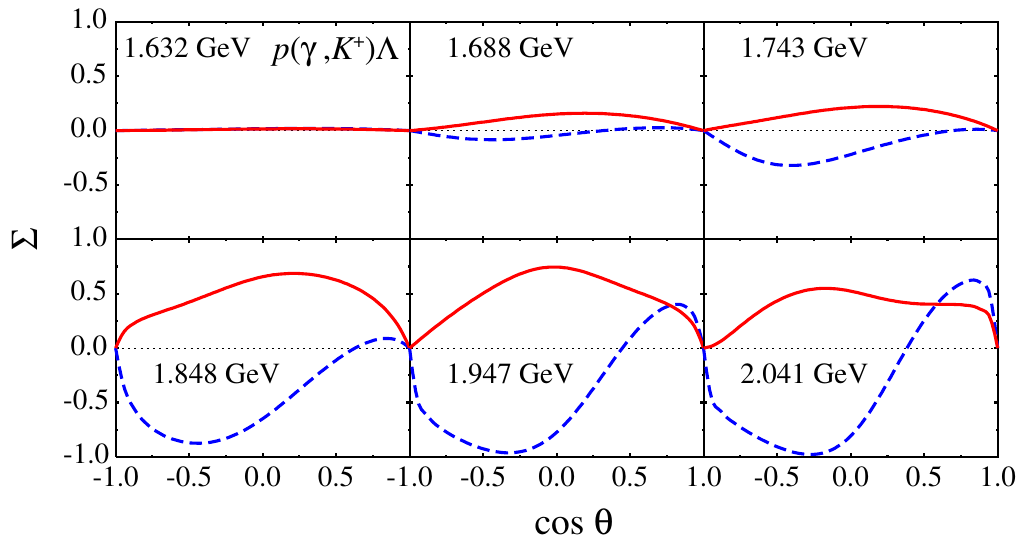} 
    \caption{Photon beam spin asymmetry $\Sigma$ obtained from fits without (dashed blue lines) and with (solid red lines) inclusion of the SAPHIR data from Ref.~\cite{SAPHIR:1998fev}. Figure from Ref.~\cite{Mart:1999ed}, with color added for clarity.}
    \label{fig:missing_photon} 
    \end{figure}

This surprising result, shown in both Table~\ref{tab:fit_to_mis_res} and Fig.~\ref{fig:mart99_SAPHIR}, could be fortuitous. Therefore, it is essential to examine additional observables that can clearly discriminate between models with and without missing resonances. One such example is presented in Fig.~\ref{fig:missing_photon}, which illustrates that including or excluding a missing resonance leads to significantly different predictions for the photon beam spin asymmetry, particularly in the higher-energy region.

%%%%%%%FIGURE 1%%%%%
%%%%%%%%%%%%%%%%%%%%

In the algebraic model~\cite{Bijker:2000gq}, both the internal (spin-flavor-color) and spatial degrees of freedom of hadrons are described within a unified algebraic framework. Within this approach, the $N(1895)3/2^-$ state observed in Ref.~\cite{Mart:1999ed} can be interpreted in terms of several possible configurations. The lowest of these corresponds to the $^{2}8_{3/2}[56, 1^-]$ state with a predicted mass of 1847~MeV, which lies in the same vibrational band as the $N(1710)1/2^+$. Alternatively, the $N(1895)3/2^-$ may be assigned to either the $^{2}8_{3/2}[70, 2^-]$ state with a mass of 1874~MeV or the $^{2}8_{3/2}[70, 1^-]$ state with a mass of 1909~MeV. If one considers only the masses, the algebraic classification of baryon excitations yields a very rich spectrum of resonance states. Nevertheless, the coupling strengths offer a more decisive means of determining the correct assignment, as exemplified by the results obtained in the isobar model (see Table~\ref{tab:fit_to_mis_res}).

Further support for the existence of the $N(1895)3/2^-$ state was provided by the relativistic quark model~\cite{Loring:2001kx}, hypercentral constituent quark model~\cite{Giannini:2001kb}, and by studies of $\omega$-meson photoproduction~\cite{Oh:2000zi} and scattering~\cite{Penner:2001fu}, although the predicted masses differ slightly among the models. By varying the hadronic form factors and including hyperon resonances in the $u$-channel, it has also been shown that the $N(1895)3/2^-$ can successfully account for the $K^+\Lambda$ photoproduction data. Nevertheless, the observed structure in the cross section at $W \approx 1900~\text{MeV}$ can be reproduced equally well by including either the $N(1895)3/2^-$ or $N(1895)3/2^+$~\cite{Janssen:2001wk}.

A subsequent study that incorporated a genetic algorithm into the fitting strategy of the $K^+\Lambda$ photoproduction data found that the available experimental data favor a $P_{11}$ resonance with a mass of 1895 MeV, whereas the $S_{11}$ and $D_{13}$ resonances with the same mass are only weakly supported \cite{Ireland:2004kp}. In contrast, the coupled-channel analysis performed by the Giessen group reported that the second peak in the $K^+\Lambda$ differential cross sections near 1900 MeV observed by SAPHIR and CLAS, can be reproduced by a coherent sum of resonance and background contributions \cite{Shklyar:2005xg}. A similar study, but focused on the forward-angle data, demonstrated that including either a $N(1900)3/2^-$ or $N(1900)1/2^+$ ``missing'' resonance together with the two-star $N(1900)3/2^+$ significantly improves the agreement with the data, with the $N(1900)1/2^+$ emerging as the more likely missing-resonance candidate \cite{Corthals:2005ce}. The importance of the $N(1900)3/2^+$ in $K^+\Lambda$ photoproduction was also emphasized by the dynamical coupled-channel model \cite{Julia-Diaz:2006rvy}, despite the fact that the Bonn-Gatchina multi-channel partial-wave analysis identified the $D_{13}$ resonance as more significant, but at a mass of 1875 MeV \cite{Anisovich:2011ye,Anisovich:2011sv}.

An analysis employing a Bayesian inference method indicated that for $K^+\Lambda$ photoproduction the “missing” resonance $N(1900)1/2^+$ plays a decisive role, whereas the “missing” $N(1900)3/2^-$ remains significant but not decisive \cite{DeCruz:2012bv}. This work was a follow-up to Ref.~\cite{DeCruz:2011xi}, which applied the Bayesian inference method and demonstrated that both the $N(1900)1/2^+$ and $N(1900)3/2^+$ are of comparable importance in describing the $K^+\Lambda$ photoproduction data. 

\begin{figure}[htb]
    \centering
    \includegraphics[width=0.6\columnwidth]{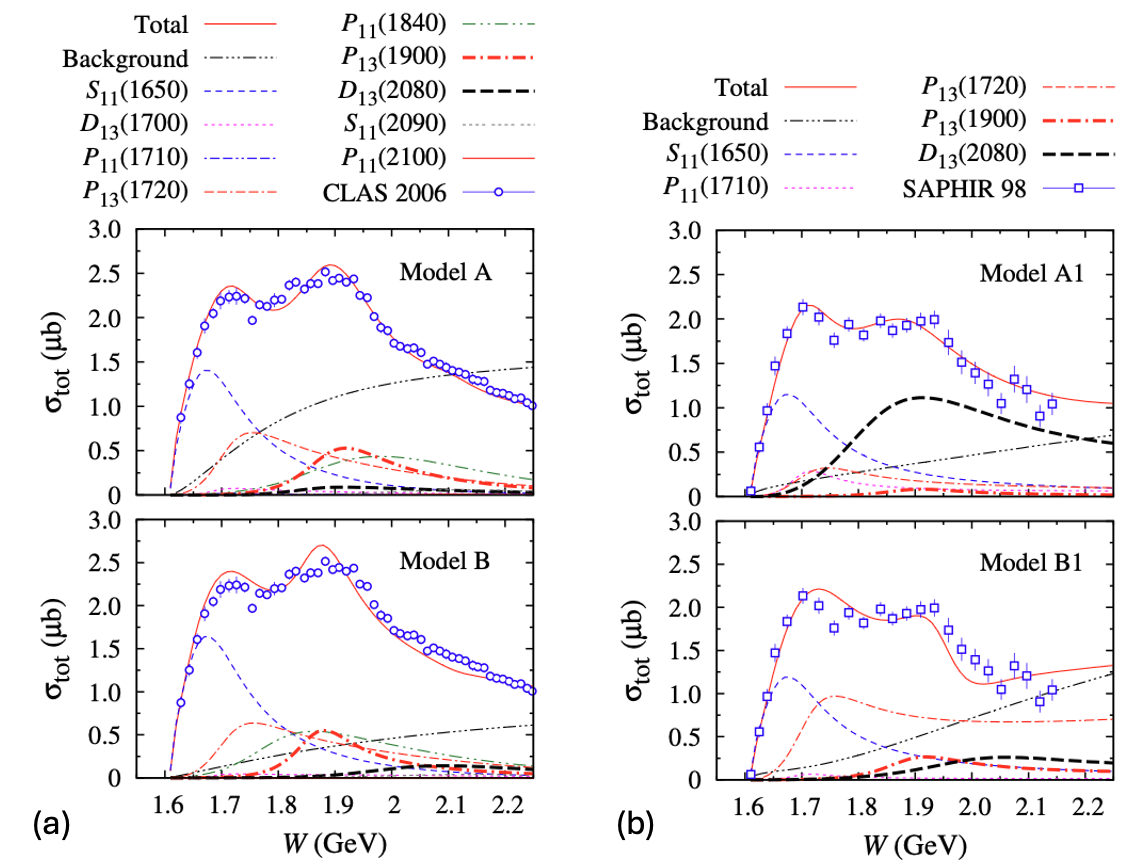} 
    \caption{(a) Contributions of the background and resonance terms to the $K^+\Lambda$ photoproduction total cross section, obtained by fitting the mass and width of either the $N(2080)3/2^-$ (top, model A) or the $N(1900)3/2^+$ (bottom). Experimental data are from CLAS \cite{Bradford:2005hq}. (b) Same as the left panels, but obtained by refitting the Kaon-MAID model. Experimental data are from SAPHIR \cite{SAPHIR:1998fev}. Note that the resonances included in models A and B differ from those in Kaon-MAID (models A1 and B1). Figures from Ref.~\cite{Mart:2012fa}.}
    \label{fig:p13_contrib_kpl}
    \end{figure}

Although the existence of the missing $N(1895)3/2^-$ remained controversial for nearly a decade, several subsequent studies on $K^+\Lambda$ photoproduction indicated that the second peak observed in the differential and total cross sections near 1900~MeV originates from the $N(1900)3/2^+$ rather than the $N(1895)3/2^-$ \cite{Anisovich:2017ygb,Nikonov:2007br,Hunt:2018wqz}. A reanalysis of $K^+\Lambda$ photoproduction using a substantially larger database than that employed in Kaon-MAID, which also includes the double-polarization observables $C_x$ and $C_z$, confirmed these findings \cite{Mart:2012fa}. The result is shown in Fig.~\ref{fig:p13_contrib_kpl}.

%%%%%%%FIGURE 1%%%%%
%%%%%%%%%%%%%%%%%%%%

It is important to emphasize that the total cross section data shown in Fig.~\ref{fig:p13_contrib_kpl} are presented solely for clarity of explanation. In reality, the total number of data points used in the fitting procedure is much larger, encompassing differential cross section, single-, and double-polarization observables. As a consequence, the database used in Ref.~\cite{Mart:2012fa} is significantly more extensive than that employed in Kaon-MAID~\cite{Mart:1999ed}. In Fig.~\ref{fig:p13_contrib_kpl}(a), the background and resonance contributions obtained when the mass and width of the $N(2080)3/2^-$ (model A) or the $N(1900)3/2^+$ (model B) are fitted to the data are compared. In model A (B), the mass and width of the $N(1900)3/2^+$ ($N(2080)3/2^-$) are fixed to their PDG values~\cite{ParticleDataGroup:2010dbb}. Model A yields a slightly smaller $\chi^2/N$, namely 2.57, compared to 2.68 for model B, indicating that the data favor the PDG parameters of the $N(1900)3/2^+$. Nevertheless, as shown in Fig.~\ref{fig:p13_contrib_kpl}(a), both models consistently indicate that the second peak arises predominantly from the $N(1900)3/2^+$. This finding is rather unexpected and contradicts the conclusion of Ref.~\cite{Mart:1999ed}.

To explain the origin of the “missing” $N(1895)3/2^-$ reported in Ref.~\cite{Mart:1999ed}, Fig.~\ref{fig:p13_contrib_kpl}(b) can be examined, where the same fitting procedure as in Fig.~\ref{fig:p13_contrib_kpl}(a) is applied but using the same resonance configuration and experimental database as in Kaon-MAID. The resulting models are denoted as A1 and B1, respectively. It is evident that by fitting the mass and width of the $N(2080)3/2^-$ while fixing those of the $N(1900)3/2^+$ (model A1), the $N(1895)3/2^-$ reported in Ref.~\cite{Mart:1999ed} is reproduced precisely, where the second peak is dominated by the $N(2080)3/2^-$ contribution. Conversely, when the $N(1900)3/2^+$ parameters are fitted and those of the $N(2080)3/2^-$ are fixed (model B1), neither resonance provides a dominant contribution to the second peak.

Therefore, there are two possible explanations for why the “missing” $N(1895)3/2^-$ reported in Ref.~\cite{Mart:1999ed} was not confirmed by subsequent studies. The first concerns the database used in Kaon-MAID, as discussed above. The second lies in the criterion employed to identify a “missing” resonance in Kaon-MAID, which was based on matching the resonance decay widths to the $K\Lambda$ and $\gamma p$ channels predicted by the constituent quark model \cite{Capstick:1998uh} with those derived directly from the coupling constants extracted through fits to the $K^+\Lambda$ photoproduction data. In contrast, subsequent analyses determined the relevance of missing resonances solely by minimizing the overall $\chi^2$ value. 

In the Kaon-MAID analysis, fitting the $S_{11}$, $P_{11}$, $P_{13}$, and $D_{13}$ resonances yielded extracted masses of 1847, 1934, 1853, and 1895~MeV, with the corresponding $\chi^2/N$ values of 2.70, 3.29, 3.15, and 3.36, respectively \cite{Mart:1999ed,Mart:2012fa}. Thus, if the Kaon-MAID criterion had been based solely on $\chi^2/N$, the $N(1847)1/2^-$ state would have been identified as the origin of the second peak. Interestingly, under this same criterion, the $N(1853)3/2^+$ state would have been a more likely candidate for the “missing” resonance than the $N(1895)3/2^-$.

\subsection{Pentaquark and Narrow Resonances}
\label{sec:penta-narrow}

\begin{figure*}[htb]
\centering
\includegraphics[width=0.75\textwidth]{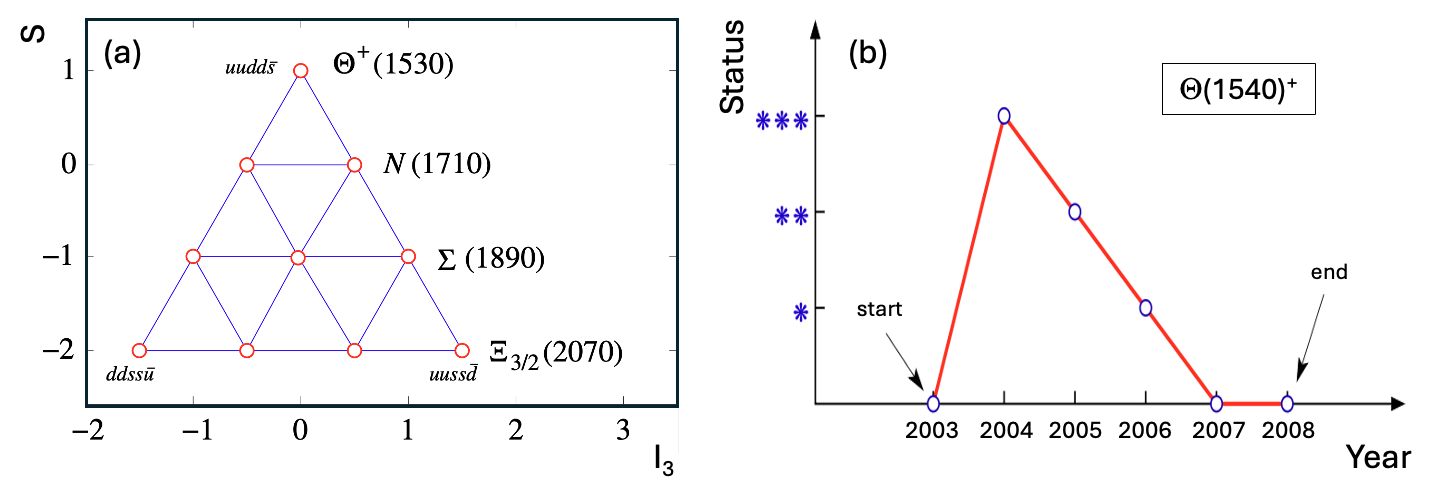}
\caption{(a) Antidecuplet of baryons predicted by the chiral soliton model~\cite{Diakonov:1997mm} in terms of strangeness vs. the third component of isospin. The states located at the corners of the antidecuplet correspond to exotic baryons, and the quark content of these exotic states is shown. (b) The history of the low-energy pentaquark $\Theta(1540)^+$ status, as documented by the Review of Particle Physics (RPP) of the Particle Data Group (PDG) between 2003 and 2008, began with the publication of the first observation by the LEPS Collaboration \cite{LEPS:2003wug}, which was first tabulated by the PDG in the following year \cite{ParticleDataGroup:2004fcd}, and concluded with the summary written by Wohl in 2008 \cite{ParticleDataGroup:2008zun}.}
\label{fig:antidecuplet} 
\end{figure*}

The quest for exotic particles can be traced back more than five decades, to a period preceding the first appearance of such candidates in the PDG listings. The PDG's 1976 edition tabulated the observed candidates $Z_0(1780)P_{01}$, $Z_0(1865)D_{03}$, and $Z_1(1900)3/2^+$, all carrying strangeness $S=1$~\cite{ParticleDataGroup:1976hcu}. However, due to a lack of strong theoretical motivation at the time, both experimental and theoretical efforts in this direction remained largely dormant until 2003, when the LEPS Collaboration reported the observation of the exotic pentaquark $\Theta^+(1540)$ in the $\gamma n \to K^-K^+n$ reaction on a $^{12}$C target~\cite{LEPS:2003wug}. This state had been anticipated in the chiral quark soliton model ($\chi$QSM) of Diakonov, Petrov, and Polyakov~\cite{Diakonov:1997mm}, where it appears as an exotic member of the antidecuplet of baryons illustrated in Fig.~\ref{fig:antidecuplet}(a), containing the $uudd{\bar s}$ quarks similar to the bound $K^+n$ system. 

%%%%%%%FIGURE 1%%%%%
%%%%%%%%%%%%%%%%%%%%

Shortly thereafter, the LEPS observation was supported by at least six independent experimental results from different laboratories~\cite{CLAS:2003yuj,DIANA:2003uet,CLAS:2003wfm,SAPHIR:2003lnh,Asratyan:2003cb,HERMES:2003phe}. These studies involved photo- and electroproduction on carbon, deuterium, and hydrogen targets, $K^+n$ charge-exchange reactions in a liquid-xenon bubble chamber, as well as analyses of compiled neutrino bubble-chamber data on hydrogen, deuterium, and other nuclei. As a result, the 2004 edition of the Review of Particle Physics (RPP) by the PDG assigned a three-star status to the pentaquark $\Theta^+(1540)$~\cite{ParticleDataGroup:2004fcd}. However, as illustrated in Fig.~\ref{fig:antidecuplet}(b), following numerous null results in subsequent experiments, the status of the $\Theta^+(1540)$ underwent a steady decline, and by 2007 it no longer carried any star rating. In 2008, the PDG removed this pentaquark from RPP and concluded that the entire episode surrounding the $\Theta^+(1540)$, from the initial discovery claims and the ensuing wave of theoretical and phenomenological studies to the eventual null findings, constitutes a remarkable chapter in the history of scientific exploration~\cite{ParticleDataGroup:2008zun}. 

%%%%%%%FIGURE 1%%%%%
%    \begin{figure}[htb]
%    \centering
%    \includegraphics[width=0.5\columnwidth]{penta_status.eps} 
%    \caption{The history of the low-energy pentaquark $\Theta(1540)^+$ status, as documented by the Review of Particle Physics (RPP) of the Particle Data Group (PDG) between 2003 and 2008, began with the publication of the first observation by the LEPS Collaboration \cite{LEPS:2003wug}, which was first tabulated by the PDG in the following year \cite{ParticleDataGroup:2004fcd}, and concluded with the summary written by C. G. Wohl in 2008 \cite{ParticleDataGroup:2008zun}.}
%    \label{fig:penta_status} 
%    \end{figure}
%%%%%%%%%%%%%%%%%%%%

Since then, research on pentaquarks became dormant once again until the LHCb Collaboration reported a significant $J/\psi p$ structure in the $\Lambda_b^0 \to J/\psi p K^-$ decay channel~\cite{LHCb:2015yax}. This structure could only be reproduced by including two Breit-Wigner amplitudes with masses of approximately 4380 and 4450~MeV, corresponding to statistical significances of 9 and 12 standard deviations, respectively. As these enhancements could not be explained by reflections from $J/\psi\Lambda^*$ resonances or other known sources, they were interpreted as genuine resonance states with a minimal quark content of $c\bar{c}uud$, and were named the charmonium-pentaquark states $P_{c\bar{c}}(4380)^+$ and $P_{c\bar{c}}(4450)^+$. To date, the PDG lists six hidden-charm pentaquark candidates observed at high energies, all of which are assigned one-star status~\cite{ParticleDataGroup:2024cfk}, while on the theoretical side, predictions (see, e.g., Ref.~\cite{Clymton:2024fbf}) have extended to hidden-charm pentaquarks with single $P_{c\bar{c}s}$ \cite{Clymton:2025hez}, double $P_{c\bar{c}ss}$ \cite{Clymton:2025zer}, and even triple $P_{c\bar{c}sss}$ \cite{Clymton:2025dzt} strangeness.

%For the purpose of present discussion it is important to remind that the photoproduction process of low-energy pentaquark $\gamma n \to K^-K^+n$ reported by the LEPS Collaboration can be considered as the photoproduction of $\Theta^+$ from the neutron, $\gamma n \to K^-\Theta^+$, which is similar to the kaon photoproduction on the neutron $\gamma n\to K^-\Sigma^+$. Since, as a target, proton is easier to handle, the process $\gamma p \to \bar{K}^0\Theta^+$ is certainly preferred. The corresponding Feynman diagrams for this process are depicted in Fig.~\ref{fig:penta_feynman}. Since, most of the coupling constants related to $\Theta^+(1540)$ are less known, only the background terms could be calculated for this process. 

%The main coupling constant is $g_{K\Theta N}$, which can be calculated from the reported decay width and obtained to be $g_{K\Theta N}/\sqrt{4\pi}=0.39$ \cite{Mart:2004at}. 

%%%%%%%FIGURE 1%%%%%
%    \begin{figure}[htb]
%    \centering
%    \includegraphics[width=0.6\columnwidth]{penta_feynman.eps} 
%    \caption{\label{fig:penta_feynman} Feynman diagrams for photoproduction of pentaquark $\Theta^+$ on the proton, $\gamma+p\to {\ba K}^0+\Theta^+$. Figure adapted from Ref.~\cite{Mart:2004at}.}
%    \end{figure}
%%%%%%%%%%%%%%%%%%%%

The low-energy pentaquark photoproduction process $\gamma n \to K^-K^+n$ reported by LEPS can be interpreted as the photoproduction of the $\Theta^+$ from the neutron, i.e., $\gamma n \to K^-\Theta^+$. This reaction is analogous to the kaon photoproduction process on the neutron, $\gamma n \to K^-\Sigma^+$, and therefore can be straightforwardly calculated using the same formalism as that employed in kaon photoproduction discussed earlier~\cite{Mart:2004at,Liu:2003dq,Rekalo:2004it,Nam:2004xt,Close:2004da,Roberts:2004rh,Yu:2004wy,He:2004tj,Hao:2005sc,Laget:2006ym, Mart:2011ss}. In view of the current PDG assessment, which considers the evidence against the existence of the $\Theta^+(1540)$ to be highly convincing, it is perhaps more relevant to focus on the non-strange member of the baryon antidecuplet shown in Fig.~\ref{fig:antidecuplet}(a), the $N(1710)$, which had previously been associated with the $N(1710)1/2^+$ resonance reported by the PDG, as its total width at the time was highly uncertain~\cite{ParticleDataGroup:1996dqp}. It should be noted that the non-observation of the $\Theta^+(1540)$ pentaquark does not necessarily imply the absence of the other members of the baryon antidecuplet.

Interestingly, the Chiral Quark Soliton model ($\chi$QSM) predicts the $N(1710)$ state with a total width of about 40~MeV, which is considerably narrower than the range suggested by the PDG (50-250 MeV). Owing to this, the $N(1710)$ was regarded as a narrow resonance. In connection with this width, the model predicts the relative branching ratios to the $\pi N$, $\eta N$, and $K\Lambda$ channels to be 0.13, 0.28, and 0.13, respectively. The relatively large branching ratio to the $\eta N$ channel subsequently stimulated significant interest in re-examining $\eta N$ photoproduction in the energy region around $W \approx 1700$~MeV. Subsequent measurements by the GRAAL Collaboration revealed that the cross section for $\eta$ photoproduction off a free neutron exhibits a pronounced enhancement around $W = 1680$ MeV~\cite{GRAAL:2006gzl}, which was later confirmed by independent experiments conducted by other collaborations~\cite{Miyahara:2007zz,CBELSA:2008epm}. Since such an enhancement is either absent or significantly weaker in the case of a free proton, it can be naturally interpreted as evidence for the presence of a narrow $P_{11}$ resonance~\cite{Fix:2007st}, although alternative interpretations have also been proposed~\cite{Doring:2009qr,Shklyar:2006xw,Choi:2005ki}.

%%%%%%%FIGURE 1%%%%%
    \begin{figure}[htb]
    \centering
    \includegraphics[width=0.45\columnwidth]{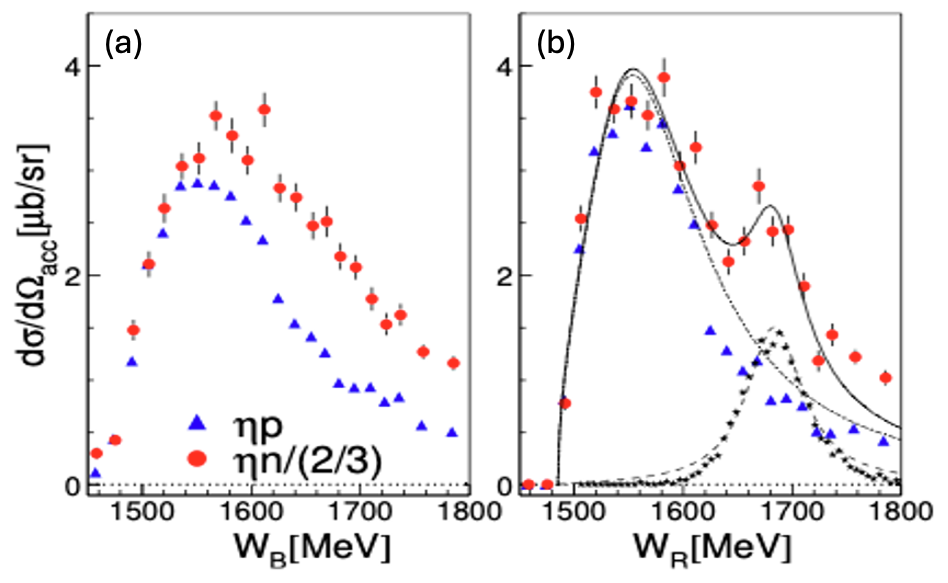} 
    \caption{Differential cross sections for $\eta$ photoproduction on the proton (solid triangles) and neutron (solid circles) as a function of the total c.m. energy, obtained from (a) the photon energy ($W_B$) and (b) reconstructed from the four-vectors of the $\eta$ meson and the recoiling nucleon ($W_R$). The stars represent the response to a $\delta$ function, illustrating the effect of the finite energy resolution. The solid, dash-dotted, and dashed curves correspond to the results of the full fit, the Breit-Wigner (BW) curve for the $N(1535)1/2^-$ resonance, and the BW curve for the second structure, respectively. Figure from Ref.~\cite{CBELSA:2008epm}.}
    \label{fig:narrow_exp_cbelsa} 
    \end{figure}
%%%%%%%%%%%%%%%%%%%%

Figure~\ref{fig:narrow_exp_cbelsa} presents the CB-ELSA measurement~\cite{CBELSA:2008epm} of the differential cross sections for $\eta$ photoproduction on the proton and neutron. To account for the effects of Fermi motion, the differential cross section is also shown as a function of the total c.m. energy reconstructed from the four-vectors of the $\eta$ meson and the recoil nucleon, denoted as $W_R$ (see Fig.~\ref{fig:narrow_exp_cbelsa}(b)), in addition to the conventional representation based on the photon energy, $W_B$ (see Fig.~\ref{fig:narrow_exp_cbelsa}(a)). A distinct second structure is clearly visible in the neutron channel near 1700~MeV in the $W_R$ spectrum, whereas no such feature appears in the proton channel. The fitted position of this structure, at 1683~MeV, is consistent with the result reported by GRAAL~\cite{GRAAL:2006gzl}.

%%%%%%%FIGURE 1%%%%%
    \begin{figure}[htb]
    \centerline{
    \includegraphics[height=0.3\textwidth, angle=90]{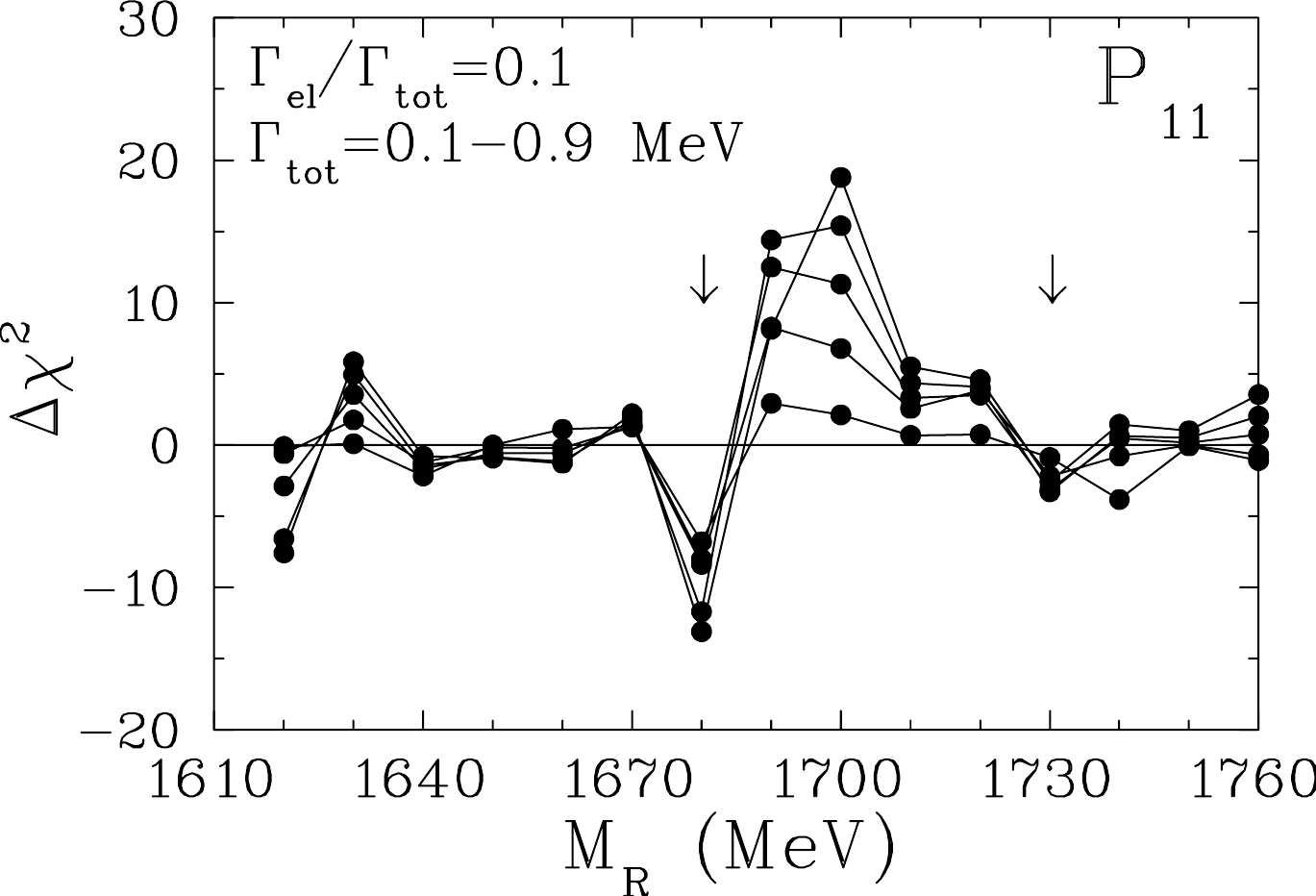}\hspace{0.1cm}
    \includegraphics[height=0.3\textwidth, angle=90]{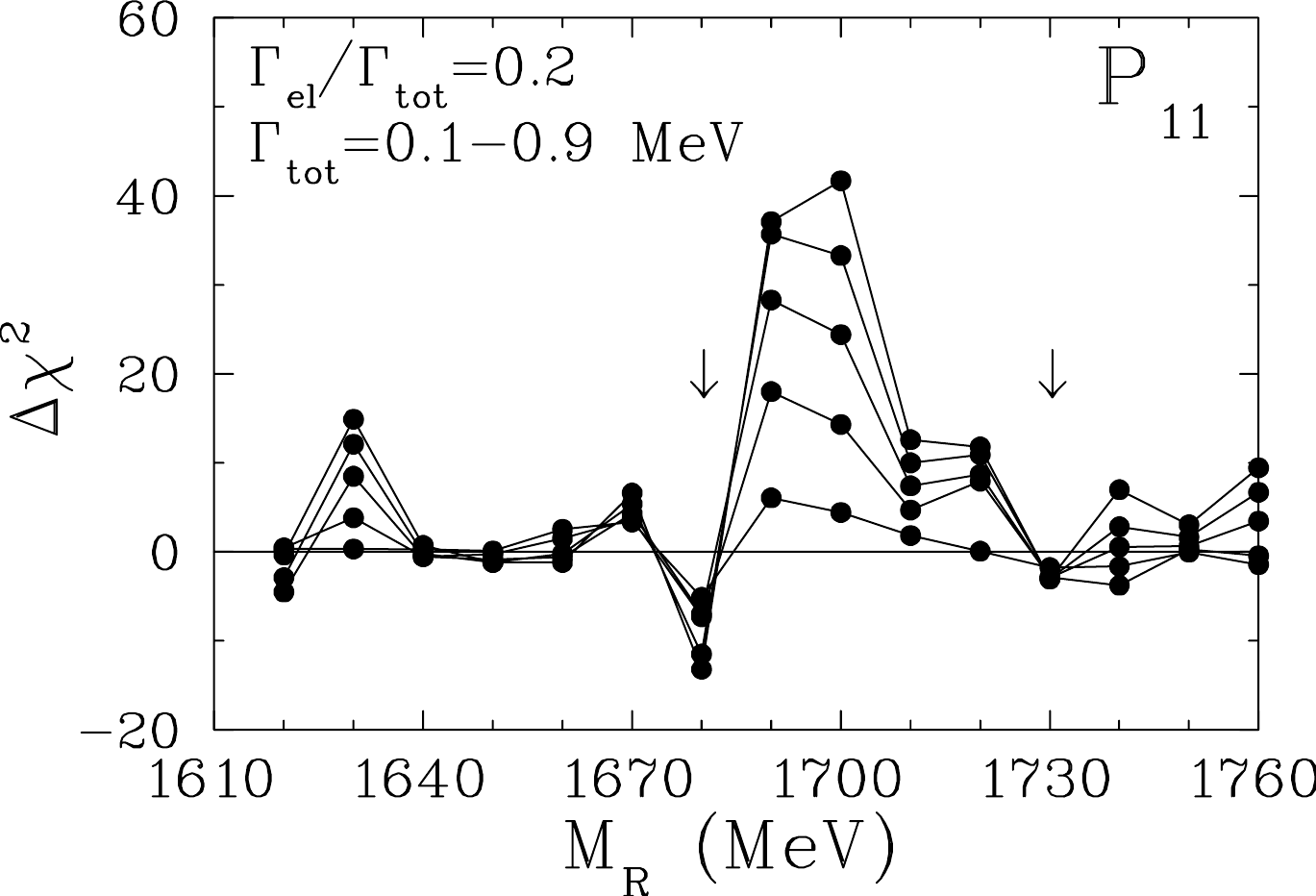}\hspace{0.1cm}
    \includegraphics[height=0.3\textwidth, angle=90]{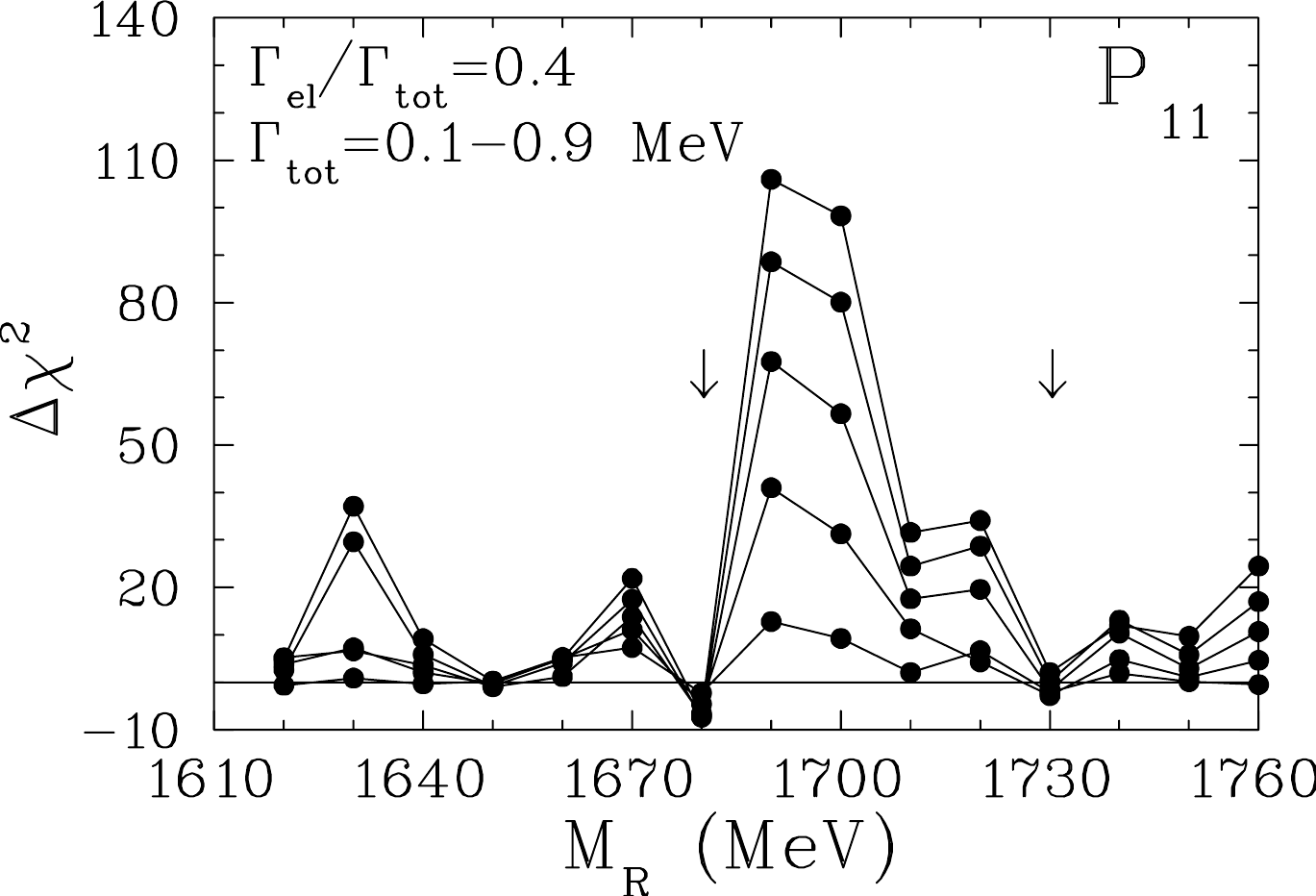}}
    \centerline{
    \includegraphics[height=0.3\textwidth, angle=90]{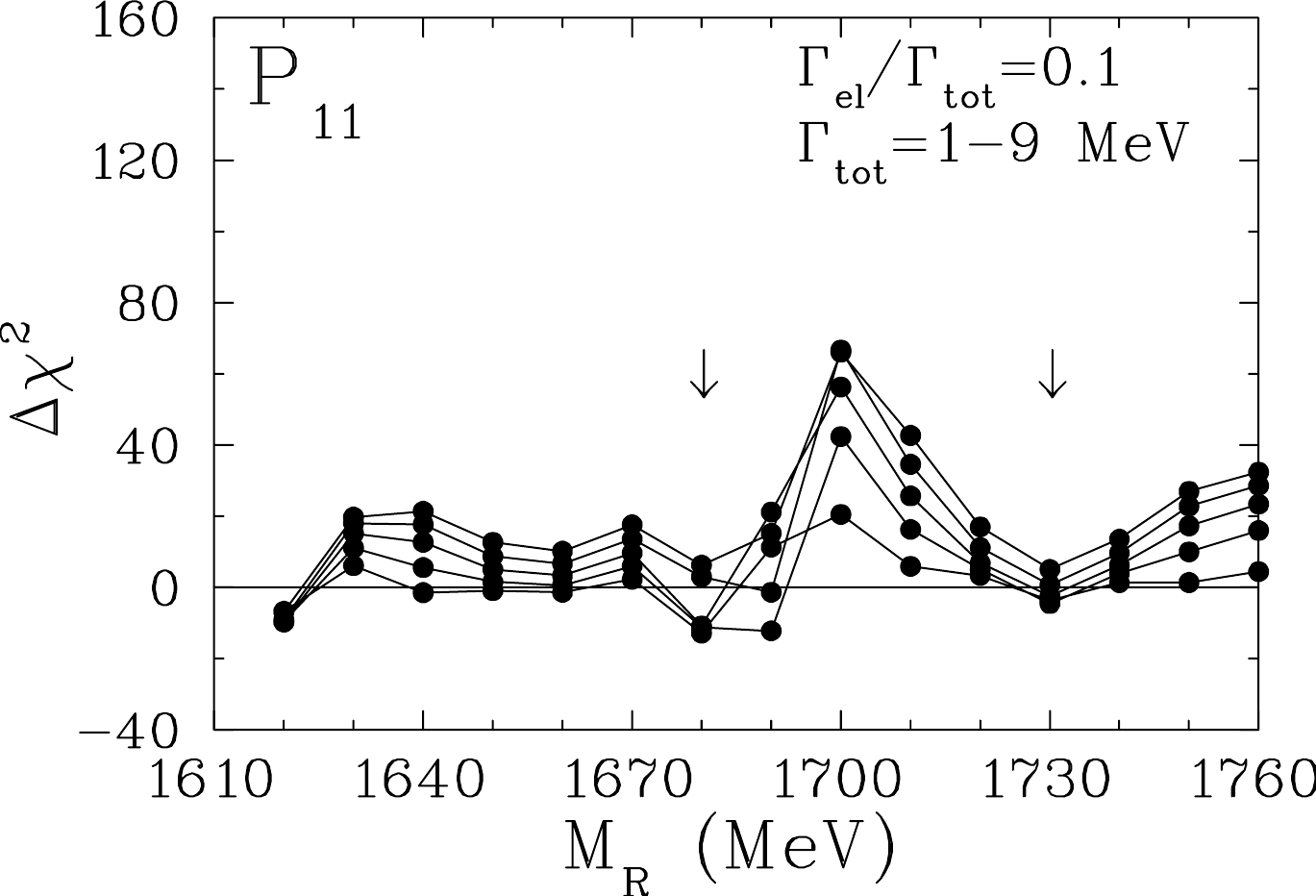}\hspace{0.1cm}
    \includegraphics[height=0.3\textwidth, angle=90]{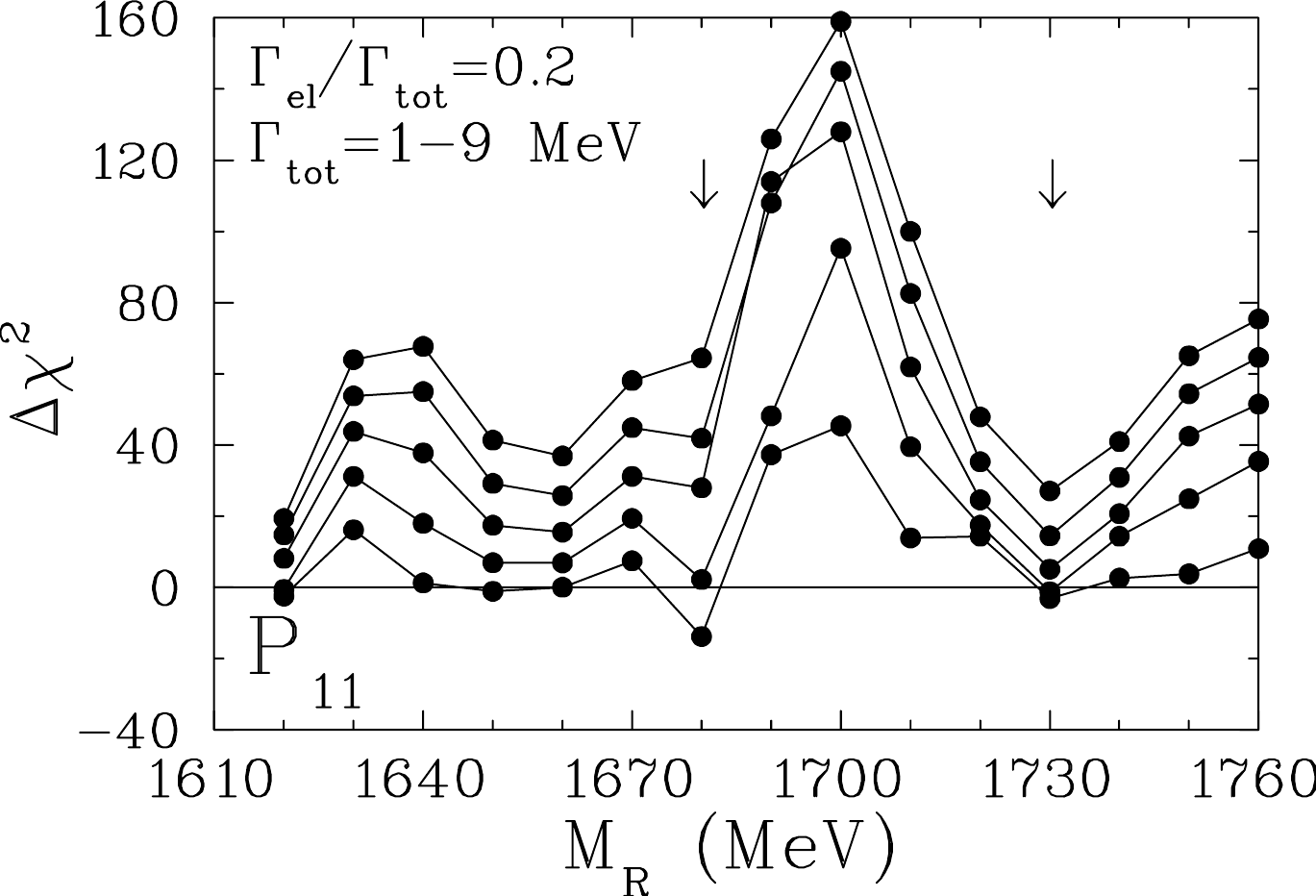}\hspace{0.1cm}
    \includegraphics[height=0.3\textwidth, angle=90]{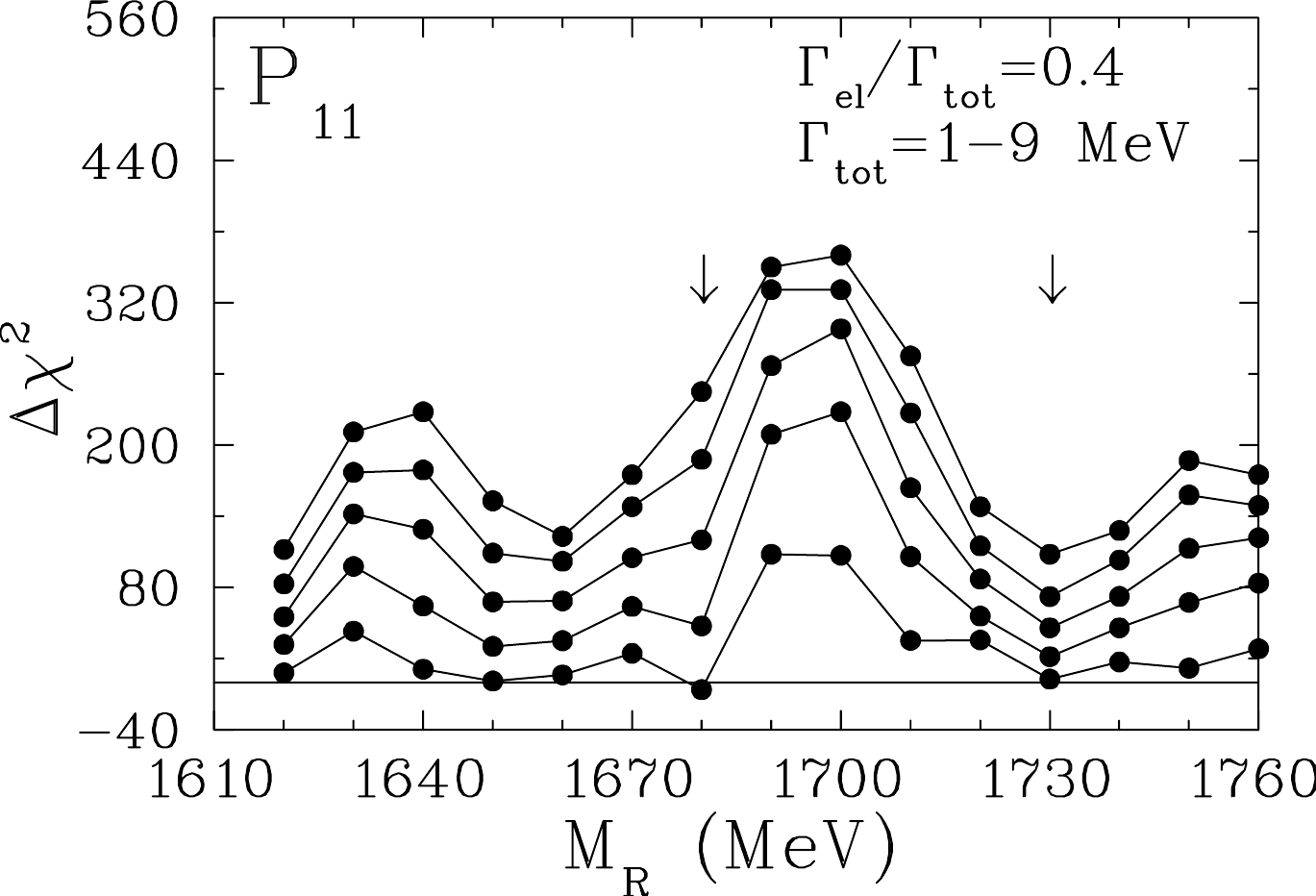}}
    \caption{Dependence of the change of overall $\chi^2$ on the resonance mass $M_R$, relative elastic width $\Gamma_{\mathrm{el}}/\Gamma_{\mathrm{tot}}$, and total width $\Gamma_{\mathrm{tot}}$ for the narrow resonance $P_{11}$. The two vertical arrows mark $M_R = 1680$ and 1730~MeV. Figures from Ref.~\cite{Arndt:2003ga}.}
    \label{fig:narrow_pion} 
    \end{figure}
%%%%%%%%%%%%%%%%%%%%

The finite branching ratio of the narrow $N(1710)$ resonance to the $\pi N$ channel was investigated by Arndt et al. and reported in Ref.~\cite{Arndt:2003ga}. In their modified partial-wave analysis~\cite{Arndt:2003if}, they introduced narrow $P_{11}$, $S_{11}$, and $P_{13}$ resonances and examined, individually, the resulting changes in the overall $\chi^2$. The resonance mass was varied from 1620 to 1760~MeV in 10~MeV steps, the relative elastic width from 0.1 to 0.4, and the total width from 0.1 to 0.9~MeV as well as from 1 to 9~MeV. The result for the $P_{11}$ state is shown in Fig.~\ref{fig:narrow_pion}. Distinct minima in $\Delta\chi^2$ appear at 1680 and 1730~MeV, with the first resonance being the more promising of the two. In the $S_{11}$ and $P_{13}$ cases, no dip was found at 1680~MeV, while a shallow dip appeared at 1730~MeV. However, since this feature differs qualitatively from the 1680~MeV signal, it was not considered a candidate for a narrow resonance. Moreover, the vicinity of this energy corresponds to the opening thresholds of the $\omega N$ and $\rho N$ channels, which can generate significant threshold effects.

The branching ratio to the $K\Lambda$ channel is comparable to that of the $\pi N$ channel. However, it was not thoroughly investigated until 2011, when a detailed study was performed using an isobar model fit to a substantial set of $K\Lambda$ photoproduction data accumulated over the previous three decades~\cite{Mart:2011ey,Mart:2013fia}. Since the earlier study of $\pi N$ photoproduction covered the energy range from threshold up to 1760~MeV, the present model was constructed as an extension of the previous isobar model developed for $K\Lambda$ photoproduction near threshold~\cite{Mart:2010ch}. The extension was limited to $W=1730$~MeV, as beyond this energy inconsistencies were observed between the SAPHIR~\cite{SAPHIR:1998fev,Glander:2003jw} and CLAS~\cite{CLAS:2005lui} datasets. Furthermore, at higher energies the inclusion of hadronic form factors becomes necessary to reproduce the data, which complicates the analysis due to the associated violation of gauge invariance. Two models were proposed in Ref.~\cite{Mart:2011ey}. In the first, the coupling parameters were fitted under SU(3) and PDG constraints, while in the second all parameters were allowed to vary freely. Although the first model provided a better description of the data, the second served to assess model dependence. As both yielded consistent results, only those from the first model are discussed here.

%%%%%%%FIGURE 1%%%%%
\begin{figure}[htb]
\centering
\includegraphics[width=0.8\textwidth]{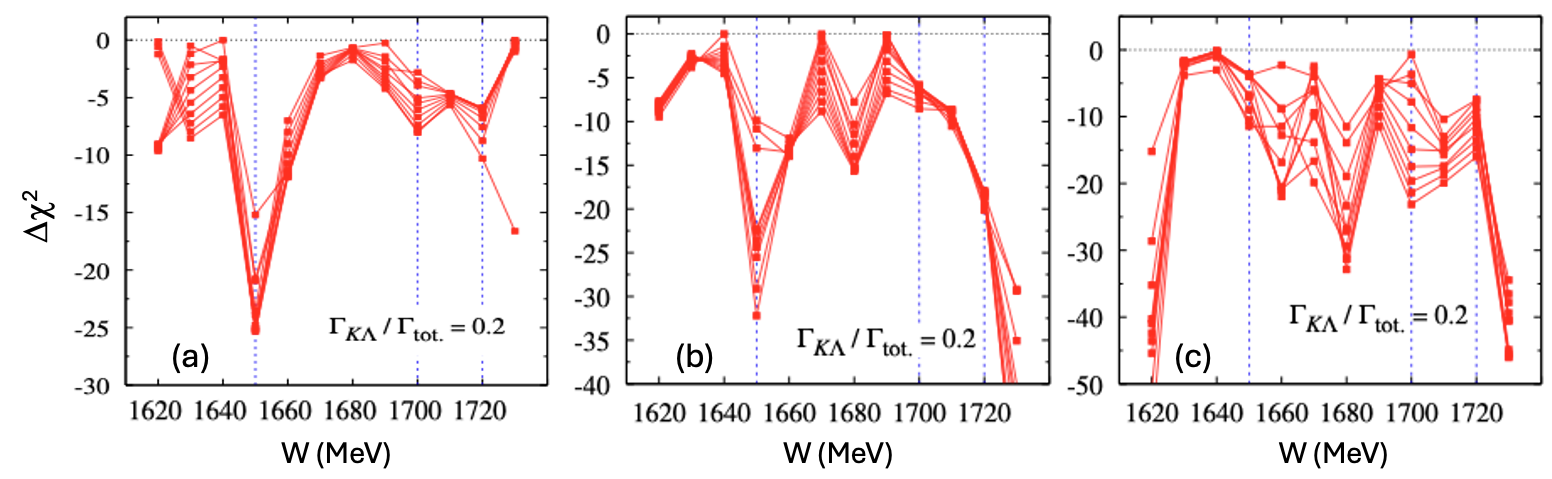}
\caption{Change of overall $\chi^2$ as a function of the resonance mass $W$ after the insertion of the (a) $P_{11}$, (b) $S_{11}$, (c) $P_{13}$ resonances in Model 1 for different values of total width $\Gamma_{\rm tot}$ (from 1 to 10~MeV with 1~MeV step). Figures from Ref.~\cite{Mart:2011ey}.}
\label{fig:narrow_kaon} 
\end{figure}
%%%%%%%%%%%%%%%%%%%%

The analysis method followed that used for the $\pi N$ channel. By scanning the variation of the total $\chi^2$ as a function of the resonance mass $W$ from 1620 to 1730~MeV in 10~MeV steps, using a partial width ratio of $\Gamma_{K\Lambda}/\Gamma_{\rm tot}=0.2$, and varying the total resonance width $\Gamma_{\rm tot}$ from 1 to 10~MeV in 1~MeV steps, the results shown in Fig.~\ref{fig:narrow_kaon}(a) were obtained. As shown, three minima appear when a $P_{11}$ state is introduced into the model, located at $W=1650$, 1700, and 1720~MeV, with the most pronounced minimum at $W=1650$~MeV. This minimum reappears when an $S_{11}$ resonance is inserted instead of the $P_{11}$, whereas the other two minima are much less distinct in this case, as shown in Fig.~\ref{fig:narrow_kaon}(b). When a narrow $P_{13}$ resonance is used, the corresponding result (see Fig.~\ref{fig:narrow_kaon}(c)) exhibits no clear minimum, unlike the $P_{11}$ case. These features remain essentially unchanged when the $K\Lambda$ branching ratio is varied between 0.1 and 0.4. A similar pattern is also obtained when the total resonance width $\Gamma_{\rm tot}$ is varied from 0.1 to 1~MeV in 0.1~MeV steps.

From the $K\Lambda$ photoproduction analysis, two possible candidates for a narrow resonance emerge: $N(1650)1/2^+$ and $N(1650)1/2^-$. However, the latter is disfavored, since introducing a narrow $N(1650)1/2^-$ state produces a pronounced dip in the total cross section at $W=1650$ MeV, which is not observed experimentally (see Fig.~14 of Ref.~\cite{Mart:2011ey}). Therefore, the $N(1650)1/2^+$ remains the most credible narrow state indicated by the kaon photoproduction data. The origin of this feature was examined in detail in Ref.~\cite{Mart:2011ey}, where it was found that the $\chi^2$ minimum arises from a pronounced structure in the recoil polarization $P$ data, consistently appears across all kaon angle bins around $W=1650$~MeV (see Fig.~16 of Ref.~\cite{Mart:2011ey}).

\subsection{Contribution of Kaon Photoproduction to the GDH Sum Rule}

The Gerasimov-Drell-Hearn (GDH) sum rule~\cite{Gerasimov:1965et,Drell:1966jv} establishes a fundamental connection between the ground-state properties of the nucleon, its anomalous magnetic moment and mass, and its excited-state properties, characterized by the helicity-dependent cross sections for the photoabsorption process $\gamma  p \to X$, where $X$ denotes all possible hadronic final states (such as $\pi N$, $\eta N$, $K\Lambda$, $\pi\pi N$, etc). The GDH sum rule is of great significance for several reasons. In addition to providing a direct link between the static properties of the nucleon and its dynamical response to electromagnetic excitation, its experimental verification offers a stringent test of fundamental principles in the non-perturbative regime of QCD, since its derivation relies solely on very general tenets of quantum field theory, i.e., Lorentz and gauge invariance, analyticity, and unitarity. Furthermore, since the GDH integral can be decomposed into contributions from various reaction channels, it offers a valuable framework for testing phenomenological and theoretical models that describe the underlying photoabsorption processes. Finally, the GDH sum rule for real photons provides the foundation for its generalization to virtual photons, thereby linking it to the spin structure functions $g_1(x,Q^2)$ and $g_2(x,Q^2)$, where $x$ denotes the Bjorken variable and $Q^2$ represents the photon virtuality. Thus, the GDH sum rule plays a crucial role in advancing our understanding of the nucleon spin structure.

For the proton, the sum rule reads
\begin{equation}
    \label{eq:GDH1}
    \frac{2\pi^2\alpha}{m_p^2}\,\kappa_p^2 = \int_0^\infty \frac{dk}{k}\Bigl[ \sigma_{3/2}(k)- \sigma_{1/2}(k)\Bigr] \equiv I_{\rm GDH}\,,
\end{equation}
where $\alpha = e^2/4\pi = 1/137$, and $m_p$ and $\kappa_p$ denote the proton mass and anomalous magnetic moment, respectively. The quantities $\sigma_{3/2}(k)$ and $\sigma_{1/2}(k)$ represent the total photoabsorption cross sections corresponding to total spin projections of $3/2$ and $1/2$, respectively, with $k$ being the photon energy. As shown in Ref.~\cite{Drechsel:2002ar}, the GDH sum rule can be derived by comparing the dispersion relation for the antisymmetric (odd) part of the forward Compton scattering amplitude,
\begin{equation}
    \label{eq:GDH3}
    {\rm Re}\, g(k) = \sum_n \left( \frac{1}{4\pi^2} \int \frac{\sigma_{1/2}(k') - \sigma_{3/2}(k')}{(k')^{2n-1}} \, dk' \right) k^{2n-1},
\end{equation}
which represents the spin-dependent component, with the corresponding low-energy expansion of the spin-flip amplitude,
\begin{equation}
    \label{eq:GDH4}
    g(k) = -\frac{e^2 \kappa_p^2}{8\pi m_p^2}\, k + \gamma_0 \, k^3 + {\cal O}(k^5).
\end{equation}
Evidently, Eq.~(\ref{eq:GDH1}) follows directly from matching the leading terms of these two expressions.

In principle, the GDH integral $I_{\rm GDH}$ can be directly evaluated from Eq.~(\ref{eq:GDH1}). In practice, however, the integration is restricted to the energy range above the reaction threshold and up to a finite photon energy, since the cross section vanishes below threshold and experimental data at asymptotically high energies are unavailable. The practical form of the GDH sum rule is therefore written as
\begin{equation}
    \label{eq:GDH2}
    I_{\rm GDH} \equiv \int_{k_{\rm thr}}^{k_{\rm max}} \frac{dk}{k}\Bigl[ \sigma_{3/2}(k)- \sigma_{1/2}(k)\Bigr],
\end{equation}
where $k_{\rm max}$ denotes the maximum photon energy. The integrand in Eq.~(\ref{eq:GDH2}) is expected to be convergent, as the cross sections tend to become approximately constant or decrease at high energies. Using the current PDG values for the proton mass and anomalous magnetic moment \cite{ParticleDataGroup:2024cfk}, the GDH integral yields $I_{\rm GDH}={2\pi^2\alpha\,\kappa_p^2}/{m_p^2}=204.78$ $\mu$b. 

It is interesting to note that the combinations of the total cross sections $\sigma_{3/2}$ and $\sigma_{1/2}$, appearing in Eqs.~(\ref{eq:GDH1}) and (\ref{eq:GDH2}), can be expressed as
\begin{equation}
\label{eq:sigma_T_TTP_GDH}
\sigma_{\rm T} = \frac{1}{2}\bigl( \sigma_{3/2} + \sigma_{1/2} \bigr)  ~~~~~~ {\rm and} ~~~~~~
%\end{equation}
%and
%\begin{equation}
%\label{eq:sigma_TTP_GDH}
\sigma_{\rm TT'} = \frac{1}{2}\bigl( \sigma_{3/2} - \sigma_{1/2} \bigr) ,
\end{equation}
which can be related to the response functions $R_{\rm T}^{00}$ and $R_{\rm TT'}^{0z}$, given in Eqs.~(\ref{eq:RT00}) and (\ref{eq:RTTP0z}) of Appendix~\ref{app:response_function}, respectively. These response functions, in turn, are connected to the differential cross section $d\sigma/d\Omega$ and the polarization observable $E$, as discussed in Section~\ref{dcs-formalism}. 

With the knowledge of the helicity-dependent cross section $\sigma_{\rm TT'}$, the forward spin polarizability $\gamma_0$ can also be evaluated. Its expression can be obtained by comparing Eq.~(\ref{eq:GDH3}) and the second term of Eq.~(\ref{eq:GDH4}), yielding
\begin{equation}
    \gamma_0 = -\frac{1}{4\pi^2} \int_0^\infty \frac{dk}{k^3} \bigl[ \sigma_{3/2}(k) - \sigma_{1/2}(k) \bigr].
\end{equation}

It should be noted that the observable $E$ corresponds to measurements with a circularly polarized photon beam and a longitudinally polarized target, i.e., a target polarized along the beam direction. As a consequence, the experimental determination of the GDH sum rule requires a dedicated setup, which has only become feasible in recent decades~\cite{GDH:2001zzk,GDH:2003xhc,Dutz:2004zz}. 

\begin{figure}[htb]
\centering
\includegraphics[width=0.8\textwidth]{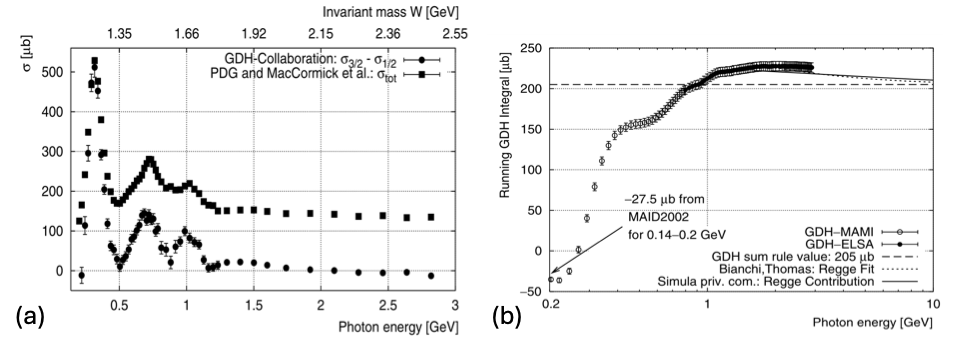}
\caption{(a) Total photoabsorption cross section (solid squares) and helicity-dependent cross-section difference, $\Delta\sigma = \sigma_{3/2} - \sigma_{1/2}$ (solid circles), for the proton as a function of the incident photon energy, as measured by the GDH Collaboration at MAMI and ELSA. (b) The corresponding running GDH integral, $I_{\rm GDH}$, is shown as a function of photon energy in the range from 200 to 2900~MeV. Figures from Ref.~\cite{Dutz:2004zz}.}
\label{fig:GDH-Dutz} 
\end{figure}

The first direct measurement was performed at MAMI by determining the total photoabsorption cross section on the proton for photon energies between 200 and 800~MeV~\cite{GDH:2001zzk}. Within this range, the GDH integral was found to be 
$226 \pm 5~{\text{(stat)}} \pm 12~{\text{(sys)}}~\mu\text{b}$. Two years later, a complementary measurement at ELSA extended the photon energy range to 680-1820~MeV, yielding $49.9 \pm 2.4~{\text{(stat)}} \pm 2.2~{\text{(sys)}}~\mu\text{b}$~\cite{GDH:2003xhc}. This result indicated that the dominant contribution to the GDH integral arises from photon energies below 680~MeV, consistent with the earlier MAMI findings. 
Finally, a combined measurement covering the full photon energy range from 200 to 2900~MeV was carried out at both MAMI and ELSA. The results of this measurement, together with theoretical analyses, are presented in Fig.~\ref{fig:GDH-Dutz}. By combining the data from both facilities, the GDH integral was determined to be $I_{\rm GDH} = 254 \pm 5~{\text{(stat)}} \pm 12~{\text{(sys)}}~\mu\text{b}$, providing a clear experimental confirmation of the GDH sum rule~\cite{Dutz:2004zz}.

%%%%%%%FIGURE 1%%%%%
%%%%%%%%%%%%%%%%%%%%

Prior to the dedicated measurements of the GDH sum rule, the GDH integral had been estimated from the sum of the total cross sections, $\sigma_{\rm TT'}$, calculated for the dominant final states $\pi N$, $\eta N$, $K\Lambda$, and $K\Sigma$ using phenomenological models. The presence of the photon energy $k$ in the denominator of Eq.~(\ref{eq:GDH2}) emphasizes the importance of the low-energy region in determining the GDH integral, as clearly demonstrated by the experimental results of the GDH Collaboration discussed above and a recent phenomenological calculation \cite{Strakovsky:2022tvu}. Consequently, earlier theoretical studies focused primarily on the single-pion photoproduction process~\cite{Karliner:1973em,Workman:1991xs,Burkert:1992yk,Sandorfi:1994ku,Drechsel:1994az,Drechsel:1998fka}, while the contributions from other channels were estimated through various assumptions and modeling approaches. As shown in Ref.~\cite{Drechsel:1998fka}, most of these theoretical predictions, unfortunately, significantly overestimated the standard GDH integral value of 204.78~$\mu$b.

\begin{figure}[htb]
    \centerline{
    \includegraphics[width=\textwidth]{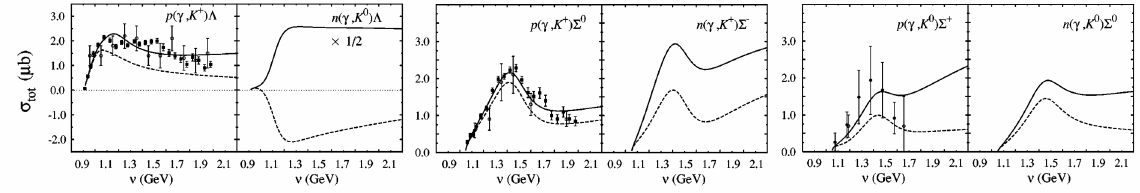}}
    
    \caption{Total cross sections $\sigma_{\rm tot}$ (solid lines) and $-\sigma_{\rm TT'}$ (dashed lines) as a function of the photon laboratory energy $\nu$ for the six isospin channels of kaon photoproduction. The $K^0\Lambda$ total cross sections are multiplied by 1/2 to fit within the same scale. Figure from Ref.~\cite{Sumowidagdo:1999yp}.}
    \label{fig:GDH-Sumowidagdo} 
    \end{figure}

However, the contribution from kaon channels was not evaluated until Ref.~\cite{Hammer:1997kw}, although that work was conducted for a slightly different purpose. The calculation was later refined in Ref.~\cite{Sumowidagdo:1999yp}, which reported the kaon-channel contributions to the GDH integral on the proton as $-1.657$, $-1.529$, and $-0.828~\mu$b for the $K^+\Lambda$, $K^+\Sigma^0$, and $K^0\Sigma^+$ channels, respectively. Consequently, if these contributions are summed coherently, the total kaon contribution to the GDH integral for the proton amounts to $-4.014~\mu$b. Furthermore, for the neutron, the kaon channels $K^0\Lambda$, $K^0\Sigma^0$, and $K^+\Sigma^-$ yield a total contribution to the GDH integral of $1.969~\mu$b. In contrast to the neutron case, the corresponding kaon-channel contributions on the proton side are negative, as the calculated total cross sections $\sigma_{\rm TT'}$ for all three proton channels are negative~\cite{Sumowidagdo:1999yp}, as shown in Fig.~\ref{fig:GDH-Sumowidagdo}. By including these contributions, the GDH integral calculated using the MAID model amounts to $202\pm 10~\mu$b~\cite{Drechsel:2000ct}, which is very close to the standard value of $204.78~\mu$b. 

%%%%%%%FIGURE 1%%%%%
%%%%%%%%%%%%%%%%%%%%

In spite of the good agreement with the standard GDH integral, contributions from vector-meson photoproduction were not included in Ref.~\cite{Drechsel:2000ct}. These were later incorporated by employing a quark model with an effective Lagrangian~\cite{Zhao:2002wf}. In our convention, the contributions of the vector channels $\omega p$, $\rho^0 p$, and $\rho^+ n$ to the proton GDH integral obtained from this model are $2.01$, $-0.05$, and $-2.22~\mu$b, respectively, resulting in a total of $-0.26~\mu$b, considerably smaller than the corresponding contributions from the kaon channels. Furthermore, the total GDH integral for the proton reported in Ref.~\cite{Zhao:2002wf} amounts to $196.72~\mu$b. On the neutron side, the total contribution from vector mesons is $1.05~\mu$b, yielding a total GDH integral of $168.03~\mu$b, which is substantially smaller than the standard value of $232.5~\mu$b.

For the contribution from the kaon channels, further refinement was carried out using the Kaon-MAID model~\cite{kaonmaid}. The resulting analysis, reported in Ref.~\cite{Mart:2008zz}, yielded a smaller contribution to the GDH integral, $I_{\rm GDH}$, of $-3.45~\mu$b for the proton. In addition, the corresponding contribution to the forward spin polarizability $\gamma_0$ was estimated to be $0.191 \times 10^{-6}~\text{fm}^4$. For comparison, the most recent evaluations of the total forward spin polarizability give $\gamma_0 = -92.9~(10.5) \times 10^{-6}~\text{fm}^4$~\cite{Gryniuk:2016gnm} and $\gamma_0 = (-90 \pm 8~{\text{(stat)}} \pm 11~{\text{(sys)}}) \times 10^{-6}~\text{fm}^4$~\cite{Pasquini:2010zr}.

It is also noteworthy that the GDH sum rule has been investigated in relation to the issue of data consistency in the $K^+\Lambda$ channel~\cite{Mart:2008ik}. It is well known that for $W \gtrsim 1500$~MeV the CLAS cross section is systematically higher than that from SAPHIR. Separate fits to each dataset lead to different predictions for the helicity-dependent cross section $\sigma_{\rm TT'}$ and, consequently, to different estimates of the $K^+\Lambda$ contribution to the GDH integral, $I_{\rm GDH}$. Determining the correct contribution in this case would therefore help to resolve this longstanding discrepancy. The investigation showed that, within the Kaon-MAID model, the contribution of $K^+\Lambda$ channel was found to be $-1.247~\mu$b, while separate fits to the CLAS and SAPHIR data yielded $-1.309~\mu$b and $0.845~\mu$b, respectively~\cite{Mart:2008ik}. The latter result is particularly interesting, as it exhibits an opposite sign when fit to the SAPHIR data.

The most recent proton study was carried out within a coupled-channels framework~\cite{Schneider:2025ndq}. The analysis yields contributions of $0.80\pm0.05$, $1.42\pm0.05$, and $-0.12\pm0.05~\mu$b for the $K^+\Lambda$, $K^+\Sigma^0$, and $K^0\Sigma^+$ channels, respectively. Notably, the first two channels exhibit signs opposite to those of previous estimates.

\subsection{Electromagnetic Form Factor of the Kaon}
\label{subsec:Kaon_form_factor}

\begin{figure}[htb]
\centerline{
\includegraphics[width=0.5\textwidth]{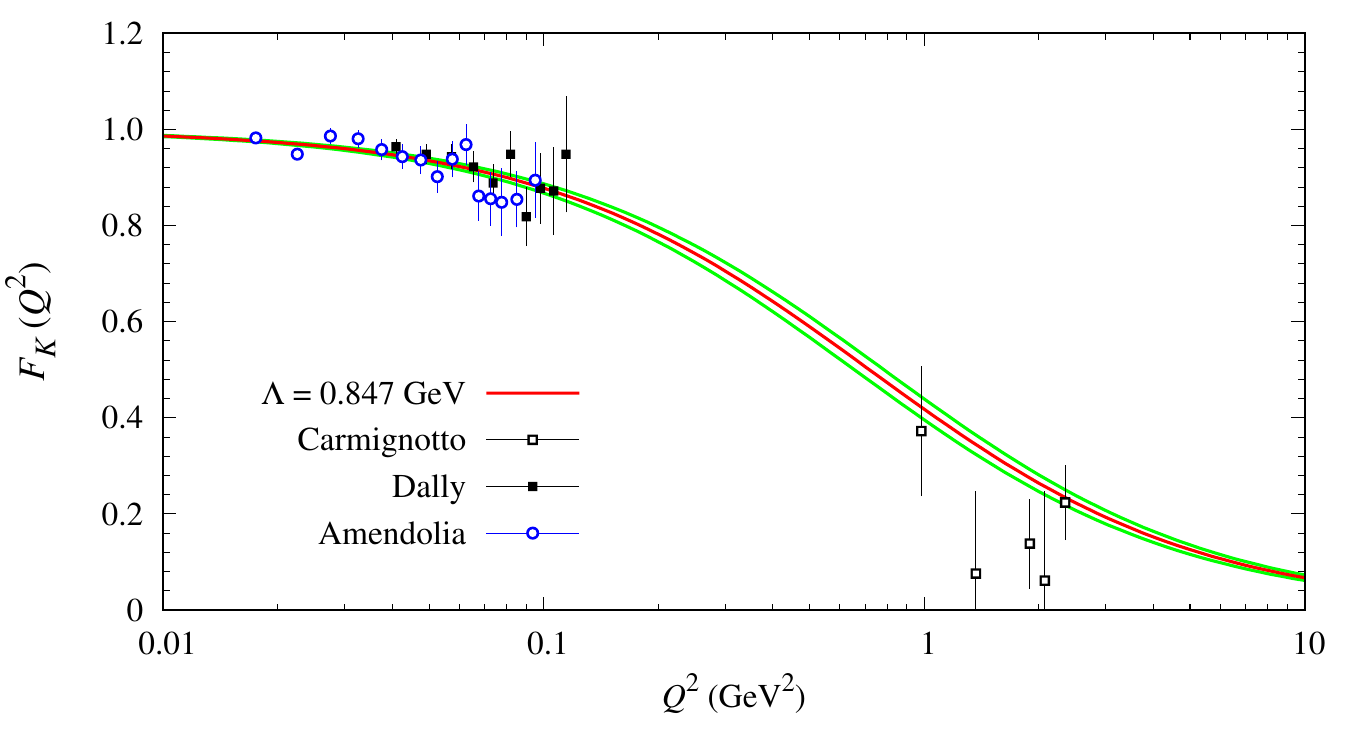}}
\caption{Electromagnetic form factor of the charged kaon as a function of $Q^2$. The same data as shown in Fig.~\ref{kaonff-hallc}, previously presented in the form of $Q^2F_K(Q^2)$, are plotted here as $F_K(Q^2)$ on a logarithmic $Q^2$ scale. Experimental data are taken from Refs.~\cite{Amendolia:1986ui} (open circles), \cite{Dally:1980dj} (solid squares), and \cite{Carmignotto:2018uqj} (open squares). The solid red curve represents a monopole fit of the form factor, $F_K(Q^2) = (1 + Q^2/\Lambda^2)^{-1}$, to the data, while the solid green curves denote the corresponding fit uncertainties. The extracted cutoff parameter is $\Lambda = 0.847 \pm 0.038$~GeV, yielding an r.m.s. radius of $0.570 \pm 0.026$~fm.}
\label{fig:kaon_em_FF} 
\end{figure}

The pion electromagnetic form factor was successfully measured nearly four decades ago in the range of $0.014 < Q^2 < 0.260$~GeV$^2$ through high-energy pion scattering on a hydrogen target. Analogous to the case of elastic electron-proton scattering, the differential cross section can be expressed as the product of the corresponding form factor and the point-like particle cross section~\cite{NA7:1986vav},
\begin{eqnarray}
\label{eq:CS_pion_FF}
\frac{d\sigma}{dQ^2} = |F_\pi(Q^2)|^2\,\frac{k}{Q^4}\,\left(1 - \frac{Q^2}{Q_{\rm max}^2}\right),
\end{eqnarray}
where $Q^2_{\rm max}$ corresponds to the value of $Q^2$ at backward scattering in the center-of-mass frame. Applying a similar technique, the electromagnetic form factor of the negative kaon was also measured in the range of $0.015 < Q^2 < 0.100$~GeV$^2$~\cite{Amendolia:1986ui}. The result is shown in Fig.~\ref{fig:kaon_em_FF} along with those obtained from the later measurements with different techniques.

%%%%%%%FIGURE 1%%%%%
%%%%%%%%%%%%%%%%%%%%

\begin{figure}[htb]
\centerline{
\includegraphics[width=0.7\textwidth]{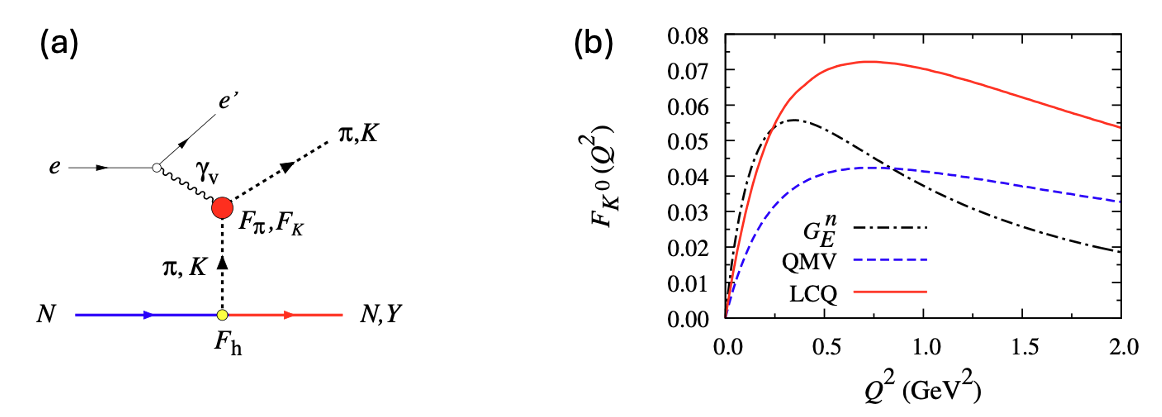}}
\caption{(a) Contribution from the $t$-channel diagram in pion (kaon) electroproduction. The amplitude is proportional to the propagator $1/(t - m^2)$, where $m$ is the pion (kaon) mass, and is enhanced when $|t - m^2|$ is minimized. The electromagnetic form factors $F_\pi(Q^2)$ and $F_K(Q^2)$ describe the meson charge distribution as a function of $Q^2$, while the hadronic form factor $F_{\rm h}(t)$ accounts for the extended structure of the hadronic vertex. (b) The neutral kaon electromagnetic form factor obtained from the quark-meson vertex (QMV) and light-cone quark (LCQ) models are shown in comparison with the neutron electric form factor $G_E^n(Q^2)$. Figures adapted from Ref.~\cite{Mart:2011ez}.}
\label{fig:t-channel_with_kaonFF} 
\end{figure}

The charged pion electromagnetic form factor can also be extracted from pion electroproduction, since the corresponding $t$-channel diagram (see Fig.~\ref{fig:t-channel_with_kaonFF}(a)) explicitly contains the electromagnetic form factor. A model-independent extraction, however, is practically impossible because contributions from other channels—particularly the resonance terms—cannot be neglected and may even be substantial. Nevertheless, by choosing kinematic conditions corresponding to small values of $|t - m_\pi^2|$, the $t$-channel contribution can be significantly enhanced. This strategy was employed, for example, in Ref.~\cite{Mart:2008sw}.

%%%%%%%FIGURE 1%%%%%
%%%%%%%%%%%%%%%%%%%%

In the kinematic region where the $t$-channel dominates the reaction amplitude, the longitudinal cross section of pion electroproduction can be expressed as~\cite{Mart:2008sw}
\begin{equation}
    (t-m_\pi^2)^2\frac{d\sigma_{\rm L}}{dt} = -t\,F_\pi^2(Q^2) F_{\rm h}^2(t)\,\frac{e^2g_{\pi NN}^2Q^2}{8\pi W(s-m^2)|\vec{k}|^3}\,\left(q_0-{\textstyle \frac{1}{2}}k_0\right)^2,
\label{eq:dsL/dt*(t-m2)}
\end{equation}
where $F_{\rm h}(t)$ denotes the hadronic form factor that accounts for the extended structure of the $\pi NN$ vertex. For each fixed value of $Q^2$, the right-hand side of Eq.~(\ref{eq:dsL/dt*(t-m2)}) can be evaluated, and the equation can be fit to the data in the form $y = F_\pi^2 ~ f(t) + g$, allowing the extraction of the pion form factor $F_\pi$. The extraction was performed using differential cross section data in the range $t = -0.025$ to $-0.3$~GeV$^2$. As shown in Ref.~\cite{Mart:2008sw}, the resulting pion form factor is in good agreement with those obtained from analyses based on Regge models~\cite{JeffersonLabFpi-2:2006ysh,JeffersonLabFpi:2007vir}.

In the case of the kaon electromagnetic form factor, the situation is somewhat more challenging because the experimental values of $|t - m_K^2|$ are considerably larger, owing to the higher kaon mass. Consequently, the extraction of the form factor can only be performed through phenomenological approaches, such as Regge and isobar models, as demonstrated in the analyses of the E01-004 and E93-018 experiments conducted in Hall~C at JLab with a 6 GeV beam~\cite{Carmignotto:2018uqj}, shown as the open squares in Fig.~\ref{fig:kaon_em_FF} (see also Section~\ref{halla-c-6gev}). At this stage, it is worth noting that the situation becomes even more complicated for form factors in the nuclear medium. As is well known, medium effects can have a significant impact on the kaon electromagnetic form factor~\cite{Gifari:2024ssz,Hutauruk:2025bjd}.

The extraction of the neutral kaon electromagnetic form factor is considerably more challenging because the $K^0$, having no net electric charge, cannot couple directly to the virtual photon in the one-photon-exchange process, as occurs in $K^- e^-$ scattering~\cite{Amendolia:1986ui}. Nevertheless, a virtual photon with non-zero four-momentum transfer ($Q^2 > 0$) can probe the internal charge distribution of the $K^0$ through its longitudinal polarization component. At finite $Q^2$, the spatial charge distributions of the constituent $d$ and $\bar{s}$ quarks do not perfectly cancel, resulting in a small but finite electromagnetic form factor. In the electroproduction reaction $e  n \to e'  K^0  \Lambda$, this form factor contributes to the longitudinal part of the virtual-photon cross section, $\sigma_{\rm L}$, via the $t$-channel $K^0$-exchange amplitude, $\mathcal{M}_t \propto {F_{K^0}(Q^2)}/{(t - m_{K^0}^2)}$ (see Fig.~\ref{fig:t-channel_with_kaonFF}(a)). Although this contribution is expected to be much smaller than in the charged-kaon case, precise measurements of the longitudinal response could, in principle, provide valuable insight into the electromagnetic structure of the neutral kaon.

An example of this effort was presented in Ref.~\cite{Mart:2011ez}, where two selected parameterizations of the $K^0$ electromagnetic form factor, derived from the quark-meson vertex (QMV)~\cite{Buck:1994hf,Ito:1993av} and light-cone quark (LCQ)~\cite{Bennhold:1995bp} models, were investigated within an isobar model of $K^0$ electroproduction. The corresponding form factor behaviors as a function of $Q^2$ are shown in Fig.~\ref{fig:t-channel_with_kaonFF}(b), together with the neutron electric form factor $G_E^n(Q^2)$ for comparison. It is evident that the $K^0$ form factor is significantly smaller than that of the charged kaon, as discussed previously.

\begin{figure}[htb]
\centerline{
\includegraphics[width=0.85\textwidth]{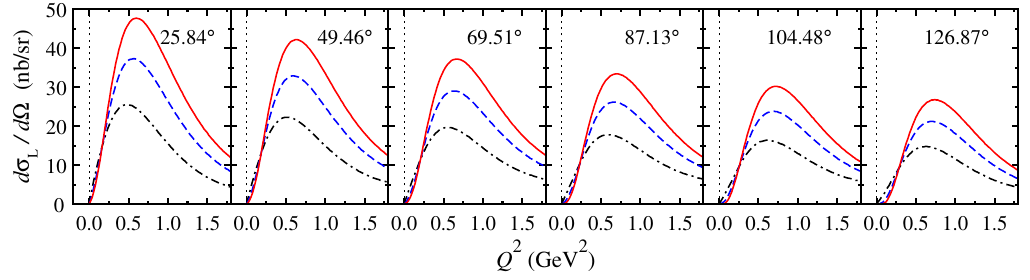}}
\caption{Longitudinal differential cross sections for neutral kaon electroproduction, $e  n \to e'  K^0  \Lambda$ as a function of $Q^2$ at $W = 1.65~\text{GeV}$, for various kaon scattering angles $\theta_K^{\mathrm{c.m.}}$ indicated within the panels. The solid and dashed curves represent calculations using the $K^0$ form factor obtained from the LCQ and QMV models, respectively, while the dash-dotted curves correspond to the calculation without the $K^0$ intermediate state, i.e., excluding the diagram in Fig.~\ref{fig:t-channel_with_kaonFF}(a). Figure from Ref.~\cite{Mart:2011ez}.}
\label{fig:sigma_L_K0FF} 
\end{figure}

Since no experimental data were available for the $en \to e'K^0\Lambda$ reaction, the model was developed as an isospin extension of the existing isobar model for the $ep \to e'K^+\Lambda$ process \cite{Mart:2011ez,Mart:2017xtf,Mart:2019add}. To minimize uncertainties arising from possible resonance contributions, the model was constrained to describe data only near the production threshold. The sensitivity of the separated differential cross sections $\sigma_{\rm T}$, $\sigma_{\rm L}$, $\sigma_{\rm TT}$, and $\sigma_{\rm LT}$ to the $K^0$ form factor was then examined for various form factor models and kaon scattering angles. It was found that the longitudinal cross section $\sigma_{\rm L}$ is the most sensitive observable to this effect, where the inclusion of the $K^0$ form factor could enhance the longitudinal cross section by up to $50\%$~\cite{Mart:2011ez}, as shown in Fig.~\ref{fig:sigma_L_K0FF}.

%%%%%%%FIGURE 1%%%%%
%%%%%%%%%%%%%%%%%%%%

The result shown in Fig.~\ref{fig:sigma_L_K0FF} clearly demonstrates that the extraction of the neutral kaon electromagnetic form factor is, in principle, feasible, provided that sufficiently precise measurements of the longitudinal cross section can be carried out. On average, Fig.~\ref{fig:sigma_L_K0FF} suggests that an accuracy of about 10~nb is required to reveal the effect of the form factor within the model. It is well known, however, that form factor extractions from indirect processes, such as electromagnetic meson production, are inherently model dependent. This issue can be mitigated by employing a relatively simple model, for instance one constrained to work only near threshold, as described above. Alternatively, the use of a Regge model offers another viable approach, as it naturally reduces model complexity through the duality principle, provided the analysis is restricted to sufficiently high energies.

\subsection{Weak Production of Strange Particles}
\label{subsec:weak_production}

The neutral-kaon system, consisting of $K^0=d\bar{s}$ and $\bar{K}^0=\bar{d}s$, exhibits weak-interaction-induced mixing between the two flavor eigenstates, giving rise to the physical states $K_S$ and $K_L$ and providing one of the earliest examples of meson oscillations \cite{Lee:1966hp}. The discovery of charge-conjugation and parity (CP) violation in neutral-kaon decays in 1964 \cite{Christenson:1964fg} revealed that the weak interaction does not exactly conserve the combined CP symmetry, a finding that ultimately led to the development of the Cabibbo-Kobayashi-Maskawa (CKM) mechanism for quark flavor mixing and CP violation \cite{Cabibbo:1963yz,Kobayashi:1973fv}. Within this framework, the Cabibbo angle and the CKM matrix element $V_{us}$ govern strangeness-changing weak transitions and account for the observed suppression of processes involving strange quarks \cite{Cabibbo:1963yz,Kobayashi:1973fv}. 

Meanwhile, the electromagnetic processes that constitute the main focus of the present review are mediated exclusively by vector currents and conserve parity at the production vertex. Complementary information can be obtained from weak kaon production induced by neutrino beams, $\nu N\rightarrow \mu K Y$, where $Y$ denotes a $\Lambda$ or $\Sigma$ hyperon. In contrast to electromagnetic production, these reactions are mediated by the charged weak current and therefore provide direct access to both the vector and axial-vector components of the hadronic current through the underlying $V\!-\!A$ interaction. 

%%%%%%%FIGURE 1%%%%%
\begin{figure}[htb]
\centerline{
\includegraphics[width=0.95\textwidth]{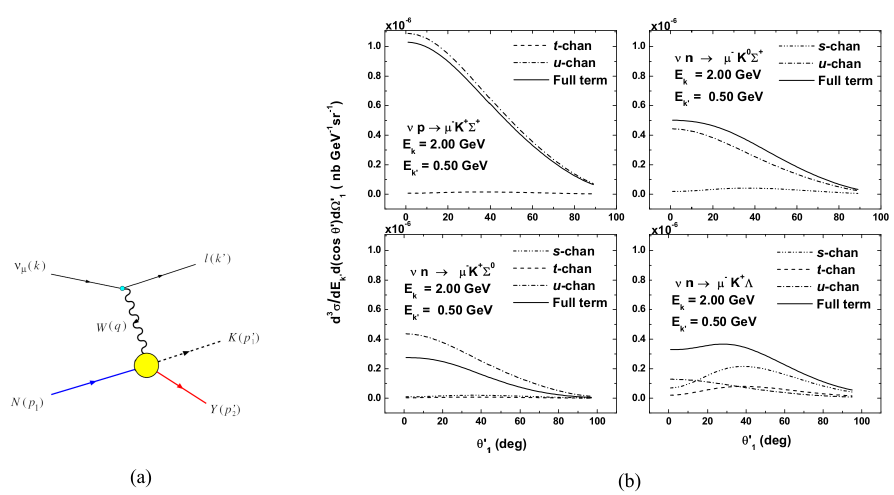}}
\caption{(a) Tree-level Feynman diagram for weak associated production of strange particles induced by neutrinos. (b) Contributions of the $s$-, $u$-, and $t$-channel amplitudes to the differential cross sections for $\nu p\to \mu^-K^+\Sigma^+$, $\nu n\to \mu^-K^0\Sigma^+$, $\nu n\to \mu^-K^+\Sigma^0$, and $\nu n\to \mu^-K^+\Lambda$ as a function of the outgoing kaon angle. Results are shown for an incoming neutrino energy of 2~GeV and an outgoing muon energy of 0.5~GeV. The $u$-channel provides the dominant contribution in all cases except in the $\nu n\to \mu^-K^+\Lambda$ process. Figures adapted from Ref.~\cite{Adera:2010zz}.}
\label{fig:weak_production} 
\end{figure}
%%%%%%%%%%%%%%%%%%%%

A notable study in this direction was carried out by Adera \emph{et al.} \cite{Adera:2010zz}, who developed a formalism for charged-current weak production of strange particles (shown in Fig.~\ref{fig:weak_production}(a)) in the threshold region based on the Born approximation, Cabibbo theory, and SU(3) flavor symmetry. Their model employed a general invariant-amplitude decomposition of the weak hadronic current and was used to investigate both polarized and unpolarized strangeness-production channels. To constrain the hadronic sector, the strong coupling constants and other hadronic parameters were adopted from the kaon-production model of Lee \emph{et al.} \cite{Lee:1999kd}, thereby establishing a direct connection between electromagnetic and weak strangeness-production studies. In particular, the model incorporates the isospin relations among all reaction channels discussed in Section~\ref{sec:isospin}. Owing to the intrinsic $V\!-\!A$ structure of the weak interaction, parity violation enters directly into the production mechanism, leading to the production of polarized hyperons whose polarization can subsequently be extracted through parity-violating decays such as $\Lambda\rightarrow p\pi^-$. 

Representative results from Ref.~\cite{Adera:2010zz} are included in Fig.~\ref{fig:weak_production}(b), which shows the differential cross sections for the four possible reaction channels as a function of the kaon scattering angle. The figure reveals that the cross sections are largely dominated by the $u$-channel contribution in the first three reactions, giving rise to their pronounced forward-peaking behavior. In contrast, the $\nu n\to \mu^-K^+\Lambda$ channel is dominated by the $s$-channel contribution, resulting in a shift of the cross-section maximum away from the forward direction. 

Consequently, weak kaon production provides access not only to the hadronic dynamics familiar from electromagnetic kaon production, but also to axial-vector currents, parity-violating observables, and strangeness-changing weak interactions governed by the CKM matrix element $V_{us}$, all of which are inaccessible in purely electromagnetic reactions. These processes have attracted increasing attention in modern neutrino programs, with experiments such as MINERvA, T2K, and the future DUNE expected to provide valuable data for testing theoretical descriptions of weak strangeness production. More recently, the theoretical framework has been improved through the implementation of unitarity constraints based on Watson's theorem \cite{RafiAlam:2010kf,RafiAlam:2012bro,Saul-Sala:2021swb}, leading to a more consistent description of the interplay between weak and strong interaction dynamics.
%=================================================

%=== UNSETTLED PROBLEMS ==========================
%  \input{unsettled_problems}
\section{Unsettled Problems}
\label{sec:unsettled}
  
Despite significant efforts over the past decades, kaon photo- and electroproduction remain among the least well-understood meson production processes, both experimentally and theoretically. This limited understanding stems from the increased complexity of the reaction mechanisms, owing to higher production thresholds compared with $\pi$ and $\eta$ production, the relative scarcity of high-precision experimental data, and the pronounced model dependence of theoretical descriptions. As a consequence, several key issues remain unsettled and warrant further investigation in future studies.

One of the most intricate challenges is the strongly model-dependent identification of the baryon resonances contributing to kaon photo- and electroproduction. Nevertheless, across a wide range of theoretical approaches, including isobar models such as Kaon-MAID~\cite{Mart:1999ed,Bennhold:1999mt}, Regge-plus-Resonance frameworks~\cite{Guidal:1997hy,Corthals:2005ce,Corthals:2006nz}, $K$-matrix models~\cite{Usov:2005wy}, and dynamical coupled-channel (DCC) analyses developed by the ANL-Osaka~\cite{Kamano:2013iva} and J\"ulich-Bonn groups~\cite{Ronchen:2018ury,Mai:2023cbp}, a consistent picture emerges in which four nucleon resonances play an important role in $K\Lambda$ production, i.e., the $N(1650)1/2^-$, $N(1710)1/2^+$, $N(1720)3/2^+$, and $N(1900)3/2^+$. Their inclusion is essential for reproducing the overall magnitude and energy dependence of the total and differential cross sections, as well as polarization observables. In particular, both DCC and multi-channel analyses consistently attribute the second peak in the $K\Lambda$ total photoproduction cross section to the $N(1900)3/2^+$~\cite{Shklyar:2005xg,Mart:2011ey,Kamano:2013iva}, thereby superseding earlier interpretations based on a “missing” $N(1895)3/2^-$~\cite{Mart:1999ed}.

Beyond these four well-established states, the resonance contribution to kaon photoproduction is less certain. Additional resonances, such as the $N(1895)1/2^-$ or higher-spin states, appear in some analyses but not in others, and their inclusion often correlates with the background terms, hadronic form factors, or interference effects. This ambiguity has been emphasized in global model comparisons and coupled-channel studies~\cite{Anisovich:2011fc,Ronchen:2018ury}, reflecting both limitations of the existing data and the difficulty of separating resonant from non-resonant mechanisms in strangeness production.

In this context, Bayesian inference has emerged as one of the promising tools for addressing the resonance-identification problem. Bayesian methods allow for quantitative model comparison through the calculation of model evidences, providing a statistically well-defined criterion for assessing whether the inclusion of specific resonances is warranted by the data. This approach has been systematically developed and applied to kaon photoproduction within the Regge-plus-Resonance framework by Ryckebusch and collaborators~\cite{DeCruz:2010ry,DeCruz:2011xi,DeCruz:2012bv}. These works demonstrated that Bayesian model selection can decisively discriminate between competing resonance sets, even when traditional $\chi^2$-based fits yield comparable descriptions of the data. Similar Bayesian strategy has been advocated in pentaquark photoproduction by Ireland and collaborators~\cite{CLAS:2007bah}. Together, these studies highlight Bayesian inference as a powerful and still underutilized framework for reducing model ambiguity in kaon production.

Another issue concerns experimental inconsistencies in the photoproduction data, particularly at forward kaon angles. This kinematic region is dominated by $t$-channel exchanges and is therefore crucial for constraining background amplitudes and Reggeized descriptions~\cite{Guidal:1997hy}. Forward-angle data are also indispensable for applications such as hypernuclear production, where maximal cross sections occur almost exclusively at small kaon angles. Despite this importance, measurements from different collaborations, most notably CLAS~\cite{CLAS:2005lui,CLAS:2009rdi}, LEPS~\cite{LEPS:2005hji}, and SAPHIR~\cite{SAPHIR:1998fev,Glander:2003jw}, exhibit significant discrepancies in both normalization and angular dependence. In particular, the long-standing inconsistencies associated with these datasets have been shown to strongly affect the extraction of resonance parameters and background contributions~\cite{Mart:2006dk}, and they continue to limit the reliability of global analyses.

On the theoretical side, the treatment of higher-spin resonances represents another open challenge. Conventional Rarita-Schwinger descriptions of spin-$3/2$ and higher fields suffer from off-shell ambiguities and the propagation of unphysical lower-spin components~\cite{Benmerrouche:1989uc,Vrancx:2011qv}. A consistent formulation was proposed by Pascalutsa, in which gauge-invariant couplings eliminate these unphysical degrees of freedom~\cite{Pascalutsa:1998pw,Pascalutsa:2000kd}. Building on this framework, consistent higher-spin interactions were further developed by Delgado Acosta et al.~\cite{DelgadoAcosta:2015ypa}. Applications of consistent treatment of spin-$3/2$ and $5/2$ interactions in kaon photoproduction have been explored in isobar frameworks~\cite{Kristiano:2017qjq,Mart:2019jtb,Mujirin:2020xwr}, indicating that a proper treatment of higher-spin degrees of freedom can significantly affect the extracted resonance content. However, such approaches have not yet been embedded in comprehensive coupled-channel analyses, and their impact on pole extractions remains an open question.

Another topic relevant to the electromagnetic production of kaons in the future is the role of lattice QCD in explaining the spectrum, structure, and interactions of hadrons in such processes. Lattice QCD is a numerical approach within QCD that discretizes space-time into a lattice, allowing quark and gluon interactions to be computed non-perturbatively. In this way, hadronic properties can be calculated directly from first principles by employing Monte Carlo simulations of correlation functions. An example is the kaon electromagnetic form factor discussed in Section~\ref{subsec:Studies-in-Hall-C} and shown in Fig.~\ref{kaonff-proj12}. In addition, the scalar, vector, and tensor form factors of the kaon have been computed within the lattice QCD framework~\cite{Alexandrou:2021ztx}, while the corresponding pion sector has been analyzed in Ref.~\cite{Chandra:2025pqs}. The connection between lattice QCD and dynamical coupled-channel (DCC) models of hadron dynamics, as well as its broader relevance to meson electromagnetic production, has been explored in, e.g., Refs.~\cite{Doring:2025sgb,Lang:2016hnn,Mai:2022eur}.

A further limitation arises from the incomplete isospin decomposition of kaon production amplitudes. Although the $K\Lambda$ final state acts as an isospin filter for $I=1/2$ resonances, thereby excluding contributions from $\Delta$ states, a complete separation of isospin amplitudes requires data on neutron targets, i.e., $\gamma n \to K^0 \Lambda$, $\gamma n \to K^+\Sigma^-$, and $\gamma n \to K^0 \Sigma^0$. Experimentally, such measurements are challenging owing to nuclear binding effects and limited statistical precision, which introduce additional uncertainties in the extracted cross sections~\cite{CLAS:2017gsu,A2:2018doh}. The situation is even more complex in the $K\Sigma$ channel, where both $I=1/2$ and $I=3/2$ amplitudes may contribute, allowing a large number of $N$ and $\Delta$ resonances to enter the model. Analyses based on isobar, $K$-matrix, and DCC frameworks consistently report a much less stable resonance content than in $K\Lambda$ production~\cite{Mart:2014eoa,Usov:2005wy,Shklyar:2005xg,Kamano:2013iva}. While some $\Delta$ resonances appear relatively robust, strong non-resonant mechanisms and interference effects obscure clear resonance signals.

In strangeness physics experiments, kaon electroproduction remains significantly less well explored than photoproduction as detailed in Section~\ref{sec:intro}. Existing data from JLab and earlier experiments (see Section~\ref{expt-measurements}) are still limited in both kinematic coverage and precision, particularly at higher $Q^2$ and $\cos \theta_K^{\rm c.m.} < 0$. However, expansive new electroproduction data from CLAS12 are expected to significantly impact the experimental landscape for $Q^2 > 0$ in the years ahead.

Electroproduction studies are complementary to photoproduction in the search for new “missing” $s$-channel baryon resonances in that the ratio of resonant to non-resonant contributions evolves strongly with $Q^2$. The fit parameters for the $N^*$ states derived from advanced theoretical models must be described by $Q^2$-independent resonance masses and hadronic decay widths. To map out the full $N^*$ spectrum and unravel the distinct structural features of these resonance states, it is necessary to consider the electroproduction data through various exclusive final states, including the non-strange channels $\pi^0 p$, $\pi^+ n$, $\eta p$, and $\pi^+ \pi^- p$, as well as the dominant strangeness channels $K^+\Lambda$ and $K^+\Sigma^0$.

The development of a reaction model that describes the full set of available cross sections and polarization observables still awaits us. Therefore, the full potential of what can be learned from the electroproduction data has yet to be realized and is a central unsettled issue in the field of strangeness physics. However, that advancement should, in time, be realized based on recent developments of both single-channel and coupled-channel approaches~\cite{Djaja:2025hzz,Wang:2024byt}. It should be stressed that a unified treatment of both photo- and electroproduction data within a consistent reaction framework is essential for probing the internal structure of baryon resonances and the role of strangeness in nucleon excitations.%=================================================

%=== CONCLUSION & OUTLOOK =====================
%    \input{conclusion_outlook}
\section{Summary Overview}
\label{sec:conclusion}

This review paper was conceived and designed to provide a comprehensive and balanced overview of the history, current developments, and future plans in the field of the electromagnetic production of strangeness. This has been a challenge to ensure that all important topics were considered from the complementary viewpoints of experiment, phenomenology, and theory. We have labored in an attempt to ensure our coverage was inclusive and our reference list as complete as possible. Looking back over the past 70 years in the history of this field has been a considerable challenge but also a labor of love as we have devoted so much of our scientific careers to this very interesting and still eminently promising area of research. These investigations provide insights into strong interaction dynamics that are relevant regarding baryon spectroscopy of strange systems, structure studies of nucleons and their excited states, investigations of the production mechanism of ground state and excited state hyperons with strangeness $S = -1, -2, -3$, hypernuclear systems that access the $YN$ and $YNN$ interactions, pseudoscalar and vector kaon production, and baryon and meson form factors. Of course, there is much more that could be explored in each of these different areas than the space we have been allowed! For example, we have not been able to cover in great depth studies associated with the electromagnetic production of $K^*Y$ and $KY^*$, even though these final states are currently under active investigation on both the theoretical and experimental fronts and provide essential insights into this field. Nevertheless, all of the topics reviewed here are relevant across all energy and distance scales from the regime of dressed quarks, to neutron stars, to galaxies, to the Universe itself, with far-reaching and fundamental implications to the fields of nuclear and particle physics. It must also be appreciated that to understand the dynamics and to account for the rich structure of strongly interacting systems across the full range of distance scales from the non-perturbative to perturbative regimes of QCD, the data from experiments with electromagnetic probes, as well as hadron- and neutrino-induced processes, must be properly understood and taken into account. This is a colossal task spanning generations of scientists!

An important goal of our efforts was to provide information on this area of research that would be helpful and relevant for undergraduate and graduate students, postdoctoral researchers, university professors, and research scientists interested in this field. It is our sincere hope that our efforts are appreciated and valuable to our colleagues, young and old.

\vspace{2mm}

\centerline{{\it ``There is no excellent beauty that hath not some strangeness in the proportion.''} (Francis Bacon, 1625)}%==============================================
	
%end of the core of the manuscript
	
%\newpage

\section*{CRediT authorship Contribution Statement}

{\bf Terry Mart}: Writing - review and editing, Writing - original draft, Conceptualization. {\bf Jovan Alfian Djaja}: Writing - review and editing, Writing - original draft, Conceptualization. {\bf Daniel S. Carman}: Writing - review and editing, Writing - original draft, Conceptualization.

\section*{Declaration of Competing Interest}

The authors declare that they have no known competing financial interests or personal relationships that could have appeared to influence the work reported in this paper. The authors acknowledge that some portions of this article (Sections 2, 5, 6) employed ChatGPT (OpenAI) to assist with polishing the English language and improving clarity of expression. However, all scientific content and interpretations are the authors' own.

\section*{Acknowledgements}

D.S.C. thanks his many collaborators over the years and, with particular regard, Victor Mokeev, Brian Raue, and Reinhard Schumacher for many seasons of productive work, motivation, and enlightening discussions. T.M. gratefully acknowledges Michael D\"oring and Maxim Mai for the fruitful discussions. He also expresses his deep appreciation for the extensive and valuable knowledge on kaon electromagnetic production acquired through decades of discussions with all of his collaborators, which has made this review possible.
This work was supported in part by the PUTI Q1 Grant from University of Indonesia under contract PKS-206/UN2.RST/HKP.05.00/2025 (T.M. and J.A.D.)  and by the U.S. Department of Energy, Office of Science, Office of Nuclear Physics under contract DE-AC05-06OR23177 (D.S.C.).

%\section*{Author's Contributions \textit{(optional section)}}

%Detailing here the contributions of the authors of the review.
	
	\newpage
    \section*{Appendix}
%=== Appendices =====================
%    \input{appendix}
%%%%%%%%%%%%%%%%%%%%%%%%%%%%%%%%%%%%%%%%%%%%%%%%%%%%%%%%%%%%%%%%%%%
\appendix
\renewcommand*{\thesection}{\Alph{section}}
	
\section{Response Functions}
\label{app:response_function}
	
For completeness, all possible response functions $R^{\nu\mu}_i$ defined in the cross section of Eq.~(\ref{eq:most_general_cs}) are listed in Table~\ref{tab:response_functions}. The definitions follow the convention of Kn\"{o}chlein et al.~\cite{Knochlein:1995qz}. Within this framework, not all response functions are independent, as certain functions are related to others. Consequently, the total number of independent response functions is reduced to 36 \cite{Knochlein:1995qz}.

%%%%%%%%%%%%%%%%%%%%%%%%%%%%%%%%%%%%%%%%%%%%%%%%%%%%%%%%%%%%%%%%%%%%%%%%%%%%%%%%%%%%%%%%%%%%%%%%%%%%%%%%%%%%%%%%%%%%%%%%%%%%%%%%%%
\begin{table}[htbp]
\begin{center}
\caption{Response functions for kaon electroproduction adapted from Kn\"ochlein et al.~\cite{Knochlein:1995qz}. The target (recoil) polarization is related to the coordinate axes denoted by $\mu$ ($\nu$) (see
Fig.~\ref{fig:frames_convention}). The last three columns are for a polarized electron beam. The symbol $\ddag$ indicates a response function that does not vanish but is related to other response functions.}
\begin{tabular} {c c || c c c c c c c c c} \hline\hline
%===ROW 1
$\nu$ & $\mu$ & ${\rm T}$ & ${\rm L}$ & ${\rm ^cTL}$ & ${\rm ^sTL}$ & ${\rm ^cTT}$ & ${\rm ^sTT}$ & ${\rm ^cTL'}$ & ${\rm ^sTL'}$ & ${\rm TT'}$ \\ \hline 
%===ROW 2
0    & 0   & $R_{\rm T}^{00}$ & $R_{\rm L}^{00}$ & $^{\rm c}\! R_{\rm TL}^{00}$ & 0 & $^{\rm c}\! R_{\rm TT}^{00}$ & 0 & 0 & $^{\rm s}\! R_{\rm TL'}^{00}$ & 0  \\ \hline
%===ROW 3
0    & $x$ & 0 & 0 & 0 & $^{\rm s}\! R_{\rm TL}^{0x}$ & 0 & $^{\rm s}\! R_{\rm TT}^{0x}$ & $^{\rm c}\! R_{\rm TL'}^{0x}$ & 0 & $R_{\rm TT'}^{0x} $ \\
%===ROW 4
0    & $y$ & $R_{\rm T}^{0y}$ & $R_{\rm L}^{0y}$ & $^{\rm c}\! R_{\rm TL}^{0y}$ & 0 & $\ddag$ & 0 & 0 & $^{\rm s}\! R_{\rm TL'}^{0y}$ & 0 \\
%===ROW 5
0    & $z$ & 0 & 0 & 0 & $^{\rm s}\! R_{\rm TL}^{0z}$ & 0 & $^{\rm s}\! R_{\rm TT}^{0z}$ & $^{\rm c}\! R_{\rm TL'}^{0z}$ & 0 & $R_{\rm TT'}^{0z} $ \\ \hline
%===ROW 6
$x'$ & 0   & 0 & 0 & 0 & $^{\rm s}\! R_{\rm TL}^{x'0}$ & 0 & $^{\rm s}\! R_{\rm TT}^{x'0}$ & $^{\rm c}\! R_{\rm TL'}^{x'0}$ & 0 & $R_{\rm TT'}^{x'0}$  \\
%===ROW 7
$y'$ & 0   & $R_{\rm T}^{y'0}$ & $\ddag$ & $\ddag$ & 0 & $\ddag$ & 0 & 0 & $\ddag$ & 0 \\
%===ROW 8
$z'$ & 0   & 0 & 0 & 0 & $^{\rm s}\! R_{\rm TL}^{z'0}$ & 0 & $^{\rm s}\! R_{\rm TT}^{z'0}$ & $^{\rm c}\! R_{\rm TL'}^{z'0}$ & 0 & $R_{\rm TT'}^{z'0}$ \\ \hline 
%===ROW 9
$x'$ & $x$ & $R_{\rm T}^{x'x}$ & $R_{\rm L}^{x'x}$ & $^{\rm c}\! R_{\rm TL}^{x'x}$ & 0 & $\ddag$ &  0 & 0 & $^{\rm s}\! R_{\rm TL'}^{x'x}$ & 0 \\
%===ROW 10
$x'$ & $y$ & 0 & 0 & 0 & $\ddag$ & 0 & $\ddag$ & $\ddag$ & 0 & $\ddag$ \\
%===ROW 11
$x'$ & $z$ & $R_{\rm T}^{x'z}$ & $R_{\rm L}^{x'z}$ & $\ddag$ & 0 & $\ddag$ & 0 & 0 & $\ddag$  & 0 \\ \hline
%===ROW 12
$y'$ & $x$ & 0 & 0 & 0 & $\ddag$ & 0 & $\ddag$ & $\ddag$ & 0 & $\ddag$ \\
%===ROW 13
$y'$ & $y$ & $\ddag$ & $\ddag$ & $\ddag$ & 0 & $\ddag$ & 0 & 0 & $\ddag$ & 0 \\
%===ROW 14
$y'$ & $z$ & 0 & 0 & 0 & $\ddag$ & 0 & $\ddag$ & $\ddag$ & 0 & $\ddag$ \\ \hline
%===ROW 15
$z'$ & $x$ & $R_{\rm T}^{z'x}$ & $\ddag$ & $^{\rm c}\! R_{\rm TL}^{z'x}$ & 0 & $\ddag$ & 0 & 0 & $^{\rm s}\! R_{\rm TL'}^{z'x}$ & 0 \\
%===ROW 16
$z'$ & $y$ & 0 & 0 & 0 & $\ddag$ & 0 & $\ddag$ & $\ddag$ & 0 & $\ddag$ \\
%===ROW 17
$z'$ & $z$ & $R_{\rm T}^{z'z}$ & $\ddag$ & $\ddag$ & 0 & $\ddag$ & 0 & 0 & $\ddag$ & 0 \\ \hline \hline
\end{tabular}
\label{tab:response_functions}
\end{center}
\end{table}
%%%%%%%%%%%%%%%%%%%%%%%%%%%%%%%%%%%%%%%%%%%%%%%%%%%%%%%%%%%%%%%%%%%%%%%%%%%%%%%%%%%%%%%%%%%%%%%%%%%%%%%%%%%%%%%%%%%%%%%%%%%%%%%%%%

The following equations specify the relations between each response function marked with a $\ddagger$ in Table~\ref{tab:response_functions} and the corresponding related function \cite{Knochlein:1995qz}.
\begin{align}
R^{00}_{\rm T}&=- ^{\rm c}\! R^{y'y}_{\rm TT},  &  R^{0y}_{\rm T}&=- ^{\rm c}\! R^{y'0}_{\rm TT},  &  R^{y'0}_{\rm T}&=- ^{\rm c}\! R^{0y}_{\rm TT},  &  R^{x'x}_{\rm T}&=- ^{\rm c}\! R^{z'z}_{\rm TT},\\
%===============================
R^{x'z}_{\rm T}&=~^{\rm c}\! R^{z'x}_{\rm TT},  &  R^{z'x}_{\rm T}&=~^{\rm c}\! R^{x'z}_{\rm TT},  &  R^{z'z}_{\rm T}&=-^{\rm c}\! R^{x'x}_{\rm TT},  &  R^{00}_{\rm L}&=-R^{y'y}_{\rm L},\\
%===============================
R^{0y}_{\rm L}&=-R^{y'0}_{\rm L},  &  R^{x'x}_{\rm L}&=-R^{z'z}_{\rm L},  &  R^{x'z}_{\rm L}&=R^{z'x}_{\rm L},  &  ^{\rm c}\! R^{00}_{\rm TL}&=-^{\rm c}\! R^{y'y}_{\rm TL},\\
%===============================
^{\rm c}\! R^{0y}_{\rm TL}&=-^{\rm c}\! R^{y'0}_{\rm TL},  &  ^{\rm c}\! R^{x'x}_{\rm TL}&=-^{\rm c}\! R^{z'z}_{\rm TL},  &  ^{\rm c}\! R^{z'x}_{\rm TL}&=~^{\rm c}\! R^{x'z}_{\rm TL},  &  ^{\rm s}\! R^{0x}_{\rm TL}&=-^{\rm c}\! R^{y'z}_{\rm TL'},\\
%===============================
^{\rm s}\! R^{0z}_{\rm TL}&=-^{\rm c}\! R^{y'x}_{\rm TL'},  &  ^{\rm s}\! R^{x'0}_{\rm TL}&=-^{\rm c}\! R^{z'y}_{\rm TL'},  &  ^{\rm s}\! R^{z'0}_{\rm TL}&=~^{\rm c}\! R^{x'y}_{\rm TL'},  &  ^{\rm c}\! R^{00}_{\rm TT}&=-R^{y'y}_{\rm T},\\
%===============================
^{\rm s}\! R^{0x}_{\rm TT}&=R^{y'z}_{\rm TT'},  &  ^{\rm s}\! R^{z'0}_{\rm TT}&=R^{x'y}_{\rm TT'},  &  ^{\rm s}\! R^{0z}_{\rm TT}&=-R^{y'x}_{\rm TT'},  &  ^{\rm s}\! R^{x'0}_{\rm TT}&=-R^{z'y}_{\rm TT'},\\
%===============================
^{\rm c}\! R^{0x}_{\rm TL'}&=-^{\rm s}\! R^{y'z}_{\rm TL},  &  ^{\rm c}\! R^{0z}_{\rm TL'}&=^{\rm s}\! R^{y'x}_{\rm TL},  &  ^{\rm c}\! R^{x'0}_{\rm TL'}&=^{\rm s}\! R^{z'y}_{\rm TL},  &  ^{\rm c}\! R^{z'0}_{\rm TL'}&=-^{\rm s}\! R^{x'y}_{\rm TL},\\
%===============================
^{\rm s}\! R^{00}_{\rm TL'}&=-^{\rm s}\! R^{y'y}_{\rm TL'},  &  ^{\rm s}\! R^{0y}_{\rm TL'}&=-^{\rm s}\! R^{y'0}_{\rm TL'},  &  ^{\rm s}\! R^{x'x}_{\rm TL'}&=-^{\rm s}\! R^{z'z}_{\rm TL'},  &  ^{\rm s}\! R^{z'x}_{\rm TL'}&=~^{\rm s}\! R^{x'z}_{\rm TL'},\\
%===============================
R^{0x}_{\rm TT'}&=-^{\rm s}\! R^{y'z}_{\rm TT},  &  R^{0z}_{\rm TT'}&=~^{\rm s}\! R^{y'x}_{\rm TT},  &  R^{x'0}_{\rm TT'}&=~^{\rm s}\! R^{z'y}_{\rm TT},  &  R^{z'0}_{\rm TT'}&=-^{\rm s}\! R^{x'y}_{\rm TT}.
\end{align}

The response functions can be expressed in terms of the CGLN amplitudes \cite{Knochlein:1995qz}. It is important to note that Ref.~\cite{Knochlein:1995qz} employs the Pauli amplitudes defined in Ref.~\cite{Dennery:1961zz}, i.e., those given in Eq.~(\ref{eq:F_in_terms_of_F1_F6}).
{\allowdisplaybreaks
\begin{align}
%===========
\label{eq:RT00}
R^{00}_{\rm T}=~&|F_1|^2+|F_2|^2+\frac{\sin^2\theta}{2}(|F_3|^2+|F_4|^2)+\mathrm{Re}~[\sin^2\theta(F_2^*F_3+F_1^*F_4+\cos{\theta}F_3^*F_4)-2\cos{\theta}F_1^*F_2],\\
%===========
R^{0y}_{\rm T}=~&\mathrm{Im}~[\sin{\theta}\{F_1^*F_3-F_2^*F_4+\cos(F_1^*F_4-F_2^*F_3)-\sin^2\theta F_3^*F_4\}],\\
%===========
R^{y'0}_{\rm T}=~&\mathrm{Im}~[\sin{\theta}\{-2F_1^*F_2-F_1^*F_3+F_2^*F_4+\cos{\theta}(F_2^*F_3-F_1^*F_4)+\sin^2\theta F_3^*F_4\}],\\
%===========
R^{x'x}_{\rm T}=~&\mathrm{Re}\left[\sin^2\theta \left\{-F_1^*F_3-F_2^*F_4-F_3^*F_4-\frac{1}{2}\cos{\theta}(|F_3|^2+|F_4|^2)\right\}\right],\\
%===========
R^{x'z}_{\rm T}=~&\sin{\theta}\left[|F_1|^2-|F_2|^2+\frac{1}{2}\sin^2\theta(|F_4|^2-|F_5|^2)-F_2^*F_3+F_1^*F_4+\cos{\theta}(F_1^*F_3-F_2^*F_4)\right],\\
%===========
R^{z'x}_{\rm T}=~&\mathrm{Re}\left[\sin{\theta}\left\{-F_2^*F_3+F_1^*F_4+\cos{\theta}(F_1^*F_3-F_2^*F_4)+\frac{1}{2}\sin^2\theta(|F_4|^2-|F_3|^2)\right\}\right],\\
%===========
R^{z'z}_{\rm T}=~&\mathrm{Re}\left[2F_1^*F_2-\cos{\theta}(|F_1|^2+|F_2|^2)+\sin^2\theta(F_1^*F_3+F_2^*F_4+F_3^*F_4)+\frac{1}{2}\cos{\theta}\sin^2\theta(|F_3|^2+|F_4|^2)\right],\\
%===========
R^{00}_{\rm L}=~&\mathrm{Re}~(|F_5|^2+|F_6|^2+2\cos{\theta}F_5^*F_6),\\
%===========
R^{0y}_{\rm L}=~&-2\sin{\theta}~\mathrm{Im}(F_5^*F_6),\\
%===========
R^{x'x}_{\rm L}=~&\mathrm{Re}~[-2F_5^*F_6-\cos{\theta}(|F_5|^2+|F_6|^2)],\\
%===========
R^{z'x}_{\rm L}=~&\sin{\theta}(|F_6|^2-|F_5|^2),\\
%===========
^{\rm c}\! R^{00}_{\rm TL}=~&\sin{\theta}~\mathrm{Re}~[-F_2^*F_5-F_3^*F_5-F_1^*F_6-F_4^*F_6-\cos{\theta}(F_4^*F_5+F_3^*F_6)],\\
%===========
^{\rm s}\! R^{0x}_{\rm TL}=~&\mathrm{Im}~[-F_1^*F_5+F_2^*F_6+\cos{\theta}(F_2^*F_5-F_1^*F_6)],\\
%===========
^{\rm c}\! R^{0y}_{\rm TL}=~&\mathrm{Im}~[-F_1^*F_5+F_2^*F_6+\cos{\theta}(F_2^*F_5-F_1^*F_6)+\sin^2\theta(F_3^*F_6-F_4^*F_5)],\\
%===========
^{\rm s}\! R^{0z}_{\rm TL}=~&\sin\theta~\mathrm{Im}~(F_2^*F_5+F_1^*F_6),\\
%===========
^{\rm s}\! R^{x'0}_{\rm TL}=~&\mathrm{Im}~[-F_2^*F_5+F_1^*F_6+\cos{\theta}(F_1^*F_5-F_2^*F_6)],\\
%===========
^{\rm s}\! R^{z'0}_{\rm TL}=~&\sin{\theta}~\mathrm{Im}~(F_1^*F_5+F_2^*F_6),\\
%===========
^{\rm c}\! R^{x'x}_{\rm TL}=~&\sin{\theta}~\mathrm{Re}~[F_1^*F_5+F_4^*F_5+F_2^*F_6+F_3^*F_6+\cos{\theta}(F_3^*F_5+F_4^*F_6)],\\
%===========
^{\rm c}\! R^{z'x}_{\rm TL}=~&\mathrm{Re}~[F_2^*F_5-F_1^*F_6+\cos{\theta}(F_2^*F_6-F_1^*F_5)+\sin^2\theta(F_3^*F_5-F_4^*F_6)],\\
%===========
^{\rm c}\! R^{00}_{\rm TT}=~&\frac{1}{2}\sin^2\theta(|F_3|^2+|F_4|^2)+\sin^2\theta~\mathrm{Re}~(F_2^*F_3+F_1^*F_4+\cos{\theta}F_3^*F_4),\\
%===========
^{\rm s}\! R^{0x}_{\rm TT}=~&\sin{\theta}~\mathrm{Im}~[2F_1^*F_2+F_1^*F_3-F_2^*F_4+\cos{\theta}(F_1^*F_4-F_2^*F_3)],\\
%===========
^{\rm s}\! R^{0z}_{\rm TT}=~&-\sin^2\theta~\mathrm{Im}~[F_2^*F_3+F_1^*F_4],\\
%===========
^{\rm s}\! R^{x'0}_{\rm TT}=~&\sin\theta~\mathrm{Im}~[F_2^*F_3-F_1^*F_4+\cos{\theta}(F_2^*F_4-F_1^*F_3)],\\
%===========
^{\rm s}\! R^{z'0}_{\rm TT}=~&-\sin^2\theta~\mathrm{Im}~(F_1^*F_3+F_2^*F_4),\\
%===========
^{\rm s}\! R^{00}_{\rm TL'}=~&-\sin{\theta}~\mathrm{Im}~[F_2^*F_5+F_3^*F_5+F_1^*F_6+F_4^*F_6+\cos{\theta}(F_4^*F_5+F_3^*F_6)],\\
%===========
^{\rm c}\! R^{0x}_{\rm TL'}=~&\mathrm{Re}~[-F_1^*F_5+F_2^*F_6+\cos{\theta}(F_2^*F_5-F_1^*F_6)],\\
%===========
^{\rm s}\! R^{0y}_{\rm TL'}=~&\mathrm{Re}~[F_1^*F_5-F_2^*F_6+\cos{\theta}(F_1^*F_6-F_2^*F_5)+\sin^2\theta(F_4^*F_5-F_3^*F_6)],\\
%===========
^{\rm c}\! R^{0z}_{\rm TL'}=~&\sin{\theta}~\mathrm{Re}~(F_2^*F_5+F_1^*F_6),\\
%===========
^{\rm c}\! R^{x'0}_{\rm TL'}=~&\mathrm{Re}~[-F_2^*F_5+F_1^*F_6+\cos{\theta}(F_1^*F_5-F_2^*F_6)],\\
%===========
^{\rm c}\! R^{z'0}_{\rm TL'}=~&\sin{\theta}~\mathrm{Re}~(F_1^*F_5+F_2^*F_6),\\
%===========
^{\rm s}\! R^{x'x}_{\rm TL'}=~&\sin{\theta}~\mathrm{Im}~[F_1^*F_5+F_4^*F_5+F_2^*F_6+F_3^*F_6+\cos{\theta}(F_3^*F_5+F_4^*F_6)],\\
%===========
^{\rm s}\! R^{z'x}_{\rm TL'}=~&\mathrm{Im}[F_2^*F_5-F_1^*F_6+\cos{\theta}(-F_1^*F_5+F_2^*F_6)+\sin^2\theta(F_3^*F_5-F_4^*F_6)],\\
%===========
R^{0x}_{\rm TT'}=~&\sin{\theta}~\mathrm{Re}[F_1^*F_3-F_2^*F_4+\cos{\theta}(-F_2^*F_3+F_1^*F_4)],\\
%===========
\label{eq:RTTP0z}
R^{0z}_{\rm TT'}=~&-|F_1|^2-|F_2|^2+\mathrm{Re}~[2\cos{\theta}F_1^*F_2-\sin^2\theta(F_2^*F_3+F_1^*F_4)],\\
%===========
R^{x'0}_{\rm TT'}=~&\sin{\theta}~\mathrm{Re}~[-|F_1|^2+|F_2|^2+F_2^*F_3-F_1^*F_4+\cos{\theta}(F_2^*F_4-F_1^*F_3)],\\
%===========
R^{z'0}_{\rm TT'}=~&\mathrm{Re}[-2F_1^*F_2+\cos\theta(|F_1|^2+|F_2|^2)-\sin^2\theta(F_1^*F_3+F_2^*F_4)].
%===========
\end{align}
}

\section{Non-Relativistic Amplitudes}
\label{app:non-rel-amplitudes}

The non-relativistic electroproduction operator in Eq.~(\ref{eq:gen_nonrel_op}) is expressed in terms of the amplitudes ${\cal F}_i$. The explicit relations between these amplitudes and the relativistic amplitudes $A_i$ defined in Eq.~(\ref{eq:Ai_Mi}) are given by \cite{Mart:2008gq}
\begin{eqnarray}
{\cal{F}}_{1} & = & k_{0}A_{1} + k \cdot q_K A_{3} + \bigl\{2P \cdot k - 
k_{0}(m_{N} + m_{Y})\bigr\}A_{4} - k^{2}A_{6} ~\!,\\
{\cal{F}}_{2} & = & -A_{1} - E_K A_{3} - (E_{N} + E_{Y} - m_{N} - 
m_{Y})A_{4} + k_{0}A_{6} ~\!, \\
{\cal{F}}_{3} & = & A_{3} - A_{6} ~\!, \\
{\cal{F}}_{4} & = & A_{3} + A_{4} ~\!, \\
{\cal{F}}_{5} & = & -A_{3} + A_{4} ~\!, \\
{\cal{F}}_{6} & = & \frac{1}{E_{N} + m_{N}}~\Bigl[~\bigl\{2P \cdot 
k\,(E_{N} - E_{Y})+({\textstyle \frac{1}{2}} k^{2} - k \cdot q_K)
(E_{N} + E_{Y})+ P \cdot kk_{0}\bigr\}A_{2} + ( k_0 E_K - k \cdot q_K ) A_{3}  
\nonumber\\
 & & \hspace{2.2cm} + \,
\bigl\{k_{0}(E_{N} + E_{Y}) - 2P \cdot k \bigr\}A_{4} - ( k_0 k \cdot q_K - 
k^2 E_K ) A_{5} 
+ (k^2 - k_0^2) A_{6}~\Bigr] ~\!, \\
{\cal{F}}_{7} & = & \frac{1}{E_{N} + m_{N}}~\Bigl[\,A_{1} - P \cdot k A_{2} 
- k_{0}A_{3} - (m_{N} + m_{Y})A_{4} - (k^{2} - k \cdot q_K)A_{5} 
 + k_{0}A_{6}\,\Bigr] , \\
{\cal{F}}_{8} & = & \frac{1}{E_{N} + m_{N}}~\Bigl[-(2P \cdot k + 
{\textstyle \frac{1}{2}} k^{2} - k \cdot q_K)A_{2} - k_{0}(A_{3} + A_{4}) - 
k^{2}A_{5}~\Bigr] , \\
{\cal{F}}_{9} & = & \frac{1}{E_{N}+m_{N}}~\Bigl[~(2P \cdot k - 
{\textstyle \frac{1}{2}} k^{2} + k \cdot q_K)A_{2} + k_{0}(A_{3} - A_{4}) + 
k^{2}A_{5}~\Bigr] , \\
{\cal{F}}_{10} & = & \frac{1}{E_{Y} + m_{Y}}~\Bigl[-\bigl\{2P \cdot 
k(E_{N} - E_{Y}) + ({\textstyle \frac{1}{2}} k^{2} - k \cdot q_K)
(E_{N} + E_{Y}) + P \cdot k k_{0} \bigr\}A_{2} + (k_0E_K - k \cdot q_K) A_{3}\nonumber\\
 & & \hspace{2.1cm}   + 
\bigl\{k_{0}(E_{N} + E_{Y}) - 2P \cdot k \bigr\}A_{4}   + ( k_0 k \cdot q_K - 
k^2 E_K ) A_{5} + (k^2-k_0^2) A_{6}~\Bigr] , \\
{\cal{F}}_{11} & = & \frac{1}{E_{Y} + m_{Y}}~\Bigl[-A_{1}+
P\cdot kA_{2} - k_{0}A_{3} + (m_{N} + m_{Y})A_{4} + (k^{2} - 
k \cdot q_K)A_{5} + k_{0}A_{6}\,\Bigr] ,\\
{\cal{F}}_{12} & = & \frac{1}{E_{Y} + m_{Y}}~\Bigl[~(2P \cdot k + 
{\textstyle \frac{1}{2}} k^{2} - k \cdot q_K)A_{2} - k_{0}(A_{3} + A_{4}) + 
k^{2}A_{5}~\Bigr] , \\
{\cal{F}}_{13} & = & \frac{1}{E_{Y} + m_{Y}}~\Bigl[-(2P \cdot k 
+ k \cdot q_K - {\textstyle \frac{1}{2}} k^{2})A_{2} + k_{0}(A_{3} - A_{4}) 
- k^{2}A_{5}~\Bigr] , \\
{\cal{F}}_{14} & = & \frac{1}{E_{N} + m_{N}}~\Bigl[-A_{1}
 + (m_{N} + m_{Y})A_{4}~\Bigr] , \\
{\cal{F}}_{15} & = & \frac{1}{E_{Y} + m_{Y}}~\Bigl[~A_{1}
 - (m_{N} + m_{Y})A_{4}~\Bigr] , \\
{\cal{F}}_{16} & = & \frac{1}{(E_{N} + m_{N})(E_{Y} + m_{Y})} 
\Bigl[-k_{0}A_{1} + k \cdot q_K A_{3} + \bigl\{2P \cdot k + k_{0}(m_{N} + 
m_{Y})\bigr\}A_{4} - k^{2}A_{6}\Bigr] ,\\
{\cal{F}}_{17} & = & \frac{1}{(E_{N} + m_{N})(E_{Y} + m_{Y})}\,
\Bigl[\,A_{1} - E_K A_{3} - (E_{N} + E_{Y}  
+ m_{N} + m_{Y})A_{4} + k_{0}A_{6}\,\Bigr] ,\\
{\cal{F}}_{18} & = & \frac{1}{(E_{N} + m_{N})(E_{Y} + m_{Y})} 
~\Bigl[~A_{3} - A_{6}~\Bigr] ~\!, \\
{\cal{F}}_{19} & = & \frac{1}{(E_{N} + m_{N})(E_{Y} + m_{Y})} 
~\Bigl[~A_{3} + A_{4}~\Bigr] ~\!, \\
{\cal{F}}_{20} & = & \frac{1}{(E_{N} + m_{N})(E_{Y} + m_{Y})} 
~\Bigl[-A_{3} + A_{4}~\Bigr] ~\!.
\end{eqnarray}

Note that the conventions for all momenta and energies are defined in Section~\ref{subsec:kinematics} and $P=p_N+p_Y$.
%==============================================
    \newpage
	\bibliography{ref}
	%Please use Bib\TeX\ to generate your bibliography and include DOIs whenever available. Example of bib file: 
	
	%%%%%%%%%%%%%%%%%%%%%%%%%%%%%%%%%%%%%%%%%%%%%%%%%%%%%%%%%%%%%%%%%%%
	% Encoding: ISO-8859-1

	%@Article{Eichmann:2016yit,
	%author        = {Eichmann, Gernot and Sanchis-Alepuz, Helios and Williams, Richard and Alkofer, Reinhard and Fischer, Christian S.},
	%title         = {{Baryons as relativistic three-quark bound states}},
	%journal       = {Prog. Part. Nucl. Phys.},
	%year          = {2016},
	%volume        = {91},
	%pages         = {1-100},
	%archiveprefix = {arXiv},
	%doi           = {10.1016/j.ppnp.2016.07.001},
	%eprint        = {1606.09602},
	%owner         = {chfi},
	%primaryclass  = {hep-ph},
	%slaccitation  = {%%CITATION = ARXIV:1606.09602;%%},
	%timestamp     = {2018.08.02},
	%}

	%@Comment{jabref-meta: databaseType:bibtex;}
	%%%%%%%%%%%%%%%%%%%%%%%%%%%%%%%%%%%%%%%%%%%%%%%%%%%%%%%%%%%%%%%%%%%

\end{document}